\documentclass[fleqn,a4paper,11pt]{ta-thesis}
\usepackage{times,mathptm,latexsym}
\usepackage{amssymb,amsfonts}
\usepackage{pstricks,textpath}
\usepackage{ta-chicago}
\usepackage{fancyheadings}
\usepackage{epsfig}
\usepackage{ta-basics}
\renewcommand{\partmark}[1]{\markboth{#1}{}}
\renewcommand{\chaptermark}[1]{\markright{Chapter~\thechapter : #1}}
\renewcommand{\sectionmark}[1]{}
\begin{document}
\frontmatter
\setcounter{page}{-1}
\thispagestyle{empty}
\setlength{\unitlength}{1mm}
\psset{unit=1mm}
\begin{pspicture}(18,12)(153,216)
\put(82,12){\makebox(0,0)[b]{Institute of Physics, Aalborg University}}
\put(52.5,16){%
 \multiput(0,0)(40,0){2}{%
  \qdisk(10,10){1.5}
  \psarc[linewidth=0.6]{-}(10,10){4}{22.5}{67.5}
  \psarc[linewidth=0.6]{-}(10,10){4}{112.5}{157.5}
  \psarc[linewidth=0.6]{-}(10,10){4}{202.5}{247.5}
  \psarc[linewidth=0.6]{-}(10,10){4}{292.5}{337.5}
  \psarc[linewidth=0.5]{-}(10,10){6.5}{22.5}{67.5}
  \psarc[linewidth=0.5]{-}(10,10){6.5}{112.5}{157.5}
  \psarc[linewidth=0.5]{-}(10,10){6.5}{202.5}{247.5}
  \psarc[linewidth=0.5]{-}(10,10){6.5}{292.5}{337.5}
  \psarc[linewidth=0.4]{-}(10,10){8.5}{22.5}{67.5}
  \psarc[linewidth=0.4]{-}(10,10){8.5}{112.5}{157.5}
  \psarc[linewidth=0.4]{-}(10,10){8.5}{202.5}{247.5}
  \psarc[linewidth=0.4]{-}(10,10){8.5}{292.5}{337.5}
 }
 \psline[linewidth=0.25]{-}(15.25,10)(17.5,10)
 \pscustom[linewidth=0.25,fillstyle=solid,fillcolor=black]{
  \pscurve[curvature=.4 .1 0]{-}(17.5,10)(20,16)(25,4)(30,16)(35,4)(40,16)(42.5,10)
  \pscurve[curvature=.6 .1 0,liftpen=1]{-}(42.5,10)(40,14)(35,6)(30,14)(25,6)(20,14)(17.5,10)
 }
 \psline[linewidth=0.25]{-}(42.5,10)(44.75,10)
 \pscustom[linewidth=0.25,fillstyle=solid,fillcolor=black]{
  \psline[linewidth=0.25](43.5,10.75)(46,10)(43.5,9.25)
 }
}
\put(82.5,152.5){
 \pstextpath[c]{\psarcn[linestyle=none](0,0){46}{220}{-40}}
  {\curly{Theoretical Study of Phase Conjugation}}
 \pstextpath[c]{\psarc[linestyle=none](0,0){48}{180}{0}}
  {\curly{in Mesoscopic Interaction Volumes}}
 \pstextpath[c]{\psarcn[linestyle=none](0,0){27}{220}{-40}}
  {\mynamefont{Torsten Andersen}}
 \psarc[linewidth=0.75mm]{*-*}(0,0){27}{144}{36}
 \put(-15,-15){%
  \setlength{\unitlength}{0.15mm}%
  \psset{unit=0.15mm}%
  \begin{pspicture}(0,0)(200,200)%
  \psarc[fillstyle=solid,fillcolor=black,linewidth=0](15,165){15}{90}{270}
  \psarc[fillstyle=solid,fillcolor=white,linewidth=0](15,160){10}{90}{270}
  \psline[linewidth=10](15,175)(185,175)
  \psarc[fillstyle=solid,fillcolor=black,linewidth=0](185,185){15}{270}{90}
  \psarc[fillstyle=solid,fillcolor=white,linewidth=0](185,190){10}{270}{90}
  \psarc[fillstyle=solid,fillcolor=black,linewidth=0](70,55){15}{90}{270}
  \psarc[fillstyle=solid,fillcolor=white,linewidth=0](70,50){10}{90}{270}
  \psline[linewidth=10](70,65)(160,65)
  \psarc[fillstyle=solid,fillcolor=black,linewidth=0](160,75){15}{270}{90}
  \psarc[fillstyle=solid,fillcolor=white,linewidth=0](160,80){10}{270}{90}
  \psarc[fillstyle=solid,fillcolor=black,linewidth=0](15,15){15}{90}{270}
  \psarc[fillstyle=solid,fillcolor=white,linewidth=0](15,20){10}{90}{270}
  \psline[linewidth=10,linearc=15](15,5)(25,5)(85,175)
  \psline[linewidth=10,linearc=15](15,5)(55,5)(115,175)
  \psline[linewidth=10,linearc=15](185,5)(175,5)(115,175)
  \psarc[fillstyle=solid,fillcolor=black,linewidth=0](185,15){15}{270}{90} 
  \psarc[fillstyle=solid,fillcolor=white,linewidth=0](185,20){10}{270}{90}
  \end{pspicture}
 }
}
\end{pspicture}
\newpage
\thispagestyle{empty}
\noindent
{\large\bf Authors note to the electronic version}
\vskip 5mm
\noindent
The author hereby grant you to redistribute and print this file for
your own personal use, provided that no modifications are made. If you
would like to cite any of the results published herein you are free to
do so, but if it is possible please make your citation to one of the
papers below:
\begin{description}
\setlength{\itemsep}{0pt plus 0.5pt}%
\setlength{\parsep}{0pt plus 0.5pt}%
\setlength{\parskip}{2.5pt plus 2.5pt}%
\item[{\rm Andersen, T. and O.~Keller (1998a).}] Local-field theory
  for optical phase conjugation by degenerate four wave mixing in
  mesoscopic interaction volumes of condensed media.  {\em Phys.
    Scr.\/}~{\em 58}, 132--144.
\item[{\rm Andersen, T. and O.~Keller (1998b).}] Optical phase
  conjugation in a single-level metallic quantum well. {\em Phys. Rev.
    B\/}~{\em 57}, 14793--14808.
\item[{\rm Andersen, T. and O.~Keller (1998c).}] Two-dimensional
  confinement of light in front of a single level quantum well phase
  conjugator.  {\em Opt. Commun.\/}~{\em 155}, 317--322.
\item[{\rm Andersen, T. and O.~Keller (1999).}] Local-field study of
  phase conjugation in nonmagnetic multi-level metallic quantum wells
  with probe fields of both propagating and evanescent character. {\em
  Phys. Rev. B\/}, (to be published). 
\end{description}
The first of these articles contains most of Chapter 3 and Parts II
and III. The second contains most of Chapters 10 and 11 and part of
Appendix B. The third article consists of the results presented in
Chapter 12. The last of the four articles is based on the analysis
presented in Part V and Appendices B and C. It is, however, greatly
expanded with numerical results.

If you plan to include some of this work (figures, tables, equations)
into your own work, please contact the author. Current (as of
October 1998) contact information is:\\
\\
Dr.~Torsten Andersen\\
Max-Planck-Institut f{\"u}r Mikrostrukturphysik\\
Weinberg 2\\
D--06120 Halle/Saale\\
Germany\\
\\
e-mail: thor@mpi-halle.mpg.de\\
www:~http://www.geocities.com/CapeCanaveral/Lab/9700/index.html\\
\\
A limited number of printed copies of the thesis are available. Please
contact the author or the publisher if you are interested.
\newpage
\thispagestyle{empty}
\begin{center}
\setlength{\unitlength}{1mm}
\psset{unit=1mm}
\begin{pspicture}(-60,-50)(60,65)
 \pstextpath[c]{\psarcn[linestyle=none](0,0){46}{220}{-40}}
  {\curly{Theoretical Study of Phase Conjugation}}
 \pstextpath[c]{\psarc[linestyle=none](0,0){48}{180}{0}}
  {\curly{in Mesoscopic Interaction Volumes}}
\end{pspicture}
\end{center}
\newpage
\thispagestyle{empty}
\noindent
{\sc
Theoretical Study of Phase Conjugation \\
in Mesoscopic Interaction Volumes.\\

\noindent
Teoretisk unders{\o}gelse af fasekonjugation \\
i mesoskopiske vekselvirkningsrumfang.\\
}

\noindent
Copyright \copyright\ 1998 by Torsten Andersen\\
and the Institute of Physics, Aalborg University.\\

\noindent
{\it Published and distributed by}\\
Institute of Physics, Aalborg University,\\
Pontoppidanstr{\ae}de 103, DK--9220 Aalborg {\O}st.\\
Phone $+45$ 96358080. Fax $+45$ 98156502.\\

\vfill

\noindent
Typeset in \LaTeXe{} by the author.\\

\noindent
Printed in Denmark by Centertrykkeriet, Aalborg University.\\

\noindent
All rights reserved. No part of this publication may be reproduced,
transmitted or translated in any form or by any means, electronic
or mechanical, including photocopy, recording, or any information
storage and retrieval system, without prior permission in writing from
the author.\\ 

\noindent
ISBN 87-89195-16-7

\newpage
\thispagestyle{empty}
\vbox to50mm{
 \centering
 \vfil
 {\Huge Torsten Andersen}
 \vskip 1\baselineskip
 \hrule width\hsize height1mm}
 \vskip 1\baselineskip 
 \begin{center}
  {{\huge\it Theoretical Study of Phase Conjugation \\
   in Mesoscopic Interaction Volumes}}
 \end{center}
\newpage
\thispagestyle{plain}
\chead[\fancyplain{}{\leftmark}]{\fancyplain{}{\leftmark}}
\chapter*{Preface}
\noindent 
The present monograph describes results of my work carried out at the
Institute of Physics at Aalborg University, Aalborg, Denmark in the
period from February 1995 to May 1998 and at the Department of Physics
and Astronomy at Mississippi State University, Starkville,
Mississippi, United States of America in the period from August 1996
to January 1997. This monograph is submitted as a Ph.D.~thesis to the
Faculty of Engineering and Science at Aalborg University.

With this dissertation, I intend to describe in a unified fashion the
work done in the time frame of the programme. It is my intention that
the contents of this monograph should go deeper and broader into the
material that has been processed for publishing in articles, and thus
include results and comments on material not suitable for publication
as parts of an article. The work is divided into six main parts, each
consisting of some separate chapters. The outline of this monograph is
presented in the following.

\section*{Outline of the dissertation}
The motivation for carrying out the present study is given in Part I,
divided into three separate chapters. In Chapter 1 a brief summary of
the contributions to scientific progress within the last century,
which I believe are the most important for my work, is presented. The
historical summary includes remarks on nonlinear optics in general,
optical phase conjugation in particular, near-field optics, and
mesoscopic optics. Chapter 2 presents the theoretical model usually
adopted in optical phase conjugation (standard theory). Chapter 3
consists of a discussion of the limitations of the standard theory as
well as the requirements to a theoretical model that can be used when
interaction takes place on small length scales and/or in the optical
near-field region.

In Part II the task of developing a nonlocal theoretical description
of phase conjugation of optical near-fields by degenerate four wave
mixing is undertaken. It consists of three chapters. In Chapter 4 the
basic working frame for the present treatment is established, starting
from Maxwell's equations. Chapter 5 sets up the first and third order
responses of an electron using the density matrix formalism starting
from the Liouville equation of motion. In Chapter 6 the general
conductivity response tensors for degenerate four-wave mixing
excluding spin-effects are established, and their symmetries are
discussed. Part II is concluded with a small discussion.

Part III discusses degenerate four-wave mixing in quantum-well
structures on a somewhat general level. For this purpose, two chapters
are written. In Chapter 8 the conductivity response tensors in the
case where a system has broken translational invariance in one spatial
direction are established and discussed, and Chapter 9 is devoted to a
discussion of the consequences of scattering in a plane using
polarized light.

In Part IV the optical phase conjugation response of a single-level
quantum well is studied in four chapters. In Chapter 10 the
theoretical considerations necessary to describe the phase conjugation
response from a single-level quantum well are discussed, and in
Chapter 11 the numerical results for the phase conjugated response
from a copper quantum well are discussed. Chapter 12 consists of a
discussion of two-dimensional confinement of light in front of the
single-level quantum-well phase conjugator considered in Chapters 10
and 11. Chapter 13 concludes this part with a short discussion.

Part V takes a similar point of view as Part IV, but for the two-level
quantum well. Theoretical considerations are presented in Chapter 14,
while the numerical results are discussed in Chapter 15. A short
discussion concludes this part in Chapter 16.

Part VI contains a concluding discussion on the developed theory and
the numerical work followed by an outlook.

There are five Appendices included, consisting of calculations not
suitable for the main text. Appendix A is a calculation of the linear
and nonlinear conductivity tensors relevant for studying degenerate
four-wave mixing in quantum well structures. Appendix B contains the
principal analytic solutions to the integrals over the states parallel
to the plane of translational invariance in the quantum well
structures. In Appendix C, the absolute solution to the integrals over
the states parallel to the plane of translational invariance in the
quantum well structures are presented in terms of the principal
solutions given in Appendix B. Appendix D contains a small calculation
of the Fermi energy for a quantum well in the low-temperature limit
and a calculation of the minimal and maximal values of the thickness
of a quantum well given the desired number of occupied eigenstates
across the quantum well. Appendix E contains some intermediate results
in the calculation of the integrals over the source region in Chapter
14.

References used in this work are listed in the bibliography at the end
of the dissertation according to the recommendations by the thirteenth
edition of {\it The Chicago Manual of Style}\/ with author(s), title,
and publication data in alphabetic order after the first authors
surname.

\section*{Notation}
Footnotes are marked using a superscript number in the text, and the
footnote itself is found at the bottom of the page. Citations to other
people's work are made with reference to the authors surname(s)
followed by the year of publication. The international system of
units (SI) has been adopted throughout the work, except that the unit
{\AA}ngstr{\"o}m ({\AA}) is used to denote certain distances
($1${\AA}$=10^{-10}$m).

Vector quantities are denoted with a unidirectional arrow above them,
i.e., $\vec{\kappa}$. Likewise, tensor quantities are denoted using a
bidirectional arrow, i.e., $\stensor{\sigma}$. Integrations over
vector quantities are denoted ``$\int{f(\vec{\kappa})}d^n\kappa$'',
where $n$ is the number of elements in the vector $\vec{\kappa}$, and
$f(\vec{\kappa})$ is an arbitrary function of the integration variable
$\vec{\kappa}$. Unit vectors are denoted $\vec{e}$ with an index
indicating which direction is taken. The unit tensor is denoted by
$\tensor{\openone}$, and is usually a $3\times3$ tensor.

Latin indices $\{i,j,k,h\}$ generally refers to the three spatial
coordinate labels $\{x,y,z\}$, and the latin indices $\{n,m,v,l\}$
generally refers to quantum states.  Exceptions from this are (i) when
the letter ``i'' appears in formulae and is not an index, it is the
complex number ${\rm{i}}^2=-1$, (ii) when the letter ``$k$'' appears
in formulae and is not an index, it is the wavenumber $k=|\vec{k}|$,
(iii) the letter ``$m$'' with an index ``$e$'' is the electron mass.
Summations over repeated indices are stated explicitly whenever it
should be performed.

To avoid confusion regarding the placement of $2\pi$'s in the Fourier
integral representation, the Fourier transform pair
\begin{displaymath}
 {\cal{F}}(t)={1\over2\pi}\int{}{\cal{F}}(\omega)e^{-{\rm{i}}\omega{}t}d\omega
 \qquad\Longleftrightarrow\qquad
 {\cal{F}}(\omega)=\int{}{\cal{F}}(t)e^{{\rm{i}}\omega{}t}dt
\end{displaymath}
is adopted between the time- and frequency domains, and thus to be
consistent the transform pair
\begin{displaymath}
 {\cal{F}}(\vec{r})
 ={1\over(2\pi)^3}\int{}{\cal{F}}(\vec{k})e^{{\rm{i}}\vec{k}\cdot\vec{r}}d^3k
 \qquad\Longleftrightarrow\qquad
 {\cal{F}}(\vec{k})=
 \int{}{\cal{F}}(\vec{r})e^{-{\rm{i}}\vec{k}\cdot\vec{r}}d^3r
\end{displaymath}
is adopted between real space and $k$-space, here shown in three
dimensions.

Furthermore, the complex conjugate and the Hermitian adjoint of a
quantity $A$ are denoted $A^{*}$ and $A^{\dag}$, respectively. The
phrases ``+\mbox{c.c.}'' and ``$+\mbox{H.a.}$'' at the end of an
equation indicates the addition of the complex conjugate or the
Hermitian adjoint of the foregoing terms. The phrase
``$+\mbox{i.t.}$'' at the end of an equation denotes the addition of a
term in which the wave-vector $\vec{k}$ is replaced by $-\vec{k}$.
The Laplacian is denoted $\nabla^2$. The Heaviside unit step function
$\Theta(x)$ has the value $+1$ for $x>0$ and $0$ for $x<0$, and the
Kronecker delta $\delta_{ij}$ has the value $+1$ for $i=j$ and $0$ for
$i\neq{}j$. The Ludolphine number $3.14159265\dots$ is denoted by the
greek letter $\pi$.

\section*{Scientific papers and presentations based on this work}
Parts of the work presented in this dissertation has been or will be
published separately in the form of proceedings papers, articles, and
letters. They are as follows:
\begin{description}
\setlength{\itemsep}{0pt plus 0.5pt}%
\setlength{\parsep}{0pt plus 0.5pt}%
\setlength{\parskip}{2.5pt plus 2.5pt}%
\item[{\rm Andersen, T.}] and O.~Keller (1995a). Optical near-field
  phase conjugation: A nonlocal DFWM response tensor.  In E.~G.
  Bortchagovsky (Ed.), {\em Proceedings of the International Autumn
    School-Conference for Young Scientists ``Solid State Physics:
    Fundamentals \&\ Applications'' (SSPFA'95)}, Kiev, pp.\ R5--R6.
  Institute of Semiconductor Physics of NASU. ISBN 5-7702-1199-7.
\item[{\rm Andersen, T. and O.~Keller (1996a).}] Phase Conjugation of
  Optical Near Fields: A new Nonlocal Microscopic Response Tensor.\ In
  O.~Keller (Ed.), {\em Notions and Perspectives of Nonlinear Optics},
  pp.\ 566--573.\ Singapore: World Scientific.\ ISBN 981-02-2627-6.
\item[{\rm Andersen, T. and O.~Keller (1998a).}] Local-field theory
  for optical phase conjugation by degenerate four wave mixing in
  mesoscopic interaction volumes of condensed media.  {\em Phys.
    Scr.\/}~{\em 58}.  In press.
\item[{\rm Andersen, T. and O.~Keller (1998b).}] Optical phase
  conjugation in a single-level metallic quantum well. {\em Phys. Rev.
    B\/}~{\em 57}, 14793--14808.
\item[{\rm Andersen, T. and O.~Keller (1998c).}] Two-dimensional
  confinement of light in front of a single level quantum well phase
  conjugator.  {\em Opt. Commun.\/}  Submitted.
\end{description}
Furthermore, an article on optical phase conjugation in a two-level
(resonant) metallic quantum well is in preparation. In addition to
the above-mentioned publications, presentations of abstracts has been
given at conferences. They are:
\begin{description}
\setlength{\itemsep}{0pt plus 0.5pt}%
\setlength{\parsep}{0pt plus 0.5pt}%
\setlength{\parskip}{2.5pt plus 2.5pt}%
\item[{\rm Keller, O., M.~Xiao and T.~Andersen (1994).}] Phase
  conjugation and Near-Field Micro\-scopy. Poster presented at the
  annual meeting of the Danish Optical Society, Lyngby, Denmark, November 24.
\item[{\rm Andersen, T.}] and O.~Keller (1995b)
  Random-phase-approximation study of the response function describing
  phase conjugation by degenerate four wave mixing. Poster presented
  at the annual meeting of the Danish Physical Society, Odense,
  Denmark, May 31--June 2.
\item[{\rm Andersen, T.}] and O.~Keller (1995c) Phase conjugation of
  optical near-fields: A new nonlocal response tensor allowing
  degenerate four wave mixing studies with probe beams strongly
  decaying in space. Talk given at the Third International Aalborg
  Summer School on Nonlinear Optics, Aalborg, Denmark, August 7--12.
\item[{\rm Andersen, T.}] and O.~Keller (1995d) Optical near-field
  phase conjugation: A nonlocal DFWM response tensor. Talk given at
  the International Autumn School-Conference for Young Scientists
  ``Solid State Physics: Fundamentals \&\ Applications'' (SSPFA'95) in
  Uzhgorod, Ukraine, September 19--26.
\item[{\rm Andersen, T.}] and O.~Keller (1996b) Optical Phase
  Conjugation by Degenerate Four Wave Mixing in a Single Level Quantum
  Well. Poster presented at the annual meeting of the Danish Physical
  Society, Nyborg, Denmark, May 23--24.
\item[{\rm Andersen, T.}] and O.~Keller (1996c) Microscopic
  Description of Optical Near-Field  Pha\-se Conjugation. Talk given at
  the South Eastern Section Meeting of The American Physical Society,
  Atlanta-Decatur, Georgia (USA), November 14--16. {\em Bulletin of
    the American Physical Society\/}~{\em 41}, p.\ 1660.
\item[{\rm Andersen, T.}] and O.~Keller (1997a) Optical Near-Field
  Phase Conjugation: A Microscopic Description. Poster presented at
  the Fourth International Conference on Near-Field Optics (NFO-4)
  Jerusalem, Israel, February 9--13.
\item[{\rm Andersen, T.}] and O.~Keller (1997b) Focusing of classical
  light beyond the diffraction limit. Poster presented at the annual
  meeting of the Danish Optical Society, Lyngby, Denmark,
  November 18--19.
\end{description}
The abstracts are printed in the relevant meeting programmes.

\section*{Software used in this project}
Creation of the numerical results presented in this work has been done
through development of computer programs, mainly in Fortran~90
\cite{Metcalf:96:1}. The final set of programs consists of
approximately 6000 lines of code developed by the present author.
Because of the size I have chosen not to include a reprint of the code
in this monograph. The presentation of the calculated data is done
using gnuplot pre-3.6 with some 400 lines of code to generate the
plots as encapsulated PostScript files. This dissertation has been
typed entirely in \LaTeXe\ \cite{Goossens:94:1,Goossens:97:1}, an
enhanced version of the typesetting program \TeX, originally developed
by \citeN{Knuth:84:1}.

\section*{Acknowledgements}
The topic of this work has been very fascinating to explore, and I
would therefore like to thank professor, Dr.~Scient.~Ole Keller for
introducing me to the world of mesoscopic physics, his huge support
and the many inspiring discussions we have had throughout the period
of this work. I would also like to thank professor, Dr.~H.~F.~Arnoldus
and the Department of Physics and Astronomy at Mississippi State
University for kindly providing facilities and inspiration during my
stay there.  Additionally, I would like to thank the scientific staff
at the Institute of Physics, Aalborg University for providing an
inspiring atmosphere in general, and Dr.~Brian Vohnsen in particular
for many discussions on near-field optics and for reviewing parts of
the text appearing in this dissertation.

My deepest thanks goes to my parents Maren and Agner Andersen for
their moral and financial support, and to my brother Jens and my
sisters Else, Karen, and Birgit. In addition, I would like to thank my
friends around the globe who---together with my family---have enriched
my life outside the academic world.

Finally, I wish to thank (i) the Faculty of Engineering and Science
and the Institute of Physics at Aalborg University for providing the
basic frame for my research work, (ii) the Mississippi Center for
Supercomputing Research at the University of Mississippi, Jackson,
Mississippi, for providing computer resources during my stay at
Mississippi State University, and (iii) the people at Aalborg
University Library for their excellent assistance in obtaining
articles from far away in time and space.

\vskip 1.5cm
\noindent
\hbox to\hsize{{\it Aalborg, Denmark, June 1998}\hfil{\bf Torsten Andersen}}
\newpage
\thispagestyle{plain}
\newpage
\tableofcontents
\listoffigures
\listoftables
\section*{List of acronyms}
\begin{tabular}{@{}p{3cm}@{}p{10.5cm}@{}}
DC & Direct current.\\
DFWM & Degenerate four-wave mixing. \\
ED & Electric dipole.\\
IB & Infinite barrier.\\
PCDFWM & Phase conjugation (by) degenerate four-wave mixing. \\
SVE & Slowly varying envelope.\\
SVEA & Slowly varying envelope approximation.\\
YAG & Y$_3$Al$_5$O$_{12}$ \cite{Yariv:84:1}.\\
\end{tabular}
\newpage
\thispagestyle{empty}
\begin{center} 
\setlength{\unitlength}{1mm}
\psset{unit=1mm}
\begin{pspicture}(-60,-50)(60,65)
 \pstextpath[c]{\psarcn[linestyle=none](0,0){46}{220}{-40}}
  {\curly{Theoretical Study of Phase Conjugation}}
 \pstextpath[c]{\psarc[linestyle=none](0,0){48}{180}{0}}
  {\curly{in Mesoscopic Interaction Volumes}}
\end{pspicture}
\end{center}
\newpage
\thispagestyle{plain}
\cleardoublepage
\mainmatter
\chead[\fancyplain{}{\leftmark}]{\fancyplain{}{\rightmark}}
\part{Motivation}\label{Part:I}
\vfill
\begin{center}
Natura inest\\
in mentibus nostrum\\
insatiabilis quaedam\\
cupiditas veri videndi\\
({\it Marcus Tullius Cicero}\/)
\end{center}
\newpage
\thispagestyle{plain}
\newpage

\chapter{Historical perspective}\label{Ch:1}
Indeed, as the great orator expressed it more than two millenia ago,
nature has planted in our minds an insatiable longing to see the
truth. This natural curiosity, I believe, has been the driving force
behind scientific investigations in the history of mankind, and thus
also behind the evolution of electromagnetic theory. But since the
electrodynamic theory as we know it was initiated by
\citeANP{Maxwell:64:1} \citeyear{Maxwell:64:1,Maxwell:91:1}, I will in
the following historical remarks concentrate on the physics of the
past century. Readers who want an overview of the evolution of
electromagnetic theories before this century are referred to
\citeN{Born:80:1}, and the comprehensive survey of
\citeN{Whittaker:51:1}.\nocite{Whittaker:53:1}

The work described in the present dissertation is mainly concerned
with a theoreti\-cal description of a nonlinear type of electromagnetic
interactions called degenerate four-wave mixing (DFWM), particularly
in the case where phase conjugation is obtained. The main interest
behind this study is to model the behaviour of the DFWM interaction in
mesoscopic volumes and in the optical near-field zone. Thus, in
relation to established branches of modern optics, this work belongs
to the fields of nonlinear optics (especial\-ly four-wave mixing),
near-field optics, and mesoscopic systems. In the remaining of this
chapter I therefore intend to describe briefly the contributions to
scientific progress within the last century, which I believe are the
most significant for the present study.

Although Einstein already in 1916 predicted the existence of
stimulated emission (\citeANP{Einstein:16:1}
\citeyearNP{Einstein:16:1,Einstein:17:1}), the main objective of
optics remained linear observations until after the development of the
maser in the early 1950's, where Townes and co-workers at Columbia
University used stimulated emission for amplification of an
electromagnetic field in combination with a resonator
(\citeNP{Gordon:54:1}, \citeyearNP{Gordon:55:1}). An application of
the principles of the maser in the optical region of the
electromagnetic spectrum was proposed in 1958 by
\citeANP{Schawlow:58:1}, and in 1960 \citeANP{Maiman:60:1} constructed
the first laser---a pulsed ruby laser. The first laser delivering a
continous-wave output was constructed in 1961 using a mixture of
Helium and Neon gasses \cite{Javan:61:1}. The laser rapidly became of
significant importance in optical physics, where the field of
nonlinear optics was ignited by the successful observation by
\citeN{Franken:61:1} of radiation of light at the second harmonic
frequency (with a wavelength $\lambda$ of $3472${\AA}) generated by a
quartz crystal illuminated with light from a ruby laser
($\lambda=6943${\AA}). Since then nonlinear optics has been of
interest to many researchers around the globe exploring a large number
of different nonlinear phenomena, such as second-, third-, and higher
order harmonic generation, optical rectification, sum and difference
frequency generation, three-, four-, six-, and higher-number
wave-mixing, laser cooling, laser induced atomic fusion, stimulated
Raman- and Brillouin scattering, to mention a few [see, e.g.,
\citeN{Bloembergen:65:1}, \citeN{Boyd:92:1}, \citeN{Mandel:95:1},
\citeN{Yariv:84:1}, \citeN{Shen:84:1}, \citeN{Schubert:86:1}, and
\citeN{Mukamel:95:1}].

The optical effect of interest in this work, optical phase
conjugation, is nowadays usually produced by means of nonlinear
optics, although the problem of reconstructing electromagnetic
wavefronts started in the linear optical regime. The pioneering work
on optical wavefront reconstruction (holography) was carried out
several years before the invention of the laser by Gabor
(\citeyearNP{Gabor:48:1,Gabor:49:1}) with the purpose of improving the
resolving power of the electron microscope [see also
\citeN{Bragg:50:1}]. But only with the high intensities and with the
degree of temporal and spatial coherence provided by the laser,
holographic imaging became of practical importance. Such experiments
were first reported by \citeANP{Leith:62:1}
(\citeyearNP{Leith:62:1,Leith:64:1}). Soon thereafter
\citeN{Kogelnik:65:1} used a hologram to correct static phase
distortions introduced onto an optical wavefront. In this experiment a
photosensitive film was used for holographic recording of an image,
and the film had to be developed prior to its application for phase
correction. This experiment of \citeANP{Kogelnik:65:1} appears to be
the first account on optical phase conjugation. However, since a new
film has to be developed every time the phase distortion changes, this
technique becomes rather cumbersome if the phase distortions changes
frequently. A key discovery of \citeN{Gerritsen:67:1} made it possible
to store holograms dynamically in crystals with an intensity-dependent
refractive index, thereby extending the applicability domain of
optical phase conjugation to cover descriptions where phase
distortions are varying in time. Experimentally, the first real-time
optical phase conjugation are cre\-dited to Zel'dovich and co-workers
\cite{Zeldovich:72:1,Nosach:72:1}, in an experiment based on
stimulated Brillouin scattering. In the late 1970's,
\citeN{Hellwarth:77:1} suggested the use of a degenerate four-wave
mixing process to produce the phase conjugated field. Immediately
thereafter \citeN{Yariv:77:1}, and independently, \citeN{Bloom:77:1},
further analyzed the optical phase conjugation via DFWM, resulting in
predictions of amplified reflection, coherent image amplification and
oscillation. Over the past twentyfive years, thousands of scientific
papers, several books and review articles describing different aspects
and applications of optical phase conjugation have been published, and
phase conjugation in the form of DFWM is now an established discipline
in modern experimental optics. The theoretical treatments of optical
phase conjugation are usually based upon the work of
\citeN{Yariv:78:1}, using the phase conjugating system as a device in
studies of other processes. A comprehensive and coherent introduction
to the field of optical phase conjugation can be found in the books by
\citeN{Zeldovich:85:1} and \citeN{Sakai:92:1}, while a more
specialized introduction can be achieved through collections of review
papers appearing in books by \citeN{Fisher:83:1} and
\citeN{Gower:94:1}, or separately, by \citeANP{Pepper:82:1}
(\citeyearNP{Pepper:82:1,Pepper:85:1}), \citeN{Hellwarth:82:1}, and
\citeN{Knoester:91:1}. Other collections of papers can be found in,
e.g., \citeN{Goodman:83:1}, and \citeN{Brueck:89:1}. Within the last
few years, DFWM has been used for creation of optical phase
conjugation in configurations where the probe field and the detector
are within subwavelength distances from the phase-conjugating medium
\cite{Bozhevolnyi:94:1,Vohnsen:97:1,Bozhevolnyi:97:1}.

The first account of attention to subwavelength (optical near-field)
interaction of light with matter seems to be \citeANP{Synge:28:1}
(\citeyearNP{Synge:28:1,Synge:32:1}), who proposed an apparatus, in
which a sample is illuminated through a small aperture in an opaque
screen, the area of the aperture being substantially smaller than the
diffraction limit of the light used for the illumination (of
subwavelength size). The aperture should be moved in small increments
(scanned) over the sample by use of a piezo-electric crystal. At every
step of the scanning procedure the light transmitted through the
sample should be collected and the intensity measured. The resolving
power of such an instrument should be limited by the size of the
aperture and the distance from the aperture to the sample rather than
by the wavelength of the illuminating light. For whatever reason, the
proposal of \citeANP{Synge:28:1} was forgotten, and even though
\citeN{Bethe:44:1} and \citeANP{Bouwkamp:50:1}
(\citeyearNP{Bouwkamp:50:1,Bouwkamp:50:2}) discussed the problem of
diffraction by small holes, the idea of an optical near-field
microscope remained forgotten until \citeN{OKeefe:56:1} made the
proposal, apparently without any knowledge of the instrument proposed
by \citeANP{Synge:28:1}. The first demonstration of an image obtained
with scanning in the electromagnetic near-field zone was given by
\citeN{Ash:72:1}, who used microwaves of wavelength of 3cm to resolve
metallic gratings with linewidths down to 0.5mm, corresponding to
$1/60$-th of a wavelength.  Another twelve years should pass before
near-field electrodynamics was adressed again. Near-field optics
evolved in the mid-eighties in the wake of the experimental works by
the groups of Pohl, Lewis, and Fischer
\cite{Pohl:84:1,Lewis:84:1,Fischer:85:1,Fischer:89:1}. The main
efforts of this new branch of modern optics is concentrated on the
original idea of subwavelength imaging [see, for example, the recently
published book by \citeN{Paesler:96:1}, the proceedings of the first
conference on near-field optics \cite{Pohl:93:1}, or proceedings from
later conferences in near-field optics
\cite{Isaacson:95:1,Paesler:95:1,Nieto-Vesperinas:96:1,Hulst:98:1}].

The appearance of microscopes with subwavelength resolution inevitably
poses the questions of the resolution limit and the degree of spatial
confinement of light---two inseparable questions in near-field optics.
Fundamentally, the spatial confinement problem is linked to the
field-matter interaction in the vicinity of the source emitting the
field and in the near-field region of the detector. In classical
optics, near-field effects traditionally have played a minor role, and
the possibilities for studying material properties on a small length
scale usually are judged in relation to the diffraction limit
criterion attributed to Ernst \citeN{Abbe:73:1} and the third baron
Rayleigh \citeyear{Rayleigh:96:1}. The Rayleigh criterion, though
mainly invoked in the context of spatial resolution, also sets the
limit for the possibilities of light compression in far-field studies.
As already emphasized by \citeANP{Rayleigh:96:1} and later discussed,
for instance, by \citeN{Ronchi:61:1}, the resolution problem is not a
simple one, even in classical optics. A recent survey of the
resolution problem within the framework of classical optics has been
given by \citeN{Dekker:97:1}.

When the interaction length of an electromagnetic field across the
individual structures in a condensed matter system is on the order of
an optical wavelength (typically a few atomic distances), the
theoretical description of the field-matter interaction belongs to the
field of mesoscopic electrodynamics. Within the last two decades
studies of the optical properties of mesoscopic systems, such as
quantum wells (single and multiple), -wires and -dots, surfaces,
interfaces, and more exotic geometries have drawn the attention of
many researchers. Because of the immediate potential for industrial
application many of these studies have been concentrated on the
properties of semiconductors (see, e.g., \citeN{Weisbuch:91:1}, and
references herein). In recent years in particular investigations of
the nonlinear electrodynamics have been in focus. Among the many
nonlinear phenomena studies of second harmonic generation
\cite{Sipe:82:1,Richmond:88:1,Heinz:91:1,Reider:95:1,Liebsch:95:1,Pedersen:95:1},
sum- and difference frequency generation
\cite{Reider:95:1,Bavli:91:1}, photon drag
\cite{Keller:93:1,Vasko:96:1,Chen:97:1,Keller:97:2}, DC-electric-field
induced second harmonic generation
\cite{Aktsipetrov:95:1,Aktsipetrov:96:1}, the Kerr effect
\cite{Pustogowa:94:1,Liu:95:1,Rasing:95:1,Rasing:96:1}, electronic and
vibrational surface Raman scattering
\cite{Nkoma:89:1,Mischenko:95:1,Garcia-Vidal:96:1}, two-photon
photoemission
\cite{Haight:95:1,Fauster:95:1,Georges:95:1,Shalaev:96:1,Tergiman:97:1},
and generation of higher harmonics
\cite{vonderLinde:96:1,Garvila:92:1} have played a prominent role.
Among the more exotic phenomena, studies of the Aharonov-Bohm effect
in mesoscopic rings \cite{Wang:97:1} and whispering-gallery modes in
microspheres \cite{Knight:95:1} have also been carried out lately.

From a theoretical point of view the refractive index concept becomes
meaningless for structures of mesoscopic size. Therefore, macroscopic
approaches to describe the field-matter interaction have to be
abandoned from the outset, and the theoretical analyses have to be
based on the microscopic Maxwell equations combined with the
Schr{\"o}dinger equation. The Schr{\"o}dinger equation describes the
quantum state of the condensed matter system, and is a fundamental
part of the quantum mechanics initiated in the beginning of the 20th
century by such scientists as Planck, Einstein, Bohr, Heisenberg,
Born, Jordan, de Broglie, Schr{\"o}dinger and Dirac. Even an attempt
to give a satisfactory historical survey of the development of quantum
mechanics at this point will fail because of the almost universal
status quantum mechanics has reached in the description of modern
physics. Instead, for the history of quantum mechanics including a
description of the mathematical foundation, please consult for example
\citeN{Neumann:32:1} or \citeN{Bohm:51:1}. A modern and comprehensive
description of quantum mechanics is given by
\citeN{Cohen-Tannoudji:77:0}, where also a comprehensive list of
references to key papers can be found. An example of interesting
papers is the series of articles by \citeANP{Schrodinger:26:0}
\citeyear{Schrodinger:26:0,Schrodinger:26:5}.

\chapter[Standard theory of optical phase conjugation by DFWM]
{Standard theory of \\ optical phase conjugation by degenerate
  four-wave mixing}\label{Ch:2} Optical phase conjugation is a
nonlinear optical phenomenon, in which an incoming optical field is
reflected in such a manner that the wavefronts of the reflected field
coincide with the incoming field, hence also the name ``wavefront
inversion'', frequently used in the literature. The principle of
optical phase conjugation has gained widespread attention because of
its ability to correct for distortions introduced in a path traversed
by an optical signal. In principle, it works like this: An optical
source is placed on one side of a distorting medium (crystal,
waveguide, atmosphere, etc.). A system in which phase conjugation
takes place (called the phase conjugator) is placed on the other side
of the distorting medium. A field emitted from the source in the
direction of the phase conjugator then travel through the distorting
medium, and is reflected by the phase conjugator. The phase conjugator
reverses the wavefront of the incoming (probe) field, and when the
reflected light comes back through the distorting medium, the
wavefront is (ideally) exactly reversed, compared to that originally
emitted by the source. Since it is possible to see how the light was
originally emitted by the source by looking at the phase conjugated
replica, it is sometimes also given the somewhat misleading term
``time reversal'' \cite{Yariv:78:1}. Several schemes exist to achieve
phase conjugation, the most widely used called ``degenerate four-wave
mixing'' (DFWM).  Optical phase conjugation in the form of degenerate
four-wave mixing (DFWM) is a nonlinear third order effect, where
mixing of two counterpropagating ``pump'' fields and a ``probe'' (or
``signal'') field---all with the same frequency $\omega$---results in,
among other signals, a generated field (the ``conjugate'') with
frequency $\omega=\omega+\omega-\omega$, which is counterpropagating
to the probe field.

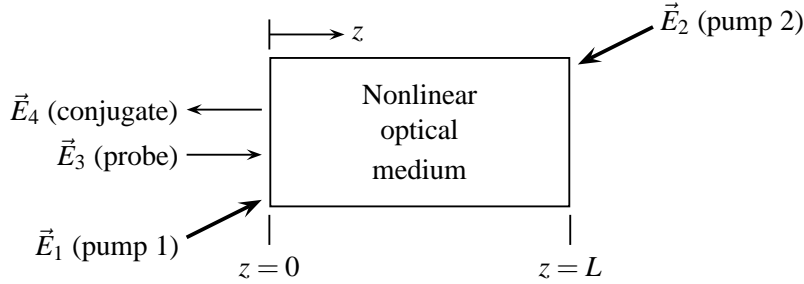
\begin{figure}[tb]
\setlength{\unitlength}{1mm}
\psset{unit=1mm}
\begin{center}
\begin{pspicture}(0,0)(120,40)
\psframe[linewidth=0.25](40,10)(80,30)
\psline[linewidth=0.25]{-}(40,5)(40,9)
\psline[linewidth=0.25]{-}(40,31)(40,35)
\psline[linewidth=0.25]{-}(80,5)(80,9)
\psline[linewidth=0.25]{->}(40,33)(50,33)
\put(40,0){\makebox(0,4)[c]{$z=0$}}
\put(80,0){\makebox(0,4)[c]{$z=L$}}
\put(51,31){\makebox(0,4)[l]{$z$}}
\put(60,22.5){\makebox(0,5)[c]{Nonlinear}}
\put(60,17.5){\makebox(0,5)[c]{optical}}
\put(60,12.5){\makebox(0,5)[c]{medium}}
\psline[linewidth=0.5]{->}(29,6)(39,11)
\psline[linewidth=0.5]{<-}(81,29)(91,34)
\psline[linewidth=0.25]{->}(29,17)(39,17)
\psline[linewidth=0.25]{<-}(29,23)(39,23)
\put(28,15){\makebox(0,4)[r]{$\vec{E}_{3}$~(probe)}}
\put(28,21){\makebox(0,4)[r]{$\vec{E}_{4}$~(conjugate)}}
\put(28,3){\makebox(0,4)[r]{$\vec{E}_{1}$~(pump~1)}}
\put(92,33){\makebox(0,4)[l]{$\vec{E}_{2}$~(pump~2)}}
\end{pspicture}
\end{center}
\caption{Geometry of phase conjugation by DFWM in the standard
  model.\label{fig:2.1}}
\end{figure}

In the following, I present the theoretical model usually adopted in
studies of optical phase conjugation by degenerate four-wave mixing,
and consequently this chapter will consist mainly of textbook material.
The treatment roughly follows that of \citeN{Yariv:83:1} and of
\citeN{Boyd:92:1}. The DFWM geometry suggested by \citeN{Yariv:77:1}
is shown in Fig.~\ref{fig:2.1}. In this configuration, a lossless
nonlinear optical medium is illuminated by two strong
counterpropagating pump fields $\vec{E}_{1}$ and $\vec{E}_{2}$ and by
a weak signal (probe) wave $\vec{E}_{3}$. The pump fields are usually
taken to be plane waves, although they in principle are allowed to
have any kind of wavefront as long as their amplitudes are complex
conjugates of each other. The probe field can have a more complex
wavefront. Resulting from the mixing process in the medium a conjugate
field appears, propagating in the direction oppositely to the probe.

In order to describe the electromagnetic field we first establish the
wave equation for the interacting fields from the macroscopic Maxwell
equations. It is thereafter reduced to its slowly varying envelope
approximation (SVEA) form. The four macroscopic Maxwell equations are
\begin{eqnarray}
 \vec{\nabla}\times\vec{E}(\vec{r},t)&=&
 -{\partial\vec{B}(\vec{r},t)\over\partial{}t},
\label{eq:macromax1}
\\
 \vec{\nabla}\times\vec{H}(\vec{r},t)&=&\vec{J}(\vec{r},t)
 +{\partial\vec{D}(\vec{r},t)\over\partial{}t},
\label{eq:macromax2}
\\
 \vec{\nabla}\cdot\vec{D}(\vec{r},t)&=&\rho(\vec{r},t),
\label{eq:macromax3}
\\
 \vec{\nabla}\cdot\vec{B}(\vec{r},t)&=&0.
\label{eq:macromax4}
\end{eqnarray}
We now assume that the material is homogeneous, nonmagnetic
($\vec{B}=\mu_0\vec{H}$), and nonconducting ($\vec{J}=\vec{0}$) and
that there are no free charges ($\rho=0$). We write the displacement
vector $\vec{D}(\vec{r},t)$ as
\begin{equation}
 \vec{D}(\vec{r},t)=\epsilon_0\vec{E}(\vec{r},t)
 +\epsilon_0\vec{P}(\vec{r},t),
\end{equation}
where $\vec{P}(\vec{r},t)$ is the polarization, which we split into
its linear, $\vec{P}_{\rm{L}}$, and nonlinear, $\vec{P}_{\rm{NL}}$,
components
\begin{equation}
 \vec{P}(\vec{r},t)=\vec{P}_{\rm{L}}(\vec{r},t)+\vec{P}_{\rm{NL}}(\vec{r},t).
\end{equation}
Above, the linear polarization describes the material response due to
interaction with the field of first order. We thus define the linear
susceptibility $\btensor{\chi}^{(1)}$ from the linear polarization in
the manner $\vec{P}_{\rm{L}}=\btensor{\chi}^{(1)}\cdot\vec{E}$. The
linear permittivity $\stensor{\epsilon}_{\!r}$ is then found from the
linear part of the displacement, giving
$\stensor{\epsilon}_{\!r}=\tensor{\openone}+\btensor{\chi}^{(1)}$. Taking
the curl of Eq.~(\ref{eq:macromax1}) and inserting
Eq.~(\ref{eq:macromax2}) into the resulting equation, we obtain (by
use of the operator identity
$\vec{\nabla}\times\vec{\nabla}\times=-\nabla^2+\vec{\nabla}\vec{\nabla}\cdot$)
the following wave equation
\begin{equation}
 \left[\tensor{\openone}\nabla^2-{\stensor{\epsilon}_{\!r}\over{}c^2}
 {\partial^2\over\partial{}t^2}\right]\cdot\vec{E}(\vec{r},t)
 ={1\over{}c^2}{\partial^2\vec{P}_{\rm{NL}}(\vec{r},t)\over\partial{}t^2},
\end{equation}
where we have assumed that the electric fields are perpendicular to
their corresponding wavevector (transversality). The nonlinear
polarization is usually described as a power series in the electric
field,
\begin{equation}
 \vec{P}=\btensor{\chi}^{(1)}\cdot\vec{E}+\btensor{\chi}^{(2)}:\vec{E}\vec{E}
 +\btensor{\chi}^{(3)}\vdots\vec{E}\vec{E}\vec{E}+\dots,
\end{equation}
where $\btensor{\chi}^{(1)}$ is the linear susceptibility,
$\btensor{\chi}^{(2)}$ is the second-order nonlinear susceptibility
tensor, $\btensor{\chi}^{(3)}$ is the third-order nonlinear
susceptibility tensor, etc. This expansion is of course only of
interest if we can assume that
$\vec{P}^{(1)}\gg\vec{P}^{(2)}\gg\vec{P}^{(3)}\gg\cdots$ (the
parametric approximation). Since we have assumed the medium to be
lossless, the susceptibility tensors are real time-independent
quantities, and hence also $\stensor{\epsilon}_{\!r}$ is a real
quantity. The linear susceptibility tensor is included in the linear
polarization ($\vec{P}_{\rm{L}}$) above, and the lowest order
nonlinear polarization of interest to DFWM is the third-order one,
i.e.,
\begin{equation}
 \vec{P}_{\rm{NL}}(\vec{r},t)=\btensor{\chi}^{(3)}\vdots
 \vec{E}(\vec{r},t)\vec{E}(\vec{r},t)\vec{E}(\vec{r},t).
\label{eq:2.9}
\end{equation}
Above, the sum-product operator ``$\vdots$'' is defined such that
element $i$ of the nonlinear polarization is
\begin{equation}
 P_{\rm{NL},i}(\vec{r},t)=\sum_{jkh}\chi_{ijkh}^{(3)}
 E_{h}(\vec{r},t)E_{k}(\vec{r},t)E_{j}(\vec{r},t).
\label{eq:2.10}
\end{equation}
The total electric field is a sum of the four individual fields in the
DFWM process,
\begin{equation}
 \vec{E}(\vec{r},t)=\sum_{\alpha=1}^{4}\vec{E}_{\alpha}(\vec{r},t)=
 {1\over2}\sum_{\alpha=1}^{4}\vec{E}_{\alpha}(\vec{r})
 e^{{\rm{i}}(\vec{k}_{\alpha}\cdot\vec{r}-\omega{}t)}+\mbox{c.c.},
\label{eq:2.11}
\end{equation}
where $\vec{E}_{\alpha}(\vec{r})$ are slowly varying quantities, and
the wavevector $\vec{k}_{\alpha}$ is real. Since we assumed that the
pump fields $\vec{E}_{1}(\vec{r},t)$ and $\vec{E}_{2}(\vec{r},t)$ are
counterpropagating, the sum of their wavevectors is zero, i.e.,
$\vec{k}_{1}+\vec{k}_{2}=\vec{0}$. Inserting Eq.~(\ref{eq:2.11}) into
Eq.~(\ref{eq:2.9}), a large number of terms are generated. In the
phase conjugation configuration we are particularly interested in the
terms related to the first harmonic in the cyclic frequency $\omega$.
Among these terms are terms that can act as phase-matched source terms
for the conjugate wave $\vec{E}_{4}(\vec{r},t)$ when the probe and
conjugate fields are counterpropagating, i.e., when
$\vec{k}_{3}+\vec{k}_{4}=\vec{0}$. Using these two properties of the
wavevectors, terms with a spatial dependence of the form
$e^{{\rm{i}}\vec{k}_{\alpha}\cdot\vec{r}}$ are particularly important
because they produce the phase-matched terms for the four interacting
electric fields. The polarizations associated with these phase-matched
contributions (at $\omega$) become
\begin{eqnarray}
 \vec{P}_{\rm{NL}}^{(1)}(\vec{r},t)&=&{3\over8}\btensor{\chi}^{(3)}\vdots
 \left[\vec{E}_{1}\vec{E}_{1}\vec{E}_{1}^{*}
 +2\sum_{\alpha\in\{2,3,4\}}\vec{E}_{\alpha}\vec{E}_{1}\vec{E}_{\alpha}^{*}
 +2\vec{E}_{2}^{*}\vec{E}_{3}\vec{E}_{4}\right]
 e^{{\rm{i}}(\vec{k}_{1}\cdot\vec{r}-\omega{}t)}+\mbox{c.c.},
\label{eq:2.12}
\\
 \vec{P}_{\rm{NL}}^{(2)}(\vec{r},t)&=&{3\over8}\btensor{\chi}^{(3)}\vdots
 \left[\vec{E}_{2}\vec{E}_{2}\vec{E}_{2}^{*}
 +2\sum_{\alpha\in\{1,3,4\}}\vec{E}_{\alpha}\vec{E}_{2}\vec{E}_{\alpha}^{*}
 +2\vec{E}_{1}^{*}\vec{E}_{3}\vec{E}_{4}\right]
 e^{{\rm{i}}(\vec{k}_{2}\cdot\vec{r}-\omega{}t)}+\mbox{c.c.},
\label{eq:2.13}
\\
 \vec{P}_{\rm{NL}}^{(3)}(\vec{r},t)&=&{3\over8}\btensor{\chi}^{(3)}\vdots
 \left[\vec{E}_{3}\vec{E}_{3}\vec{E}_{3}^{*}
 +2\sum_{\alpha\in\{1,2,4\}}\vec{E}_{\alpha}\vec{E}_{3}\vec{E}_{\alpha}^{*}
 +2\vec{E}_{1}\vec{E}_{2}\vec{E}_{4}^{*}\right]
 e^{{\rm{i}}(\vec{k}_{3}\cdot\vec{r}-\omega{}t)}+\mbox{c.c.},
\label{eq:2.14}
\\
 \vec{P}_{\rm{NL}}^{(4)}(\vec{r},t)&=&{3\over8}\btensor{\chi}^{(3)}\vdots
 \left[\vec{E}_{4}\vec{E}_{4}\vec{E}_{4}^{*}
 +2\sum_{\alpha\in\{1,2,3\}}\vec{E}_{\alpha}\vec{E}_{4}\vec{E}_{\alpha}^{*}
 +2\vec{E}_{1}\vec{E}_{2}\vec{E}_{3}^{*}\right]
 e^{{\rm{i}}(\vec{k}_{4}\cdot\vec{r}-\omega{}t)}+\mbox{c.c.},
\label{eq:2.15}
\end{eqnarray}
in a short notation where $\vec{E}\equiv\vec{E}(\vec{r})$. With this
splitting of the nonlinear polarization, the wave equation is
satisfied, when each of the four fields and their related
polarizations satisfy the wave equation separately. Next, assuming
that the pump fields are much stronger than the probe and the
conjugate fields, we can drop the terms in
Eqs.~(\ref{eq:2.12})--(\ref{eq:2.15}) containing more than one
weak-field component, thus obtaining
\begin{eqnarray}
 \vec{P}_{\rm{NL}}^{(1)}(\vec{r},t)&=&{3\over8}\btensor{\chi}^{(3)}\vdots
 \left[\vec{E}_{1}\vec{E}_{1}\vec{E}_{1}^{*}
 +2\vec{E}_{2}\vec{E}_{1}\vec{E}_{2}^{*}\right]
 e^{{\rm{i}}(\vec{k}_{1}\cdot\vec{r}-\omega{}t)}+\mbox{c.c.},
\label{eq:2.16}
\\
 \vec{P}_{\rm{NL}}^{(2)}(\vec{r},t)&=&{3\over8}\btensor{\chi}^{(3)}\vdots
 \left[\vec{E}_{2}\vec{E}_{2}\vec{E}_{2}^{*}
 +2\vec{E}_{1}\vec{E}_{2}\vec{E}_{1}^{*}\right]
 e^{{\rm{i}}(\vec{k}_{2}\cdot\vec{r}-\omega{}t)}+\mbox{c.c.},
\label{eq:2.17}
\\
 \vec{P}_{\rm{NL}}^{(3)}(\vec{r},t)&=&{3\over4}\btensor{\chi}^{(3)}\vdots
 \left[\vec{E}_{1}\vec{E}_{3}\vec{E}_{1}^{*}
 +\vec{E}_{2}\vec{E}_{3}\vec{E}_{2}^{*}
 +\vec{E}_{1}\vec{E}_{2}\vec{E}_{4}^{*}\right]
 e^{{\rm{i}}(\vec{k}_{3}\cdot\vec{r}-\omega{}t)}+\mbox{c.c.},
\label{eq:2.18}
\\
 \vec{P}_{\rm{NL}}^{(4)}(\vec{r},t)&=&{3\over4}\btensor{\chi}^{(3)}\vdots
 \left[\vec{E}_{1}\vec{E}_{4}\vec{E}_{1}^{*}
 +\vec{E}_{2}\vec{E}_{4}\vec{E}_{2}^{*}
 +\vec{E}_{1}\vec{E}_{2}\vec{E}_{3}^{*}\right]
 e^{{\rm{i}}(\vec{k}_{4}\cdot\vec{r}-\omega{}t)}+\mbox{c.c.},
\label{eq:2.19}
\end{eqnarray}
again in short notation. Note that by this approximation the
polarizations associated with the pump fields have been decoupled from
the probe and conjugate fields. Then we may first solve the wave
equations for the pump fields, and thereafter insert the result into
the wave equations for the probe and conjugate fields. Following this
insertion, the probe and conjugate fields can be found. Since
$\nabla^2\{\vec{E}(\vec{r})e^{{\rm{i}}(\vec{k}\cdot\vec{r}-\omega{}t)}\}=\{(\nabla^2+2{\rm{i}}[\vec{k}\cdot\vec{\nabla}]-k^2)\vec{E}(\vec{r})\}e^{{\rm{i}}(\vec{k}\cdot\vec{r}-\omega{}t)}$,
the wave equation for pump field 1 can be written
\begin{equation}
 \left[\tensor{\openone}\left(\nabla^2+2{\rm{i}}[\vec{k}_{1}\cdot\vec{\nabla}]
 -k^2\right)+{\omega^2\over{}c^2}\stensor{\epsilon}_{\!r}\right]
 \cdot\vec{E}_{1}=-{3\omega^2\over8c^2}\btensor{\chi}^{(3)}\vdots
 \left[\vec{E}_{1}\vec{E}_{1}\vec{E}_{1}^{*}
 +2\vec{E}_{2}\vec{E}_{1}\vec{E}_{2}^{*}\right],
\label{eq:2.20}
\end{equation}
still in the short notation from above. Assuming now that we have an
isotropic medium, the susceptibility tensors must be invariant to
inversion and rotation around any axis in the chosen Cartesian
coordinate system. The demand of inversion symmetry leaves all tensor
elements with an odd number of $x$'s, $y$'s, or $z$'s zero, and thus
only diagonal elements survive in the linear susceptibility tensor,
and only the $21$ elements in the nonlinear susceptibility tensor of
the form $\chi_{iijj}^{(3)}$, $\chi_{ijji}^{(3)}$, and
$\chi_{ijij}^{(3)}$ are nonzero, $i$ and $j$ being any $x$, $y$, or
$z$. The demand of invariance to rotational transformations results in
the demand that the three remaining nonzero elements of the linear
susceptibility tensor are equal, and thus we find that
$\stensor{\epsilon}_{\!r}=\tensor{\openone}\epsilon_r$. In terms of
the refractive index $n$ of the medium, that is $\epsilon_r=n^2$. For
the nonlinear susceptibility tensor this demand implies that the
nonzero elements can be written
\begin{equation}
 \chi_{ijkh}^{(3)}=
 \chi_{xxyy}^{(3)}\delta_{ij}\delta_{kh}
 +\chi_{xyxy}^{(3)}\delta_{ik}\delta_{jh}
 +\chi_{xyyx}^{(3)}\delta_{ih}\delta_{jk}.
\label{eq:2.21}
\end{equation}
In the DFWM case, permutation symmetry between the two fields without
the complex conjugation makes $k$ and $h$ interchangeable, and thus
$\chi_{xyxy}^{(3)}$ is equal to $\chi_{xyyx}^{(3)}$, leaving the
nonzero elements
\begin{equation}
 \chi_{ijkh}^{(3)}=\chi_{xxyy}^{(3)}\delta_{ij}\delta_{kh}
 +\chi_{xyyx}^{(3)}(\delta_{ih}\delta_{jk}+\delta_{ik}\delta_{jh})
\label{eq:2.22}
\end{equation}
in the nonlinear susceptibility tensor. Furthermore, inside the medium
the modulus of the wavevectors are the same,
$k_1=k_2=k_3=k_4=k=n\omega/c$. Under the above assumptions,
Eq.~(\ref{eq:2.20}) takes the form
\begin{eqnarray}
\lefteqn{
 \left(\nabla^2+2{\rm{i}}[\vec{k}_{1}\cdot\vec{\nabla}]\right)
 \vec{E}_{1}(\vec{r})=-{3\omega^2\over8c^2}\left\{
 \chi_{xxyy}^{(3)}
 \left[[\vec{E}_{1}(\vec{r})\cdot\vec{E}_{1}(\vec{r})]\vec{E}_{1}^{*}(\vec{r})
 +2[\vec{E}_{2}(\vec{r})\cdot\vec{E}_{1}(\vec{r})]\vec{E}_{2}^{*}(\vec{r})
 \right]
\right.}\nonumber\\ &\quad&\left.\!
 +2\chi_{xyyx}^{(3)}
 \left[\vec{E}_{1}(\vec{r})[\vec{E}_{1}(\vec{r})\cdot\vec{E}_{1}^{*}(\vec{r})]
 +2\vec{E}_{2}(\vec{r})[\vec{E}_{1}(\vec{r})\cdot\vec{E}_{2}^{*}(\vec{r})]
 \right]
 \right\}.
\label{eq:2.23}
\end{eqnarray}
Now the simplest assumption is that the four fields travel in a direction
almost parallel to the $z$-axis (the paraxial approximation), that
they have the same state of polarization, and that the pump waves have
plane wavefronts (independent of $x$ and $y$). Then instead of
Eq.~(\ref{eq:2.23}), Eq.~(\ref{eq:2.20}) is rewritten into the form
\begin{equation}
 \left({d^2\over{}dz^2}+2{\rm{i}}k{d\over{}dz}\right) 
 E_{1}(z)=-{3\omega^2\over8c^2}\chi_{xxxx}^{(3)}
 \left[|E_{1}(z)|^2+2|E_{2}(z)|^2\right]E_{1}(z),
\label{eq:2.24}
\end{equation}
still for an isotropic medium, and now having $k_1=k_z$. Introducing
into Eq.~(\ref{eq:2.24}) the slowly varying envelope approximation
(SVEA), in which it is assumed that $|k(dE/dz)|$ $\gg|d^2E/dz^2|$, we
obtain
\begin{equation}
 {dE_1(z)\over{}dz}={3{\rm{i}}\omega\over16nc}\chi_{xxxx}^{(3)}
 \left[|E_1(z)|^2+2|E_2(z)|^2\right]E_1(z)\equiv{\rm{i}}\kappa_1E_1(z).
\label{eq:2.25}
\end{equation}
In a similar fashion we find that the pump field going in the
negative $z$-direction is descibed by the equation
\begin{equation}
 {dE_2(z)\over{}dz}=-{3{\rm{i}}\omega\over16nc}\chi_{xxxx}^{(3)}
 \left[|E_2(z)|^2+2|E_1(z)|^2\right]E_2(z)\equiv-{\rm{i}}\kappa_2E_2(z),
\label{eq:2.26}
\end{equation}
since $k_2=-k_z$. Since $\chi_{xxxx}^{(3)}$ and $n$ are real
quantities (from the assumption of a lossless medium), $\kappa_1$ and
$\kappa_2$ are also real quantities. Eqs.~(\ref{eq:2.25}) and
(\ref{eq:2.26}) have solutions on the form
$E_1(z)=E_1(0)e^{{\rm{i}}\kappa_1z}$ and
$E_2(z)=E_2(0)e^{-{\rm{i}}\kappa_2z}$, respectively.

Next, we consider the probe and conjugate fields. If we assume that
the incident probe wave can be decomposed into plane waves we can for
simplicity consider only one of these at a time. Under this
assumption, and keeping the approximations mentioned before,
the wave equations for the probe and conjugate fields are
\begin{eqnarray}
\lefteqn{
 {dE_3(z)\over{}dz}={3{\rm{i}}\omega\over8nc}\chi_{xxxx}^{(3)}
 \left\{\left[|E_1(0)|^2+2|E_2(0)|^2\right]E_3(z)+E_1(0)E_2(0)E_4^*(z)
 e^{{\rm{i}}(\kappa_1-\kappa_2)z}\right\}
}\nonumber\\ &\quad&\equiv
 {\rm{i}}\kappa_3E_3(z)+{\rm{i}}\kappa{}E_4^*(z),
\label{eq:2.27}
\\
\lefteqn{
 {dE_4(z)\over{}dz}=-{3{\rm{i}}\omega\over8nc}\chi_{xxxx}^{(3)}
 \left\{\left[|E_1(0)|^2+2|E_2(0)|^2\right]E_4(z)+E_1(0)E_2(0)E_3^*(z)
 e^{{\rm{i}}(\kappa_1-\kappa_2)z}\right\}
}\nonumber\\ &&\equiv
 -{\rm{i}}\kappa_3E_4(z)-{\rm{i}}\kappa{}E_3^*(z).
\label{eq:2.28}
\end{eqnarray}
To achieve perfect phase matching between the probe and the conjugate
field, $\kappa$ has to be constant along $z$, requiring that
$\kappa_1=\kappa_2$, which means that the intensity of the two pump
fields must be the same ($|E_1(z)|^2=|E_2(z)|^2$). If we additionally
introduce a change of variables by letting
$E_3(z)=E_3'(z)e^{{\rm{i}}\kappa_3z}$ and
$E_4(z)=E_4'(z)e^{-{\rm{i}}\kappa_3z}$, Eqs.~(\ref{eq:2.27}) and
(\ref{eq:2.28}) become
\begin{eqnarray}
 {dE_3'(z)\over{}dz}&=&{\rm{i}}\kappa{}E_4^{\prime{}*}(z),
\label{eq:2.29}\\
 {dE_4'(z)\over{}dz}&=&-{\rm{i}}\kappa{}E_3^{\prime{}*}(z),
\label{eq:2.30}
\end{eqnarray}
and we notice in passing that the primed and the unprimed variables
coincide in the input plane of the interaction region, i.e., at $z=0$.
Eqs.~(\ref{eq:2.29}) and (\ref{eq:2.30}) shows why degenerate
four-wave mixing leads to phase conjugation, since the generated field
$E_4'(z)$ is driven only by the complex conjugate of the probe field
amplitude. Differentiation of Eq.~(\ref{eq:2.29}) and insertion of
Eq.~(\ref{eq:2.30}), an vice versa, we get
\begin{eqnarray}
 {d^2E_3'(z)\over{}dz^2}+\kappa^2E_3'(z)&=&0,
\label{eq:2.31}\\
 {d^2E_4'(z)\over{}dz^2}+\kappa^2E_4'(z)&=&0.
\label{eq:2.32}
\end{eqnarray}
The characteristic equation is $\lambda^2+\kappa^2=0$, which has
solutions $\lambda=\pm{\rm{i}}\kappa$. The general solution to
Eqs.~(\ref{eq:2.31}) and (\ref{eq:2.32}) is then
\begin{eqnarray}
 E_3'(z)&=&C_1e^{{\rm{i}}\kappa{}z}+C_2e^{-{\rm{i}}\kappa{}z},
\label{eq:2.33}\\
 E_4'(z)&=&C_3e^{{\rm{i}}\kappa{}z}+C_4e^{-{\rm{i}}\kappa{}z}.
\label{eq:2.34}
\end{eqnarray}
Assuming that we know the values $E_3'(0)$ and $E_4'(L)$, we can then
find $E_3'(z)$ and $E_4'(z)$ as a function of these two boundary
values. Then the solutions to the coupled differential equations,
Eqs.~(\ref{eq:2.29}) and (\ref{eq:2.30}), describing the electric
field inside the phase conjugating medium, become
\begin{eqnarray}
 E_3'(z)&=&E_3'(0){\cos[\kappa(L-z)]\over\cos(\kappa{}L)}
 -{\rm{i}}E_4^{\prime{}*}(L){\sin(\kappa{}z)\over\cos(\kappa{}L)},
\label{eq:2.35}\\
 E_4'(z)&=&E_4'(L){\cos(\kappa{}z)\over\cos(\kappa{}L)}
 -{\rm{i}}E_3^{\prime{}*}(0){\sin[\kappa(L-z)]\over\cos(\kappa{}L)}.
\label{eq:2.36}
\end{eqnarray}
In the practical case, $E_3'(0)$ (the probe field coming into the
medium) is finite and $E_4'(L)$ (the phase conjugated field at the
other end of the medium) is zero. The phase conjugated field coming
out of the medium at $z=0$ is then
\begin{equation}
 E_4'(0)=-{\rm{i}}E_3^{\prime{}*}(0)\tan(\kappa{}L).
\label{eq:2.37}
\end{equation}
Thus the phase conjugated field depends on (i) the intensity of the
pump fields, (ii) the length of the active medium, and (iii) the
incoming probe field, and we notice that the magnitude of the phase
conjugated field can be larger than the magnitude of the incoming
probe field.

\chapter{Discussion}\label{Ch:3}
The theoretical description given in the preceding chapter is not the
only existing description of phase conjugation by DFWM in the
macroscopic sense, but it illustrates quite well the usual line of
thought when considering optical phase conjugation. As examples on
theoretical papers going beyond the description in Chapter~\ref{Ch:2},
let us mention that (i) polarization properties have been studied by
\citeN{Ducloy:84:1}, (ii) descriptions taking into account the
vectorial properties [see Eq.~(\ref{eq:2.20})] have been given, e.g.,
by \citeN{Syed:96:1}, (iii) improvements to the standard theory in the
form of abandoning the slowly varying envelope approximation (SVEA)
have also been discussed [see, e.g., \citeN{Marburger:83:1} and
\citeN{Farzad:97:1}]. A feature of the standard theory [see
Eq.~(\ref{eq:2.37})] is that the phase conjugated response depends on
the length of the nonlinear crystal used (infinite at
$\kappa{}L=(2p+1)\pi/2$ for any integer value of $p$). The standard
theory has proven to be a satisfactory description for spatially
nondecaying fields containing no evanescent components.

Though the overwhelming majority of optical phase conjugation
experiments can be described without inclusion of evanescent
components of the electromagnetic field, the possible phase
conjugation of these components has been discussed from time to time.
With the experimental observation of \citeANP{Bozhevolnyi:94:1}
\citeyear{Bozhevolnyi:94:1,Bozhevolnyi:95:2},
\nocite{Bozhevolnyi:95:4} the need for inclusion of near-field
components and thus evanescent modes in the description of optical
phase conjugation has drawn renewed attention.

In an important paper by \citeN{Agarwal:95:1} the treatment was
focused on an analysis of the phase conjugated replica produced by a
so-called ideal phase conjugator, characterized phenomenologically by
a polarization- and angle of incidence independent nonlinear amplitude
reflection coefficient, and in recent articles by
\citeANP{Keller:96:2} \citeyear{Keller:96:3,Keller:96:2} attention was
devoted to an investigation of the spatial confinement problem of the
phase conjugated field. Macroscopic theories including near-field
components in the optical phase conjugation process have also appeared
recently \cite{Bozhevolnyi:95:3,Arnoldus:95:2}.

In their work \citeANP{Bozhevolnyi:94:1} used degenerate four-wave
mixing (DFWM) produced by a 10mW HeNe laser with a wavelength of 633nm
in an iron-doped lithium-niobate (Fe:LiNbO$_3$) crystal and an
external-reflection near-field optical microscope to achieve phase
conjugated light foci, which with a diameter of $\sim180$nm were well
below the classical diffraction limit. The main conclusion of their
experiments was that to achieve a spot size as small as 180nm phase
conjugation of at least parts of the optical near-field emitted from
the source must have taken place.

In the present work we go one step further in the theoretical study of
the phase conjugation of optical signals which include near-field
components by abandoning the ideal phase conjugator assumption. For
simplicity our description is limited to cover only the degenerate
four-wave mixing configuration for which the interacting optical
fields all have the same cyclic frequency $\omega$.

\begin{figure}[tb]
\setlength{\unitlength}{25mm}
\psset{unit=25mm}
\begin{center}
\begin{pspicture}(0,0)(4,2.5)
\put(0,-0.5){
 \qdisk(0.25,2.55){0.05}
 \psarc[linewidth=0.25mm]{-}(0.25,2.55){0.15}{10}{35}
 \psarc[linewidth=0.25mm]{-}(0.25,2.55){0.15}{55}{80}
 \psarc[linewidth=0.25mm]{-}(0.25,2.55){0.15}{100}{125}
 \psarc[linewidth=0.25mm]{-}(0.25,2.55){0.15}{145}{170}
 \psarc[linewidth=0.25mm]{-}(0.25,2.55){0.15}{190}{215}
 \psarc[linewidth=0.25mm]{-}(0.25,2.55){0.15}{235}{260}
 \psarc[linewidth=0.25mm]{-}(0.25,2.55){0.15}{280}{305}
 \psarc[linewidth=0.25mm]{-}(0.25,2.55){0.15}{325}{350}
 \psarc[linewidth=0.25mm]{-}(0.25,2.55){0.25}{10}{35}
 \psarc[linewidth=0.25mm]{-}(0.25,2.55){0.25}{55}{80}
 \psarc[linewidth=0.25mm]{-}(0.25,2.55){0.25}{100}{125}
 \psarc[linewidth=0.25mm]{-}(0.25,2.55){0.25}{145}{170}
 \psarc[linewidth=0.25mm]{-}(0.25,2.55){0.25}{190}{215}
 \psarc[linewidth=0.25mm]{-}(0.25,2.55){0.25}{235}{260}
 \psarc[linewidth=0.25mm]{-}(0.25,2.55){0.25}{280}{305}
 \psarc[linewidth=0.25mm]{-}(0.25,2.55){0.25}{325}{350}
 \put(0.7,2.25){\makebox(0,0.4)[l]{$=$\large$\displaystyle\sum_{q_{\|}\leq{\omega\over{}c_0}}$}}
 \psline[linewidth=0.5mm]{->}(1.7,2.75)(1.7,2.25)
 \psline[linewidth=0.5mm]{->}(1.6875,2.75)(2.2,2.75)
 \psline[linewidth=0.5mm]{->}(1.7,2.75)(2.2,2.25)
 \psline[linewidth=0.2mm,linestyle=dashed]{-}(1.7,2.25)(2.2,2.25)
 \psline[linewidth=0.2mm,linestyle=dashed]{-}(2.2,2.25)(2.2,2.75)
 \psline[linewidth=0.2mm]{-}(1.75,2.60)(1.85,2.70)
 \psline[linewidth=0.2mm]{-}(1.80,2.55)(1.90,2.65)
 \psline[linewidth=0.2mm]{-}(1.85,2.50)(1.95,2.60)
 \psline[linewidth=0.2mm]{-}(1.90,2.45)(2.00,2.55)
 \psline[linewidth=0.2mm]{-}(1.95,2.40)(2.05,2.50)
 \psline[linewidth=0.2mm]{-}(2.00,2.35)(2.10,2.45)
 \put(2.5,2.25){\makebox(0,0.4)[l]{$+$\large$\displaystyle\sum_{q_{\|}>{\omega\over{}c_0}}$}}
 \psline[linewidth=0.5mm]{->}(3.5,2.75)(3.5,2.25)
 \psline[linewidth=0.5mm]{->}(3.4875,2.75)(4.0,2.75)
 \psline[linewidth=0.2mm]{-}(3.60,2.65)(3.60,2.85)
 \psline[linewidth=0.2mm]{-}(3.65,2.65)(3.65,2.85)
 \psline[linewidth=0.2mm]{-}(3.70,2.65)(3.70,2.85)
 \psline[linewidth=0.2mm]{-}(3.75,2.65)(3.75,2.85)
 \psline[linewidth=0.2mm]{-}(3.80,2.65)(3.80,2.85)
 \psline[linewidth=0.3mm,linestyle=dotted,dotsep=0.5mm]{-}(3.4,2.65)(3.6,2.65)
 \psline[linewidth=0.3mm,linestyle=dotted,dotsep=0.5mm]{-}(3.4,2.60)(3.6,2.60)
 \psline[linewidth=0.3mm,linestyle=dotted,dotsep=0.5mm]{-}(3.4,2.55)(3.6,2.55)
 \psline[linewidth=0.3mm,linestyle=dotted,dotsep=0.5mm]{-}(3.4,2.50)(3.6,2.50)
 \psline[linewidth=0.3mm,linestyle=dotted,dotsep=0.5mm]{-}(3.4,2.45)(3.6,2.45)
}
 \pscurve[linestyle=solid,linewidth=0.25mm](2,1)(1.61,0.9)(1.37,0.8)(1.22,0.7)(1.14,0.6)(1.082,0.5)(1.050,0.4)(1.030,0.3)(1.018,0.2)(1.011,0.1)(1.007,0.0)
 \psline[linewidth=0.5mm]{->}(1,1.1)(1,0)
 \psline[linewidth=0.5mm]{->}(0.8,1)(2.2,1)
 \psline[linewidth=0.25mm]{-}(0,0.75)(4,0.75)
 \psline[linewidth=0.25mm]{-}(0,0.35)(4,0.35)
 \qdisk(1,1){0.05}
 \psline[linewidth=0.25mm]{->}(1.3,1.3)(1.1,1.1)
 \put(1.325,1.20){\makebox(0,0.4)[l]{Mesoscopic source}}
 \put(3.9,0.8){\makebox(0,0.4)[r]{Vacuum}}
 \put(3.9,0.35){\makebox(0,0.4)[r]{Quantum well/Thin film}}
 \put(3.9,0){\makebox(0,0.4)[r]{Substrate}}
\end{pspicture}
\end{center}
\caption[Weyl expansion in plane waves of a spherical wave over a
plane]{The upper part is a schematic illustration showing the Weyl
  representation of a spherical wave-field from a point (mesoscopic)
  source. In this representation the field is expanded in plane waves
  over a plane, in practice the surface in consideration. The
  two-dimensional wavevector ($\vec{q}_{\|}$) expansion consists of
  those terms for which $q_{\|}\leq\omega/c_0$ ($c_0$ being the vacuum
  speed of light) plus those having $q_{\|}>\omega/c_0$. In the first
  group of terms the component of the wavevector perpendicular to the
  surface (vertical arrow) is real so that the individual plane-wave
  modes are propagating, and in the second group, consisting of
  evanescent modes, this component is purely imaginary. The solid
  lines attached to two of the arrows indicate planes of constant
  phase, and the dotted lines attached to the evanescent modes
  indicate lines of constant amplitude. The lower part is a schematic
  illustration showing an exponentially decaying mode from a
  mesoscopic source placed near a thin film (quantum well) phase
  conjugator. To phase conjugate an evanescent mode in an effective
  manner the near field of the source must overlap the phase
  conjugator, and as indicated it is not always correct to assume that
  the selfconsistently determined evanescent field is constant across
  the thin film.
\label{fig:PS1}}
\end{figure}
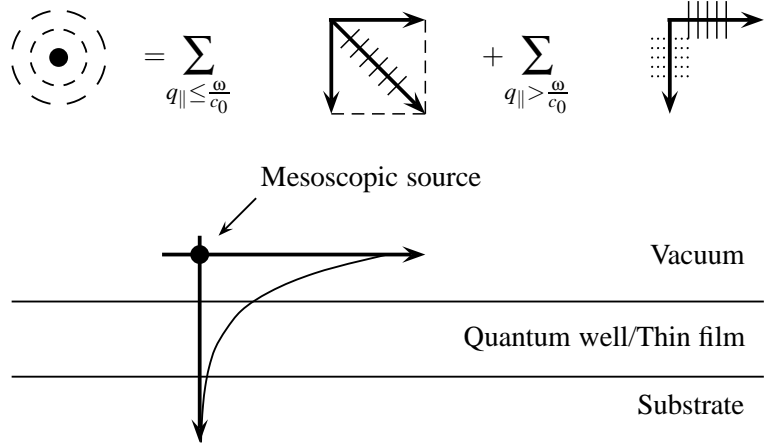

Because of the small range of the optical evanescent fields from the
(mesoscopic) source a substantial part of the near-field phase
conjugation process is bound to take place in the surface region of
the phase conjugating medium. It is thus from the very outset
necessary to focus the attention on the surface region of the
nonlinear mirror and investigate the phase conjugation process on a
length scale (much) smaller than the optical wavelength. This fact in
itself makes use of the ideal phase conjugator assumption doubtful.
For a bulk phase conjugator it may furthermore be difficult to assure
an effective nonlinear mixing in a surface layer as thin as the field
penetration depth. Thus, experimentally it might be advantageous to
use a thin film or even a quantum well as the nonlinear medium (see
Fig.~\ref{fig:PS1}). From a different perspective the use of a thin
film as the nonlinear medium has already drawn attention
\cite{Montemezzani:96:1}.

The present theory has been constructed in such a manner that it
offers a framework for microscopic studies of degenerate four-wave
mixing at surfaces of bulk media, in thin films and quantum wells, and
in small particles. To carry out in detail a rigorous microscopic
numerical analysis of the DFWM process it is, however, necessary to
consider mesoscopic media with a particularly simple electronic
structure, and we shall demonstrate later how the present theory can
be applied to a simple quantum well structure.

In conventional descriptions of optical phase conjugation by DFWM it
is assumed that the interaction length is long compared to the
wavelength of the probe fields, thus building up pictorially speaking
from one of the pump beams and the probe beam a grating, from which
the other pump beam is scattered into a phase conjugated replica (the
`real-time holography' picture). Furthermore it is assumed that the
amplitudes of the fields are slowly varying on the optical wavelength
scale [Slowly varying envelope (SVE) approximation] and thus also
constant across the individual scattering units (atoms, molecules,
\dots) [Electric dipole (ED) approximation] of the phase conjugating
medium. Consider\-ing optical near fields, which contain components
varying rapidly in space, the aforementioned approximations do not
hold and we thus exclude them in the present formalism. We also avoid
other approximations often made in the literature, namely (i) the
assumption of a lossless medium, (ii) the {\it ab initio}\/
requirement of phase matching between the interacting optical signals,
and (iii) the assumption that the probe field is weak compared to the
pump fields.

\begin{figure}[tb]
\setlength{\unitlength}{20mm}
\psset{unit=20mm}
\begin{center}
\begin{pspicture}(-0.5,-1.0)(4,4)
 \psline[linestyle=dotted,dotsep=1mm]{-}(0,0)(4,4)
 \pswedge[linestyle=dashed,linewidth=0.25mm](0,0){1.51}{0}{90}
 \pscurve[linestyle=dashed,linewidth=0.25mm](1.51,0.00)(1.60,0.53)(2.00,1.31)(2.50,1.99)(3.00,2.59)(3.50,3.16)(4.00,3.70)
 \pswedge[linestyle=solid,linewidth=0.25mm](0,0){1.0}{0}{90}
 \pscurve[linestyle=solid,linewidth=0.25mm](1.00,0.00)(1.10,0.46)(1.30,0.83)(1.50,1.12)(2.00,1.73)(2.50,2.29)(3.00,2.83)(3.50,3.35)(4.00,3.87)
 \psline[linestyle=solid,linewidth=0.5mm]{->}(-0.05,0)(4,0)
 \psline[linestyle=solid,linewidth=0.5mm]{->}(0,-0.05)(0,4)
 \put(-0.1,3.75){\makebox(0.0,0.0)[r]{$|q_{\perp}|$}}
 \psline[linestyle=solid,linewidth=0.5mm]{-}(-0.05,1.51)(0.05,1.51)
 \put(-0.1,1.51){\makebox(0.0,0.0)[r]{${n\omega\over{}c_0}$}}
 \psline[linestyle=solid,linewidth=0.5mm]{-}(-0.05,1.0)(0.05,1.0)
 \put(-0.1,1.0){\makebox(0.0,0.0)[r]{${\omega\over{}c_0}$}}
 \put(-0.1,0.0){\makebox(0.0,0.0)[r]{$0$}}
 \put(0.0,-0.1){\makebox(0.0,0.0)[t]{$0$}}
 \psline[linestyle=solid,linewidth=0.5mm]{-}(1.0,-0.05)(1.0,0.05)
 \put(1.0,-0.1){\makebox(0.0,0.0)[t]{${\omega\over{}c_0}$}}
 \psline[linestyle=solid,linewidth=0.5mm]{-}(1.51,-0.05)(1.51,0.05)
 \put(1.51,-0.1){\makebox(0.0,0.0)[t]{${n\omega\over{}c_0}$}}
 \put(3.75,-0.1){\makebox(0.0,0.0)[t]{$q_{\|}$}}
 \put(0.25,1.0){\makebox(0.0,0.0)[b]{$q_{\perp}^{0}$}}
 \put(0.5,1.51){\makebox(0.0,0.0)[b]{$q_{\perp}^{n}$}}
 \put(1.67,1.0){\makebox(0.0,0.0)[b]{$\alpha_{\perp}^{0}$}}
 \put(2.35,1.5){\makebox(0.0,0.0)[b]{$\alpha_{\perp}^{n}$}}
 \psline[linewidth=0.25mm]{<->}(0,-0.65)(1,-0.65)
 \psline[linewidth=0.25mm]{<-}(1,-0.65)(4,-0.65)
 \psline[linewidth=0.25mm]{-}(0,-0.7)(0,-0.6)
 \psline[linewidth=0.25mm]{-}(1,-0.7)(1,-0.6)
 \put(0.5,-0.625){\makebox(0,0)[b]{$q_{\perp}^{0}$}}
 \put(1.5,-0.625){\makebox(0,0)[bl]{$q_{\perp}^{0}={\rm{i}}\alpha_{\perp}^{0}$}}
 \psline[linewidth=0.25mm]{<->}(0,-1)(1.51,-1)
 \psline[linewidth=0.25mm]{<-}(1.51,-1)(4,-1)
 \psline[linewidth=0.25mm]{-}(0,-1.05)(0,-0.95)
 \psline[linewidth=0.25mm]{-}(1.51,-1.05)(1.51,-0.95)
 \put(0.75,-0.975){\makebox(0,0)[b]{$q_{\perp}$}}
 \put(2.0,-0.975){\makebox(0,0)[bl]{$q_{\perp}^{n}={\rm{i}}\alpha_{\perp}^{n}$}}
 \psline[linestyle=solid,linewidth=0.5mm]{->}(0.4875,3.50)(1.5,3.50)
 \psline[linestyle=solid,linewidth=0.5mm]{->}(0.5,3.50)(1.5,3.00)
 \psline[linestyle=solid,linewidth=0.5mm]{->}(0.5,3.50)(0.5,3.00)
 \put(1.55,3.50){\makebox(0.0,0.0)[l]{$q_{\|}$}}
 \put(1.55,3.00){\makebox(0.0,0.0)[l]{$q$}}
 \put(0.45,3.00){\makebox(0.0,0.25)[r]{$q_{\perp}^{n}$}}
 \put(0.45,3.25){\makebox(0.0,0.25)[r]{$q_{\perp}^{0}$}}
 \psline[linestyle=dashed,dash=0.05 0.05,linewidth=0.25mm]{-}(0.5,3.00)(1.5,3.00)
 \psline[linestyle=dashed,dash=0.05 0.05,linewidth=0.25mm]{-}(1.5,3.00)(1.5,3.50)
 \psline[linestyle=solid,linewidth=0.5mm]{->}(2.4875,1.0)(3.5,1.0)
 \psline[linestyle=solid,linewidth=0.5mm]{->}(2.5,1.0)(2.5,0.5)
 \put(3.55,1.0){\makebox(0.0,0.0)[l]{$q_{\|}$}}
 \put(2.45,0.5){\makebox(0.0,0.25)[r]{$\alpha_{\perp}^{n}$}}
 \put(2.45,0.75){\makebox(0.0,0.25)[r]{$\alpha_{\perp}^{0}$}}
\end{pspicture}
\end{center}
\caption[The probe wavevector component perpendicular to a surface as
a function of the parallel component]{The component of the probe
  wavevector perpendicular to ($q_{\perp}$) a vacuum/bulk phase
  conjugator interface as a function of its real parallel component
  (${q}_{\|}$) in vacuum (solid line, $q_{\perp}^{0}$) and in the
  substrate (dashed line, $q_{\perp}^{n}$). For
  $q_{\|}\leq\omega/c_0$,
  $q_{\perp}^{0}=[(\omega/c_0)^2-q_{\|}^2]^{1/2}$, and
  $q_{\perp}^{n}=[(n\omega/c_0)^2-q_{\|}^2]^{1/2}$ are both real, and
  the associated plane waves are thus propagating (and nondecaying) in
  both the vacuum and the phase conjugator, neglecting absorption. In
  the region $\omega/c_0<q_{\|}\leq{}n\omega/c_0$ ($n$ being the
  linear (real) refractive index of the phase conjugator),
  $q_{\perp}^{0}={\rm{i}}\alpha_{\perp}^{0}$ becomes purely imaginary
  (we plot $\alpha_{\perp}^{0}=[q_{\|}^2-(\omega/c_0)^2]^{1/2}$), but
  $q_{\perp}$ is still real. The field in the vacuum is thus
  evanescent in this region. In the region $q_{\|}>n\omega/c_0$, also
  $q_{\perp}={\rm{i}}\alpha_{\perp}$ is a purely imaginary number (we
  plot $\alpha_{\perp}^{n}=[q_{\|}^2-(n\omega/c_0)^2]^{1/2}$), so that
  also the field in the phase conjugator is evanescent.
\label{fig:PS2}}
\end{figure}
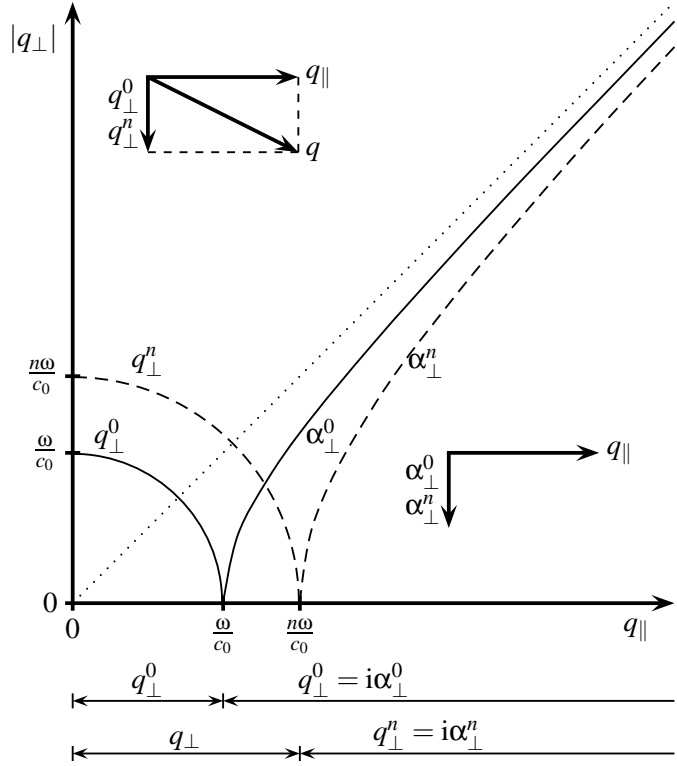

To illustrate the need for a theory going beyond the SVE and ED
approximations, we have in Fig.~\ref{fig:PS2} shown the component of
the probe wavevector perpendicular to the surface, inside as well as
outside the phase conjugator, as a function of its parallel component.
When the parallel component of the probe wavevector becomes larger
than $\omega/c_0$, the perpendicular component of the wavevector
becomes purely imaginary in the vacuum, but it is still real inside
the phase conjugating mirror. A purely imaginary wavevector component
means that the electromagnetic field is evanescent, whereas a real
component indicates that the field is propagating and nondecaying (in
the absense of absorption). When the parallel component becomes larger
than $n\omega/c_0$ (where $n$ is the refractive index of the
substrate) the perpendicular component of the probe wavevector
becomes evanescent also inside the phase conjugating mirror, and the
larger the parallel component, the more wrong the SVE and ED
approximations become. Thus to study, for instance, the phase
conjugation of all field components possibly emitted from a mesoscopic
source in the vicinity of the phase conjugator it is necessary to
abandon these approximations.

As already mentioned, the present theory not only allows one to
investigate the optical phase conjugation of evanescent waves with
small penetration depths, it also enables one to investigate the
possibility of achieving DFWM in mesoscopic films (quantum wells), a
subject of interest in its own right. The main reason that the present
formulation may be used in near-field optics as well as in
mesoscopic-film electrodynamics originates in the fact that in both
cases the microscopic local-field calculation is the crucial quantity.
A further advantage of the present theory is that it allows us to
study phase conjugation when one or more of the interacting fields are
surface-wave fields.

The construction of such a theoretical model begins with the
microscopic Maxwell--Lorentz equations, which combined with the
nonlocal linear and third-order nonlinear constitutive equations are
used to set up the basic wave equation for the phase conjugated field.
The linear and nonlinear conductivity responses of the electrons will
be calculated within the framework of the random-phase-approximation
theory, the starting point being the Liouville equation of motion for
the density matrix operator. In the description we include in the
interaction Hamiltonian not only the standard $\vec{p}\cdot\vec{A}$
term ($\vec{p}$ being the momentum operator and $\vec{A}$ being the
vector potential) but also the term proportional to the square of the
vector potential, i.e., $\vec{A}\cdot\vec{A}$.  For a monochromatic
driving field (of cyclic frequency $\omega$), this term contains
$2\omega$- and DC-parts, and both of these are in general important
for the description of the microscopic phase conjugation process. In
the current density operator we include the term containing the vector
potential. This term, needed in order to ensure the gauge invariance
of quantum electrodynamics, also turns out to be of importance in some
cases. Starting from a dipolar interaction Hamiltonian the first
explicit microscopic derivation of the third-order conductivity
(susceptibility) response appears to be due to
\citeN{Bloembergen:78:1}. The result of \citeANP{Bloembergen:78:1} is
based on a $\vec{r}\cdot\vec{E}$ calculation and only the
vector-potential independent part of the current density operator is
kept. Apart from a single study dealing with the electromagnetic
self-action in a BCS-paired superconductor \cite{Keller:95:7}, it
seems that in all theoretical investigations of the DFWM-process in
which microscopic considerations have appeared, the
\citeANP{Bloembergen:78:1} expression has been used. We cannot use
this expression here, however, since we need to address a local-field
problem when dealing with mesoscopic interaction volumes, and such a
problem necessitates that we take into account values of $q_{\|}$ much
larger than $\omega/c_0$, and thus that the calculation goes beyond
the ED approximation. To account for local-field effects it is
necessary to perform a spatially nonlocal calculation of the
third-order conductivity, and this is most adequately done beginning
with the minimal coupling interaction Hamiltonian which contains both
the $\vec{p}\cdot\vec{A}$ and $\vec{A}\cdot\vec{A}$ terms. In the
local limit where the vector potential only depends on time our
expression for the nonlinear conductivity and the
$\vec{r}\cdot\vec{E}$ based one of \citeANP{Bloembergen:78:1} are
physically equivalent, {\it provided}\/ the terms stemming from the
gauge conserving vector potential dependent part of the current
density operator are neglected. Though physically equivalent, the
explicit forms of the relation between the nonlinear current density
and the electric field only coincides after having performed a
relevant unitary transformation on the minimal coupling Hamiltonian
and the related electronic wave functions. In stead of using the
minimal coupling Hamiltonian to describe the nonlocal dynamics one
could in principle have used the multipolar Hamiltonian. In practice
this is less convenient for the present purpose due to the fact that
the pronounced nonlocality we sometimes are facing in mesoscopic media
would require that many multipole terms were kept in the Hamiltonian.
The essentially nonlocal terms in the nonlinear conductivity are
included in our treatment because they in certain cases---especially
for very small interaction volumes---are the only contributing ones,
and in other cases they dominate the phase conjugated response. Since
we deal with a spatially nonlocal description it is important to
characterize the spatial structure involved in the physical processes
behind the phase conjugation, and the various physical processes
hidden in the nonlinear and nonlocal constitutive equation are
therefore identified.  Following the identification of the physical
processes, an expression for the so-called conductivity response
tensor describing the nonlinear material response in the DFWM process
is established, and the eigensymmetries of the conductivity tensors
belonging to each of the processes occuring in the DFWM process are
discussed. Rather than solving the full spatial problem (which would
be cumbersome, if not impossible), we consider a simplified system
possessing infinitesimal translational invariance in two directions.
For such a system the potential of the related Schr{\"o}dinger
equation only varies in the direction perpendicular to the plane of
translational invariance. The fundamental solutions to the
time-independent Schr{\"o}dinger equation are inserted into the linear
and nonlinear conductivity tensors thus giving us the framework for a
theoretical description of the DFWM process in mesoscopic films
(quantum wells) as well as for evanescent waves. Compared to
conventional descriptions of optical phase conjugation in bulk media
the concept of phase matching (momentum conservation) now appears only
in {\it two}\/ dimensions.  The lack of translational invariance in
the third dimension implies that no phase matching occurs in this
dimension. Phase matching (in one, two or three dimensions) is not a
precondition set on our theory, it follows to the extent that the
phase conjugating medium exhibits infinitesimal translational
invariance.  Despite the fact that the phase matching is lost in the
third dimension, phase conjugation may still take place in quantum
wells and thin films, and with evanescent fields, just as second
harmonic generation can occur in quantum-well systems, at metallic
(and semiconducting) surfaces and from nonlinear (sub)monolayer films
deposited on linear substrates \cite{Richmond:88:1}. To complete our
local-field calculation of the optical phase conjugation by degenerate
four-wave mixing in mesoscopic interaction volumes, we use a Green's
function formalism to establish new integral equations for the phase
conjugated field in the general case, and in the case where the
nonlinear medium exhibits translational invariance in two dimensions.
The microscopic local-field theory thus established is then used to
describe the DFWM in one- and two-level quantum-well phase
conjugators.

\part[Microscopic model for DFWM]{Microscopic model for \\ degenerate
  four-wave mixing}\label{part:II}
\newpage
\thispagestyle{plain}
\newpage

\chapter{The electromagnetic field}\label{ch:4}
In this chapter a description of the electromagnetic field from a
phase conjugating medi\-um is established, starting from the microscopic
Maxwell equations. First, we derive the relevant wave equation for the
phase conjugated field. The field--matter interaction is then
described through the use of constitutive relations. In the final step
of this basic framework, a self-consistent description of the phase
conjugated field is established.

\section{Wave equation for the phase conjugated field}
As a starting point we take the microscopic Maxwell-Lorentz equations,
in which the material response at the space-time point $(\vec{r},t)$
is completely described via the microscopic current density
$\vec{J}(\vec{r},t)$, and the related charge density,
$\rho(\vec{r},t)$. They are
\begin{eqnarray}
 \vec{\nabla}\times\vec{E}(\vec{r},t)&=&
  -{\partial\vec{B}(\vec{r},t)\over\partial{}t},
\label{eq:Max1}\\
 \vec{\nabla}\times\vec{B}(\vec{r},t)&=&\mu_{0}\vec{J}(\vec{r},t)
  +{1\over{}c_0^2}{\partial\vec{E}(\vec{r},t)\over\partial{}t},
\label{eq:Max2}\\
\vec{\nabla}\cdot\vec{E}(\vec{r},t)&=&{1\over\epsilon_{0}}\rho(\vec{r},t),
\label{eq:Max3}\\
 \vec{\nabla}\cdot\vec{B}(\vec{r},t)&=&0,
\label{eq:Max4}
\end{eqnarray}
$\vec{E}(\vec{r},t)$ and $\vec{B}(\vec{r},t)$ being the electric and
magnetic fields prevailing at the space point $\vec{r}$ at the time
$t$.

Taking the curl of Eq.~(\ref{eq:Max1}) and inserting the result into
Eq.~(\ref{eq:Max2}) we obtain the following wave equation for the
prevailing local electric field $\vec{E}(\vec{r},t)$:
\begin{equation}
\left({\tensor{\openone}}{\Box}
  +\vec{\nabla}\otimes\vec{\nabla}\right)\cdot\vec{E}(\vec{r},t)
=-\mu_0{\partial\vec{J}(\vec{r},t)\over\partial{}t},\label{eq:Wave}
\end{equation}
where ${\Box}={1\over{}c^{2}}{\partial^2\over\partial{}t^2}-\nabla^2$
is the d'Alembertian operator, $\tensor{\openone}$ is the ($3\times3$)
unit tensor, and $\otimes$ is the outer (dyadic) product operator.

Introducing the electric field as a Fourier series in the cyclic
frequency $\omega$, viz.
\begin{equation}
 \vec{E}(\vec{r},t)={1\over2}\sum_{\alpha=0}^{\infty}
 \vec{E}_{-\alpha\omega}(\vec{r}\,)e^{-{\rm{i}}\alpha\omega{}t}+\mbox{c.c.},
\label{eq:Ert}
\end{equation}
where $\alpha$ is an integer and ``c.c.''  denotes the complex
conjugate of the first terms, we subsequently can limit ourselves to a
harmonic analysis. Since $\vec{E}(\vec{r},t)$ is a real quantity,
$\vec{E}^{*}_{-\alpha\omega}(\vec{r}\,)=\vec{E}_{\alpha\omega}(\vec{r}\,)$.

Likewise, we write the current density as a Fourier series in
$\omega$, in which each component implicitly is expressed as a power
series in the electric field. Thus
\begin{equation}
 \vec{J}(\vec{r},t)=
 {1\over2}\sum_{\alpha=0}^{\infty}\sum_{\beta=0}^{\infty}
 \left(\vec{J}_{-\alpha\omega}^{\,(\alpha+2\beta)}(\vec{r}\,)
 e^{-{\rm{i}}\alpha\omega{}t}+\mbox{c.c.}\right),
\label{eq:J-power}
\end{equation}
where $\alpha$ and $\beta$ are integers. Looking for solutions at the
cyclic frequency $\omega$, only fields and current densities with
$\alpha=1$ in Eqs.~(\ref{eq:Ert}) and (\ref{eq:J-power}) contributes.
Accordingly we in the following write the phase conjugated (PC)
electric field without the reference to the cyclic frequency, i.e.,
$\vec{E}_{-\omega}(\vec{r}\,)\equiv\vec{E}_{\rm{PC}}(\vec{r}\,)$. In
the case of DFWM we will assume that the lowest order nonlinear
interaction dominates over higher order mixing processes. Thus, in
order to describe the DFWM response of our medium we retain only the
two currents of lowest order in $\beta$, namely the linear
contribution $\vec{J}_{-\omega}^{\,(1)}(\vec{r}\,)$ and the lowest
order nonlinear contribution $\vec{J}_{-\omega}^{\,(3)}(\vec{r}\,)$.
The wave equation for the negative frequency part of the phase
conjugated response hence takes the form
\begin{equation}
 \left[\tensor{\openone}\left({\omega^2\over{}c^2}+{\nabla}^2\right)
 -\vec{\nabla}\otimes\vec{\nabla}\right]
 \cdot\vec{E}_{\rm{PC}}(\vec{r}\,)
 =-i\mu_0\omega\left(\vec{J}_{-\omega}^{\,(1)}(\vec{r}\,)
 +\vec{J}_{-\omega}^{\,(3)}(\vec{r}\,)\right).
\label{eq:Wave2}
\end{equation}

\section{Constitutive relations for the current densities}
To close the loop for the calculation of the phase conjugated field,
the microscopic current densities
$\vec{J}_{-\omega}^{\,(1)}(\vec{r}\,)$ and
$\vec{J}_{-\omega}^{\,(3)}(\vec{r}\,)$ are given in terms of the local
electric field through constitutive relations describing the
field--matter interaction in a perturbative manner. Choosing a gauge
where the time-dependent part of the scalar potential is zero, the
electric field is related to the vector potential via
$\vec{E}(\vec{r}\,)={\rm{i}}\omega\vec{A}(\vec{r}\,)$. Thus the
microscopic current densities can be related via the constitutive
relations to the vector potentials of the phase conjugated field
($\vec{A}_{\rm{PC}}$) and the fields driving the process
($\vec{A}\,$). The linear constitutive relation we therefore write in
the form
\begin{equation}
 \vec{J}_{-\omega}^{\,(1)}(\vec{r}\,)={i\omega}\int
 \stensor{\sigma}(\vec{r},\vec{r}\,')
 \cdot\vec{A}_{\rm{PC}}(\vec{r}\,')d^3r',
\label{eq:Jr}
\end{equation}
where $\stensor{\sigma}(\vec{r},\vec{r}\,')\equiv
\stensor{\sigma}(\vec{r},\vec{r}\,';\omega)$ is the linear conductivity
tensor. The $i$'th element of the first order current density is
proportional to the integral of
$[\stensor{\sigma}\cdot\vec{A}_{\rm{PC}}]_{i}=
\sum_{j}{\sigma}_{ij}{A}_{{\rm{PC}},j}$.  The nonlinear DFWM
constitutive relation is written in a similar fashion, i.e.,
\begin{equation}
 \vec{J}_{-\omega}^{\,(3)}(\vec{r}\,)=(i\omega)^3\int\int\int
 \tensor{\Xi}(\vec{r},\vec{r}\,',\vec{r}\,'',\vec{r}\,''')
 \vdots\,\vec{A}(\vec{r}\,''')\vec{A}(\vec{r}\,'')\vec{A}^{*}(\vec{r}\,')
 d^3r'''d^3r''d^3r',
\label{eq:J3r}
\end{equation}
where $\tensor{\Xi}(\vec{r},\vec{r}\,',\vec{r}\,'',\vec{r}\,''')\equiv
\tensor{\Xi}(\vec{r},\vec{r}\,',\vec{r}\,'',\vec{r}\,''';\omega)$ is
the nonlocal third order conductivity tensor. The three-dimensional
sum-product operator ``$\vdots$'' is here meant to be interpreted for
the $i$'th element of the third order current density in the following
way:
\begin{eqnarray}
 \left[\tensor{\Xi}(\vec{r},\vec{r}\,',\vec{r}\,'',\vec{r}\,''')\vdots
 \,\vec{A}(\vec{r}\,''')\vec{A}(\vec{r}\,'')\vec{A}^{*}(\vec{r}\,')\right]_{i}
 =
 \sum_{jkh}{\Xi}_{ijkh}(\vec{r},\vec{r}\,',\vec{r}\,'',\vec{r}\,''')
 {A}_{h}(\vec{r}\,'''){A}_{k}(\vec{r}\,''){A}^{*}_{j}(\vec{r}\,').
\nonumber\\
\label{eq:sumprod3}
\end{eqnarray}
By inserting Eqs.~(\ref{eq:Jr}) and (\ref{eq:J3r}), with
$\vec{A}_{\rm{PC}}=\vec{E}_{\rm{PC}}/({\rm{i}}\omega)$, into
Eq.~(\ref{eq:Wave2}) the loop for the phase conjugated field is
closed.

\section{The phase conjugated field}\label{sec:field}
From the outset we assume that the parametric approximation can be
adopted, i.e., we assume that the generated phase conjugated field does
not affect the dynamics of the pump and signal fields. In the present
case, where the phase conjugated field originates mainly in evanescent
modes or from a quantum well, the interaction volume is small and the
magnitude of the phase conjugated field thus very limited so that one
may expect the parametric approximation to be quite good. The inherent
spatial nonlocality of the processes which underlies the microscopic
calculation of the local fields and currents is crucial and must be
kept throughout the following analysis.

Above we used the microscopic Maxwell-Lorentz equations to establish
a wave equation [Eq.~(\ref{eq:Wave2})] for the phase conjugated
electric field. Since this equation holds not only inside the phase
conjugator but also in the medium possibly in contact with the phase
conjugator, it is adequate to divide the linear part of the induced
current density into two, i.e.,
\begin{equation}
 \vec{J}(\vec{r}\,)=\vec{J}_{\rm{cont}}(\vec{r};\omega)
 +\vec{J}_{\rm{PC}}^{\,(1)}(\vec{r};\omega),
\end{equation}
where $\vec{J}_{\rm{cont}}(\vec{r};\omega)$ is the linear current
density of the medium in contact (cont) with the phase conjugator, and
$\vec{J}_{\rm{PC}}^{\,(1)}(\vec{r};\omega)$ is the linear current
density of the phase conjugator. In setting up the above-mentioned
equation we have implicitly assumed that there is no (significant)
electronic overlap between the phase conjugator and the contact
medium. The two electron distributions can still be
electromagnetically coupled, of course. In the quantum well case,
$\vec{J}_{\rm{cont}}(\vec{r};\omega)$ is to be identified as the
current density induced in the (assumed linear) response of the
substrate. To deal with the evanescent response of a (semiinfinite)
phase conjugator one just puts
$\vec{J}_{\rm{cont}}(\vec{r};\omega)=\vec{0}$.

Instead of proceeding directly with the differential equation
[Eq.~(\ref{eq:Wave2})] for the phase conjugated local field we convert
it into an integral relation between the phase conjugated electric
field and the prevailing current density, namely
\begin{equation}
 \vec{E}_{\rm{PC}}(\vec{r};\omega)=\vec{E}^{\rm{ext}}_{\rm{PC}}(\vec{r};\omega)
 -{\rm{i}}\mu_{0}\omega
 \int\tensor{G}_{0}(\vec{r},\vec{r}\,';\omega)
 \cdot\left[\vec{J}_{\rm{cont}}(\vec{r}\,';\omega)
 +\vec{J}_{\rm{PC}}^{\,(1)}(\vec{r}\,';\omega)\right]d^3r',
\label{eq:EPCrw}
\end{equation}
where $\vec{E}^{\rm{ext}}_{\rm{PC}}(\vec{r};\omega)$ is the so-called
external (ext) field driving the phase conjugation process, and
$\tensor{G}_{0}(\vec{r},\vec{r}\,';\omega)$ is the electromagnetic
vacuum propagator. Instead of proceeding with Eq.~(\ref{eq:EPCrw}) as
it stands, if possible, it is often advantageous to eliminate the
current density of the contact medium in favour of a so-called
pseudo-vacuum (or contact-medium) propagator,
$\tensor{G}(\vec{r},\vec{r}\,';\omega)$. Doing this, one obtains
\begin{equation}
 \vec{E}_{\rm{PC}}(\vec{r};\omega)=\vec{E}^{\rm{B}}_{\rm{PC}}(\vec{r};\omega)
 -{\rm{i}}\mu_{0}\omega
 \int\tensor{G}(\vec{r},\vec{r}\,';\omega)
 \cdot\vec{J}_{\rm{PC}}^{\,(1)}(\vec{r}\,';\omega)d^3r',
\label{eq:EPCrwB}
\end{equation}
where $\vec{E}^{\rm{B}}_{\rm{PC}}(\vec{r};\omega)$ is the so-called
background (B) response of the phase conjugator. The background field
is effectively the field driving the phase conjugated response. From a
knowledge of the nonlinear part, $\vec{J}_{-\omega}^{\,(3)}(\vec{r}\,)$,
of the current density of the phase conjugator, the background field
can be calculated from the integral relation
\begin{equation}
 \vec{E}_{\rm{PC}}^{\rm{B}}(\vec{r};\omega)=-{\rm{i}}\mu_0\omega\int
 \tensor{G}(\vec{r},\vec{r}\,';\omega)\cdot
 \vec{J}_{-\omega}^{\,(3)}(\vec{r}\,')d^3r'.
\label{eq:EPCB}
\end{equation}
In the parametric approximation adopted here the background field can
be considered as a prescribed quantity. By inserting the linear
constitutive equation 
\begin{equation}
 \vec{J}_{\rm{PC}}^{\,(1)}(\vec{r};\omega)=
 \int\stensor{\sigma}(\vec{r},\vec{r}\,';\omega)\cdot
 \vec{E}_{\rm{PC}}(\vec{r}\,';\omega)d^3r'
\end{equation}
into Eq.~(\ref{eq:EPCrwB}) one obtains the following integral equation
for the phase conjugated field:
\begin{equation}
 \vec{E}_{\rm{PC}}(\vec{r};\omega)=\vec{E}_{\rm{PC}}^{\rm{B}}(\vec{r};\omega)
 -{\rm{i}}\mu_{0}\omega\int\int\tensor{G}(\vec{r},\vec{r}^{\,\prime\prime};\omega)
 \cdot\stensor{\sigma}(\vec{r}\,'',\vec{r}\,';\omega)
 \cdot\vec{E}_{\rm{PC}}(\vec{r}\,';\omega)d^3r''d^3r'.
\label{eq:Inte}
\end{equation}
The formal solution of this equation is given by
\begin{equation}
 \vec{E}_{\rm{PC}}(\vec{r};\omega)=
 \int\tensor{\Gamma}(\vec{r},\vec{r}\,';\omega)
 \cdot\vec{E}_{\rm{PC}}^{\rm{B}}(\vec{r}\,';\omega)d^3r',
\label{eq:Sol}
\end{equation}
where the nonlocal field-field response tensor
$\tensor{\Gamma}(\vec{r},\vec{r}\,';\omega)$ is to be derived from the
dyadic integral equation
\begin{equation}
 \tensor{\Gamma}(\vec{r},\vec{r}\,';\omega)=\tensor{\openone}\delta(\vec{r}-\vec{r}\,')
 +\int\tensor{K}(\vec{r},\vec{r}\,'';\omega)
 \cdot\tensor{\Gamma}(\vec{r}\,'',\vec{r}\,';\omega)d^3r''.
\label{eq:Gamma}
\end{equation}
In Eq.~(\ref{eq:Gamma}) the tensor
\begin{equation}
 \tensor{K}(\vec{r},\vec{r}\,'';\omega)=-{\rm{i}}\mu_{0}\omega
 \int\tensor{G}(\vec{r},\vec{r}\,';\omega)
 \cdot\stensor{\sigma}(\vec{r}\,',\vec{r}\,'';\omega)d^3r'
\label{eq:Kernel}
\end{equation}
is the kernel of the integral equation in Eq.~(\ref{eq:Inte}).  This
kernel formally is identical to the one playing a prominent role in
the electrodynamics of mesoscopic media and small particles [see
\citeN{Keller:96:1}, section 4].

By inserting Eq.~(\ref{eq:EPCB}) into Eq.~(\ref{eq:Sol}) and
thereafter making use of Eq.~(\ref{eq:J3r}), the phase conjugated
field may in principle be calculated from known quantities. In
practice it is not so easy, since the integral equation in
Eq.~(\ref{eq:Inte}) for the phase conjugated field in general is too
difficult to handle numerically even if rather simple linear
conductivity response tensors are used, the reason being the inherent
three-dimensional ($\vec{r}\,$) nature of the problem. One therefore
has to resort to one sort of approximation or another. Just as in
other linear and nonlinear studies of mesoscopic media, or media with
a small interaction volume, a tractable problem is obtained if the
medium in question possesses translational invariance in two
directions as discussed in Part~\ref{part:III}.

\chapter{Single-electron current density response}\label{ch:5}
In this chapter the Liouville equation of motion for the single-body
density matrix operator is used together with the single-particle
Hamiltonian to establish a more general quantum mechanical expression
for the third-order current density than those hitherto found in the
literature. The generalisation is of significant importance for the
theory of near-field phase conjugation and for DFWM in mesoscopic
films. Following the derivation of the linear and the DFWM responses,
we end this chapter by a discussion of the underlying physical
processes.

\section{Density matrix operator approach}
The starting point for this calculation is the Liouville equation of
motion for the single-body density matrix operator $\rho$, i.e.,
\begin{equation}
 {\rm{i}}\hbar{\partial\rho\over\partial{}t}=\left[{\cal{H}},\rho\right].
\end{equation}
In the equation above, the single-particle Hamiltonian ${\cal{H}}$
appearing in the commutator $\left[{\cal{H}},\rho\right]$ in the
present description is given by
\begin{equation}
 {\cal{H}}={\cal{H}}_{\,0}+{\cal{H}}_{\,\rm{R}}+{\cal{H}}_{\,0}^{(2)}
 +{1\over2}\sum_{\alpha=1}^{2}\left(
 {\cal{H}}_{-\alpha\omega}^{(\alpha)}e^{-{\rm{i}}\alpha\omega{}t}
 +\mbox{H.a.}\right),
\end{equation}
where ${\cal{H}}_{0}$ is the Hamiltonian operator for the electron in
the material when the perturbing optical field is absent,
${\cal{H}}^{(1)}$ is the interaction Hamiltonian of first order in the
vector potential $\vec{A}(\vec{r}\,)$, ${\cal{H}}^{(2)}$ is the
interaction Hamiltonian of second order in $\vec{A}(\vec{r}\,)$,
${\cal{H}}_{\,\rm{R}}$ represents the irreversible coupling to the
``surroundings'', and ``H.a.'' denotes the Hermitian adjoint. Although
the spin and spin-orbit dynamics may be included in the formalism in a
reasonably simple fashion we have omitted to do so because spin
effects are judged to be significant only for nonlinear phenomena of
even order. Hence
\begin{eqnarray}
{\cal{H}}_{\,0}&=&{1\over{2m_{e}}}\vec{p}\cdot\vec{p}+V(\vec{r}\,),
\label{eq:H0}\\
{\cal{H}}_{-\omega}^{(1)}=\left({\cal{H}}_{\,\omega}^{(1)}\right)^{\dag}&=&
 {e\over2m_{e}}\left(\vec{p}\cdot\vec{A}(\vec{r}\,)+
 \vec{A}(\vec{r}\,)\cdot\vec{p}\,\right),
\label{eq:H-+1}\\
{\cal{H}}_{-2\omega}^{(2)}=\left({\cal{H}}_{\,2\omega}^{(2)}\right)^{\dag}&=&
 {e^2\over4m_{e}}\vec{A}(\vec{r}\,)\cdot\vec{A}(\vec{r}\,),
\label{eq:H-+2}\\
{\cal{H}}_{\,0}^{(2)}&=&
 {e^2\over4m_{e}}\vec{A}(\vec{r}\,)\cdot\vec{A}^{*}(\vec{r}\,),
\label{eq:H02}
\end{eqnarray}
where $\dag$ stands for Hermitian adjugation, $V(\vec{r}\,)$ is the
scalar potential of the field-un\-per\-tur\-bed Schr{\"o}dinger equation,
$\vec{p}=-{\rm{i}}\hbar\vec{\nabla}$ denotes the momentum operator, $m_{e}$ is
the mass of the electron, and $-e$ is its electric charge.

As often is the practice in optics we assume that the irreversible
coupling to the surrounding reservoir can be described using a
phenomenological relaxation-time ansatz in the Liouville equation, so
that
\begin{equation}
 {1\over{}{\rm{i}}\hbar}\left[{\cal{H}}_{\,\rm{R}},\rho_{nm}\right]=
 {\rho_{nm}^{(0)}-\rho_{nm}\over\tau_{nm}},\qquad n\neq{}m, 
\end{equation}
$\rho_{nm}^{(0)}$ being the $nm$'th element of the thermal equilibrium
density matrix operator, and $\tau_{nm}$ the associated relaxation
time.

In the present harmonic analysis we also use a combined Fourier and
power series expansion of the density matrix operator, namely
\begin{equation}
 {\rho}={1\over2}\sum_{\alpha=0}^{\infty}\sum_{\beta=0}^{\infty}\left(
 \rho_{-\alpha\omega}^{(\alpha+2\beta)}e^{-{\rm{i}}\alpha\omega{}t}
 +\mbox{H.a.}\right),
\label{eq:rho}
\end{equation}
where $\alpha$ and $\beta$ are integers, as before. The density matrix
operator is Hermitian, i.e.,
$(\rho_{-\alpha\omega}^{(\alpha+2\beta)})^{\dag}=
\rho_{\alpha\omega}^{(\alpha+2\beta)}$, and we solve the Liouville
equation of motion in the usual iterative manner.

To determine the conductivity response tensors,
$\stensor{\sigma}(\vec{r},\vec{r}\,')$ and
$\tensor{\Xi}(\vec{r},\vec{r}\,',\vec{r}\,'',\vec{r}\,''')$,
appropriate for describing the phase conjugation process, we consider
the ensemble average $\vec{J}(\vec{r},t)$ of the microscopic
single-body current-density operator $\vec{j}(\vec{r},t)$. This
ensemble average is obtained as the trace of ${\rho}\vec{j}$,
carried out in the usual manner as a quantum mechanical double sum
over states, i.e.,
\begin{equation}
 \vec{J}(\vec{r},t)=\mbox{Tr}\left\{{\rho}\vec{j}\,\right\}
 \equiv\sum_{nm}\rho_{nm}\vec{j}_{mn}.
\label{eq:Trrhoj}
\end{equation}
In Eq.~(\ref{eq:Trrhoj}) and hereafter the $ab$'th matrix element of a
single-body operator ${\cal{O}}$ as usual is denoted by
${\cal{O}}_{ab}=\langle{a}|{\cal{O}}|{b}\rangle$. In the absence of
spin effects the microscopic current-density operator is given by
\cite{Bloembergen:65:1}
\begin{equation}
 \vec{j}(\vec{r},t)=\vec{j}^{\,(0)}(\vec{r}\,)
 +{1\over2}\left(\vec{j}_{-\omega}^{\,(1)}e^{-{\rm{i}}\omega{t}}
 +\mbox{H.a.}\right),
\label{eq:jrt}
\end{equation}
where
\begin{eqnarray}
 \vec{j}^{\,(0)}(\vec{r}\,)&=&
  -{e\over2m_{e}}\Bigl(\vec{p}(\vec{r}_e)\delta(\vec{r}-\vec{r}_e)
  +\delta(\vec{r}-\vec{r}_e)\vec{p}(\vec{r}_e)\Bigr)
\label{eq:j(0)r}\\
 \vec{j}_{-\omega}^{\,(1)}&=&
  -{e^2\over{}m_{e}}\vec{A}(\vec{r}_e)\delta(\vec{r}-\vec{r}_e).
\label{eq:j(1)r}
\end{eqnarray}

\section{Linear response}
Because of its usefulness for a subsequent comparison to the forced
DFWM current density we first present the well known result for the
linear response (\citeNP{Feibelman:75:1},
\citeyearNP{Feibelman:82:1}). Thus, by using the expressions for the
current density [Eq.~(\ref{eq:jrt})] and density matrix
[Eq.~(\ref{eq:rho})] operators it is realised that the linear current
density is to be obtained from
\begin{equation}
 \vec{J}_{-\omega}^{\,(1)}(\vec{r}\,)=
 \mbox{Tr}\left\{\rho^{(0)}\vec{j}^{\,(1)}_{-\omega}\right\}
 +\mbox{Tr}\left\{\rho^{(1)}_{-\omega}\vec{j}^{\,(0)}\right\}.
\label{eq:J1Tr}
\end{equation}
In explicit form the two traces are
\begin{eqnarray}
 \mbox{Tr}\left\{\rho^{(0)}\vec{j}^{\,(1)}_{-\omega}\right\}&=&
 \sum_{n}f_{n}\vec{j}^{\,(1)}_{-\omega,nn}.
\label{eq:J1-1}
\\
 \mbox{Tr}\left\{\rho^{(1)}_{-\omega}\vec{j}^{\,(0)}\right\}&=&
 \sum_{nm}{f_{n}-f_{m}\over\hbar}{{\cal{H}}^{(1)}_{-\omega,nm}
  \over\tilde{\omega}_{nm}-\omega}\vec{j}^{\,(0)}_{mn}.
\label{eq:J1-2}
\end{eqnarray}
In the equations above, we have introduced the complex cyclic
transition frequency
$\tilde{\omega}_{nm}=\omega_{nm}-{\rm{i}}\tau_{nm}^{-1}$ between
states $n$ and $m$. The respective energies ${\cal{E}}_{\,n}$ and
${\cal{E}}_{\,m}$ of these states appear in the usual transition
frequency
$\omega_{nm}=\left({\cal{E}}_{\,n}-{\cal{E}}_{\,m}\right)/\hbar$. The
quantity
\begin{equation}
 f_{a}=\left[1+\exp\left({{\cal{E}}_{\,a}-\mu\over{}k_{B}T}\right)\right]^{-1}
\label{eq:fn}
\end{equation}
denotes the Fermi--Dirac distribution function for state $a$
($a\in\{m,n\}$ above), $k_{B}$ being the Boltzmann constant, $\mu$ the
chemical potential of the electron system, and $T$ the absolute
temperature.

\section{DFWM response}
The nonlinear current density at $-\omega$, which originates in third
order effects in the electric field, and which is the driving source
for the DFWM process is given by
\begin{equation}
 \vec{J}_{-\omega}^{\,(3)}(\vec{r}\,)={1\over2}
 \mbox{Tr}\left\{{\rho}_{-2\omega}^{(2)}\vec{j}_{\omega}^{\,(1)}\right\}
 +\mbox{Tr}\left\{{\rho}_{0}^{(2)}\vec{j}_{-\omega}^{\,(1)}\right\}
 +\mbox{Tr}\left\{{\rho}_{-\omega}^{(3)}\vec{j}^{\,(0)}\right\},
\label{eq:J3Tr}
\end{equation}
as one readily realises from Eqs.~(\ref{eq:J-power}), (\ref{eq:rho}),
and (\ref{eq:jrt}). The tedious calculation of the three traces can
be carried out in a fashion similar to that used for the linear case,
finally leading to
\begin{eqnarray}
\lefteqn{
 {1\over2}\mbox{Tr}\left\{\rho_{-2\omega}^{(2)}\vec{j}_{\omega}^{\,(1)}\right\}
 =\sum_{nm}{f_{n}-f_{m}\over2\hbar}
 {{\cal{H}}_{-2\omega,nm}^{(2)}\vec{j}_{\omega,mn}^{\,(1)}
  \over\tilde{\omega}_{nm}-2\omega}
}\nonumber\\ &\quad&
 +\sum_{nmv}\left({f_{m}-f_{v}\over\tilde{\omega}_{v{}m}-\omega}
 +{f_{n}-f_{v}\over\tilde{\omega}_{nv}-\omega}\right)
 {{\cal{H}}_{-\omega,nv}^{(1)}{\cal{H}}_{-\omega,v{}m}^{(1)} 
 \vec{j}_{\omega,mn}^{(1)}
  \over4\hbar^2(\tilde{\omega}_{nm}-2\omega)},
\label{eq:J3-1} \\
\lefteqn{
 \mbox{Tr}\left\{\rho_{0}^{(2)}\vec{j}_{-\omega}^{\,(1)}\right\}
 =\sum_{nm}{f_{n}-f_{m}\over\hbar}
 {{\cal{H}}_{\,0,nm}^{(2)}\vec{j}_{-\omega,mn}^{\,(1)}
  \over\tilde{\omega}_{nm}}
+\sum_{nmv}\left\{
 \left({f_{m}-f_{v}\over\tilde{\omega}_{v{}m}-\omega}
 +{f_{n}-f_{v}\over\tilde{\omega}_{nv}+\omega}\right)
\right.}\nonumber\\ &&\times\left.\!
 {{\cal{H}}_{\,\omega,nv}^{(1)}{\cal{H}}_{-\omega,v{}m}^{(1)}
 \vec{j}_{-\omega,mn}^{\,(1)}\over4\hbar^2\tilde{\omega}_{nm}}
 +\left({f_{m}-f_{v}\over\tilde{\omega}_{v{}m}+\omega}
 +{f_{n}-f_{v}\over\tilde{\omega}_{nv}-\omega}\right)
 {{\cal{H}}_{-\omega,nv}^{(1)}{\cal{H}}_{\,\omega,v{}m}^{(1)}
 \vec{j}_{-\omega,mn}^{\,(1)}\over4\hbar^2\tilde{\omega}_{nm}}\right\},
\label{eq:J3-2} \\
\lefteqn{
 \mbox{Tr}\left\{\rho_{-\omega}^{(3)}\vec{j}^{\,(0)}\right\}
 =\sum_{nmv}{1\over2\hbar^2(\tilde\omega_{nm}-\omega)}
 \left\{\left({f_{m}-f_{v}\over2(\tilde{\omega}_{v{}m}-2\omega)}
 +{f_{n}-f_{v}\over2(\tilde{\omega}_{nv}+\omega)}\right)
 {\cal{H}}_{\,\omega,nv}^{(1)}{\cal{H}}_{-2\omega,v{}m}^{(2)}
\right.}\nonumber\\ &\quad&
 +\left({f_{n}-f_{v}\over2(\tilde{\omega}_{nv}-2\omega)}
 +{f_{m}-f_{v}\over2(\tilde{\omega}_{v{}m}+\omega)}\right)
 {\cal{H}}_{-2\omega,nv}^{(2)}{\cal{H}}_{\,\omega,v{}m}^{(1)}
 +\left({f_{m}-f_{v}\over\tilde{\omega}_{v{}m}}
 +{f_{n}-f_{v}\over\tilde{\omega}_{nv}-\omega}\right) 
\nonumber\\ &&\times\left.\!
 {\cal{H}}_{-\omega,nv}^{(1)}{\cal{H}}_{\,0,v{}m}^{(2)}
 +\left({f_{n}-f_{v}\over\tilde{\omega}_{nv}}
 +{f_{m}-f_{v}\over\tilde{\omega}_{v{}m}-\omega}\right)
 {\cal{H}}_{\,0,nv}^{(2)}{\cal{H}}_{-\omega,v{}m}^{(1)}
 \right\}\vec{j}_{mn}^{\,(0)}
\nonumber\\ &&
+\sum_{nmvl}{1\over2\hbar(\tilde\omega_{nm}-\omega)}
 \left\{
 \left[
 \left({f_{l}-f_{m}\over\tilde{\omega}_{l{}m}-\omega}
 +{f_{l}-f_{v}\over\tilde{\omega}_{v{}l}-\omega}\right)
 {1\over4\hbar^2(\tilde{\omega}_{v{}m}-2\omega)}
\right.\right.\nonumber\\ &&\left.
 +\left({f_{l}-f_{v}\over\tilde{\omega}_{vl}-\omega}
 +{f_{n}-f_{v}\over\tilde{\omega}_{nv}+\omega}\right)
 {1\over4\hbar^2\tilde{\omega}_{nl}}
 \right]
 {\cal{H}}_{\,\omega,nv}^{(1)}{\cal{H}}_{-\omega,v{}l}^{(1)}
 {\cal{H}}_{-\omega,l{}m}^{(1)}
\nonumber\\ &&
 +\left[
 \left({f_{l}-f_{m}\over\tilde{\omega}_{l{}m}-\omega}
 +{f_{l}-f_{v}\over\tilde{\omega}_{v{}l}+\omega}\right)
 {1\over4\hbar^2\tilde{\omega}_{v{}m}}
 +\left({f_{l}-f_{v}\over\tilde{\omega}_{vl}+\omega}
 +{f_{n}-f_{v}\over\tilde{\omega}_{nv}-\omega}\right)
 {1\over4\hbar^2\tilde{\omega}_{nl}}
 \right]
\nonumber\\ &&\times
 {\cal{H}}_{-\omega,nv}^{(1)}{\cal{H}}_{\,\omega,v{}l}^{(1)}
 {\cal{H}}_{-\omega,l{}m}^{(1)}
 +\left[
 \left({f_{l}-f_{m}\over\tilde{\omega}_{l{}m}+\omega}
 +{f_{l}-f_{v}\over\tilde{\omega}_{v{}l}-\omega}\right)
 {1\over4\hbar^2\tilde{\omega}_{v{}m}}
\right.\nonumber\\ &&\left.\!\left.\!
 +\left({f_{l}-f_{v}\over\tilde{\omega}_{vl}-\omega}
 +{f_{n}-f_{v}\over\tilde{\omega}_{nv}-\omega}\right)
 {1\over4\hbar^2(\tilde{\omega}_{nl}-2\omega)}
 \right]
 {\cal{H}}_{-\omega,nv}^{(1)}{\cal{H}}_{-\omega,v{}l}^{(1)}
 {\cal{H}}_{\,\omega,l{}m}^{(1)}
 \right\}\vec{j}_{mn}^{\,(0)}.
\label{eq:J3-3}
\end{eqnarray}
Though quite complicated in its appearence the expression for the
driving current density of the DFWM process is needed in order to
understand the near-field phase conjugation process from a general
point of view.\footnote{With respect to the result published in
  \citeN{Andersen:97:1} the last sum in Eq.~(\ref{eq:J3-3}) above is
  written in a more compact form than in Eq.~(22) of
  \citeN{Andersen:97:1}. The compact form in Eq.~(\ref{eq:J3-3}) is
  obtained by exchanging indices $v$ and $l$ in the last three terms
  of \citeN{Andersen:97:1}, Eq.~(22). As a consequence of this, the
  same difference occur between Eqs.~(\ref{eq:XiAGrz}) and
  (\ref{eq:XiAG}) and Eqs.~(34) and (51) of \citeN{Andersen:97:1},
  respectively.} Special scattering configurations of course can lead
to analytical simplifications of the general result.  The above result
also enables us to establish a microscopic theory for DFWM in quantum
wells and thin films as described in Parts~\ref{part:III},
\ref{part:IV}, and \ref{part:V}.

\section{Physical processes underlying the current densities}
To gain insight into the physics underlying the nonlinear constitutive
equation, given implicitly in Eqs.~(\ref{eq:J3-1})--(\ref{eq:J3-3}), we
next discuss the processes connecting in a nonlocal fashion the
current density at a given point in space to the field points of the
surroundings. To facilitate the understanding of the nonlinear
response we start by a brief summary of the linear response.

\subsection{Linear part}
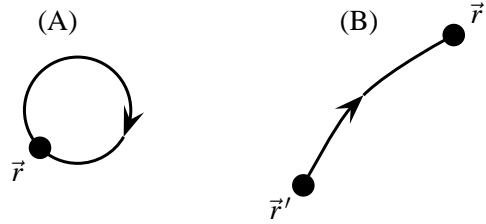
\begin{figure}[tb]
\setlength{\unitlength}{1mm}
\noindent
\begin{center}
\psset{unit=1mm}
\begin{pspicture}(0,0)(80,30)
\put(0,0){
 \put(15,25){\makebox(0,5)[br]{(A)}}
 \psarcn[linewidth=0.4,linestyle=solid,arrowscale=1.5]{->}(15,15){7.07}{-30}{330}
 \qdisk(10,10){1.5}
 \put(7,7){\makebox(0,0)[c]{$\vec{r}$}}
}
\put(40,0){
 \put(15,25){\makebox(0,5)[br]{(B)}}
 \qdisk(25,25){1.5}
 \qdisk(5,5){1.5}
 \psecurve[linewidth=0.4,linestyle=solid]{-}(5,5)(13,17)(25,25)(25,25)
 \psecurve[linewidth=0.4,linestyle=solid,arrowscale=1.5]{->}(5,5)(5,5)(13,17)(25,25)
 \put(2,2){\makebox(0,0)[c]{$\vec{r}\,'$}}
 \put(28,28){\makebox(0,0)[c]{$\vec{r}$}}
}
\end{pspicture}
\end{center}
\caption[Schematic diagrams of the two processes described by the
linear response tensor]{Schematic illustration of the two processes
  described by the linear response tensor. The process in diagram A
  is purely diamagnetic whereas the process in diagram B is purely
  paramagnetic.\label{fig:2Tr}}
\end{figure}

The electrodynamic coupling connecting a source point for the field to
an observation point for the current density associated to each of the
two linear processes underlying Eqs.~(\ref{eq:J1-1}) and
(\ref{eq:J1-2}) is adequately illustrated in diagrammatic form as
shown in Fig.~\ref{fig:2Tr}.

Hence, Fig.~\ref{fig:2Tr}.A represents a picture of the well known
diamagnetic process originating in the quantity
$\vec{j}^{\,(1)}_{-\omega,nn}$ in Eq.~(\ref{eq:J1-1}).  In this
process, a photon is absorbed at the observation point $\vec{r}$ for
the current density. Fig.~\ref{fig:2Tr}.B is a picture of the
paramagnetic process stemming from the term
${\cal{H}}^{(1)}_{-\omega,nm}$ appearing in Eq.~(\ref{eq:J1-2}). In
this case, a photon is absorbed at space point $\vec{r}\,'$, and
observation takes place at $\vec{r}$.

\subsection{Nonlinear part}
\begin{figure}[tb]
\setlength{\unitlength}{0.76mm}
\begin{center}
\psset{unit=0.76mm}
\begin{pspicture}(-4,0)(170,80)
\put(0,40){ 
 \put(15,25){\makebox(0,5)[br]{(A)}}
 \psarcn[linewidth=0.4,linestyle=dotted,dotsep=1,arrowscale=1.5]{->}(30,30){7.07}{-30}{330}
 \qdisk(25,25){1.5}
 \qdisk(5,5){1.5}
 \pscurve[linewidth=0.4,linestyle=solid]{-}(6,5)(6,5)(17,13)(26,25)
 \psecurve[linewidth=0.4,linestyle=solid,arrowscale=1.5]{->}(6,5)(6,5)(17,13)(26,25)
 \pscurve[linewidth=0.4,linestyle=solid]{-}(5,6)(5,6)(13,17)(25,26)
 \psecurve[linewidth=0.4,linestyle=solid,arrowscale=1.5]{->}(5,6)(5,6)(13,17)(25,26)
 \put(1.5,2){\makebox(0,0)[c]{$\vec{r}\,''$}}
 \put(28,28){\makebox(0,0)[c]{$\vec{r}$}}
}
\put(39,40){
 \put(15,25){\makebox(0,5)[br]{(B)}}
 \psarcn[linewidth=0.4,linestyle=dotted,dotsep=1,arrowscale=1.5]{->}(30,30){7.07}{-30}{330}
 \qdisk(25,25){1.5}
 \qdisk(5,5){1.5}
 \qdisk(38,5){1.5}
 \pscurve[linewidth=0.4,linestyle=solid]{-}(5,5)(5,5)(14,16)(25,25)
 \psecurve[linewidth=0.4,linestyle=solid,arrowscale=1.5]{->}(5,5)(5,5)(14,16)(25,25)
 \pscurve[linewidth=0.4,linestyle=solid]{-}(38,5)(38,5)(29,15)(25,25)
 \psecurve[linewidth=0.4,linestyle=solid,arrowscale=1.5]{->}(38,5)(38,5)(29,15)(25,25)
 \put(1.5,2){\makebox(0,0)[c]{$\vec{r}\,''$}}
 \put(28,28){\makebox(0,0)[c]{$\vec{r}$}}
 \put(34,2){\makebox(0,0)[c]{$\vec{r}\,'''$}}
}
\put(88,40){
 \put(15,25){\makebox(0,5)[br]{(C)}}
 \psarcn[linewidth=0.4,linestyle=solid,arrowscale=1.5]{->}(30,30){7.07}{-30}{330}
 \qdisk(25,25){1.5}
 \qdisk(5,5){1.5}
 \psecurve[linewidth=0.4,linestyle=dotted,dotsep=1]{-}(6,5)(17,13)(26,25)(26,25)
 \psecurve[linewidth=0.4,linestyle=dotted,dotsep=1,arrowscale=1.5]{->}(6,5)(6,5)(17,13)(26,25)
 \pscurve[linewidth=0.4,linestyle=solid]{-}(5,6)(5,6)(13,17)(25,26)
 \psecurve[linewidth=0.4,linestyle=solid,arrowscale=1.5]{->}(5,6)(5,6)(13,17)(25,26)
 \put(2,2){\makebox(0,0)[c]{$\vec{r}\,'$}}
 \put(28,28){\makebox(0,0)[c]{$\vec{r}$}}
}
\put(127,40){
 \put(15,25){\makebox(0,5)[br]{(D)}}
 \psarcn[linewidth=0.4,linestyle=solid,arrowscale=1.5]{->}(30,30){7.07}{-30}{330}
 \qdisk(25,25){1.5} \qdisk(5,5){1.5} \qdisk(38,5){1.5}
 \pscurve[linewidth=0.4,linestyle=solid]{-}(5,5)(5,5)(14,16)(25,25)
 \psecurve[linewidth=0.4,linestyle=solid,arrowscale=1.5]{->}(5,5)(5,5)(14,16)(25,25)
 \psecurve[linewidth=0.4,linestyle=dotted,dotsep=1]{-}(38,5)(29,15)(25,25)(25,25)
 \psecurve[linewidth=0.4,linestyle=dotted,dotsep=1,arrowscale=1.5]{->}(38,5)(38,5)(29,15)(25,25)
 \put(2,2){\makebox(0,0)[c]{$\vec{r}\,'$}}
 \put(28,28){\makebox(0,0)[c]{$\vec{r}$}}
 \put(34.5,2){\makebox(0,0)[c]{$\vec{r}\,''$}}
}
\put(0,0){
 \put(15,25){\makebox(0,5)[br]{(E)}}
 \qdisk(25,25){1.5} \qdisk(5,5){1.5} \qdisk(45,5){1.5}
 \pscurve[linewidth=0.4,linestyle=solid]{-}(6,5)(6,5)(17,13)(26,25)
 \psecurve[linewidth=0.4,linestyle=solid,arrowscale=1.5]{->}(6,5)(6,5)(17,13)(26,25)
 \pscurve[linewidth=0.4,linestyle=solid]{-}(5,6)(5,6)(13,17)(25,26)
 \psecurve[linewidth=0.4,linestyle=solid,arrowscale=1.5]{->}(5,6)(5,6)(13,17)(25,26)
 \psecurve[linewidth=0.4,linestyle=dotted,dotsep=1]{-}(45,5)(34,14)(25,25)(25,25)
 \psecurve[linewidth=0.4,linestyle=dotted,dotsep=1,arrowscale=1.5]{->}(45,5)(45,5)(34,14)(25,25)
 \put(1.5,2){\makebox(0,0)[c]{$\vec{r}\,''$}}
 \put(28,28){\makebox(0,0)[c]{$\vec{r}$}}
 \put(42,2){\makebox(0,0)[c]{$\vec{r}\,'$}}
}
\put(60,0){
 \put(15,25){\makebox(0,5)[br]{(F)}}
 \qdisk(25,25){1.5} \qdisk(5,5){1.5} \qdisk(45,5){1.5}
 \psecurve[linewidth=0.4,linestyle=dotted,dotsep=1]{-}(6,5)(17,13)(26,25)(26,25)
 \psecurve[linewidth=0.4,linestyle=dotted,dotsep=1,arrowscale=1.5]{->}(6,5)(6,5)(17,13)(26,25)
 \pscurve[linewidth=0.4,linestyle=solid]{-}(5,6)(5,6)(13,17)(25,26)
 \psecurve[linewidth=0.4,linestyle=solid,arrowscale=1.5]{->}(5,6)(5,6)(13,17)(25,26)
 \pscurve[linewidth=0.4,linestyle=solid]{-}(45,5)(45,5)(34,14)(25,25)
 \psecurve[linewidth=0.4,linestyle=solid,arrowscale=1.5]{->}(45,5)(45,5)(34,14)(25,25)
 \put(2,2){\makebox(0,0)[c]{$\vec{r}\,'$}}
 \put(28,28){\makebox(0,0)[c]{$\vec{r}$}}
 \put(41.5,2){\makebox(0,0)[c]{$\vec{r}\,''$}}
}
\put(120,0){
 \put(15,25){\makebox(0,5)[br]{(G)}}
 \qdisk(25,30){1.5} \qdisk(5,10){1.5} \qdisk(45,10){1.5} \qdisk(25,5){1.5}
 \pscurve[linewidth=0.4,linestyle=solid]{-}(25,5)(25,5)(26,18)(25,30)
 \psecurve[linewidth=0.4,linestyle=solid,arrowscale=1.5]{->}(25,5)(25,5)(26,18)(25,30)
 \pscurve[linewidth=0.4,linestyle=solid]{-}(5,10)(5,10)(13,22)(25,30)
 \psecurve[linewidth=0.4,linestyle=solid,arrowscale=1.5]{->}(5,10)(5,10)(13,22)(25,30)
 \psecurve[linewidth=0.4,linestyle=dotted,dotsep=1]{-}(45,10)(37,22)(25,30)(25,30)
 \psecurve[linewidth=0.4,linestyle=dotted,dotsep=1,arrowscale=1.5]{->}(45,10)(45,10)(37,22)(25,30)
 \put(2,7){\makebox(0,0)[c]{$\vec{r}\,'$}}
 \put(21.5,2){\makebox(0,0)[c]{$\vec{r}\,''$}}
 \put(41,7){\makebox(0,0)[c]{$\vec{r}\,'''$}}
 \put(28,32){\makebox(0,0)[c]{$\vec{r}$}}
}
\end{pspicture}
\end{center}
\caption[Process diagrams for the seven processes in the DFWM
response]{Schematic illustration of the processes underlying the DFWM
  response tensor. In this illustration, the solid paths results from
  the $-\omega$ terms, and the dotted paths from the $+\omega$ terms.
  The processes in diagrams A--D include both diamagnetic effects
  (drawn as circles) and paramagnetic effects (lines). The processes
  in diagrams E--G are purely paramagnetic. The standard theory for
  the DFWM susceptibility (conductivity) is obtained from diagram G in
  the local limit.\label{fig:3Tr}}
\end{figure}
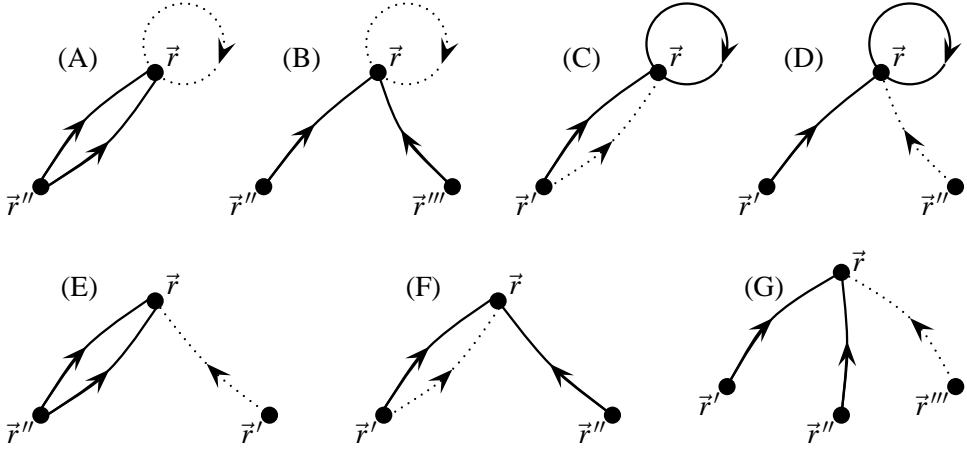

In the DFWM process, the coupling between the three source points for the
field and the observation point for the current density, described in
Eqs.~(\ref{eq:J3-1})--(\ref{eq:J3-3}), can be pictured in diagrammatic
form as shown in Fig.~\ref{fig:3Tr}.

Hence, in Fig.~\ref{fig:3Tr}.A the mixing process contained in the
product ${\cal{H}}_{-2\omega,nm}^{(2)} \vec{j}_{\omega,mn}^{\,(1)}$ in
Eq.~(\ref{eq:J3-1}) is illustrated. Here, two photons are {\it
  simultaneously}\/ absorbed at space point $\vec{r}\,''$, and one
photon is emitted at the point of observation $\vec{r}$ for the
current density. Fig.~\ref{fig:3Tr}.B pictures the other mixing
process in Eq.~(\ref{eq:J3-1}), namely that associated with the
product ${\cal{H}}_{-\omega,nv}^{(1)}
{\cal{H}}_{-\omega,v{}m}^{(1)}\vec{j}_{\omega,mn}^{\,(1)}$. In this
process, one photon is absorbed at $\vec{r}\,''$, another at
$\vec{r}\,'''$, and the last one is emitted at $\vec{r}$.
Fig.~\ref{fig:3Tr}.C gives a view of the mixing process from the terms
in Eq.~(\ref{eq:J3-2}) containing the product
${\cal{H}}_{\,0,nm}^{(2)} \vec{j}_{-\omega,mn}^{\,(1)}$. In this case
a photon is absorbed and another is emitted {\it simultaneously}\/ at
space point $\vec{r}\,'$, and the third photon is absorbed at
$\vec{r}$.  Fig.~\ref{fig:3Tr}.D shows the other type of mixing
process occuring in Eq.~(\ref{eq:J3-2}).  This process is described by
the products ${\cal{H}}_{\,\omega,nv}^{(1)}
{\cal{H}}_{-\omega,v{}m}^{(1)}\vec{j}_{-\omega,mn}^{\,(1)}$ and
${\cal{H}}_{-\omega,nv}^{(1)}{\cal{H}}_{\,\omega,v{}m}^{(1)}
\vec{j}_{-\omega,mn}^{\,(1)}$. Here, photons are absorbed at
$\vec{r}\,'$ and at the point of observation $\vec{r}$, while a photon
is emitted at $\vec{r}\,''$. Fig.~\ref{fig:3Tr}.E represents the
diagram for the mixing process appearing in the terms containing the
product ${\cal{H}}_{\,\omega,nv}^{(1)}
{\cal{H}}_{-2\omega,v{}m}^{(2)}$ (and the equivalent product
${\cal{H}}_{-2\omega,v{}m}^{(2)}{\cal{H}}_{\,\omega,nv}^{(1)}$) in
Eq.~(\ref{eq:J3-3}). Here two photons are {\it simultanously}\/
absorbed at $\vec{r}\,''$, and one is emitted at $\vec{r}\,'$.
Fig.~\ref{fig:3Tr}.F is the diagrammatic representation of the terms
containing the product ${\cal{H}}_{\,0,nv}^{(2)}
{\cal{H}}_{-\omega,v{}m}^{(1)}$ (and the equivalent one
${\cal{H}}_{-\omega,nv}^{(1)}{\cal{H}}_{\,0,v{}m}^{(2)}$) in
Eq.~(\ref{eq:J3-3}). In these terms a photon is absorbed at
$\vec{r}\,'$ and at the {\it same time}\/ one is emitted from there.
The last photon is absorbed at $\vec{r}\,''$. Finally,
Fig.~\ref{fig:3Tr}.G gives a picture of one of six equivalent products
of the last type appearing in Eq.~(\ref{eq:J3-3}). These are of the
form ${\cal{H}}_{\,\omega,nv}^{(1)}
{\cal{H}}_{-\omega,vl}^{(1)}{\cal{H}}_{-\omega,l{}m}^{(1)}$ or
equivalent forms (all six possible permutations of one
``$\omega$''-term and two ``$-\omega$''-terms). Here, a photon is
absorbed at $\vec{r}\,'$, another at $\vec{r}\,''$, and the last
photon is emitted at $\vec{r}\,'''$.

At this stage it is fruitful to compare the nonlocal result for the
DFWM current density, shown in diagrammatic form in
Fig.~\ref{fig:3Tr}, with the commonly used standard (textbook) result.
In the standard description all diamagnetic effects are neglected from
the outset. The diamagnetic process is hidden in the diagrams
containing a closed loop, cf. Fig.~\ref{fig:2Tr}. This means that all
the processes depicted in Figs.~\ref{fig:3Tr}.A--\ref{fig:3Tr}.D are
absent in the standard description. Omission of diamagnetic effects in
the nonlinear optics of quantum wells and in mesoscopic near-field
optics is known to be dangerous \cite{Keller:96:1}, and thus we {\it
  cannot}\/ omit these terms here. We shall substantiate on this point
later. Also the interaction channels given by the diagrams in
Figs.~\ref{fig:3Tr}.E and \ref{fig:3Tr}.F are absent in textbook
formulations.  This is so because simultaneous two-photon processes
originating in the $\vec{A}\cdot\vec{A}$ part of the interaction
Hamiltonian are left out from the beginning. These processes however
are known to be important in mesoscopic electrodynamics and can not be
omitted {\it a priori}.\label{Local} In the local limit the result
given by the diagram in Fig.~\ref{fig:3Tr}.G is identical to the
$\vec{r}\cdot\vec{E}$ dipolar interaction Hamiltonian, since a unitary
transformation of the form
$S=\exp(-{\rm{i}}e\vec{A}(t)\cdot\vec{r}/\hbar)$ performed on the wave
functions and the minimal coupling Hamiltonian would display the
equivalence of the two formalisms \cite{Ackerhalt:84:1,Milonni:89:1}.

\newpage
\thispagestyle{plain}
\newpage

\chapter{Conductivity tensors for DFWM response}\label{ch:6}
In the preceding two chapters, we have found expressions for the phase
conjugated field and the single-electron current density response. In
the present chapter the connection between the single-electron current
densities [Eqs.~(\ref{eq:J1Tr}) and (\ref{eq:J3Tr})] and their related
conductivity tensors [Eqs.~(\ref{eq:Jr}) and (\ref{eq:J3r})] is
established. First, the matrix elements of the Hamiltonian and the
current density operator are written in terms of the vector
potential. Then the symmetries of the various contributions to the
conductivity tensors are studied. Finally, the expressions for the
nonzero and independent elements of the conductivity tensors are
written on explicit form.

\section{General considerations}
In order to determine (i) the linear conductivity response tensor
$\stensor{\sigma}(\vec{r},\vec{r}\,')$ introduced in Eq.~(\ref{eq:Jr})
from the expression for the linear current density in
Eq.~(\ref{eq:J1Tr}) [with insertion of Eqs.~(\ref{eq:J1-1}) and
(\ref{eq:J1-2})], and (ii) the nonlinear conductivity response
function $\tensor{\Xi}(\vec{r},\vec{r}\,',\vec{r}\,'',\vec{r}\,''')$
introduced in Eq.~(\ref{eq:J3r}) from the expression for the DFWM
current density given in Eq.~(\ref{eq:J3Tr}) [with
Eqs.~(\ref{eq:J3-1})--(\ref{eq:J3-3}) inserted] we by now essentially
just need to relate the various matrix elements appearing in
Eqs.~(\ref{eq:J1-1}), (\ref{eq:J1-2}), and
(\ref{eq:J3-1})--(\ref{eq:J3-3}) to the vector potential.

Taking the $nm$ matrix element of the ``$-\omega$'' part of the part of
the Hamiltonian which is linear in the vector potential one finds on
integral form
\begin{equation}
 {\cal{H}}^{(1)}_{-\omega,nm}=\left({\cal{H}}^{(1)}_{\omega,mn}\right)^{*}=
 -\int\vec{J}_{mn}(\vec{r}\,)\cdot\vec{A}(\vec{r}\,)d^3r,
\label{eq:H-w1nm}\label{eq:Hw1nm}
\end{equation}
where we have introduced the transition current density {\it from}\/
state $m$ {\it to}\/ state $n$,
i.e., $\vec{j}^{\,(0)}_{nm}\equiv\vec{J}_{mn}$, in its explicit
form, viz.
\begin{equation}
 \vec{J}_{mn}(\vec{r}\,)={e\hbar\over2{\rm{i}}m_{e}}
 \left(\psi_{m}(\vec{r}\,)\vec{\nabla}\psi_{n}^{*}(\vec{r}\,)
 -\psi_{n}^{*}(\vec{r}\,)\vec{\nabla}\psi_{m}(\vec{r}\,)\right),
\label{eq:Jm->n}
\end{equation}
$\psi_{a}$ ($a\in\{m,n\}$) being the electronic eigenstate satisfying
the unperturbed Schr{\"o}dinger equation ${\cal{H}}_{\,0}\psi_{a}=
{\cal{E}}_{\,a}\psi_{a}$. From Eq.~(\ref{eq:Jm->n}) we note that
$\vec{J}_{nm}(\vec{r}\,)=\vec{J}_{mn}^{*}(\vec{r}\,)$.  Similarly, the
$nm$ matrix elements of the ``$-2\omega$'' part of the Hamiltonian becomes
\begin{equation}
 {\cal{H}}^{(2)}_{-2\omega,nm}
 ={e^2\over4m_{e}}\int\psi_{n}^{*}(\vec{r}\,)\psi_{m}(\vec{r}\,)
 \vec{A}(\vec{r}\,)\cdot\vec{A}(\vec{r}\,)d^3r
\label{eq:H-2w2nm}
\end{equation}
on integral form. Next, the matrix elements of the part of the
Hamiltonian which is proportional to $\vec{A}\cdot\vec{A}^{*}$ are
given by
\begin{equation}
 {\cal{H}}^{(2)}_{\,0,nm}={e^2\over4m_{e}}
 \int\psi_{n}^{*}(\vec{r}\,)\psi_{m}(\vec{r}\,)\vec{A}(\vec{r}\,)
 \cdot\vec{A}^{*}(\vec{r}\,)d^3r.
\label{eq:H02nm}
\end{equation}
Finally, the matrix elements of the current density operator
$\vec{j}^{\,(1)}$ are found to be
\begin{equation}
 \vec{j}_{-\omega,nm}^{\,(1)}=\vec{j}_{\omega,mn}^{\,(1)*}=
 -{e^2\over{}m_{e}}\psi_{n}^{*}(\vec{r}\,)\psi_{m}(\vec{r}\,)
 \vec{A}(\vec{r}\,).
\label{eq:j-w1}\label{eq:jw1}
\end{equation}

The calculation of the DFWM conductivity tensor is finalized in two
steps. Thus we start by
inserting Eqs.~(\ref{eq:H-w1nm})--(\ref{eq:j-w1}) into the three
traces in Eqs.~(\ref{eq:J3-1})--(\ref{eq:J3-3}), and thereafter we extract
the vector potential in such a manner that the result takes the
general form given in Eq.~(\ref{eq:J3r}). For convenience, we in the
following divide the nonlinear conductivity tensor into a sum
of subparts A--G referring to the processes (A)--(G) shown in
Fig.~\ref{fig:3Tr}.

Since we are using the linear response function in the description of
the phase conjugated field it is adequate for consistency again to
describe the linear process, although it is already well known. The
calculation is done in a similar manner as for the DFWM response, by
inserting Eqs.~(\ref{eq:H-w1nm}), (\ref{eq:Jm->n}) and (\ref{eq:j-w1})
into the two traces in Eqs.~(\ref{eq:J1-1}) and (\ref{eq:J1-2}), and
thereafter isolating the vector potential so that the result takes the
form of Eq.~(\ref{eq:Jr}). In the following the linear conductivity
tensor is divided into a sum of subparts A--B referring to the two
processes shown in Fig.~\ref{fig:2Tr}.

\section{Symmetry properties of the conductivity tensors}
In order to study the symmetries of the various contributions to the
conductivity tensors $\sigma(\vec{r},\vec{r}\,')$ and
$\tensor{\Xi}(\vec{r},\vec{r}\,',\vec{r}\,'',\vec{r}\,''')$ one
notices that the vector potential only appears via
${\cal{H}}^{(1)}_{-\omega}$, ${\cal{H}}^{(1)}_{\,\omega}$,
${\cal{H}}^{(2)}_{-2\omega}$, ${\cal{H}}^{(2)}_{\,0}$,
${\vec{j}}^{\,(1)}_{-\omega}$, and ${\vec{j}}^{\,(1)}_{\omega}$ of
Eqs.~(\ref{eq:H-w1nm}) and (\ref{eq:H-2w2nm})--(\ref{eq:j-w1}). One
further observes from Eq.~(\ref{eq:H-w1nm}) that the matrix elements
of ${\cal{H}}^{(1)}_{-\omega}$ and ${\cal{H}}^{(1)}_{\omega}$ contain
inner products between a transition current density and a vector
potential and that those of ${\cal{H}}^{(2)}_{-2\omega}$ and
${\cal{H}}^{(2)}_{\,0}$ involve inner products between two vector
potentials, see Eqs.~(\ref{eq:H-2w2nm}) and (\ref{eq:H02nm}). These
last inner products may conveniently be written in the form
$\tensor{\openone}:\vec{A}\vec{A}$ and
$\tensor{\openone}:\vec{A}\vec{A}^{*}$, respectively. The matrix
elements of the current densities $\vec{j}_{-\omega}^{\,(1)}$ and
$\vec{j}_{\omega}^{\,(1)}$ are directly proportional to the vector
potential and may thus for the present purpose adequately be written
in the forms $\tensor{\openone}\cdot\vec{A}$ and
$\tensor{\openone}\cdot\vec{A}^{*}$, respectively.

\subsection{Linear conductivity tensor}
In the view of the aforementioned remarks it is concluded that part
A of the linear conductivity tensor, given by Eq.~(\ref{eq:J1-1})
has the symmetry of the unit tensor $\tensor{\openone}$. Part A thus
have $3$ nonzero elements. Furthermore, of these $3$ nonzero elements
only $1$ is independent, since the Cartesian index of the linear
current density follows that of the vector potential appearing in
$\vec{j}_{\omega}^{\,(1)}$, and thus $i=j$. 
\begin{figure}[tb]
\setlength{\unitlength}{0.77mm}
\psset{unit=0.77mm}
\begin{center}
\begin{pspicture}(0,4)(70,29)
\multiput(2,0)(40,0){2}{
 \psline[linewidth=0.5]{-}(6,6)(4,6)(4,22)(6,22)
 \psline[linewidth=0.5]{-}(22,6)(24,6)(24,22)(22,22)
 {\footnotesize
  \put(0,22.5){\makebox(0,5)[l]{$[ij]$}}
 }
}
\multiput(10,8)(40,0){2}{
  \multiput(0,0)(6,0){3}{\multiput(0,0)(0,6){3}{\circle*{0.1}
}}}
\put(27,6){\makebox(0,5)[bl]{(A)}}
\put(45,6){\makebox(0,5)[br]{(B)}}
\put(10,20){\circle*{2}}
\put(16,14){\circle*{2}}
\put(22,8){\circle*{2}}
\multiput(50,8)(6,0){3}{\multiput(0,0)(0,6){3}{\circle*{2}}}
\psline[linewidth=0.25mm]{-}(10,20)(22,8)
\end{pspicture}
\end{center}
\caption[Symmetry schemes for the linear conductivity tensor]{Symmetry
  schemes for the linear conductivity tensor. Tensor elements labeled
  with a ``$\bullet$'' are nonzero, elements labeled with a
  ``$\cdot$'' are zero, and the solid line connect equal nonzero
  elements.\label{fig:Sigma}}
\end{figure}
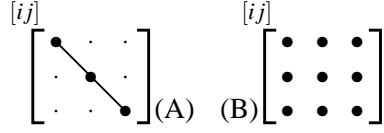
The other part of the linear conductivity tensor (part B) is
extracted from Eq.~(\ref{eq:J1-2}), and it shows a symmetry to the
outer product $\vec{J}_1\otimes\vec{J}_2$, where $\vec{J}_1$ and
$\vec{J}_2$ in general are different, and part B of the linear
conductivity tensor thus has $9$ independent nonzero elements. The
symmetry schemes of the linear conductivity tensor are shown in
Fig.~\ref{fig:Sigma}.

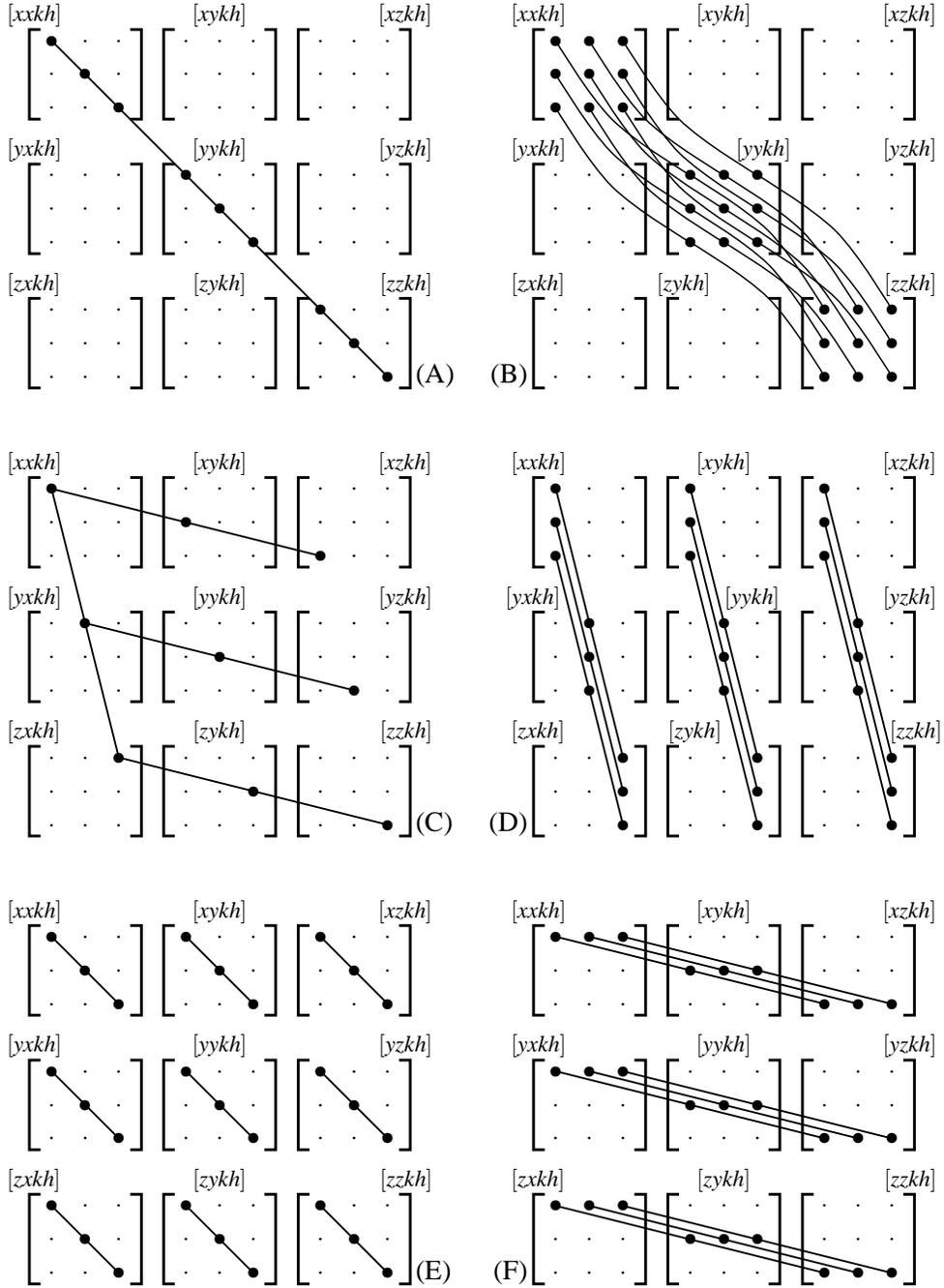
\begin{figure}[tbp]
\setlength{\unitlength}{0.77mm}
\psset{unit=0.77mm}
\begin{center}
\begin{pspicture}(0,4)(170,240)
\put(0,160){
\put(75,6){\makebox(10,5)[bl]{(A)}}
\multiput(2,0)(24,0){3}{\multiput(0,0)(0,24){3}{
  \psline[linewidth=0.5]{-}(6,6)(4,6)(4,22)(6,22)
  \psline[linewidth=0.5]{-}(22,6)(24,6)(24,22)(22,22)
}}
\multiput(10,8)(24,0){3}{\multiput(0,0)(0,24){3}{
  \multiput(0,0)(6,0){3}{\multiput(0,0)(0,6){3}{\circle*{0.1}
}}}}
{\footnotesize
 \put(2,70){\makebox(0,5)[l]{$[xxkh]$}}
 \put(2,46){\makebox(0,5)[l]{$[yxkh]$}}
 \put(2,22){\makebox(0,5)[l]{$[zxkh]$}}
 \put(40,70){\makebox(0,5)[c]{$[xykh]$}}
 \put(40,46){\makebox(0,5)[c]{$[yykh]$}}
 \put(40,22){\makebox(0,5)[c]{$[zykh]$}}
 \put(78,70){\makebox(0,5)[r]{$[xzkh]$}}
 \put(78,46){\makebox(0,5)[r]{$[yzkh]$}}
 \put(78,22){\makebox(0,5)[r]{$[zzkh]$}}
}
\put(10,68){\circle*{2}}
\put(16,62){\circle*{2}}
\put(22,56){\circle*{2}}
\put(34,44){\circle*{2}}
\put(40,38){\circle*{2}}
\put(46,32){\circle*{2}}
\put(58,20){\circle*{2}}
\put(64,14){\circle*{2}}
\put(70,8){\circle*{2}} 
\psline[linewidth=0.25mm]{-}(10,68)(70,8)
}
\put(90,160){
\put(5,6){\makebox(0,5)[br]{(B)}}
\multiput(2,0)(24,0){3}{\multiput(0,0)(0,24){3}{
  \psline[linewidth=0.5]{-}(6,6)(4,6)(4,22)(6,22)
  \psline[linewidth=0.5]{-}(22,6)(24,6)(24,22)(22,22)
}}
\multiput(10,8)(24,0){3}{\multiput(0,0)(0,24){3}{
  \multiput(0,0)(6,0){3}{\multiput(0,0)(0,6){3}{\circle*{0.1}
}}}}
{\footnotesize
 \put(2,70){\makebox(0,5)[l]{$[xxkh]$}}
 \put(2,46){\makebox(0,5)[l]{$[yxkh]$}}
 \put(2,22){\makebox(0,5)[l]{$[zxkh]$}}
 \put(40,70){\makebox(0,5)[c]{$[xykh]$}}
 \put(52,46){\makebox(0,5)[r]{$[yykh]$}}
 \put(28,22){\makebox(0,5)[l]{$[zykh]$}}
 \put(78,70){\makebox(0,5)[r]{$[xzkh]$}}
 \put(78,46){\makebox(0,5)[r]{$[yzkh]$}}
 \put(78,22){\makebox(0,5)[r]{$[zzkh]$}}
}
\multiput(10,56)(0,6){3}{\multiput(0,0)(6,0){3}{\circle*{2}}}
\multiput(34,32)(0,6){3}{\multiput(0,0)(6,0){3}{\circle*{2}}}
\multiput(58,8)(0,6){3}{\multiput(0,0)(6,0){3}{\circle*{2}}}
\pscurve[linewidth=.2mm]{-}(10,56)(20,42)(34,32) 
\pscurve[linewidth=.2mm]{-}(34,32)(48,22)(58,8) 
\pscurve[linewidth=.2mm]{-}(10,62)(20,48)(34,38) 
\pscurve[linewidth=.2mm]{-}(34,38)(48,28)(58,14) 
\pscurve[linewidth=.2mm]{-}(10,68)(20,54)(34,44) 
\pscurve[linewidth=.2mm]{-}(34,44)(48,34)(58,20) 
\pscurve[linewidth=.2mm]{-}(16,56)(26,42)(40,32) 
\pscurve[linewidth=.2mm]{-}(40,32)(54,22)(64,8) 
\pscurve[linewidth=.2mm]{-}(16,62)(26,48)(40,38) 
\pscurve[linewidth=.2mm]{-}(40,38)(54,28)(64,14) 
\pscurve[linewidth=.2mm]{-}(16,68)(26,54)(40,44) 
\pscurve[linewidth=.2mm]{-}(40,44)(54,34)(64,20) 
\pscurve[linewidth=.2mm]{-}(22,56)(32,42)(46,32) 
\pscurve[linewidth=.2mm]{-}(46,32)(60,22)(70,8) 
\pscurve[linewidth=.2mm]{-}(22,62)(32,48)(46,38) 
\pscurve[linewidth=.2mm]{-}(46,38)(60,28)(70,14) 
\pscurve[linewidth=.2mm]{-}(22,68)(32,54)(46,44) 
\pscurve[linewidth=.2mm]{-}(46,44)(60,34)(70,20) 
}
\put(0,80){
\put(75,6){\makebox(10,5)[bl]{(C)}}
\multiput(2,0)(24,0){3}{\multiput(0,0)(0,24){3}{
  \psline[linewidth=0.5]{-}(6,6)(4,6)(4,22)(6,22)
  \psline[linewidth=0.5]{-}(22,6)(24,6)(24,22)(22,22)
}}
\multiput(10,8)(24,0){3}{\multiput(0,0)(0,24){3}{
  \multiput(0,0)(6,0){3}{\multiput(0,0)(0,6){3}{\circle*{0.1}
}}}}
{\footnotesize
 \put(2,70){\makebox(0,5)[l]{$[xxkh]$}}
 \put(2,46){\makebox(0,5)[l]{$[yxkh]$}}
 \put(2,22){\makebox(0,5)[l]{$[zxkh]$}}
 \put(40,70){\makebox(0,5)[c]{$[xykh]$}}
 \put(40,46){\makebox(0,5)[c]{$[yykh]$}}
 \put(40,22){\makebox(0,5)[c]{$[zykh]$}}
 \put(78,70){\makebox(0,5)[r]{$[xzkh]$}}
 \put(78,46){\makebox(0,5)[r]{$[yzkh]$}}
 \put(78,22){\makebox(0,5)[r]{$[zzkh]$}}
}
\put(10,68){\circle*{2}}
\put(34,62){\circle*{2}}
\put(58,56){\circle*{2}}
\put(16,44){\circle*{2}}
\put(40,38){\circle*{2}}
\put(64,32){\circle*{2}}
\put(22,20){\circle*{2}}
\put(46,14){\circle*{2}}
\put(70,8){\circle*{2}} 
\psline[linewidth=0.25mm]{-}(10,68)(58,56)
\psline[linewidth=0.25mm]{-}(16,44)(64,32)
\psline[linewidth=0.25mm]{-}(22,20)(70,8)
\psline[linewidth=0.25mm]{-}(10,68)(22,20)
}
\put(90,80){
\put(5,6){\makebox(0,5)[br]{(D)}}
\multiput(2,0)(24,0){3}{\multiput(0,0)(0,24){3}{
  \psline[linewidth=0.5]{-}(6,6)(4,6)(4,22)(6,22)
  \psline[linewidth=0.5]{-}(22,6)(24,6)(24,22)(22,22)
}}
\multiput(10,8)(24,0){3}{\multiput(0,0)(0,24){3}{
  \multiput(0,0)(6,0){3}{\multiput(0,0)(0,6){3}{\circle*{0.1}
}}}}
{\footnotesize
 \put(2,70){\makebox(0,5)[l]{$[xxkh]$}}
 \put(1,46){\makebox(0,5)[l]{$[yxkh]$}}
 \put(2,22){\makebox(0,5)[l]{$[zxkh]$}}
 \put(40,70){\makebox(0,5)[c]{$[xykh]$}}
 \put(50,46){\makebox(0,5)[r]{$[yykh]$}}
 \put(30,22){\makebox(0,5)[l]{$[zykh]$}}
 \put(78,70){\makebox(0,5)[r]{$[xzkh]$}}
 \put(78,46){\makebox(0,5)[r]{$[yzkh]$}}
 \put(79,22){\makebox(0,5)[r]{$[zzkh]$}}
}
\put(10,68){\circle*{2}}
\put(10,62){\circle*{2}}
\put(10,56){\circle*{2}}
\put(34,68){\circle*{2}}
\put(34,62){\circle*{2}}
\put(34,56){\circle*{2}}
\put(58,68){\circle*{2}}
\put(58,62){\circle*{2}}
\put(58,56){\circle*{2}}
\put(16,44){\circle*{2}}
\put(16,38){\circle*{2}}
\put(16,32){\circle*{2}}
\put(40,44){\circle*{2}}
\put(40,38){\circle*{2}}
\put(40,32){\circle*{2}}
\put(64,44){\circle*{2}}
\put(64,38){\circle*{2}}
\put(64,32){\circle*{2}}
\put(22,20){\circle*{2}}
\put(22,14){\circle*{2}}
\put(22,8){\circle*{2}}
\put(46,20){\circle*{2}}
\put(46,14){\circle*{2}}
\put(46,8){\circle*{2}} 
\put(70,20){\circle*{2}}
\put(70,14){\circle*{2}}
\put(70,8){\circle*{2}} 
\multiput(10,68)(24,0){3}{%
 \multiput(0,0)(0,-6){3}{%
  \psline[linewidth=0.25mm]{-}(0,0)(12,-48)
}}
}
\put(0,0){
\put(75,6){\makebox(10,5)[bl]{(E)}}
\multiput(2,0)(24,0){3}{\multiput(0,0)(0,24){3}{
  \psline[linewidth=0.5]{-}(6,6)(4,6)(4,22)(6,22)
  \psline[linewidth=0.5]{-}(22,6)(24,6)(24,22)(22,22)
}}
\multiput(10,8)(24,0){3}{\multiput(0,0)(0,24){3}{
  \multiput(0,0)(6,0){3}{\multiput(0,0)(0,6){3}{\circle*{0.1}
}}}}
{\footnotesize
 \put(2,70){\makebox(0,5)[l]{$[xxkh]$}}
 \put(2,46){\makebox(0,5)[l]{$[yxkh]$}}
 \put(2,22){\makebox(0,5)[l]{$[zxkh]$}}
 \put(40,70){\makebox(0,5)[c]{$[xykh]$}}
 \put(40,46){\makebox(0,5)[c]{$[yykh]$}}
 \put(40,22){\makebox(0,5)[c]{$[zykh]$}}
 \put(78,70){\makebox(0,5)[r]{$[xzkh]$}}
 \put(78,46){\makebox(0,5)[r]{$[yzkh]$}}
 \put(78,22){\makebox(0,5)[r]{$[zzkh]$}}
}
\put(10,68){\circle*{2}}
\put(16,62){\circle*{2}}
\put(22,56){\circle*{2}}
\put(34,68){\circle*{2}}
\put(40,62){\circle*{2}}
\put(46,56){\circle*{2}}
\put(58,68){\circle*{2}}
\put(64,62){\circle*{2}}
\put(70,56){\circle*{2}}
\put(10,44){\circle*{2}}
\put(16,38){\circle*{2}}
\put(22,32){\circle*{2}}
\put(34,44){\circle*{2}}
\put(40,38){\circle*{2}}
\put(46,32){\circle*{2}}
\put(58,44){\circle*{2}}
\put(64,38){\circle*{2}}
\put(70,32){\circle*{2}}
\put(10,20){\circle*{2}}
\put(16,14){\circle*{2}}
\put(22,8){\circle*{2}} 
\put(34,20){\circle*{2}}
\put(40,14){\circle*{2}}
\put(46,8){\circle*{2}} 
\put(58,20){\circle*{2}}
\put(64,14){\circle*{2}}
\put(70,8){\circle*{2}} 
\multiput(10,68)(24,0){3}{%
 \multiput(0,0)(0,-24){3}{%
  \psline[linewidth=0.25mm]{-}(0,0)(12,-12)
}}
}
\put(90,0){
\put(5,6){\makebox(0,5)[br]{(F)}}
\multiput(2,0)(24,0){3}{\multiput(0,0)(0,24){3}{
  \psline[linewidth=0.5]{-}(6,6)(4,6)(4,22)(6,22)
  \psline[linewidth=0.5]{-}(22,6)(24,6)(24,22)(22,22)
}}
\multiput(10,8)(24,0){3}{
 \multiput(0,0)(0,24){3}{
  \multiput(0,0)(6,0){3}{
   \multiput(0,0)(0,6){3}{\circle*{0.1}
}}}}
{\footnotesize
 \put(2,70){\makebox(0,5)[l]{$[xxkh]$}}
 \put(2,46){\makebox(0,5)[l]{$[yxkh]$}}
 \put(2,22){\makebox(0,5)[l]{$[zxkh]$}}
 \put(40,70){\makebox(0,5)[c]{$[xykh]$}}
 \put(40,46){\makebox(0,5)[c]{$[yykh]$}}
 \put(40,22){\makebox(0,5)[c]{$[zykh]$}}
 \put(78,70){\makebox(0,5)[r]{$[xzkh]$}}
 \put(78,46){\makebox(0,5)[r]{$[yzkh]$}}
 \put(78,22){\makebox(0,5)[r]{$[zzkh]$}}
}
\multiput(10,20)(0,24){3}{
 \multiput(0,0)(24,-6){3}{
  \multiput(0,0)(6,0){3}{\circle*{2}
}}}
\multiput(10,68)(0,-24){3}{%
 \multiput(0,0)(6,0){3}{%
  \psline[linewidth=0.25mm]{-}(0,0)(48,-12)
}}
}
\end{pspicture}
\end{center}
\caption[Symmetry schemes for parts A--F of the DFWM conductivity
tensor]{The symmetry schemes for parts A--F of the DFWM conductivity
  tensor in their most general spin-less forms. Tensor elements
  labeled with a ``$\bullet$'' are nonzero, elements labeled with a
  ``$\cdot$'' are zero, and the solid lines connect nonzero elements
  of equal magnitude.\label{fig:6.2}}
\end{figure}

\subsection{DFWM conductivity tensor}
Taking our symmetry analysis to the DFWM conductivity tensor we
conclude that part A, given by the first sum on the right hand side
of Eq.~(\ref{eq:J3-1}), has a symmetry given by the outer product
$\tensor{\openone}\otimes\tensor{\openone}$. Part A thus have $9$
nonzero elements. Furthermore, since the Cartesian index of the DFWM
current density follows that of the vector potential appearing in
$\vec{j}_{\omega}^{\,(1)}$, $i=j$ in the index notation of
Eq.~(\ref{eq:sumprod3}). From the form of the
${\cal{H}}^{(2)}_{-2\omega}$ term we next conclude that $k=h$.
Altogether it is realised that the $9$ nonzero elements are identical.
To get an overview of the conclusion, we show in
Fig.~\ref{fig:6.2}.A the result in terms of a symmetry scheme.

Utilising the same type of arguments it is concluded that each term in
the second sum in Eq.~(\ref{eq:J3-1}), which gives rise to part B of
the conductivity tensor, when written in the form of
Eq.~(\ref{eq:sumprod3}) has a symmetry identical to the outer product
$\tensor{\openone}\otimes\vec{J}_{1}\otimes\vec{J}_{2}$, where
$\vec{J}_{1}$ and $\vec{J}_{2}$ are two generally different transition
current densities. The form of this outer product leaves us with $27$
nonzero elements. Also here the coordinate convention of
Eq.~(\ref{eq:sumprod3}) implies that $i=j$. Furthermore we observe
that elements with $i=x$, $i=y$, and $i=z$ are identical, since the
two ${\cal{H}}^{(1)}_{-\omega}$ terms essentially produces numbers.
Finally, we see that the independent nature of the two
${\cal{H}}^{(1)}_{-\omega}$ terms makes them interchangeable, and thus
gives us two different ways of constructing the sum in
Eq.~(\ref{eq:sumprod3}). Of the $27$ nonzero elements only $9$ are
independent, since as we have realised,
$\Xi_{xxkh}^{\rm{B}}=\Xi_{yykh}^{\rm{B}}=\Xi_{zzkh}^{\rm{B}}$ for all
permutations of $k$ and $h$ in the three Cartesian coordinates
$\{x,y,z\}$. Expressed in terms of a symmetry scheme, the deductions
above lead to the symmetry scheme shown in Fig.~\ref{fig:6.2}.B.

The first sum in Eq.~(\ref{eq:J3-2}) gives rise to part C of the
DFWM conductivity tensor, and the second sum in this equation leads to
part D. Looking at the first sum it appears that this is
proportional to $(\tensor{\openone}:\vec{A}\vec{A}^{*})\vec{A}$, a
fact which in relation to the form given in
Eq.~(\ref{eq:sumprod3}) implies that the symmetry of the conductivity
tensor is given by the outer product
$\vec{e}_{A}\otimes\tensor{\openone}\otimes\vec{e}_{A}$, with
$\vec{e}_{A}=\vec{A}/A$. This product form leaves us with $9$ nonzero
elements. As far as the Cartesian indices are concerned the above
symmetry implies that $j=h$ and $i=k$. Finally we observe that the
same constant appears in front of the vector potential indexed $k$.
This leads to the conclusion that the cases $i=x$, $i=y$, and $i=z$
are equal, leaving at the end only one independent nonzero element of
part C of the conductivity tensor. The symmetry scheme for part C
is shown in Fig.~\ref{fig:6.2}.C.

In the second sum of Eq.~(\ref{eq:J3-2}) the symmetry is proportional
to the outer product
$\vec{e}_{A}\otimes\vec{J}_{1}\otimes\vec{J}_{2}\otimes\vec{e}_{A}$
and thus we are left with $27$ nonzero elements. Then, in the form of
Eq.~(\ref{eq:sumprod3}), $i=h$, and the permutations over $i$ are seen
to be equal, so that we end up with only $9$ independent elements.
Using the fact that
$\Xi_{xjkx}^{\rm{D}}=\Xi_{yjky}^{\rm{D}}=\Xi_{zjkz}^{\rm{D}}$ one
obtains the symmetry scheme shown in Fig.~\ref{fig:6.2}.D.

\begin{table}[tb]
\begin{center}
\setlength{\tabcolsep}{0pt}
\begin{tabular}{p{63.5mm}p{63.5mm}}
\hline
\hfil Conductivity\hfil &\hfil Tensor symmetry\hfil\\ \hline\hline
\end{tabular}
\vskip 0.25mm
\begin{tabular}{p{63.5mm}p{63.5mm}}
$\hfil \stensor{\sigma}^{\rm{A}}(\vec{r},\vec{r}\,')$\hfil
&\hfil \hbox{$\tensor{\openone}$}\hfil
\\
\hfil $\stensor{\sigma}^{\rm{B}}(\vec{r},\vec{r}\,')$\hfil 
&\hfil $\vec{J}_{1}\otimes\vec{J}_{2}$\hfil 
\\ \hline
\end{tabular}
\vskip 0.25mm
\begin{tabular}{p{63.5mm}p{63.5mm}}
\hfil $\tensor{\Xi}^{\rm{A}}(\vec{r},\vec{r}\,',\vec{r}\,'',\vec{r}\,''')$\hfil
&\hfil $\tensor{\openone}\otimes\tensor{\openone}$\hfil 
\\
\hfil $\tensor{\Xi}^{\rm{B}}(\vec{r},\vec{r}\,',\vec{r}\,'',\vec{r}\,''')$\hfil
&\hfil $\tensor{\openone}\otimes\vec{J}_{1}\otimes\vec{J}_{2}$\hfil 
\\
\hfil $\tensor{\Xi}^{\rm{C}}(\vec{r},\vec{r}\,',\vec{r}\,'',\vec{r}\,''')$\hfil
&\hfil $\vec{e}_{A}\otimes\tensor{\openone}\otimes\vec{e}_{A}$\hfil 
\\
\hfil $\tensor{\Xi}^{\rm{D}}(\vec{r},\vec{r}\,',\vec{r}\,'',\vec{r}\,''')$\hfil
&\hfil $\vec{e}_{A}\otimes\vec{J}_{1}\otimes\vec{J}_{2}\otimes\vec{e}_{A}$\hfil
\\
\hfil $\tensor{\Xi}^{\rm{E}}(\vec{r},\vec{r}\,',\vec{r}\,'',\vec{r}\,''')$\hfil
&\hfil $\vec{J}_{1}\otimes\tensor{\openone}\otimes\vec{J}_{2}$\hfil 
\\
\hfil $\tensor{\Xi}^{\rm{F}}(\vec{r},\vec{r}\,',\vec{r}\,'',\vec{r}\,''')$\hfil
&\hfil $\vec{J}_{1}\otimes\vec{J}_{2}\otimes\tensor{\openone}$\hfil 
\\
\hfil $\tensor{\Xi}^{\rm{G}}(\vec{r},\vec{r}\,',\vec{r}\,'',\vec{r}\,''')$\hfil
&\hfil $\vec{J}_{1}\otimes\vec{J}_{2}\otimes\vec{J}_{3}\otimes\vec{J}_{4}$\hfil
\\ \hline
\end{tabular}
\end{center}
\caption[The tensor symmetries of the linear and the DFWM conductivity
tensors]{The tensor symmetries of the various parts A--B of the
  linear, and A--G of the DFWM conductivity. As explained in the
  text, $\vec{J}_{1}$--$\vec{J}_{4}$ are four in general different
  vectors obtained by a weighted superposition of single-particle
  transition current densities, and
  $\vec{e}_{A}=\vec{A}/A$.\label{fig:TableSym}\label{tab:1}}
\end{table}

Let us now take a closer look at the third trace in
Eq.~(\ref{eq:J3-3}). It is convenient to split the first sum in this
equation into two parts related to the two different processes that
occur. The first part of the sum, which refers to the
${\cal{H}}^{(2)}_{-2\omega,nv}{\cal{H}}^{(1)}_{\,\omega,v{}m}$-type
of terms, gives rise to part E of the third order conductivity
tensor corresponding to process (E) of Fig.~\ref{fig:3Tr}. The second
part of the first sum is related to part F of the third order
conductivity tensor [process (F) of Fig.~\ref{fig:3Tr}]. Finally, the
second sum on the right side of Eq.~(\ref{eq:J3-3}) produces part G
of the third order conductivity tensor, corresponding to process
(G) of Fig.~\ref{fig:3Tr}.

The first part of the first sum, in relation to the representation in
Eq.~(\ref{eq:sumprod3}), has a symmetry which can be represented by
the outer product
$\vec{J}_{1}\otimes\tensor{\openone}\otimes\vec{J}_{2}$, leaving $27$
nonzero elements. From the term ${\cal{H}}_{-2\omega}^{(2)}$, we see
that $k=h$ in the chosen representation of coordinate sets, and
furthermore we realise that elements with $k=x$, $k=y$, and $k=z$ are
equal. These deductions reduce the number of independent nonzero
elements to $9$, which fulfills
$\Xi_{ijxx}^{\rm{E}}=\Xi_{ijyy}^{\rm{E}}=\Xi_{ijzz}^{\rm{E}}$ for all
permutations of $i$ and $j$ in the three Cartesian coordinates
$\{x,y,z\}$. The result is shown on schematic form in terms of the
symmetry scheme in Fig.~\ref{fig:6.2}.E.

In the second part of the first sum in Eq.~(\ref{eq:J3-3}) we observe
that the symmetry of part F of the conductivity tensor, in relation
to the form of Eq.~(\ref{eq:sumprod3}), is proportional to the outer
product $\vec{J}_{1}\otimes\vec{J}_{2}\otimes\tensor{\openone}$, again
leaving $27$ nonzero elements. Due to the chosen convention of the
coordinate sets, we realise from the ${\cal{H}}_{\,0}^{(2)}$ term that
the condition $k=j$ applies. We furthermore notice from this that
terms with $j=x$, $j=y$, and $j=z$ are equal, leaving $9$ independent
nonzero elements related by
$\Xi_{ixxh}^{\rm{F}}=\Xi_{iyyh}^{\rm{F}}=\Xi_{izzh}^{\rm{F}}$ for all
permutations of $i$ and $h$ in the three Cartesian coordinates
$\{x,y,z\}$.  This means that the symmetry scheme is as shown in
Fig.~\ref{fig:6.2}.F.

Part G of the third order conductivity tensor, which originates in
the second sum in Eq.~(\ref{eq:J3-3}), obviously has the tensor form
$\vec{J}_{1}\otimes\vec{J}_{2}\otimes\vec{J}_{3}\otimes\vec{J}_{4}$,
and there will hence in general be $81$ independent nonzero elements
in the associated symmetry scheme. 

The considerations laying the foundations for the symmetry schemes of
the various parts of the linear and nonlinear conductivity tensors are
displayed in Tab.~\ref{fig:TableSym}, where the relevant combinations
of $\vec{J}$'s and $\tensor{\openone}$'s are given.

\section{Expressions for the conductivity tensors}
We end this chapter by giving the explicit expressions for the
independent tensor elements of $\stensor{\sigma}(\vec{r},\vec{r}\,')$
and $\tensor{\Xi}(\vec{r},\vec{r}\,',\vec{r}\,'',\vec{r}\,''')$.
Thus, the only independent tensor element in part A of the linear
conductivity tensor is
\begin{equation}
 {\sigma}^{\rm{A}}_{xx}(\vec{r},\vec{r}\,')=
 {2{\rm{i}}\over{}\omega}{e^2\over{}m_{e}}\sum_{n}f_{n}
 |\psi_{n}|^2\delta(\vec{r}-\vec{r}\,'),
\label{eq:SigmaAArz}
\end{equation}
and the nine independent elements of part B are
\begin{equation}
 {\sigma}^{\rm{B}}_{ij}(\vec{r},\vec{r}\,')=
 {2{\rm{i}}\over\omega}{1\over\hbar}\sum_{nm}
 {f_{n}-f_{m}\over\tilde{\omega}_{nm}-\omega}J_{j,mn}'J_{i,nm}.
\label{eq:SigmaABrz}
\end{equation}
The only independent tensor element in part A of the third order
conductivity tensor thus is
\begin{eqnarray}
\lefteqn{
 \Xi_{xxxx}^{\rm{A}}(\vec{r},\vec{r}\,',\vec{r}\,'',\vec{r}\,''')=
 {2{\rm{i}}\over\omega^3}{e^4\over8m_{e}^2\hbar}\sum_{nm}
 {f_{n}-f_{m}\over\tilde{\omega}_{nm}-2\omega}
 \psi_{n}^{*}(\vec{r}\,'')\psi_{m}(\vec{r}\,'')
 \psi_{m}^{*}(\vec{r}\,)\psi_{n}(\vec{r}\,)
}\nonumber\\ &\quad&\times
 \delta(\vec{r}-\vec{r}\,')\delta(\vec{r}\,''-\vec{r}\,'''),
\label{eq:XiAArz}
\end{eqnarray}
and the nine independent elements of part B are
\begin{eqnarray}
\lefteqn{
 \Xi_{xxkh}^{\rm{B}}(\vec{r},\vec{r}\,',\vec{r}\,'',\vec{r}\,''')=
 {2{\rm{i}}\over\omega^3}{e^2\over4m_{e}\hbar^2}\sum_{nmv}
 {1\over\tilde{\omega}_{nm}-2\omega}
 \left({f_{m}-f_{v}\over\tilde{\omega}_{v{}m}-\omega}
 +{f_{n}-f_{v}\over\tilde{\omega}_{nv}-\omega}\right)
}\nonumber\\ &\quad&\times
 J_{h,v{}n}(\vec{r}\,''')J_{k,mv}(\vec{r}\,'')
 \psi_{m}^{*}(\vec{r}\,)\psi_{n}(\vec{r}\,)
 \delta(\vec{r}-\vec{r}\,').
\label{eq:XiABrz}
\end{eqnarray}
The only independent nonzero element of part C
of the third order conductivity tensor is given by
\begin{eqnarray}
\lefteqn{
 \Xi_{xxxx}^{\rm{C}}(\vec{r},\vec{r}\,',\vec{r}\,'',\vec{r}\,''')=
 {2{\rm{i}}\over\omega^3}{e^4\over4m_{e}^{2}\hbar}\sum_{nm}
 {f_{n}-f_{m}\over\tilde{\omega}_{nm}}\psi_{n}^{*}(\vec{r}\,')\psi_{m}(\vec{r}\,')
 \psi_{m}^{*}(\vec{r}\,)\psi_{n}(\vec{r}\,)
}\nonumber\\ &\quad&\times
 \delta(\vec{r}\,'-\vec{r}\,''')\delta(\vec{r}-\vec{r}\,''),
\label{eq:XiACrz}
\end{eqnarray}
and the nine independent nonzero elements of part D of the third
order conductivity tensor are
\begin{eqnarray}
\lefteqn{
 \Xi_{xjkx}^{\rm{D}}(\vec{r},\vec{r}\,',\vec{r}\,'',\vec{r}\,''')=
 {2{\rm{i}}\over\omega^3}{e^2\over4m_{e}\hbar^2}\sum_{nmv}
 {1\over\tilde{\omega}_{nm}}\left\{
 \left({f_{m}-f_{v}\over\tilde{\omega}_{v{}m}-\omega}
 +{f_{n}-f_{v}\over\tilde{\omega}_{nv}+\omega}\right)
 J_{j,v{}n}(\vec{r}\,')J_{k,mv}(\vec{r}\,'')
\right.}\nonumber\\ &\quad&\left.\!
 +\left({f_{m}-f_{v}\over\tilde{\omega}_{v{}m}+\omega}
 +{f_{n}-f_{v}\over\tilde{\omega}_{nv}-\omega}\right)
 J_{k,v{}n}(\vec{r}\,'')J_{j,mv}(\vec{r}\,')
 \right\}
 \psi_{m}^{*}(\vec{r}\,)\psi_{n}(\vec{r}\,)
 \delta(\vec{r}-\vec{r}\,''').
\label{eq:XiADrz}
\end{eqnarray}
The nine
independent elements of part E have the explicit form
\begin{eqnarray}
\lefteqn{
 \Xi_{ijxx}^{\rm{E}}(\vec{r},\vec{r}\,',\vec{r}\,'',\vec{r}\,''')=
 {2{\rm{i}}\over\omega^3}{e^2\over16m_{e}\hbar^2}
}\nonumber\\ &\quad&\times
\sum_{nmv}
 {1\over\tilde{\omega}_{nm}-\omega}\left\{
 \left({f_{m}-f_{v}\over\tilde{\omega}_{v{}m}-2\omega}
 +{f_{n}-f_{v}\over\tilde{\omega}_{nv}+\omega}\right)
 J_{j,v{}n}(\vec{r}\,')\psi_{v}^{*}(\vec{r}\,'')\psi_{m}(\vec{r}\,'')
\right.\nonumber\\ &&\left.\!
 +\left({f_{n}-f_{v}\over\tilde{\omega}_{nv}-2\omega}
 +{f_{m}-f_{v}\over\tilde{\omega}_{v{}m}+\omega}\right)
 J_{j,mv}(\vec{r}\,')\psi_{n}^{*}(\vec{r}\,'')\psi_{v}(\vec{r}\,'')
 \right\}J_{i,nm}(\vec{r}\,)\delta(\vec{r}\,''-\vec{r}\,'''),
\label{eq:XiAErz}
\end{eqnarray}
and the nine independent elements of part F of the third order
conductivity tensor are
\begin{eqnarray}
\lefteqn{
 \Xi_{ixxh}^{\rm{F}}(\vec{r},\vec{r}\,',\vec{r}\,'',\vec{r}\,''')=
 {2{\rm{i}}\over\omega^3}{e^2\over8m_{e}\hbar^2}
}\nonumber\\ &\quad&\times
 \sum_{nmv}
 {1\over\tilde{\omega}_{nm}-\omega}\left\{
 \left({f_{m}-f_{v}\over\tilde{\omega}_{v{}m}}
 +{f_{n}-f_{v}\over\tilde{\omega}_{nv}-\omega}\right)
 J_{h,v{}n}(\vec{r}\,'')\psi_{v}^{*}(\vec{r}\,')\psi_{m}(\vec{r}\,')
\right.\nonumber\\ &&\left.
 +\left({f_{n}-f_{v}\over\tilde{\omega}_{nv}}
 +{f_{m}-f_{v}\over\tilde{\omega}_{v{}m}-\omega}\right)
 J_{h,mv}(\vec{r}\,'')\psi_{n}^{*}(\vec{r}\,')\psi_{v}(\vec{r}\,')
 \right\}J_{i,nm}(\vec{r}\,)\delta(\vec{r}\,'-\vec{r}\,''').
\label{eq:XiAFrz}
\end{eqnarray}
Finally, the eighty-one independent elements of part G of the third
order conductivity tensor are given by
\begin{eqnarray}
\lefteqn{
 \Xi_{ijkh}^{\rm{G}}(\vec{r},\vec{r}\,',\vec{r}\,'',\vec{r}\,''')=
 {2{\rm{i}}\over\omega^3}{1\over8\hbar^3}\sum_{nmvl}
 {1\over\tilde{\omega}_{nm}-\omega}
 \left\{
 \left[
 \left({f_{l}-f_{m}\over\tilde{\omega}_{l{}m}-\omega}
 +{f_{l}-f_{v}\over\tilde{\omega}_{vl}-\omega}\right)
 {1\over\tilde{\omega}_{v{}m}-2\omega}
\right.\right.}\nonumber\\ &\quad&\left.\!
 +\left({f_{l}-f_{v}\over\tilde{\omega}_{vl}-\omega}
 +{f_{n}-f_{v}\over\tilde{\omega}_{nv}+\omega}\right)
 {1\over\tilde{\omega}_{nl}}
 \right]
 J_{h,ml}(\vec{r}\,''')J_{k,lv}(\vec{r}\,'')J_{j,v{}n}(\vec{r}\,')
\nonumber\\ &&
 +\left[
 \left({f_{l}-f_{m}\over\tilde{\omega}_{l{}m}-\omega}
 +{f_{l}-f_{v}\over\tilde{\omega}_{vl}+\omega}\right)
 {1\over\tilde{\omega}_{v{}m}}
 +\left({f_{l}-f_{v}\over\tilde{\omega}_{vl}+\omega}
 +{f_{n}-f_{v}\over\tilde{\omega}_{nv}-\omega}\right)
 {1\over\tilde{\omega}_{nl}}
 \right]
\nonumber\\ &&\times
 J_{h,ml}(\vec{r}\,''')J_{k,v{}n}(\vec{r}\,'')J_{j,lv}(\vec{r}\,')
 +\left[
 \left({f_{l}-f_{m}\over\tilde{\omega}_{l{}m}+\omega}
 +{f_{l}-f_{v}\over\tilde{\omega}_{vl}-\omega}\right)
 {1\over\tilde{\omega}_{v{}m}}
\right.\nonumber\\ &&\left.\!\left.\!
 +\left({f_{l}-f_{v}\over\tilde{\omega}_{vl}-\omega}
 +{f_{n}-f_{v}\over\tilde{\omega}_{nv}-\omega}\right)
 {1\over\tilde{\omega}_{nl}-2\omega}
 \right]
 J_{h,lv}(\vec{r}\,''')J_{k,v{}n}(\vec{r}\,'')J_{j,ml}(\vec{r}\,')
\right\}J_{i,nm}(\vec{r}\,).
\label{eq:XiAGrz}
\end{eqnarray}
The number $2$ appearing in the first fraction of each part of the
conductivity tensors represents the degeneracy of the spin energies,
thus giving two electrons in each energy eigenstate.

\chapter{Discussion}\label{ch:7}
In Chapters~\ref{ch:4}--\ref{ch:6} we have established a spatially
nonlocal theoretical model for optical phase conjugation in mesoscopic
media. The comparison to the existing (local) descriptions of the
degenerate four-wave mixing response can be made by taking the local
limit of our nonlocal response tensor and abandoning the contributions
stemming from the microscopic current density of first order in
$\vec{A}$.

In the local limit, the amplitudes of the interacting vector
potentials are assumed not to vary with the spatial coordinates, and
thus the expressions for the nonlocal DFWM conductivity tensor can be
integrated over the $\vec{r}\,'''$, $\vec{r}\,''$, and $\vec{r}\,'$
spaces to obtain the local DFWM conductivity tensor, i.e.,
\begin{equation}
 \tensor{\Xi}(\vec{r})=
 \iiint\tensor{\Xi}(\vec{r},\vec{r}\,',\vec{r}\,'',\vec{r}\,''')
 d^3r'''d^3r''d^3r'.
\end{equation}
Using an orthogonal set of wave equations, parity teaches
\begin{equation}
 \int\psi_n^*(\vec{r}\,)\psi_m(\vec{r}\,)d^3r=\delta_{nm},
\end{equation}
where $\delta_{nm}$ is the Kronecker delta. The integrals over the
current densities gives zero if the two quantum numbers are identical,
otherwise they depend on the individual wave functions. The
consequences are the following: (i) Integration over the spatial
coordinates $\vec{r}\,'''$ and $\vec{r}\,''$ in Eq.~(\ref{eq:XiAArz})
gives $n=m$, and thus $f_n=f_m$, such that part A of the DFWM
conductivity tensor vanish. (ii) In part C, given by
Eq.~(\ref{eq:XiACrz}), the effect is similar, but is here obtained
after integration over $\vec{r}\,'''$, $\vec{r}\,''$, and
$\vec{r}\,'$. (iii) In part E, integration over $\vec{r}\,'''$ and
$\vec{r}\,''$ in Eq.~(\ref{eq:XiAErz}) makes two terms disappear
immediately, and an inspection of the remaining two terms shows that
they are of equal magnitude, but with opposite sign, ultimately
cancelling the rest of part E. (iv) In part F, integration over
$\vec{r}\,'''$ and $\vec{r}\,'$ in Eq.~(\ref{eq:XiAFrz}) gives a
result similar in consequences as for part E. Thus, parts A, C, E, and
F of the DFWM conductivity tensor are inherently nonlocal, while parts
B, D, and G also contributes to the response in the local limit.

Abandoning parts B and D of the DFWM conductivity tensor because they
are based on the response of the microscopic current density of first
order in $\vec{A}$, we conclude that only the local contribution from
part G is included in the previous descriptions of the DFWM response
\cite{Bloembergen:78:1}, as postulated on page \pageref{Local}.

The single-electron model for degenerate four-wave mixing established
in the previous chapters can be used to study the four-wave mixing
response from a number of different materials. For example, one could
study (i) systems built from molecules or atoms with no electronic
overlap (dielectrics), in which case the response from each molecule
(atom) can be found separately. The coupling between the individual
molecules would then be described using electromagnetic propagators.
Another approach (ii) can be taken for studies of the response from
metals, where a number of electrons from each ion in the metallic
structure is shared with the other ions in a free-electron-like cloud,
or (iii) one could study semiconductors, in which the behaviour of the
electrons are strongly coupled.

In the present work, we will concentrate on the metallic case, and we
proceed to give a simplified description of potential interest for
DFWM in mesoscopic films and in near-field optics.

\part[DFWM in quantum well structures]{Degenerate four-wave mixing \\
  in quantum well structures}\label{part:III}
\newpage
\thispagestyle{plain}
\newpage

\chapter[DFWM in 2-D translational invariant media]{DFWM in
  two-dimensionally translational invariant
  media}\label{sec:Film}\label{sec:pot}
After having established and analyzed the DFWM conductivity response
in its most general form we now turn the attention towards the
specific case in which the medium under consideration {\it
  effectively}\/ exhibits translational invariance in two directions,
say $x$ and $y$ in a Cartesian $(x,y,z)$ coordinate system. We study
such a case because it appears to be of particular importance for
optical phase conjugation (i) in mesoscopic films (quantum wells), and
(ii) related to evanescent waves in near-field optics. In neither of
these cases a microscopic theory exists today to our knowledge. For
mesoscopic films the dynamics perpendicular to the film plane (here,
the $x$-$y$-plane) has to be treated from a microscopic nonlocal point
of view, whereas the dynamics in the plane of the film often is well
modelled by a local conductivity (dielectric) function. In the
following we assume for simplicity that the electron motion in the
plane of the film is free-electron-like. It is possible to replace the
free-electron-like behaviour with extended Bloch-function (or
tight-binding) dynamics if necessary but we shall not do this here,
since after all, in the local limit only matrix elements are changed
in the oscillator model when the free-electron dynamics is replaced by
a more complicated one. In the optical near-field case where
evanescent waves with extremely small penetration depths in say the
$z$-direction appear, it is crucial to keep the microscopic dynamics
perpendicular to the surface of the phase conjugating mirror when
calculating the DFWM response. So far, four-wave mixing in media with
two-dimensional translational invariance has only been studied in the
context of phase conjugation of electromagnetic surface waves
(\citeNP{Fukui:78:1,Ujihara:82:1}, \citeyearNP{Ujihara:82:2}), and of
a bulk wave by surface waves
\cite{Zeldovich:80:1,Ujihara:83:1,Stegeman:83:1,Ninzi:84:1,Mamaev:84:1,Mukhin:85:1,Arutyunyan:87:1,Pilipetskii:87:1}.
In these investigations macroscopic approaches was used.

\section{General DFWM response}
The assumed two-dimensional translational invariance against
displacements parallel to the $x$-$y$-plane makes it natural to
express the various vector and tensor quantities in a mixed Fourier
representation. Thus, by a Fourier analysis in the $x$- and
$y$-coordinates, the vector potential is
\begin{equation}
 \vec{A}(z,\vec{r}_{\|})={1\over(2\pi)^2}\int
 \vec{A}(z;\vec{q}_{\|})e^{{\rm{i}}\vec{q}_{\|}\cdot\vec{r}_{\|}}d^2q_{\|},
\label{eq:Azr}
\end{equation}
where $\vec{q}_{\|}=(q_x,q_y,0)$ and $\vec{r}_{\|}=(x,y,0)$. Likewise,
the inverse relation reads for the current density of order $\alpha$
and linear in the cyclic frequency $\omega$
\begin{equation}
 \vec{J}^{\,(\alpha)}_{-\omega}(z;\vec{q}_{\|})=
 \int\vec{J}^{\,(\alpha)}_{-\omega}(z,\vec{r}_{\|})
 e^{-{\rm{i}}\vec{q}_{\|}\cdot\vec{r}_{\|}}d^2r_{\|}.
\label{eq:Jzq}
\end{equation}
In the mixed Fourier representation the relevant constitutive
relations takes the form
\begin{eqnarray}
\lefteqn{
 \vec{J}_{-\omega}^{\,(1)}(z;\vec{q}_{\|})={{\rm{i}}\omega}
 \int\stensor{\sigma}(z,z';\vec{q}_{\|})\cdot\vec{A}(z';\vec{q}_{\|})dz',
}\label{eq:J1zq}
\\
\lefteqn{
 \vec{J}_{-\omega}^{\,(3)}(z;\vec{q}_{\|})={({\rm{i}}\omega)^3\over(2\pi)^4}
 \int\cdots\int\tensor{\Xi}(z,z',z'',z''';\vec{q}_{\|},\vec{q}_{\|}^{\,\prime},
 \vec{q}_{\|}^{\,\prime\prime},\vec{q}_{\|}^{\,\prime\prime\prime})
}\nonumber\\ &\quad&
 \vdots\,
 \vec{A}(z''';\vec{q}_{\|}^{\,\prime\prime\prime})
 \vec{A}(z'';\vec{q}_{\|}^{\,\prime\prime})
 \vec{A}^{*}(z';\vec{q}_{\|}^{\,\prime})
 d^2q_{\|}'''d^2q_{\|}''d^2q_{\|}'
 dz'''dz''dz'.
\label{eq:J3zq}
\end{eqnarray}
Due to the manner in which the nonlinear conductivity response tensor
was constructed in Chapter~\ref{ch:5}, the various components parallel
to the $x$-$y$-plane are not completely independent but satisfy the
momentum conservation criterion
\begin{equation}
 \vec{q}_{\|}^{\,\prime\prime\prime}+\vec{q}_{\|}^{\,\prime\prime}
  -\vec{q}_{\|}^{\,\prime}-\vec{q}_{\|}=\vec{0}.
\label{eq:momentum}
\end{equation}
In passing we stress again that Eq.~(\ref{eq:momentum}) is {\it not}\/
an extra condition put on the dynamics, the equation is {\it
  derived}\/ from the general theory [see Appendix~\ref{ch:ConCalc}].
To study the phase conjugated response originating in the mixing of
three incoming waves one must choose for the fields of the two pump
waves, denoted by $(1)$ and $(2)$, the vector potentials with the
double and triple primes in Eq.~(\ref{eq:J3zq}). The incoming probe
field [indexed $(p)$] is represented via the vector potential with the
single prime.

\section{Phase conjugation DFWM response}
So far, we have not utilized the translational invariance condition on
the properties of the medium. We do this first indirectly by assuming
that each of the three incoming electromagnetic fields contains only
one plane-wave component parallel to the $x$-$y$-plane. Further
limiting our study to the case where the DFWM response becomes the
phase conjugated response, i.e., the wavevector of the response must
be counterpropagating to the probe field, conservation of
pseudomomentum requires that the two pump fields are
counterpropagating.  Thus we take for the pump fields
\begin{eqnarray}
 \vec{A}(z''';\vec{q}_{\|}^{\,\prime\prime\prime})
 &\equiv&\vec{A}(z''';-\vec{k}_{\|})
 \delta(\vec{q}_{\|}^{\,\prime\prime\prime}+\vec{k}_{\|}),
\label{eq:Apump-1a}\\
 \vec{A}(z'';\vec{q}_{\|}^{\,\prime\prime})
 &\equiv&\vec{A}(z'';\vec{k}_{\|})
 \delta(\vec{q}_{\|}^{\,\prime\prime}-\vec{k}_{\|}),
\label{eq:Apump-2a}
\end{eqnarray}
where $\vec{k}_{\|}$ is the common wavevector for the two pump
fields. With these substitutions we can perform the integrals over
$q_{\|}'''$ and $q_{\|}''$ in Eq.~(\ref{eq:J3zq}), and the
conservation of pseudo-momentum is reduced from its general degenerate
four-wave mixing form, $\vec{q}_{\|}^{\,\prime\prime\prime}
+\vec{q}_{\|}^{\,\prime\prime}-\vec{q}_{\|}^{\,\prime}-\vec{q}_{\|}=\vec{0}$,
to $\vec{q}_{\|}^{\,\prime}+\vec{q}_{\|}=\vec{0}$. This allows us also
to solve the integral over $q_{\|}'$ in Eq.~(\ref{eq:J3zq}).

In relation to the conventional theory of three-dimensional (bulk)
phase conjugation, the relation
$\vec{q}_{\|}^{\,\prime}+\vec{q}_{\|}=\vec{0}$ expresses the fact that
the two-dimensional wavevector $\vec{q}_{\|}$ of the phase conjugated
field is equal in magnitude to the two-dimensional probe wavevector
($\vec{q}_{\|}^{\,\prime}$) but points in the opposite direction.
Using the aforementioned criteria, the nonlinear constitutive equation
is reduced to the form
\begin{eqnarray}
\lefteqn{
 \vec{J}_{-\omega}^{\,(3)}(z;\vec{q}_{\|})=
 {({\rm{i}}\omega)^3\over(2\pi)^4}\int\int\int
 \tensor{\Xi}(z,z',z'',z''';\vec{q}_{\|},\vec{k}_{\|})
}\nonumber\\ &\quad&
 \vdots\,
 \vec{A}(z''';-\vec{k}_{\|})\vec{A}(z'';\vec{k}_{\|})
 \vec{A}^{*}(z';-\vec{q}_{\|})
 dz'''dz''dz'+\mbox{i.t.},
\label{eq:J3-wzq}
\end{eqnarray}
where appropriate integration over
$\vec{q}_{\|}^{\,\prime\prime\prime}$,
$\vec{q}_{\|}^{\,\prime\prime}$, and $\vec{q}_{\|}^{\,\prime}$ has
been performed. The term ``i.t.''  denotes the so-called
``interchanged term''. This term is obtained from the first one by
interchanging the two pump fields. The reason that such a term has to
be added arises from the fact that each of the vector potentials
basically consists of a sum of all three incoming fields, and that the
phase conjugated term from the product of the three vector potentials
thus must include both permutations of the pump fields. The new phase
conjugation DFWM (PCDFWM) conductivity tensor appearing after
integration over $\vec{q}_{\|}^{\,\prime\prime\prime}$,
$\vec{q}_{\|}^{\,\prime\prime}$, and $\vec{q}_{\|}^{\,\prime}$ is
denoted $\tensor{\Xi}(z,z',z'',z''';\vec{q}_{\|},\vec{k}_{\|})$.

In order to calculate the nonlinear conductivity tensor
$\tensor{\Xi}(z,z',z'',z''';\vec{q}_{\|},\vec{k}_{\|})$ in the mixed
Fourier representation [as well as the linear one,
$\stensor{\sigma}(z,z';\vec{q}_{\|})$], we begin by looking at the
energy eigenstates for the light-unperturbed Schr{\"o}dinger equation.
Hence, since the potential energy of the individual electrons is
independent of $x$ and $y$, i.e., $V(\vec{r}\,)=V(z)$ under our
translational invariance assumption, the basis set may be taken in the
generic form
\begin{equation}
 \psi_{n}(z,\vec{r}_{\|})
 \equiv\psi_{n,\vec{\kappa}_{\|}}(z,\vec{r}_{\|})
 ={1\over2\pi}\psi_{n}(z)e^{{\rm{i}}\vec{\kappa}_{\|}\cdot\vec{r}_{\|}}
\label{eq:eigenstate}
\end{equation}
where $\vec{\kappa}_{\|}=(\kappa_x,\kappa_y,0)$ is the wavevector
describing the free-particle motion perpendicular to the
$z$-direction. For a medium of macroscopic extension in the $x$- and
$y$-directions, the set of wavevectors commonly denoted by
$\vec{\kappa}_{\|}$ forms a two-dimensional quasi-continuum. Albeit
the index $n$ in the wave function $\psi_{n}(z,\vec{r}_{\|})$ stands
for a triple set of quantum numbers we also use this index to classify
the various wave function parts, $\psi_{n}(z)$, belonging to the
single indexed $z$-dynamics. In a readily understandable notation the
energy eigenstates, ${\cal{E}}_{\,n}$, associated with the generic
solution in Eq.~(\ref{eq:eigenstate}) is
\begin{equation}
 {\cal{E}}_{\,n}=\varepsilon_{n}
 +{\hbar^2\over2m_{e}}|\vec{\kappa}_{\|}|^2,
\end{equation}
where we have introduced $\varepsilon_{n}$ as the energy of state $n$
in the solution dependent on the $z$-coordinate only. In the view of
the abovementioned considerations the cyclic transition frequency
becomes
\begin{equation}
 \omega_{nm}={1\over\hbar}\left[\varepsilon_{n}-\varepsilon_{m}+
 {\hbar^2\over2m_{e}}\left(|\vec{\kappa}_{\|,\bar{n}}|^2
 -|\vec{\kappa}_{\|,\bar{m}}|^2\right)\right],
\label{eq:transition-w}
\end{equation}
in a notation where adequate subscripts $\bar{n}$ and $\bar{m}$ have
been put on the wavevectors.  In abbreviated form the complex
transition frequency, which includes the relaxation time, is for the
sake of the following analysis written in the form
\begin{equation}
 \tilde{\omega}_{nm}=\tilde{\omega}_{nm}(\vec{\kappa}_{\|,\bar{n}},
 \vec{\kappa}_{\|,\bar{m}}),
\label{eq:transitionw}
\end{equation}
omitting the reference to $\varepsilon_{n}$ and $\varepsilon_{m}$,
since this is already implicitly given by the $nm$ subscript. The
Fermi-Dirac distribution function we also present in an abbreviated
form, viz.
\begin{equation}
 f_{n}({\cal{E}}_{\,n})=f_{n}\left(\varepsilon_{n}
  +{\hbar^2\kappa_{\|,\bar{n}}^2\over2m_{e}}\right)
 \equiv f_{n}(\vec{\kappa}_{\|,\bar{n}}).
\label{eq:f_n}
\end{equation}
By inserting the generic solution in Eq.~(\ref{eq:eigenstate}) into
the expression for the transition current density in
Eq.~(\ref{eq:Jm->n}), we obtain
\begin{eqnarray}
\lefteqn{
 \vec{J}_{mn}(\vec{r}\,)=-{e\hbar\over2{\rm{i}}m_{e}}{1\over(2\pi)^2}
 \Biggl[{\rm{i}}(\vec{\kappa}_{\|,\bar{m}}
 +\vec{\kappa}_{\|,\bar{n}})\psi_{n}^{*}(z)\psi_{m}(z)
}\nonumber\\ &\quad&\left.\!
+\vec{e}_{z}\left(
 \psi_{n}^{*}(z){\partial\psi_{m}(z)\over\partial{}z}- 
 \psi_{m}(z){\partial\psi_{n}^{*}(z)\over\partial{}z}\right)\right]
 e^{{\rm{i}}(\vec{\kappa}_{\|,\bar{m}}-\vec{\kappa}_{\|,\bar{n}})\cdot\vec{r}_{\|}}
\nonumber\\ &&
 \equiv{1\over(2\pi)^2}
 \vec{j}_{mn}(z;\vec{\kappa}_{\|,\bar{m}}+\vec{\kappa}_{\|,\bar{n}})
 e^{{\rm{i}}(\vec{\kappa}_{\|,\bar{m}}-\vec{\kappa}_{\|,\bar{n}})\cdot\vec{r}_{\|}},
\label{eq:Jm->n||}
\label{eq:j_hnm}
\end{eqnarray}
where for convenience we have defined a new transition current density
$\vec{j}_{mn}(z;\vec{\kappa}_{\|,\bar{m}}+\vec{\kappa}_{\|,\bar{n}})$
to separate out the dependence on the Cartesian coordinates
$\vec{r}_{\|}$. For the various Cartesian components of this current
density, we use the notation
$j_{i,nm}(z;\vec{\kappa}_{\|,\bar{m}}+\vec{\kappa}_{\|,\bar{n}})$,
$i\in\{x,y,z\}$.

\section{Conductivity tensors}
The explicit expression for the phase conjugation degenerate four-wave
mixing (PCDFWM) conductivity tensor
$\tensor{\Xi}(z,z',z'',z''';\vec{q}_{\|},\vec{k}_{\|})$ is calculated
by insertion of (i) the solutions to the time-independent
Schr{\"o}dinger equation given in Eq.~(\ref{eq:eigenstate}), (ii) the
Fourier representation of the vector potential given by
Eq.~(\ref{eq:Azr}), and (iii) the new form of the transition current
given in Eq.~(\ref{eq:Jm->n||}) into the nonlinear DFWM constitutive
relation in Eq.~(\ref{eq:J3r}) with the phase conjugation conductivity
tensor in real space given by
Eqs.~(\ref{eq:XiAArz})--(\ref{eq:XiAGrz}), and thereafter inserting
the outcome of these steps into the expression for the nonlinear
current density in the mixed Fourier representation in
Eq.~(\ref{eq:Jzq}). Finally, we perform the integrals over the two
dimensions ($x$ and $y$) in real space and over relevant sets of
$\vec{\kappa}_{\|}$-states. Altogether we are left with an expression
on the form of Eq.~(\ref{eq:J3-wzq}). For the processes in
Figs.~\ref{fig:2Tr} and \ref{fig:3Tr}, the abovementioned calculations
are supplied in Appendix~\ref{ch:ConCalc}, where also the general DFWM
conductivity tensors are given. The nonzero element of the linear
conductivity tensor part A become
\begin{equation}
 {\sigma}_{xx}^{\,\rm{A}}(z,z';\vec{q}_{\|})=
 {2{\rm{i}}\over\omega}{e^2\over{}m_{e}}{1\over(2\pi)^2}
 \sum_{n}\int{}f_{n}(\vec{\kappa}_{\|})
 d^2\kappa_{\|}
 |\psi_{n}(z)|^2\delta(z-z'),
\label{eq:SigmaAA}
\end{equation}
and the nine nonzero elements of part B are
\begin{eqnarray}
\lefteqn{
 {\sigma}_{ij}^{\,\rm{B}}(z,z';\vec{q}_{\|})=
 {2{\rm{i}}\over\omega}{1\over\hbar}{1\over(2\pi)^2}\sum_{nm}\int
 {f_{n}(\vec{\kappa}_{\|}+\vec{q}_{\|})-f_{m}(\vec{\kappa}_{\|})
 \over\tilde{\omega}_{nm}(\vec{\kappa}_{\|}+\vec{q}_{\|},
 \vec{\kappa}_{\|})-\omega}
}\nonumber\\ &\quad&\times
 j_{j,mn}(z';2\vec{\kappa}_{\|}+\vec{q}_{\|})
 j_{i,nm}(z;2\vec{\kappa}_{\|}+\vec{q}_{\|})
 d^2\kappa_{\|}.
\label{eq:SigmaAB}
\end{eqnarray}
The nonzero element of the PCDFWM conductivity tensor part A are
\begin{eqnarray}
\lefteqn{
 {\Xi}_{xxxx}^{\rm{A}}(z,z',z'',z''';\vec{q}_{\|},\vec{k}_{\|})=
 {e^4\over8m_{e}^2\hbar}{1\over(2\pi)^2}
 {2{\rm{i}}\over\omega^3}
 \sum_{nm}
 \psi_{n}^{*}(z'')\psi_{m}(z'')\psi_{m}^{*}(z)\psi_{n}(z)
}\nonumber\\ &\quad&\times
 \delta(z-z')\delta(z''-z''')\int
 {f_{n}(\vec{\kappa}_{\|})-f_{m}(\vec{\kappa}_{\|})
 \over\tilde{\omega}_{nm}(\vec{\kappa}_{\|},\vec{\kappa}_{\|})-2\omega}
 d^2\kappa_{\|},
\label{eq:XiAA}
\end{eqnarray}
and the nine nonzero elements of part B become
\begin{eqnarray}
\lefteqn{
 {\Xi}_{xxkh}^{\rm{B}}(z,z',z'',z''';\vec{q}_{\|},\vec{k}_{\|})=
 {e^2\over4m_{e}\hbar^2}{1\over(2\pi)^2}
 {2{\rm{i}}\over\omega^3}\sum_{nmv}\psi_{m}^{*}(z)\psi_{n}(z)\delta(z-z')
}\nonumber\\ &&\times
 \int
 {1\over\tilde{\omega}_{nm}(\vec{\kappa}_{\|},\vec{\kappa}_{\|})-2\omega}
 \left({f_{m}(\vec{\kappa}_{\|})-f_{v}(\vec{\kappa}_{\|}+\vec{k}_{\|})
 \over\tilde{\omega}_{v{}m}(\vec{\kappa}_{\|}+\vec{k}_{\|},
 \vec{\kappa}_{\|})-\omega}
 +{f_{n}(\vec{\kappa}_{\|})-f_{v}(\vec{\kappa}_{\|}+\vec{k}_{\|})\over
 \tilde{\omega}_{nv}(\vec{\kappa}_{\|},
 \vec{\kappa}_{\|}+\vec{k}_{\|})-\omega}\right)
\nonumber\\ &&\times
 j_{h,v{}n}(z''';2\vec{\kappa}_{\|}+\vec{k}_{\|})
 j_{k,mv}(z'';2\vec{\kappa}_{\|}+\vec{k}_{\|})
 d^2\kappa_{\|}.
\label{eq:XiAB}
\end{eqnarray}
In part C of the nonlinear conductivity tensor the nonzero element is
\begin{eqnarray}
\lefteqn{
 {\Xi}_{xxxx}^{\rm{C}}(z,z',z'',z''';\vec{q}_{\|},\vec{k}_{\|})=
 {e^4\over4m_{e}^2\hbar}{1\over(2\pi)^2}
 {2{\rm{i}}\over\omega^3}\sum_{nm}
 \psi_{n}^{*}(z')\psi_{m}(z')\psi_{m}^{*}(z)\psi_{n}(z)
}\nonumber\\ &&\times
 \delta(z'-z''')\delta(z-z'')\int
 {f_{n}(\vec{\kappa}_{\|}-\vec{k}_{\|}+\vec{q}_{\|})
 -f_{m}(\vec{\kappa}_{\|})\over\tilde{\omega}_{nm}(\vec{\kappa}_{\|}
 -\vec{k}_{\|}+\vec{q}_{\|},\vec{\kappa}_{\|})}
 d^2\kappa_{\|},
\label{eq:XiAC}
\end{eqnarray}
and the nine nonzero tensor elements in part D become
\begin{eqnarray}
\lefteqn{
 {\Xi}_{xjkx}^{\rm{D}}(z,z',z'',z''';\vec{q}_{\|},\vec{k}_{\|})=
 {e^2\over4m_{e}\hbar^2}{1\over(2\pi)^2}
 {2{\rm{i}}\over\omega^3}\sum_{nmv} \psi_{m}^{*}(z)\psi_{n}(z)
 \delta(z-z''')
}\nonumber\\ &&\times
 \int
 {1\over\tilde{\omega}_{nm}(\vec{\kappa}_{\|}+\vec{k}_{\|}+\vec{q}_{\|}
 ,\vec{\kappa}_{\|})}
\left\{
 \left({f_{m}(\vec{\kappa}_{\|})-f_{v}(\vec{\kappa}_{\|}+\vec{k}_{\|})
 \over\tilde{\omega}_{v{}m}(\vec{\kappa}_{\|}+\vec{k}_{\|},
 \vec{\kappa}_{\|})-\omega}
\right.\right.\nonumber\\ &&\left.\!
 +{f_{n}(\vec{\kappa}_{\|}+\vec{k}_{\|}+\vec{q}_{\|})
 -f_{v}(\vec{\kappa}_{\|}+\vec{k}_{\|})\over
 \tilde{\omega}_{nv}(\vec{\kappa}_{\|}+\vec{k}_{\|}+\vec{q}_{\|},
 \vec{\kappa}_{\|}+\vec{k}_{\|})+\omega}\right)
 j_{j,v{}n}(z';2\vec{\kappa}_{\|}+2\vec{k}_{\|}+\vec{q}_{\|})
 j_{k,mv}(z'';2\vec{\kappa}_{\|}+\vec{k}_{\|})
\nonumber\\ &&
 +
 \left({f_{m}(\vec{\kappa}_{\|})-f_{v}(\vec{\kappa}_{\|}+\vec{q}_{\|})
 \over\tilde{\omega}_{v{}m}(\vec{\kappa}_{\|}+\vec{q}_{\|},
 \vec{\kappa}_{\|})+\omega}
 +{f_{n}(\vec{\kappa}_{\|}+\vec{k}_{\|}+\vec{q}_{\|})
 -f_{v}(\vec{\kappa}_{\|}+\vec{q}_{\|})\over
 \tilde{\omega}_{nv}(\vec{\kappa}_{\|}+\vec{k}_{\|}+\vec{q}_{\|},
 \vec{\kappa}_{\|}+\vec{q}_{\|})-\omega}\right)
\nonumber\\ &&\times
 j_{k,v{}n}(z'';2\vec{\kappa}_{\|}+\vec{k}_{\|}+2\vec{q}_{\|})
 j_{j,mv}(z';2\vec{\kappa}_{\|}+\vec{q}_{\|})
 \Bigr\}
 d^2\kappa_{\|}.
\label{eq:XiAD}
\end{eqnarray}
The nonlinear conductivity tensor part E has the nine nonzero elements
\begin{eqnarray}
\lefteqn{
 {\Xi}_{ijxx}^{\rm{E}}(z,z',z'',z''';\vec{q}_{\|},\vec{k}_{\|})=
 {e^2\over16m_{e}\hbar^2}{1\over(2\pi)^2}
 {2{\rm{i}}\over\omega^3}\sum_{nmv}\delta(z''-z''')\int
 {1\over\tilde{\omega}_{nm}(\vec{\kappa}_{\|}+\vec{q}_{\|},\vec{\kappa}_{\|})
 -\omega}
}\nonumber\\ &&\times
\!\left\{\!
 \left(\!{f_{m}(\vec{\kappa}_{\|})-f_{v}(\vec{\kappa}_{\|})
 \over\tilde{\omega}_{v{}m}(\vec{\kappa}_{\|},\vec{\kappa}_{\|})-2\omega}
 +{f_{n}(\vec{\kappa}_{\|}+\vec{q}_{\|})-f_{v}(\vec{\kappa}_{\|})\over
 \tilde{\omega}_{nv}(\vec{\kappa}_{\|}+\vec{q}_{\|},\vec{\kappa}_{\|})
 +\omega}\right)
 j_{j,v{}n}(z';2\vec{\kappa}_{\|}+\vec{q}_{\|})\psi_{v}^{*}(z'')\psi_{m}(z'')
\right.\nonumber\\ &&
 +
 \left({f_{m}(\vec{\kappa}_{\|})-f_{v}(\vec{\kappa}_{\|}+\vec{q}_{\|})\over
 \tilde{\omega}_{v{}m}(\vec{\kappa}_{\|}+\vec{q}_{\|},\vec{\kappa}_{\|})
 +\omega}+{f_{n}(\vec{\kappa}_{\|}+\vec{q}_{\|})
 -f_{v}(\vec{\kappa}_{\|}+\vec{q}_{\|})\over
 \tilde{\omega}_{nv}(\vec{\kappa}_{\|}+\vec{q}_{\|},\vec{\kappa}_{\|}
 +\vec{q}_{\|})-2\omega}\right)
\nonumber\\ &&\times
 j_{j,mv}(z';2\vec{\kappa}_{\|}+\vec{q}_{\|})
 \psi_{n}^{*}(z'')\psi_{v}(z'')
 \Bigr\}j_{i,nm}(z;2\vec{\kappa}_{\|}+\vec{q}_{\|})d^2\kappa_{\|}.
\label{eq:XiAE}
\end{eqnarray}
Part F also has nine nonzero elements, which are
\begin{eqnarray}
\lefteqn{
 {\Xi}_{ixxh}^{\rm{F}}(z,z',z'',z''';\vec{q}_{\|},\vec{k}_{\|})=
 {e^2\over8m_{e}\hbar^2}{1\over(2\pi)^2}
 {2{\rm{i}}\over\omega^3}\sum_{nmv}\delta(z'-z''')\int
 {1\over\tilde{\omega}_{nm}(\vec{\kappa}_{\|}+\vec{q}_{\|},\vec{\kappa}_{\|})
 -\omega}
}\nonumber\\ &&\times
\left\{
 \left({f_{m}(\vec{\kappa}_{\|})-f_{v}(\vec{\kappa}_{\|}-\vec{k}_{\|}
 +\vec{q}_{\|})\over\tilde{\omega}_{v{}m}(\vec{\kappa}_{\|}
 -\vec{k}_{\|}+\vec{q}_{\|},\vec{\kappa}_{\|})}
 +{f_{n}(\vec{\kappa}_{\|}+\vec{q}_{\|})
 -f_{v}(\vec{\kappa}_{\|}-\vec{k}_{\|}+\vec{q}_{\|})\over
 \tilde{\omega}_{nv}(\vec{\kappa}_{\|}+\vec{q}_{\|},\vec{\kappa}_{\|}
 -\vec{k}_{\|}+\vec{q}_{\|})-\omega}\right)
\right.\nonumber\\ &&\times
 j_{h,v{}n}(z'';2\vec{\kappa}_{\|}-\vec{k}_{\|}+\vec{q}_{\|})
 \psi_{v}^{*}(z')\psi_{m}(z')
\nonumber\\ &&
 +
 \left(
 {f_{m}(\vec{\kappa}_{\|})-f_{v}(\vec{\kappa}_{\|}+\vec{k}_{\|})\over
 \tilde{\omega}_{v{}m}(\vec{\kappa}_{\|}+\vec{k}_{\|},
 \vec{\kappa}_{\|})-\omega}+{f_{n}(\vec{\kappa}_{\|}+\vec{q}_{\|})
 -f_{v}(\vec{\kappa}_{\|}+\vec{k}_{\|})\over
 \tilde{\omega}_{nv}(\vec{\kappa}_{\|}+\vec{q}_{\|},\vec{\kappa}_{\|}
 +\vec{k}_{\|})}\right)
\nonumber\\ &&\times
 j_{h,mv}(z'';2\vec{\kappa}_{\|}+\vec{k}_{\|})
 \psi_{n}^{*}(z')\psi_{v}(z')
 \Bigr\}j_{i,nm}(z;2\vec{\kappa}_{\|}+\vec{q}_{\|})
 d^2\kappa_{\|}.
\label{eq:XiAF}
\end{eqnarray}
Finally, the PCDFWM conductivity tensor part G has the eightyone
nonzero elements
\begin{eqnarray}
\lefteqn{
 {\Xi}_{ijkh}^{\rm{G}}(z,z',z'',z''';\vec{q}_{\|},\vec{k}_{\|})=
 {1\over8\hbar^3}{1\over(2\pi)^2}
 {2{\rm{i}}\over\omega^3}\sum_{nmvl}\int
 {1\over\tilde{\omega}_{nm}(\vec{\kappa}_{\|}+\vec{q}_{\|},\vec{\kappa}_{\|})
 -\omega}
}\nonumber\\ &&\times
\left\{
 \left[
 \left({f_{l}(\vec{\kappa}_{\|}-\vec{k}_{\|})
 -f_{m}(\vec{\kappa}_{\|})\over\tilde{\omega}_{l{}m}(\vec{\kappa}_{\|}
 -\vec{k}_{\|},\vec{\kappa}_{\|})-\omega}
 +{f_{l}(\vec{\kappa}_{\|}-\vec{k}_{\|})-f_{v}(\vec{\kappa}_{\|})
 \over\tilde{\omega}_{vl}(\vec{\kappa}_{\|},
 \vec{\kappa}_{\|}-\vec{k}_{\|})-\omega}\right)
 {1\over\tilde{\omega}_{v{}m}(\vec{\kappa}_{\|},\vec{\kappa}_{\|})-2\omega}
\right.\right.\nonumber\\ &&\left.
 +\left({f_{l}(\vec{\kappa}_{\|}-\vec{k}_{\|})
 -f_{v}(\vec{\kappa}_{\|})\over
 \tilde{\omega}_{vl}(\vec{\kappa}_{\|},
 \vec{\kappa}_{\|}-\vec{k}_{\|})-\omega}
 +{f_{n}(\vec{\kappa}_{\|}+\vec{q}_{\|})-f_{v}(\vec{\kappa}_{\|})\over
 \tilde{\omega}_{nv}(\vec{\kappa}_{\|}+\vec{q}_{\|},\vec{\kappa}_{\|})
 +\omega}\right)
 {1\over\tilde{\omega}_{nl}(\vec{\kappa}_{\|}+\vec{q}_{\|},\vec{\kappa}_{\|}-\vec{k}_{\|})}
 \right]
\nonumber\\ &&\times
 j_{h,ml}(z''';2\vec{\kappa}_{\|}-\vec{k}_{\|})
 j_{k,lv}(z'';2\vec{\kappa}_{\|}-\vec{k}_{\|})
 j_{j,v{}n}(z';2\vec{\kappa}_{\|}+\vec{q}_{\|})
\nonumber\\ &&
 +
 \left[
 \left({f_{l}(\vec{\kappa}_{\|}-\vec{k}_{\|})
 -f_{m}(\vec{\kappa}_{\|})\over\tilde{\omega}_{l{}m}(\vec{\kappa}_{\|}
 -\vec{k}_{\|},\vec{\kappa}_{\|})-\omega}
 +{f_{l}(\vec{\kappa}_{\|}-\vec{k}_{\|})-f_{v}(\vec{\kappa}_{\|}
 -\vec{k}_{\|}+\vec{q}_{\|})\over
 \tilde{\omega}_{vl}(\vec{\kappa}_{\|}-\vec{k}_{\|}+\vec{q}_{\|},
 \vec{\kappa}_{\|}-\vec{k}_{\|})+\omega}\right)
\right.\nonumber\\ &&\times
 {1\over\tilde{\omega}_{v{}m}(\vec{\kappa}_{\|}-\vec{k}_{\|}
 +\vec{q}_{\|},\vec{\kappa}_{\|})}
 +\left({f_{l}(\vec{\kappa}_{\|}-\vec{k}_{\|})
 -f_{v}(\vec{\kappa}_{\|}-\vec{k}_{\|}+\vec{q}_{\|})\over
 \tilde{\omega}_{vl}(\vec{\kappa}_{\|}-\vec{k}_{\|}+\vec{q}_{\|},
 \vec{\kappa}_{\|}-\vec{k}_{\|})+\omega}
\right.\nonumber\\ &&\left.\!\left.\!
 +{f_{n}(\vec{\kappa}_{\|}+\vec{q}_{\|})
 -f_{v}(\vec{\kappa}_{\|}-\vec{k}_{\|}+\vec{q}_{\|})\over
 \tilde{\omega}_{nv}(\vec{\kappa}_{\|}+\vec{q}_{\|},\vec{\kappa}_{\|}
 -\vec{k}_{\|}+\vec{q}_{\|})-\omega}
 \right)
 {1\over\tilde{\omega}_{nl}(\vec{\kappa}_{\|}
 +\vec{q}_{\|},\vec{\kappa}_{\|}-\vec{k}_{\|})}
 \right]
\nonumber\\ &&\times
 j_{h,ml}(z''';2\vec{\kappa}_{\|}-\vec{k}_{\|})
 j_{k,v{}n}(z'';2\vec{\kappa}_{\|}-\vec{k}_{\|}+2\vec{q}_{\|})
 j_{j,lv}(z';2\vec{\kappa}_{\|}-2\vec{k}_{\|}+\vec{q}_{\|})
\nonumber\\ &&
 +
 \left[
 \left({f_{l}(\vec{\kappa}_{\|}+\vec{q}_{\|})-f_{m}(\vec{\kappa}_{\|})
 \over\tilde{\omega}_{l{}m}(\vec{\kappa}_{\|}+\vec{q}_{\|},
 \vec{\kappa}_{\|})+\omega}
 +{f_{l}(\vec{\kappa}_{\|}+\vec{q}_{\|})-f_{v}(\vec{\kappa}_{\|}
 -\vec{k}_{\|}+\vec{q}_{\|})\over
 \tilde{\omega}_{vl}(\vec{\kappa}_{\|}-\vec{k}_{\|}+\vec{q}_{\|},
 \vec{\kappa}_{\|}+\vec{q}_{\|})-\omega}\right)
\right.\nonumber\\ &&\times
 {1\over\tilde{\omega}_{v{}m}(\vec{\kappa}_{\|}-\vec{k}_{\|}
 +\vec{q}_{\|},\vec{\kappa}_{\|})}
 +\left({f_{l}(\vec{\kappa}_{\|}+\vec{q}_{\|})
 -f_{v}(\vec{\kappa}_{\|}-\vec{k}_{\|}+\vec{q}_{\|})\over
 \tilde{\omega}_{vl}(\vec{\kappa}_{\|}-\vec{k}_{\|}+\vec{q}_{\|},
 \vec{\kappa}_{\|}+\vec{q}_{\|})-\omega}
\right.\nonumber\\ &&\left.\!\left.\!
 +{f_{n}(\vec{\kappa}_{\|}+\vec{q}_{\|})
 -f_{v}(\vec{\kappa}_{\|}-\vec{k}_{\|}+\vec{q}_{\|})\over
 \tilde{\omega}_{nv}(\vec{\kappa}_{\|}+\vec{q}_{\|},\vec{\kappa}_{\|}
 -\vec{k}_{\|}+\vec{q}_{\|})-\omega}\right)
 {1\over\tilde{\omega}_{nl}(\vec{\kappa}_{\|}
 +\vec{q}_{\|},\vec{\kappa}_{\|}+\vec{q}_{\|})-2\omega}
 \right]
\nonumber\\ &&\times
 j_{h,lv}(z''';2\vec{\kappa}_{\|}-\vec{k}_{\|}+2\vec{q}_{\|})
 j_{k,v{}n}(z'';2\vec{\kappa}_{\|}-\vec{k}_{\|}+2\vec{q}_{\|})
 j_{j,ml}(z';2\vec{\kappa}_{\|}+\vec{q}_{\|})
\Bigr\}
\nonumber\\ &&\times
 j_{i,nm}(z;2\vec{\kappa}_{\|}+\vec{q}_{\|})
 d^2\kappa_{\|}.
\label{eq:XiAG}
\end{eqnarray}
In Eqs.~(\ref{eq:SigmaAA})--(\ref{eq:XiAG}) above we have dropped the
now superfluous index on $\vec{\kappa}_{\|}$.

\section{Phase conjugated field}
After having sketched the calculation of the nonlinear DFWM response
we turn our attention to the phase conjugated electric field. In the
present case where the main parts of the interaction takes place in
very small interaction volumes, we can expect that the generated phase
conjugated field does not affect the dynamics of the pump and probe
fields much, and thus take the parametric approximation. 

Then the loop equation in Eq.~(\ref{eq:Inte}) is reduced to the
single-coordinate form in the two-dimensional phase matching case
\cite{Keller:96:1}
\begin{eqnarray}
\lefteqn{
 \vec{E}_{\rm{PC}}(z;\vec{q}_{\|},\omega)=
 \vec{E}_{\rm{PC}}^{\rm{B}}(z;\vec{q}_{\|},\omega)
}\nonumber\\ &\quad&
 -{\rm{i}}\mu_{0}\omega\int\int
 \tensor{G}(z,z'';\vec{q}_{\|},\omega)\cdot
 \stensor{\sigma}(z'',z';\vec{q}_{\|},\omega)\cdot
 \vec{E}_{\rm{PC}}(z';\vec{q}_{\|},\omega)dz''dz',
\label{eq:Loophole}
\label{eq:loop}
\end{eqnarray}
possibly with $\tensor{G}(z,z'';\vec{q}_{\|},\omega)$ replaced by
$\tensor{G}_{0}(z,z'';\vec{q}_{\|},\omega)$. In the quantum-well case
the explicit form of $\tensor{G}(z,z'';\vec{q}_{\|},\omega)$ is known
\cite{Bagchi:79:1}, and also
$\tensor{G}_{0}(z,z'';\vec{q}_{\|},\omega)$, adequate in near-field
optics, is of course known. For few (one, two, three, \dots)-level
quantum wells several schemes exist for the handling of the integral
equation problem in Eq.~(\ref{eq:Loophole}), cf., e.g.,
\citeN{Keller:96:1}. The only factor which in the parametric
approximation makes the DFWM loop problem different from those
hitherto investigated is the background field. In the present case
this is given by
\begin{equation}
 \vec{E}_{\rm{PC}}^{\rm{B}}(z;\vec{q}_{\|},\omega)=-{\rm{i}}\mu_{0}\omega\int
 \tensor{G}(z,z';\vec{q}_{\|},\omega)\cdot
 \vec{J}_{-\omega}^{\,(3)}(z';\vec{q}_{\|},\omega)dz',
\end{equation}
with $\vec{J}_{-\omega}^{\,(3)}(z';\vec{q}_{\|},\omega)$ taken from
Eq.~(\ref{eq:J3-wzq}) in the case of simple two-dimensional plane-wave
mixing, or in general from Eq.~(\ref{eq:J3zq}).

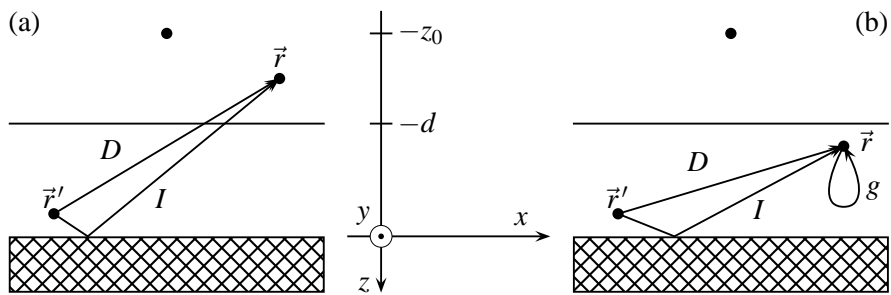
\begin{figure}[tb]
\setlength{\unitlength}{1.5mm}
\psset{unit=1.5mm}
\begin{center}
\begin{pspicture}(0,0)(80,25)
 \psline[linewidth=0.25mm]{-}(1,15)(29,15)
 \psframe[linewidth=0.25mm,fillstyle=crosshatch](1,0)(29,5)
 \psline[linewidth=0.25mm,linearc=0mm]{->}(5,7)(8,5)(25,19)
 \psline[linewidth=0.25mm]{->}(5,7)(25,19)
 \qdisk(5,7){0.5}
 \qdisk(25,19){0.5}
 \put(5,7.5){\makebox(0,0)[b]{$\vec{r}\,'$}}
 \put(25,20){\makebox(0,0)[b]{$\vec{r}$}}
 \put(11,13.5){\makebox(0,0)[tr]{$D$}}
 \put(14,7.5){\makebox(0,0)[bl]{$I$}}
 \put(1,25){\makebox(0,0)[tl]{(a)}}
 \psline[linewidth=0.25mm]{-}(51,15)(79,15)
 \psframe[linewidth=0.25mm,fillstyle=crosshatch](51,0)(79,5)
 \psline[linewidth=0.25mm,linearc=0mm]{->}(55,7)(60,5)(75,13)
 \psline[linewidth=0.25mm]{->}(55,7)(75,13)                  
 \pscurve[linewidth=0.25mm]{->}(75,13)(74,8)(76,8)(75,13)
 \qdisk(55,7){0.5}
 \qdisk(75,13){0.5}
 \put(55,7.5){\makebox(0,0)[b]{$\vec{r}\,'$}}
 \put(77,13.5){\makebox(0,0){$\vec{r}$}}
 \put(63,12.5){\makebox(0,0)[tr]{$D$}}
 \put(67,6.5){\makebox(0,0)[bl]{$I$}}
 \put(77,9){\makebox(0,0)[l]{$g$}}
 \put(79,25){\makebox(0,0)[tr]{(b)}}
 \psline[linewidth=0.25mm]{->}(31,5)(49,5)
 \psline[linewidth=0.25mm]{->}(34,25)(34,0)
 \pscircle[linewidth=0.1mm,linecolor=black,fillstyle=solid,fillcolor=white](34,5){1.0}
 \qdisk(34,5){0.25}
 \put(47,6){\makebox(0,0)[rb]{$x$}}
 \put(33,6){\makebox(0,0)[rb]{$y$}}
 \put(33,0){\makebox(0,0)[rb]{$z$}}
 \psline[linewidth=0.25mm]{-}(33,15)(35,15)
 \put(35.5,13){\makebox(0,4)[l]{$-d$}}
 \psline[linewidth=0.25mm]{-}(33,23)(35,23)
 \put(35.5,21){\makebox(0,4)[l]{$-z_0$}}
 \qdisk(15,23){0.5}
 \qdisk(65,23){0.5}
\end{pspicture}
\end{center}
\caption[Electromagnetic propagators for a
vacuum/quantum-well/substrate system, and the Cartesian coordinate
system used]{The propagators appearing in the calculation of the phase
  conjugated field in the system we consider in this communication.
  The system consists of a three layer thin film structure, namely
  vacuum, film (quantum well, extending from $0$ to $-d$) and
  substrate (crosshatched). In the vacuum may be placed different
  kinds of sources, e.g., a quantum wire with its axis along the
  $y$-direction (shown as a dot). In Fig.~(a) the propagation of the
  electromagnetic field from a source point $\vec{r}\,'$ inside the
  quantum well to an observation point $\vec{r}$ outside the quantum
  well is shown, while in (b) the propagation of the electromagnetic
  field is illustrated in the case where both source and observation
  point are inside the quantum well.  $D$ is the propagation path
  described by the direct propagator, $I$ is the propagation path
  described by the indirect propagator, and $g$ denotes the self-field
  action propagator. In the center of the figure is shown the
  Cartesian coordinate system used in our calculations.\label{fig:1}}
\end{figure}

In the quantum well case, the pseudo-vacuum propagator
$\tensor{G}(z,z'';\vec{q}_{\|},\omega)$ can be written as a sum of
three terms
\begin{equation}
 \tensor{G}(z,z';\vec{q}_{\|},\omega)=
 \tensor{D}(z-z';\vec{q}_{\|},\omega)+
 \tensor{I}(z+z';\vec{q}_{\|},\omega)+
 \btensor{g}(z-z';\omega),
\label{eq:Green1}
\end{equation}
where the first two are named after the processes they describe. Thus
the term $\tensor{D}(z-z';\vec{q}_{\|},\omega)$ describes the direct
propagation of the electromagnetic field from a source point at $z'$
to the observation point at $z$. It is given by
\begin{equation}
 \tensor{D}(z-z';\vec{q}_{\|},\omega)=
 {e^{{\rm{i}}q_{\perp}|z-z'|}\over2{\rm{i}}q_{\perp}}\left[
 \vec{e}_{y}\otimes\vec{e}_{y}+\Theta(z-z')\vec{e}_{i}\otimes\vec{e}_{i}
 +\Theta(z'-z)\vec{e}_{r}\otimes\vec{e}_{r}\right].
\label{eq:Green2}
\end{equation}
\noindent The indirect term,
$\tensor{I}(z+z';\vec{q}_{\|},\omega)$, describes the propagation from
the source point of the part of the electromagnetic field that is
going to the point of observation via the surface of the bulk medium.
The expression for the indirect term reads
\begin{equation}
 \tensor{I}(z+z';\vec{q}_{\|},\omega)=
 {e^{-{\rm{i}}q_{\perp}(z+z')}\over2{\rm{i}}q_{\perp}}\left[
 r^s\vec{e}_{y}\otimes\vec{e}_{y}+r^p\vec{e}_{r}\otimes\vec{e}_{i}\right].
\label{eq:Green3}
\end{equation}
Finally, the self-field term characterizes the field generated at the
observation point by the current density at the same point. The
self-field part of the propagator is given by
\begin{equation}
 \btensor{g}(z-z';\omega)=q^{-2}\delta(z-z')\vec{e}_{z}\otimes\vec{e}_{z},
\label{eq:Green4}
\end{equation}
where $q=\omega/c_0$ is the vacuum wavenumber.  In the above
equations, $q_{\perp}=[q^2-q_{\|}^2]^{1/2}$,
$\vec{e}_{i}=q^{-1}(q_{\perp},0,-q_{\|})$, and
$\vec{e}_{r}=q^{-1}(-q_{\perp},0,-q_{\|})$, taking
$\vec{q}_{\|}=q_{\|}\vec{e}_{x}$. The quantities $r^s$ and $r^p$ are
the amplitude reflection coefficients of the vacuum/substrate
interface in the absence of the quantum well. In general these are
functions of $\vec{q}_{\|}$. The appropriate propagators for a single
quantum well system are shown in Fig.~\ref{fig:1}.

\section{Some limits of the PCDFWM conductivity tensor}

\subsection{Local limit in the $z$-coordinates}\label{sec:local-z}
In the local limit the three interacting fields are independent of the
$z$-coordinate, and thus we may calculate the local PCDFWM response
tensor as
\begin{equation}
 \tensor{\Xi}(z;\vec{q}_{\|},\vec{k}_{\|})=
 \iiint\tensor{\Xi}(z,z',z'',z''';\vec{q}_{\|},\vec{k}_{\|})dz'''dz''dz'.
\end{equation}
Since the dependence on the three coordinates $z'''$, $z''$, and $z'$
are fairly simple we may draw some conclusions directly from looking
at Eqs.~(\ref{eq:XiAA})--(\ref{eq:XiAG}). Using an orthogonal set of
wave functions, parity teaches
\begin{equation}
 \int\psi_n(z)\psi_m(z)dz=\delta_{nm},
\end{equation}
where $\delta_{nm}$ is the Kronecker delta. By inspection of
Eq.~(\ref{eq:j_hnm}), this is the type of integral appearing when
considering the $x$ and $y$ coordinates of this current density.

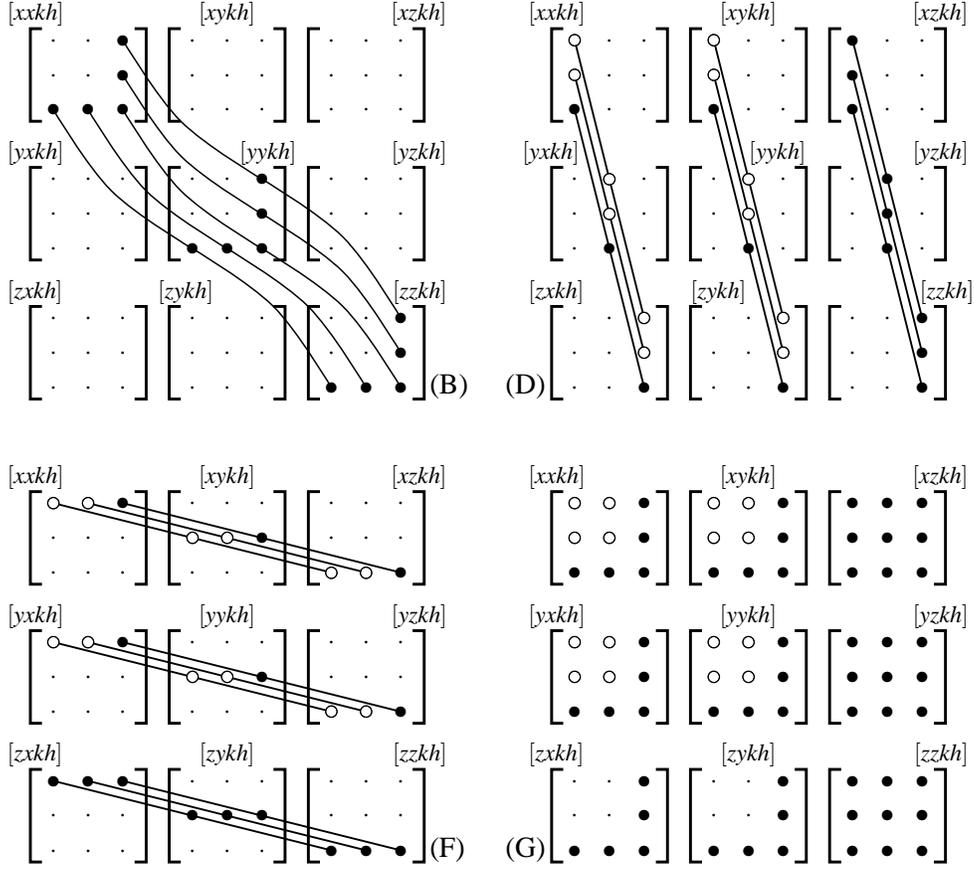
\begin{figure}[tb]
\setlength{\unitlength}{0.77mm}
\psset{unit=0.77mm}
\begin{center}
\begin{pspicture}(0,4)(170,160)
\put(0,80){
 \put(75,6){\makebox(0,5)[bl]{(B)}}
 \multiput(2,0)(24,0){3}{\multiput(0,0)(0,24){3}{
  \psline[linewidth=0.5]{-}(6,6)(4,6)(4,22)(6,22)
  \psline[linewidth=0.5]{-}(22,6)(24,6)(24,22)(22,22)
 }}
 \multiput(10,8)(24,0){3}{\multiput(0,0)(0,24){3}{
  \multiput(0,0)(6,0){3}{\multiput(0,0)(0,6){3}{\circle*{0.1}
 }}}}
 {\footnotesize
  \put(2,70){\makebox(0,5)[l]{$[xxkh]$}}
  \put(2,46){\makebox(0,5)[l]{$[yxkh]$}}
  \put(2,22){\makebox(0,5)[l]{$[zxkh]$}}
  \put(40,70){\makebox(0,5)[c]{$[xykh]$}}
  \put(52,46){\makebox(0,5)[r]{$[yykh]$}}
  \put(28,22){\makebox(0,5)[l]{$[zykh]$}}
  \put(78,70){\makebox(0,5)[r]{$[xzkh]$}}
  \put(78,46){\makebox(0,5)[r]{$[yzkh]$}}
  \put(78,22){\makebox(0,5)[r]{$[zzkh]$}}
 }
 \multiput(10,56)(0,6){3}{\multiput(0,0)(6,0){3}{\circle*{2}}}
 \multiput(34,32)(0,6){3}{\multiput(0,0)(6,0){3}{\circle*{2}}}
 \multiput(58,8)(0,6){3}{\multiput(0,0)(6,0){3}{\circle*{2}}}
 \pscurve[linewidth=.2mm]{-}(10,56)(20,42)(34,32) 
 \pscurve[linewidth=.2mm]{-}(34,32)(48,22)(58,8) 
 \pscurve[linewidth=.2mm]{-}(16,56)(26,42)(40,32) 
 \pscurve[linewidth=.2mm]{-}(40,32)(54,22)(64,8) 
 \pscurve[linewidth=.2mm]{-}(22,56)(32,42)(46,32) 
 \pscurve[linewidth=.2mm]{-}(46,32)(60,22)(70,8) 
 \pscurve[linewidth=.2mm]{-}(22,62)(32,48)(46,38) 
 \pscurve[linewidth=.2mm]{-}(46,38)(60,28)(70,14) 
 \pscurve[linewidth=.2mm]{-}(22,68)(32,54)(46,44) 
 \pscurve[linewidth=.2mm]{-}(46,44)(60,34)(70,20) 
 \multiput(10,62)(24,-24){3}{\multiput(0,0)(6,0){2}{\multiput(0,0)(0,6){2}{
  \pscircle[fillstyle=solid,fillcolor=white,linewidth=0,linecolor=white](0,0){1.5}
   \circle*{0.1}
 }}}
}
\put(90,80){
 \put(5,6){\makebox(0,5)[br]{(D)}}
 \multiput(2,0)(24,0){3}{\multiput(0,0)(0,24){3}{
  \psline[linewidth=0.5]{-}(6,6)(4,6)(4,22)(6,22)
  \psline[linewidth=0.5]{-}(22,6)(24,6)(24,22)(22,22)
 }}
 \multiput(10,8)(24,0){3}{\multiput(0,0)(0,24){3}{
  \multiput(0,0)(6,0){3}{\multiput(0,0)(0,6){3}{\circle*{0.1}
 }}}}
 {\footnotesize
  \put(2,70){\makebox(0,5)[l]{$[xxkh]$}}
  \put(1,46){\makebox(0,5)[l]{$[yxkh]$}}
  \put(2,22){\makebox(0,5)[l]{$[zxkh]$}}
  \put(40,70){\makebox(0,5)[c]{$[xykh]$}}
  \put(50,46){\makebox(0,5)[r]{$[yykh]$}}
  \put(30,22){\makebox(0,5)[l]{$[zykh]$}}
  \put(78,70){\makebox(0,5)[r]{$[xzkh]$}}
  \put(78,46){\makebox(0,5)[r]{$[yzkh]$}}
  \put(79,22){\makebox(0,5)[r]{$[zzkh]$}}
 }
 \multiput(22,8)(-6,24){3}{\multiput(0,0)(24,0){3}{
  \multiput(0,0)(0,6){3}{\circle*{2}
 }}}
\multiput(10,68)(24,0){3}{%
 \multiput(0,0)(0,-6){3}{%
  \psline[linewidth=0.25mm]{-}(0,0)(12,-48)
}}
 \multiput(22,14)(24,0){2}{\multiput(0,0)(-6,24){3}{
  \multiput(0,0)(0,6){2}{
   \pscircle[fillstyle=solid,fillcolor=white,linewidth=0](0,0){1}
   \pscircle[fillcolor=white,linecolor=black,linewidth=0.1](0,0){1}
 }}}
}
\put(0,0){
 \put(75,6){\makebox(0,5)[bl]{(F)}}
 \multiput(2,0)(24,0){3}{\multiput(0,0)(0,24){3}{
  \psline[linewidth=0.5]{-}(6,6)(4,6)(4,22)(6,22)
  \psline[linewidth=0.5]{-}(22,6)(24,6)(24,22)(22,22)
 }}
 \multiput(10,8)(24,0){3}{\multiput(0,0)(0,24){3}{
  \multiput(0,0)(6,0){3}{\multiput(0,0)(0,6){3}{\circle*{0.1}
 }}}}
 {\footnotesize
  \put(2,70){\makebox(0,5)[l]{$[xxkh]$}}
  \put(2,46){\makebox(0,5)[l]{$[yxkh]$}}
  \put(2,22){\makebox(0,5)[l]{$[zxkh]$}}
  \put(40,70){\makebox(0,5)[c]{$[xykh]$}}
  \put(40,46){\makebox(0,5)[c]{$[yykh]$}}
  \put(40,22){\makebox(0,5)[c]{$[zykh]$}}
  \put(78,70){\makebox(0,5)[r]{$[xzkh]$}}
  \put(78,46){\makebox(0,5)[r]{$[yzkh]$}}
  \put(78,22){\makebox(0,5)[r]{$[zzkh]$}}
 }
 \multiput(10,20)(0,24){3}{\multiput(0,0)(24,-6){3}{
  \multiput(0,0)(6,0){3}{\circle*{2}
 }}}
\multiput(10,68)(0,-24){3}{%
 \multiput(0,0)(6,0){3}{%
  \psline[linewidth=0.25mm]{-}(0,0)(48,-12)
}}
 \multiput(10,44)(0,24){2}{\multiput(0,0)(24,-6){3}{
  \multiput(0,0)(6,0){2}{
   \pscircle[fillstyle=solid,fillcolor=white,linewidth=0](0,0){1}
   \pscircle[fillcolor=white,linecolor=black,linewidth=0.1](0,0){1}
 }}}
}
\put(90,0){
 \put(5,6){\makebox(0,5)[br]{(G)}}
 \multiput(2,0)(24,0){3}{\multiput(0,0)(0,24){3}{
  \psline[linewidth=0.5]{-}(6,6)(4,6)(4,22)(6,22)
  \psline[linewidth=0.5]{-}(22,6)(24,6)(24,22)(22,22)
 }}
 \multiput(10,8)(24,0){3}{\multiput(0,0)(0,24){3}{
  \multiput(0,0)(6,0){3}{\circle*{2}
 }}}
 \multiput(10,8)(24,0){3}{\multiput(12,6)(0,24){3}{
  \multiput(0,0)(0,6){2}{\circle*{2}
 }}}
 \multiput(58,8)(0,24){3}{\multiput(0,6)(0,6){2}{
  \multiput(0,0)(6,0){2}{\circle*{2}
 }}}
 \multiput(10,32)(24,0){2}{\multiput(0,6)(0,24){2}{\multiput(0,0)(6,0){2}{
  \multiput(0,0)(0,6){2}{
   \pscircle[fillstyle=solid,fillcolor=white,linewidth=0](0,0){1}
   \pscircle[fillcolor=white,linecolor=black,linewidth=0.1](0,0){1}
 }}}}
 \multiput(10,8)(24,0){2}{\multiput(0,6)(0,6){2}{
  \multiput(0,0)(6,0){3}{\circle*{0.1}
 }}}
 {\footnotesize
  \put(2,70){\makebox(0,5)[l]{$[xxkh]$}}
  \put(2,46){\makebox(0,5)[l]{$[yxkh]$}}
  \put(2,22){\makebox(0,5)[l]{$[zxkh]$}}
  \put(40,70){\makebox(0,5)[c]{$[xykh]$}}
  \put(40,46){\makebox(0,5)[c]{$[yykh]$}}
  \put(40,22){\makebox(0,5)[c]{$[zykh]$}}
  \put(78,70){\makebox(0,5)[r]{$[xzkh]$}}
  \put(78,46){\makebox(0,5)[r]{$[yzkh]$}}
  \put(78,22){\makebox(0,5)[r]{$[zzkh]$}}
 }
}
\end{pspicture}
\end{center}
\caption[Symmetry schemes for the PCDFWM conductivity tnesor in the
local limit and in the single level quantum well case]{Symmetry
  schemes for the PCDFWM conductivity tensor parts B, D, F, and G in
  (i) the local limit in the $z$ coordinates, and (ii) the single
  level quantum well case. In the local limit, tensor elements labeled
  with a ``$\bullet$'' gives nonzero mixed interband/intraband
  contributions, elements labeled with a ``$\circ$'' gives nonzero
  pure intraband contributions, and elements labeled with a
  ``$\cdot$'' are zero. In the single level quantum well case, only
  elements labeled with a ``$\circ$'' contributes to the solution. The
  solid lines connect equal nonzero elements.\label{fig:8.2}}
\end{figure}

Then we may conclude that (i) the only independent element of part A
of the PCDFWM conductivity tensor is zero in the local limit, since
the two Fermi-Dirac distribution functions in Eq.~(\ref{eq:XiAA})
becomes identical for $n=m$. This occurs when taking the local limit
in the coordinate $z''$.  In addition, (ii) the five independent
elements of part E of the PCDFWM conductivity tensor also becomes
zero, since the two pure interband terms are zero by themselves, and
the two other terms are of the same magnitude but with opposite sign.
This occurs when taking the local limit in the coordinates $z'''$ and
$z''$. Furthermore, (iii) for part G of the PCDFWM conductivity
tensor, elements with the Cartesian index $i=z$ and the other indices
different from $z$ becomes zero, since the other indices implies that
all quantum numbers in the summation become identical.  Finally, (iv)
the only independent element of part C of the PCDFWM conductivity
tensor is reduced to a pure intraband contribution. The same reduction
appears in tensor elements of parts D, F, and G with no Cartesian
coordinate index $z$ in indices $jk$, $ih$, and $ijkh$, respectively.
In part B of the PCDFWM conductivity tensor, the elements with no
Cartesian index $z$ in $kh$ apparantly gives the same result, but the
following integration over $\vec{\kappa}_{\|}$ makes them vanish.
These conclusions are shown in the form of symmetry schemes in
Fig.~\ref{fig:8.2}.

\subsection{Local limit along the surface}\label{sec:local-xy}
Taking the local limit along the surface coordinates, the wavenumbers
are considered to be much less than the Fermi wavenumber, i.e., we
take the limit where $\vec{q}_{\|}\rightarrow\vec{0}$ and
$\vec{k}_{\|}\rightarrow\vec{0}$ (the dipole limit).

Then from Eq.~(\ref{eq:transition-w}) we observe that the transition
frequencies become independent of $\vec{\kappa}_{\|}$, and thus we
conclude that this approximation makes the integration over
$\vec{\kappa}_{\|}$ particularly simple in the low temperature limit,
since no integration variables appears in any of the denominators in
Eqs.~(\ref{eq:XiAA})--(\ref{eq:XiAG}). We further observe that only
interband contributions are left compared to the full description.

\subsection{Local limit in three coordinates}\label{sec:local-xyz}
Locality in all three coordinates is achieved by a combination of the
two limits mentioned in Secs.~\ref{sec:local-z} and \ref{sec:local-xy}
above. Thus, in this limit (i) parts A and E of the DFWM
conductivity tensor does not contribute for the same reasons as
before. Furthermore (ii) the integration over $z'$ in part C makes
this part vanish, (iii) the integration over $z'$ in part F of the
DFWM conductivity tensor cancels two terms, and the other two are of
the same magnitude, but with opposite signs, resulting in the fact
that part F does not contribute to the response in this limit.
Finally (iv), all purely intraband contributing tensor elements found
in the local limit vanish. All in all we are left with five
independent nonzero elements of parts B and D, and fiftyseven
independent nonzero elements of G, all labeled with a ``$\bullet$'' in
Fig.~\ref{fig:8.2}.

\subsection{Single level quantum well}\label{sec:OneLevel}
A substantial simplification of Eqs.~(\ref{eq:XiAA})--(\ref{eq:XiAG})
occur in {\it one}\/ special case, namely in the case where the thin
film is a single level quantum well in the $z$-direction. In the
single level quantum well, the summation indices are all equal to $1$.

Special attention is in this case devoted to the current density
defined in Eq.~(\ref{eq:j_hnm}), which in a single level quantum well
is reduced to
\begin{equation}
 j_{h,11}(z;{\kappa}_{h})=-{e\hbar\over2m_{e}}
 (\delta_{hx}+\delta_{hy})
 {\kappa}_{h}|\psi_{1}(z)|^2,
\end{equation}
for $h\in\{x,y\}$, since the $z$-dependent part vanish for any
$n=m$. This observation leads to a drastic reduction of the number of
contributing elements in most of the symmetry schemes associated with
the occuring processes.

The only nonzero element in part A of the nonlinear conductivity
tensor vanish for $n=m$. In part B of the conductivity tensor all
elements with $k$ or $h$ equal to $z$ vanish for $n=m=v$, and the rest
of the elements vanish by integration over $\vec{\kappa}_{\|}$. The
only nonzero element of part C of the nonlinear conductivity tensor is
conserved, but simplified. Part D of the nonlinear conductivity tensor
is reduced somewhat, since either combination of $j=z$ or $k=z$ gives
zero. Then we are left with four nonzero independent elements, as
shown in Fig.~\ref{fig:8.2}.D. Part E does not give any contributions
to the intraband transitions, since two terms in the sum gives no
intraband contributions in general, and the other two terms cancel
each other.  Part F is reduced in a manner similar to parts B and D,
since any combination of $i=z$ or $h=z$ gives zero. The resulting four
nonzero independent elements are shown in Fig.~\ref{fig:8.2}.F. In the
last part (G) of the nonlinear conductivity tensor any combination of
$i=z$, $j=z$, $k=z$, or $h=z$ gives zero. As a consequence of this
rather drastic reduction we are left with sixteen nonzero elements, as
shown in Fig.~\ref{fig:8.2}.G.

\chapter{Polarized light in the $x$-$z$-plane}\label{ch:9}\label{ch:system}
Restricting ourselves to consider light propagating in the
$x$-$z$-plane, which furthermore is polarized either in the
$x$-$z$-plane ($p$-polarized) or perpendicular to the $x$-$z$-plane
($s$-polarized), the treatment can be split into eight separate parts
related to the possible combinations of polarization of the three
different incident fields. In this scattering geometry $\vec{q}_{\|}$
(and $\vec{k}_{\|}$) lie along the $x$-axis, giving a mirror plane at
$y=0$. Consequently, only tensor elements in the
($3\times3\times3\times3$) PCDFWM response tensor with a Cartesian
index even numbered in $y$ contributes, and the $81$ tensor elements
generally appearing are reduced to $41$.

\begin{figure}[tb]
\setlength{\unitlength}{1mm}
\psset{unit=1mm}
\begin{center}
\begin{picture}(80,76)(0,4)
\newgray{mygray}{0.95}
\multiput(2,0)(24,0){3}{\multiput(0,0)(0,24){3}{
  \psline[linewidth=0.385]{-}(6,6)(4,6)(4,22)(6,22)
  \psline[linewidth=0.385]{-}(22,6)(24,6)(24,22)(22,22)
}}
{\small
 \put(2,70){\makebox(0,5)[l]{$[xxkh]$}}
 \put(2,46){\makebox(0,5)[l]{$[yxkh]$}}
 \put(2,22){\makebox(0,5)[l]{$[zxkh]$}}
 \put(40,70){\makebox(0,5)[c]{$[xykh]$}}
 \put(40,46){\makebox(0,5)[c]{$[yykh]$}}
 \put(40,22){\makebox(0,5)[c]{$[zykh]$}}
 \put(78,70){\makebox(0,5)[r]{$[xzkh]$}}
 \put(78,46){\makebox(0,5)[r]{$[yzkh]$}}
 \put(78,22){\makebox(0,5)[r]{$[zzkh]$}}
}
\put(40,38){\circle*{2}}
\put(10,68){\circle{2}}
\put(10,56){\circle{2}}
\put(10,20){\circle{2}}
\put(10,8){\circle{2}} 
\put(22,68){\circle{2}}
\put(22,56){\circle{2}}
\put(22,20){\circle{2}}
\put(22,8){\circle{2}}
\put(58,68){\circle{2}}
\put(58,56){\circle{2}}
\put(58,20){\circle{2}}
\put(58,8){\circle{2}}
\put(70,68){\circle{2}}
\put(70,56){\circle{2}}
\put(70,20){\circle{2}}
\put(70,8){\circle{2}}
\put(16,62){\makebox(0,0)[c]{$\blacksquare$}}
\put(16,14){\makebox(0,0)[c]{$\blacksquare$}}
\put(64,62){\makebox(0,0)[c]{$\blacksquare$}}
\put(64,14){\makebox(0,0)[c]{$\blacksquare$}}
\put(34,32){\makebox(0,0)[c]{$\square$}}
\put(34,44){\makebox(0,0)[c]{$\square$}}
\put(46,32){\makebox(0,0)[c]{$\square$}}
\put(46,44){\makebox(0,0)[c]{$\square$}}
\put(40,8){\makebox(0,0)[c]{$\clubsuit$}}
\put(40,20){\makebox(0,0)[c]{$\clubsuit$}}
\put(40,56){\makebox(0,0)[c]{$\clubsuit$}}
\put(40,68){\makebox(0,0)[c]{$\clubsuit$}}
\put(16,32){\makebox(0,0)[c]{$\diamondsuit$}}
\put(16,44){\makebox(0,0)[c]{$\diamondsuit$}}
\put(64,32){\makebox(0,0)[c]{$\diamondsuit$}}
\put(64,44){\makebox(0,0)[c]{$\diamondsuit$}}
\put(34,14){\makebox(0,0)[c]{$\heartsuit$}}
\put(46,14){\makebox(0,0)[c]{$\heartsuit$}}
\put(34,62){\makebox(0,0)[c]{$\heartsuit$}}
\put(46,62){\makebox(0,0)[c]{$\heartsuit$}}
\put(10,38){\makebox(0,0)[c]{$\spadesuit$}}
\put(22,38){\makebox(0,0)[c]{$\spadesuit$}}
\put(58,38){\makebox(0,0)[c]{$\spadesuit$}}
\put(70,38){\makebox(0,0)[c]{$\spadesuit$}}
\put(10,14){\circle*{0.1}}
\put(10,32){\circle*{0.1}}
\put(10,44){\circle*{0.1}}
\put(10,62){\circle*{0.1}}
\put(16,8){\circle*{0.1}}
\put(16,20){\circle*{0.1}}
\put(16,38){\circle*{0.1}}
\put(16,56){\circle*{0.1}}
\put(16,68){\circle*{0.1}}
\put(22,14){\circle*{0.1}}
\put(22,32){\circle*{0.1}}
\put(22,44){\circle*{0.1}}
\put(22,62){\circle*{0.1}}
\put(34,8){\circle*{0.1}}
\put(34,20){\circle*{0.1}}
\put(34,38){\circle*{0.1}}
\put(34,56){\circle*{0.1}}
\put(34,68){\circle*{0.1}}
\put(40,14){\circle*{0.1}}
\put(40,32){\circle*{0.1}}
\put(40,44){\circle*{0.1}}
\put(40,62){\circle*{0.1}}
\put(46,8){\circle*{0.1}}
\put(46,20){\circle*{0.1}}
\put(46,38){\circle*{0.1}}
\put(46,56){\circle*{0.1}}
\put(46,68){\circle*{0.1}}
\put(58,14){\circle*{0.1}}
\put(58,32){\circle*{0.1}}
\put(58,44){\circle*{0.1}}
\put(58,62){\circle*{0.1}}
\put(64,8){\circle*{0.1}}
\put(64,20){\circle*{0.1}}
\put(64,38){\circle*{0.1}}
\put(64,56){\circle*{0.1}}
\put(64,68){\circle*{0.1}}
\put(70,14){\circle*{0.1}}
\put(70,32){\circle*{0.1}}
\put(70,44){\circle*{0.1}}
\put(70,62){\circle*{0.1}}
\end{picture}
\end{center}
\caption[The contributing matrix elements of the DFWM conductivity
tensor when using $s$- and $p$-polarized light in the
$x$-$z$-plane]{The contributing matrix elements of the third order
  conductivity tensor in the cases where $(\bullet)$ both the pump
  fields and the probe field are $s$-polarized, $(\circ)$ both the
  pump fields and the probe field are $p$-polarized, $(\blacksquare)$
  both pump fields are $s$-polarized, and the probe field is
  $p$-polarized, $(\square)$ both pump fields are $p$-polarized, and
  the probe field is $s$-polarized, $(\clubsuit)$ (and $\heartsuit$
  for the ``interchanged term'') pump $1$ is $s$-polarized, pump $2$
  is $p$-polarized, and the probe is $s$-polarized, $(\heartsuit)$
  (and $\clubsuit$ for the ``interchanged term'') pump $1$ is
  $p$-polarized, pump $2$ is $s$-polarized, and the probe is
  $s$-polarized, $(\diamondsuit)$ (and $\spadesuit$ for the
  ``interchanged term'') pump $1$ is $s$-polarized, pump $2$ is
  $p$-polarized, and the probe is $p$-polarized, and $(\spadesuit)$
  (and $\diamondsuit$ for the ``interchanged term'') pump $1$ is
  $p$-polarized, pump $2$ is $s$-polarized, and the probe is
  $p$-polarized.\label{fig:polarize}}
\end{figure}

Applying the two polarization states $s$ and $p$ chosen above to the
three interacting fields, the resulting eight different combinations
uses different matrix elements in the nonlinear conductivity tensor,
and (as would be expected) these eight combinations together make
use of {\it all}\/ elements of the nonlinear conductivity tensor.
This division is shown in Fig.~\ref{fig:polarize} for the $41$
contributing tensor elements as described in the following. The
noncontributing elements of the nonlinear conductivity tensor is
denoted using the symbol ``$\cdot$'' in Fig.~\ref{fig:polarize}.

\begin{figure}[tb]
\setlength{\unitlength}{1mm}
\psset{unit=1mm}
\begin{center}
\begin{pspicture}(0,0)(115,20)
\multiput(0,0)(30,0){4}{
 \psline[linewidth=0.25mm]{-}(0,10)(25,10)
 \psline[linewidth=0.5mm]{->}(0,5)(11.5,5)
 \psline[linewidth=0.5mm]{->}(25,5)(13.5,5)
 \put(25,0){\makebox(0,4)[r]{(1)}}
 \put(0,0){\makebox(0,4)[l]{(2)}}
 \put(1,13){\makebox(0,4)[l]{(p)}}
 \psline[linewidth=0.5mm]{->}(3.4,19.1)(11.5,11)
}
\put(0,0){
 \pscircle[linewidth=0.1mm,linecolor=black,fillstyle=solid,fillcolor=white](6.5,5){1.0}
 \qdisk(6.5,5){0.25}
 \pscircle[linewidth=0.1mm,linecolor=black,fillstyle=solid,fillcolor=white](18.5,5){1.0}
 \qdisk(18.5,5){0.25}
 \put(8,14.5){%
  \pscircle[linewidth=0.1mm,linecolor=black,fillstyle=solid,fillcolor=white](0,0){1.0}
  \qdisk(0,0){0.25}
 }
 \put(25,20){\makebox(0,0)[tr]{(a)}}
}
\put(30,0){
 \psline[linewidth=0.25mm]{<->}(6.5,2)(6.5,8)
 \psline[linewidth=0.25mm]{<->}(18.5,2)(18.5,8)
 \put(8,14.5){%
  \pscircle[linewidth=0.1mm,linecolor=black,fillstyle=solid,fillcolor=white](0,0){1.0}
  \qdisk(0,0){0.25}
 }
 \put(25,20){\makebox(0,0)[tr]{(b)}}
}
\put(60,0){
 \pscircle[linewidth=0.1mm,linecolor=black,fillstyle=solid,fillcolor=white](6.5,5){1.0}
 \qdisk(6.5,5){0.25}
 \pscircle[linewidth=0.1mm,linecolor=black,fillstyle=solid,fillcolor=white](18.5,5){1.0}
 \qdisk(18.5,5){0.25}
 \psline[linewidth=0.25mm]{<->}(5.9,12.4)(10.1,16.6)
 \put(25,20){\makebox(0,0)[tr]{(c)}}
}
\put(90,0){
 \psline[linewidth=0.25mm]{<->}(6.5,2)(6.5,8)
 \psline[linewidth=0.25mm]{<->}(18.5,2)(18.5,8)
 \psline[linewidth=0.25mm]{<->}(5.9,12.4)(10.1,16.6)
 \put(25,20){\makebox(0,0)[tr]{(d)}}
}
\end{pspicture}
\end{center}
\caption[Polarization combinations for $s$ to $s$ and $p$ to $p$
responses]{Schematic illustration showing the four possible field
  polarization combinations giving rise to the same polarization of
  the phase conjugated response as the probe field. Figs.~a and b
  shows the $s$ to $s$ response for $s$-polarized and $p$-polarized
  pump fields, respectively, while Figs.~c and d shows the $p$ to $p$
  response corresponding to these pump field polarizations. In
  Figs.~a--d the pump fields are denoted (1) and (2) and the probe
  field is denoted (p).\label{fig:equalpumps}\label{fig:9.2}}
\setlength{\unitlength}{1mm}
\psset{unit=1mm}
\begin{center}
\begin{pspicture}(0,0)(115,20)
\put(0,0){
 \psline[linewidth=0.25mm]{-}(0,10)(25,10)
 \psline[linewidth=0.5mm]{->}(0,5)(11.5,5)
 \psline[linewidth=0.5mm]{->}(25,5)(13.5,5)
 \psline[linewidth=0.25mm]{<->}(6.5,2)(6.5,8)
 \pscircle[linewidth=0.1mm,linecolor=black,fillstyle=solid,fillcolor=white](18.5,5){1.0}
 \qdisk(18.5,5){0.25}
 \put(25,0){\makebox(0,4)[r]{(1)}}
 \put(0,0){\makebox(0,4)[l]{(2)}}
 \put(1,13){\makebox(0,4)[l]{(p)}}
 \psline[linewidth=0.5mm]{->}(3.4,19.1)(11.5,11)
 \put(8,14.5){%
  \pscircle[linewidth=0.1mm,linecolor=black,fillstyle=solid,fillcolor=white](0,0){1.0}
  \qdisk(0,0){0.25}
 }
 \put(25,20){\makebox(0,0)[tr]{(a)}}
}
\put(30,0){
 \psline[linewidth=0.25mm]{-}(0,10)(25,10)
 \psline[linewidth=0.5mm]{->}(0,5)(11.5,5)
 \psline[linewidth=0.5mm]{->}(25,5)(13.5,5)
 \pscircle[linewidth=0.1mm,linecolor=black,fillstyle=solid,fillcolor=white](6.5,5){1.0}
 \qdisk(6.5,5){0.25}
 \psline[linewidth=0.25mm]{<->}(18.5,2)(18.5,8)
 \put(25,0){\makebox(0,4)[r]{(1)}}
 \put(0,0){\makebox(0,4)[l]{(2)}}
 \put(1,13){\makebox(0,4)[l]{(p)}}
 \psline[linewidth=0.5mm]{->}(3.4,19.1)(11.5,11)
 \put(8,14.5){%
  \pscircle[linewidth=0.1mm,linecolor=black,fillstyle=solid,fillcolor=white](0,0){1.0}
  \qdisk(0,0){0.25}
 }
 \put(25,20){\makebox(0,0)[tr]{(b)}}
}
\put(60,0){
 \psline[linewidth=0.25mm]{-}(0,10)(25,10)
 \psline[linewidth=0.5mm]{->}(0,5)(11.5,5)
 \psline[linewidth=0.5mm]{->}(25,5)(13.5,5)
 \psline[linewidth=0.25mm]{<->}(6.5,2)(6.5,8)
 \pscircle[linewidth=0.1mm,linecolor=black,fillstyle=solid,fillcolor=white](18.5,5){1.0}
 \qdisk(18.5,5){0.25}
 \put(20,0){\makebox(0,4)[r]{(1)}}
 \put(0,0){\makebox(0,4)[l]{(2)}}
 \put(1,13){\makebox(0,4)[l]{(p)}}
 \psline[linewidth=0.5mm]{->}(3.4,19.1)(11.5,11)
 \psline[linewidth=0.25mm]{<->}(5.9,12.4)(10.1,16.6)
 \put(25,20){\makebox(0,0)[tr]{(c)}}
}
\put(90,0){
 \psline[linewidth=0.25mm]{-}(0,10)(25,10)
 \psline[linewidth=0.5mm]{->}(0,5)(11.5,5)
 \psline[linewidth=0.5mm]{->}(25,5)(13.5,5)
 \pscircle[linewidth=0.1mm,linecolor=black,fillstyle=solid,fillcolor=white](6.5,5){1.0}
 \qdisk(6.5,5){0.25}
 \psline[linewidth=0.25mm]{<->}(18.5,2)(18.5,8)
 \put(25,0){\makebox(0,4)[r]{(1)}}
 \put(0,0){\makebox(0,4)[l]{(2)}}
 \put(1,13){\makebox(0,4)[l]{(p)}}
 \psline[linewidth=0.5mm]{->}(3.4,19.1)(11.5,11)
 \psline[linewidth=0.25mm]{<->}(5.9,12.4)(10.1,16.6)
 \put(25,20){\makebox(0,0)[tr]{(d)}}
}
\end{pspicture}
\end{center}
\caption[Polarization combinations for $s$ to $p$ and $p$ to $s$
responses]{Schematic illustration showing the four possible field
  polarization combinations giving rise to different polarization of
  the phase conjugated response seen from the point of view of the
  probe field. Figs.~a and b shows the combinations of the fields
  giving a $s$ to $p$ response, while Figs.~c and d shows the $p$ to
  $s$ response configurations. In all four cases the two pump fields
  are differently polarized with respect to each other. In Figs.~a--d
  the pump fields are denoted (1) and (2) and the probe field is
  denoted (p).\label{fig:diff.pumps}\label{fig:9.3}}
\end{figure}

\section{Eight sets of contributing matrix elements}
From the point of view of the probe, the eight different combinations
of polarized light can be divided into two groups of four, namely four
giving a PCDFWM response with the same polarization as the probe and
four giving the other polarization as the PCDFWM response. In the four
combinations giving response of the same polarization as the probe,
the two pump fields have the same polarization states. These
configurations are sketched in Fig.~\ref{fig:equalpumps}. The other
four combinations, where the two pump fields are differently polarized
are sketched in Fig.~\ref{fig:diff.pumps}.

Two out of the first four combinations describe $s$ to $s$
transitions, seen from the point of view of the probe. (i) The
simplest combination arises when both pump fields and the probe field
are $s$-polarized, as shown in Fig.~\ref{fig:equalpumps}.a.  In this
case, only the $yyyy$ element of the nonlinear conductivity tensor is
present. In Fig.~\ref{fig:polarize} it is marked with a ``$\bullet$''.
It should be noted that this is the {\it only}\/ case of the eight, in
which a single matrix element can be determined independently in an
actual experiment. (ii) When both pump fields are $p$-polarized and
the probe field is $s$-polarized (see Fig.~\ref{fig:equalpumps}.b),
the four contributing matrix elements in the nonlinear conductivity
tensor have indices $i$ and $j$ equal to $y$ and indices $k$ and $h$
different from $y$. Each of these four elements is marked with a
``$\square$'' in Fig.~\ref{fig:polarize}.

From the same point of view the other two of the first four
combinations describe $p$ to $p$ transitions. (iii) If both pump
fields are $s$-polarized and the probe field is $p$-polarized the
configuration is sketched in Fig.~\ref{fig:equalpumps}.c, and four
matrix elements in the nonlinear conductivity tensor contribute to
the solution. They have indices $k$ and $h$ equal to $y$ and indices
$i$ and $j$ different from $y$. In Fig.~\ref{fig:polarize}, each of
these elements is marked with a ``$\blacksquare$''. (iv) The other
extreme case [the simple extreme has been described in item (i)] is
the combination where both pump fields and the probe field are
$p$-polarized, as shown in Fig.~\ref{fig:equalpumps}.d. In order to
obtain the solution for this combination as many as sixteen elements
of the nonlinear conductivity tensor are required, since every element
with an index without $y$'s in it contributes.  Each of these elements
is marked in Fig.~\ref{fig:polarize} with a ``$\circ$''.

Still taking the ``probe to response'' point of view, two of the
remaining four cases represent a probe to response transition from
$s$ to $p$.  (v) If pump field $1$ is $s$-polarized, pump field $2$ is
$p$-polarized, and the probe field is $s$-polarized (the corresponding
diagram is showed in Fig.~\ref{fig:diff.pumps}.a), the four
contributing matrix elements have indices $j$ and $k$ equal to $y$ and
indices $i$ and $h$ different from $y$. In Fig.~\ref{fig:polarize}
each of these elements is marked with the symbol ``$\clubsuit$''. In
the other of these cases, (vi), pump field $1$ is $p$-polarized and
pump field $2$ is $s$-polarized, and we take the probe field to be
$s$-polarized. This combination is sketched in
Fig.~\ref{fig:diff.pumps}.b, and the four contributing elements in the
nonlinear conductivity tensor then have indices $j$ and $k$ equal to
$y$ and indices $i$ and $h$ different from $y$. In
Fig.~\ref{fig:polarize} the symbol ``$\heartsuit$'' is used to show
these elements. As a direct consequence of the conservation of
momentum criterion these two combinations are equivalent, since by
replacing $\vec{q}_{\|}$ with $-\vec{q}_{\|}$ and $\vec{k}_{\|}$ with
$-\vec{k}_{\|}$ the situation in (vi) changes to the situation in (v).

The last two combinations represent a transition from $p$ to $s$ in
the picture from probe to response.  (vii) If pump field $1$ is
$s$-polarized and pump field $2$ is $p$-polarized, but the probe field
is $p$-polarized, the situation is as sketched in
Fig.~\ref{fig:diff.pumps}.c. Then again four elements of the nonlinear
conductivity tensor contribute to the solution. These four elements
have indices $i$ and $h$ equal to $y$ and indices $j$ and $k$
different from $y$, and each element is marked using the symbol
``$\diamondsuit$'' in Fig.~\ref{fig:polarize}. Finally, (viii), when
pump field $1$ is $p$-polarized and pump field $2$ is $s$-polarized,
but the probe field is $p$-polarized, the configuration appears as
shown in Fig.~\ref{fig:diff.pumps}.d, which again gives four elements
of the nonlinear conductivity tensor contributing to the solution, the
indices $i$ and $k$ being equal to $y$ and the indices $j$ and $h$
being different from $y$. We mark each of these elements with the
symbol ``$\spadesuit$'' in Fig.~\ref{fig:polarize}. For the same
reason as before, cases (vii) and (viii) are equivalent.

\section{Simplified description by choice of pump fields}
Although the two pump fields has been drawn parallel to the interface
in Figs.~\ref{fig:equalpumps} and \ref{fig:diff.pumps}, this is not a
requirement, as long as they are counterpropagating to each other.  If
we are looking for a simplification in the treatment of optical phase
conjugation from a quantum well structure, a reduction in the number
of tensor elements to be calculated could be one alternative.  Two
immediate possibilities comes to mind. In the first case, the pump
fields are taken to be parallel to the $x$-axis. The second case has
the pump fields parallel to the $z$-axis. The consequences of these
two cases are described in the following.

If the pump fields propagate in a direction parallel to the $x$-axis,
the number of contributing tensor elements in the nonlinear
conductivity tensor
$\tensor{\Xi}(z,z',z'',z''';\vec{q}_{\|},\vec{k}_{\|})$ is reduced
from $41$ to $18$ when considering $s$- and $p$-polarized light only.
The surviving elements can be divided into four cases following the
four possible combinations of polarization of the pump fields. When
(i) both pump fields are $p$-polarized, their respective electric
fields have only a $z$-component, and hence $h=k=z$. Similarly (ii),
when both pump fields are $s$-polarized, their electric fields only
have a $y$-component, that is, $h=k=y$. In case (iii) pump field 1 is
$p$-polarized while pump field 2 is $s$-polarized, giving $h=z$ and
$k=y$. In the final case (iv) the pump fields are polarized oppositely
to those in case (iii), i.e., $h=y$ and $k=z$.

If we choose the pump fields to propagate in a direction parallel to
the $z$ axis instead, the number of contributing tensor elements in
the nonlinear conductivity tensor
$\tensor{\Xi}(z,z',z'',z''';\vec{q}_{\|},\vec{k}_{\|})$ is again
reduced from $41$ to $18$ when considering $s$- and $p$-polarized
light, and again the surviving elements are divided into four groups
following the four possible combinations of polarization the pump
fields can have. Thus, (i) when both pump fields are $p$-polarized,
the electric fields representing them have only $x$-components, i.e.,
$h=k=x$, (ii) for the pump fields both being $s$-polarized, the same
elements as when the pump fields are parallel to the $x$-axis
contributes to the solution, giving again $h=k=y$. In the cases of
differently polarized pump fields, (iii) $h=x$ and $k=y$ when pump
field 1 is $p$-polarized and pump field 2 is $s$-polarized, and (iv)
$h=y$ and $k=x$ when the opposite polarizations occur.

In conclusion, these two possibilities of choice have five common
contributing elements, namely the ones where $k=h=y$. At the same
time, ten elements of the nonlinear conductivity tensor does not
contribute to either simplification. They have $kh\in\{xz,zx\}$.

\part[Optical phase conjugation in a single-level metallic quantum
  well]{Optical phase conjugation \\ in a single-level metallic
  quantum well}\label{part:IV}
\newpage
\thispagestyle{plain}
\newpage

\chapter{Theoretical considerations}\label{ch:10}
Having discussed the properties of optical phase conjugation in
quantum well structures in general, let us consider here the simplest
configuration of a mesoscopic metallic optical quantum-well phase
conjugator. In this case only a single bound state exists below the
Fermi level and it is assumed that no levels above the Fermi level can
be reached with the applied optical field. Such a quantum well is
called a single-level quantum well.

\section{Phase conjugated field}
In a mesoscopic film the electric field generated via the direct and
indirect processes at a given point is roughly speaking of the order
$(\mu_0\omega/q_{\perp})\int\vec{J}_{-\omega}^{\,(3)}dz'$, whereas the
self-field has the magnitude
$(\mu_0\omega/q^{2})\vec{J}_{-\omega}^{\,(3)}$. Since $qd\ll1$, where
$d$ is the thickness of the film, we judge the self-field term to
dominate the phase conjugated field inside the quantum well, at least
for single-level metallic quantum wells which have thicknesses on the
atomic length scale. In the following we therefore use the so-called
self-field (electrostatic) approximation to calculate the phase
conjugated field inside the quantum well. With the propagator
$\tensor{G}(z,z'';\vec{q}_{\|},\omega)$ replaced by
$\btensor{g}(z-z';\omega)$, the phase conjugated field fulfills the
integral equation
\begin{equation}
 \vec{E}_{\rm{PC}}(z;\vec{q}_{\|},\omega)=
 \vec{E}_{\rm{PC}}^{\rm{B}}(z;\vec{q}_{\|},\omega)
 +{\vec{e}_{z}\otimes\vec{e}_{z}\over{}{\rm{i}}\varepsilon_0\omega}
 \cdot\int\stensor{\sigma}(z,z';\vec{q}_{\|},\omega)\cdot
 \vec{E}_{\rm{PC}}(z';\vec{q}_{\|},\omega)dz'
\label{eq:ping}
\end{equation}
inside the well, and the background field is now
\begin{equation}
 \vec{E}_{\rm{PC}}^{\rm{B}}(z;\vec{q}_{\|},\omega)=
 {\vec{e}_{z}\otimes\vec{e}_{z}\over{}{\rm{i}}\varepsilon_0\omega}
 \vec{J}_{-\omega}^{\,(3)}(z;\vec{q}_{\|},\omega).
\label{eq:pong}
\end{equation}
In the self-field approach the phase conjugated field has only a
component perpendicular to the surface (the $z$-component) inside the
well and only the $z$-component of the nonlinear current density
$\vec{J}_{-\omega}^{\,(3)}$ drives the process.

Once the phase conjugated field inside the quantum well has been
determined in a self-consistent manner from Eq.~(\ref{eq:ping}), it can
be determined outside using Eq.~(\ref{eq:loop}). The self-field does
of course not contribute to the exterior field, and no loop problem is
involved. All that need to be done is to integrate known quantities in
the $z$-direction over the well.

\section{Nonlinear conductivity tensor} 
As we may recall, the nonlinear conductivity tensor appearing in
Eq.~(\ref{eq:J3-wzq}) may in general be written as a sum of seven
parts (A--G) after the physical processes they describe. These have
the tensor symmetries shown in Tab.~\ref{tab:1}. In this chapter we use
this conductivity tensor in the form it takes for media with
two-dimensional translational invariance as it was developed in
Part~\ref{part:II}, but for quantum wells so thin that only a single
bound level exists.  The quantum well may be free standing, or it may
be deposited on a substrate that can be described by a refractive
index $n$ relative to the vacuum on the other side of the film. The
surface of the film is parallel to the $x$-$y$-plane in a Cartesian
coordinate system, and the interface between the film and the
substrate is placed at $z=0$ as shown in Fig.~\ref{fig:1}. We further
limit our study to the case where (i) all scattering takes place in
the $x$-$z$-plane, (ii) the interacting fields are linearly polarized
in ($p$) or perpendicular to ($s$) the scattering plane, (iii) the
pump fields in the phase conjugating system are counterpropagating
monochromatic plane waves with a uniform amplitude along the $z$-axis
and propagating in a direction parallel to the $x$-axis, and (iv) the
field is calculated within the self-field approximation.

From (i) above we get a mirror plane at $y=0$, leaving only tensor
elements of the conductivity tensors with an even number $(0,2,4)$ of
$y$'s in the Cartesian index nonzero. Condition (iii) implies as a
consequence of condition (ii) that no tensor elements of the nonlinear
conductivity tensor with one or both of the last two Cartesian indices
as $x$ contributes to the phase conjugated response.  Requirement (iv)
above implies that the first Cartesian index of a tensor element
should be $z$ in order to contribute to the phase conjugated response.
The choice of a single level quantum well in itself restricts the
transition current density to contain $x$- and $y$- components only.
Together with the fact that part B gives zero after integration over
$\vec{\kappa}_{\|}$ and E gives pure interband contributions, these
choices leave two nonzero elements of the nonlinear conductivity
tensor, namely
\begin{eqnarray}
\lefteqn{
{\Xi}_{zyyz}^{\rm{C}}(z,z',z'',z''';q_{\|}-k_{\|})=
{\Xi}_{zzzz}^{\rm{C}}(z,z',z'',z''';q_{\|}-k_{\|})=
}\nonumber\\ &\quad&
 {e^4\over2^4\pi^2{\rm{i}}\hbar\omega^3m_{e}^2}
 {\cal{C}}(q_{\|}-k_{\|})
 \delta(z'-z''')\delta(z-z'')
 |\psi(z')|^2|\psi(z)|^2,
\label{eq:XiEC}
\\
\lefteqn{
{\Xi}_{zyyz}^{\rm{D}}(z,z',z'',z''';q_{\|},k_{\|})=
 {e^4\over2^6\pi^2i\omega^3m_e^3}
 {\cal{D}}(q_{\|},k_{\|})
 \delta(z-z''')
 |\psi(z'')|^2|\psi(z')|^2|\psi(z)|^2,
}\nonumber\\
\label{eq:XiED}
\end{eqnarray}
where
\begin{eqnarray}
\lefteqn{
 {\cal{C}}(q_{\|}-k_{\|})=2
 \int{f(\vec{\kappa}_{\|}+[{q}_{\|}-{k}_{\|}]\vec{e}_{x})-f(\vec{\kappa}_{\|})
 \over\hbar(q_{\|}-k_{\|})[2\kappa_x+q_{\|}-k_{\|}]/(2m_e)-{\rm{i}}/\tau}
 d^2\kappa_{\|},
}\label{eq:C}
\\
\lefteqn{
 {\cal{D}}(q_{\|},k_{\|})=2\int
 {\kappa_y^2\over\hbar(q_{\|}+k_{\|})[2\kappa_x+q_{\|}+k_{\|}]/(2m_e)-{\rm{i}}/\tau}
}\nonumber\\ &\quad&\times
 \left({f(\vec{\kappa}_{\|})-f(\vec{\kappa}_{\|}+{k}_{\|}\vec{e}_{x})
 \over\hbar{}k_{\|}[2\kappa_x+k_{\|}]/(2m_e)-{\rm{i}}/\tau-\omega}
 +{f(\vec{\kappa}_{\|}+[{k}_{\|}+{q}_{\|}]\vec{e}_{x})
 -f(\vec{\kappa}_{\|}+{k}_{\|}\vec{e}_{x})
 \over\hbar{}q_{\|}[2\kappa_x+q_{\|}+2k_{\|}]/(2m_e)-{\rm{i}}/\tau+\omega}
\right.\nonumber\\ &&\left.
 +{f(\vec{\kappa}_{\|})-f(\vec{\kappa}_{\|}+{q}_{\|}\vec{e}_{x})
 \over\hbar{}q_{\|}[2\kappa_x+q_{\|}]/(2m_e)-{\rm{i}}/\tau+\omega}
 +{f(\vec{\kappa}_{\|}+[{k}_{\|}+{q}_{\|}]\vec{e}_{x})
 -f(\vec{\kappa}_{\|}+{q}_{\|}\vec{e}_{x})
 \over\hbar{}k_{\|}[2\kappa_x+k_{\|}+2q_{\|}]/(2m_e)-{\rm{i}}/\tau-\omega}
 \right)
 d^2\kappa_{\|}.
\nonumber\\
\label{eq:D}
\end{eqnarray}
\noindent
The number $2$ appearing in front of the integrals above represents
the summation over the degenerate spin energies.

The free-particle character of the electron motion in the plane of the
quantum well enables us to write the solutions to the
light-unperturbed Schr{\"o}dinger equation in the form
$\Psi(\vec{r}\,)=(2\pi)^{-1}\psi(z)\exp({\rm{i}}\vec{\kappa}_{\|}\cdot\vec{r}\,)$,
where $\vec{\kappa}_{\|}=(\kappa_x,\kappa_y,0)$ is the wavevector of
the electron in consideration and $\psi(z)$, appearing in
Eqs.~(\ref{eq:XiEC}) and (\ref{eq:XiED}), is the $z$-dependent part of
the wave function, common to all electrons. The $x$-$y$-dependent
parts of the wave functions,
$(2\pi)^{-1}\exp({\rm{i}}\vec{\kappa}_{\|}\cdot\vec{r}\,)$, are
orthonormalized in the Dirac sense, i.e., they obey the equation
$(2\pi)^{-2}\int\exp[{\rm{i}}(\vec{\kappa}_{\|}-\vec{\kappa}_{\|}')\cdot\vec{r}\,]d^2r=\delta(\vec{\kappa}_{\|}-\vec{\kappa}_{\|}')$,
and the $z$-dependent part fulfills the separate normalization
condition $\int|\psi(z)|^2dz=1$. In Eqs.~(\ref{eq:C}) and (\ref{eq:D})
the response of all electrons is taken into account by integrating
over all possible $\vec{\kappa}_{\|}$ wavevectors. The eigenenergy
${\cal{E}}(\vec{\kappa}_{\|})$ belonging to the state
$\Psi(\vec{r}\,)$ is obtained by adding to the common bound-state
energy $\varepsilon$, the kinetic energy in the parallel motion. Thus
\begin{equation}
 {\cal{E}}(\vec{\kappa}_{\|})=\varepsilon+{\hbar^2\over2m_{e}}\kappa_{\|}^2.
\end{equation}
The quantity
$f(\vec{\kappa}_{\|})=[1+\exp\{({\cal{E}}(\vec{\kappa}_{\|})-\mu)/(k_{\rm{B}}T)\}]^{-1}$
denotes the Fermi-Dirac distribution function for this eigenstate,
$\mu$ being the chemical potential of the electron system,
$k_{\rm{B}}$ the Boltzmann constant, and $T$ the absolute temperature.

\section{Probe with single Fourier component}

In the following we calculate the phase conjugated field generated by
a probe field which consists of only one plane-wave component of
wavevector $\vec{q}=(q_{\|},0,q_{\perp})$. A probe field of the form
$\vec{E}(z;\vec{q}_{\|})=\vec{E}e^{{\rm{i}}q_{\perp}z}$ is hence
inserted in Eq.~(\ref{eq:J3-wzq}).

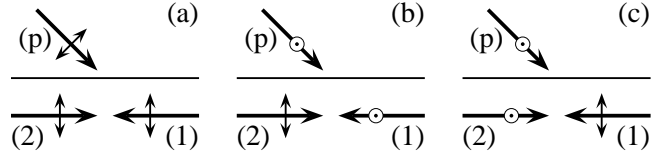
\begin{figure}[tb]
\setlength{\unitlength}{1mm}
\psset{unit=1mm}
\begin{center}
\begin{pspicture}(0,0)(85,22)
\put(0,0){
 \psline[linewidth=0.25mm]{-}(0,10)(25,10)
 \psline[linewidth=0.5mm]{->}(0,5)(11.5,5)
 \psline[linewidth=0.5mm]{->}(25,5)(13.5,5)
 \psline[linewidth=0.25mm]{<->}(6.5,2)(6.5,8)
 \psline[linewidth=0.25mm]{<->}(18.5,2)(18.5,8)
 \put(25,0){\makebox(0,4)[r]{(1)}}
 \put(0,0){\makebox(0,4)[l]{(2)}}
 \put(1,13){\makebox(0,4)[l]{(p)}}
 \psline[linewidth=0.5mm]{->}(3.4,19.1)(11.5,11)
 \psline[linewidth=0.25mm]{<->}(5.9,12.4)(10.1,16.6)
 \put(25,20){\makebox(0,0)[tr]{(a)}}
}
\put(30,0){
 \psline[linewidth=0.25mm]{-}(0,10)(25,10)
 \psline[linewidth=0.5mm]{->}(0,5)(11.5,5)
 \psline[linewidth=0.5mm]{->}(25,5)(13.5,5)
 \psline[linewidth=0.25mm]{<->}(6.5,2)(6.5,8)
 \pscircle[linewidth=0.1mm,linecolor=black,fillstyle=solid,fillcolor=white](18.5,5){1.0}
 \qdisk(18.5,5){0.25}
 \put(25,0){\makebox(0,4)[r]{(1)}}
 \put(0,0){\makebox(0,4)[l]{(2)}}
 \put(1,13){\makebox(0,4)[l]{(p)}}
 \psline[linewidth=0.5mm]{->}(3.4,19.1)(11.5,11)
 \put(8.0,14.5){%
  \pscircle[linewidth=0.1mm,linecolor=black,fillstyle=solid,fillcolor=white](0,0){1.0}
  \qdisk(0,0){0.25}
 }
 \put(25,20){\makebox(0,0)[tr]{(b)}}
}
\put(60,0){
 \psline[linewidth=0.25mm]{-}(0,10)(25,10)
 \psline[linewidth=0.5mm]{->}(0,5)(11.5,5)
 \psline[linewidth=0.5mm]{->}(25,5)(13.5,5)
 \pscircle[linewidth=0.1mm,linecolor=black,fillstyle=solid,fillcolor=white](6.5,5){1.0}
 \qdisk(6.5,5){0.25}
 \psline[linewidth=0.25mm]{<->}(18.5,2)(18.5,8)
 \put(25,0){\makebox(0,4)[r]{(1)}}
 \put(0,0){\makebox(0,4)[l]{(2)}}
 \put(1,13){\makebox(0,4)[l]{(p)}}
 \psline[linewidth=0.5mm]{->}(3.4,19.1)(11.5,11)
 \put(8.0,14.5){%
  \pscircle[linewidth=0.1mm,linecolor=black,fillstyle=solid,fillcolor=white](0,0){1.0}
  \qdisk(0,0){0.25}
 }
 \put(25,20){\makebox(0,0)[tr]{(c)}}
}
\end{pspicture}
\end{center}
\caption[Possible combinations of polarized light in a single-level
quantum well]{Schematic illustration showing three of the possible
  field polarization combinations which may give rise to a phase
  conjugated response in a single-level quantum well, viz.~(a) the
  purely $p$-polarized configuration, and (b--c) the mixed
  polarization configurations where the pump fields are differently
  polarized while the probe is $s$-polarized. The two mixed
  polarization states are closely related, since replacing
  $\vec{q}_{\|}$ with $-\vec{q}_{\|}$ in one of them yields the other.
  In both (a and b--c) cases, the phase conjugated response is
  $p$-polarized. The schemes are shown in the Cartesian coordinate
  system given in Fig.~\ref{fig:1}, such that the small arrows in the
  plane represents $p$-polarized states and the circles represents
  $s$-polarized states. The large arrows show the direction of (one
  Fourier component of) the wavevectors of the pump fields (1 and 2)
  and the probe field (p).\label{fig:2}}
\end{figure}

Then, when using linearly polarized light, three different
combinations of polarization gives a nonlinear current density, namely
(i) the one in which all participating fields are $p$-polarized
($ppp$), and (ii) the two combinations where the pump fields are
differently polarized and the probe field is $s$-polarized ($sps$ and
$pss$). In all cases, the phase conjugated response is $p$-polarized,
and thus characterized in terms of the polarization states of the
probe and phase conjugated fields, case (i) may be classified as a $p$
to $p$ transition, and cases (ii) as $s$ to $p$ transitions. A
schematic illustration of these interaction configurations is shown in
Fig.~\ref{fig:2}. Defining the $z$-independent quantity
\begin{equation} 
 {\cal{J}}_{-\omega,z}^{(3)}(\vec{q}_{\|})
 \equiv{J_{-\omega,z}^{(3)}(z;\vec{q}_{\|})\over|\psi(z)|^2},
\end{equation} 
the above conditions yields for the $p$ to $p$ transition 
\begin{eqnarray}
\lefteqn{
 {\cal{J}}_{-\omega,z}^{(3)}(\vec{q}_{\|})=
 {e^4\over2^8\pi^6{\rm{i}}\hbar\omega^3m_{e}^2}
 \left[{\cal{C}}(q_{\|}-k_{\|})
 +{\cal{C}}(q_{\|}+k_{\|})\right]
 E_z^{(1)}E_z^{(2)}E_z^{*}
}\nonumber\\ &\quad&\times
 \int|\psi(z')|^2e^{-{\rm{i}}q_{\perp}^{*}z'}dz'
\label{eq:J3-ppp-0}
\end{eqnarray}
and for the $s$ to $p$ transitions
\begin{eqnarray}
\lefteqn{
 {\cal{J}}_{-\omega,z}^{(3)}(\vec{q}_{\|})=
 {e^4\over2^8\pi^6{\rm{i}}\hbar\omega^3m_e^2}
 \left[{\cal{C}}(q_{\|}+k_{\|})
 +{\hbar\over4m_e}{\cal{D}}(\vec{q}_{\|},-\vec{k}_{\|})\right]
 E_y^{(1)}E_z^{(2)}E_y^{*}
}\nonumber\\ &\quad&\times
 \int|\psi(z')|^2 e^{-{\rm{i}}q_{\perp}^{*}z'}dz',
\label{eq:J3-sps-0}
\\
\lefteqn{
 {\cal{J}}_{-\omega,z}^{(3)}(\vec{q}_{\|})=
 {e^4\over2^8\pi^6{\rm{i}}\hbar\omega^3m_e^2}
 \left[{\cal{C}}(q_{\|}-k_{\|})
 +{\hbar\over4m_e}{\cal{D}}(q_{\|},k_{\|})\right]
 E_z^{(1)}E_y^{(2)}E_y^{*}
}\nonumber\\ &&\times
 \int|\psi(z')|^2 e^{-{\rm{i}}q_{\perp}^{*}z'}dz'.
\label{eq:J3-pss-0}
\end{eqnarray}
\noindent
In the above three equations, the superscript $(1)$ refers to the pump
field propagating along the $x$-axis in the positive direction
($\vec{k}_{\|}=k_{\|}\vec{e}_{x}$), and the superscript $(2)$ refers
to the other pump field.  The $s$ to $p$ transitions are symmetric in
the sense that if the probe wavevector $\vec{q}_{\|}$ is replaced by
$-\vec{q}_{\|}$ in Eq.~(\ref{eq:J3-sps-0}), then the result of
Eq.~(\ref{eq:J3-pss-0}) is obtained, and vice versa. The $p$ to $p$
transition is symmetric to itself in this sense.

For a single-level quantum well, the $zz$-component of the linear
conductivity tensor is given by \cite{Feibelman:82:1}
\begin{equation}
 \sigma_{zz}(z,z';\vec{q}_{\|})={{\rm{i}}e^2{\cal{N}}\over{}m_e(\omega+{\rm{i}}/\tau)}
 |\psi(z)|^2\delta(z-z'),
\label{eq:sigma-zz}
\end{equation}
where
\begin{equation}
 {\cal{N}}={2\over(2\pi)^{2}}\int{}f(\vec{\kappa}_{\|})d^2\kappa_{\|}.
\label{eq:N}
\end{equation}
In order to take into account the coupling to surroundings we have
introduced a phenomenological relaxation time $\tau$ in the
diamagnetic expression for $\sigma_{zz}$ [Eq.~(\ref{eq:sigma-zz})]
\cite{Feibelman:82:1}. A factor of two in this equation again stems
from the spin summation, and the quantity ${\cal{N}}|\psi(z)|^2$ is
the conduction electron density. The phase conjugated field inside the
quantum well has a $z$-component, $E_{{\rm{PC}},z}(z;\vec{q}_{\|})$,
only, and by combining Eqs.~(\ref{eq:ping}), (\ref{eq:pong}), and
(\ref{eq:sigma-zz}) it appears that this is given by
\begin{equation}
 E_{{\rm{PC}},z}(z;\vec{q}_{\|})=
 {im_e(\omega+{\rm{i}}/\tau)\over{}e^2{\cal{N}}|\psi(z)|^2
  -\epsilon_0m_e\omega(\omega+{\rm{i}}/\tau)}
 J_{-\omega,z}^{(3)}(z;\vec{q}_{\|}).
\end{equation}
Using now Eq.~(\ref{eq:loop}), the $z$-components of the phase
conjugated field outside the quantum well can be calculated, and the
result is
\begin{eqnarray}
 E_{{\rm{PC}},z}(z;\vec{q}_{\|})=
 {\cal{J}}_{-\omega,z}^{(3)}(\vec{q}_{\|})
 e^{-{\rm{i}}q_{\perp}z}
 {q_{\|}^2m_e(\omega+{\rm{i}}/\tau)\over2q_{\perp}}
 \int{(e^{{\rm{i}}q_{\perp}z'}+r^pe^{-{\rm{i}}q_{\perp}z'})|\psi(z')|^2\over
 e^2{\cal{N}}|\psi(z')|^2-\epsilon_0m_e\omega(\omega+{\rm{i}}/\tau)}dz',
\nonumber\\
\label{eq:EPC-IInt-na}
\end{eqnarray}
where the relevant expression for
${\cal{J}}_{-\omega,z}^{(3)}(\vec{q}_{\|})$ is taken from
Eq.~(\ref{eq:J3-ppp-0}), (\ref{eq:J3-sps-0}), or (\ref{eq:J3-pss-0}).
Given the $z$-component of the phase conjugated field, the
$x$-component may be found from
\begin{equation}
 {E}_{{\rm{PC}},x}(z;\vec{q}_{\|})
 ={q_{\perp}\over{}q_{\|}}{E}_{{\rm{PC}},z}(z;\vec{q}_{\|}),
\label{eq:EPCx-zq}
\end{equation}
which follows from the expression for the electromagnetic propagator,
or equivalently from the demand that the phase conjugated field must
be transverse in vacuum.

The integral in Eq.~(\ref{eq:EPC-IInt-na}) is different from zero only
in the region of the quantum well [from around $z'=-d$ to around
$z'=0$ in the chosen coordinate system, the exact domain depending on
the extent of the electronic wave function $\psi(z')$]. Since the
width ($\sim{}d$) of a single-level metallic quantum well is in the
{\AA}ngstr{\"o}m range, and $q_{\perp}$ is typically in the micrometer
range for optical signals such that $q_{\perp}d\ll1$, it is a good
approximation to put $\exp(\pm{}{\rm{i}}q_{\perp}z')=1$ in
Eq.~(\ref{eq:EPC-IInt-na}). For electromagnetic frequencies so high
that $q_{\perp}\sim{}d^{-1}$, the present theory would anyway be too
simple to rely on [the Bloch function character of the wave functions
along the surface and excitation to the continuum (photoemission)
should be incorporated at least]. With the above-mentioned
approximation, Eq.~(\ref{eq:EPC-IInt-na}) is reduced to
\begin{equation}
 E_{{\rm{PC}},z}(z;\vec{q}_{\|})=
 {\cal{J}}_{-\omega,z}^{(3)}(\vec{q}_{\|})
 e^{-{\rm{i}}q_{\perp}z}
 {(1+r^p)q_{\|}^2\over2\epsilon_0\omega{}q_{\perp}}
 \int{|\psi(z')|^2\over{}\gamma|\psi(z')|^2-1}dz',
\label{eq:EPC-IInt}
\end{equation}
where $\gamma=e^2{\cal{N}}/[\epsilon_0m_e\omega(\omega+{\rm{i}}/\tau)]$. Using the
approximation $\exp({\rm{i}}q_{\perp}z')=1$ and the normalization condition
on $\psi(z')$, Eqs.~(\ref{eq:J3-ppp-0})--(\ref{eq:J3-pss-0}) are
reduced to
\begin{eqnarray}
 {\cal{J}}_{-\omega,z}^{(3)}(\vec{q}_{\|})&=&
 {e^4\over2^8\pi^6{\rm{i}}\hbar\omega^3m_{e}^2}
 \left[{\cal{C}}(q_{\|}-k_{\|})
 +{\cal{C}}(q_{\|}+k_{\|})\right]
 E_z^{(1)}E_z^{(2)}E_z^{*},
\label{eq:J3-ppp-0-a}
\\
 {\cal{J}}_{-\omega,z}^{(3)}(\vec{q}_{\|})&=&
 {e^4\over2^8\pi^6{\rm{i}}\hbar\omega^3m_e^2}
 \left[{\cal{C}}(q_{\|}+k_{\|})
 +{\hbar\over4m_e}{\cal{D}}(\vec{q}_{\|},-\vec{k}_{\|})\right]
 E_y^{(1)}E_z^{(2)}E_y^{*},
\label{eq:J3-sps-0-a}
\end{eqnarray}
and
\begin{eqnarray}
 {\cal{J}}_{-\omega,z}^{(3)}(\vec{q}_{\|})&=&
 {e^4\over2^8\pi^6{\rm{i}}\hbar\omega^3m_e^2}
 \left[{\cal{C}}(q_{\|}-k_{\|})
 +{\hbar\over4m_e}{\cal{D}}(q_{\|},k_{\|})\right]
 E_z^{(1)}E_y^{(2)}E_y^{*},
\label{eq:J3-pss-0-a}
\end{eqnarray}
respectively.

Thus the phase conjugated field from a single-level quantum well is
described in the mixed Fourier space by Eq.~(\ref{eq:EPC-IInt}) with
insertion of Eq.~(\ref{eq:J3-ppp-0-a}), (\ref{eq:J3-sps-0-a}), or
(\ref{eq:J3-pss-0-a}), the expressions for ${\cal{C}}$
[Eq.~(\ref{eq:C})] and ${\cal{D}}$ [Eq.~(\ref{eq:D})] carrying the
information on the two-dimensional electron dynamics.

So far, the description of the phase conjugated response has been
independent of the actual wave functions in the active medium, and
thus independent of the form of the quantum well potential. In order
to prepare our theory for a numerical study we now introduce a model
potential in our quantum well system, namely the infinite barrier
potential.

\section{Infinite barrier model}
To achieve a qualitative impression of the phase conjugation from a
single-level metallic quantum well it is sufficient to carry out
numerical calculations on the basis of the simple infinite barrier
(IB) model. In this model the one-dimensional potential $V(z)$ is
taken to be zero in the interval $-d\leq{}z\leq{}0$ (inside the
quantum well) and infinite elsewhere. The stationary state wave
function now is given by $\psi(z)=\sqrt{2/d}\sin(\pi{}z/d)$ inside the
well and $\psi(z)=0$ outside, and the associated energy is
$\varepsilon=(\pi\hbar)^2/(2m_ed^2)$. In the IB model the number of
bound states is of course infinite, and to use this model in the
context of a single level calculation, one must be sure that only one
of the bound states (the ground state) has an energy below the Fermi
energy, and that the optical frequency is so low that interlevel
excitations are negligible.

For a metallic quantum well one may even at room temperature
approximate the Fermi-Dirac distribution function appearing in the
expressions for ${\cal{C}}$, ${\cal{D}}$, and ${\cal{N}}$ in
Eqs.~(\ref{eq:C}), (\ref{eq:D}), and (\ref{eq:N}) by its value at zero
temperature, i.e.,
\begin{equation}
 \lim_{T\rightarrow0}f(\vec{\kappa}_{\|})=
 \Theta\left\{{\cal{E}}_{F}-{\hbar^2\over2m_e}
 \left[\left(\pi\over{}d\right)^2+\kappa_{\|}^2\right]\right\},
\label{eq:fermi-T=0}
\end{equation} 
where $\Theta$ is the Heaviside step function and ${\cal{E}}_{F}$ is
the Fermi energy of the system. In the low temperature limit it is
possible to find analytical solutions to the integrals over
$\vec{\kappa}_{\|}$ appearing in Eqs.~(\ref{eq:C}) and (\ref{eq:D}).
This is adequately achieved by performing a coordinate transformation
into cylindrical coordinates, since each Heaviside step function gives
nonzero values in the $\kappa_x$-$\kappa_y$-space only inside a circle
with radius, say, $\alpha$. The explicit calculations are tedious but
trivial to carry out, and since the final expressions for ${\cal{C}}$
and ${\cal{D}}$ are rather long we do not present them here. For the
interested reader these calculations are reproduced in
Appendices~\ref{ch:Solve-Q} and \ref{app:C} [specifically,
Section~\ref{sec:CDN}].

The Fermi energy is calculated from the global charge neutrality
condition [see \citeN{Keller:96:1} and the calculation performed in
Appendix~\ref{app:D}], which for a single level quantum well takes the
form
\begin{equation}
 {\cal{N}}=ZN_+d,
\end{equation}
where $N_+$ is the number of positive ions per unit volume and $Z$ is
the valence of these ions. Since
${\cal{N}}=m_e({\cal{E}}_{F}-\varepsilon)/(\pi\hbar^2)$, cf.~the
calculation in the Section~\ref{sec:CDN}, one gets
\begin{equation}
 {\cal{E}}_{F}={\pi\hbar^2\over{}m_e}\left[ZN_+d+{\pi\over2d^2}\right].
\label{eq:Fermi}
\end{equation}
In order that just the ground state (energy $\varepsilon$) has an
energy less than the Fermi energy, the film thickness must be less
than a certain maximum value $d_{\rm{max}}$. When the thickness of the
well becomes so large that the Fermi energy equals the energy
$\varepsilon_2=(2\pi\hbar)^2/(2m_ed^2)$ of the first excited state a
second bound state of energy less than ${\cal{E}}_{F}$ will appear.
From the condition
${\cal{E}}_{F}(d_{\rm{max}})=\varepsilon_2(d_{\rm{max}})$,
$d_{\rm{max}}$ can be calculated, and one gets by means of
Eq.~(\ref{eq:Fermi})
\begin{equation}
 d_{\rm{max}}=\sqrt[3]{3\pi/(2ZN_+)}\,,
\label{eq:dmax}
\end{equation}
i.e., a result which depends on the number of conduction electrons in
the film. The minimum thickness is in the IB model zero, but in
reality the smallest thickness is a single monolayer.

Inserting the IB model into the integral over the source region
appearing in Eq.~(\ref{eq:EPC-IInt}) we get
\begin{eqnarray}
 \int{|\psi(z')|^2\over{}\gamma|\psi(z')|^2-1}dz'
 =\int_{-d}^{0}{2\sin^2({\pi{}z'/d})
  \over2\gamma\sin^2({\pi{}z'/d})-d}dz',
\end{eqnarray}
which by substitution of $\theta=\pi{}z'/d$, addition and
subtraction of $d$ in the nominator of the integral, and use of
$2\gamma\sin^2\theta-d=2\gamma[\sqrt{1-d/(2\gamma)}-\cos\theta][\sqrt{1-d/(2\gamma)}+\cos\theta]$
gives
\begin{eqnarray}
 {d\over\pi\gamma}\left[\pi-{d\over4\gamma}{1\over\sqrt{1-d/(2\gamma)}}
 \int_{0}^{2\pi}{d\theta\over\sqrt{1-d/(2\gamma)}+\cos\theta}\right]
 ={d\over\gamma}\left[1-{1\over\sqrt{2\gamma/d-1}}\right]\approx{d\over\gamma}.
\nonumber\\
\label{eq:intdg}
\end{eqnarray}
The solution to the integral in Eq.~(\ref{eq:intdg}) is obtained by
use of Eq.~(\ref{eq:Ang-nocos}), and since $2|\gamma|/d\gg1$ [for
metals, $|\gamma|$ lies typically between $1$ and $100$ in the optical
region (e.g., for copper $|\gamma|\approx85$ in the present study) and
$d$ is in the {\AA}ngstr{\"o}m range]. Using this result and the
expression for the Fermi energy given in Eq.~(\ref{eq:Fermi}), we
obtain by insertion into Eq.~(\ref{eq:EPC-IInt}) the result
\begin{equation}
 E_{{\rm{PC}},z}(z;\vec{q}_{\|})=
 {q_{\|}^2m_e(\omega+{\rm{i}}/\tau)(1+r^p)\over2q_{\perp}e^2ZN_+}
 {\cal{J}}_{-\omega,z}^{(3)}(\vec{q}_{\|})
 e^{-{\rm{i}}q_{\perp}z}.
\end{equation}
By insertion of the relevant expressions for
${\cal{J}}_{-\omega,z}^{(3)}(\vec{q}_{\|})$ we finally obtain the following
results for the $z$-component of the phase conjugated field outside
the quantum well:
\begin{eqnarray}
\lefteqn{
 E_{{\rm{PC}},z}(z;\vec{q}_{\|})=
 {e^2(\omega+{\rm{i}}/\tau)(1+r^p)\over2^9\pi^6\hbar\omega^3ZN_+m_e}
 {q_{\|}^2\over{}{\rm{i}}q_{\perp}}
 \left[{\cal{C}}(q_{\|}-k_{\|})
 +{\cal{C}}(q_{\|}+k_{\|})\right]
}\nonumber\\ &\quad&\times
 E_z^{(1)}E_z^{(2)}E_z^{*}e^{-{\rm{i}}q_{\perp}z},
\label{eq:PC-ppp}
\end{eqnarray}
for the purely $p$-polarized configuration, and 
\begin{eqnarray}
\lefteqn{
 E_{{\rm{PC}},z}(z;\vec{q}_{\|})=
 {e^2(\omega+{\rm{i}}/\tau)(1+r^p)\over2^9\pi^6\hbar\omega^3ZN_+m_e}
 {q_{\|}^2\over{}{\rm{i}}q_{\perp}}
 \left[{\cal{C}}(q_{\|}+k_{\|})
 +{\hbar\over4m_e}{\cal{D}}(q_{\|},-k_{\|})\right]
}\nonumber\\ &\quad&\times
 E_y^{(1)}E_z^{(2)}E_y^{*}
 e^{-{\rm{i}}q_{\perp}z},
\label{eq:PC-sps}
\\
\lefteqn{
 E_{{\rm{PC}},z}(z;\vec{q}_{\|})=
 {e^2(\omega+{\rm{i}}/\tau)(1+r^p)\over2^9\pi^6\hbar\omega^3ZN_+m_e}
 {q_{\|}^2\over{}{\rm{i}}q_{\perp}}
 \left[{\cal{C}}(q_{\|}-k_{\|})
 +{\hbar\over4m_e}{\cal{D}}(q_{\|},k_{\|})\right]
}\nonumber\\ &&\times
 E_z^{(1)}E_y^{(2)}E_y^{*}
 e^{-{\rm{i}}q_{\perp}z}
\label{eq:PC-pss}
\end{eqnarray}
\noindent 
for the configurations with mixed polarization of the pump fields.
The $x$-component of the phase conjugated field is obtained using
Eq.~(\ref{eq:EPCx-zq}).

\chapter{Numerical results}\label{sec:num}\label{Ch:11}
The theoretical description presented in the previous chapter resulted
in expressions for the phase conjugated field from a single level
quantum well. Thus for the numerical work, the phase conjugated field
is given completely by Eqs.~(\ref{eq:PC-ppp})--(\ref{eq:PC-pss}) and
(\ref{eq:EPCx-zq}) with the insertion of the expressions for the
electron dynamics parallel to the surface plane, given by
Eqs.~(\ref{eq:C-solved})--(\ref{eq:D-solved}) in
Appendix~\ref{app:C}. In the following we will present the phase
conjugation reflection coefficient, succeeded by a discussion of a
possible excitation scheme which might be adequate for studies of
phase conjugation of optical near fields \cite{Bozhevolnyi:94:1}.

\section{Phase conjugation reflection coefficient}\label{sec:11.1} 
To estimate the amount of light we get back through the phase
conjugated channel, we define the phase conjugation (energy)
reflection coefficient as
\begin{equation}
 R_{\rm{PC}}(z;\vec{q}_{\|})={I_{\rm{PC}}(z;\vec{q}_{\|})
  \over{}I^{(1)}I^{(2)}I_{\rm{Probe}}(-d;\vec{q}_{\|})},
\label{eq:RPC}
\end{equation}
in which $I^{(1)}$, $I^{(2)}$, $I_{\rm{Probe}}$, and $I_{\rm{PC}}$ are
the intensities of the two pump beams, the probe and the phase
conjugated field, respectively. Each of the intensities are given by
\begin{equation}
 I={\epsilon_0c_0\over2}{\vec{E}\cdot\vec{E}^{*}\over(2\pi)^4},
\label{eq:RPC-I}
\end{equation}
where the factor of $(2\pi)^{-4}$ originates from the manner in which
we have introduced the Fourier amplitudes of the fields. If the probe
field is evanescent the intensity of the phase conjugated field,
$I_{\rm{PC}}(z;\vec{q}_{\|})$, will depend on the distance from the
surface, and consequently the reflection coefficient is $z$-dependent
in such a case.

For the remaining part of this work we choose a copper quantum well
with $N_+=8.47\times10^{28}$m$^{-3}$ and $Z=1$ [data taken from
\citeN{Ashcroft:76:1}]. Then from Eq.~(\ref{eq:dmax}), the maximal
thickness becomes $d_{\rm{max}}=3.82${\AA}, which is more than two
monolayers and less than three. Thus we have two obvious choices for
the thickness of the quantum well, namely a single monolayer or two
monolayers. We thus take a look at both possibilities in the
following, corresponding to a thickness of $d=1.8${\AA} for one
monolayer and $d=3.6${\AA} for two monolayers. The Cu quantum well can
adequately be deposited on a glass substrate for which we use a
refractive index $n$ of 1.51. With this substrate, a reasonable
description of the linear vaccum/substrate amplitude reflection
coefficient $r^p$ is obtained by use of the classical Fresnel formula
\begin{equation}
 r^p={n^2q_{\perp}-({n^2q^2-q_{\|}^2})^{1\over2}\over
      n^2q_{\perp}+({n^2q^2-q_{\|}^2})^{1\over2}},
\label{eq:rp}
\end{equation}
$q=\omega/c_0$ being the vacuum wavenumber, as before. Then, having
the pump fields parallel to the $x$-axis gives a pump wavenumber
$k_{\|}=1.51q$. The wavelength $\lambda$ of the light is chosen to be
$\lambda=1061$nm.

\begin{figure}[tb]
\setlength{\unitlength}{1mm}
\psset{unit=1mm}
\begin{center}
\begin{pspicture}(0,0)(127,96)
\put(-8,2){\epsfig{file=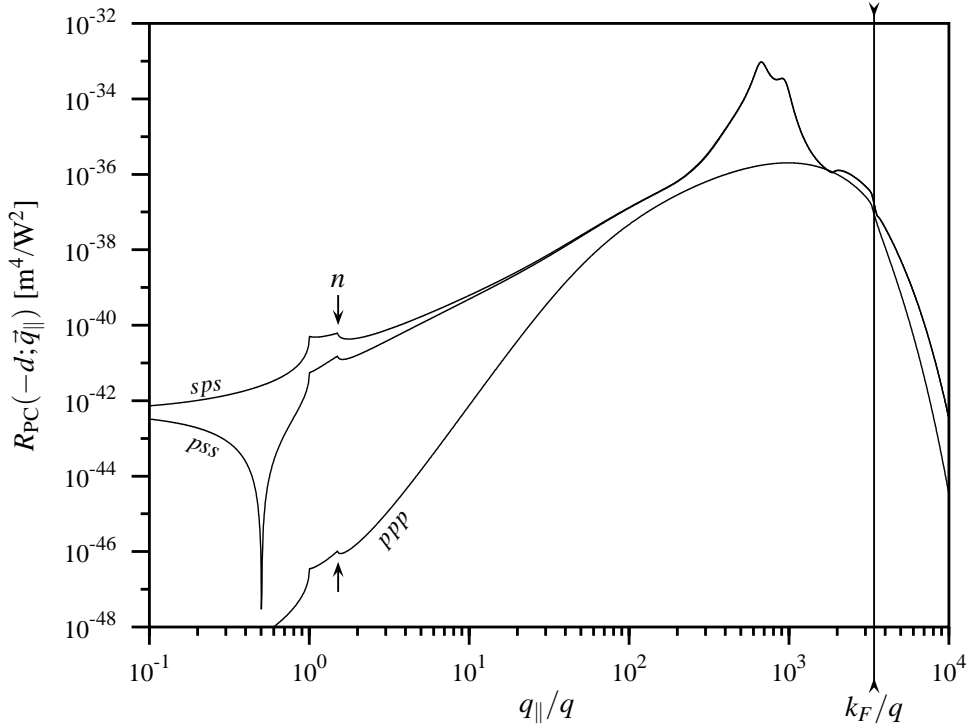,width=135\unitlength}}
\rput[t]{90}(0,53){$R_{\rm{PC}}(-d;\vec{q}_{\|})$~[m$^4$/W$^2$]}
\put(71.5,0){\makebox(0,0)[b]{$q_{\|}/q$}}
\pstextpath[l]{\psline[linestyle=none](24,45)(31,46)}{\footnotesize$sps$}
\pstextpath[l]{\psline[linestyle=none](24,38)(29,36)}{\footnotesize$pss$}
\pstextpath[l]{\psline[linestyle=none](49,24)(53,29)}{\footnotesize$ppp$}
\psline[linewidth=0.25mm]{->}(43.75,57)(43.75,53)
\psline[linewidth=0.25mm]{->}(43.75,17.5)(43.75,21.5)
\put(43.75,58){\makebox(0,0)[b]{$n$}}
\psline[linewidth=0.25mm]{>-<}(115,4)(115,96)
\put(115,0){\makebox(0,0)[b]{$k_F/q$}}
\end{pspicture}
\end{center}
\caption[The phase conjugation reflection coefficient at the
vacuum/film interface (one monolayer Cu quantum well)]{The phase
  conjugation reflection coefficient at the vacuum/film interface of a
  single monolayer copper quantum well,
  $R_{\rm{PC}}(-d;\vec{q}_{\|})$, is plotted for ($ppp$) the $p$ to
  $p$ transition (corresponding to diagram (a) in Fig.~\ref{fig:2}),
  ($sps$) one of the $s$ to $p$ transitions (corresponding to diagram
  (b) in Fig.~\ref{fig:2}), and ($pss$) the other $s$ to $p$
  transition (corresponding to diagram (c) in Fig.~\ref{fig:2}), as a
  function of the normalized component of the probe wavevector along
  the interface, $q_{\|}/q$. The normalized Fermi wavenumber is
  indicated by the vertical line. It is for a single monolayer of
  copper $k_F/q=3.38\times10^3$. The set of arrows labeled $n$ are
  placed at $q_{\|}=nq$.\label{fig:11.1}}
\end{figure}

\begin{figure}[tb]
\setlength{\unitlength}{1mm}
\psset{unit=1mm}
\begin{center}
\begin{pspicture}(0,0)(127,96)
\put(-8,2){\epsfig{file=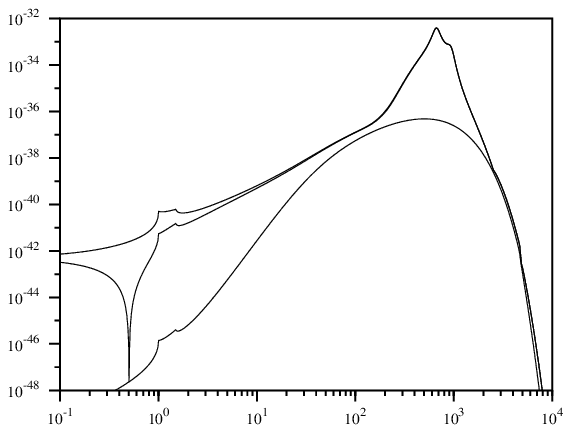,width=135\unitlength}}
\rput[t]{90}(0,53){$R_{\rm{PC}}(-d;\vec{q}_{\|})$~[m$^4$/W$^2$]}
\put(71.5,0){\makebox(0,0)[b]{$q_{\|}/q$}}
\pstextpath[l]{\psline[linestyle=none](24,45)(31,46)}{\footnotesize$sps$}
\pstextpath[l]{\psline[linestyle=none](24,38)(29,36)}{\footnotesize$pss$}
\pstextpath[l]{\psline[linestyle=none](49,27)(53,32)}{\footnotesize$ppp$}
\psline[linewidth=0.25mm]{->}(43.75,57)(43.75,53)
\psline[linewidth=0.25mm]{->}(43.75,20.5)(43.75,24.5)
\put(43.75,58){\makebox(0,0)[b]{$n$}}
\psline[linewidth=0.25mm]{>-<}(113,4)(113,96)
\put(113,0){\makebox(0,0)[b]{$k_F/q$}}
\end{pspicture}
\end{center}
\caption[The phase conjugation reflection coefficient at the
vacuum/film interface (two monolayer Cu quantum well)]{The phase
  conjugation reflection coefficient at the vacuum/film interface of a
  two-monolayer copper quantum well, $R_{\rm{PC}}(-d;\vec{q}_{\|})$,
  is plotted for ($ppp$) the $p$ to $p$ transition (corresponding to
  diagram (a) in Fig.~\ref{fig:2}), ($sps$) one of the $s$ to $p$
  transitions (corresponding to diagram (b) in Fig.~\ref{fig:2}), and
  ($pss$) the other $s$ to $p$ transition (corresponding to diagram
  (c) in Fig.~\ref{fig:2}), as a function of the normalized component
  of the probe wavevector along the interface, $q_{\|}/q$. The
  normalized Fermi wavenumber is indicated by the vertical line. For a
  two-monolayer copper film it is $k_F/q=2.78\times10^3$. The set of
  arrows labeled $n$ are placed at
  $q_{\|}=nq$.\label{fig:3}\label{fig:11.2}}
\end{figure}

\begin{figure}[tb]
\setlength{\unitlength}{1mm}
\psset{unit=1mm}
\begin{center}
\begin{pspicture}(0,0)(127,143)
\put(-8,2){\epsfig{file=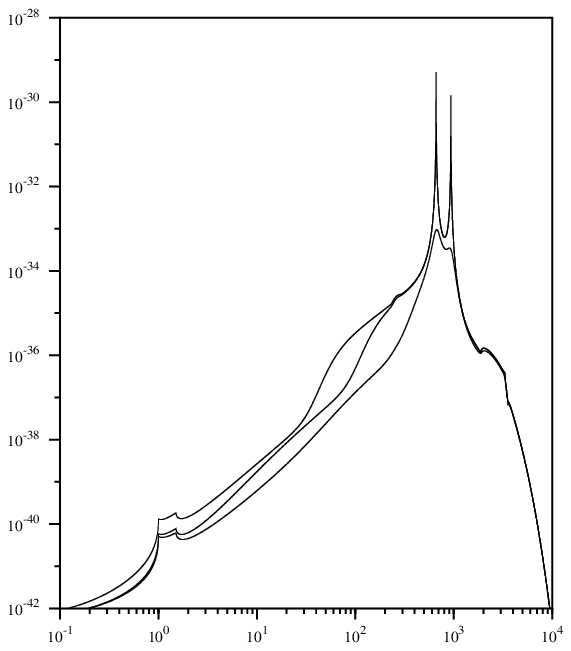,width=135\unitlength}}
\rput[t]{90}(0,76.5){$R_{\rm{PC}}(-d;\vec{q}_{\|})$~[m$^4$/W$^2$]}
\put(71.5,0){\makebox(0,0)[b]{$q_{\|}/q$}}
\put(10,85){\epsfig{file=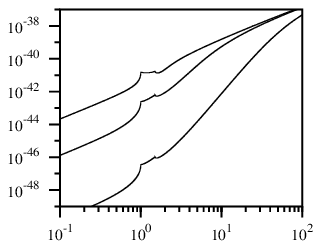,width=75\unitlength}}
\psline[linewidth=0.25mm]{->}(43.75,38.5)(43.75,34.5)
\psline[linewidth=0.25mm]{->}(43.75,23)(43.75,27)
\psline[linewidth=0.25mm]{->}(53.75,100)(53.75,104)
\psline[linewidth=0.25mm]{->}(53.75,127)(53.75,123)
\put(43.75,39.5){\makebox(0,0)[b]{$n$}}
\put(53.75,128){\makebox(0,0)[b]{$n$}}
\psline[linewidth=0.25mm]{->}(99.75,134)(99.75,130)
\psline[linewidth=0.25mm]{->}(99.75,87)(99.75,91)
\psline[linewidth=0.25mm]{->}(103,129)(103,125)
\psline[linewidth=0.25mm]{->}(103,82)(103,86)
\put(123,14){\makebox(0,0)[br]{$sps$}}
\put(82,96){\makebox(0,0)[br]{$ppp$}}
\pstextpath[bc]{\pscurve[linestyle=none](71,57)(74,64)(78,71)(83,76)}{\footnotesize$\tau=200$fs}
\pstextpath[bc]{\pscurve[linestyle=none](76,59)(80,64)(83,71)}{\footnotesize$\tau=30$fs}
\pstextpath[bc]{\pscurve[linestyle=none](85,65)(89,69)(91,73)}{\footnotesize$\tau=3$fs}
\pstextpath[bc]{\pscurve[linestyle=none](36,115)(41,117)(47,120)}{\footnotesize$\tau=200$fs}
\pstextpath[bc]{\pscurve[linestyle=none](36,108)(41,110)(47,113.5)}{\footnotesize$\tau=30$fs}
\pstextpath[bc]{\pscurve[linestyle=none](41,97)(46,100)(49,103.5)}{\footnotesize$\tau=3$fs}
\end{pspicture}
\end{center}
\caption[The phase conjugation reflection coefficient at the surface
of the phase conjugator for different values of the relaxation time
(single monolayer)]{The phase conjugation reflection coefficient at
  the surface of the phase conjugator (single monolayer Cu film) is
  plotted for different values ($\tau\in\{200,30,3\}$ femtoseconds) of
  the relaxation time. The main figure shows the result for the $sps$
  configuration, while the inserted picture shows the $ppp$ result.
  The two sets of arrows labeled $n$ are placed at $q_{\|}=nq$. The
  other two sets of arrows are explained in the main
  text.\label{fig:11.3}}
\end{figure}

\begin{figure}[tb]
\setlength{\unitlength}{1mm}
\psset{unit=1mm}
\begin{center}
\begin{pspicture}(0,0)(127,143)
\put(-8,2){\epsfig{file=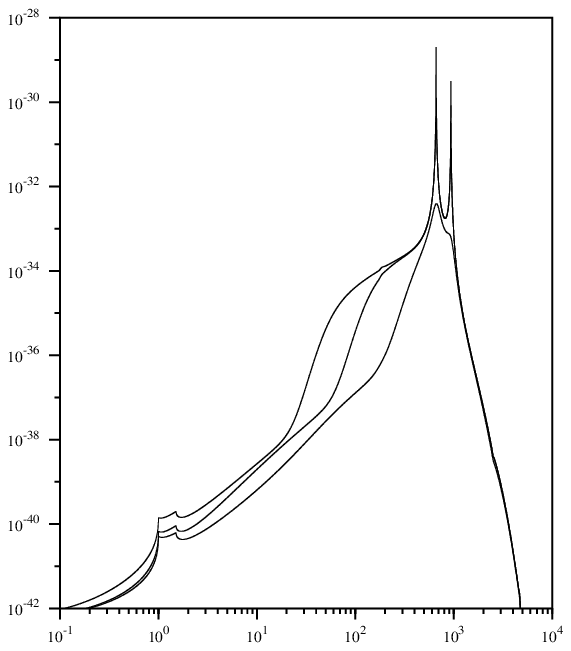,width=135\unitlength}}
\rput[t]{90}(0,76.5){$R_{\rm{PC}}(-d;\vec{q}_{\|})$~[m$^4$/W$^2$]}
\put(71.5,0){\makebox(0,0)[b]{$q_{\|}/q$}}
\put(10,85){\epsfig{file=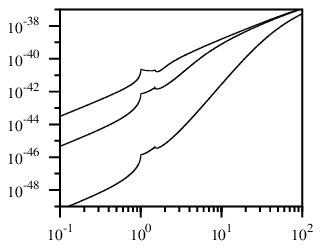,width=75\unitlength}}
\psline[linewidth=0.25mm]{->}(43.75,39)(43.75,35)
\psline[linewidth=0.25mm]{->}(43.75,23)(43.75,27)
\psline[linewidth=0.25mm]{->}(53.75,101.5)(53.75,105.5)
\psline[linewidth=0.25mm]{->}(53.75,127)(53.75,123)
\put(43.75,40){\makebox(0,0)[b]{$n$}}
\put(53.75,128){\makebox(0,0)[b]{$n$}}
\psline[linewidth=0.25mm]{->}(99.75,139)(99.75,135)
\psline[linewidth=0.25mm]{->}(99.75,92)(99.75,96)
\psline[linewidth=0.25mm]{->}(103,132)(103,128)
\psline[linewidth=0.25mm]{->}(103,84)(103,88)
\put(124,14){\makebox(0,0)[br]{$sps$}}
\put(82,96){\makebox(0,0)[br]{$ppp$}}
\pstextpath[bc]{\pscurve[linestyle=none](69,58)(71,65)(73,72)(75,76)}{\footnotesize$\tau=200$fs}
\pstextpath[bc]{\pscurve[linestyle=none](76,60)(79,68)(82,78)}{\footnotesize$\tau=30$fs}
\pstextpath[bc]{\pscurve[linestyle=none](87,68)(90,77)(92,83)}{\footnotesize$\tau=3$fs}
\pstextpath[bc]{\pscurve[linestyle=none](36,116)(41,118)(47,121)}{\footnotesize$\tau=200$fs}
\pstextpath[bc]{\pscurve[linestyle=none](36,110)(43,113)(48,116)}{\footnotesize$\tau=30$fs}
\pstextpath[bc]{\pscurve[linestyle=none](38,98)(44,101)(48,104)}{\footnotesize$\tau=3$fs}
\end{pspicture}
\end{center}
\caption[The phase conjugation reflection coefficient at the surface
of the phase conjugator for different values of the relaxation time
(two monolayers)]{The phase conjugation reflection coefficient at the
  surface of the phase conjugator (two-monolayer Cu film) is plotted
  for different values ($\tau\in\{200,30,3\}$ femtoseconds) of the
  relaxation time. The main figure shows the result for the $sps$
  configuration, while the inserted picture shows the $ppp$ result.
  The two sets of arrows labeled $n$ are placed at $q_{\|}=nq$. The
  other two sets of arrows are explained in the main
  text.\label{fig:4}\label{fig:11.4}}
\end{figure}

The phase conjugation reflection coefficient at the vacuum/film
interface, $R_{\rm{PC}}(-d;\vec{q}_{\|})$ is plotted in
Figs.~\ref{fig:11.1} and \ref{fig:3} as a function of the parallel
component ($q_{\|}$) of the wavevector for both the $p$ to $p$
transition and the two $s$ to $p$ transitions. The reason that the two
curves for the $s$ to $p$ transitions appear the same in the high end
of the $q_{\|}/q$ spectrum is due to the fact that for
$k_{\|}\ll{}q_{\|}$ we have
${\cal{C}}(q_{\|}-k_{\|})\simeq{\cal{C}}(q_{\|}+k_{\|})$ and
${\cal{D}}(q_{\|},k_{\|})\simeq{\cal{D}}(q_{\|},-k_{\|})$. The
``bubble'' appearing on the $sps$ and $pss$ curves from around
$q_{\|}/q\sim100$ to $q_{\|}/q\sim{}k_F/q$ is due to the
two-dimensional electron dynamics hidden in
${\cal{D}}(q_{\|},k_{\|})$. To be a little more specific, the left of
the two peaks stems from the second term, while the peak to the right
in the bubble stems from the third term.

To illustrate the similarity between the two possible $s$ to $p$
transitions, we can take Eq.~(\ref{eq:PC-sps}) to describe the phase
conjugated field, which for positive values of $q_{\|}/q$ gives the
result in Fig.~\ref{fig:3}~($sps$). Using the other $s$ to $p$
transition, given by Eq.~(\ref{eq:PC-pss}), instead we get the result
in Fig.~\ref{fig:3}~($pss$) for positive values of $q_{\|}/q$. The
symmetry between the two configurations is obtained by looking at the
negative values of $q_{\|}/q$, since Eq.~(\ref{eq:PC-sps}) plotted for
negative values of $q_{\|}/q$ gives the ($pss$) curve in
Fig.~\ref{fig:3}. Similarly, by starting with Eq.~(\ref{eq:PC-pss}),
the resulting curve for negative values of $q_{\|}/q$ gives the
($sps$) result in Fig.~\ref{fig:3}.

The choice of an adequate relaxation time $\tau$ is a difficult
problem and it appears from Figs.~\ref{fig:11.3} and \ref{fig:4} that
the value of the relaxation time has a great impact on the phase
conjugation reflection coefficient. We have plotted the reflection
coefficent for three values of the relaxation time, namely (i) 30fs
and (ii) 200fs, which are typical values one would find for bulk
copper \cite{Ashcroft:76:1} at (i) room temperature and (ii) at 77K,
and (iii) 3fs. The value in case (iii) is obtained by a conjecture
based on the difference between measured data for a lead quantum well
\cite{Jalochowski:97:1} and the bulk value for lead at room
temperature. The difference between the relaxation time measured by
\citeN{Jalochowski:97:1} is for two monolayers approximately one order
of magnitude.  Based on the results of \citeN{Jalochowski:97:1} we
have for the data presented in this work chosen the value of the
relaxation time to be 3fs. As it can be seen from Fig.~\ref{fig:4},
the bubble in the curve corresponding to the $sps$ configuration
appears earlier in the $q_{\|}/q$-spectrum for higher values of
$\tau$. For the $ppp$ configuration the lower end of the spectrum is
damped when $\tau$ becomes smaller.

So what is the difference between using a single monolayer or two
monolayers in the quantum well? In the single monolayer quantum well,
the distance between the occupied energy level and the first free
energy level in the infinite barrier model, and between the occupied
energy level and the continuum states in a finite barrier model is
larger than for a two monolayer well. Thus the single-monolayer well
should behave more ideally like a single-level quantum well at higher
frequencies than the two-monolayer well. If we take a look at
Figs.~\ref{fig:11.1} and \ref{fig:11.2} we observe that the bubble in
the $sps$ and $pss$ curves has the highest maximal magnitude for the
two-monolayer well, and the earliest falloff in the high end of the
$q_{\|}/q$-spectrum.  The value of each of the two peaks in the bubble
is reached at the same $q_{\|}/q$-value in the two cases, as is also
evident from Figs.~\ref{fig:11.3} and \ref{fig:11.4} (shown using a
set of arrows for each peak). From these two figures we also observe
that the relaxation-time dependent low-$q_{\|}/q$ beginning of the
bubble occurs a little earlier and is increasing faster in the
two-monolayer well compared to the other. In the low end of the
$q_{\|}/q$-spectrum the $sps$ and $pss$ curves are of equal magnitude.
Looking at the $ppp$ curve, we observe that it is damped roughly by a
factor of two in the low end of the $q_{\|}/q$-spectrum using a
single-monolayer film in stead of two monolayers. In the high end it
takes its maximal value for the single-monolayer well at rougly twice
the value of $q_{\|}/q$ than for the two-monolayer film. In
conclusion, the differences between the phase conjugated response for
a single-monolayer film and a two-monolayer film will probably be very
difficult, if not impossible, to observe in an experiment with single
mode excitation. In the rest of this chapter we thus present results for
the two-monolayer film only.

\begin{figure}[tb]
\setlength{\unitlength}{1mm}
\psset{unit=1mm}
\begin{center}
\begin{pspicture}(0,0)(127,117)
\put(0,60){
\put(-8,2){\epsfig{file=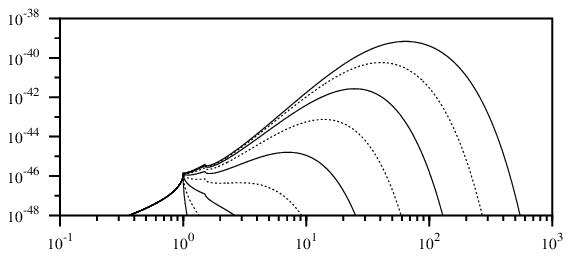,width=135\unitlength}}
\rput[t]{90}(0,34){$R_{\rm{PC}}(z;\vec{q}_{\|})$~[m$^4$/W$^2$]}
\put(71.5,0){\makebox(0,0)[b]{$q_{\|}/q$}}
\put(20,54){\makebox(0,0)[tl]{$ppp$}}
\pstextpath[r]{\pscurve[linestyle=none](97,52)(103,49)(108,44)(112,38)(115,31)(117,24)}{\footnotesize$|z+d|=\lambda/256$}
\pstextpath[r]{\pscurve[linestyle=none](94,47)(101,41)(106,32)(109,23)}{\footnotesize$\lambda/128$}
\pstextpath[r]{\pscurve[linestyle=none](86,42)(93,37)(98,29)(101,22)}{\footnotesize$\lambda/64$}
\pstextpath[r]{\pscurve[linestyle=none](82,34)(88,28)(92,21)}{\footnotesize$\lambda/32$}
\pstextpath[r]{\pscurve[linestyle=none](71,28)(77,25)(82,19)}{\footnotesize$\lambda/16$}
\pstextpath[r]{\pscurve[linestyle=none](61,22)(66,20)(70,17)}{\footnotesize$\lambda/8$}
\pstextpath[r]{\pscurve[linestyle=none](49,20)(53,17)(57,15)}{\footnotesize$\lambda/4$}
}
\put(0,0){
\put(-8,2){\epsfig{file=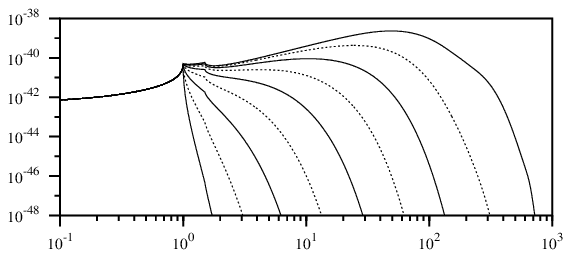,width=135\unitlength}}
\rput[t]{90}(0,34){$R_{\rm{PC}}(z;\vec{q}_{\|})$~[m$^4$/W$^2$]}
\put(71.5,0){\makebox(0,0)[b]{$q_{\|}/q$}}
\put(20,54){\makebox(0,0)[tl]{$sps$}}
\pstextpath[r]{\pscurve[linestyle=none](105,50)(109,46)(114,41)(117,35)(120,28)}{\footnotesize$|z+d|=\lambda/256$}
\pstextpath[r]{\pscurve[linestyle=none](100,44)(104,38)(107,32)(110,26)}{\footnotesize$\lambda/128$}
\pstextpath[r]{\pscurve[linestyle=none](95,36)(98,31)(101,24)}{\footnotesize$\lambda/64$}
\pstextpath[r]{\pscurve[linestyle=none](85,36)(90,28)(93,22)}{\footnotesize$\lambda/32$}
\pstextpath[r]{\pscurve[linestyle=none](78,32)(81,26)(84,21)}{\footnotesize$\lambda/16$}
\pstextpath[r]{\pscurve[linestyle=none](68,32)(72,26)(75.5,19)}{\footnotesize$\lambda/8$}
\pstextpath[r]{\pscurve[linestyle=none](59,32)(63,26)(66.5,17)}{\footnotesize$\lambda/4$}
\pstextpath[r]{\psline[linestyle=none](55.5,25)(59.5,16)}{\footnotesize$\lambda/2$}
\pstextpath[r]{\psline[linestyle=none](51,24)(53,14)}{\footnotesize$\lambda$}
}
\end{pspicture}
\end{center}
\caption[The phase conjugated reflection coefficient at different
distances from the phase conjugator]{The $q_{\|}/q$-dependence of the
  phase conjugation reflection coefficient,
  $R_{\rm{PC}}(z;\vec{q}_{\|})$, is plotted at different distances
  $|z+d|\in\{\lambda, \lambda/2, \lambda/4, \lambda/8, \lambda/16,
  \lambda/32, \lambda/64, \lambda/128, \lambda/256\}$ from the
  vacuum/film interface. The upper figure shows the results for the
  $p$ to $p$ transition. The lower figure shows the results for the
  $s$ to $p$ transition which corresponds to configuration (b) in
  Fig.~\ref{fig:2}.\label{fig:5}}
\end{figure}

\begin{figure}[tbp]
\setlength{\unitlength}{1mm}
\psset{unit=1mm}
\begin{center}
\begin{pspicture}(0,1)(127,177)
\put(0,120){
\put(-8,2){\epsfig{file=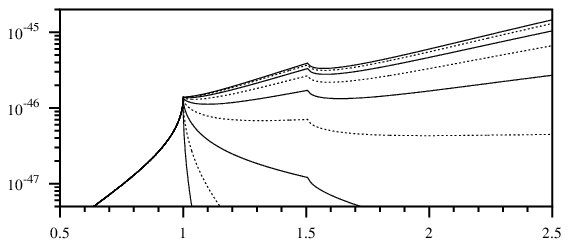,width=135\unitlength}}
\rput[t]{90}(0,34){$R_{\rm{PC}}(z;\vec{q}_{\|})$~[m$^4$/W$^2$]}
\put(71.5,0){\makebox(0,0)[b]{$q_{\|}/q$}}
\put(20,54){\makebox(0,0)[tl]{$ppp$}}
\pstextpath[r]{\psline[linestyle=none](89,47)(120,54)}{\footnotesize$|z+d|=\lambda/256$}
\pstextpath[r]{\psline[linestyle=none](108,46)(120,48)}{\footnotesize$\lambda/32$}
\pstextpath[r]{\psline[linestyle=none](100,40)(120,42.5)}{\footnotesize$\lambda/16$}
\pstextpath[r]{\psline[linestyle=none](100,30)(120,30)}{\footnotesize$\lambda/8$}
\pstextpath[r]{\psline[linestyle=none](69,22)(81,16)}{\footnotesize$\lambda/4$}
\pstextpath[r]{\psline[linestyle=none](50,23)(54,15)}{\footnotesize$\lambda/2$}
\pstextpath[r]{\psline[linestyle=none](48,21)(49,15)}{\footnotesize$\lambda$}
}
\put(0,60){
\put(-8,2){\epsfig{file=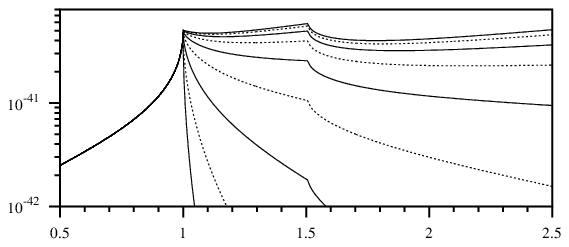,width=135\unitlength}}
\rput[t]{90}(0,34){$R_{\rm{PC}}(z;\vec{q}_{\|})$~[m$^4$/W$^2$]}
\put(71.5,0){\makebox(0,0)[b]{$q_{\|}/q$}}
\put(20,54){\makebox(0,0)[tl]{$sps$}}
\pstextpath[r]{\psline[linestyle=none](90,51)(120,53)}{\footnotesize$|z+d|=\lambda/256$}
\pstextpath[r]{\psline[linestyle=none](100,45)(120,45)}{\footnotesize$\lambda/32$}
\pstextpath[r]{\psline[linestyle=none](96,39)(120,37)}{\footnotesize$\lambda/16$}
\pstextpath[r]{\psline[linestyle=none](100,25)(120,20)}{\footnotesize$\lambda/8$}
\pstextpath[r]{\psline[linestyle=none](71,23)(77,15)}{\footnotesize$\lambda/4$}
\pstextpath[r]{\psline[linestyle=none](52,24)(56,15)}{\footnotesize$\lambda/2$}
\pstextpath[r]{\psline[linestyle=none](48,23)(49,15)}{\footnotesize$\lambda$}
}
\put(0,0){
\put(-8,2){\epsfig{file=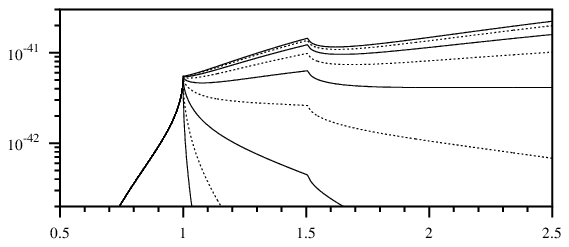,width=135\unitlength}}
\rput[t]{90}(0,34){$R_{\rm{PC}}(z;\vec{q}_{\|})$~[m$^4$/W$^2$]}
\put(71.5,0){\makebox(0,0)[b]{$q_{\|}/q$}}
\put(20,54){\makebox(0,0)[tl]{$pss$}}
\pstextpath[r]{\psline[linestyle=none](90,50)(120,54)}{\footnotesize$|z+d|=\lambda/256$}
\pstextpath[r]{\psline[linestyle=none](100,46)(120,48)}{\footnotesize$\lambda/32$}
\pstextpath[r]{\psline[linestyle=none](100,40.5)(120,40.5)}{\footnotesize$\lambda/16$}
\pstextpath[r]{\psline[linestyle=none](99,29)(120,26)}{\footnotesize$\lambda/8$}
\pstextpath[r]{\psline[linestyle=none](70,23)(80,15)}{\footnotesize$\lambda/4$}
\pstextpath[r]{\psline[linestyle=none](51,22)(55,15)}{\footnotesize$\lambda/2$}
\pstextpath[r]{\psline[linestyle=none](48,21)(49,15)}{\footnotesize$\lambda$}
}
\end{pspicture}
\end{center}
\caption[The phase conjugated reflection coefficient in the range
where we expect single-mode excitation is experimentally feasible]{The
  phase conjugation reflection coefficient,
  $R_{\rm{PC}}(z;\vec{q}_{\|})$, is plotted at different distances
  $|z+d|\in\{\lambda, \lambda/2, \lambda/4, \lambda/8, \lambda/16,
  \lambda/32, \lambda/64, \lambda/128, \lambda/256\}$ from the
  vacuum/film interface as a function of the normalized probe wave
  number $q_{\|}/q$. Results for a two-monolayer thick quantum-well
  phase conjugator are here shown for the three polarization
  combinations $ppp$, $sps$, and $pss$ in the range where we expect
  single mode excitation to be experimentally feasible.\label{fig:6}}
\end{figure}

We have in Fig.~\ref{fig:5} plotted the phase conjugation reflection
coefficient for the $p$ to $p$ transition and one of the $s$ to $p$
transitions, respectively, for different distances from the surface of
the phase conjugator. Due to our particular interest in the phase
conjugation of the evanescent modes in the Fourier spectrum the chosen
distances are fractions of the vacuum wavelength. In Fig.~\ref{fig:6}
we have plotted the part of the Fourier spectrum for all three
configurations which is judged to be the most easily accessible to
single-mode excitation in experimental investigations.

It appears from Fig.~\ref{fig:5}~($ppp$) that the phase
conjugation reflection coefficient is independent of the distance from
the metal film in the region where $q_{\|}/q\leq1$. This is so because
the probe field, and hence also the phase conjugated field, are of
propagating character ($q_{\perp}=[q^2-q_{\|}^2]^{1/2}$ is
real). In the region where $q_{\|}/q>1$, both the probe field and the
phase conjugated field are evanescent
($q_{\perp}={\rm{i}}[q_{\|}^2-q^2]^{1/2}$ is a purely imaginary
quantity), and in consequence the reflection coefficient decreases
rapidly with the distance from the phase conjugator. Already a single
wavelength away from the surface of the phase conjugator the
evanescent modes of the phase conjugated field have essentially
vanished and only propagating modes are detectable.  Although the
evanescent Fourier components of the phase conjugated field are
present only less than an optical wavelength from the surface, this
{\it does not}\/ imply that the nonlinear mixing of the
electromagnetic waves is less effective in the regime of the
evanescent modes. It is in fact opposite, as may be seen for instance
from Fig.~\ref{fig:3}. The maximum coupling for the $p$ to $p$
transition is obtained for $q_{\|}/q\simeq500$, and in comparison with
$R_{\rm{PC}}$ at $q_{\|}/q\simeq1$, the maximum in $R_{\rm{PC}}$ is
nine orders of magnitude larger, and seven, respectively eight orders
of magnitude larger for the two $s$ to $p$ transitions, which have
their maxima at around $q_{\|}/q\simeq700$. As we observe from
Fig.~\ref{fig:5}, as the distance from the film increases the
maximum value decreases and is shifted downwards in the $q_{\|}/q$
spectrum.  But only when the distance from the phase conjugator
becomes larger than $\sim\lambda/10$ ($ppp$) respectively
$\sim\lambda/60$ ($sps$), the phase conjugated signal is largest at
$q_{\|}/q\approx1$.

The absolute value of the reflection coefficients may seem very small,
but utilizing a high-power Nd:YAG laser with, say an energy of 100mJ
per pulse available for each of the three incoming fields, a pulse
(assumed square for simplicity) duration of 4ns and an interaction
area of 25mm$^2$, the intensity of each of these fields will be in the
order of $1$TW/m$^2$, and the phase conjugated intensity lies between
100pW/m$^2$ and 1W/m$^2$ in the full range of $q_{\|}/q$ for which
the reflection coefficient has been plotted in Fig.~\ref{fig:5}~($ppp$),
and between 1$\mu$W/m$^2$ and 1kW/m$^2$ in relation to the data in
Fig.~\ref{fig:5}~($sps$).

In many theoretical studies of the properties of phase conjugated
fields it is assumed that the phase conjugator is ideal
\cite{Hendriks:89:1,Agarwal:95:1,Keller:96:2}. By this is meant that
the phase conjugation reflection coefficient is independent of the
angle of incidense of the (propagating) probe field (and maybe also of
the state of polarization). In the present case, the ideal phase
conjugator assumption is certainly not good. Prior to the observation
that evanescent fields could be phase conjugated
\cite{Bozhevolnyi:94:1} it was often assumed in theory
\cite{Yariv:82:1} \nocite{Wolf:82:1} that $R_{\rm{PC}}=0$ in the
region $q_{\|}/q>1$, and in later studies
\cite{Agarwal:95:1,Keller:96:2} it has been assumed that also the
phase conjugation of evanescent waves is ideal, i.e., independent of
$q_{\|}/q$ ($\gtrsim1$). When it comes to the phase conjugation from
quantum well systems our analysis indicates that use of an energy
reflection coefficient independent of $q_{\|}/q$ in general is bad.
Only at specific distances the ideal phase conjugator assumption might
be justified, see, e.g., the results representing $R_{\rm{PC}}$ at
$|z+d|=\lambda/8$ in Fig.~\ref{fig:6}.  The kink in the reflection
coefficient (which is most pronounced close to the metal/vacuum
interface) found at $q_{\|}/q=n$ ($=1.51$) appears when the probe
field changes from being propagating to being evanescent inside the
substrate.

Above we have discussed the nonlinear reflection coefficient for the
$p$ to $p$ configuration. It appears from Figs.~\ref{fig:5} and
\ref{fig:6} that the quantitative picture is the same for the
$s$ to $p$ cases, though the reflection coefficient for the $s$ to $p$
transitions roughly speaking are five orders of magnitude larger in
the experimentally most adequate evanescent region of the Fourier
spectrum ($1\leq{}q_{\|}/q\lesssim2.5$) for single mode excitation.

The IB model only offers a crude description of the electronic
properties of a quantum well.\label{kF} Among other things, the
electron density profile at the ion/vacuum edge is poorly accounted
for in this model, which gives too sharp a profile and underestimates
the spill-out of the wave function. Altogether one should be careful
to put too much reality into the IB model when treating local-field
variations (related to, say, $q_{\|}$ or $q_{\perp}$) on the atomic
length scale.  Also the neclect of the Bloch character of the wave
functions accounting for the dynamics in the plane of the well is
doubtful in investigations of the local field among the atoms of the
quantum well.  The crucial quantity in the above-mentioned context is
the Fermi wavenumber $k_F=(2m_e{\cal{E}}_{F})^{1/2}/\hbar$, and in
relation to Fig.~\ref{fig:5}, only results for $q_{\|}/q$ ratios less
than approximately
\begin{equation}
 {k_F\over{}q}=\lambda\sqrt{{ZN_+d\over2\pi}+{1\over4d^2}},
\end{equation}
appears reliable. Insertion of the appropriate values for copper:
$ZN_+=8.47\times10^{28}$, $d=1.8${\AA} (single monolayer film) or
$d=3.6${\AA} (two-monolayer film), and the wavelength $\lambda=1061$nm
gives $k_F=3.38\times10^3q$ for a single monolayer of copper, and
$k_F=2.76\times10^3q$ for two monolayers of copper, respectively. The
data presented in Fig.~\ref{fig:5} should therefore be well within
this limit of our model.

Returning to the curve in Fig.~\ref{fig:5}~($ppp$) which
represents the reflection coefficient closest to the surface of the
phase conjugator ($|z+d|=\lambda/256$) one finds approximately a
relation of the form $R_{\rm{PC}}=b(q_{\|}/q)^a$ with $a\simeq5$ in
the lower wavenumber end of the evanescent region. The falloff of
$R_{\rm{PC}}$ with $q_{\|}/q$ after the maximum (located at
$q_{\|}/q\sim50$) is much stronger than the increase towards the
maximum. As the distance from the phase conjugator is increased the
value of $a$ gradually decreases. In Fig.~\ref{fig:5}~(b) we
observe a similar behaviour, but this time the value of $a$ in the
approximate relation in the low end of the evanescent part of the
Fourier spectrum is smaller, namely $a\simeq1.5$.

The energy reflection coefficient calculated at the vacuum/quantum
well interface, $R_{\rm{PC}}(-d;\vec{q}_{\|})$, characterizes the
effectiveness with which a given ($q_{\|}$) plane-wave probe field
(propagating or evanescent) may be phase conjugated, and the results
presented in Fig.~\ref{fig:3} indicate that this effectiveness
(nonlinear coupling) is particularly large for (part of the)
evanescent modes. The maximum in the effectivity is reached for a
value of $q_{\|}/q$ as large as $\sim500$--$700$. The strong coupling
in part of the evanescent region does not necessarily reflect itself
in any easy manner experimentally. First of all, one must realize that
the strong coupling effect only may be observed close to the quantum
well, i.e., at distances $z\lesssim\lambda$. Secondly, one must be able
to produce evanescent probe fields with relatively large values of
$q_{\|}/q$. This is in itself by no means simple outside the range
when the standard Otto \citeyear{Otto:68:1,Otto:76:1} [or possibly
Kretschmann \cite{Kretschmann:68:1,Raether:88:1}] techniques can be
adopted. Roughly speaking, this range coincides with the ones shown in
Fig.~\ref{fig:6}. To create probe fields with larger $q_{\|}/q$ values
other kinds of experimental techniques must be used, and in the
following we shall consider a particular example and in a qualitative
manner discuss the resulting Fourier spectrum of the phase conjugated
field.

\section{Phase conjugated response using a wire source}
In near-field optics evanescent fields with relatively large values of
$q_{\|}/q$ are produced by various methods, all aiming at compressing
the source field to subwavelength spatial extension [see,
e.g., \citeN{Pohl:93:1} and \citeN{Nieto-Vesperinas:96:1}]. From a
theoretical point of view the radiation from a subwavelength source
may in some cases be modelled by the radiation from an (electric)
point-dipole source, or an assembly of such sources. It is a
straightforward matter to decompose an electric point-dipole field
into its relevant evanescent and propagating modes, and thereby
estimate the intensity of the phase conjugated field in each of the
$q_{\|}$-components. However, in order to determine the
characteristics of the phase conjugated light focus generated by the
quantum well one would have to calculate the four-wave mixing also for
probe fields with wavevectors not confined to the $x$-$z$-plane, and
to do this our theory must first be generalized to non-planar phase
conjugation.

Within the framework of the present theory, it is possible, however,
to study the spatial confinement (focusing) of the phase conjugated
field generated by a quantum wire adequately placed above the surface
of the quantum well \cite{Keller:98:1}, and let us therefore as an
example consider the case where the source of the probe field is a
(quantum) wire. We imagine that the axis of the wire is placed
parallel to the $y$-axis and cuts the $x$-$z$-plane in the point
$(0,-z_0)$, cf.~Fig.~\ref{fig:1}. Under the assumption that the
spatial electron confinement in the wire is perfect (complete) and the
wire current density is the same all along the wire at a given time,
the harmonic source current density is given by
\begin{equation}
 \vec{J}(\vec{r};\omega)=\vec{J}_{0}(\omega)\delta(x)\delta(z+z_0),
\label{eq:Jrw}
\end{equation}
where $\vec{J}_{0}(\omega)$ is its possibly frequency dependent
vectorial amplitude. The spatial distribution of the field from this
source is
\begin{eqnarray}
\lefteqn{
 \vec{E}(x,z;\omega)={1\over(2\pi)^2}\int_{-\infty}^{\infty}
 \vec{E}(z;\vec{q}_{\|},\omega)e^{{\rm{i}}\vec{q}_{\|}\cdot\vec{r}}
 \delta(q_{\|,y})d^2q_{\|},
}\nonumber\\ &\quad&=
 {1\over(2\pi)^2}\int_{-\infty}^{\infty}
 \vec{E}(z;\vec{q}_{\|},\omega)e^{{\rm{i}}q_{\|}x}dq_{\|},
\end{eqnarray}
where
\begin{equation}
 \vec{E}(z;\vec{q}_{\|},\omega)=
 -{e^{{\rm{i}}q_{\perp}(z+z_0)}\over2\epsilon_0\omega{}q_{\perp}}
 \left[\begin{array}{ccc}
 q_{\perp}^2 & 0 & -q_{\|}q_{\perp} \\
 0 & q^2 & 0 \\
 -q_{\|}q_{\perp} & 0 & q_{\|}^2
 \end{array}\right]
 \cdot\vec{J}_{0}(\omega),
\label{eq:Wire-z}
\end{equation}
where as hitherto $q_{\|}^2+q_{\perp}^2=q^2$. At the phase conjugating
mirror, the Fourier components of the wire probe are
$\vec{E}(-d;\vec{q}_{\|},\omega)$.

\begin{figure}[tb]
\setlength{\unitlength}{1mm}
\psset{unit=1mm}
\begin{center}
\begin{pspicture}(0,0)(127,95)
\put(-3,2){\epsfig{file=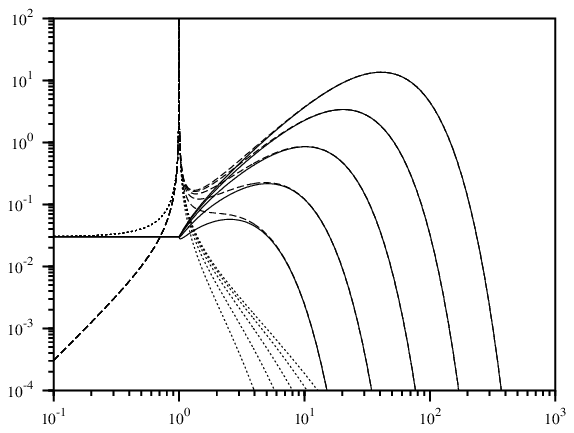,width=130\unitlength}}
\rput[t]{90}(0,51){$\displaystyle{\Delta{}I_{\rm{probe}}(-d;\vec{q}_{\|})\over|J_{0}(\omega)|^2}$ [V/A]}
\put(73,0){\makebox(0,0)[b]{$q_{\|}/q$}}
\pstextpath[l]{\psline[linestyle=none](23,42)(38,42)}{\footnotesize$J_{0}(\omega)\vec{e}_{x}$}
\pstextpath[l]{\pscurve[linestyle=none](53,45)(58,38)(64,30)(72,20)}{\footnotesize$J_{0}(\omega)\vec{e}_{y}$}
\pstextpath[l]{\psline[linestyle=none](26,20)(37,32)}{\footnotesize$J_{0}(\omega)\vec{e}_{z}$}
\pstextpath[r]{\pscurve[linestyle=none](109,53)(112,40)(114,30)(115,20)}{\footnotesize$z_0-d=\lambda/256$}
\pstextpath[r]{\pscurve[linestyle=none](99,52)(102,40)(104,30)(106,19)}{\footnotesize$\lambda/128$}
\pstextpath[r]{\pscurve[linestyle=none](88,53)(92,42)(95,30)(97,18)}{\footnotesize$\lambda/64$}
\pstextpath[r]{\pscurve[linestyle=none](75,52)(81,43)(85,31)(88,17)}{\footnotesize$\lambda/32$}
\pstextpath[r]{\pscurve[linestyle=none](67,46)(71,40)(75,31)(79,16)}{\footnotesize$\lambda/16$}
\pstextpath[r]{\pscurve[linestyle=none](53,45)(58,38)(64,30)(73,20)}{\footnotesize$\lambda/256$}
\pstextpath[r]{\psline[linestyle=none](48,41)(59,15)}{\footnotesize$\lambda/16$}
\end{pspicture}
\end{center}
\caption[Quantum wire $q_{\|}/q$-spectrum at the phase conjugator]{The
  angular Fourier spectrum reaching the surface of the phase
  conjugating medium when the probe field is radiated from a (quantum)
  wire. The dotted curves ($J_{0}(\omega)\vec{e}_{y}$) show the
  Fourier components when the wire current density is polarized along
  the $y$ axis. Similarly, the dashed curves
  ($J_{0}(\omega)\vec{e}_{z}$) and the fully drawn curves
  ($J_{0}(\omega)\vec{e}_{x}$) show the Fourier components from a wire
  source with its current density oscillating along the $z$-axis and
  the $x$-axis, respectively. The angular Fourier spectrum is for all
  three cases shown for five different distances
  $z_0-d\in\{\lambda/16$, $\lambda/32$, $\lambda/64$, $\lambda/128$,
  $\lambda/256\}$ of the wire from the phase conjugator.\label{fig:7}}
\end{figure}

To illustrate the angular spectral distribution of the field from this
kind of wire source at the phase conjugator, we look closer at the
cases, where (i) the current density is polarized along the $x$-axis,
and (ii) along the $y$-axis. Thus, in case (i) we use
$\vec{J}_{0}(\omega)=J_{0}(\omega)\vec{e}_{x}$, and by normalizing the
electric fields to the amplitude of the current density, the
corresponding normalized differential intensity
[$\Delta{}I_{\rm{Probe}}\equiv{1\over2}\epsilon_0c_0\vec{E}(-d;\vec{q}_{\|},\omega)\cdot\vec{E}^{*}(-d;\vec{q}_{\|},\omega)/(2\pi)^4$]
becomes
\begin{eqnarray}
\lefteqn{
 {\Delta{}I_{\rm{Probe}}(-d;\vec{q}_{\|})\over|{J}_{0}(\omega)|^2}=
 {1\over2^7\pi^4\epsilon_0c_0}\left\{\Theta\left({1-(q_{\|}/q)}\right)
 +\Theta\left((q_{\|}/q)-1\right)
 \left[2({q_{\|}/q})^2-1\right]
\right.}\nonumber\\ &\quad&\times\left.
 \exp\left({-2(z_0-d)q\sqrt{(q_{\|}/q)^2-1}}\right)\right\}
\label{eq:Ip-i}
\end{eqnarray}
is shown in Fig.~\ref{fig:7} for different values of the distance
$z_0-d$ from the wire to the vacuum/film interface. In case (ii),
$\vec{J}_{0}(\omega)=J_{0}(\omega)\vec{e}_{y}$, and the associated
normalized intensity which is given by
\begin{eqnarray}
\lefteqn{
 {\Delta{}I_{\rm{Probe}}(-d;\vec{q}_{\|})\over|{J}_{0}(\omega)|^2}=
 {1\over2^7\pi^4\epsilon_0c_0}\left\{
 {\Theta\left({1-(q_{\|}/q)}\right)\over1-(q_{\|}/q)^2}
 +{\Theta\left((q_{\|}/q)-1\right)\over({q_{\|}/q})^2-1}
\right.}\nonumber\\ &\quad&\left.\!\times
 \exp\left({-2(z_0-d)q\sqrt{(q_{\|}/q)^2-1}}\right)\right\},
\label{eq:Ip-ii}
\end{eqnarray}
\noindent 
is also presented in Fig.~\ref{fig:7}, for the same distances as in
case (i). The third curve in Fig.~\ref{fig:7} represents the case
where $\vec{J}_{0}(\omega)=J_{0}(\omega)\vec{e}_{z}$, and is shown for
reference.

Looking at the curve in Fig.~\ref{fig:7} corresponding to
$\vec{J}_{0}(\omega)=J_{0}(\omega)\vec{e}_{y}$ (and the curve
corresponding to $\vec{J}_{0}(\omega)=J_{0}(\omega)\vec{e}_{z}$), we
notice that a singularity occurs when $q_{\|}/q=1$, or equivalently
where $q_{\perp}=0$. The presence of this singularity is an artifact
originating in the (model) assumption that the electron confinement is
complete in the $x$- and $z$-directions (see Eq.~(\ref{eq:Jrw})). If
we had started from a quantum wire current density of finite (but
small) extension in $x$ and $z$ the singularity would have been
replaced by a (narrow) peak of finite height. Not only in quantum wire
optics, but also in optical studies of quantum dots and wells
singularities would appear if complete electron confinement was
assumed (in 3D and 1D, respectively). In the present context the
assumption of perfect electron confinement works well because we only
consider the generated field outside the self-field region of the wire
[see, e.g., \citeN{Keller:97:3}]. In an experiment one would always end
up integrating over some finite interval of $q_{\|}$ around the
singularity, and this integral can in all cases be proven finite.  At
each distance of the wire from the phase conjugator the two curves
$J_{0}\vec{e}_{x}$ and $J_{0}\vec{e}_{z}$ in Fig.~\ref{fig:7} becomes
identical when $(q_{\|}/q)^2\gg{}1$, since from Eq.~(\ref{eq:Wire-z})
we may draw the relation $E_z=-(q_{\|}/q_{\perp})E_x$, and since
$q_{\|}/q_{\perp}\simeq1$ when $(q_{\|}/q)^2\gg{}1$.

When the current oscillates in the direction of the wire, it appears
that the field intensity in the evanescent probe modes is very small.
An appreciable amount of the radiated energy is stored in components
in the region $q_{\|}/q\sim1$ (and in the propagating modes). To study
the phase conjugation of evanescent modes it is therefore better to
start from $\vec{J}_{0}(\omega)=J_{0}(\omega)\vec{e}_{x}$ or from
$\vec{J}_{0}(\omega)=J_{0}(\omega)\vec{e}_{z}$ because these two probe
current densities give rise to significant probe intensities in the
evanescent regime. If we look at the curve in Fig.~\ref{fig:7}
representing the field at the surface of the phase conjugator when the
probe is placed at $z_0-d=\lambda/256$, $I_{\rm{Probe}}$ peaks in both
these cases at $q_{\|}/q\sim50$ in the evanescent regime. When the
current density oscillates along the surface (in the $x$-direction)
there is no singularity (and no peak) at $q_{\|}/q\sim1$, and the
maximum value of $I_{\rm{Probe}}$, occuring at $q_{\|}/q\sim50$, is
three orders of magnitude larger than the probe intensities of
every one of the propagating modes. Above $q_{\|}/q\approx50$ the
amplitude of the $q_{\|}$ components descends rapidly and has lost six
orders of magnitude within the next order of magnitude of $q_{\|}/q$.
At larger probe to surface distances the maximum in the $q_{\|}/q$
spectrum of the probe field at the vacuum/film interface is shifted
downwards, and the magnitude becomes smaller, too.  That is, compared
to the raw $p$ to $p$ reflection coefficient, the intensity of each of
the Fourier components available from the probe field begin their own
falloff about one to two orders of magnitude before the reflection
coefficient descends, depending on the distance from the probe to the
surface of the phase conjugator. The $s$-polarized probe field starts
the descending tendency already where the character of the Fourier
components shifts from being propagating to evanescent ($q_{\perp}$
becoming imaginary), cf.\ the remarks above.

\begin{figure}[tb]
\setlength{\unitlength}{1mm}
\psset{unit=1mm}
\begin{center}
\begin{pspicture}(0,0)(127,117)
\put(0,60){
\put(-3,2){\epsfig{file=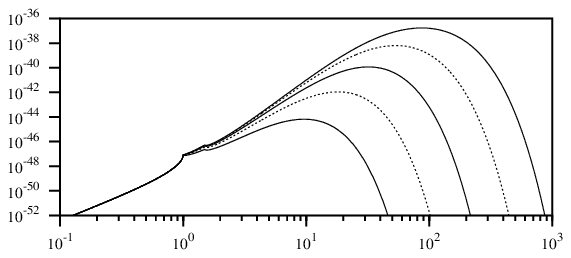,width=130\unitlength}}
\rput[t]{90}(0,34){$\displaystyle{I_{\rm{PC}}(-d;\vec{q}_{\|})\over{}I^{(1)}I^{(2)}|J_{0}(\omega)|^2}$~[m$^2$/WA$^2$]}
\put(71.5,0){\makebox(0,0)[b]{$q_{\|}/q$}}
\put(24,52){\makebox(0,0)[tl]{$J_{0}\vec{e}_{x}$}}
\pstextpath[r]{\pscurve[linestyle=none](109,49)(116,41)(121,30)(123,22)}{\footnotesize$z_0-d=\lambda/256$}
\pstextpath[r]{\pscurve[linestyle=none](108,40)(113,30)(116,20)}{\footnotesize$\lambda/128$}
\pstextpath[r]{\pscurve[linestyle=none](98,40)(104,30)(108.5,18)}{\footnotesize$\lambda/64$}
\pstextpath[r]{\pscurve[linestyle=none](88,38)(95,30)(101,16)}{\footnotesize$\lambda/32$}
\pstextpath[r]{\pscurve[linestyle=none](82,32)(87,27)(92,14)}{\footnotesize$\lambda/16$}
}
\put(0,0){
\put(-3,2){\epsfig{file=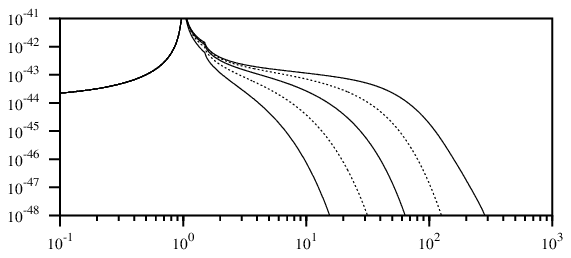,width=130\unitlength}}
\rput[t]{90}(0,34){$\displaystyle{I_{\rm{PC}}(-d;\vec{q}_{\|})\over{}I^{(1)}I^{(2)}|J_{0}(\omega)|^2}$~[m$^2$/WA$^2$]}
\put(71.5,0){\makebox(0,0)[b]{$q_{\|}/q$}}
\put(24,52){\makebox(0,0)[tl]{$J_{0}\vec{e}_{y}$}}
\pstextpath[r]{\pscurve[linestyle=none](82,43)(92,40)(98,36)(103,30)(106.5,25)}{\footnotesize$z_0-d=\lambda/256$}
\pstextpath[r]{\pscurve[linestyle=none](87,38)(95,31)(100,22)}{\footnotesize$\lambda/128$}
\pstextpath[r]{\pscurve[linestyle=none](80,37)(87,31)(94,20)}{\footnotesize$\lambda/64$}
\pstextpath[r]{\pscurve[linestyle=none](73,37)(79,31)(87,17)}{\footnotesize$\lambda/32$}
\pstextpath[r]{\pscurve[linestyle=none](67,36)(72.5,30)(80.5,14)}{\footnotesize$\lambda/16$}
}
\end{pspicture}
\end{center}
\caption[The phase conjugated response when using a quantum wire
source]{The convolution of the probe field from a wire source with the
  phase conjugation reflection coefficient at the vacuum/film
  interface is shown for different distances between the wire source
  and the vacuum/film interface, namely $z_0-d\in\{\lambda/16,
  \lambda/32, \lambda/64, \lambda/128, \lambda/256\}$ as a function of
  the normalized probe wavenumber $q_{\|}/q$. In the top figure the
  current density of the wire oscillates along the $x$-axis, and in
  the bottom figure along the $y$-axis.\label{fig:8}}
\end{figure}

Using a quantum wire as the source for the probe field, the angular
spectrum of the phase conjugated response, normalized to the pump
fields and the absolute square of the amplitude of the wire current
density, is given by 
\begin{equation}
 R_{\rm{PC}}(z;\vec{q}_{\|})
 {I_{\rm{Probe}}(-d;\vec{q}_{\|})\over|J_{0}(\omega)|^2}=
 {I_{\rm{PC}}(z;\vec{q}_{\|})\over{}I^{(1)}I^{(2)}|J_{0}(\omega)|^2},
\label{eq:RI/J}
\end{equation}
and is obtained numerically by multiplying the energy reflection
coefficient, $R_{\rm{PC}}(z;\vec{q}_{\|})$, with the normalized probe
intensity, $I_{\rm{Probe}}(-d;\vec{q}_{\|})/|J_{0}(\omega)|^2$. In
Fig.~\ref{fig:8}, the angular spectrum at the vacuum/quantum-well
interface ($z=-d$) given by Eq.~(\ref{eq:RI/J}) is shown for the cases
where $\vec{J}_{0}(\omega)=J_{0}(\omega)\vec{e}_{x}$ and
$\vec{J}_{0}(\omega)=J_{0}(\omega)\vec{e}_{y}$. It is plotted for the
two-monolayer film, but since the main contribution is in the low end
of the $q_{\|}/q$-spectrum, the similar curves for the
single-monolayer film would be indistinguishable from the ones plotted
(apart from a factor of two in the $ppp$ case). In both cases data are
presented for the wire placed at different distances from the
vacuum/film interface. By comparison with the raw reflection data in
Fig.~\ref{fig:3} it appears that the high end of the reflected
$q_{\|}$-spectrum is strongly damped. For the $s$ to $p$ transition we
see that the energy of the phase conjugated signal is concentrated
around $q_{\|}/q=1$, which is mainly due to the fact that the
concentration of the radiated energy spectrum from the wire lies
around that same point. In the $p$ to $p$ transition the evanescent
components are still by far dominating the response at the place of
the wire compared to the propagating components.

\chapter[Two-dimensional confinement of light]{Two-dimensional
  confinement of light \\ in front of a single-level quantum-well
  phase conjugator}\label{ch:12}
The possibility of compressing light in space to a degree (much)
better than predictable by classical diffraction theory has gained
widespread attention only with the birth of near-field optics
\cite{Pohl:93:1,Nieto-Vesperinas:96:1}.

As stated in Chapter~\ref{Ch:1}, sub-wavelength electrodynamics was
discussed only sporadically until near-field optics evolved in the
mid-eighties in the wake of the experimental works by the groups of
Pohl, Lewis, and Fischer \cite{Pohl:84:1,Lewis:84:1,Fischer:85:1}. The
first investigations are usually attributed to \citeANP{Synge:28:1},
who presented a proposal for sub-wavelength microscopy as early as in
1928. The subject was studied again in 1944 by \citeANP{Bethe:44:1},
by \citeANP{Bouwkamp:50:1} in 1950, and a proposal much similar to
that of Synge was made by \citeANP{OKeefe:56:1} in 1956. Using
microwaves, \citeANP{Ash:72:1} resolved a grating with a linewidth of
$\lambda/60$ in 1972.

In the wake of theoretical studies of the possibility for phase
conjugating the field emitted from a mesoscopic object carried out by
\citeN{Keller:92:1}, creation of light foci with a diameter below the
classical diffraction limit was demonstrated experimentally by
\citeANP{Bozhevolnyi:94:1}
\citeyear{Bozhevolnyi:94:1,Bozhevolnyi:95:2}, who used the fibre tip
of a near field optical microscope to create source spots of red light
(633nm) in front of a photorefractive Fe:LiNbO$_3$ crystal, which
acted as a phase conjugator.  After creation, the phase conjugated
replica of these light spots could be detected using the near field
microscope, since they were maintained for approximately ten minutes
because of the long memory of the phase conjugation process in the
crystal. The resulting phase conjugated light foci had diameters of
around 180nm, and the conclusion of their work was therefore that at
least some of the evanescent field components of the source also must
have been phase conjugated in order to achieve the observed size of
the phase conjugated image.

The above-mentioned observation drew renewed attention to the
description of focusing of electromagnetic fields in front of phase
conjugating mirrors, and required inclusion of evanescent modes in the
description of the optical phase conjugation process. In an important
paper, \citeN{Agarwal:95:1}, extending an original idea of
\citeN{Agarwal:82:1}, undertook an analysis of the phase conjugated
replica of the field from a point particle as it is produced by a
so-called ideal phase conjugator, and in recent articles by
\citeANP{Keller:96:2} \citeyear{Keller:96:3,Keller:96:2} attention was
devoted to an investigation of microscopic aspects of the spatial
confinement problem of the phase conjugated field.

In the previous chapters, we developed a microscopic theory for
optical phase conjugation by degenerate four-wave mixing in mesoscopic
interaction volumes, with the aim of establishing a theoretical
framework for inclusion of near field components in the analysis.  In
order for near field components to give a significant contribution to
the phase conjugated response, the phase conjugation process must be
effective in a surface layer of thickness (much) less than the optical
wavelength. This makes quantum well systems particularly adequate
candidates for the nonlinear mixing process.

In this chapter we employ the developed microscopic theory to a study
of the spatial confinement of an electromagnetic field emitted from an
ideal line source (quantum wire with complete electron confinement),
paying particular attention to the evanescent part of the angular
spectrum.  As phase conjugator a single-level metallic quantum well,
particularly effective in phase conjugating evanescent field
components, albeit with an overall small conversion efficiency, is
used. The relevant expression for the phase conjugated field is given
and the result of a numerical calculation of the nonlinear energy
reflection coefficient for a copper well presented. Finally, the
intensity distribution of the phase conjugated field in the region
between the line source and the phase conjugator is calculated and the
two-dimensional spatial focusing investigated.

\begin{figure}[tb]
\setlength{\unitlength}{1mm}
\psset{unit=1mm}
\begin{center}
\begin{pspicture}(0,0)(127,95)
\put(-3,2){\epsfig{file=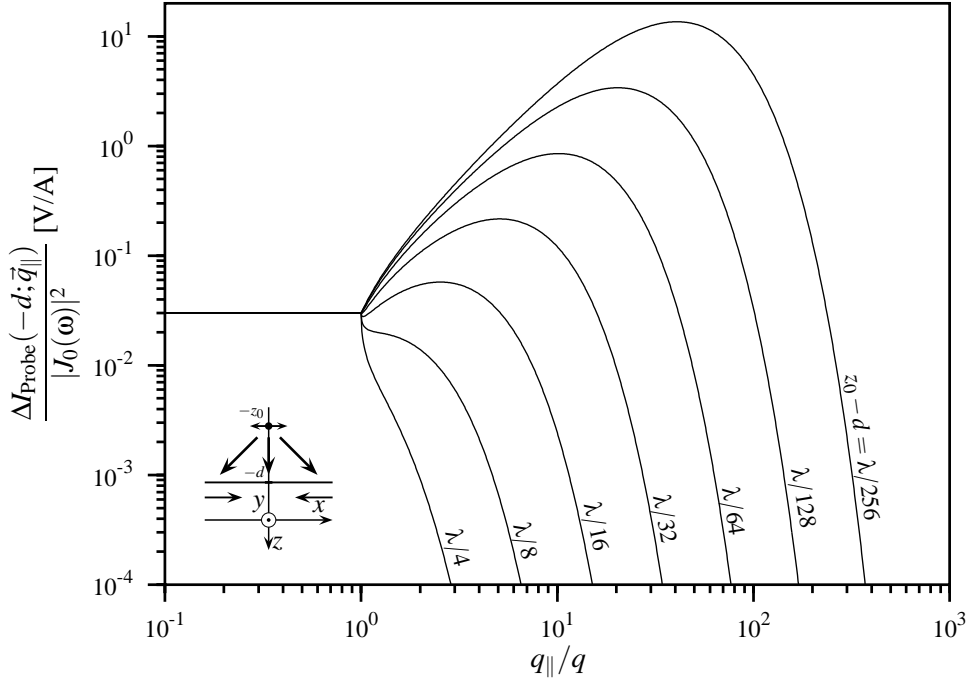,width=130\unitlength}}
\rput[t]{90}(0,51){$\displaystyle{\Delta{}I_{\rm{Probe}}(-d;\vec{q}_{\|})\over|J_{0}(\omega)|^2}$ [V/A]}
\put(73,0){\makebox(0,0)[b]{$q_{\|}/q$}}
\pstextpath[r]{\pscurve[linestyle=none](108,63)(111,49)(113,37)(115,21)}{\footnotesize$z_0-d=\lambda/256$}
\pstextpath[r]{\pscurve[linestyle=none](101,50)(104,35)(106.5,20)}{\footnotesize$\lambda/128$}
\pstextpath[r]{\pscurve[linestyle=none](93,42)(95,32)(97,19)}{\footnotesize$\lambda/64$}
\pstextpath[r]{\pscurve[linestyle=none](83,41)(86,29)(88,18)}{\footnotesize$\lambda/32$}
\pstextpath[r]{\pscurve[linestyle=none](74,36)(77,26)(79,17)}{\footnotesize$\lambda/16$}
\pstextpath[r]{\pscurve[linestyle=none](64,34)(67,25)(69.5,16)}{\footnotesize$\lambda/8$}
\pstextpath[r]{\pscurve[linestyle=none](55,31)(58.5,22)(60.5,15)}{\footnotesize$\lambda/4$}
\put(22,21){
 \psline[linewidth=0.25mm]{-}(4,5)(21,5)
 \psline[linewidth=0.25mm]{->}(4,3)(9,3)
 \psline[linewidth=0.25mm]{->}(21,3)(16,3)
 \psline[linewidth=0.35mm]{->}(12.5,11)(12.5,6)
 \psline[linewidth=0.35mm]{->}(11,11)(6,6)
 \psline[linewidth=0.35mm]{->}(14,11)(19,6)
 \qdisk(12.5,12.5){0.5}
 \psline[linewidth=0.25mm,arrowscale=0.7]{<->}(10,12.5)(15,12.5)
 \psline[linewidth=0.2mm]{->}(4,0)(21,0)
 \psline[linewidth=0.2mm]{->}(12.5,15)(12.5,-4)
 \pscircle[linewidth=0.1mm,linecolor=black,fillstyle=solid,fillcolor=white](12.5,0){1.0}
 \qdisk(12.5,0){0.25}
 \put(20,1){\makebox(0,0)[rb]{\small $x$}}
 \put(12,1){\makebox(0,0)[rb]{\small $y$}}
 \put(13,-4){\makebox(0,0)[lb]{\small $z$}}
 \psline[linewidth=0.35mm]{-}(12,5)(13,5)
 \put(12,5.5){\makebox(0,0)[rb]{\tiny $-d$}}
 \put(12,13.5){\makebox(0,0)[rb]{\tiny $-z_0$}}
}
\end{pspicture}
\end{center}
\caption[The angular spectral distribution of the field generated by a
quantum wire at the surface of the phase conjugator]{The angular
  spectral distribution of the field generated by a quantum wire at
  the surface of the phase conjugator is shown for the seven different
  distances $z_0-d\in\{\lambda/4, \lambda/8,$ $\lambda/16, \lambda/32,
  \lambda/64, \lambda/128, \lambda/256\}$ of the wire from the phase
  conjugating film. Inserted is shown the scattering geometry. The
  film/substrate and vacuum/film interfaces are at $z=0$ and $z=-d$,
  respectively, and the wire crosses the $x$-$z$-plane in the point
  $(x,z)=(0,-z_0)$. As indicated by the double arrow the current
  oscillations of the wire is along the $x$-direction. The arrows
  between the vacuum/film and film/substrate interfaces indicate the
  two pump wavevectors. The arrows drawn from the source towards the
  vacuum/film interface indicate the angular spreading of the source
  field.\label{fig:wire}}
\end{figure}

\section{Quantum wire as a two-dimensional point source}
As starting point, we consider a line source (quantum wire) placed
parallel to the $y$-axis of a Cartesian coordinate system and cutting
the $x$-$z$-plane in the point $(0,-z_0)$. In the description of the
source we assume (i) perfect spatial electron confinement in the wire
and (ii) constant current density along the wire at a given time.
Choosing the wire current to oscillate along the $x$-direction, the
above-mentioned assumptions lead to a harmonic source current density
given by
\begin{equation}
 \vec{J}(\vec{r};\omega)={J}_{0}(\omega)\delta(x)\delta(z+z_0)\vec{e}_{x},
\end{equation}
where $\vec{J}_{0}(\omega)={J}_{0}(\omega)\vec{e}_{x}$ is its possibly
frequency dependent vectorial amplitude, $\vec{e}_{x}$ being a unit
vector in the $x$-direction. In order to calculate the phase
conjugated response using the developed microscopic model we perform a
Fourier analysis of the source field along the $x$-axis. The electric
field of the quantum wire at the surface ($z=-d$) of the phase
conjugating mirror thus becomes
\begin{equation}
 \vec{E}(x,-d;\omega)={1\over(2\pi)^2}\int_{-\infty}^{\infty}
 \vec{E}(-d;\vec{q}_{\|},\omega)e^{iq_{\|}x}dq_{\|},
\label{eq:Wire-xz}
\end{equation}
where the parallel component of the probe wavevector lies along the
$x$-axis, i.e., $\vec{q}_{\|}=q_{\|}\vec{e}_{x}$, and \cite{Keller:98:1}
\begin{equation}
 \vec{E}(-d;\vec{q}_{\|},\omega)=
 -{e^{iq_{\perp}(z_0-d)}\over2\epsilon_0\omega}
 \left[\begin{array}{c} q_{\perp} \\ 0 \\ -q_{\|} \end{array}\right]
 {J}_{0}(\omega).
\label{eq:Wire-z2}
\end{equation}
In Eq.~(\ref{eq:Wire-z2}), $q_{\perp}$ is determined from the vacuum
dispersion relation for the field,
i.e., $(q_{\perp}^2+q_{\|}^2)^{1/2}=q$, where $q=\omega/c_0$ is the
vacuum wavenumber. For propagating modes, satisfying the inequality
$|q_{\|}|<q$, $q_{\perp}=(q^2-q_{\|}^2)^{1/2}$ is real (and positive),
whereas for evanescent modes having $|q_{\|}|>q$,
$q_{\perp}=i(q_{\|}^2-q^2)^{1/2}$ is purely imaginary. To illustrate
the angular spectral distribution of the field from the wire at the
phase conjugator, we calculate the magnitude of the differential
source (probe) field intensity, i.e., $\Delta
I_{\rm{probe}}(-d;\vec{q}_{\|},\omega)= {1\over2}\epsilon_0c_0
\vec{E}(-d;\vec{q}_{\|},\omega)\cdot
\vec{E}^{*}(-d;\vec{q}_{\|},\omega)/(2\pi)^4$.  From Eq.~(\ref{eq:Wire-z2})
one obtains [Eq.~(\ref{eq:Ip-i})]
\begin{eqnarray}
\lefteqn{
 {\Delta I_{\rm{probe}}(-d;\vec{q}_{\|})\over|{J}_{0}(\omega)|^2}=
 {1\over2^7\pi^4\epsilon_0c_0}
 \Bigl\{\Theta\left({1-(q_{\|}/q)}\right)
 +\Theta\left((q_{\|}/q)-1\right)
 \left[2({q_{\|}/q})^2-1\right]
}\nonumber\\ &\quad&\times
 \exp\left({-2(z_0-d)q\sqrt{(q_{\|}/q)^2-1}}\right)\Bigr\},
\label{eq:Ip-i2}
\end{eqnarray}
where $\Theta$ is the Heaviside unit step function. It appears from
Eq.~(\ref{eq:Ip-i2}), that $\Delta{}I_{\rm{probe}}$ is independent of
$q_{\|}$ for the propagating modes. In Fig.~\ref{fig:wire}, the
normalized differential probe distribution
$\Delta{}I_{\rm{probe}}/|J_0(\omega)|^2$ is shown as a function of
$q_{\|}/q$ for various distances $z_0-d$ between the wire and the
vaccum/phase-conjugator interface. It is seen that the evanescent
components tend to dominate the angular spectrum when
$z_0-d\lesssim\lambda/4$.

\section{Single-level metallic quantum-well phase conjugator}
We take as the active medium a metallic quantum well, and we describe
the conduction electron dynamics using the infinite barrier model
potential. While such a model potential from a quantitative point of
view of course is too na{\"\i}ve, in particular in cases where the
conduction electrons of the well are allowed to mix with those of a
semiconducting or metallic substrate, it suffices in the present
context. We further assume that only the lowest lying band is below
the Fermi energy and that the photon energy is so small that interband
transitions do not contribute to the electrodynamics. The quantum well
is deposited on a substrate that can be described alone by its
refractive index $n$. Because of the chosen polarization of the wire
current density we have limited the description to cover only the case
where all interacting electric fields are polarized in the scattering
plane ($p$-polarization). Then, within the limits of a self-field
approximation, the $z$-component of the phase conjugated field becomes
[Eq.~(\ref{eq:PC-ppp})]
\begin{eqnarray}
\lefteqn{
 E_{{\rm{PC}},z}(z;\vec{q}_{\|},\omega)=
 {e^2(\omega+i/\tau)(1+r^p)\over2^9\pi^6\hbar\omega^3ZN_+m_e}
 {q_{\|}^2\over{}iq_{\perp}}
 \left[{\cal{C}}(q_{\|}-k_{\|})
 +{\cal{C}}(q_{\|}+k_{\|})\right]
}\nonumber\\ &\quad&\times
 E_z^{(1)}E_z^{(2)}E_z^{*}e^{-iq_{\perp}z},
\label{eq:PC-ppp2}
\end{eqnarray}
where
\begin{equation}
 {\cal{C}}(q_{\|}\pm{}k_{\|})=
 {4\pi\over{}a^2}\left\{\sqrt{b^2-a^2\alpha^2}-a(q_{\|}\pm{}k_{\|})
 -\sqrt{\left[b-a(q_{\|}\pm{}k_{\|})\right]^2-a^2\alpha^2}\right\},
\end{equation}

\begin{figure}[tb]
\setlength{\unitlength}{1mm}
\psset{unit=1mm}
\begin{center}
\begin{pspicture}(0,0)(127,117)
\put(0,60){
 \put(-1,2){\epsfig{file=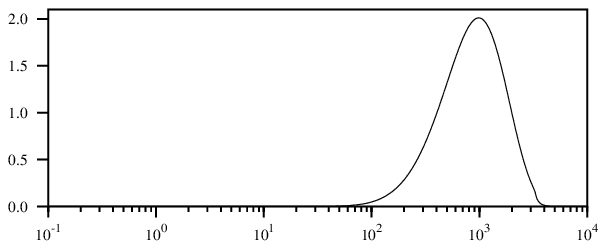,width=128\unitlength}}
 \rput[t]{90}(0,32){$R_{\rm{PC}}({q}_{\|})$~[$10^{-36}$m$^4$/W$^2$]}
 \put(71.5,0){\makebox(0,0)[b]{$q_{\|}/q$}}
 \psline[linewidth=0.25mm]{>-<}(115,4)(115,56)
 \put(115,0){\makebox(0,0)[b]{$k_F/q$}}
 \put(16,51){\makebox(0,0)[tl]{1ML~Cu}}
}
\put(0,0){
 \put(-1,2){\epsfig{file=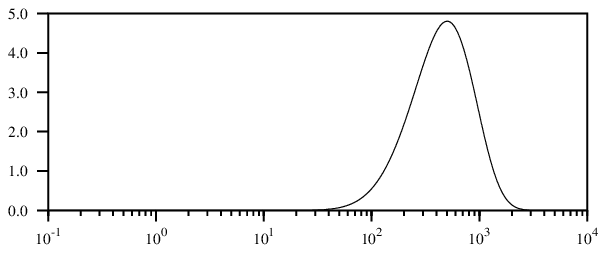,width=128\unitlength}}
 \rput[t]{90}(0,32){$R_{\rm{PC}}({q}_{\|})$~[$10^{-37}$m$^4$/W$^2$]}
 \put(71.5,0){\makebox(0,0)[b]{$q_{\|}/q$}}
 \psline[linewidth=0.25mm]{>-<}(113,4)(113,56)
 \put(113,0){\makebox(0,0)[b]{$k_F/q$}}
 \put(16,51){\makebox(0,0)[tl]{2ML~Cu}}
}
\end{pspicture}
\end{center}
\caption[Linear plot of the phase conjugation energy reflection
coefficient in Figs. \ref{fig:11.1}~(ppp) and \ref{fig:3}~(ppp)]{The
  phase conjugation energy reflection coefficient
  $R_{\rm{PC}}({q}_{\|})\equiv{}R_{\rm{PC}}(-d,\vec{q}_{\|})$ is shown
  on a linear scale as a function of the probe wavevector component
  parallel to the film plane, normalized to the vacuum wavenumber.
  The upper figure shows the result for a single monolayer (1ML)
  copper film [linear plot of Fig.~\ref{fig:11.1}~(ppp)]. The lower
  figure gives the result for the two-monolayer (2ML) film [linear
  plot of Fig.~\ref{fig:3}~(ppp)].\label{fig:RPC}}
\end{figure}

\noindent
with $a=\hbar(q_{\|}\pm{}k_{\|})/m_e$ and
$b=\hbar(q_{\|}\pm{}k_{\|})^2/(2m_e)-i/\tau$. The quantity $\alpha$ is
the radius of the (two-dimensional) Fermi circle, given by
$\alpha=[k_F^2-(\pi/d)^2]^{1/2}$. Given the $z$-component of the phase
conjugated field, the $x$-component is also known, since the electric
field must be transverse in vacuum. Above, $-e$ and $m_e$ are the
electron charge and mass, respectively, $\omega$ and $\tau$ are the
cyclic frequency of the optical field and the electron relaxation
time, and $k_{\|}>0$ is the parallel component of the wavevector of
pump field number $1$ having a $z$-component $E_z^{(1)}$. The
corresponding quantities for pump field number $2$ are $-k_{\|}$ and
$E_z^{(2)}$. When the two pump fields have numerically equal
wavevector components in the plane of the phase conjugator,
conservation of momentum parallel to the surface implies that given
angular components of the probe and phase conjugated fields are
counterpropagating along the surface. Moreover, $Z$ is the number of
conduction electrons each atom in the quantum well contributes to the
assumed free-electron gas, $N_+$ is the number of atoms per unit
volume in the quantum well, and $k_F$ is the Fermi wavenumber. Since
the two pump fields are counterpropagating, the wavevectors of the
probe field ($\vec{q}_{\|}^{\,\prime}$) and the phase conjugated field
($\vec{q}_{\|}$) are related through the conservation of momentum,
$\vec{q}_{\|}^{\,\prime}+\vec{q}_{\|}=\vec{0}$. This property was used
in the derivation of Eq.~(\ref{eq:PC-ppp2}). Thus, the $z$-component
of the probe field in Eq.~(\ref{eq:Wire-z2}), $E_z$, is given by
$E_z=\int{}E_z(-d;\vec{q}_{\|}^{\,\prime},\omega)
\delta(\vec{q}_{\|}^{\,\prime}+\vec{q}_{\|})d^2q_{\|}'$, where
$E_z(-d;\vec{q}_{\|}^{\,\prime},\omega)$ is taken from
Eq.~(\ref{eq:Wire-z2}).

\section{Numerical results and discussion}
For the numerical calculation we consider a copper quantum well
[$d=1.8${\AA} (for single-monolayer film), respectively $3.6${\AA} (two
monolayers), $N_+=8.47\times10^{28}$m$^{-3}$ and $Z=1$
\cite{Ashcroft:76:1}] deposited on a glass substrate with $n=1.51$,
giving $k_{\|}=1.51q$ when the wavevectors of the pump fields are
parallel to the $x$-axis.  The wavelength of the light is chosen to be
$\lambda=1061$nm. For a glass substrate it is adequate to calculate
the linear vaccum/substrate amplitude reflection coefficient $r^p$ by
means of the classical Fresnel formula
$r^p=[{n^2q_{\perp}-({n^2q^2-q_{\|}^2})^{1\over2}}]/
[{n^2q_{\perp}+({n^2q^2-q_{\|}^2})^{1\over2}}]$. In the view of the
recent experimental data discussed in Section~\ref{sec:11.1}
\cite{Jalochowski:97:1}, we have chosen an intraband relaxation time
of $\tau=3$fs for the electrons in the ultrathin quantum-well film.

To give an impression of the efficiency of the phase conjugation
process for the various evanescent modes, the nonlinear energy
reflection coefficient of the phase conjugator at the vacuum/film
interface, $R_{\rm{PC}}({q}_{\|})$, is shown in Fig.~\ref{fig:RPC} as
a function of the parallel component ($q_{\|}$) of the probe
wavevector for the two possible single-level quantum wells. It appears
from this figure that in particular high spatial frequency components
($10^2\lesssim{}q_{\|}/q\lesssim{}10^3$) are phase conjugated in an
effective manner. This is associated with the fact that in a
single-level quantum well the two-dimensional intraband electron
dynamics along the plane of the well is responsible for the main part
of the phase conjugation process.

Using a square-potential barrier model to describe the quantum well
the integration limits should not extend beyond the (two-dimensional)
Fermi wavenumber $k_F=[2\pi{}ZN_+d+(\pi/d)^2]^{1/2}$ for the single
level quantum well. Looking at the phase conjugation reflection
coefficient shown in Fig.~\ref{fig:RPC}, we not only notice that the
main contribution to the phase conjugated signal is well above the
point where the probe field becomes evanescent in vaccum, but also
does not extend beyond the Fermi wavenumber. The phase conjugated
image of our quantum-wire source field is then given by
\begin{equation}
 \vec{E}_{\rm{PC}}(x,z;\omega)={1\over2\pi}\int_{-k_F}^{k_F}
 \vec{E}_{\rm{PC}}(z;\vec{q}_{\|},\omega)e^{iq_{\|}x}dq_{\|},
\end{equation}
in the $x$-$z$-plane.

\begin{figure}[tbp]
\setlength{\unitlength}{0.977mm}
\psset{unit=0.977mm}
\begin{center}
\begin{pspicture}(-2,-4)(128,145)
\newcounter{puppet}
\put(0,0){
 \put(0,120){\rput[lt]{90}(0,0){\epsfig{file=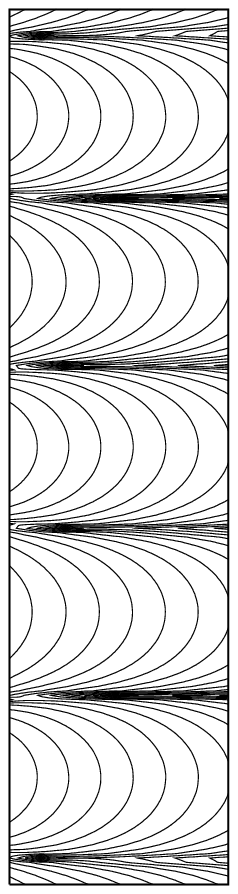,height=80\unitlength}}}
 \put(0,90){\rput[lt]{90}(0,0){\epsfig{file=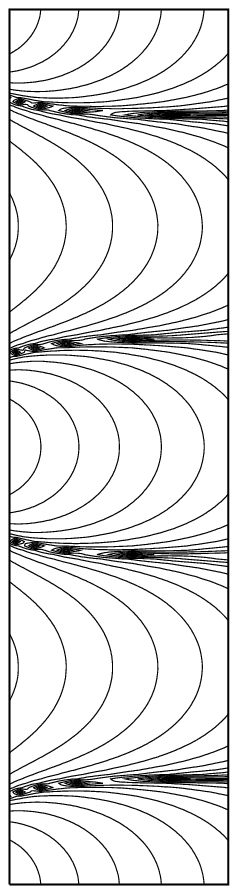,height=80\unitlength}}}
 \put(0,60){\rput[lt]{90}(0,0){\epsfig{file=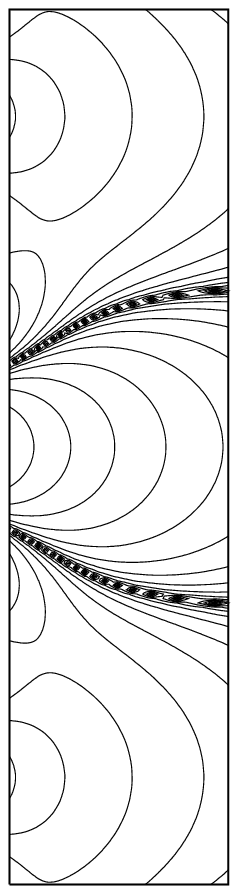,height=80\unitlength}}}
 \put(0,30){\rput[lt]{90}(0,0){\epsfig{file=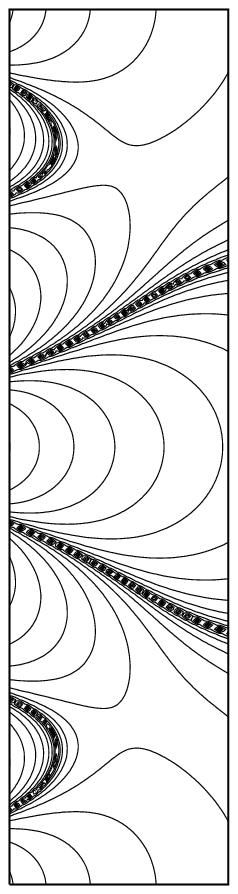,height=80\unitlength}}}
 \put(0,0){\rput[lt]{90}(0,0){\epsfig{file=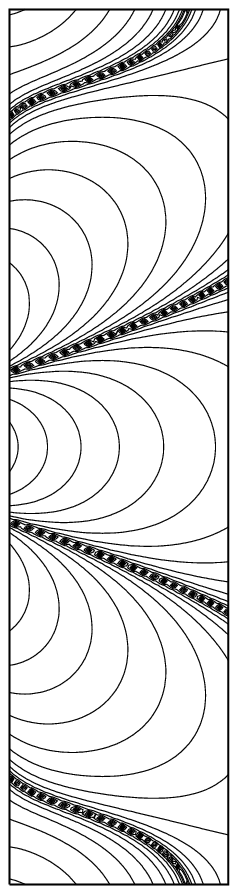,height=80\unitlength}}}
 \multiput(0,120)(0,-30){5}{
  \psline[linewidth=0.25mm]{->}(40,-2)(40,23)
  \psline[linewidth=0.25mm]{->}(-2,0)(83,0)
  \multiput(0,0)(10,0){9}{\psline[linewidth=0.25mm]{-}(0,-1.25)(0,0)}
  \multiput(0,0)(5,0){17}{\psline[linewidth=0.25mm]{-}(0,-0.75)(0,0)}
  \put(83,1.5){\makebox(0,0)[br]{$x$}}
  \put(38.5,23){\makebox(0,0)[tr]{$z$}}
  \put(40.5,-2){\makebox(0,0)[tl]{\small{$0$}}}
 }
 \setcounter{puppet}{2}
 \multiput(0,120)(0,-60){3}{
  \addtocounter{puppet}{\value{puppet}}
  \put(60,-1.5){\makebox(0,0)[t]{\tiny{${\lambda\over\arabic{puppet}}$}}}
  \put(19,-1.5){\makebox(0,0)[t]{\tiny{$-{\lambda\over\arabic{puppet}}$}}}
  \put(42,20.5){\makebox(0,0)[bl]{\footnotesize{${\lambda/\arabic{puppet}}$}}}
 }
 \setcounter{puppet}{8}
 \multiput(0,90)(0,-60){2}{
  \addtocounter{puppet}{\value{puppet}}
  \put(60,-1.5){\makebox(0,0)[t]{\tiny{${3\lambda\over\arabic{puppet}}$}}}
  \put(19,-1.5){\makebox(0,0)[t]{\tiny{$-{3\lambda\over\arabic{puppet}}$}}}
  \put(42,20.5){\makebox(0,0)[bl]{\footnotesize{${3\lambda/\arabic{puppet}}$}}}
 }
 \setcounter{puppet}{4}
 \multiput(0,120)(0,-60){3}{
  \addtocounter{puppet}{\value{puppet}}
  \put(50,-1.5){\makebox(0,0)[t]{\tiny{${\lambda\over\arabic{puppet}}$}}}
  \put(29,-1.5){\makebox(0,0)[t]{\tiny{$-{\lambda\over\arabic{puppet}}$}}}
  \put(70,-1.5){\makebox(0,0)[t]{\tiny{${3\lambda\over\arabic{puppet}}$}}}
  \put(9,-1.5){\makebox(0,0)[t]{\tiny{$-{3\lambda\over\arabic{puppet}}$}}}
 }
 \setcounter{puppet}{16}
 \multiput(0,90)(0,-60){2}{
  \addtocounter{puppet}{\value{puppet}}
  \put(50,-1.5){\makebox(0,0)[t]{\tiny{${3\lambda\over\arabic{puppet}}$}}}
  \put(29,-1.5){\makebox(0,0)[t]{\tiny{$-{3\lambda\over\arabic{puppet}}$}}}
  \put(70,-1.5){\makebox(0,0)[t]{\tiny{${9\lambda\over\arabic{puppet}}$}}}
  \put(9,-1.5){\makebox(0,0)[t]{\tiny{$-{9\lambda\over\arabic{puppet}}$}}}
 }
 \setcounter{puppet}{1}
 \multiput(0,120)(0,-60){3}{
  \addtocounter{puppet}{\value{puppet}}
  \put(80,-1.5){\makebox(0,0)[t]{\tiny{${\lambda\over\arabic{puppet}}$}}}
  \put(-1,-1.5){\makebox(0,0)[t]{\tiny{$-{\lambda\over\arabic{puppet}}$}}}
 }
 \setcounter{puppet}{4}
 \multiput(0,90)(0,-60){2}{
  \addtocounter{puppet}{\value{puppet}}
  \put(80,-1.5){\makebox(0,0)[t]{\tiny{${3\lambda\over\arabic{puppet}}$}}}
  \put(-1,-1.5){\makebox(0,0)[t]{\tiny{$-{3\lambda\over\arabic{puppet}}$}}}
 }
 \setcounter{puppet}{0}
 \multiput(80,120)(0,-30){5}{
  \addtocounter{puppet}{1}
  \put(1,22){\makebox(0,0)[tl]{(\alph{puppet})}}
 }
}
\put(88,0){
 \put(0,120){\epsfig{file=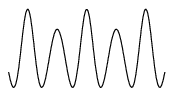,width=36\unitlength}}
 \put(0,90){\epsfig{file=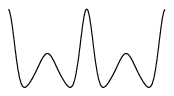,width=36\unitlength}}
 \put(0,60){\epsfig{file=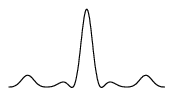,width=36\unitlength}}
 \put(0,30){\epsfig{file=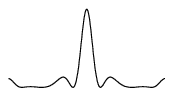,width=36\unitlength}}
 \put(0,0){\epsfig{file=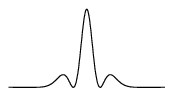,width=36\unitlength}}
 \put(0,120){\epsfig{file=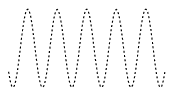,width=36\unitlength}}
 \put(0,90){\epsfig{file=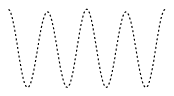,width=36\unitlength}}
 \put(0,60){\epsfig{file=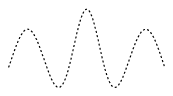,width=36\unitlength}}
 \put(0,30){\epsfig{file=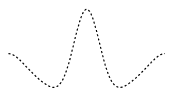,width=36\unitlength}}
 \put(0,0){\epsfig{file=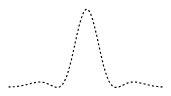,width=36\unitlength}}
 \multiput(0,120)(0,-30){5}{
  \psline[linewidth=0.25mm]{->}(18,-2)(18,22)
  \psline[linewidth=0.25mm]{->}(-2,0)(40,0)
  \multiput(0,0)(9,0){5}{\psline[linewidth=0.25mm]{-}(0,-1.25)(0,0)}
  \multiput(0,0)(2.25,0){17}{\psline[linewidth=0.25mm]{-}(0,-0.75)(0,0)}
  \put(40,1.5){\makebox(0,0)[br]{$x$}}
  \put(16.5,22){\makebox(0,0)[tr]{$I_{\rm{PC}}$}}
  \put(18.5,-2){\makebox(0,0)[tl]{\small{$0$}}}
 }
 \setcounter{puppet}{2}
 \multiput(0,120)(0,-60){3}{
  \addtocounter{puppet}{\value{puppet}}
  \put(27,-1.5){\makebox(0,0)[t]{\tiny{${\lambda\over\arabic{puppet}}$}}}
  \put(8.5,-1.5){\makebox(0,0)[t]{\tiny{$-{\lambda\over\arabic{puppet}}$}}}
 }
 \setcounter{puppet}{8}
 \multiput(0,90)(0,-60){2}{
  \addtocounter{puppet}{\value{puppet}}
  \put(27,-1.5){\makebox(0,0)[t]{\tiny{${3\lambda\over\arabic{puppet}}$}}}
  \put(8.5,-1.5){\makebox(0,0)[t]{\tiny{$-{3\lambda\over\arabic{puppet}}$}}}
 }
 \setcounter{puppet}{1}
 \multiput(0,120)(0,-60){3}{
  \addtocounter{puppet}{\value{puppet}}
  \put(36,-1.5){\makebox(0,0)[t]{\tiny{${\lambda\over\arabic{puppet}}$}}}
  \put(-0.5,-1.5){\makebox(0,0)[t]{\tiny{$-{\lambda\over\arabic{puppet}}$}}}
 }
 \setcounter{puppet}{4}
 \multiput(0,90)(0,-60){2}{
  \addtocounter{puppet}{\value{puppet}}
  \put(36,-1.5){\makebox(0,0)[t]{\tiny{${3\lambda\over\arabic{puppet}}$}}}
  \put(-0.5,-1.5){\makebox(0,0)[t]{\tiny{$-{3\lambda\over\arabic{puppet}}$}}}
 }
}
\end{pspicture}
\end{center}
\caption[The intensity of the phase conjugated field from a quantum
wire (1ML Cu phase conjugator)]{The intensity of the phase conjugated
  field from a quantum wire is plotted for different distances between
  the wire and the surface of a single-monolayer copper quantum-well
  phase conjugator. The figures to the left show lines of equal
  intensity on a logarithmic scale of the phase conjugated intensity
  in the $x$-$z$-plane between the surface of the phase conjugator and
  the wire. The figures to the right show the phase conjugated
  intensity (i) at the surface of the phase conjugator (solid lines)
  and (ii) at the height of the wire, $z=-z_0$ (dashed lines).  The
  intensity in these figures is plotted on a linear scale (arbitrary
  units). In order to make the two curves in these plots comparable,
  the curve associated with case (ii) has been multiplied by a factor
  of (a) $2590$, (b) $545$, (c) $271$, (d) $353$, and (e) $586$. The
  sets of figures are shown for distances $z_0-d$ of (a) $\lambda/4$,
  (b) $3\lambda/16$, (c) $\lambda/8$, (d) $3\lambda/32$, and (e)
  $\lambda/16$ between the quantum wire and the surface of the phase
  conjugating mirror.\label{fig:intensity-ptop-1.8}\label{fig:12.3}}
\end{figure}

\begin{figure}[t]
\setlength{\unitlength}{0.977mm}
\psset{unit=0.977mm}
\begin{center}
\begin{pspicture}(-2,-4)(128,145)
\put(0,0){
 \put(0,120){\rput[lt]{90}(0,0){\epsfig{file=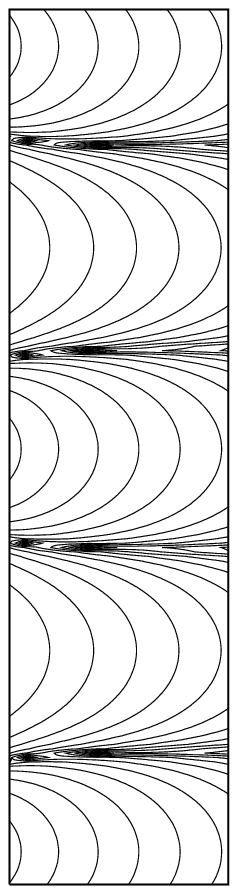,height=80\unitlength}}}
 \put(0,90){\rput[lt]{90}(0,0){\epsfig{file=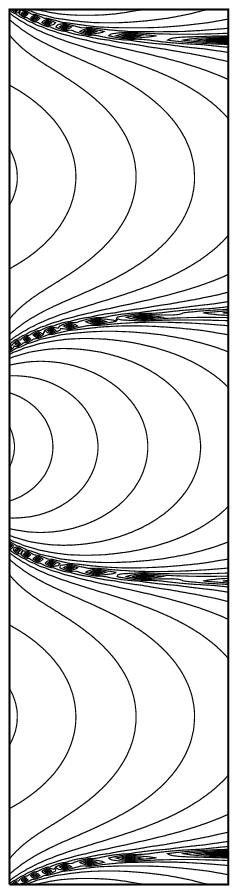,height=80\unitlength}}}
 \put(0,60){\rput[lt]{90}(0,0){\epsfig{file=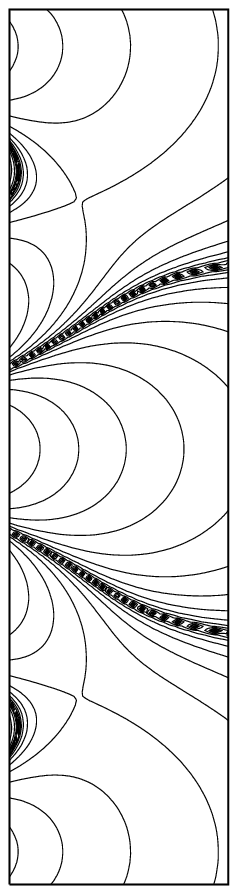,height=80\unitlength}}}
 \put(0,30){\rput[lt]{90}(0,0){\epsfig{file=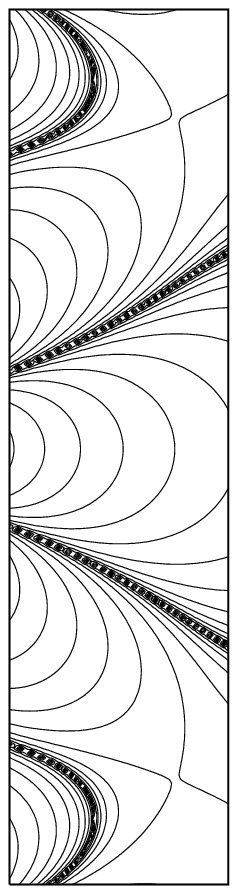,height=80\unitlength}}}
 \put(0,0){\rput[lt]{90}(0,0){\epsfig{file=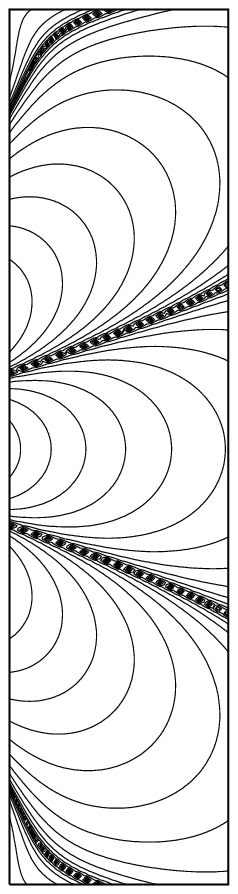,height=80\unitlength}}}
 \multiput(0,120)(0,-30){5}{
  \psline[linewidth=0.25mm]{->}(40,-2)(40,23)
  \psline[linewidth=0.25mm]{->}(-2,0)(83,0)
  \multiput(0,0)(10,0){9}{\psline[linewidth=0.25mm]{-}(0,-1.25)(0,0)}
  \multiput(0,0)(5,0){17}{\psline[linewidth=0.25mm]{-}(0,-0.75)(0,0)}
  \put(83,1.5){\makebox(0,0)[br]{$x$}}
  \put(38.5,23){\makebox(0,0)[tr]{$z$}}
  \put(40.5,-2){\makebox(0,0)[tl]{\small{$0$}}}
 }
 \setcounter{puppet}{2}
 \multiput(0,120)(0,-60){3}{
  \addtocounter{puppet}{\value{puppet}}
  \put(60,-1.5){\makebox(0,0)[t]{\tiny{${\lambda\over\arabic{puppet}}$}}}
  \put(19,-1.5){\makebox(0,0)[t]{\tiny{$-{\lambda\over\arabic{puppet}}$}}}
  \put(42,20.5){\makebox(0,0)[bl]{\footnotesize{${\lambda/\arabic{puppet}}$}}}
 }
 \setcounter{puppet}{8}
 \multiput(0,90)(0,-60){2}{
  \addtocounter{puppet}{\value{puppet}}
  \put(60,-1.5){\makebox(0,0)[t]{\tiny{${3\lambda\over\arabic{puppet}}$}}}
  \put(19,-1.5){\makebox(0,0)[t]{\tiny{$-{3\lambda\over\arabic{puppet}}$}}}
  \put(42,20.5){\makebox(0,0)[bl]{\footnotesize{${3\lambda/\arabic{puppet}}$}}}
 }
 \setcounter{puppet}{4}
 \multiput(0,120)(0,-60){3}{
  \addtocounter{puppet}{\value{puppet}}
  \put(50,-1.5){\makebox(0,0)[t]{\tiny{${\lambda\over\arabic{puppet}}$}}}
  \put(29,-1.5){\makebox(0,0)[t]{\tiny{$-{\lambda\over\arabic{puppet}}$}}}
  \put(70,-1.5){\makebox(0,0)[t]{\tiny{${3\lambda\over\arabic{puppet}}$}}}
  \put(9,-1.5){\makebox(0,0)[t]{\tiny{$-{3\lambda\over\arabic{puppet}}$}}}
 }
 \setcounter{puppet}{16}
 \multiput(0,90)(0,-60){2}{
  \addtocounter{puppet}{\value{puppet}}
  \put(50,-1.5){\makebox(0,0)[t]{\tiny{${3\lambda\over\arabic{puppet}}$}}}
  \put(29,-1.5){\makebox(0,0)[t]{\tiny{$-{3\lambda\over\arabic{puppet}}$}}}
  \put(70,-1.5){\makebox(0,0)[t]{\tiny{${9\lambda\over\arabic{puppet}}$}}}
  \put(9,-1.5){\makebox(0,0)[t]{\tiny{$-{9\lambda\over\arabic{puppet}}$}}}
 }
 \setcounter{puppet}{1}
 \multiput(0,120)(0,-60){3}{
  \addtocounter{puppet}{\value{puppet}}
  \put(80,-1.5){\makebox(0,0)[t]{\tiny{${\lambda\over\arabic{puppet}}$}}}
  \put(-1,-1.5){\makebox(0,0)[t]{\tiny{$-{\lambda\over\arabic{puppet}}$}}}
 }
 \setcounter{puppet}{4}
 \multiput(0,90)(0,-60){2}{
  \addtocounter{puppet}{\value{puppet}}
  \put(80,-1.5){\makebox(0,0)[t]{\tiny{${3\lambda\over\arabic{puppet}}$}}}
  \put(-1,-1.5){\makebox(0,0)[t]{\tiny{$-{3\lambda\over\arabic{puppet}}$}}}
 }
 \setcounter{puppet}{0}
 \multiput(80,120)(0,-30){5}{
  \addtocounter{puppet}{1}
  \put(1,22){\makebox(0,0)[tl]{(\alph{puppet})}}
 }
}
\put(88,0){
 \put(0,120){\epsfig{file=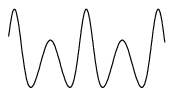,width=36\unitlength}}
 \put(0,90){\epsfig{file=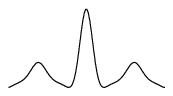,width=36\unitlength}}
 \put(0,60){\epsfig{file=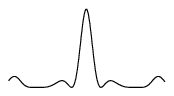,width=36\unitlength}}
 \put(0,30){\epsfig{file=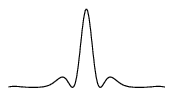,width=36\unitlength}}
 \put(0,0){\epsfig{file=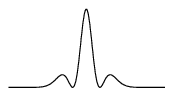,width=36\unitlength}}
 \put(0,120){\epsfig{file=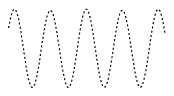,width=36\unitlength}}
 \put(0,90){\epsfig{file=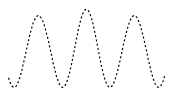,width=36\unitlength}}
 \put(0,60){\epsfig{file=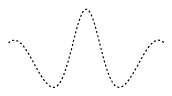,width=36\unitlength}}
 \put(0,30){\epsfig{file=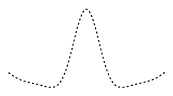,width=36\unitlength}}
 \put(0,0){\epsfig{file=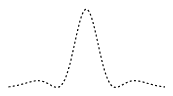,width=36\unitlength}}
 \multiput(0,120)(0,-30){5}{
  \psline[linewidth=0.25mm]{->}(18,-2)(18,22)
  \psline[linewidth=0.25mm]{->}(-2,0)(40,0)
  \multiput(0,0)(9,0){5}{\psline[linewidth=0.25mm]{-}(0,-1.25)(0,0)}
  \multiput(0,0)(2.25,0){17}{\psline[linewidth=0.25mm]{-}(0,-0.75)(0,0)}
  \put(40,1.5){\makebox(0,0)[br]{$x$}}
  \put(16.5,22){\makebox(0,0)[tr]{$I_{\rm{PC}}$}}
  \put(18.5,-2){\makebox(0,0)[tl]{\small{$0$}}}
 }
 \setcounter{puppet}{2}
 \multiput(0,120)(0,-60){3}{
  \addtocounter{puppet}{\value{puppet}}
  \put(27,-1.5){\makebox(0,0)[t]{\tiny{${\lambda\over\arabic{puppet}}$}}}
  \put(8.5,-1.5){\makebox(0,0)[t]{\tiny{$-{\lambda\over\arabic{puppet}}$}}}
 }
 \setcounter{puppet}{8}
 \multiput(0,90)(0,-60){2}{
  \addtocounter{puppet}{\value{puppet}}
  \put(27,-1.5){\makebox(0,0)[t]{\tiny{${3\lambda\over\arabic{puppet}}$}}}
  \put(8.5,-1.5){\makebox(0,0)[t]{\tiny{$-{3\lambda\over\arabic{puppet}}$}}}
 }
 \setcounter{puppet}{1}
 \multiput(0,120)(0,-60){3}{
  \addtocounter{puppet}{\value{puppet}}
  \put(36,-1.5){\makebox(0,0)[t]{\tiny{${\lambda\over\arabic{puppet}}$}}}
  \put(-0.5,-1.5){\makebox(0,0)[t]{\tiny{$-{\lambda\over\arabic{puppet}}$}}}
 }
 \setcounter{puppet}{4}
 \multiput(0,90)(0,-60){2}{
  \addtocounter{puppet}{\value{puppet}}
  \put(36,-1.5){\makebox(0,0)[t]{\tiny{${3\lambda\over\arabic{puppet}}$}}}
  \put(-0.5,-1.5){\makebox(0,0)[t]{\tiny{$-{3\lambda\over\arabic{puppet}}$}}}
 }
}
\end{pspicture}
\end{center}
\caption[The intensity of the phase conjugated field from a quantum
wire (2ML Cu phase conjugator)]{The intensity of the phase conjugated
  field from a quantum wire is plotted for different distances between
  the wire and the surface of a two-monolayer copper quantum-well
  phase conjugator. The figures to the left show lines of equal
  intensity on a logarithmic scale of the phase conjugated intensity
  in the $x$-$z$-plane between the surface of the phase conjugator and
  the wire. The figures to the right show the phase conjugated
  intensity (i) at the surface of the phase conjugator (solid lines)
  and (ii) at the height of the wire, $z=-z_0$ (dashed lines).  The
  intensity in these figures is plotted on a linear scale (arbitrary
  units). In order to make the two curves in these plots comparable,
  the curve associated with case (ii) has been multiplied by a factor
  of (a) $555$, (b) $212$, (c) $204$, (d) $322$, and (e) $528$. The
  sets of figures are shown for distances $z_0-d$ of (a) $\lambda/4$,
  (b) $3\lambda/16$, (c) $\lambda/8$, (d) $3\lambda/32$, and (e)
  $\lambda/16$ between the quantum wire and the surface of the phase
  conjugating mirror.\label{fig:intensity-ptop}\label{fig:12.4}}
\end{figure}

It appears from Eq.~(\ref{eq:PC-ppp2}), that the individual angular
components of the phase conjugated field decay exponentially with the
distance from the phase conjugator in the evanescent part of the
Fourier spectrum. The evanescent components of the source likewise
decay exponentially with the distance from the quantum wire.
Therefore, the contribution from the evanescent components to the
total phase conjugated field is expected to increase significantly
when the distance between the source and the phase conjugator becomes
smaller. Experimentally, it is feasible presently to carry out
measurements at distances from the surface down to $\sim40${\AA}
(using near-field microscopes). For the chosen system this leads to an
intensity distribution in the $x$-$z$-plane between the probe and the
phase conjugator as shown in Figs.~\ref{fig:12.3} and \ref{fig:12.4}
for different distances between the probe and the film,
$z_0-d\in\{\lambda/4$, $3\lambda/16$, $\lambda/8$, $3\lambda/32$,
$\lambda/16\}$, Fig.~\ref{fig:12.3} corresponding to the
single-monolayer film and Fig.~\ref{fig:12.4} to the two-monolayer
film. For $z_0-d\gtrsim\lambda/2$, the effect of the near-field
components are negligible. In both Figs.~\ref{fig:12.3} and
\ref{fig:12.4}, the figures to the left show by equal-intensity
contours the intensity distribution of the phase conjugated field in
the area of the $x$-$z$-plane between the quantum wire and the surface
of the phase conjugator. The width (along the $x$-axis) of the shown
area is in all cases twice the height (along the $z$-axis) on both
sides of $x=0$. The contours are drawn in an exponential sequence, so
that if the first contour corresponds to the intensity
$I_{\rm{PC}}^{(1)}$ the $n$-th contour is associated with the
intensity $I_{\rm{PC}}^{(n)}=I_{\rm{PC}}^{(1)}\exp[(1-n)\alpha]$,
$\alpha$ varying from figure to figure. To further illustrate the
capabilities of light focusing the chosen system possess, we have to
the right on a linear scale shown the phase conjugated intensity at
(i) the surface of the phase conjugator (solid lines) and (ii) along
an axis parallel to the $x$-axis placed at the same distance from the
phase conjugator as the wire (dashed lines). The two curves in each of
the plots to the right are adjusted by multiplication of the curve in
case (ii) by a factor of (a) $2590$, (b) $545$, (c) $271$, (d) $353$,
and (e) $586$ in Fig.~\ref{fig:12.3}, and by a factor of (a) $555$,
(b) $212$, (c) $204$, (d) $322$, and (e) $528$ in Fig.~\ref{fig:12.4},
respectively, so that the maximum values coincide in the plots.

It is seen from Figs.~\ref{fig:12.3} and \ref{fig:12.4} that the width
of the focus created by the phase conjugated field in all cases is
smallest at the surface of the phase conjugator. Furthermore one
observe that the focus becomes narrower when the distance between the
source and the film becomes smaller, as one should expect when
evanescent field components give a significant contribution to the
process. At the surface the width of the main peak decreases roughly
by a factor of two every time the source-film distance decreases by
the same factor.  This tendency continues closer to the surface of the
phase conjugator, at least down to around $\lambda/256\approx4$nm,
where the structure of the intensity distribution looks more or less
like Fig.~\ref{fig:12.3}.e [and Fig.~\ref{fig:12.4}.e], scaled
appropriately with the distance.

The difference between the ability of the single-monolayer film and
the two-mono\-layer film to focus the field emitted from the source is
small, as we would expect from the analysis in the previous chapter.
We observe from Fig.~\ref{fig:12.3} that when the source is far away
from the film (a--b), the intensity of the phase conjugated field
decays faster in the case of a single-monolayer film than in the other
case. Furthermore, we observe that the width of the centre peak is
somewhat smaller for the focus in front of the single-monolayer phase
conjugator, and that the height of the first sidelobe is slightly
higher compared to the centre peak. Taking the source closer to the
surface (c--e) we observe that the distance between minima in
Fig.~\ref{fig:12.3} is still smaller than in Fig.~\ref{fig:12.4}, but
that the width of the centre peak becomes more and more equal in the
two cases.

To give an impression of the size of the phase conjugated focus, let
us take a look at Fig.~\ref{fig:intensity-ptop}.e, where the
distance from the probe to the surface is $\lambda/16\approx66$nm. The
distance between the two minima at the surface of the phase conjugator
is in this case around $40$nm (approximately $\lambda/25$).  In the
plane of the probe the distance between the two minima is around
$100$nm ($\sim\lambda/10$).

\newpage

\clearpage
\thispagestyle{plain}
\newpage

\chapter{Discussion}
We have in the previous three chapters discussed the phase conjugation
response of a single-level quantum well, where only intraband
transitions are possible. It is evident from our analysis that in this
case the phase conjugated response depends strongly on the component
of the probe wavevector that is parallel to the surface of the phase
conjugator ($q_{\|}$). Consequently, the assumption of an ideal
phase conjugator with constant reflection coefficient throughout the
full $q_{\|}$-spectrum must be abandoned, at least when a single-level
quantum-well phase conjugator is considered. The nonlinear coupling is
strongest in the evanescent part of the $q_{\|}$-spectrum above the
point up to which the probe field is propagating in the substrate
($q_{\|}/q=n$).  As a consequence, if one wants to observe the phase
conjugation of a broad Fourier spectrum of evanescent modes, both the
observation and the excitation are required to take place near the
surface of the phase conjugator.

As a possible method to excite the Fourier components in the high end
of the $q_{\|}$-spectrum we have analyzed the consequences of using a
quantum wire. When the quantum wire is placed close to the phase
conjugator the phase conjugated response contains a broad range of
evanescent components. This property made it a good candidate for
investigations of the problem of focusing light to a spatial extent
less than the Rayleigh limit. The spatial focusing of the phase
conjugated response from a quantum wire was studied, and the problem
of the resolution limit has been addressed. The conclusion of this
study is in agreement with previous studies and mainly shows that the
focus gets narrower when the distance from the quantum wire to the
phase conjugator gets shorter.  Judging from this we may conclude that
in order to establish a better estimate of the limit of resolution the
present model has to be improved, since continuing to get closer
becomes meaningless at some point.

In examining the phase conjugation response of the single-level
quantum well we have chosen a specific frequency of the interacting
electromagnetic fields, which should be feasible for an experiment. It
would be interesting to give a more detailed account of how (i) the
phase conjugated response and (ii) the focusing of the phase
conjugated field varies with the frequency. These properties will not
be discussed in detail in this dissertation, but let us at this point
just mention that (i) it seems like the evanescent components in the
high end of the Fourier spectrum becomes better phase conjugated when
the interacting fields move to longer wavelengths, and (ii) the
spatial distance between minima in the intensity of the phase
conjugated response when using a quantum wire becomes smaller when
measured in fractions of the wavelength. Going to shorter wavelengths,
the tendency goes in the opposite direction.

\part[Optical phase conjugation in multi-level metallic quantum
  wells]{Optical phase conjugation in \\ multi-level metallic quantum
  wells}\label{part:V}
\newpage
\thispagestyle{plain}
\newpage

\chapter{Theoretical properties}\label{ch:14}
From the simple description of a quantum well where only intraband
transitions contributed to the phase conjugated response we now turn
our attention to the case where transitions between energy levels in
the quantum well (interband transitions) can take place. In this
chapter we therefore give the theoretical description that is
necessary to describe the phase conjugated response from a quantum
well where both interband and intraband transitions contribute to the
response. In the following we adopt the same scattering geometry as in
the previous treatment, i.e., scattering takes place in the
$x$-$z$-plane and we use light that is polarized either in
($p$-polarized) or perpendicular to ($s$-polarized) the scattering
plane.

\section{Phase conjugated field}
Unlike in the case of a single-level quantum well, we cannot rely on
the self-field approximation when considering multi-level quantum
wells (with resonances). We therefore begin this treatment with the
loop equation for the phase conjugated field in the two-dimensional
Fourier space [Eq.~(\ref{eq:loop})]. It is repeated here for
convenience:
\begin{eqnarray}
\lefteqn{
 \vec{E}_{\rm{PC}}(z;\vec{q}_{\|},\omega)=
 \vec{E}_{\rm{PC}}^{\rm{B}}(z;\vec{q}_{\|},\omega)
}\nonumber\\ &\quad&
 -{\rm{i}}\mu_0\omega
 \iint
 \tensor{G}(z,z'';\vec{q}_{\|},\omega)
 \cdot\stensor{\sigma}(z'',z';\vec{q}_{\|},\omega)
 \cdot \vec{E}_{\rm{PC}}(z';\vec{q}_{\|},\omega)dz''dz'.
\label{eq:PC-loop}
\end{eqnarray}
The background field in Eq.~(\ref{eq:PC-loop}) is given by
\begin{equation}
\vec{E}_{\rm{PC}}^{\rm{B}}(z;\vec{q}_{\|},\omega)=
 -{\rm{i}}\mu_0\omega\int
 \tensor{G}(z,z';\vec{q}_{\|},\omega)
 \cdot\vec{J}_{-\omega}^{\,(3)}(z';\vec{q}_{\|},\omega)dz',
\end{equation}
and can be determined from the previous analysis. The linear
conductivity tensor consists in general of a diamagnetic and a
paramagnetic part (see the discussion in Chapters~\ref{ch:5} and
\ref{ch:6}). It is, however, possible to combine the two parts in such
a way that the diamagnetic conductivity tensor can be written as a
correction to the paramagnetic one (\citeNP{Keller:96:1},
\citeyearNP{Keller:97:1,Keller:97:3}). Then the expression for the
total linear conductivity tensor becomes
\begin{eqnarray}
 \stensor{\sigma}(\vec{r},\vec{r}\,';\omega)=
 -{2\over{}i\omega}{1\over\hbar}\sum_{nm}{\omega\over\tilde{\omega}_{nm}}
 {f_n-f_m\over\tilde{\omega}_{nm}-\omega}
 \vec{J}_{nm}(\vec{r})\otimes\vec{J}_{mn}(\vec{r}\,'),
\end{eqnarray}
the correction from the diamagnetic term to the paramagnetic
response being the factor $\omega/\tilde{\omega}_{nm}$, which close
to resonance becomes $1$. In the two-dimensional mixed Fourier space
this is
\begin{eqnarray}
\lefteqn{
 \stensor{\sigma}(z,z';\vec{q}_{\|},\omega)=
 -{2\over{}i\omega}{1\over\hbar}{1\over(2\pi)^2}\sum_{nm}\int{\omega\over
  \tilde{\omega}_{nm}(\vec{\kappa}_{\|}+\vec{q}_{\|},\vec{\kappa}_{\|})}
 {f_{n}(\vec{\kappa}_{\|}+\vec{q}_{\|})-f_{m}(\vec{\kappa}_{\|})\over
  \tilde{\omega}_{nm}(\vec{\kappa}_{\|}+\vec{q}_{\|},\vec{\kappa}_{\|})-\omega}
}\nonumber\\ &\quad&\times
 \vec{j}_{nm}(z;2\vec{\kappa}_{\|}+\vec{q}_{\|})\otimes
 \vec{j}_{mn}(z';2\vec{\kappa}_{\|}+\vec{q}_{\|})
 d^2\kappa_{\|}.
\label{eq:sigma-zz'qw}
\end{eqnarray}
The transition frequency appearing in Eq.~(\ref{eq:sigma-zz'qw}) is
\begin{equation}
 \tilde{\omega}_{nm}(\vec{\kappa}_{\|}+\vec{q}_{\|},\vec{\kappa}_{\|})=
 {1\over\hbar}\left[\varepsilon_n-\varepsilon_m+{\hbar^2\over2m_{e}}
 \left(2\kappa_xq_x+q_x^2\right)\right]-{i\over\tau_{nm}},
\end{equation}
which by insertion into Eq.~(\ref{eq:sigma-zz'qw}) gives the five
nonzero elements of the linear conductivity tensor
\begin{eqnarray}
 \sigma_{xx}(z,z';\vec{q}_{\|},\omega)&=&
 \sum_{nm}{\cal{Q}}_{\,\,\,nm}^{xx}(\vec{q}_{\|},\omega)
 {\cal{Z}}^{x}_{nm}(z){\cal{Z}}^{x}_{mn}(z'),
\label{eq:s-xx}\\
 \sigma_{xz}(z,z';\vec{q}_{\|},\omega)&=&-i
 \sum_{nm}{\cal{Q}}_{\,\,\,nm}^{xz}(\vec{q}_{\|},\omega)
 {\cal{Z}}^{x}_{nm}(z){\cal{Z}}^{z}_{mn}(z'),
\label{eq:s-xz}\\
 \sigma_{yy}(z,z';\vec{q}_{\|},\omega)&=&
 \sum_{nm}{\cal{Q}}_{\,\,\,nm}^{yy}(\vec{q}_{\|},\omega)
 {\cal{Z}}^{x}_{nm}(z){\cal{Z}}^{x}_{mn}(z'),
\label{eq:s-yy}\\
 \sigma_{zx}(z,z';\vec{q}_{\|},\omega)&=&-i
 \sum_{nm}{\cal{Q}}_{\,\,\,nm}^{xz}(\vec{q}_{\|},\omega)
 {\cal{Z}}^{z}_{nm}(z){\cal{Z}}^{x}_{mn}(z'),
\label{eq:s-zx}\\
 \sigma_{zz}(z,z';\vec{q}_{\|},\omega)&=&-
 \sum_{nm}{\cal{Q}}_{\,\,\,nm}^{zz}(\vec{q}_{\|},\omega)
 {\cal{Z}}^{z}_{nm}(z){\cal{Z}}^{z}_{mn}(z'),
\label{eq:s-zz}
\end{eqnarray}
where we for the sake of notational simplicity have divided the total
expression for each element into a $z$-dependent part and a
$z$-independent part, the $z$-independent quantities being
\begin{eqnarray}
\lefteqn{ {\cal{Q}}_{\,\,\,nm}^{xx}(\vec{q}_{\|},\omega)={2i\hbar\over(2\pi)^2}
 \left({e\hbar\over2m_{e}}\right)^2
 \int
 {4\kappa_x^2+4\kappa_xq_x+q_x^2
  \over\varepsilon_n-\varepsilon_m+\hbar(2\kappa_xq_x+q_x^2)/(2m_{e})
 -i\hbar/\tau_{nm}}
}\nonumber\\ &\quad&\times
 {f_{n}(\vec{\kappa}_{\|}+\vec{q}_{\|})-f_{m}(\vec{\kappa}_{\|})
  \over\varepsilon_n-\varepsilon_m+\hbar(2\kappa_xq_x+q_x^2)/(2m_{e})
  -i\hbar/\tau_{nm}-\hbar\omega}d^2\kappa_{\|},
\label{eq:A-xx}\\
\lefteqn{ {\cal{Q}}_{\,\,\,nm}^{xz}(\vec{q}_{\|},\omega)={2i\hbar\over(2\pi)^2}
 \left({e\hbar\over2m_{e}}\right)^2
 \int
 {2\kappa_x+q_x
  \over\varepsilon_n-\varepsilon_m+\hbar(2\kappa_xq_x+q_x^2)/(2m_{e})
 -i\hbar/\tau_{nm}}
}\nonumber\\ &&\times
 {f_{n}(\vec{\kappa}_{\|}+\vec{q}_{\|})-f_{m}(\vec{\kappa}_{\|})
  \over\varepsilon_n-\varepsilon_m+\hbar(2\kappa_xq_x+q_x^2)/(2m_{e})
  -i\hbar/\tau_{nm}-\hbar\omega}d^2\kappa_{\|},
\label{eq:A-xz}\\
\lefteqn{ {\cal{Q}}_{\,\,\,nm}^{yy}(\vec{q}_{\|},\omega)={2i\hbar\over(2\pi)^2}
 \left({e\hbar\over2m_{e}}\right)^2
 \int
 {4\kappa_y^2
  \over\varepsilon_n-\varepsilon_m+\hbar(2\kappa_xq_x+q_x^2)/(2m_{e})
 -i\hbar/\tau_{nm}}
}\nonumber\\ &&\times
 {f_{n}(\vec{\kappa}_{\|}+\vec{q}_{\|})-f_{m}(\vec{\kappa}_{\|})
  \over\varepsilon_n-\varepsilon_m+\hbar(2\kappa_xq_x+q_x^2)/(2m_{e})
  -i\hbar/\tau_{nm}-\hbar\omega}d^2\kappa_{\|},
\label{eq:A-yy}\\
\lefteqn{ {\cal{Q}}_{\,\,\,nm}^{zz}(\vec{q}_{\|},\omega)={2i\hbar\over(2\pi)^2}
 \left({e\hbar\over2m_{e}}\right)^2
 \int
 {1\over\varepsilon_n-\varepsilon_m+\hbar(2\kappa_xq_x+q_x^2)/(2m_{e})
 -i\hbar/\tau_{nm}}
}\nonumber\\ &&\times
 {f_{n}(\vec{\kappa}_{\|}+\vec{q}_{\|})-f_{m}(\vec{\kappa}_{\|})
  \over\varepsilon_n-\varepsilon_m+\hbar(2\kappa_xq_x+q_x^2)/(2m_{e})
  -i\hbar/\tau_{nm}-\hbar\omega}d^2\kappa_{\|},
\label{eq:A-zz}
\end{eqnarray}
since ${\cal{Q}}_{\,\,\,nm}^{xz}(\vec{\kappa}_{\|},\omega)=
{\cal{Q}}_{\,\,\,nm}^{zx}(\vec{\kappa}_{\|},\omega)$. The
$z$-dependent quantities in Eqs.~(\ref{eq:s-xx})--(\ref{eq:s-zz})
above are
\begin{eqnarray}
 {\cal{Z}}^{x}_{nm}(z)&=&{\cal{Z}}^{\,y}_{nm}(z)=\psi_m^*(z)\psi_n(z), \\
 {\cal{Z}}^{z}_{nm}(z)&=&\psi_m^*(z){\partial\psi_n(z)\over\partial{}z}
 -\psi_n(z){\partial\psi_m^*(z)\over\partial{}z}.
\end{eqnarray}
Eqs.~(\ref{eq:A-xx})--(\ref{eq:A-zz}) has the solutions given in
Appendix~\ref{app:C}, section \ref{sec:Q} in terms of the analytic
solution to the integrals given in Appendix~\ref{ch:Solve-Q}.
Inserting this solution into Eq.~(\ref{eq:PC-loop}), we get
\begin{equation}
 \vec{E}_{\rm{PC}}(z;\vec{q}_{\|},\omega)=
 \vec{E}_{\rm{PC}}^{\rm{B}}(z;\vec{q}_{\|},\omega)+\sum_{nm}
 \tensor{F}_{nm}(z;\vec{q}_{\|},\omega)
 \cdot \vec{\Gamma}_{mn}(\vec{q}_{\|},\omega),
\label{eq:loop-FG}
\end{equation}
in which we have introduced the $3\times3$ tensor
$\tensor{F}_{nm}(z;\vec{q}_{\|},\omega)$ with the nonzero elements
\begin{eqnarray}
\lefteqn{ 
 F_{nm}^{xx}(z;\vec{q}_{\|},\omega)=
 -{\rm{i}}\mu_0\omega
 \left\{ 
 {\cal{Q}}_{\,\,\,nm}^{xx}(\vec{q}_{\|},\omega)
 \int G_{xx}(z,z'';\vec{q}_{\|},\omega){\cal{Z}}^{x}_{nm}(z'')dz''
\right.}\nonumber\\ &\quad&\left.\!
 -i{\cal{Q}}_{\,\,\,nm}^{xz}(\vec{q}_{\|},\omega)
 \int G_{xz}(z,z'';\vec{q}_{\|},\omega){\cal{Z}}^{z}_{nm}(z'')dz''
 \right\},
\label{eq:F-xx}\\
\lefteqn{ 
 F_{nm}^{xz}(z;\vec{q}_{\|},\omega)=
 {\rm{i}}\mu_0\omega
 \left\{ 
 i{\cal{Q}}_{\,\,\,nm}^{xz}(\vec{q}_{\|},\omega)
 \int G_{xx}(z,z'';\vec{q}_{\|},\omega){\cal{Z}}^{x}_{nm}(z'')dz''
\right.}\nonumber\\ &&\left.\!
 +{\cal{Q}}_{\,\,\,nm}^{zz}(\vec{q}_{\|},\omega)
 \int G_{xz}(z,z'';\vec{q}_{\|},\omega){\cal{Z}}^{z}_{nm}(z'')dz''
 \right\},
\label{eq:F-xz}\\
\lefteqn{
 F_{nm}^{yy}(z;\vec{q}_{\|},\omega)=
 -{\rm{i}}\mu_0\omega {\cal{Q}}_{\,\,\,nm}^{yy}(\vec{q}_{\|},\omega)
 \int G_{yy}(z,z'';\vec{q}_{\|},\omega){\cal{Z}}^{x}_{nm}(z'')dz'',
}
\label{eq:F-yy}\\
\lefteqn{
 F_{nm}^{zx}(z;\vec{q}_{\|},\omega)=
 {q_{\|}\over{}q_{\perp}}F_{nm}^{xx}(z;\vec{q}_{\|},\omega),
}\label{eq:F-zx}\\
\lefteqn{
 F_{nm}^{zz}(z;\vec{q}_{\|},\omega)=
 {q_{\|}\over{}q_{\perp}}F_{nm}^{xz}(z;\vec{q}_{\|},\omega),
}\label{eq:F-zz}
\end{eqnarray}
and the vector
\begin{equation}
 \vec{\Gamma}_{mn}(\vec{q}_{\|},\omega)=\left(
 \begin{array}{c}
 \displaystyle 
 \int{\cal{Z}}^{x}_{mn}(z')E_{{\rm{PC}},x}(z';\vec{q}_{\|},\omega)dz' \\[1mm]
 \displaystyle 
 \int{\cal{Z}}^{x}_{mn}(z')E_{{\rm{PC}},y}(z';\vec{q}_{\|},\omega)dz' \\[1mm]
 \displaystyle
 \int{\cal{Z}}^{z}_{mn}(z')E_{{\rm{PC}},z}(z';\vec{q}_{\|},\omega)dz'
 \end{array}\right)
\end{equation}
can be determined from the following set of algebraic equations:
\begin{equation}
 \vec{\Gamma}_{mn}(\vec{q}_{\|},\omega)
 -\sum_{vl}\tensor{K}_{mn}^{vl}(\vec{q}_{\|},\omega)
 \cdot\vec{\Gamma}_{vl}(\vec{q}_{\|},\omega)
 =\vec{\Omega}_{mn}(\vec{q}_{\|},\omega).
\label{eq:Gammafind}
\end{equation}
Since we may now determine the different $\Gamma$ values independently
of their dependence on the phase conjugated field,
$\vec{E}_{\rm{PC}}(z;\vec{q}_{\|},\omega)$, we have by this operation
kept the self-consistency in Eq.~(\ref{eq:loop-FG}), but the problem
of solution has been reduced to a problem of solving a linear
algebraic set of equations with just as many unknowns. This problem
can be treated as a matrix problem and is thus in principle fairly
easy to solve numerically. In Eq.~(\ref{eq:Gammafind}) above, the
vectorial quantity $\vec{\Omega}_{mn}$ is given by
\begin{equation}
 \vec{\Omega}_{mn}(\vec{q}_{\|},\omega)=
 \left(\begin{array}{c}
 \displaystyle
 \int{\cal{Z}}^{x}_{mn}(z)E_{{\rm{PC}},x}^{\rm{B}}(z;\vec{q}_{\|},\omega)dz \\[1mm]
 \displaystyle
 \int{\cal{Z}}^{x}_{mn}(z)E_{{\rm{PC}},y}^{\rm{B}}(z;\vec{q}_{\|},\omega)dz \\[1mm]
 \displaystyle
 \int{\cal{Z}}^{z}_{mn}(z)E_{{\rm{PC}},z}^{\rm{B}}(z;\vec{q}_{\|},\omega)dz
 \end{array}\right)
\end{equation}
and the $3\times3$ tensorial quantity
$\tensor{K}_{mn}^{vl}(\vec{q}_{\|},\omega)$ has the five nonzero
elements
\begin{eqnarray}
 K_{xx,mn}^{vl}(\vec{q}_{\|},\omega)&=&
 \int{\cal{Z}}^{x}_{mn}(z)F_{lv}^{xx}(z;\vec{q}_{\|},\omega)dz,
\label{eq:Kxx}
\\
 K_{xz,mn}^{vl}(\vec{q}_{\|},\omega)&=&
 \int{\cal{Z}}^{x}_{mn}(z)F_{lv}^{xz}(z;\vec{q}_{\|},\omega)dz,
\\
 K_{yy,mn}^{vl}(\vec{q}_{\|},\omega)&=&
 \int{\cal{Z}}^{x}_{mn}(z)F_{lv}^{yy}(z;\vec{q}_{\|},\omega)dz,
\\
 K_{zx,mn}^{vl}(\vec{q}_{\|},\omega)&=&{q_{\|}\over{}q_{\perp}}
 \int{\cal{Z}}^{z}_{mn}(z)F_{lv}^{xx}(z;\vec{q}_{\|},\omega)dz,
\\
 K_{zz,mn}^{vl}(\vec{q}_{\|},\omega)&=&{q_{\|}\over{}q_{\perp}}
 \int{\cal{Z}}^{z}_{mn}(z)F_{lv}^{xz}(z;\vec{q}_{\|},\omega)dz.
\label{eq:Kzz}
\end{eqnarray}
If we limit our treatment to polarized light perpendicular to the
scattering plane ($s$) and in the scattering plane ($p$), we
get for $s$-polarized light the set of equations
\begin{equation}
 \Gamma_{y,mn}-\sum_{vl}K_{yy,mn}^{vl}\Gamma_{y,vl}
 =\Omega_{y,mn},
\label{eq:Gamma-y}
\end{equation}
which is $m\times n$ equations with just as many unknowns, and
for $p$-polarized light the set of equations
\begin{eqnarray}
 \Gamma_{x,mn}-\sum_{vl}\left(K_{xx,mn}^{vl}\Gamma_{x,vl}
 +K_{xz,mn}^{vl}\Gamma_{z,vl}\right)&=&\Omega_{x,mn},
\label{eq:Gamma-x}
\\
 \Gamma_{z,mn}-\sum_{vl}\left(K_{zx,mn}^{vl}\Gamma_{x,vl}
 +K_{zz,mn}^{vl}\Gamma_{z,vl}\right)&=&\Omega_{z,mn},
\label{eq:Gamma-z}
\end{eqnarray}
which is $2m\times n$ equations with just as many unknowns.

\section{Infinite barrier quantum well}
Applying the infinite barrier quantum well to the above formalism, we
are able to determine the integrals over the Cartesian coordinates in
explicit form. The wave function constructs ${\cal{Z}}(z)$
becomes in the infinite barrier model
\begin{eqnarray}
 {\cal{Z}}^{x,\rm{IB}}_{nm}(z)&=&
 {1\over{}d}\left[\cos\left({(n-m)\pi{}z\over{}d}\right)
 -\cos\left({(n+m)\pi{}z\over{}d}\right)\right],
\label{eq:ZxIB}
\\
 {\cal{Z}}^{z,\rm{IB}}_{nm}(z)&=&
 {\pi\over{}d^2}\left[(n-m)\sin\left({(n+m)\pi{}z\over{}d}\right)
 -(n+m)\sin\left({(n-m)\pi{}z\over{}d}\right)\right].
\label{eq:ZzIB}
\end{eqnarray}
With this result the integrals over the source region apperaing in
Eqs.~(\ref{eq:F-xx})-- (\ref{eq:F-zz}) and
(\ref{eq:Kxx})--(\ref{eq:Kzz}) can be solved (see
Appendix~\ref{app:E}), and the $K$ quantities thus become
\begin{eqnarray}
\lefteqn{K_{xx,mn}^{vl}(\vec{q}_{\|},\omega)=
 -{8\pi^4nmlv{}q_{\perp}d\left[1-e^{iq_{\perp}d}(-1)^{n+m}\right]
 \over\epsilon_0\omega
 [(iq_{\perp}d)^2+\pi^2(n-m)^2][(iq_{\perp}d)^2+\pi^2(n+m)^2]}
}\nonumber\\ &\quad&\times
 {1+r^p-\left(e^{-iq_{\perp}d}+r^pe^{iq_{\perp}d}\right)(-1)^{l+v}
 \over[(iq_{\perp}d)^2+\pi^2(l-v)^2][(iq_{\perp}d)^2+\pi^2(l+v)^2]}
\nonumber\\ &&\times
 \left\{ 
 {\cal{Q}}_{\,\,\,lv}^{xx}(\vec{q}_{\|},\omega)
 q_{\perp}^2d
 +{\cal{Q}}_{\,\,\,lv}^{xz}(\vec{q}_{\|},\omega)
 {\pi^2(l^2-v^2)q_{\|}\over{}d}
 \right\},
\\
\lefteqn{K_{xz,mn}^{vl}(\vec{q}_{\|},\omega)=
 {8\pi^4inmlv{}q_{\perp}d\left[1-e^{iq_{\perp}d}(-1)^{n+m}\right]
 \over\epsilon_0\omega
 [(iq_{\perp}d)^2+\pi^2(n-m)^2][(iq_{\perp}d)^2+\pi^2(n+m)^2]}
}\nonumber\\ &&\times
 {1+r^p-\left(e^{-iq_{\perp}d}+r^pe^{iq_{\perp}d}\right)(-1)^{l+v}
 \over[(iq_{\perp}d)^2+\pi^2(l-v)^2][(iq_{\perp}d)^2+\pi^2(l+v)^2]}
\nonumber\\ &&\times
 \left\{ 
 {\cal{Q}}_{\,\,\,lv}^{xz}(\vec{q}_{\|},\omega)
 q_{\perp}^2d
 +{\cal{Q}}_{\,\,\,lv}^{zz}(\vec{q}_{\|},\omega)
 {\pi^2(l^2-v^2)q_{\|}\over{}d}
 \right\},
\\
\lefteqn{K_{yy,mn}^{vl}(\vec{q}_{\|},\omega)=
 {\cal{Q}}_{\,\,\,lv}^{yy}(\vec{q}_{\|},\omega)
 {8\pi^4\mu_0nmlv\omega{}q_{\perp}d^2[e^{iq_{\perp}d}(-1)^{n+m}-1]
 \over[(iq_{\perp}d)^2+\pi^2(n-m)^2][(iq_{\perp}d)^2+\pi^2(n+m)^2]}
}\nonumber\\ &&\times
 {1-r^s-\left(e^{-iq_{\perp}d}-r^se^{iq_{\perp}d}\right)(-1)^{l+v}
 \over[(iq_{\perp}d)^2+\pi^2(l-v)^2][(iq_{\perp}d)^2+\pi^2(l+v)^2]}
\\
\lefteqn{K_{zx,mn}^{vl}(\vec{q}_{\|},\omega)=
 -{2\pi^2ilv\over\epsilon_0\omega}
 {4\pi^4nm(n^2-m^2)[e^{iq_{\perp}d}(-1)^{n+m}-1]
 \over{}d[(iq_{\perp}d)^2+\pi^2(n-m)^2][(iq_{\perp}d)^2+\pi^2(n+m)^2]}
}\nonumber\\ &\quad&\times
 {1+r^p-\left(e^{-iq_{\perp}d}+r^pe^{iq_{\perp}d}\right)(-1)^{l+v}
 \over[(iq_{\perp}d)^2+\pi^2(l-v)^2][(iq_{\perp}d)^2+\pi^2(l+v)^2]}
\nonumber\\ &&\times
 \left\{ 
 {\cal{Q}}_{\,\,\,lv}^{xx}(\vec{q}_{\|},\omega)
 {q_{\perp}^2d}
 +{\cal{Q}}_{\,\,\,lv}^{xz}(\vec{q}_{\|},\omega)
 {\pi^2(l^2-v^2)q_{\|} \over{}d}
 \right\} 
\\
\lefteqn{K_{zz,mn}^{vl}(\vec{q}_{\|},\omega)=
 {2\pi^2ilv\over\epsilon_0\omega}
 {4\pi^4nm(n^2-m^2)[e^{iq_{\perp}d}(-1)^{n+m}-1]
 \over{}d[(iq_{\perp}d)^2+\pi^2(n-m)^2][(iq_{\perp}d)^2+\pi^2(n+m)^2]}
}\nonumber\\ &&\times
 {1+r^p-\left(e^{-iq_{\perp}d}+r^pe^{iq_{\perp}d}\right)(-1)^{l+v}
 \over[(iq_{\perp}d)^2+\pi^2(l-v)^2][(iq_{\perp}d)^2+\pi^2(l+v)^2]}
\nonumber\\ &&\times
 \left\{ 
 i{\cal{Q}}_{\,\,\,lv}^{xz}(\vec{q}_{\|},\omega)
 q_{\perp}^2d
 -{\cal{Q}}_{\,\,\,lv}^{zz}(\vec{q}_{\|},\omega)
 {\pi^2(l^2-v^2)q_{\|}\over{}id}
 \right\}.
\end{eqnarray}
To find $\vec{\Omega}_{mn}(\vec{q}_{\|},\omega)$ is in general a much
more difficult task, but insertion of the expression for
$\vec{E}_{\rm{PC}}^{\rm{B}}(z;\vec{q}_{\|},\omega)$ gives
\begin{eqnarray}
\lefteqn{
 {\Omega}_{x,mn}(\vec{q}_{\|},\omega)=-{\rm{i}}\mu_0\omega
 \int_{-d}^{0}{\cal{Z}}^{x}_{mn}(z)\int_{-d}^{0}
 \left[G_{xx}(z,z';\vec{q}_{\|})J_{-\omega,x}^{(3)}(z';\vec{q}_{\|})
\right.}\nonumber\\ &\quad&\left.\!
 +G_{xz}(z,z';\vec{q}_{\|})J_{-\omega,z}^{(3)}(z';\vec{q}_{\|})\right]dz'dz,
\\
\lefteqn{
 {\Omega}_{y,mn}(\vec{q}_{\|},\omega)=-{\rm{i}}\mu_0\omega
 \int_{-d}^{0}{\cal{Z}}^{x}_{mn}(z)\int_{-d}^{0}
 G_{yy}(z,z';\vec{q}_{\|})J_{-\omega,y}^{(3)}(z';\vec{q}_{\|})dz'dz,
}
\\
\lefteqn{
 {\Omega}_{z,mn}(\vec{q}_{\|},\omega)=-{\rm{i}}\mu_0\omega
 \int_{-d}^{0}{\cal{Z}}^{z}_{mn}(z)\int_{-d}^{0}
 {q_{\|}\over{}q_{\perp}}
 \left[G_{xx}(z,z';\vec{q}_{\|})J_{-\omega,x}^{(3)}(z';\vec{q}_{\|})
\right.}\nonumber\\ &\quad&\left.\!
 +G_{xz}(z,z';\vec{q}_{\|})J_{-\omega,z}^{(3)}(z';\vec{q}_{\|})\right]dz'dz.
\end{eqnarray}
These integrals can by insertion of the propagators and the wave
functions be solved for the integral over $z$, and thus we find
\begin{eqnarray}
\lefteqn{ {\Omega}_{x,mn}(\vec{q}_{\|},\omega)=
 -{2\pi^2inmq_{\perp}d\left[e^{iq_{\perp}d}(-1)^{n+m}-1\right]
 \over\epsilon_0\omega
 [(iq_{\perp}d)^2+\pi^2(n+m)^2][(iq_{\perp}d)^2+\pi^2(n-m)^2]}
}\nonumber\\ &\quad&\times
 \!\int_{-d}^{0}\left[q_{\perp}
 \left(e^{iq_{\perp}z}-r^pe^{-iq_{\perp}z}\right)
 J_{-\omega,x}^{(3)}(z;\vec{q}_{\|})
 +q_{\|}
 \left(e^{iq_{\perp}z}+r^pe^{-iq_{\perp}z}\right)
 J_{-\omega,z}^{(3)}(z;\vec{q}_{\|})\right]dz,
\nonumber\\
\label{eq:14.44}
\\
\lefteqn{ {\Omega}_{y,mn}(\vec{q}_{\|},\omega)=
 -{2\pi^2{\rm{i}}\mu_0\omega{}nmd\left[e^{iq_{\perp}d}(-1)^{n+m}-1\right]
 \over[(iq_{\perp}d)^2+\pi^2(n+m)^2][(iq_{\perp}d)^2+\pi^2(n-m)^2]}
}\nonumber\\ &&\times
 \int_{-d}^{0}
 \left(e^{iq_{\perp}z}+r^se^{-iq_{\perp}z}\right)
 J_{-\omega,y}^{(3)}(z;\vec{q}_{\|})dz,
\\
\lefteqn{ {\Omega}_{z,mn}(\vec{q}_{\|},\omega)=
-{2\pi^4q_{\|}nm(n^2-m^2)\left[e^{iq_{\perp}d}(-1)^{m+n}-1\right]
 \over\epsilon_0\omega{}d
 [(iq_{\perp}d)^2+\pi^2(n+m)^2][(iq_{\perp}d)^2+\pi^2(n-m)^2]}
}\nonumber\\ &&\times
 \int_{-d}^{0}\left[
 \left(e^{iq_{\perp}z}-r^pe^{-iq_{\perp}z}\right)
 J_{-\omega,x}^{(3)}(z;\vec{q}_{\|})
 +{q_{\|}\over{}q_{\perp}}
 \left(e^{iq_{\perp}z}+r^pe^{-iq_{\perp}z}\right)
 J_{-\omega,z}^{(3)}(z;\vec{q}_{\|})\right]dz,
\nonumber\\
\label{eq:14.46}
\end{eqnarray}
where we have dropped the now superfluous marking $z'$ in favor of a
new $z$. Since the $z$-dependence of
$\vec{J}_{-\omega}^{\,(3)}(z;\vec{q}_{\|})$ is expressed via the
interacting fields and the wave functions, and we are limiting
ourselves to studies where (i) the pump fields are parallel to either
the $x$-axis or the $z$-axis and with uniform amplitude profile along
that axis, and (ii) the probe field has only one plane-wave component
on the form $\vec{E}(z;\vec{q}_{\|})=\vec{E}e^{iq_{\perp}z}$, the last
integral above can be solved. This solution is discussed in
Appendix~\ref{app:C}, sections \ref{sec:Z} and \ref{sec:Ps}. Thus, in
Eq.~(\ref{eq:Gammafind}), all $K$'s and $\Omega$'s are numbers with no
inline integrals to solve numerically.

\chapter{Numerical results for a two-level quantum well}\label{ch:15}
Besides calculation of the nonlinear current densities, the main
numerical work consists of finding the solution to the appropriate
sets of equations, given by Eq.~(\ref{eq:Gamma-y}) for processes with
$s$-polarized response, and Eqs.~(\ref{eq:Gamma-x}) and
(\ref{eq:Gamma-z}) for processes with $p$-polarized response.
Computational procedures to solve this kind of problems are well known
(see, e.g., \citeNP{Press:92:1}, \citeyearNP{Press:96:1,Press:96:2}
for description and Fortran routines).

\section{Phase conjugation reflection coefficient} 
To estimate the amount of light we get back through the phase
conjugated channel, we use the phase conjugation reflection
coefficient $R_{\rm{PC}}(z;\vec{q}_{\|})$ defined in Eq.~(\ref{eq:RPC})
together with the expression for the intensities given by
Eq.~(\ref{eq:RPC-I}). As before, the reflection coefficient at the
surface of the quantum well is thus $R_{\rm{PC}}(-d;\vec{q}_{\|})$.

In order to give an impression of the difference between the
calculation where only intraband contributions were taken into account
(chapter \ref{Ch:11}) the present calculation is also based on the
data for a two-monolayer thick copper quantum well
[$N_+=8.47\times10^{28}$m$^{-3}$, $Z=1$, and $d=3.8${\AA}
\cite{Ashcroft:76:1}]. As was the case for the single-level Cu quantum
well, the two-level Cu quantum well can adequately be deposited on a
glass substrate for which we use a refractive index $n$ of 1.51. With
this substrate, a reasonable description of the linear
vaccum/substrate amplitude reflection coefficients ($r^p$ for the
$p$-polarized light and $r^s$ for the $s$-polarized light) can be
obtained by use of the classical Fresnel formulae, given by
Eq.~(\ref{eq:rp}) and
\begin{eqnarray}
 r^s&=&{q_{\perp}-({n^2q^2-q_{\|}^2})^{1\over2}\over
        q_{\perp}+({n^2q^2-q_{\|}^2})^{1\over2}}.
\label{eq:rs}
\end{eqnarray}
Keeping the pump fields parallel to the $x$-axis, we get a pump
wavenumber $k_{\|}=1.51q$.

\begin{figure}[tb]
\setlength{\unitlength}{1mm}
\psset{unit=1mm}
\begin{center}
\begin{pspicture}(0,0)(127,96)
\put(-8,2){\epsfig{file=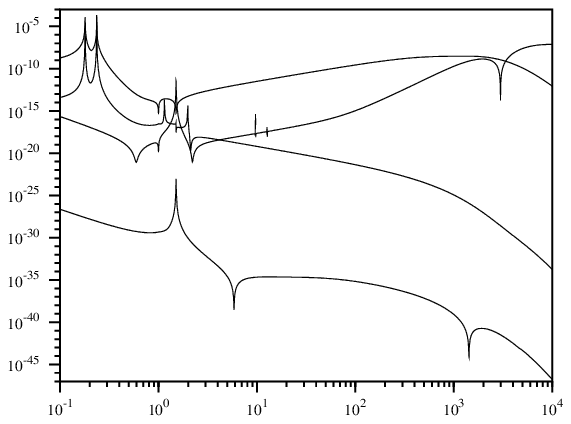,width=135\unitlength}}
\rput[t]{90}(0,53){$R_{\rm{PC}}(-d;\vec{q}_{\|})$~[m$^4$/W$^2$]}
\put(71.5,0){\makebox(0,0)[b]{$q_{\|}/q$}}
\pstextpath[l]{\pscurve[linestyle=none](80,83)(87,84)(97,85)}{\footnotesize$ppp$}
\pstextpath[l]{\pscurve[linestyle=none](80,73)(86,75)(95,79)}{\footnotesize$ssp$}
\pstextpath[l]{\pscurve[linestyle=none](80,62)(86,61)(96,58)}{\footnotesize$pps$}
\pstextpath[l]{\pscurve[linestyle=none](80,37)(90,35)(97,33)}{\footnotesize$sss$}
\psline[linewidth=0.25mm]{->}(43.75,84)(43.75,80)
\psline[linewidth=0.25mm]{->}(43.75,41)(43.75,45)
\put(43.75,85){\makebox(0,0)[b]{$n$}}
\psline[linewidth=0.25mm]{>-<}(113,4)(113,96)
\put(113,0){\makebox(0,0)[b]{$k_F/q$}}
\end{pspicture}
\end{center}
\caption[Phase conjugated response for a two-level quantum well as a
function of $q_{\|}/q$ , $s$ to $s$ and $p$ to $p$ responses]{The
  phase conjugation reflection coefficient at the vaccum/quantum-well
  interface is plotted as a function of the normalized component of
  the probe wavevector along the surface, $q_{\|}/q$, for the four
  combinations of polarization of the three interacting fields, in
  which the two pump fields have the same polarization ($ppp$, $sss$,
  $ssp$, and $pps$), corresponding to the four diagrams shown in
  Fig.~\ref{fig:9.2}. The vertical line indicates the normalized Fermi
  wavenumber, which for the two-monolayer Cu quantum well is
  $2.78\times10^{3}$. The set of arrows labeled $n$ are placed at
  $q_{\|}=nq$.\label{fig:15.1}}
\end{figure}

\begin{figure}[tbp]
\setlength{\unitlength}{1mm}
\psset{unit=1mm}
\begin{center}
\begin{pspicture}(0,6.55)(127,187.9)
\put(0,94){
 \put(-8,2){\epsfig{file=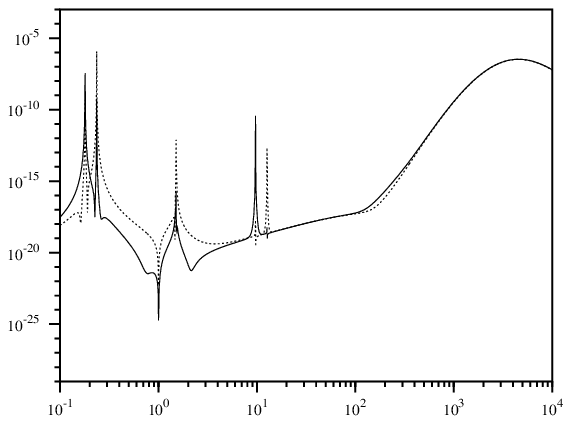,width=135\unitlength}}
 \rput[t]{90}(0,53){$R_{\rm{PC}}(-d;\vec{q}_{\|})$~[m$^4$/W$^2$]}
 \put(71.5,2){\makebox(0,0)[b]{$q_{\|}/q$}}
 \psline[linewidth=0.25mm]{->}(43.75,70)(43.75,66)
 \psline[linewidth=0.25mm]{->}(43.75,38)(43.75,42)
 \put(43.75,71){\makebox(0,0)[b]{$n$}}
 \psline[linewidth=0.25mm]{>-<}(113,6)(113,95)
 \put(113,2){\makebox(0,0)[b]{$k_F/q$}}
}
\put(0,0){
 \put(-8,2){\epsfig{file=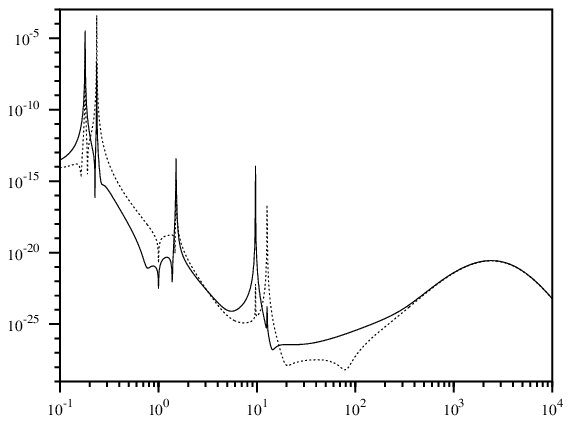,width=135\unitlength}}
 \rput[t]{90}(0,53){$R_{\rm{PC}}(-d;\vec{q}_{\|})$~[m$^4$/W$^2$]}
 \put(71.5,2){\makebox(0,0)[b]{$q_{\|}/q$}}
 \psline[linewidth=0.25mm]{->}(43.75,66)(43.75,62)
 \psline[linewidth=0.25mm]{->}(43.75,30)(43.75,34)
 \put(43.75,67){\makebox(0,0)[b]{$n$}}
 \psline[linewidth=0.25mm]{>-<}(113,6)(113,95)
 \put(113,2){\makebox(0,0)[b]{$k_F/q$}}
}
\end{pspicture}
\end{center}
\caption[Phase conjugated response for a two-level quantum well as a
function of $q_{\|}/q$, $s$ to $p$ and $p$ to $s$ responses]{Same
  parameters as in Fig.~\ref{fig:15.1}, but for the other four
  polarization combinations (see Fig.~\ref{fig:9.3}). The upper figure
  shows the responses where $s$-polarized probe gives $p$-polarized
  response [$pss$ (solid line) and $sps$ (dotted line)]. The lower
  figure shows the opposite cases [$psp$ (solid line) and $psp$
  (dotted line)]. Further explanation can be found in the main
  text.\label{fig:15.2}}
\end{figure}

In Figs.~\ref{fig:15.1} and \ref{fig:15.2} we have plotted the phase
conjugation reflection coefficient at the interface between the vacuum
and the quantum well as a function of the parallel component of the
wavevector normalized to the vacuum wavenumber, $q_{\|}/q$. The
wavelength has in these plots been fixed to $\lambda=1061$nm (the same
as in the single-level case). The plots have been divided into two
sets, together covering all eight different combinations of
polarization of the interacting fields. In Fig.~\ref{fig:15.1} is
plotted the four combinations leading to a response with the same
state of polarization as the probe, i.e., (i) the purely $p$-polarized
case where all interacting fields are polarized in the scattering
plane (denoted $ppp$), (ii) the purely $s$-polarized ($sss$) case
where all three interacting fields are polarized perpendicular to the
scattering plane, (iii) the case where the two pump fields both are
$p$-polarized and the probe field is $s$-polarized ($pps$), and (iv)
the case where the probe field is $p$-polarized and the two pump
fields are $s$-polarized ($ssp$). The results for the other four
combinations of polarization has been plotted in pairs in
Fig.~\ref{fig:15.2}. The upper figure in Fig.~\ref{fig:15.2} shows the
two cases where the pump fields are differently polarized and the
probe field is $s$-polarized, while in the lower figure, the probe
field is $p$-polarized, still with differently polarized pump fields.
The vertical line inserted into Figs.~\ref{fig:15.1} and
\ref{fig:15.2} indicates the normalized Fermi wavenumber, which for
the two-monolayer Cu quantum well is $2.78\times10^{3}$. The
discussion of this quantity has been given in chapter \ref{Ch:11} (in
the paragraph starting at the end of page \pageref{kF}).

\begin{figure}[tb]
\setlength{\unitlength}{1mm}
\psset{unit=1mm}
\begin{center}
\begin{pspicture}(0,3)(127,96)
\put(-8,2){\epsfig{file=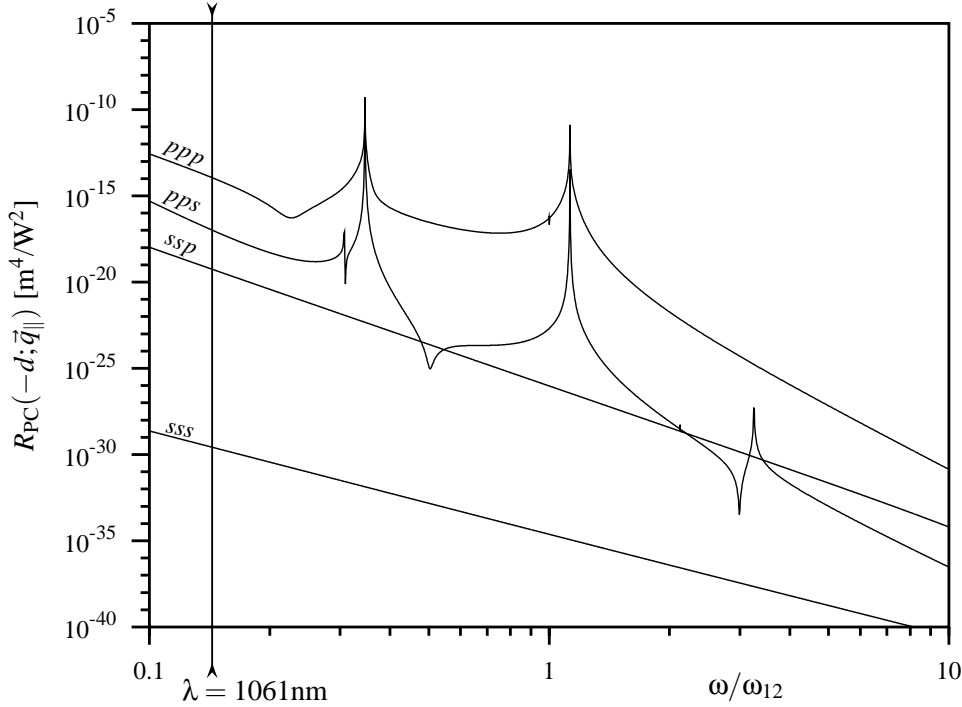,width=135\unitlength}}
\rput[t]{90}(0,53){$R_{\rm{PC}}(-d;\vec{q}_{\|})$~[m$^4$/W$^2$]}
\put(98,3){\makebox(0,0)[b]{$\omega/\omega_{12}$}}
 \pstextpath[l]{\psline[linestyle=none](21,77.5)(30,73.5)}{\footnotesize$ppp$}
 \pstextpath[l]{\psline[linestyle=none](21,71)(30,67)}{\footnotesize$pps$}
 \pstextpath[l]{\psline[linestyle=none](21,65)(30,61.5)}{\footnotesize$ssp$}
 \pstextpath[l]{\psline[linestyle=none](21,40)(30,38)}{\footnotesize$sss$}
 \psline[linewidth=0.25mm]{>-<}(27,6)(27,96)
 \put(23,3){\makebox(0,0)[bl]{$\lambda=1061$nm}}
\end{pspicture}
\end{center}
\caption[Phase conjugated response for a two-level quantum well as a
function of $\omega/\omega_{12}$, $s$ to $s$ and $p$ to $p$
responses]{The phase conjugation reflection coefficient at the
  vaccum/quantum-well interface is plotted as a function of the
  optical frequency normalized to the transition frequency of the
  two-level quantum well, $\omega/\omega_{12}$, for the four
  combinations of polarization of the three interacting fields which
  have equal pump field polarization ($ppp$, $sss$, $ssp$, and $pps$).
  The transition frequency $\omega_{12}$ in the present case is
  $\omega_{12}=1.32\times10^{16}$rad/s, corresponding to a
  wavelength of $\lambda=142.4$nm. The vertical line indicates
  $\lambda=1061$nm, the point in the frequency spectrum where
  Figs.~\ref{fig:11.1} and \ref{fig:11.2} have been
  drawn.\label{fig:15.3}}
\end{figure}

\begin{figure}[tbp]
\setlength{\unitlength}{1mm}
\psset{unit=1mm}
\begin{center}
\begin{pspicture}(0,6.55)(127,187.9)
\put(0,94){
 \put(-8,2){\epsfig{file=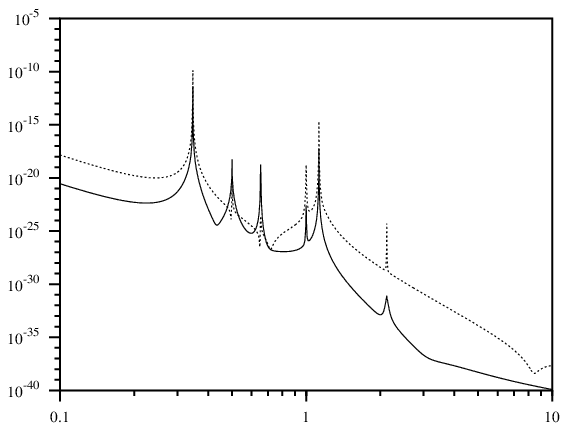,width=135\unitlength}}
 \rput[t]{90}(0,53){$R_{\rm{PC}}(-d;\vec{q}_{\|})$~[m$^4$/W$^2$]}
 \put(98,3){\makebox(0,0)[b]{$\omega/\omega_{12}$}}
 \psline[linewidth=0.25mm]{>-<}(27,6)(27,95)
 \put(23,3){\makebox(0,0)[bl]{$\lambda=1061$nm}}
}
\put(0,0){
 \put(-8,2){\epsfig{file=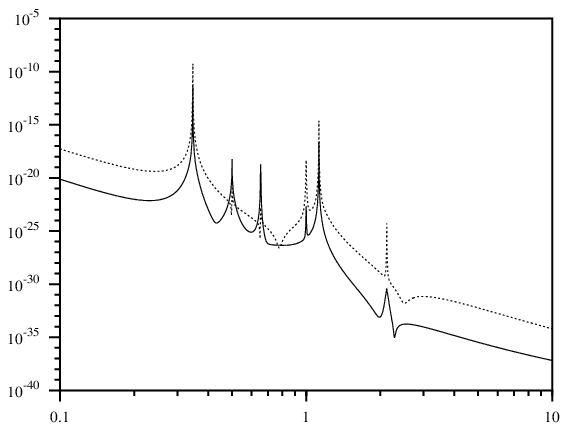,width=135\unitlength}}
 \rput[t]{90}(0,53){$R_{\rm{PC}}(-d;\vec{q}_{\|})$~[m$^4$/W$^2$]}
 \put(98,3){\makebox(0,0)[b]{$\omega/\omega_{12}$}}
 \psline[linewidth=0.25mm]{>-<}(27,6)(27,95)
 \put(23,3){\makebox(0,0)[bl]{$\lambda=1061$nm}}
}
\end{pspicture}
\end{center}
\caption[Phase conjugated response for a two-level quantum well as a
function of $\omega/\omega_{12}$, $s$ to $s$ and $p$ to $p$
responses]{Same parameters as in Fig.~\ref{fig:15.3}, but for the
  other four polarization combinations. The upper figure shows the
  responses where $s$-polarized probe gives $p$-polarized response
  [$pss$ (solid line) and $sps$ (dotted line)]. The lower figure shows
  the opposite cases [$psp$ (solid line) and $spp$ (dotted line)]. For
  further explanation, please consult the main text.\label{fig:15.4}}
\end{figure}

In addition to the plots in Figs.~\ref{fig:15.1} and \ref{fig:15.2},
where the phase conjugation reflection coefficient was plotted as a
function of $q_{\|}/q$, we have in Figs.~\ref{fig:15.3} and
\ref{fig:15.4} plotted the phase conjugation reflection coefficient as
a function of the optical frequency normalized to the interband
transition frequency, $\omega/\omega_{12}$. The parallel component of
the wavevector has in this case been fixed at $q_{\|}=0.8q$ (in the
propagating regime). Again, the four cases of pump fields having the
same polarization are plotted in the first of the two figures
(Fig.~\ref{fig:15.3}), and the remaining four in the other figure
(Fig.~\ref{fig:15.4}).

All plots in Figs.~\ref{fig:15.1}--\ref{fig:15.4} have been plotted
using a relaxation time of $200$fs in the interband transition from
the occupied state to the unoccupied state and a relaxation time of
$3$fs within the intraband transitions of the occupied state. Unlike
in the case of pure intraband response, the choice of adequate
relaxation times seem less important in the two-level quantum-well
case. Changing either of the relaxation times (or both) an order of
magnitude up or down doesn't change the results shown in
Figs.~\ref{fig:15.1}--\ref{fig:15.4} so much that the two curves for
the respective choices of relaxation times would differ from each
other, as it was the case in the single-level quantum well (see
Fig.~\ref{fig:11.4}).

Returning our attention to Fig.~\ref{fig:15.1}, we observe that the
purely $p$-polarized combination of polarization gives the strongest
phase conjugated response while the purely $s$-polarized combination
gives the weakest response of the four. Looking at the $ppp$ curve, we
see that the maximum value is reached in the propagating regime of the
$q_{\|}/q$-spectrum, where two peaks occur approximately at the values
of $q_{\|}=0.18q$ and $q_{\|}=0.22q$. These peaks must be due to the
pump waves being $p$-polarized, since they also occur in the $pps$
configuration, but in neither of the $sss$ and $ssp$ configurations.
In the evanescent regime the $ppp$ curve is increasing from the point
$q_{\|}=nq$ until it reaches its maximal value at around
$q_{\|}=10^3q$. Above $q_{\|}=10^3q$, the $ppp$ response starts
decaying again. The $pps$ curve in Fig.~\ref{fig:15.1} has, apart from
the two peaks discussed above, two additional peaks occuring
symmetrically around the point $nq$ in the $q_{\|}$-spectrum, at
approximately $q_{\|}=1.1q$ and $q_{\|}=1.9q$, respectively. After the
second of these peaks, the amplitude of the response fades away with
growing values of $q_{\|}$. The $ssp$ curve has a maximum when the
probe field is perpendicularly incident on the phase conjugator
($q_{\|}/q=0$), and another one where the probe field becomes
evanescent in the substrate, i.e., at $q_{\|}=nq$. In the evanescent
regime of the $q_{\|}/q$-spectrum the response is increasing, with two
small narrow peaks occuring at $q_{\|}\approx10q$ and
$q_{\|}\approx13q$, and it reaches a maximum at
$q_{\|}\approx2\times10^3$, and after going down to a minimum right
after $k_F/q$ it increases again. This indicates that if we are able
to produce probe fields with a significant amount of evanescent modes
above $k_F/q$, the present model is probably not sufficient to
describe the $ssp$ response (and maybe not sufficient to describe the
$ppp$ response either). The last of the curves in Fig.~\ref{fig:15.1}
represents the purely $s$-polarized case ($sss$). It has maxima at
$q_{\|}/q=0$, and again at $q_{\|}/q=n$. Above $q_{\|}=nq$ it falls
off rapidly.

Looking at the $q_{\|}/q$-spectrum of the phase conjugated response in
the other four combinations of polarization, depicted in
Fig.~\ref{fig:15.2}, we see that both pairs have peaks in the
propagating regime of the $q_{\|}/q$-spectrum at the same places as
the $ppp$ and $pps$ curves of Fig.~\ref{fig:15.1} had. Another peak
appears when $q_{\|}=nq$, and in the purely evanescent part of the
$q_{\|}/q$-spectrum peaks appear at $q_{\|}\approx10q$ and
$q_{\|}\approx13q$, the same places as in the $ssp$ response shown in
Fig.~\ref{fig:15.1}. After these two peaks the responses of the
$pss/sps$-pair increases until they reach their maximum at around
$q_{\|}=5\times10^{3}$, after which they decrease again. This maximum
is comparable in magnitude to the peaks in the propagating end of the
$q_{\|}/q$-spectrum. We observe that the two curves in each of the
pairs shown in Fig.~\ref{fig:15.2} becomes identical for high values
of $q_{\|}$, as they should from the previous analysis. The
$spp/psp$-pair of curves also starts increasing in magnitude in the
high end of the $q_{\|}/q$-spectrum showed. They reach their maximum
at around $q_{\|}=2.5\times10^3q$, after which the magnitude decreases
again. In this case, however, the magnitude of this maximum is some
fifteen orders of magnitude less than the magnitude of the peaks in
the propagating regime. The problem in the $sps/pss$-pair of curves is
that the maximal value is reached after the point $q_{\|}=k_F/q$, and
the conclusion must therefore be the same as in the $ssp$ case, namely
that if the probe has components of significance in the high end of
the $q_{\|}/q$-spectrum, then the model should probably be extended in
one way or another.

Continuing to the frequency plots, we observe from Fig.~\ref{fig:15.3}
that the two cases where both pump fields are $s$-polarized has no
resonances at all. Their decrease in magnitude as the frequency
increases is mainly due to the factor of $\omega^{-3}$ occuring in the
nonlinear conductivity tensor. The $ppp$ curve has a peak of high
magnitude at $\omega\approx\omega_{12}/3$, a small one at
$\omega=\omega_{12}$, and a large one again at
$\omega\approx1.1\omega_{12}$. The $pps$ curve has a peak of small
magnitude at $\omega=0.3\omega_{12}$, two large ones at
$\omega\approx\omega_{12}/3$ and at $\omega\approx1.1\omega_{12}$, and
finally a small one at $\omega\approx3.2\omega_{12}$. The peaks around
$\omega\approx\omega_{12}$ probably arise from the combination of the
denominators in the nonlinear conductivity tensors, but none of the
peaks have been clearly identified from the formulas yet. Going to any
of the two sides in the frequency spectrum away from this group of
resonances, the curves behave like the $sss$ and $ssp$ curves, with
the magnitude proportional to $\omega^{-3}$. The frequency plots for
the two pairs of polarization combinations where the pump fields are
differently polarized (Fig.~\ref{fig:15.4}) have resonances with the
approximate values of $\omega/\omega_{12}$ of $1/3$, $1/2$, $2/3$,
$1$, $1.1$, and $2$. Like in the previous case, these peaks have not
been clearly identified from the formulas yet. Again one might assume
that the peaks at $1$ and $1.1$ arise directly from (some of) the
denominators in the nonlinear conductivity tensors. As before, outside
of the shown frequency range, the behaviour of the response is
proportional to $\omega^{-3}$. The results presented in this chapter
will be treated more thoroughly in a forthcoming paper.

\newpage
\thispagestyle{plain}
\newpage

\chapter{Discussion}
In the past two chapter we have briefly shown how to calculate the
phase conjugated response from a multilevel quantum well and given
numerical results for a two-level quantum well. As we concluded in the
single-level quantum-well case, the assumption of an ideal phase
conjugator also does not hold for a two-level quantum-well phase
conjugator. In this case, however, it is not so much because of the
efficiency in the high end of the $q_{\|}$-spectrum, since in most of
the cases shown in chapter \ref{ch:15} the efficiency in the
evanescent regime is not so much larger than in the propagating
regime. It is more because the two-level phase conjugator is much more
efficient for certain values of $q_{\|}$ than for the rest of the
spectrum.

However, before we can give a full description of the phase conjugated
response from multi-level quantum wells, some aspects has to be
addressed. Among the important ones are the fact that we need to
identify (i) which terms of the nonlinear conductivity tensor that are
dominating the phase conjugation response, and if it is possible by a
careful choice of the system to make different terms dominate.
Furthermore (ii) it is desirable to find out more precisely which
individual terms in the nonlinear conductivity tensor gives rise to
each of the peaks occuring in the curves in
Figs.~\ref{fig:15.1}--\ref{fig:15.4}. Many other things has to be
investigated, for example (i) the behaviour of the response close to
the peaks in Figs.~\ref{fig:15.1}--\ref{fig:15.4}, both in the
frequency spectrum and in the $q_{\|}$-spectrum, (ii) the frequency
dependence in general, and (iii) the response to sources with a large
number of Fourier components in the $q_{\|}$-spectrum, such as the
quantum wire discussed in Chapter \ref{ch:12}.

\newpage
\thispagestyle{plain}
\newpage

\chead[\fancyplain{}{\leftmark}]{\fancyplain{}{\leftmark}}
\part{Closing remarks}
\newpage
\section*{Summary of conclusions}
We have developed a spatially nonlocal theoretical model of degenerate
four-wave mixing of electromagnetic fields on the mesoscopic length
scale. We have analyzed the physical processes involved in creating
the DFWM response and identified the independent nonzero elements of
the related conductivity tensor for each type of process. Following
the more general treatment in real space we have specialized the
treatment to take into account only cases where translational
invariance against displacements in two of the three spatial
dimensions occur, thereby favouring a description in which the optical
processes occur in surfaces and thin films of condensed matter.

As a consequence of this choice we have transformed the response
function into Fourier space in two spatial coordinates, keeping the
real-space coordinate in the third dimension. From there, the emphasis
has been laid on phase conjugation, although the more general DFWM
response tensor has been carried out in this mixed Fourier space as
well. The emphasis on phase conjugation was realized by the choice of
letting two of the interacting fields be spatially counterpropagating.
Letting the two counterpropagating fields act as pump fields in the
phase conjugation process, the third of the interacting fields became
what we have referred to as the probe field.

The choice of a scattering geometry in which the pump fields were
taken to be undamped plane waves traveling parallel to the
translationally invariant plane resulted in a description where the
main effort could be concentrated on studying the response due to the
probe field, thus letting the pump fields effectively being a part of
the phase conjugator. We concluded that using different combinations
of light polarized in the scattering plane or perpendicular to this
plane lead to different properties of the phase conjugated field
compared to the incoming probe field, including changes in
polarization in some cases.

Using the developed model on a single-level metallic quantum-well
phase conjugator we have shown that the phase-conjugation reflection
coefficient behaves quite differently from the uniform reflection
coefficient that has often been assumed in previous studies where
evanescent components have been included
\cite{Agarwal:95:1,Keller:92:1}. The response in the high end of the
$q_{\|}$-spectrum turned out to be as much as ten orders of magnitude
larger than in the propagating regime.  Subsequently, it was shown
that by use of the single-level phase conjugator it was possible to
phase conjugate light emitted from a subwavelength source in the
vicinity of the phase conjugator, and that the phase conjugated light
at the plane parallel to the phase conjugator, where the source has
been placed, has a subwavelength distance between the minima in the
intensity.  Consistent with another recent prediction
\cite{Bozhevolnyi:97:1,Bozhevolnyi:97:2} we have observed that the
smallest distance between the two minima surrounding the main lobe in
the phase conjugated field occur at the surface of the phase
conjugator.

The theoretical model was concluded with a description of a quantum
well with an arbitrary number of bound states, followed by a numerical
calculation of the response from a two-level quantum well.  We have
shown that also the two-level quantum well does not come close to an
ideal phase conjugator with a uniform reflection coefficient in the
$q_{\|}$-spectrum. Furthermore, it does not behave the same way as
the single-level quantum well, even though the combinations of
polarization for the interacting fields that gives a $p$-polarized
response lead to similar results in the high end of the
$q_{\|}$-spectrum. In the low end of the $q_{\|}$-spectrum the phase
conjugated response from a two-level quantum well is several orders
of magnitude stronger in a small number of very narrow ranges in
$q_{\|}$ than in the rest.

Finally, we concluded that if one is able to excite the two-level
quantum well in the $q_{\|}$-range around the point of the Fermi wave
number the present model could prove insufficient, because the maximum
value of the phase conjugation reflection coefficient in the high end
of the $q_{\|}$-spectrum in several cases is above the Fermi wave
number.

\section*{Discussion and outlook}
With respect to the single-level quantum well several properties would
be interesting to examine from a fundamental point of view, including
(i) the response in the far-infrared and ultraviolet parts of the
frequency spectrum, (ii) how the response can be divided into an
electrostatic and an electrodynamic part, (iii) how the width of the
phase conjugated focus from a quantum wire scales with the wavelength
of the electromagnetic field used in the interaction, and (iv) the
problem of a three-dimensional source.

For the two-level quantum well plenty of work remains to be done
before it would be wise to take up some of the above-mentioned
properties. First of all, we have to determine how much each of the
terms in the nonlinear conductivity tensor contributes to the phase
conjugated response. Also, the problem of phase conjugating a broad
angular band should be addressed in order to study, for instance,
focusing of light in front of a two-level phase conjugator.

The problem of focusing has to be addressed more carefully, since the
present study has revealed only that when the mesoscopic source is
moved closer to the phase conjugator the focus is narrowed. I imagine
that this problem could be addressed properly using a pure engineering
approach to make an adjustment of the present model by (i) abandoning
the infinite barrier model by insertion of a more sophisticated
potential across the barrier, and (ii) abandoning the point-source
description of the probe field.  Thereby one would also be able to
discuss the problem of resolution in a near-field optical microscope.

Furthermore, it could be interesting to establish a model which
provides a temporal resolution. It could be used to study, for example,
the time delay and distortion of an electromagnetic pulse (wave
packet) being phase conjugated, initially using, for example, plane
waves as pumps, and ultimately to give an understanding of four-wave
mixing using pulsed interacting fields. Such a model would provide a
framework for a description of time-resolved optical phase conjugation
in, for example, communications systems.

On the more sophisticated fundamental side it could be interesting to
investigate the phase conjugated response when the phase conjugator
is, for example, a mesoscopic ring, cylinder, sphere, or a quantum
wire. Another possibility is to take into account spin effects in
order to treat the phase conjugation response from magnetic materials.
We also believe that there is a connection between the model for
electromagnetic phase conjugation presented in this work and phase
conjugation of electrons and atoms [for an introduction to atomic
phase conjugation and nonlinear atom optics, see \citeANP{Lenz:93:1}
\citeyear{Lenz:93:1,Lenz:94:1} and \citeN{Goldstein:95:1}].

A problem that has to be taken into account when using the present
formulation to describe nonlinear optical processes is the apparantly
divergent behaviour in the long-wavelength (low-frequency) limit
stemming (in third-order problems) from the $\omega^{-3}$-term in the
beginning of the nonlinear conductivity tensor. This problem is a
general one in the theoretical model, and although the linear problem
has been solved \cite{Keller:96:1}, the problem has still not been
solved for any nonlinear case, including DFWM.

If time permits, I shall substantiate on some of these points in future
work. Otherwise, it will be left for others to do.

\vfill
\hbox to\hsize{\hfil We are very lucky to live in an age in which we are still making discoveries.}
\hbox to\hsize{\hfil {\it Richard P. Feynman}\/ in {\it The Character of
    Physical Law, p. 172}\/.}

\part*{Appendices}
\renewcommand{\partmark}[1]{\markboth{#1}{}}
\renewcommand{\chaptermark}[1]{\markright{Appendix~\thechapter : #1}}
\appendix
\chead[\fancyplain{}{\leftmark}]{\fancyplain{}{\rightmark}}
\newpage
\thispagestyle{plain}
\newpage

\chapter[Conductivity tensors in the
$(z;\vec{q}_{\|})$-space]{Calculation of linear and nonlinear
  conductivity tensors in two-dimensionally translational invariant
  systems [$(z;\vec{q}_{\|})$-space]}\label{ch:ConCalc}
In this appendix I present a calculation of linear and nonlinear
conductivity tensors suitable for calculation of the linear and
nonlinear current densities in a physical system with translational
invariance only in the $x$ and $y$ directions of the Cartesian
$x$-$y$-$z$-coordinate system. The basic ingredients in this
calculation consists of (i) the Fourier integral representation of the
vector potential in the $x$- and $y$-coordinates given by
Eq.~(\ref{eq:Azr}), (ii) the inverse relation for the current
densities linear in the cyclic frequency $\omega$ (appeared as
Eq.~(\ref{eq:Jzq})), (iii) the basis set of the wave functions taken
on the form of Eq.~(\ref{eq:eigenstate}), and (iv) the corresponding
transition current density in Eq.~(\ref{eq:Jm->n||}). Using these
ingredients we start from the three-dimensional expressions in real
space, which in the linear case are given by Eq.~(\ref{eq:Jr}) with
insertion of Eqs.~(\ref{eq:SigmaAArz})--(\ref{eq:SigmaABrz}) and in
the nonlinear case are given by Eq.~(\ref{eq:J3r}) with insertion of
Eqs.~(\ref{eq:XiAArz})--(\ref{eq:XiAGrz}). The results of these
calculations are presented as expressions for the individual matrix
elements according to the definitions given by Eq.~(\ref{eq:J1zq}) for
the linear conductivity tensor and Eq.~(\ref{eq:J3-wzq}) for the
nonlinear conductivity tensor, the cyclic transition frequencies being
expressed in the form of Eq.~(\ref{eq:transitionw}).

For convenience, we in the following treatment divide the linear
current density into two parts following the two processes shown in
Fig.~\ref{fig:2Tr}. If we define the linear current density as
$J_{i}(z;\vec{q}_{\|})=\sum_jJ_{ij}(z;\vec{q}_{\|})$, the result of
calculating the linear conductivity tensor is presented as the
individual nonzero matrix elements corresponding to the symmetry
analysis presented in Chapter~\ref{ch:6}. Like in the linear case, it
is convenient in the nonlinear case to define the nonlinear current
density as $J_{i}(\vec{r}\,)\equiv\sum_{jkh}J_{ijkh}(\vec{r}\,)$, with
$J_{ijkh}(\vec{r}\,)=\Xi_{ijkh}(\vec{r},\vec{r}\,',\vec{r}\,'',\vec{r}\,''')
A_h(\vec{r}\,''')A_k(\vec{r}\,'')A_j(\vec{r}\,')$, and then split the
treatment of the nonlinear current density in such a way that each of
the processes mentioned in Fig.~\ref{fig:3Tr} is treated separately.

\section{Linear process A}
From Eq.~(\ref{eq:SigmaAArz}) we have the $xx$ element of part A of
the linear current, in which we insert the expressions for the wave
function and the vector potential in the two-dimensional Fourier
representation, giving the result
\begin{eqnarray}
\lefteqn{
 J_{xx}^{\rm{A}}(z;\vec{q}_{\|})=
 -{e^2\over{}m_{e}}
 {2\over(2\pi)^4}
 \iiint
 \sum_{n}f_{n}|\psi_{n}(z)|^2\delta(\vec{r}-\vec{r}\,')
 A_{x}(z';\vec{q}_{\|}^{\,\prime})e^{{\rm{i}}\vec{q}_{\|}^{\,\prime}
 \cdot\vec{r}_{\|}^{\,\prime}}d^2q_{\|}'d^3r'
}\nonumber\\ &\quad&\times
 e^{-{\rm{i}}\vec{q}_{\|}\cdot\vec{r}_{\|}}d^2r_{\|}.
\end{eqnarray}
Solving the integrals over the real-space coordinates in the
$x$-$y$-directions, $\int{}d^2r_{\|}'$ and $\int{}d^2r_{\|}$, we
obtain
\begin{eqnarray}
\lefteqn{ J_{xx}^{\rm{A}}(z;\vec{q}_{\|})=
 -{e^2\over{}m_{e}}{2\over(2\pi)^2}
 \iiint
 \sum_{n}f_{n}(\vec{\kappa}_{\|,\bar{n}})
 |\psi_{n}(z)|^2\delta(z-z')
 \delta(\vec{q}_{\|}^{\,\prime}-\vec{q}_{\|})
}\nonumber\\ &\quad&\times
 A_{x}(z';\vec{q}_{\|}^{\,\prime})
 d^2\kappa_{\|,\bar{n}}
 d^2q_{\|}'dz',
\label{eq:J1xx-A}
\end{eqnarray}
where the (infinite) sum over the $\kappa_{\|}$ coordinates has been
replaced by an integral. From this expression we extract part A of
the linear conductivity tensor as
\begin{equation}
 {\sigma}_{xx}^{\,\rm{A}}(z,z';\vec{q}_{\|},\vec{q}_{\|}^{\,\prime})=
 -{2\over{}{\rm{i}}\omega}{e^2\over{}m_{e}}{1\over(2\pi)^2}
 \sum_{n}\int{}f_{n}(\vec{\kappa}_{\|})
 d^2\kappa_{\|}
 |\psi_{n}(z)|^2\delta(z-z')
 \delta(\vec{q}_{\|}^{\,\prime}-\vec{q}_{\|}),
\end{equation}
where we have omitted the now superfluous reference to $n$ from
$\vec{\kappa}_{\|}$. Taking into account the conservation of momentum
given by the Dirac delta function
$\delta(\vec{q}_{\|}^{\,\prime}-\vec{q}_{\|})$ we may integrate over
$q_{\|}'$ in Eq.~(\ref{eq:J1xx-A}), and thereafter extract part A of
the linear conductivity tensor as presented in Eq.~(\ref{eq:SigmaAA}).

\section{Linear process B}
Taking from Eq.~(\ref{eq:SigmaABrz}) element $ij$ of part B of the
linear current density and inserting the expressions for the wave
function [Eq.~(\ref{eq:eigenstate})] and the transition current
density [Eq.~(\ref{eq:j_hnm})] in the two-dimensional Fourier
representation, we get
\begin{eqnarray}
\lefteqn{
 J_{ij}^{\rm{B}}(z;\vec{q}_{\|})=
 -{1\over\hbar}{2\over(2\pi)^6}\iiint
 \sum_{nm}
 {f_{n}-f_{m}\over\tilde{\omega}_{nm}-\omega}
 j_{j,mn}(z';\vec{\kappa}_{\|,\bar{m}}+\vec{\kappa}_{\|,\bar{n}})
 j_{i,nm}(z;\vec{\kappa}_{\|,\bar{n}}+\vec{\kappa}_{\|,\bar{m}})
}\nonumber\\ &\quad&\times
 e^{{\rm{i}}(\vec{\kappa}_{\|,\bar{m}}-\vec{\kappa}_{\|,\bar{n}})
  \cdot\vec{r}_{\|}^{\,\prime}}
 e^{{\rm{i}}(\vec{\kappa}_{\|,\bar{n}}-\vec{\kappa}_{\|,\bar{m}})
  \cdot\vec{r}_{\|}}
 A_{j}(z';\vec{q}_{\|}^{\,\prime})
 e^{{\rm{i}}\vec{q}_{\|}^{\,\prime}\cdot\vec{r}_{\|}^{\,\prime}}
 d^2q_{\|}'d^3r'e^{-{\rm{i}}\vec{q}_{\|}\cdot\vec{r}_{\|}}d^2r_{\|}.
\end{eqnarray}
Solving the Cartesian integrals $\int{}d^2r_{\|}'$ and
$\int{}d^2r_{\|}$ along the surface and replacing the infinite sum
over $\vec{\kappa}_{\|}$ with an integral, this is
\begin{eqnarray}
\lefteqn{
 J_{ij}^{\rm{B}}(z;\vec{q}_{\|})=
 -{1\over\hbar}
 {2\over(2\pi)^2}
 \idotsint
 \sum_{nm}
 {f_{n}(\vec{\kappa}_{\|,\bar{n}})-f_{m}(\vec{\kappa}_{\|,\bar{m}})
 \over\tilde{\omega}_{nm}(\vec{\kappa}_{\|,\bar{n}},\vec{\kappa}_{\|,\bar{m}})-\omega}
 j_{j,mn}(z';\vec{\kappa}_{\|,\bar{m}}+\vec{\kappa}_{\|,\bar{n}})
}\nonumber\\ &\quad&\times
 j_{i,nm}(z;\vec{\kappa}_{\|,\bar{n}}+\vec{\kappa}_{\|,\bar{m}})
 \delta(\vec{\kappa}_{\|,\bar{m}}-\vec{\kappa}_{\|,\bar{n}}+\vec{q}_{\|}^{\,\prime})
 \delta(\vec{\kappa}_{\|,\bar{n}}-\vec{\kappa}_{\|,\bar{m}}-\vec{q}_{\|})
\nonumber\\ &&\times
 A_{j}(z';\vec{q}_{\|}^{\,\prime})
 d^2\kappa_{\|,\bar{n}}d^2\kappa_{\|,\bar{m}}
 d^2q_{\|}'dz'.
\end{eqnarray}
Of these two integrals over the surface states, we can solve one
because of the coupling between the surface states and the wavevectors
introduced by the Dirac delta functions appearing. Keeping the
$\vec{\kappa}_{\|,\bar{m}}$ set, and thus we solve for the $n$ set.
Solving for this set, we find that $\vec{\kappa}_{\|,\bar{n}}$ is
replaced by $\vec{\kappa}_{\|,\bar{m}}+\vec{q}_{\|}^{\,\prime}$, thus
giving
\begin{eqnarray}
\lefteqn{
 J_{ij}^{\rm{B}}(z;\vec{q}_{\|})=
 -{1\over\hbar}
 {2\over(2\pi)^2}
 \iiint
 \sum_{nm}
 {f_{n}(\vec{\kappa}_{\|,\bar{m}}+\vec{q}_{\|}^{\,\prime})-f_{m}(\vec{\kappa}_{\|,\bar{m}})
 \over\tilde{\omega}_{nm}(\vec{\kappa}_{\|,\bar{m}}+\vec{q}_{\|}^{\,\prime},
 \vec{\kappa}_{\|,\bar{m}})-\omega}
 j_{j,mn}(z';2\vec{\kappa}_{\|,\bar{m}}+\vec{q}_{\|}^{\,\prime})
}\nonumber\\ &\quad&\times
 j_{i,nm}(z;2\vec{\kappa}_{\|,\bar{m}}+\vec{q}_{\|}^{\,\prime})
 \delta(\vec{q}_{\|}^{\,\prime}-\vec{q}_{\|})
 A_{j}(z';\vec{q}_{\|}^{\,\prime})
 d^2\kappa_{\|,\bar{m}}
 d^2q_{\|}'dz'.
\label{eq:J1ij-B}
\end{eqnarray}
From the above expression we extract part B of the linear
conductivity tensor as
\begin{eqnarray}
\lefteqn{
 {\sigma}_{ij}^{\,\rm{B}}(z,z';\vec{q}_{\|},\vec{q}_{\|}^{\,\prime})=
 -{2\over{}{\rm{i}}\omega}{1\over\hbar}{1\over(2\pi)^2}\sum_{nm}\int
 {f_{n}(\vec{\kappa}_{\|}+\vec{q}_{\|}^{\,\prime})-f_{m}(\vec{\kappa}_{\|})
 \over\tilde{\omega}_{nm}(\vec{\kappa}_{\|}+\vec{q}_{\|}^{\,\prime},
 \vec{\kappa}_{\|})-\omega}
 j_{j,mn}(z';2\vec{\kappa}_{\|}+\vec{q}_{\|}^{\,\prime})
}\nonumber\\ &\quad&\times
 j_{i,nm}(z;2\vec{\kappa}_{\|}+\vec{q}_{\|}^{\,\prime})
 \delta(\vec{q}_{\|}^{\,\prime}-\vec{q}_{\|})
 d^2\kappa_{\|},
\end{eqnarray}
where we have omitted the now superfluous index $m$ from the surface
states $\vec{\kappa}_{\|}$. Again we take into account the
conservation of pseudo-momentum,
$\delta(\vec{q}_{\|}^{\,\prime}-\vec{q}_{\|})$, letting us perform the
integration over $q_{\|}'$ in Eq.~(\ref{eq:J1ij-B}). From this result
the linear conductivity tensor part B is extracted on the form shown
in Eq.~(\ref{eq:SigmaAB}).

\section{Nonlinear process A}
Inserting Eq.~(\ref{eq:XiAArz}) into Eq.~(\ref{eq:J3r}), we take
element $xxxx$ of part A of the nonlinear current density. In the
result we insert the expressions for the wave function and the vector
potential in the two-dimensional Fourier representation
[Eqs.~(\ref{eq:eigenstate}) and (\ref{eq:Azr}), respectively]. Then by
use of Eq.~(\ref{eq:Jzq}) we find
\begin{eqnarray}
\lefteqn{
 J_{xxxx}^{\rm{A}}(z;\vec{q}_{\|})=
 -{e^4\over8m_{e}^2\hbar}
 {2\over(2\pi)^{10}}\idotsint
 \sum_{nm}
 {f_{n}-f_{m}\over\tilde{\omega}_{nm}-2\omega}
 \psi_{n}^{*}(z'')\psi_{m}(z'')\psi_{m}^{*}(z)\psi_{n}(z)
}\nonumber\\ &\quad&\times
 A_{x}(z''';\vec{q}_{\|}^{\,\prime\prime\prime})
 A_{x}(z'';\vec{q}_{\|}^{\,\prime\prime})
 A_{x}^{*}(z';\vec{q}_{\|}^{\,\prime})
 e^{{\rm{i}}(\vec{\kappa}_{\|,\bar{m}}-\vec{\kappa}_{\|,\bar{n}})\cdot\vec{r}_{\|}^{\,\prime\prime}}
 e^{{\rm{i}}(\vec{\kappa}_{\|,\bar{n}}-\vec{\kappa}_{\|,\bar{m}})\cdot\vec{r}_{\|}}
 e^{{\rm{i}}\vec{q}_{\|}^{\,\prime\prime\prime}\cdot\vec{r}_{\|}^{\,\prime\prime\prime}}
 e^{{\rm{i}}\vec{q}_{\|}^{\,\prime\prime}\cdot\vec{r}_{\|}^{\,\prime\prime}}
 e^{-{\rm{i}}\vec{q}_{\|}^{\,\prime}\cdot\vec{r}_{\|}^{\,\prime}}
\nonumber\\ &&\times
 \delta(\vec{r}-\vec{r}\,')
 \delta(\vec{r}\,''-\vec{r}\,''')
 d^2q_{\|}'''d^2q_{\|}''d^2q_{\|}'
 d^3r'''d^3r''d^3r'e^{-{\rm{i}}\vec{q}_{\|}\cdot\vec{r}_{\|}}d^2r_{\|}.
\end{eqnarray}
In this equation, we first solve the integrals $\int{}d^2r_{\|}'$ and
$\int{}d^2r_{\|}'''$, thereafter the remaining Cartesian integrals
$\int{}d^2r_{\|}''$ and $\int{}d^2r_{\|}$, and finally replace the
infinite sums over the different $\vec{\kappa}_{\|}$ coordinates with
integrals, thereby obtaining
\begin{eqnarray}
\lefteqn{
 J_{xxxx}^{\rm{A}}(z;\vec{q}_{\|})=
 -{e^4\over8m_{e}^2\hbar}
 {2\over(2\pi)^6}
 \idotsint
 \sum_{nm}
 {f_{n}(\vec{\kappa}_{\|,\bar{n}})-f_{m}(\vec{\kappa}_{\|,\bar{m}})
 \over\tilde{\omega}_{nm}(\vec{\kappa}_{\|,\bar{n}},\vec{\kappa}_{\|,\bar{m}})-2\omega}
 \psi_{n}^{*}(z'')\psi_{m}(z'')\psi_{m}^{*}(z)\psi_{n}(z)
}\nonumber\\ &\quad&\times
 A_{x}(z''';\vec{q}_{\|}^{\,\prime\prime\prime})
 A_{x}(z'';\vec{q}_{\|}^{\,\prime\prime})
 A_{x}^{*}(z';\vec{q}_{\|}^{\,\prime})
 \delta(\vec{\kappa}_{\|,\bar{m}}-\vec{\kappa}_{\|,\bar{n}}+\vec{q}_{\|}^{\,\prime\prime\prime}
 +\vec{q}_{\|}^{\,\prime\prime})
 \delta(\vec{\kappa}_{\|,\bar{n}}-\vec{\kappa}_{\|,\bar{m}}-\vec{q}_{\|}^{\,\prime}-\vec{q}_{\|})
\nonumber\\ &&\times
 \delta(z-z')
 \delta(z''-z''')
 d^2\kappa_{\|,\bar{n}}d^2\kappa_{\|,\bar{m}}
 d^2q_{\|}'''d^2q_{\|}''d^2q_{\|}'
 dz'''dz''dz'.
\end{eqnarray}
Of the two integrals over the $\kappa_{\|}$ quantities, we can solve
one because of the coupling of these to the wavevectors introduced by
the Dirac delta functions appearing. Keeping the
$\vec{\kappa}_{\|,\bar{m}}$ set we thus solve the integrals for the
$n$ set. Solving for this set, we find that
$\vec{\kappa}_{\|,\bar{n}}$ is replaced by
$\vec{\kappa}_{\|,\bar{m}}+\vec{q}_{\|}^{\,\prime}+\vec{q}_{\|}$,
which gives
\begin{eqnarray}
\lefteqn{
 J_{xxxx}^{\rm{A}}(z;\vec{q}_{\|})=
 -{e^4\over8m_{e}^2\hbar}
 {2\over(2\pi)^6}
 \idotsint
 \sum_{nm}
 {f_{n}(\vec{\kappa}_{\|,\bar{m}}+\vec{q}_{\|}^{\,\prime}+\vec{q}_{\|})
 -f_{m}(\vec{\kappa}_{\|,\bar{m}})
 \over\tilde{\omega}_{nm}(\vec{\kappa}_{\|,\bar{m}}+\vec{q}_{\|}^{\,\prime}+\vec{q}_{\|},
 \vec{\kappa}_{\|,\bar{m}})-2\omega}
 \psi_{n}^{*}(z'')\psi_{m}(z'')
}\nonumber\\ &\quad&\times
 \psi_{m}^{*}(z)\psi_{n}(z)
 A_{x}(z''';\vec{q}_{\|}^{\,\prime\prime\prime})
 A_{x}(z'';\vec{q}_{\|}^{\,\prime\prime})
 A_{x}^{*}(z';\vec{q}_{\|}^{\,\prime})
 \delta(\vec{q}_{\|}^{\,\prime\prime\prime}+\vec{q}_{\|}^{\,\prime\prime}-\vec{q}_{\|}^{\,\prime}-\vec{q}_{\|})
 \delta(z-z')
\nonumber\\ &&\times
 \delta(z''-z''')
 d^2\kappa_{\|,\bar{m}}
 d^2q_{\|}'''d^2q_{\|}''d^2q_{\|}'
 dz'''dz''dz'.
\label{eq:J3A}
\end{eqnarray}
From this we may extract part A of the nonlinear conductivity tensor
as defined in Eq.~(\ref{eq:J3zq}) as
\begin{eqnarray}
\lefteqn{
{\Xi}_{xxxx}^{\rm{A}}(z,z',z'',z''';\vec{q}_{\|},\vec{q}_{\|}^{\,\prime},
 \vec{q}_{\|}^{\,\prime\prime},\vec{q}_{\|}^{\,\prime\prime\prime})=
 {2{\rm{i}}\over\omega^3}{e^4\over8m_{e}^2\hbar}{1\over(2\pi)^2}
 \sum_{nm}\int
 {f_{n}(\vec{\kappa}_{\|}+\vec{q}_{\|}^{\,\prime}+\vec{q}_{\|})
 -f_{m}(\vec{\kappa}_{\|})
 \over\tilde{\omega}_{nm}(\vec{\kappa}_{\|}+\vec{q}_{\|}^{\,\prime}+\vec{q}_{\|},
 \vec{\kappa}_{\|})-2\omega}
}\nonumber\\ &\quad&\times
 \psi_{n}^{*}(z'')\psi_{m}(z'')\psi_{m}^{*}(z)\psi_{n}(z)
 \delta(\vec{q}_{\|}^{\,\prime\prime\prime}+\vec{q}_{\|}^{\,\prime\prime}-\vec{q}_{\|}^{\,\prime}-\vec{q}_{\|})
 \delta(z-z')\delta(z''-z''')
 d^2\kappa_{\|},
\end{eqnarray}
where we have omitted the now superfluous index $m$ from the surface
states $\vec{\kappa}_{\|}$. Taking into account the fact that we look
for the phase conjugation response we restrict ourselves to the case
where the pump fields are counterpropagating, thus taking
\begin{eqnarray}
 \vec{A}(z''';\vec{q}_{\|}^{\,\prime\prime\prime})&\equiv&\vec{A}(z''';-\vec{k}_{\|})
 \delta(\vec{q}_{\|}^{\,\prime\prime\prime}+\vec{k}_{\|}),
\label{eq:Apump-1}\\
 \vec{A}(z'';\vec{q}_{\|}^{\,\prime\prime})&\equiv&\vec{A}(z'';\vec{k}_{\|})
 \delta(\vec{q}_{\|}^{\,\prime\prime}-\vec{k}_{\|}),
\label{eq:Apump-2}
\end{eqnarray}
where $\vec{k}_{\|}$ is the common wavevector for the two pump
fields. With these substitutions we can perform the integrals over
$q_{\|}'''$ and $q_{\|}''$ in Eq.~(\ref{eq:J3A}), and the conservation
of pseudo-momentum is reduced from its general degenerate four-wave
mixing form,
$\delta(\vec{q}_{\|}^{\,\prime\prime\prime}+\vec{q}_{\|}^{\,\prime\prime}-\vec{q}_{\|}^{\,\prime}-\vec{q}_{\|})$,
to $\delta(\vec{q}_{\|}^{\,\prime}+\vec{q}_{\|})$. This allows us also
to solve the integral over $q_{\|}'$ in Eq.~(\ref{eq:J3A}), and on the
form of Eq.~(\ref{eq:J3-wzq}) we can extract the PCDFWM conductivity
tensor part A, appearing as Eq.~(\ref{eq:XiAA}).

\section{Nonlinear process B}
Inserting Eq.~(\ref{eq:XiABrz}) into Eq.~(\ref{eq:J3r}), we take
element $xxkh$ of part B of the nonlinear current density. In the
result we insert the expressions for the wave function, the vector
potential and the transition current density in the two-dimensional
Fourier representation [Eqs.~(\ref{eq:eigenstate}), (\ref{eq:Azr}) and
(\ref{eq:Jm->n||}), respectively]. Then by use of Eq.~(\ref{eq:Jzq})
we find
\begin{eqnarray}
\lefteqn{
 J_{xxkh}^{\rm{B}}(z;\vec{q}_{\|})=
 -{e^2\over4m_{e}\hbar^2}
 {2\over(2\pi)^{12}}
 \idotsint
 \sum_{nmv}
 {1\over\tilde{\omega}_{nm}-2\omega}
 \left({f_{m}-f_{v}\over\tilde{\omega}_{v{}m}-\omega}
 +{f_{n}-f_{v}\over\tilde{\omega}_{nv}-\omega}\right)
}\nonumber\\ &\quad&\times
 j_{h,v{}n}(z''';\vec{\kappa}_{\|,\bar{v}}+\vec{\kappa}_{\|,\bar{n}})
 j_{k,mv}(z'';\vec{\kappa}_{\|,\bar{m}}+\vec{\kappa}_{\|,\bar{v}})
 \psi_{m}^{*}(z)
 \psi_{n}(z)
\nonumber\\ &&\times
 A_{h}(z''';\vec{q}_{\|}^{\,\prime\prime\prime})
 A_{k}(z'';\vec{q}_{\|}^{\,\prime\prime})
 A_{x}^{*}(z';\vec{q}_{\|}^{\,\prime})
 e^{{\rm{i}}(\vec{\kappa}_{\|,\bar{n}}-\vec{\kappa}_{\|,\bar{m}})\cdot\vec{r}_{\|}}
 e^{{\rm{i}}(\vec{\kappa}_{\|,\bar{v}}-\vec{\kappa}_{\|,\bar{n}})\cdot\vec{r}_{\|}^{\,\prime\prime\prime}}
 e^{{\rm{i}}(\vec{\kappa}_{\|,\bar{m}}-\vec{\kappa}_{\|,\bar{v}})\cdot\vec{r}_{\|}^{\,\prime\prime}}
\nonumber\\ &&\times
 e^{{\rm{i}}\vec{q}_{\|}^{\,\prime\prime\prime}\cdot\vec{r}_{\|}^{\,\prime\prime\prime}}
 e^{{\rm{i}}\vec{q}_{\|}^{\,\prime\prime}\cdot\vec{r}_{\|}^{\,\prime\prime}}
 e^{-{\rm{i}}\vec{q}_{\|}^{\,\prime}\cdot\vec{r}_{\|}^{\,\prime}}
 d^2q_{\|}'''d^2q_{\|}''d^2q_{\|}'
 \delta(\vec{r}-\vec{r}\,')
 d^3r'''d^3r''d^3r'e^{-{\rm{i}}\vec{q}_{\|}\cdot\vec{r}_{\|}}d^2r_{\|}.
\end{eqnarray}
Solving first the integral $\int{}d^2r_{\|}'$, then the integrals
$\int{}d^2r_{\|}'''$, $\int{}d^2r_{\|}''$, and $\int{}d^2r_{\|}$, and
finally replacing the sums over the $\vec{\kappa}_{\|}$ quantities
with integrals, we obtain
\begin{eqnarray}
\lefteqn{
 J_{xxkh}^{\rm{B}}(z;\vec{q}_{\|})=
 -{e^2\over4m_{e}\hbar^2}
 {2\over(2\pi)^6}
 \idotsint
 \sum_{nmv}
 {1\over\tilde{\omega}_{nm}(\vec{\kappa}_{\|,\bar{n}},\vec{\kappa}_{\|,\bar{m}})-2\omega}
}\nonumber\\ &\quad&\times
 \left({f_{m}(\vec{\kappa}_{\|,\bar{m}})-f_{v}(\vec{\kappa}_{\|,\bar{v}})\over
 \tilde{\omega}_{v{}m}(\vec{\kappa}_{\|,\bar{v}},\vec{\kappa}_{\|,\bar{m}})-\omega}
 +{f_{n}(\vec{\kappa}_{\|,\bar{n}})-f_{v}(\vec{\kappa}_{\|,\bar{v}})\over
 \tilde{\omega}_{nv}(\vec{\kappa}_{\|,\bar{n}},\vec{\kappa}_{\|,\bar{v}})-\omega}
 \right)
 j_{h,v{}n}(z''';\vec{\kappa}_{\|,\bar{v}}+\vec{\kappa}_{\|,\bar{n}})
\nonumber\\ &&\times
 j_{k,mv}(z'';\vec{\kappa}_{\|,\bar{m}}+\vec{\kappa}_{\|,\bar{v}})
 \psi_{m}^{*}(z)
 \psi_{n}(z)
 A_{h}(z''';\vec{q}_{\|}^{\,\prime\prime\prime})
 A_{k}(z'';\vec{q}_{\|}^{\,\prime\prime})
 A_{x}^{*}(z';\vec{q}_{\|}^{\,\prime})
\nonumber\\ &&\times
 \delta(\vec{\kappa}_{\|,\bar{v}}-\vec{\kappa}_{\|,\bar{n}}+\vec{q}_{\|}^{\,\prime\prime\prime})
 \delta(\vec{\kappa}_{\|,\bar{m}}-\vec{\kappa}_{\|,\bar{v}}+\vec{q}_{\|}^{\,\prime\prime})
 \delta(\vec{\kappa}_{\|,\bar{n}}-\vec{\kappa}_{\|,\bar{m}}-\vec{q}_{\|}^{\,\prime}-\vec{q}_{\|})
\nonumber\\ &&\times
 \delta(z-z')
 d^2\kappa_{\|,\bar{n}}d^2\kappa_{\|,\bar{m}}d^2\kappa_{\|,\bar{v}}
 d^2q_{\|}'''d^2q_{\|}''d^2q_{\|}'
 dz'''dz''dz'.
\end{eqnarray}
Of these three integrals over $\vec{\kappa}_{\|}$, we can solve two
because of the coupling to the wave\-vectors introduced by the Dirac
delta functions appearing. Keeping the $\vec{\kappa}_{\|,\bar{m}}$ set
of surface states, we thus solve the integrals for the $v$ and $n$
sets (in that order). Solving for the $v$ set, we find that
$\vec{\kappa}_{\|,\bar{v}}$ is replaced by
$\vec{\kappa}_{\|,\bar{m}}+\vec{q}_{\|}^{\,\prime\prime}$, which then
allows us to solve the $n$ set by replacing
$\vec{\kappa}_{\|,\bar{n}}$ by
$\vec{\kappa}_{\|,\bar{m}}+\vec{q}_{\|}^{\,\prime\prime\prime}+\vec{q}_{\|}^{\,\prime\prime}$.
Then we get
\begin{eqnarray}
\lefteqn{
 J_{xxkh}^{\rm{B}}(z;\vec{q}_{\|})=
 -{e^2\over4m_{e}\hbar^2}
 {2\over(2\pi)^6}
 \idotsint
 \sum_{nmv}
 {1\over\tilde{\omega}_{nm}(\vec{\kappa}_{\|,\bar{m}}+\vec{q}_{\|}^{\,\prime\prime\prime}+\vec{q}_{\|}^{\,\prime\prime},
 \vec{\kappa}_{\|,\bar{m}})-2\omega}
}\nonumber\\ &\quad&\times
 \left({f_{m}(\vec{\kappa}_{\|,\bar{m}})-f_{v}(\vec{\kappa}_{\|,\bar{m}}+\vec{q}_{\|}^{\,\prime\prime})
 \over\tilde{\omega}_{v{}m}(\vec{\kappa}_{\|,\bar{m}}+\vec{q}_{\|}^{\,\prime\prime},
 \vec{\kappa}_{\|,\bar{m}})-\omega}
 +{f_{n}(\vec{\kappa}_{\|,\bar{m}}+\vec{q}_{\|}^{\,\prime\prime\prime}+\vec{q}_{\|}^{\,\prime\prime})
 -f_{v}(\vec{\kappa}_{\|,\bar{m}}+\vec{q}_{\|}^{\,\prime\prime})\over
 \tilde{\omega}_{nv}(\vec{\kappa}_{\|,\bar{m}}+\vec{q}_{\|}^{\,\prime\prime\prime}+\vec{q}_{\|}^{\,\prime\prime},
 \vec{\kappa}_{\|,\bar{m}}+\vec{q}_{\|}^{\,\prime\prime})-\omega}\right)
\nonumber\\ &&\times
 j_{h,v{}n}(z''';2\vec{\kappa}_{\|,\bar{m}}+\vec{q}_{\|}^{\,\prime\prime\prime}+2\vec{q}_{\|}^{\,\prime\prime})
 j_{k,mv}(z'';2\vec{\kappa}_{\|,\bar{m}}+\vec{q}_{\|}^{\,\prime\prime})
 \psi_{m}^{*}(z)\psi_{n}(z)
\nonumber\\ &&\times
 A_{h}(z''';\vec{q}_{\|}^{\,\prime\prime\prime})
 A_{k}(z'';\vec{q}_{\|}^{\,\prime\prime})
 A_{x}^{*}(z';\vec{q}_{\|}^{\,\prime})
 \delta(\vec{q}_{\|}^{\,\prime\prime\prime}+\vec{q}_{\|}^{\,\prime\prime}-\vec{q}_{\|}^{\,\prime}-\vec{q}_{\|})
\nonumber\\ &&\times
 \delta(z-z')d^2\kappa_{\|,\bar{m}}
 d^2q_{\|}'''d^2q_{\|}''d^2q_{\|}'dz'''dz''dz'.
\label{eq:J3B}
\end{eqnarray}
On the form of Eq.~(\ref{eq:J3zq}) we thus get part B of the
conductivity tensor as
\begin{eqnarray}
\lefteqn{
 {\Xi}_{xxkh}^{\rm{B}}(z,z',z'',z''';\vec{q}_{\|},\vec{q}_{\|}^{\,\prime},
 \vec{q}_{\|}^{\,\prime\prime},\vec{q}_{\|}^{\,\prime\prime\prime})=
}\nonumber\\ &\quad&
 {2{\rm{i}}\over{}\omega^3}
 {1\over(2\pi)^2}{e^2\over4m_{e}\hbar^2}
 \sum_{nmv}\int
 {1\over\tilde{\omega}_{nm}(\vec{\kappa}_{\|}+\vec{q}_{\|}^{\,\prime\prime\prime}+\vec{q}_{\|}^{\,\prime\prime},
 \vec{\kappa}_{\|})-2\omega}
 \left({f_{m}(\vec{\kappa}_{\|})-f_{v}(\vec{\kappa}_{\|}+\vec{q}_{\|}^{\,\prime\prime})\over
 \tilde{\omega}_{v{}m}(\vec{\kappa}_{\|}+\vec{q}_{\|}^{\,\prime\prime},\vec{\kappa}_{\|})
 -\omega}
\right.\nonumber\\ &&\left.\!
 +{f_{n}(\vec{\kappa}_{\|}+\vec{q}_{\|}^{\,\prime\prime\prime}+\vec{q}_{\|}^{\,\prime\prime})
 -f_{v}(\vec{\kappa}_{\|}+\vec{q}_{\|}^{\,\prime\prime})\over
 \tilde{\omega}_{nv}(\vec{\kappa}_{\|}+\vec{q}_{\|}^{\,\prime\prime\prime}+\vec{q}_{\|}^{\,\prime\prime},
 \vec{\kappa}_{\|}+\vec{q}_{\|}^{\,\prime\prime})-\omega}\right)
 j_{h,v{}n}(z''';2\vec{\kappa}_{\|}+\vec{q}_{\|}^{\,\prime\prime\prime}+2\vec{q}_{\|}^{\,\prime\prime})
\nonumber\\ &&\times
 j_{k,mv}(z'';2\vec{\kappa}_{\|}+\vec{q}_{\|}^{\,\prime\prime})
 \psi_{m}^{*}(z)\psi_{n}(z)
 \delta(\vec{q}_{\|}^{\,\prime\prime\prime}+\vec{q}_{\|}^{\,\prime\prime}-\vec{q}_{\|}^{\,\prime}-\vec{q}_{\|})
 \delta(z-z')d^2\kappa_{\|},
\end{eqnarray}
where we have omitted the now superfluous index $m$ from the surface
states $\vec{\kappa}_{\|}$. Looking for the phase conjugation response
the pump fields take the form of
Eqs.~(\ref{eq:Apump-1})--(\ref{eq:Apump-2}), and integration over
$q_{\|}'''$ and $q_{\|}''$ in Eq.~(\ref{eq:J3B}) can be performed.
Thereby the Dirac delta function accounting for conservation of
pseudo-momentum is reduced from its general DFWM form,
$\delta(\vec{q}_{\|}^{\,\prime\prime\prime}+\vec{q}_{\|}^{\,\prime\prime}-\vec{q}_{\|}^{\,\prime}-\vec{q}_{\|})$,
to $\delta(\vec{q}_{\|}^{\,\prime}+\vec{q}_{\|})$. Thus performing
also the integral over $q_{\|}'$, the PCDFWM conductivity tensor on
the form of Eq.~(\ref{eq:J3-wzq}) can be extracted, and
Eq.~(\ref{eq:XiAB}) appear.

\section{Nonlinear process C}
Inserting Eq.~(\ref{eq:XiACrz}) into Eq.~(\ref{eq:J3r}), we take
element $xxxx$ of part C of the nonlinear current density. In the
result we insert the expressions for the wave function and the vector
potential in the two-dimensional Fourier representation
[Eqs.~(\ref{eq:eigenstate}) and (\ref{eq:Azr}), respectively]. Then by
use of Eq.~(\ref{eq:Jzq}) we find
\begin{eqnarray}
\lefteqn{
 J_{xxxx}^{\rm{C}}(z;\vec{q}_{\|})=
 -{e^4\over4m_{e}^2\hbar}
 {2\over(2\pi)^{10}}
 \idotsint
 \sum_{nm}
 {f_{n}-f_{m}\over\tilde{\omega}_{nm}}
 \psi_{n}^{*}(z')\psi_{m}(z')
 \psi_{m}^{*}(z)\psi_{n}(z)
 A_{x}(z''';\vec{q}_{\|}^{\,\prime\prime\prime})
}\nonumber\\ &\quad&\times
 A_{x}(z'';\vec{q}_{\|}^{\,\prime\prime})
 A_{x}^{*}(z';\vec{q}_{\|}^{\,\prime})
 e^{{\rm{i}}(\vec{\kappa}_{\|,\bar{m}}-\vec{\kappa}_{\|,\bar{n}})\cdot\vec{r}_{\|}^{\,\prime}}
 e^{{\rm{i}}(\vec{\kappa}_{\|,\bar{n}}-\vec{\kappa}_{\|,\bar{m}})\cdot\vec{r}_{\|}}
 e^{{\rm{i}}\vec{q}_{\|}^{\,\prime\prime\prime}\cdot\vec{r}_{\|}^{\,\prime\prime\prime}}
 e^{{\rm{i}}\vec{q}_{\|}^{\,\prime\prime}\cdot\vec{r}_{\|}^{\,\prime\prime}}
 e^{-{\rm{i}}\vec{q}_{\|}^{\,\prime}\cdot\vec{r}_{\|}^{\,\prime}}
 \delta(\vec{r}\,'-\vec{r}\,''')
\nonumber\\ &&\times
 \delta(\vec{r}-\vec{r}\,'')
 d^2q_{\|}'''d^2q_{\|}''d^2q_{\|}'
 d^3r'''d^3r''d^3r'e^{-{\rm{i}}\vec{q}_{\|}\cdot\vec{r}_{\|}}d^2r_{\|}.
\end{eqnarray}
Solving first the integrals $\int{}d^2r_{\|}'''$ and
$\int{}d^2r_{\|}''$, then the integrals $\int{}d^2r_{\|}'$
and $\int{}d^2r_{\|}$, and finally replacing the sum over the
$\vec{\kappa}_{\|}$ quantities with integrals, we get
\begin{eqnarray}
\lefteqn{
 J_{xxxx}^{\rm{C}}(z;\vec{q}_{\|})=
 -{e^4\over4m_{e}^2\hbar}
 {2\over(2\pi)^6}
 \idotsint
 \sum_{nm}
 {f_{n}(\vec{\kappa}_{\|,\bar{n}})-f_{m}(\vec{\kappa}_{\|,\bar{m}})\over
 \tilde{\omega}_{nm}(\vec{\kappa}_{\|,\bar{n}},\vec{\kappa}_{\|,\bar{m}})}
 \psi_{n}^{*}(z')\psi_{m}(z')
 \psi_{m}^{*}(z)\psi_{n}(z)
}\nonumber\\ &\quad&\times
 A_{x}(z''';\vec{q}_{\|}^{\,\prime\prime\prime})
 A_{x}(z'';\vec{q}_{\|}^{\,\prime\prime})
 A_{x}^{*}(z';\vec{q}_{\|}^{\,\prime})
 \delta(\vec{\kappa}_{\|,\bar{m}}-\vec{\kappa}_{\|,\bar{n}}+\vec{q}_{\|}^{\,\prime\prime\prime}-\vec{q}_{\|}^{\,\prime})
 \delta(\vec{\kappa}_{\|,\bar{n}}-\vec{\kappa}_{\|,\bar{m}}+\vec{q}_{\|}^{\,\prime\prime}-\vec{q}_{\|})
\nonumber\\ &&\times
 \delta(z'-z''')
 \delta(z-z'')
 d^2\kappa_{\|,\bar{n}}d^2\kappa_{\|,\bar{m}}
 d^2q_{\|}'''d^2q_{\|}''d^2q_{\|}'
 dz'''dz''dz'.
\end{eqnarray}
Of the two integrals over the $\vec{\kappa}_{\|}$ quantities, we can
solve one because of the coupling to the wavevectors introduced by
the Dirac delta functions appearing. We aim at keeping the
$\vec{\kappa}_{\|,\bar{m}}$ set, and thus we solve the integrals for
the $n$ set. Solving for this set, we find that
$\vec{\kappa}_{\|,\bar{n}}$ is replaced by
$\vec{\kappa}_{\|,\bar{m}}-\vec{q}_{\|}^{\,\prime\prime}+\vec{q}_{\|}$,
thus leading to the result
\begin{eqnarray}
\lefteqn{
 J_{xxxx}^{\rm{C}}(z;\vec{q}_{\|})=
 -{e^4\over4m_{e}^2\hbar}
 {2\over(2\pi)^6}
 \idotsint
 \sum_{nm}
 {f_{n}(\vec{\kappa}_{\|,\bar{m}}-\vec{q}_{\|}^{\,\prime\prime}+\vec{q}_{\|})
 -f_{m}(\vec{\kappa}_{\|,\bar{m}})\over\tilde{\omega}_{nm}(\vec{\kappa}_{\|,\bar{m}}
 -\vec{q}_{\|}^{\,\prime\prime}+\vec{q}_{\|},\vec{\kappa}_{\|,\bar{m}})}
 \psi_{n}^{*}(z')\psi_{m}(z')
}\nonumber\\ &\quad&\times
 \psi_{m}^{*}(z)\psi_{n}(z)
 A_{x}(z''';\vec{q}_{\|}^{\,\prime\prime\prime})A_{x}(z'';\vec{q}_{\|}^{\,\prime\prime})
 A_{x}^{*}(z';\vec{q}_{\|}^{\,\prime})
 \delta(\vec{q}_{\|}^{\,\prime\prime\prime}+\vec{q}_{\|}^{\,\prime\prime}-\vec{q}_{\|}^{\,\prime}-\vec{q}_{\|})
 \delta(z'-z''')
\nonumber\\ &&\times
 \delta(z-z'')
 d^2\kappa_{\|,\bar{m}}d^2q_{\|}'''d^2q_{\|}''d^2q_{\|}'dz'''dz''dz'.
\label{eq:J3C}
\end{eqnarray}
On the form of Eq.~(\ref{eq:J3zq}) we thus get part C of the
conductivity tensor as
\begin{eqnarray}
\lefteqn{
 {\Xi}_{xxxx}^{\rm{C}}(z,z',z'',z''';\vec{q}_{\|},
 \vec{q}_{\|}^{\,\prime},\vec{q}_{\|}^{\,\prime\prime},
 \vec{q}_{\|}^{\,\prime\prime\prime})=
 {2{\rm{i}}\over\omega^3}{e^4\over4m_{e}^2\hbar}{1\over(2\pi)^2}
 \sum_{nm}\int
 {f_{n}(\vec{\kappa}_{\|}-\vec{q}_{\|}^{\,\prime\prime}+\vec{q}_{\|})
 -f_{m}(\vec{\kappa}_{\|})\over\tilde{\omega}_{nm}(\vec{\kappa}_{\|}
 -\vec{q}_{\|}^{\,\prime\prime}+\vec{q}_{\|},\vec{\kappa}_{\|})}
}\nonumber\\ &\quad&\times
 \psi_{n}^{*}(z')\psi_{m}(z')
 \psi_{m}^{*}(z)\psi_{n}(z)
 \delta(\vec{q}_{\|}^{\,\prime\prime\prime}+\vec{q}_{\|}^{\,\prime\prime}-\vec{q}_{\|}^{\,\prime}-\vec{q}_{\|})
 \delta(z'-z''')\delta(z-z'')
 d^2\kappa_{\|}.
\end{eqnarray}
where we have omitted the now superfluous index $m$ from the surface
states $\vec{\kappa}_{\|}$. The phase conjugation response is found
using the same procedure as before, since using the pump fields
defined in Eqs.~(\ref{eq:Apump-1})--(\ref{eq:Apump-2}) the integrals
over $q_{\|}'''$ and $q_{\|}''$ in Eq.~(\ref{eq:J3C}) can be
performed. Then (again) the conservation of pseudo-momentum,
$\delta(\vec{q}_{\|}^{\,\prime\prime\prime}+\vec{q}_{\|}^{\,\prime\prime}-\vec{q}_{\|}^{\,\prime}-\vec{q}_{\|})$,
is reduced to $\delta(\vec{q}_{\|}^{\,\prime}+\vec{q}_{\|})$, and
after integration over $q_{\|}'$ we obtain on the form of
Eq.~(\ref{eq:J3-wzq}) the PCDFWM conductivity tensor part C, appearing
as Eq.~(\ref{eq:XiAC}).

\section{Nonlinear process D}
Inserting Eq.~(\ref{eq:XiADrz}) into Eq.~(\ref{eq:J3r}), we take
element $xjkx$ of part D of the nonlinear current density. In the
result we insert the expressions for the wave function, the vector
potential and the transition current density in the two-dimensional
Fourier representation [Eqs.~(\ref{eq:eigenstate}), (\ref{eq:Azr}) and
(\ref{eq:Jm->n||}), respectively]. Then by use of Eq.~(\ref{eq:Jzq})
we find
\begin{eqnarray}
\lefteqn{
 J_{xjkx}^{\rm{D}}(z;\vec{q}_{\|})=
 -{e^2\over4m_{e}\hbar^2}
 {2\over(2\pi)^{12}}
 \idotsint
 \sum_{nmv}
 {1\over\tilde{\omega}_{nm}}\left\{
 \left({f_{m}-f_{v}\over\tilde{\omega}_{v{}m}-\omega}
 +{f_{n}-f_{v}\over\tilde{\omega}_{nv}+\omega}\right)
\right.}\nonumber\\ &\quad&\times
 j_{j,v{}n}(z';\vec{\kappa}_{\|,\bar{v}}+\vec{\kappa}_{\|,\bar{n}})
 j_{k,mv}(z'';\vec{\kappa}_{\|,\bar{m}}+\vec{\kappa}_{\|,\bar{v}})
 e^{{\rm{i}}(\vec{\kappa}_{\|,\bar{v}}-\vec{\kappa}_{\|,\bar{n}})\cdot\vec{r}_{\|}^{\,\prime}}
 e^{{\rm{i}}(\vec{\kappa}_{\|,\bar{m}}-\vec{\kappa}_{\|,\bar{v}})\cdot\vec{r}_{\|}^{\,\prime\prime}}
\nonumber\\ &&
 +\left({f_{m}-f_{v}\over\tilde{\omega}_{v{}m}+\omega}
 +{f_{n}-f_{v}\over\tilde{\omega}_{nv}-\omega}\right)
 j_{k,v{}n}(z'';\vec{\kappa}_{\|,\bar{v}}+\vec{\kappa}_{\|,\bar{n}})
 j_{j,mv}(z';\vec{\kappa}_{\|,\bar{m}}+\vec{\kappa}_{\|,\bar{v}})
 e^{{\rm{i}}(\vec{\kappa}_{\|,\bar{v}}-\vec{\kappa}_{\|,\bar{n}})\cdot\vec{r}_{\|}^{\,\prime\prime}}
\nonumber\\ &&\times\left.\!
 e^{{\rm{i}}(\vec{\kappa}_{\|,\bar{m}}-\vec{\kappa}_{\|,\bar{v}})\cdot\vec{r}_{\|}^{\,\prime}}
 \right\}
 \psi_{m}^{*}(z)
 \psi_{n}(z)
 \delta(\vec{r}-\vec{r}\,''')
 A_{x}(z''';\vec{q}_{\|}^{\,\prime\prime\prime})
 A_{k}(z'';\vec{q}_{\|}^{\,\prime\prime})
 A_{j}^{*}(z';\vec{q}_{\|}^{\,\prime})
 e^{{\rm{i}}(\vec{\kappa}_{\|,\bar{n}}-\vec{\kappa}_{\|,\bar{m}})\cdot\vec{r}_{\|}}
\nonumber\\ &&\times
 e^{{\rm{i}}\vec{q}_{\|}^{\,\prime\prime\prime}\cdot\vec{r}_{\|}^{\,\prime\prime\prime}}
 e^{{\rm{i}}\vec{q}_{\|}^{\,\prime\prime}\cdot\vec{r}_{\|}^{\,\prime\prime}}
 e^{-{\rm{i}}\vec{q}_{\|}^{\,\prime}\cdot\vec{r}_{\|}^{\,\prime}}
 d^2q_{\|}'''d^2q_{\|}''d^2q_{\|}'
 d^3r'''d^3r''d^3r'e^{-{\rm{i}}\vec{q}_{\|}\cdot\vec{r}_{\|}}d^2r_{\|}.
\end{eqnarray}
Solving in this equation the integral $\int{}d^2r_{\|}'''$, and then
the integrals $\int{}d^2r_{\|}''$, $\int{}d^2r_{\|}'$, and
$\int{}d^2r_{\|}$ we get, after having replaced the sums over the
various $\vec{\kappa}_{\|}$ quantities with integrals as before,
\begin{eqnarray}
\lefteqn{
 J_{xjkx}^{\rm{D}}(z;\vec{q}_{\|})=
 -{e^2\over4m_{e}\hbar^2}
 {2\over(2\pi)^6}
 \idotsint
 \sum_{nmv}
 {1\over\tilde{\omega}_{nm}(\vec{\kappa}_{\|,\bar{n}},\vec{\kappa}_{\|,\bar{m}})}\left\{
 \left({f_{m}(\vec{\kappa}_{\|,\bar{m}})-f_{v}(\vec{\kappa}_{\|,\bar{v}})\over
 \tilde{\omega}_{v{}m}(\vec{\kappa}_{\|,\bar{v}},\vec{\kappa}_{\|,\bar{m}})-\omega}
\right.\right.}\nonumber\\ &\quad&\left.\!
 +{f_{n}(\vec{\kappa}_{\|,\bar{n}})-f_{v}(\vec{\kappa}_{\|,\bar{v}})\over
 \tilde{\omega}_{nv}(\vec{\kappa}_{\|,\bar{n}},\vec{\kappa}_{\|,\bar{v}})+\omega}
 \right)
 j_{j,v{}n}(z';\vec{\kappa}_{\|,\bar{v}}+\vec{\kappa}_{\|,\bar{n}})
 j_{k,mv}(z'';\vec{\kappa}_{\|,\bar{m}}+\vec{\kappa}_{\|,\bar{v}})
 \delta(\vec{\kappa}_{\|,\bar{m}}-\vec{\kappa}_{\|,\bar{v}}+\vec{q}_{\|}^{\,\prime\prime})
\nonumber\\ &&\times
 \delta(\vec{\kappa}_{\|,\bar{v}}-\vec{\kappa}_{\|,\bar{n}}-\vec{q}_{\|}^{\,\prime})
 +\left({f_{m}(\vec{\kappa}_{\|,\bar{m}})-f_{v}(\vec{\kappa}_{\|,\bar{v}})\over
 \tilde{\omega}_{v{}m}(\vec{\kappa}_{\|,\bar{v}},\vec{\kappa}_{\|,\bar{m}})+\omega}
 +{f_{n}(\vec{\kappa}_{\|,\bar{n}})-f_{v}(\vec{\kappa}_{\|,\bar{v}})\over
 \tilde{\omega}_{nv}(\vec{\kappa}_{\|,\bar{n}},\vec{\kappa}_{\|,\bar{v}})-\omega}
 \right)
\nonumber\\ &&\times\left.
 j_{k,v{}n}(z'';\vec{\kappa}_{\|,\bar{v}}+\vec{\kappa}_{\|,\bar{n}})
 j_{j,mv}(z';\vec{\kappa}_{\|,\bar{m}}+\vec{\kappa}_{\|,\bar{v}})
 \delta(\vec{\kappa}_{\|,\bar{v}}-\vec{\kappa}_{\|,\bar{n}}+\vec{q}_{\|}^{\,\prime\prime})
 \delta(\vec{\kappa}_{\|,\bar{m}}-\vec{\kappa}_{\|,\bar{v}}-\vec{q}_{\|}^{\,\prime})
 \right\}
\nonumber\\ &&\times
 \psi_{m}^{*}(z)\psi_{n}(z)
 \delta(z-z''')
 \delta(\vec{\kappa}_{\|,\bar{n}}-\vec{\kappa}_{\|,\bar{m}}+\vec{q}_{\|}^{\,\prime\prime\prime}-\vec{q}_{\|})
 A_{x}(z''';\vec{q}_{\|}^{\,\prime\prime\prime})
 A_{k}(z'';\vec{q}_{\|}^{\,\prime\prime})
\nonumber\\ &&\times
 A_{j}^{*}(z';\vec{q}_{\|}^{\,\prime})
 d^2\kappa_{\|,\bar{n}}d^2\kappa_{\|,\bar{m}}d^2\kappa_{\|,\bar{v}}
 d^2q_{\|}'''d^2q_{\|}''d^2q_{\|}'
 dz'''dz''dz'.
\end{eqnarray}
Of the three integrals over $\vec{\kappa}_{\|}$ quantities, we can
solve two because of the coupling to the wavevectors introduced by
the Dirac delta functions appearing. We aim at keeping the
$\vec{\kappa}_{\|,\bar{m}}$ set, and thus we solve the integrals for
the $v$ and $n$ sets (in that order). Solving for the $v$ set, we find
that $\vec{\kappa}_{\|,\bar{v}}$ is replaced by
$\vec{\kappa}_{\|,\bar{m}}+\vec{q}_{\|}^{\,\prime\prime}$ in the first
part of the sum and by
$\vec{\kappa}_{\|,\bar{m}}-\vec{q}_{\|}^{\,\prime}$ in the second part
of the sum, which then allows us to solve the $n$ set by replacing
$\vec{\kappa}_{\|,\bar{n}}$ with
$\vec{\kappa}_{\|,\bar{m}}+\vec{q}_{\|}^{\,\prime\prime}-\vec{q}_{\|}^{\,\prime}$
in general, giving the result
\begin{eqnarray}
\lefteqn{
 J_{xjkx}^{\rm{D}}(z;\vec{q}_{\|})=
 -{e^2\over4m_{e}\hbar^2}
 {2\over(2\pi)^6}
 \idotsint
 \sum_{nmv}
 {1\over\tilde{\omega}_{nm}(\vec{\kappa}_{\|,\bar{m}}+\vec{q}_{\|}^{\,\prime\prime}-\vec{q}_{\|}^{\,\prime}
 ,\vec{\kappa}_{\|,\bar{m}})}
}\nonumber\\ &\quad&\times
\left\{
 \left({f_{m}(\vec{\kappa}_{\|,\bar{m}})-f_{v}(\vec{\kappa}_{\|,\bar{m}}+\vec{q}_{\|}^{\,\prime\prime})
 \over\tilde{\omega}_{v{}m}(\vec{\kappa}_{\|,\bar{m}}+\vec{q}_{\|}^{\,\prime\prime},
 \vec{\kappa}_{\|,\bar{m}})-\omega}
 +{f_{n}(\vec{\kappa}_{\|,\bar{m}}+\vec{q}_{\|}^{\,\prime\prime}-\vec{q}_{\|}^{\,\prime})
 -f_{v}(\vec{\kappa}_{\|,\bar{m}}+\vec{q}_{\|}^{\,\prime\prime})\over
 \tilde{\omega}_{nv}(\vec{\kappa}_{\|,\bar{m}}+\vec{q}_{\|}^{\,\prime\prime}-\vec{q}_{\|}^{\,\prime},
 \vec{\kappa}_{\|,\bar{m}}+\vec{q}_{\|}^{\,\prime\prime})+\omega}\right)
\right.\nonumber\\ &&\times
 j_{j,v{}n}(z';2\vec{\kappa}_{\|,\bar{m}}+2\vec{q}_{\|}^{\,\prime\prime}-\vec{q}_{\|}^{\,\prime})
 j_{k,mv}(z'';2\vec{\kappa}_{\|,\bar{m}}+\vec{q}_{\|}^{\,\prime\prime})
\nonumber\\ &&
 +\left({f_{m}(\vec{\kappa}_{\|,\bar{m}})-f_{v}(\vec{\kappa}_{\|,\bar{m}}-\vec{q}_{\|}^{\,\prime})
 \over\tilde{\omega}_{v{}m}(\vec{\kappa}_{\|,\bar{m}}-\vec{q}_{\|}^{\,\prime},
 \vec{\kappa}_{\|,\bar{m}})+\omega}
 +{f_{n}(\vec{\kappa}_{\|,\bar{m}}+\vec{q}_{\|}^{\,\prime\prime}-\vec{q}_{\|}^{\,\prime})
 -f_{v}(\vec{\kappa}_{\|,\bar{m}}-\vec{q}_{\|}^{\,\prime})\over
 \tilde{\omega}_{nv}(\vec{\kappa}_{\|,\bar{m}}+\vec{q}_{\|}^{\,\prime\prime}-\vec{q}_{\|}^{\,\prime},
 \vec{\kappa}_{\|,\bar{m}}-\vec{q}_{\|}^{\,\prime})-\omega}\right)
\nonumber\\ &&\times\left.\!
 j_{k,v{}n}(z'';2\vec{\kappa}_{\|,\bar{m}}+\vec{q}_{\|}^{\,\prime\prime}-2\vec{q}_{\|}^{\,\prime})
 j_{j,mv}(z';2\vec{\kappa}_{\|,\bar{m}}-\vec{q}_{\|}^{\,\prime})
 \right\}
 \psi_{m}^{*}(z)\psi_{n}(z)
 \delta(z-z''')
\nonumber\\ &&\times
 \delta(\vec{q}_{\|}^{\,\prime\prime\prime}+\vec{q}_{\|}^{\,\prime\prime}-\vec{q}_{\|}^{\,\prime}-\vec{q}_{\|})
 A_{x}(z''';\vec{q}_{\|}^{\,\prime\prime\prime})
 A_{k}(z'';\vec{q}_{\|}^{\,\prime\prime})
 A_{j}^{*}(z';\vec{q}_{\|}^{\,\prime})
 d^2\kappa_{\|,\bar{m}}
 d^2q_{\|}'''d^2q_{\|}''d^2q_{\|}'
\nonumber\\ &&\times
 dz'''dz''dz'.
\label{eq:J3D}
\end{eqnarray}
On the form of Eq.~(\ref{eq:J3zq}) we thus get part D of the
conductivity tensor as
\begin{eqnarray}
\lefteqn{
 {\Xi}_{xjkx}^{\rm{D}}(z,z',z'',z''';\vec{q}_{\|},\vec{q}_{\|}^{\,\prime},
 \vec{q}_{\|}^{\,\prime\prime},\vec{q}_{\|}^{\,\prime\prime\prime})=
 {2{\rm{i}}\over\omega^3}{e^2\over4m_{e}\hbar^2}{1\over(2\pi)^2}\sum_{nmv}\int
 {1\over\tilde{\omega}_{nm}(\vec{\kappa}_{\|}+\vec{q}_{\|}^{\,\prime\prime}-\vec{q}_{\|}^{\,\prime}
 ,\vec{\kappa}_{\|})}
}\nonumber\\ &\quad&\times
\left\{
 \left({f_{m}(\vec{\kappa}_{\|})-f_{v}(\vec{\kappa}_{\|}+\vec{q}_{\|}^{\,\prime\prime})
 \over\tilde{\omega}_{v{}m}(\vec{\kappa}_{\|}+\vec{q}_{\|}^{\,\prime\prime},
 \vec{\kappa}_{\|})-\omega}
 +{f_{n}(\vec{\kappa}_{\|}+\vec{q}_{\|}^{\,\prime\prime}-\vec{q}_{\|}^{\,\prime})
 -f_{v}(\vec{\kappa}_{\|}+\vec{q}_{\|}^{\,\prime\prime})\over
 \tilde{\omega}_{nv}(\vec{\kappa}_{\|}+\vec{q}_{\|}^{\,\prime\prime}-\vec{q}_{\|}^{\,\prime},
 \vec{\kappa}_{\|}+\vec{q}_{\|}^{\,\prime\prime})+\omega}\right)
\right.\nonumber\\ &&\times
 j_{j,v{}n}(z';2\vec{\kappa}_{\|}+2\vec{q}_{\|}^{\,\prime\prime}-\vec{q}_{\|}^{\,\prime})
 j_{k,mv}(z'';2\vec{\kappa}_{\|}+\vec{q}_{\|}^{\,\prime\prime})
\nonumber\\ &&
 +\left({f_{m}(\vec{\kappa}_{\|})-f_{v}(\vec{\kappa}_{\|}-\vec{q}_{\|}^{\,\prime})
 \over\tilde{\omega}_{v{}m}(\vec{\kappa}_{\|}-\vec{q}_{\|}^{\,\prime},
 \vec{\kappa}_{\|})+\omega}
 +{f_{n}(\vec{\kappa}_{\|}+\vec{q}_{\|}^{\,\prime\prime}-\vec{q}_{\|}^{\,\prime})
 -f_{v}(\vec{\kappa}_{\|}-\vec{q}_{\|}^{\,\prime})\over
 \tilde{\omega}_{nv}(\vec{\kappa}_{\|}+\vec{q}_{\|}^{\,\prime\prime}-\vec{q}_{\|}^{\,\prime},
 \vec{\kappa}_{\|}-\vec{q}_{\|}^{\,\prime})-\omega}\right)
\nonumber\\ &&\times\left.\!
 j_{k,v{}n}(z'';2\vec{\kappa}_{\|}+\vec{q}_{\|}^{\,\prime\prime}-2\vec{q}_{\|}^{\,\prime})
 j_{j,mv}(z';2\vec{\kappa}_{\|}-\vec{q}_{\|}^{\,\prime})
 \right\}
 \psi_{m}^{*}(z)\psi_{n}(z)
 \delta(z-z''')
\nonumber\\ &&\times
 \delta(\vec{q}_{\|}^{\,\prime\prime\prime}+\vec{q}_{\|}^{\,\prime\prime}-\vec{q}_{\|}^{\,\prime}-\vec{q}_{\|})
 d^2\kappa_{\|},
\end{eqnarray}
where we have omitted the now superfluous index $m$ from the surface
states $\vec{\kappa}_{\|}$. Again, when looking for the DFWM response
tensor we insert the pump fields defined in
Eqs.~(\ref{eq:Apump-1})--(\ref{eq:Apump-2}) and integrate over
$q_{\|}'''$ and $q_{\|}''$ in Eq.~(\ref{eq:J3D}), again reducing the
Dirac delta function accounting for conservation of pseudo-momentum to
$\delta(\vec{q}_{\|}^{\,\prime}+\vec{q}_{\|})$. After integration over
$q_{\|}'$ and separation according to Eq.~(\ref{eq:J3-wzq}),
Eq.~(\ref{eq:XiAD}) appear as the PCDFWM conductivity tensor part D.

\section{Nonlinear Process E}
Inserting Eq.~(\ref{eq:XiAErz}) into Eq.~(\ref{eq:J3r}), we take
element $ijxx$ of part E of the nonlinear current density. In the
result we insert the expressions for the wave function, the vector
potential and the transition current density in the two-dimensional
Fourier representation [Eqs.~(\ref{eq:eigenstate}), (\ref{eq:Azr}) and
(\ref{eq:Jm->n||}), respectively]. Then by use of Eq.~(\ref{eq:Jzq})
we find
\begin{eqnarray}
\lefteqn{
 J_{ijxx}^{\rm{E}}(z;\vec{q}_{\|})=
 -{e^2\over16m_{e}\hbar^2}
 {2\over(2\pi)^{12}}
 \idotsint
 \sum_{nmv}
 {1\over\tilde{\omega}_{nm}-\omega}\left\{
 \left({f_{m}-f_{v}\over\tilde{\omega}_{v{}m}-2\omega}
 +{f_{n}-f_{v}\over\tilde{\omega}_{nv}+\omega}\right)
\right.}\nonumber\\ &&\times
 j_{j,v{}n}(z';\vec{\kappa}_{\|,\bar{v}}+\vec{\kappa}_{\|,\bar{n}})
 \psi_{v}^{*}(z'')
 \psi_{m}(z'')
 e^{{\rm{i}}(\vec{\kappa}_{\|,\bar{v}}-\vec{\kappa}_{\|,\bar{n}})\cdot\vec{r}_{\|}^{\,\prime}}
 e^{{\rm{i}}(\vec{\kappa}_{\|,\bar{m}}-\vec{\kappa}_{\|,\bar{v}})\cdot\vec{r}_{\|}^{\,\prime\prime}}
\nonumber\\ &&
 +\left({f_{n}-f_{v}\over\tilde{\omega}_{nv}-2\omega}
 +{f_{m}-f_{v}\over\tilde{\omega}_{v{}m}+\omega}\right)
 j_{j,mv}(z';\vec{\kappa}_{\|,\bar{m}}+\vec{\kappa}_{\|,\bar{v}})
 \psi_{n}^{*}(z'')
 \psi_{v}(z'')
 e^{{\rm{i}}(\vec{\kappa}_{\|,\bar{m}}-\vec{\kappa}_{\|,\bar{v}})\cdot\vec{r}_{\|}^{\,\prime}}
\nonumber\\ &&\times\left.\!
 e^{{\rm{i}}(\vec{\kappa}_{\|,\bar{v}}-\vec{\kappa}_{\|,\bar{n}})\cdot\vec{r}_{\|}^{\,\prime\prime}}
 \right\}
 j_{i,nm}(z;\vec{\kappa}_{\|,\bar{n}}+\vec{\kappa}_{\|,\bar{m}})
 \delta(\vec{r}\,''-\vec{r}\,''')
 A_{x}(z''';\vec{q}_{\|}^{\,\prime\prime\prime})
 A_{x}(z'';\vec{q}_{\|}^{\,\prime\prime})
 A_{j}^{*}(z';\vec{q}_{\|}^{\,\prime})
\nonumber\\ &&\times
 e^{{\rm{i}}(\vec{\kappa}_{\|,\bar{n}}-\vec{\kappa}_{\|,\bar{m}})\cdot\vec{r}_{\|}}
 e^{{\rm{i}}\vec{q}_{\|}^{\,\prime\prime\prime}\cdot\vec{r}_{\|}^{\,\prime\prime\prime}}
 e^{{\rm{i}}\vec{q}_{\|}^{\,\prime\prime}\cdot\vec{r}_{\|}^{\,\prime\prime}}
 e^{-{\rm{i}}\vec{q}_{\|}^{\,\prime}\cdot\vec{r}_{\|}^{\,\prime}}
 d^2q_{\|}'''d^2q_{\|}''d^2q_{\|}'
 d^3r'''d^3r''d^3r'e^{-{\rm{i}}\vec{q}_{\|}\cdot\vec{r}_{\|}}d^2r_{\|}.
\end{eqnarray}
In this equation, we first solve the integral $\int{}d^2r_{\|}'''$,
and then the integrals $\int{}d^2r_{\|}''$, $\int{}d^2r_{\|}'$, and
$\int{}d^2r_{\|}$, which together with replacement of the sums over
the different $\vec{\kappa}_{\|}$ quantities with integrals yields the
result
\begin{eqnarray}
\lefteqn{
 J_{ijxx}^{\rm{E}}(z;\vec{q}_{\|})=
 -{e^2\over16m_{e}\hbar^2}
 {2\over(2\pi)^6}
 \idotsint
 \sum_{nmv}
 {1\over\tilde{\omega}_{nm}(\vec{\kappa}_{\|,\bar{n}},\vec{\kappa}_{\|,\bar{m}})-\omega}
\left\{\!\!\!
 \left({f_{m}(\vec{\kappa}_{\|,\bar{m}})-f_{v}(\vec{\kappa}_{\|,\bar{v}})\over
 \tilde{\omega}_{v{}m}(\vec{\kappa}_{\|,\bar{v}},\vec{\kappa}_{\|,\bar{m}})-2\omega}
\right.\right.}\nonumber\\ &\quad&\left.\!
 +{f_{n}(\vec{\kappa}_{\|,\bar{n}})-f_{v}(\vec{\kappa}_{\|,\bar{v}})\over
 \tilde{\omega}_{nv}(\vec{\kappa}_{\|,\bar{n}},\vec{\kappa}_{\|,\bar{v}})+\omega}
 \right)
 j_{j,v{}n}(z';\vec{\kappa}_{\|,\bar{v}}+\vec{\kappa}_{\|,\bar{n}})
 \psi_{v}^{*}(z'')
 \psi_{m}(z'')
 \delta(\vec{\kappa}_{\|,\bar{v}}-\vec{\kappa}_{\|,\bar{n}}-\vec{q}_{\|}^{\,\prime})
\nonumber\\ &&\times
 \delta(\vec{\kappa}_{\|,\bar{m}}-\vec{\kappa}_{\|,\bar{v}}
 +\vec{q}_{\|}^{\,\prime\prime\prime}+\vec{q}_{\|}^{\,\prime\prime})
 +\left({f_{n}(\vec{\kappa}_{\|,\bar{n}})-f_{v}(\vec{\kappa}_{\|,\bar{v}})\over
 \tilde{\omega}_{nv}(\vec{\kappa}_{\|,\bar{n}},\vec{\kappa}_{\|,\bar{v}})-2\omega}
 +{f_{m}(\vec{\kappa}_{\|,\bar{m}})-f_{v}(\vec{\kappa}_{\|,\bar{v}})\over
 \tilde{\omega}_{v{}m}(\vec{\kappa}_{\|,\bar{v}},\vec{\kappa}_{\|,\bar{m}})+\omega}
 \right)
\nonumber\\ &&\times\left.\!
 j_{j,mv}(z';\vec{\kappa}_{\|,\bar{m}}+\vec{\kappa}_{\|,\bar{v}})
 \psi_{n}^{*}(z'')
 \psi_{v}(z'')
 \delta(\vec{\kappa}_{\|,\bar{m}}-\vec{\kappa}_{\|,\bar{v}}-\vec{q}_{\|}^{\,\prime})
 \delta(\vec{\kappa}_{\|,\bar{v}}-\vec{\kappa}_{\|,\bar{n}}+\vec{q}_{\|}^{\,\prime\prime\prime}
 +\vec{q}_{\|}^{\,\prime\prime})
 \right\}
\nonumber\\ &&\times
 j_{i,nm}(z;\vec{\kappa}_{\|,\bar{n}}+\vec{\kappa}_{\|,\bar{m}})
 \delta(z''-z''')
 A_{x}(z''';\vec{q}_{\|}^{\,\prime\prime\prime})
 A_{x}(z'';\vec{q}_{\|}^{\,\prime\prime})
 A_{j}^{*}(z';\vec{q}_{\|}^{\,\prime})
\nonumber\\ &&\times
 \delta(\vec{\kappa}_{\|,\bar{n}}-\vec{\kappa}_{\|,\bar{m}}-\vec{q}_{\|})
 d^2\kappa_{\|,\bar{n}}d^2\kappa_{\|,\bar{m}}d^2\kappa_{\|,\bar{v}}
 d^2q_{\|}'''d^2q_{\|}''d^2q_{\|}'
 dz'''dz''dz'.
\end{eqnarray}
Of the three integrals over $\vec{\kappa}_{\|}$ quantities, we can
solve two because of the coupling to the wavevectors introduced by
the Dirac delta functions appearing. We aim at keeping the
$\vec{\kappa}_{\|,\bar{m}}$ set, and thus we solve the integrals for
the $v$ and $n$ sets (in that order). Solving for the $v$ set, we find
that $\vec{\kappa}_{\|,\bar{v}}$ is replaced by
$\vec{\kappa}_{\|,\bar{m}}+\vec{q}_{\|}^{\,\prime\prime\prime}+\vec{q}_{\|}^{\,\prime\prime}$
in the first part of the sum and by
$\vec{\kappa}_{\|,\bar{m}}-\vec{q}_{\|}^{\,\prime}$ in the second part
of the sum, which then allows us to solve the $n$ set by replacing
$\vec{\kappa}_{\|,\bar{n}}$ with
$\vec{\kappa}_{\|,\bar{m}}+\vec{q}_{\|}^{\,\prime\prime\prime}+\vec{q}_{\|}^{\,\prime\prime}-\vec{q}_{\|}^{\,\prime}$
in general, giving
\begin{eqnarray}
\lefteqn{
 J_{ijxx}^{\rm{E}}(z;\vec{q}_{\|})=
 -{e^2\over16m_{e}\hbar^2}
 {2\over(2\pi)^6}
 \idotsint
 \sum_{nmv}
 {1\over\tilde{\omega}_{nm}(\vec{\kappa}_{\|,\bar{m}}+\vec{q}_{\|}^{\,\prime\prime\prime}+\vec{q}_{\|}^{\,\prime\prime}
 -\vec{q}_{\|}^{\,\prime},\vec{\kappa}_{\|,\bar{m}})-\omega}
}\nonumber\\ &\quad&\times
\left\{
 \left({f_{m}(\vec{\kappa}_{\|,\bar{m}})-f_{v}(\vec{\kappa}_{\|,\bar{m}}+\vec{q}_{\|}^{\,\prime\prime\prime}
 +\vec{q}_{\|}^{\,\prime\prime})\over\tilde{\omega}_{v{}m}(\vec{\kappa}_{\|,\bar{m}}
 +\vec{q}_{\|}^{\,\prime\prime\prime}+\vec{q}_{\|}^{\,\prime\prime},\vec{\kappa}_{\|,\bar{m}})-2\omega}
\right.\right.\nonumber\\ &&\left.\!
 +{f_{n}(\vec{\kappa}_{\|,\bar{m}}+\vec{q}_{\|}^{\,\prime\prime\prime}+\vec{q}_{\|}^{\,\prime\prime}-\vec{q}_{\|}^{\,\prime})
 -f_{v}(\vec{\kappa}_{\|,\bar{m}}+\vec{q}_{\|}^{\,\prime\prime\prime}+\vec{q}_{\|}^{\,\prime\prime})\over
 \tilde{\omega}_{nv}(\vec{\kappa}_{\|,\bar{m}}+\vec{q}_{\|}^{\,\prime\prime\prime}+\vec{q}_{\|}^{\,\prime\prime}
 -\vec{q}_{\|}^{\,\prime},\vec{\kappa}_{\|,\bar{m}}+\vec{q}_{\|}^{\,\prime\prime\prime}+\vec{q}_{\|}^{\,\prime\prime})+\omega}
 \right)
\nonumber\\ &&\times
 j_{j,v{}n}(z';2\vec{\kappa}_{\|,\bar{m}}+2\vec{q}_{\|}^{\,\prime\prime\prime}+2\vec{q}_{\|}^{\,\prime\prime}
 -\vec{q}_{\|}^{\,\prime})
 \psi_{v}^{*}(z'')
 \psi_{m}(z'')
\nonumber\\ &&
 +\left({f_{n}(\vec{\kappa}_{\|,\bar{m}}+\vec{q}_{\|}^{\,\prime\prime\prime}+\vec{q}_{\|}^{\,\prime\prime}
 -\vec{q}_{\|}^{\,\prime})-f_{v}(\vec{\kappa}_{\|,\bar{m}}-\vec{q}_{\|}^{\,\prime})\over
 \tilde{\omega}_{nv}(\vec{\kappa}_{\|,\bar{m}}+\vec{q}_{\|}^{\,\prime\prime\prime}+\vec{q}_{\|}^{\,\prime\prime}
 -\vec{q}_{\|}^{\,\prime},\vec{\kappa}_{\|,\bar{m}}-\vec{q}_{\|}^{\,\prime})-2\omega}
 +{f_{m}(\vec{\kappa}_{\|,\bar{m}})-f_{v}(\vec{\kappa}_{\|,\bar{m}}-\vec{q}_{\|}^{\,\prime})\over
 \tilde{\omega}_{v{}m}(\vec{\kappa}_{\|,\bar{m}}-\vec{q}_{\|}^{\,\prime},\vec{\kappa}_{\|,\bar{m}})
 +\omega}\right)
\nonumber\\ &&\times\left.
 j_{j,mv}(z';2\vec{\kappa}_{\|,\bar{m}}-\vec{q}_{\|}^{\,\prime})
 \psi_{n}^{*}(z'')
 \psi_{v}(z'')
 \right\}
 j_{i,nm}(z;2\vec{\kappa}_{\|,\bar{m}}+\vec{q}_{\|}^{\,\prime\prime\prime}+\vec{q}_{\|}^{\,\prime\prime}-\vec{q}_{\|}^{\,\prime})
 \delta(z''-z''')
\nonumber\\ &&\times
 A_{x}(z''';\vec{q}_{\|}^{\,\prime\prime\prime})
 A_{x}(z'';\vec{q}_{\|}^{\,\prime\prime})
 A_{j}^{*}(z';\vec{q}_{\|}^{\,\prime})
 \delta(\vec{q}_{\|}^{\,\prime\prime\prime}+\vec{q}_{\|}^{\,\prime\prime}-\vec{q}_{\|}^{\,\prime}-\vec{q}_{\|})
 d^2\kappa_{\|,\bar{m}}
 d^2q_{\|}'''d^2q_{\|}''d^2q_{\|}'
\nonumber\\ &&\times
 dz'''dz''dz'.
\label{eq:J3E}
\end{eqnarray}
On the form of Eq.~(\ref{eq:J3zq}) we thus get part E of the
conductivity tensor as
\begin{eqnarray}
\lefteqn{
 {\Xi}_{ijxx}^{\rm{E}}(z,z',z'',z''';\vec{q}_{\|},\vec{q}_{\|}^{\,\prime},
 \vec{q}_{\|}^{\,\prime\prime},\vec{q}_{\|}^{\,\prime\prime\prime})=
}\nonumber\\ &\quad&
 {2{\rm{i}}\over\omega^3}{e^2\over16m_{e}\hbar^2}{1\over(2\pi)^2}
 \sum_{nmv}\int
 {1\over\tilde{\omega}_{nm}(\vec{\kappa}_{\|}+\vec{q}_{\|}^{\,\prime\prime\prime}+\vec{q}_{\|}^{\,\prime\prime}
 -\vec{q}_{\|}^{\,\prime},\vec{\kappa}_{\|})-\omega}
\nonumber\\ &&\times
\!\left\{\!\!
 \left(\!{f_{m}(\vec{\kappa}_{\|})-f_{v}(\vec{\kappa}_{\|}+\vec{q}_{\|}^{\,\prime\prime\prime}
 +\vec{q}_{\|}^{\,\prime\prime})\over\tilde{\omega}_{v{}m}(\vec{\kappa}_{\|}
 +\vec{q}_{\|}^{\,\prime\prime\prime}+\vec{q}_{\|}^{\,\prime\prime},\vec{\kappa}_{\|})-2\omega}
 +{f_{n}(\vec{\kappa}_{\|}+\vec{q}_{\|}^{\,\prime\prime\prime}+\vec{q}_{\|}^{\,\prime\prime}-\vec{q}_{\|}^{\,\prime})
 -f_{v}(\vec{\kappa}_{\|}+\vec{q}_{\|}^{\,\prime\prime\prime}+\vec{q}_{\|}^{\,\prime\prime})\over
 \tilde{\omega}_{nv}(\vec{\kappa}_{\|}+\vec{q}_{\|}^{\,\prime\prime\prime}+\vec{q}_{\|}^{\,\prime\prime}
 -\vec{q}_{\|}^{\,\prime},\vec{\kappa}_{\|}+\vec{q}_{\|}^{\,\prime\prime\prime}+\vec{q}_{\|}^{\,\prime\prime})+\omega}
 \!\right)
\right.\nonumber\\ &&\times
 j_{j,v{}n}(z';2\vec{\kappa}_{\|}+2\vec{q}_{\|}^{\,\prime\prime\prime}+2\vec{q}_{\|}^{\,\prime\prime}
 -\vec{q}_{\|}^{\,\prime})
 \psi_{v}^{*}(z'')\psi_{m}(z'')
\nonumber\\ &&
 +\left({f_{n}(\vec{\kappa}_{\|}+\vec{q}_{\|}^{\,\prime\prime\prime}+\vec{q}_{\|}^{\,\prime\prime}
 -\vec{q}_{\|}^{\,\prime})-f_{v}(\vec{\kappa}_{\|}-\vec{q}_{\|}^{\,\prime})\over
 \tilde{\omega}_{nv}(\vec{\kappa}_{\|}+\vec{q}_{\|}^{\,\prime\prime\prime}+\vec{q}_{\|}^{\,\prime\prime}
 -\vec{q}_{\|}^{\,\prime},\vec{\kappa}_{\|}-\vec{q}_{\|}^{\,\prime})-2\omega}
 +{f_{m}(\vec{\kappa}_{\|})-f_{v}(\vec{\kappa}_{\|}-\vec{q}_{\|}^{\,\prime})\over
 \tilde{\omega}_{v{}m}(\vec{\kappa}_{\|}-\vec{q}_{\|}^{\,\prime},\vec{\kappa}_{\|})
 +\omega}\right)
\nonumber\\ &&\times\left.\!
 j_{j,mv}(z';2\vec{\kappa}_{\|}-\vec{q}_{\|}^{\,\prime})
 \psi_{n}^{*}(z'')\psi_{v}(z'')
 \right\}
 j_{i,nm}(z;2\vec{\kappa}_{\|}+\vec{q}_{\|}^{\,\prime\prime\prime}+\vec{q}_{\|}^{\,\prime\prime}-\vec{q}_{\|}^{\,\prime})
 \delta(z''-z''')
\nonumber\\ &&\times
 \delta(\vec{q}_{\|}^{\,\prime\prime\prime}+\vec{q}_{\|}^{\,\prime\prime}-\vec{q}_{\|}^{\,\prime}-\vec{q}_{\|})
 d^2\kappa_{\|},
\end{eqnarray}
where we have omitted the now superfluous index $m$ from the surface
states $\vec{\kappa}_{\|}$. The PCDFWM response tensor part E we find
by insertion of the pump fields defined by Eqs.~(\ref{eq:Apump-1}) and
(\ref{eq:Apump-2}) into Eq.~(\ref{eq:J3E}) and perform the integrals
over $q_{\|}'''$ and $q_{\|}''$, followed by integration over
$q_{\|}'$ because of the reduction in the Dirac delta function
accounting for conservation of pseudo-momentum. After these
operations, the resulting expression is separated in the form of
Eq.~(\ref{eq:J3-wzq}), and Eq.~(\ref{eq:XiAE}) is obtained.

\section{Nonlinear process F}
Inserting Eq.~(\ref{eq:XiAFrz}) into Eq.~(\ref{eq:J3r}), we take
element $ixxh$ of part F of the nonlinear current density. In the
result we insert the expressions for the wave function, the vector
potential and the transition current density in the two-dimensional
Fourier representation [Eqs.~(\ref{eq:eigenstate}), (\ref{eq:Azr}) and
(\ref{eq:Jm->n||}), respectively]. Then by use of Eq.~(\ref{eq:Jzq})
we find
\begin{eqnarray}
\lefteqn{
 J_{ixxh}^{\rm{F}}(z;\vec{q}_{\|})=
 -{e^2\over8m_{e}\hbar^2}
 {2\over(2\pi)^{12}}
 \idotsint
 \sum_{nmv}
 {1\over\tilde{\omega}_{nm}-\omega}\left\{
 \left({f_{m}-f_{v}\over\tilde{\omega}_{v{}m}}
 +{f_{n}-f_{v}\over\tilde{\omega}_{nv}-\omega}\right)
\right.}\nonumber\\ &&\times
 j_{h,v{}n}(z'';\vec{\kappa}_{\|,\bar{v}}+\vec{\kappa}_{\|,\bar{n}})
 \psi_{v}^{*}(z')
 \psi_{m}(z')
 e^{{\rm{i}}(\vec{\kappa}_{\|,\bar{v}}-\vec{\kappa}_{\|,\bar{n}})\cdot\vec{r}_{\|}^{\,\prime\prime}}
 e^{{\rm{i}}(\vec{\kappa}_{\|,\bar{m}}-\vec{\kappa}_{\|,\bar{v}})\cdot\vec{r}_{\|}^{\,\prime}}
\nonumber\\ &&\left.\!
 +\!\left(\!{f_{n}-f_{v}\over\tilde{\omega}_{nv}}
 +{f_{m}-f_{v}\over\tilde{\omega}_{v{}m}-\omega}\right)
 j_{h,mv}(z'';\vec{\kappa}_{\|,\bar{m}}+\vec{\kappa}_{\|,\bar{v}})
 \psi_{n}^{*}(z')
 \psi_{v}(z')
 e^{{\rm{i}}(\vec{\kappa}_{\|,\bar{m}}-\vec{\kappa}_{\|,\bar{v}})\cdot\vec{r}_{\|}^{\,\prime\prime}}
 e^{{\rm{i}}(\vec{\kappa}_{\|,\bar{v}}-\vec{\kappa}_{\|,\bar{n}})\cdot\vec{r}_{\|}^{\,\prime}}
 \!\right\}\!
\nonumber\\ &&\times
 j_{i,nm}(z;\vec{\kappa}_{\|,\bar{n}}+\vec{\kappa}_{\|,\bar{m}})
 \delta(\vec{r}\,'-\vec{r}\,''')
 A_{h}(z''';\vec{q}_{\|}^{\,\prime\prime\prime})
 A_{x}(z'';\vec{q}_{\|}^{\,\prime\prime})
 A_{x}^{*}(z';\vec{q}_{\|}^{\,\prime})
 e^{{\rm{i}}(\vec{\kappa}_{\|,\bar{n}}-\vec{\kappa}_{\|,\bar{m}})\cdot\vec{r}_{\|}}
\nonumber\\ &&\times
 e^{{\rm{i}}\vec{q}_{\|}^{\,\prime\prime\prime}\cdot\vec{r}_{\|}^{\,\prime\prime\prime}}
 e^{{\rm{i}}\vec{q}_{\|}^{\,\prime\prime}\cdot\vec{r}_{\|}^{\,\prime\prime}}
 e^{-{\rm{i}}\vec{q}_{\|}^{\,\prime}\cdot\vec{r}_{\|}^{\,\prime}}
 d^2q_{\|}'''d^2q_{\|}''d^2q_{\|}'
 d^3r'''d^3r''d^3r'e^{-{\rm{i}}\vec{q}_{\|}\cdot\vec{r}_{\|}}d^2r_{\|}.
\end{eqnarray}
Solving in this equation first the integral $\int{}d^2r_{\|}'''$, and
then the integrals $\int{}d^2r_{\|}''$, $\int{}d^2r_{\|}'$, and
$\int{}d^2r_{\|}$, followed by a replacement of the sums over the
various $\vec{\kappa}_{\|}$ quantities with integrals, we get
\begin{eqnarray}
\lefteqn{
 J_{ixxh}^{\rm{F}}(z;\vec{q}_{\|})=
 -{e^2\over8m_{e}\hbar^2}
 {2\over(2\pi)^6}
 \idotsint
 \sum_{nmv}
 {1\over\tilde{\omega}_{nm}(\vec{\kappa}_{\|,\bar{n}},\vec{\kappa}_{\|,\bar{m}})-\omega}
\left\{
 \left({f_{m}(\vec{\kappa}_{\|,\bar{m}})-f_{v}(\vec{\kappa}_{\|,\bar{v}})\over
 \tilde{\omega}_{v{}m}(\vec{\kappa}_{\|,\bar{v}},\vec{\kappa}_{\|,\bar{m}})}
\right.\right.}\nonumber\\ &\quad&\left.\!
 +{f_{n}(\vec{\kappa}_{\|,\bar{n}})-f_{v}(\vec{\kappa}_{\|,\bar{v}})\over
 \tilde{\omega}_{nv}(\vec{\kappa}_{\|,\bar{n}},\vec{\kappa}_{\|,\bar{v}})-\omega}
 \right)
 j_{h,v{}n}(z'';\vec{\kappa}_{\|,\bar{v}}+\vec{\kappa}_{\|,\bar{n}})
 \psi_{v}^{*}(z')
 \psi_{m}(z')
 \delta(\vec{\kappa}_{\|,\bar{m}}-\vec{\kappa}_{\|,\bar{v}}+\vec{q}_{\|}^{\,\prime\prime\prime}
 -\vec{q}_{\|}^{\,\prime})
\nonumber\\ &&\times
 \delta(\vec{\kappa}_{\|,\bar{v}}-\vec{\kappa}_{\|,\bar{n}}+\vec{q}_{\|}^{\,\prime\prime})
 +\left({f_{n}(\vec{\kappa}_{\|,\bar{n}})-f_{v}(\vec{\kappa}_{\|,\bar{v}})\over
 \tilde{\omega}_{nv}(\vec{\kappa}_{\|,\bar{n}},\vec{\kappa}_{\|,\bar{v}})}
 +{f_{m}(\vec{\kappa}_{\|,\bar{m}})-f_{v}(\vec{\kappa}_{\|,\bar{v}})\over
 \tilde{\omega}_{v{}m}(\vec{\kappa}_{\|,\bar{v}},\vec{\kappa}_{\|,\bar{m}})-\omega}
 \right)
\nonumber\\ &&\times\left.
 j_{h,mv}(z'';\vec{\kappa}_{\|,\bar{m}}+\vec{\kappa}_{\|,\bar{v}})
 \psi_{n}^{*}(z')
 \psi_{v}(z')
 \delta(\vec{\kappa}_{\|,\bar{v}}-\vec{\kappa}_{\|,\bar{n}}+\vec{q}_{\|}^{\,\prime\prime\prime}
 -\vec{q}_{\|}^{\,\prime})
 \delta(\vec{\kappa}_{\|,\bar{m}}-\vec{\kappa}_{\|,\bar{v}}+\vec{q}_{\|}^{\,\prime\prime})
 \right\}
\nonumber\\ &&\times
 j_{i,nm}(z;\vec{\kappa}_{\|,\bar{n}}+\vec{\kappa}_{\|,\bar{m}})
 \delta(z'-z''')
 A_{h}(z''';\vec{q}_{\|}^{\,\prime\prime\prime})
 A_{x}(z'';\vec{q}_{\|}^{\,\prime\prime})
 A_{x}^{*}(z';\vec{q}_{\|}^{\,\prime})
\nonumber\\ &&\times
 \delta(\vec{\kappa}_{\|,\bar{n}}-\vec{\kappa}_{\|,\bar{m}}-\vec{q}_{\|})
 d^2\kappa_{\|,\bar{n}}d^2\kappa_{\|,\bar{m}}d^2\kappa_{\|,\bar{v}}
 d^2q_{\|}'''d^2q_{\|}''d^2q_{\|}'
 dz'''dz''dz'.
\end{eqnarray}
Of the three integrals over $\vec{\kappa}_{\|}$ quantities, we can
solve two because of the coupling to the wavevectors introduced by
the Dirac delta functions appearing. We aim at keeping the
$\vec{\kappa}_{\|,\bar{m}}$ set, and thus we solve
the integrals for the $v$ and $n$ sets (in that order). Solving
for the $v$ set, we find that $\vec{\kappa}_{\|,\bar{v}}$ is
replaced by $\vec{\kappa}_{\|,\bar{m}}+\vec{q}_{\|}^{\,\prime\prime\prime}-\vec{q}_{\|}^{\,\prime}$
in the first part of the sum and by
$\vec{\kappa}_{\|,\bar{m}}+\vec{q}_{\|}^{\,\prime\prime}$ in the second part of the
sum, which afterwards allows us to solve the $n$ set by replacing
$\vec{\kappa}_{\|,\bar{n}}$ with
$\vec{\kappa}_{\|,\bar{m}}+\vec{q}_{\|}^{\,\prime\prime\prime}+\vec{q}_{\|}^{\,\prime\prime}-\vec{q}_{\|}^{\,\prime}$ in
general, giving
\begin{eqnarray}
\lefteqn{
 J_{ixxh}^{\rm{F}}(z;\vec{q}_{\|})=
 -{e^2\over8m_{e}\hbar^2}
 {2\over(2\pi)^6}
 \idotsint
 \sum_{nmv}
 {1\over\tilde{\omega}_{nm}(\vec{\kappa}_{\|,\bar{m}}+\vec{q}_{\|}^{\,\prime\prime\prime}+\vec{q}_{\|}^{\,\prime\prime}
 -\vec{q}_{\|}^{\,\prime},\vec{\kappa}_{\|,\bar{m}})-\omega}
}\nonumber\\ &\quad&\times
\left\{
 \left({f_{m}(\vec{\kappa}_{\|,\bar{m}})-f_{v}(\vec{\kappa}_{\|,\bar{m}}+\vec{q}_{\|}^{\,\prime\prime\prime}
 -\vec{q}_{\|}^{\,\prime})\over\tilde{\omega}_{v{}m}(\vec{\kappa}_{\|,\bar{m}}
 +\vec{q}_{\|}^{\,\prime\prime\prime}-\vec{q}_{\|}^{\,\prime},\vec{\kappa}_{\|,\bar{m}})}
\right.\right.\nonumber\\ &&\left.\!
 +{f_{n}(\vec{\kappa}_{\|,\bar{m}}+\vec{q}_{\|}^{\,\prime\prime\prime}+\vec{q}_{\|}^{\,\prime\prime}-\vec{q}_{\|}^{\,\prime})
 -f_{v}(\vec{\kappa}_{\|,\bar{m}}+\vec{q}_{\|}^{\,\prime\prime\prime}-\vec{q}_{\|}^{\,\prime})\over
 \tilde{\omega}_{nv}(\vec{\kappa}_{\|,\bar{m}}+\vec{q}_{\|}^{\,\prime\prime\prime}+\vec{q}_{\|}^{\,\prime\prime}
 -\vec{q}_{\|}^{\,\prime},\vec{\kappa}_{\|,\bar{m}}+\vec{q}_{\|}^{\,\prime\prime\prime}-\vec{q}_{\|}^{\,\prime})-\omega}
 \right)
\nonumber\\ &&\times
 j_{h,v{}n}(z'';2\vec{\kappa}_{\|,\bar{m}}+2\vec{q}_{\|}^{\,\prime\prime\prime}+\vec{q}_{\|}^{\,\prime\prime}
 -\vec{q}_{\|}^{\,\prime})
 \psi_{v}^{*}(z')
 \psi_{m}(z')
\nonumber\\ &&
 +\left({f_{n}(\vec{\kappa}_{\|,\bar{m}}+\vec{q}_{\|}^{\,\prime\prime\prime}+\vec{q}_{\|}^{\,\prime\prime}
 -\vec{q}_{\|}^{\,\prime})-f_{v}(\vec{\kappa}_{\|,\bar{m}}+\vec{q}_{\|}^{\,\prime\prime})\over
 \tilde{\omega}_{nv}(\vec{\kappa}_{\|,\bar{m}}+\vec{q}_{\|}^{\,\prime\prime\prime}+\vec{q}_{\|}^{\,\prime\prime}
 -\vec{q}_{\|}^{\,\prime},\vec{\kappa}_{\|,\bar{m}}+\vec{q}_{\|}^{\,\prime\prime})}
 +{f_{m}(\vec{\kappa}_{\|,\bar{m}})-f_{v}(\vec{\kappa}_{\|,\bar{m}}+\vec{q}_{\|}^{\,\prime\prime})\over
 \tilde{\omega}_{v{}m}(\vec{\kappa}_{\|,\bar{m}}+\vec{q}_{\|}^{\,\prime\prime},
 \vec{\kappa}_{\|,\bar{m}})-\omega}
 \right)
\nonumber\\ &&\times\left.\!
 j_{h,mv}(z'';2\vec{\kappa}_{\|,\bar{m}}+\vec{q}_{\|}^{\,\prime\prime})
 \psi_{n}^{*}(z')
 \psi_{v}(z')
 \right\}
 j_{i,nm}(z;2\vec{\kappa}_{\|,\bar{m}}+\vec{q}_{\|}^{\,\prime\prime\prime}+\vec{q}_{\|}^{\,\prime\prime}-\vec{q}_{\|}^{\,\prime})
 \delta(z'-z''')
\nonumber\\ &&\times
 A_{h}(z''';\vec{q}_{\|}^{\,\prime\prime\prime})
 A_{x}(z'';\vec{q}_{\|}^{\,\prime\prime})
 A_{x}^{*}(z';\vec{q}_{\|}^{\,\prime})
 \delta(\vec{q}_{\|}^{\,\prime\prime\prime}+\vec{q}_{\|}^{\,\prime\prime}-\vec{q}_{\|}^{\,\prime}-\vec{q}_{\|})
 d^2\kappa_{\|,\bar{m}}
 d^2q_{\|}'''d^2q_{\|}''d^2q_{\|}'
\nonumber\\ &&\times
 dz'''dz''dz'.
\label{eq:J3F}
\end{eqnarray}
On the form of Eq.~(\ref{eq:J3zq}) we thus get part F of the
conductivity tensor as
\begin{eqnarray}
\lefteqn{ 
 {\Xi}_{ixxh}^{\rm{F}}(z,z',z'',z''';\vec{q}_{\|},\vec{q}_{\|}^{\,\prime},
 \vec{q}_{\|}^{\,\prime\prime},\vec{q}_{\|}^{\,\prime\prime\prime})=
}\nonumber\\ &\quad&
 {2{\rm{i}}\over\omega^3}{e^2\over8m_{e}\hbar^2}{1\over(2\pi)^2}\sum_{nmv}\int
 {1\over\tilde{\omega}_{nm}(\vec{\kappa}_{\|}+\vec{q}_{\|}^{\,\prime\prime\prime}+\vec{q}_{\|}^{\,\prime\prime}
 -\vec{q}_{\|}^{\,\prime},\vec{\kappa}_{\|})-\omega}
\nonumber\\ &&\times
\left\{
 \left({f_{m}(\vec{\kappa}_{\|})-f_{v}(\vec{\kappa}_{\|}+\vec{q}_{\|}^{\,\prime\prime\prime}
 -\vec{q}_{\|}^{\,\prime})\over\tilde{\omega}_{v{}m}(\vec{\kappa}_{\|}
 +\vec{q}_{\|}^{\,\prime\prime\prime}-\vec{q}_{\|}^{\,\prime},\vec{\kappa}_{\|})}
 +{f_{n}(\vec{\kappa}_{\|}+\vec{q}_{\|}^{\,\prime\prime\prime}+\vec{q}_{\|}^{\,\prime\prime}-\vec{q}_{\|}^{\,\prime})
 -f_{v}(\vec{\kappa}_{\|}+\vec{q}_{\|}^{\,\prime\prime\prime}-\vec{q}_{\|}^{\,\prime})\over
 \tilde{\omega}_{nv}(\vec{\kappa}_{\|}+\vec{q}_{\|}^{\,\prime\prime\prime}+\vec{q}_{\|}^{\,\prime\prime}
 -\vec{q}_{\|}^{\,\prime},\vec{\kappa}_{\|}+\vec{q}_{\|}^{\,\prime\prime\prime}-\vec{q}_{\|}^{\,\prime})-\omega}
 \right)
\right.\nonumber\\ &&\times
 j_{h,v{}n}(z'';2\vec{\kappa}_{\|}+2\vec{q}_{\|}^{\,\prime\prime\prime}+\vec{q}_{\|}^{\,\prime\prime}
 -\vec{q}_{\|}^{\,\prime})
 \psi_{v}^{*}(z')\psi_{m}(z')
\nonumber\\ &&
 +\left({f_{n}(\vec{\kappa}_{\|}+\vec{q}_{\|}^{\,\prime\prime\prime}+\vec{q}_{\|}^{\,\prime\prime}
 -\vec{q}_{\|}^{\,\prime})-f_{v}(\vec{\kappa}_{\|}+\vec{q}_{\|}^{\,\prime\prime})\over
 \tilde{\omega}_{nv}(\vec{\kappa}_{\|}+\vec{q}_{\|}^{\,\prime\prime\prime}+\vec{q}_{\|}^{\,\prime\prime}
 -\vec{q}_{\|}^{\,\prime},\vec{\kappa}_{\|}+\vec{q}_{\|}^{\,\prime\prime})}
 +{f_{m}(\vec{\kappa}_{\|})-f_{v}(\vec{\kappa}_{\|}+\vec{q}_{\|}^{\,\prime\prime})\over
 \tilde{\omega}_{v{}m}(\vec{\kappa}_{\|}+\vec{q}_{\|}^{\,\prime\prime},
 \vec{\kappa}_{\|})-\omega}
 \right)
\nonumber\\ &&\times\left.\!
 j_{h,mv}(z'';2\vec{\kappa}_{\|}+\vec{q}_{\|}^{\,\prime\prime})
 \psi_{n}^{*}(z')\psi_{v}(z')
 \right\}
 j_{i,nm}(z;2\vec{\kappa}_{\|}+\vec{q}_{\|}^{\,\prime\prime\prime}+\vec{q}_{\|}^{\,\prime\prime}-\vec{q}_{\|}^{\,\prime})
 \delta(z'-z''')
\nonumber\\ &&\times
 \delta(\vec{q}_{\|}^{\,\prime\prime\prime}+\vec{q}_{\|}^{\,\prime\prime}-\vec{q}_{\|}^{\,\prime}-\vec{q}_{\|})
 d^2\kappa_{\|},
\end{eqnarray}
where we have omitted the now superfluous index $m$ from the surface
states $\vec{\kappa}_{\|}$. Inserting the DFWM pump fields defined by
Eqs.~(\ref{eq:Apump-1}) and (\ref{eq:Apump-2}) into
Eq.~(\ref{eq:J3F}), the integrals over $q_{\|}'''$ and $q_{\|}''$ can
be solved. The resulting expression can then be solved for $q_{\|}'$
for the same reason as before, and on the form of
Eq.~(\ref{eq:J3-wzq}), the PCDFWM conductivity tensor part F appears
as Eq.~(\ref{eq:XiAF}).

\section{Nonlinear process G}
Inserting Eq.~(\ref{eq:XiAGrz}) into Eq.~(\ref{eq:J3r}), we take
element $ijkh$ of part G of the nonlinear current density. In the
result we insert the expressions for the wave function, the vector
potential and the transition current density in the two-dimensional
Fourier representation [Eqs.~(\ref{eq:eigenstate}), (\ref{eq:Azr}) and
(\ref{eq:Jm->n||}), respectively]. Then by use of Eq.~(\ref{eq:Jzq})
we find
\begin{eqnarray}
\lefteqn{
 J_{ijkh}^{\rm{G}}(z;\vec{q}_{\|})=
 -{1\over8\hbar^3}
 {2\over(2\pi)^{14}}
 \idotsint
 \sum_{nmvl}
 {1\over\tilde{\omega}_{nm}-\omega}
\left\{
\left[
 \left({f_{l}-f_{m}\over\tilde{\omega}_{l{}m}-\omega}
 +{f_{l}-f_{v}\over\tilde{\omega}_{vl}-\omega}\right)
 {1\over\tilde{\omega}_{v{}m}-2\omega}
\right.\right.}\nonumber\\ &\quad&\left.\!
 +\left({f_{l}-f_{v}\over\tilde{\omega}_{vl}-\omega}
 +{f_{n}-f_{v}\over\tilde{\omega}_{nv}+\omega}\right)
 {1\over\tilde{\omega}_{nl}}
\right]
 j_{h,ml}(z''';\vec{\kappa}_{\|,\bar{m}}+\vec{\kappa}_{\|,\bar{l}})
 j_{k,lv}(z'';\vec{\kappa}_{\|,\bar{l}}+\vec{\kappa}_{\|,\bar{v}})
\nonumber\\ &&\times
 j_{j,v{}n}(z';\vec{\kappa}_{\|,\bar{v}}+\vec{\kappa}_{\|,\bar{n}})
 e^{{\rm{i}}(\vec{\kappa}_{\|,\bar{m}}-\vec{\kappa}_{\|,\bar{l}})\cdot\vec{r}_{\|}^{\,\prime\prime\prime}}
 e^{{\rm{i}}(\vec{\kappa}_{\|,\bar{l}}-\vec{\kappa}_{\|,\bar{v}})\cdot\vec{r}_{\|}^{\,\prime\prime}}
 e^{{\rm{i}}(\vec{\kappa}_{\|,\bar{v}}-\vec{\kappa}_{\|,\bar{n}})\cdot\vec{r}_{\|}^{\,\prime}}
\nonumber\\ &&
 +
\left[
 \left({f_{l}-f_{m}\over\tilde{\omega}_{l{}m}-\omega}
 +{f_{l}-f_{v}\over\tilde{\omega}_{vl}+\omega}\right)
 {1\over\tilde{\omega}_{v{}m}}
 +\left({f_{l}-f_{v}\over\tilde{\omega}_{vl}+\omega}
 +{f_{n}-f_{v}\over\tilde{\omega}_{nv}-\omega}\right)
 {1\over\tilde{\omega}_{nl}}
\right]
\nonumber\\ &&\times
 j_{h,ml}(z''';\vec{\kappa}_{\|,\bar{m}}+\vec{\kappa}_{\|,\bar{l}})
 j_{k,v{}n}(z'';\vec{\kappa}_{\|,\bar{v}}+\vec{\kappa}_{\|,\bar{n}})
 j_{j,lv}(z';\vec{\kappa}_{\|,\bar{l}}+\vec{\kappa}_{\|,\bar{v}})
 e^{{\rm{i}}(\vec{\kappa}_{\|,\bar{m}}-\vec{\kappa}_{\|,\bar{l}})\cdot\vec{r}_{\|}^{\,\prime\prime\prime}}
\nonumber\\ &&\times
 e^{{\rm{i}}(\vec{\kappa}_{\|,\bar{v}}-\vec{\kappa}_{\|,\bar{n}})\cdot\vec{r}_{\|}^{\,\prime\prime}}
 e^{{\rm{i}}(\vec{\kappa}_{\|,\bar{l}}-\vec{\kappa}_{\|,\bar{v}})\cdot\vec{r}_{\|}^{\,\prime}}
 +
\left[
 \left({f_{l}-f_{m}\over\tilde{\omega}_{l{}m}+\omega}
 +{f_{l}-f_{v}\over\tilde{\omega}_{vl}-\omega}\right)
 {1\over\tilde{\omega}_{v{}m}}
\right.\nonumber\\ &&\left.\!
 +\left({f_{l}-f_{v}\over\tilde{\omega}_{vl}-\omega}
 +{f_{n}-f_{v}\over\tilde{\omega}_{nv}-\omega}\right)
 {1\over\tilde{\omega}_{nl}-2\omega}
\right]
 j_{h,lv}(z''';\vec{\kappa}_{\|,\bar{l}}+\vec{\kappa}_{\|,\bar{v}})
 j_{k,v{}n}(z'';\vec{\kappa}_{\|,\bar{v}}+\vec{\kappa}_{\|,\bar{n}})
\nonumber\\ &&\times\left.\!
 j_{j,ml}(z';\vec{\kappa}_{\|,\bar{m}}+\vec{\kappa}_{\|,\bar{l}})
 e^{{\rm{i}}(\vec{\kappa}_{\|,\bar{l}}-\vec{\kappa}_{\|,\bar{v}})\cdot\vec{r}_{\|}^{\,\prime\prime\prime}}
 e^{{\rm{i}}(\vec{\kappa}_{\|,\bar{v}}-\vec{\kappa}_{\|,\bar{n}})\cdot\vec{r}_{\|}^{\,\prime\prime}}
 e^{{\rm{i}}(\vec{\kappa}_{\|,\bar{m}}-\vec{\kappa}_{\|,\bar{l}})\cdot\vec{r}_{\|}^{\,\prime}}
\right\}
\nonumber\\ &&\times
 j_{i,nm}(z;\vec{\kappa}_{\|,\bar{n}}+\vec{\kappa}_{\|,\bar{m}})
 A_{h}(z''';\vec{q}_{\|}^{\,\prime\prime\prime})
 A_{k}(z'';\vec{q}_{\|}^{\,\prime\prime})
 A_{j}^{*}(z';\vec{q}_{\|}^{\,\prime})
 e^{{\rm{i}}(\vec{\kappa}_{\|,\bar{n}}-\vec{\kappa}_{\|,\bar{m}})\cdot\vec{r}_{\|}}
\nonumber\\ &&\times
 e^{{\rm{i}}\vec{q}_{\|}^{\,\prime\prime\prime}\cdot\vec{r}_{\|}^{\,\prime\prime\prime}}
 e^{{\rm{i}}\vec{q}_{\|}^{\,\prime\prime}\cdot\vec{r}_{\|}^{\,\prime\prime}}
 e^{-{\rm{i}}\vec{q}_{\|}^{\,\prime}\cdot\vec{r}_{\|}^{\,\prime}}
 d^2q_{\|}'''d^2q_{\|}''d^2q_{\|}'
 d^3r'''d^3r''d^3r'e^{-{\rm{i}}\vec{q}_{\|}\cdot\vec{r}_{\|}}d^2r_{\|}.
\end{eqnarray}
In the above equation, we may immediately solve the integrals
$\int{}d^2r_{\|}'''$, $\int{}d^2r_{\|}''$, $\int{}d^2r_{\|}'$, and
$\int{}d^2r_{\|}$, and by replacing the sums over the various
$\vec{\kappa}_{\|}$ quantities with integrals, as before, we get
\begin{eqnarray}
\lefteqn{
 J_{ijkh}^{G}(z;\vec{q}_{\|})=
 -{1\over8\hbar^3}{2\over(2\pi)^6}
 \idotsint\sum_{nmvl}
 {1\over\tilde{\omega}_{nm}(\vec{\kappa}_{\|,\bar{n}},
  \vec{\kappa}_{\|,\bar{m}})-\omega}
}\nonumber\\ &&\times
\left\{
\left[
 \left({f_{l}(\vec{\kappa}_{\|,\bar{l}})-f_{m}(\vec{\kappa}_{\|,\bar{m}})
  \over\tilde{\omega}_{l{}m}(\vec{\kappa}_{\|,\bar{l}},
  \vec{\kappa}_{\|,\bar{m}})-\omega}
 +{f_{l}(\vec{\kappa}_{\|,\bar{l}})-f_{v}(\vec{\kappa}_{\|,\bar{v}})
  \over\tilde{\omega}_{vl}(\vec{\kappa}_{\|,\bar{v}},
  \vec{\kappa}_{\|,\bar{l}})-\omega}
 \right)
 {1\over\tilde{\omega}_{v{}m}(\vec{\kappa}_{\|,\bar{v}},
  \vec{\kappa}_{\|,\bar{m}})-2\omega}
\right.\right.\nonumber\\ &\quad&\left.\!
 +\left({f_{l}(\vec{\kappa}_{\|,\bar{l}})
  -f_{v}(\vec{\kappa}_{\|,\bar{v}})\over
  \tilde{\omega}_{vl}(\vec{\kappa}_{\|,\bar{v}},
  \vec{\kappa}_{\|,\bar{l}})-\omega}
 +{f_{n}(\vec{\kappa}_{\|,\bar{n}})-f_{v}(\vec{\kappa}_{\|,\bar{v}})\over
  \tilde{\omega}_{nv}(\vec{\kappa}_{\|,\bar{n}},\vec{\kappa}_{\|,\bar{v}})
  +\omega}
 \right)
 {1\over\tilde{\omega}_{nl}(\vec{\kappa}_{\|,\bar{n}},
  \vec{\kappa}_{\|,\bar{l}})}
\right]
\nonumber\\ &&\times
 j_{h,ml}(z''';\vec{\kappa}_{\|,\bar{m}}+\vec{\kappa}_{\|,\bar{l}})
 j_{k,lv}(z'';\vec{\kappa}_{\|,\bar{l}}+\vec{\kappa}_{\|,\bar{v}})
 j_{j,v{}n}(z';\vec{\kappa}_{\|,\bar{v}}+\vec{\kappa}_{\|,\bar{n}})
\nonumber\\ &&\times
 \delta(\vec{\kappa}_{\|,\bar{m}}-\vec{\kappa}_{\|,\bar{l}}+\vec{q}_{\|}^{\,\prime\prime\prime})
 \delta(\vec{\kappa}_{\|,\bar{l}}-\vec{\kappa}_{\|,\bar{v}}
  +\vec{q}_{\|}^{\,\prime\prime})
 \delta(\vec{\kappa}_{\|,\bar{v}}-\vec{\kappa}_{\|,\bar{n}}-\vec{q}_{\|}^{\,\prime})
\nonumber\\ &&
 +
\left[
 \left({f_{l}(\vec{\kappa}_{\|,\bar{l}})-f_{m}(\vec{\kappa}_{\|,\bar{m}})
  \over\tilde{\omega}_{l{}m}(\vec{\kappa}_{\|,\bar{l}},
  \vec{\kappa}_{\|,\bar{m}})-\omega}
 +{f_{l}(\vec{\kappa}_{\|,\bar{l}})-f_{v}(\vec{\kappa}_{\|,\bar{v}})
  \over\tilde{\omega}_{vl}(\vec{\kappa}_{\|,\bar{v}},
  \vec{\kappa}_{\|,\bar{l}})+\omega}
 \right)
 {1\over\tilde{\omega}_{v{}m}(\vec{\kappa}_{\|,\bar{v}},
  \vec{\kappa}_{\|,\bar{m}})}
\right.\nonumber\\ &&\left.
 +\left({f_{l}(\vec{\kappa}_{\|,\bar{l}})
  -f_{v}(\vec{\kappa}_{\|,\bar{v}})\over
  \tilde{\omega}_{vl}(\vec{\kappa}_{\|,\bar{v}},
  \vec{\kappa}_{\|,\bar{l}})+\omega}
 +{f_{n}(\vec{\kappa}_{\|,\bar{n}})-f_{v}(\vec{\kappa}_{\|,\bar{v}})\over
  \tilde{\omega}_{nv}(\vec{\kappa}_{\|,\bar{n}},\vec{\kappa}_{\|,\bar{v}})
  -\omega}
 \right)
 {1\over\tilde{\omega}_{nl}(\vec{\kappa}_{\|,\bar{n}},
  \vec{\kappa}_{\|,\bar{l}})}
\right]
\nonumber\\ &&\times
 j_{h,ml}(z''';\vec{\kappa}_{\|,\bar{m}}+\vec{\kappa}_{\|,\bar{l}})
 j_{k,v{}n}(z'';\vec{\kappa}_{\|,\bar{v}}+\vec{\kappa}_{\|,\bar{n}})
 j_{j,lv}(z';\vec{\kappa}_{\|,\bar{l}}+\vec{\kappa}_{\|,\bar{v}})
\nonumber\\ &&\times
 \delta(\vec{\kappa}_{\|,\bar{m}}-\vec{\kappa}_{\|,\bar{l}}+\vec{q}_{\|}^{\,\prime\prime\prime})
 \delta(\vec{\kappa}_{\|,\bar{v}}-\vec{\kappa}_{\|,\bar{n}}+\vec{q}_{\|}^{\,\prime\prime})
 \delta(\vec{\kappa}_{\|,\bar{l}}-\vec{\kappa}_{\|,\bar{v}}-\vec{q}_{\|}^{\,\prime})
\nonumber\\ &&
 +
\left[
 \left({f_{l}(\vec{\kappa}_{\|,\bar{l}})-f_{m}(\vec{\kappa}_{\|,\bar{m}})
  \over\tilde{\omega}_{l{}m}(\vec{\kappa}_{\|,\bar{l}},
  \vec{\kappa}_{\|,\bar{m}})+\omega}
 +{f_{l}(\vec{\kappa}_{\|,\bar{l}})-f_{v}(\vec{\kappa}_{\|,\bar{v}})
  \over\tilde{\omega}_{vl}(\vec{\kappa}_{\|,\bar{v}},
  \vec{\kappa}_{\|,\bar{l}})-\omega}
 \right)
 {1\over\tilde{\omega}_{v{}m}(\vec{\kappa}_{\|,\bar{v}},
  \vec{\kappa}_{\|,\bar{m}})}
\right.\nonumber\\ &&\left.
 +\left({f_{l}(\vec{\kappa}_{\|,\bar{l}})
  -f_{v}(\vec{\kappa}_{\|,\bar{v}})\over
 \tilde{\omega}_{vl}(\vec{\kappa}_{\|,\bar{v}},
  \vec{\kappa}_{\|,\bar{l}})-\omega}
 +{f_{n}(\vec{\kappa}_{\|,\bar{n}})-f_{v}(\vec{\kappa}_{\|,\bar{v}})\over
  \tilde{\omega}_{nv}(\vec{\kappa}_{\|,\bar{n}},\vec{\kappa}_{\|,\bar{v}})
  -\omega}
 \right)
 {1\over\tilde{\omega}_{nl}(\vec{\kappa}_{\|,\bar{n}},
  \vec{\kappa}_{\|,\bar{l}})-2\omega}
\right]
\nonumber\\ &&\times
 j_{h,lv}(z''';\vec{\kappa}_{\|,\bar{l}}+\vec{\kappa}_{\|,\bar{v}})
 j_{k,v{}n}(z'';\vec{\kappa}_{\|,\bar{v}}+\vec{\kappa}_{\|,\bar{n}})
 j_{j,ml}(z';\vec{\kappa}_{\|,\bar{m}}+\vec{\kappa}_{\|,\bar{l}})
\nonumber\\ &&\times\left.
 \delta(\vec{\kappa}_{\|,\bar{l}}-\vec{\kappa}_{\|,\bar{v}}+\vec{q}_{\|}^{\,\prime\prime\prime})
 \delta(\vec{\kappa}_{\|,\bar{v}}-\vec{\kappa}_{\|,\bar{n}}+\vec{q}_{\|}^{\,\prime\prime})
 \delta(\vec{\kappa}_{\|,\bar{m}}-\vec{\kappa}_{\|,\bar{l}}-\vec{q}_{\|}^{\,\prime})
\right\}
 j_{i,nm}(z;\vec{\kappa}_{\|,\bar{n}}+\vec{\kappa}_{\|,\bar{m}})
\nonumber\\ &&\times
 A_{h}(z''';\vec{q}_{\|}^{\,\prime\prime\prime})
 A_{k}(z'';\vec{q}_{\|}^{\,\prime\prime})
 A_{j}^{*}(z';\vec{q}_{\|}^{\,\prime})
 \delta(\vec{\kappa}_{\|,\bar{n}}-\vec{\kappa}_{\|,\bar{m}}-\vec{q}_{\|})
 d^2\kappa_{\|,\bar{n}}d^2\kappa_{\|,\bar{m}}d^2\kappa_{\|,\bar{v}}d^2\kappa_{\|,\bar{l}}
\nonumber\\ &&\times
 d^2q_{\|}'''d^2q_{\|}''d^2q_{\|}'
 dz'''dz''dz'.
\end{eqnarray}
Of the four integrals over $\vec{\kappa}_{\|}$ quantities, we can
solve three because of the coupling to the wavevectors introduced by
the Dirac delta functions appearing. We aim at keeping the
$\vec{\kappa}_{\|,\bar{m}}$ set, and thus we solve the integrals for
the $l$, $v$, and $n$ sets. Thus (i), in the first part of the sum, we
find that $\vec{\kappa}_{\|,\bar{l}}$ can be replaced by
$\vec{\kappa}_{\|,\bar{m}}+\vec{q}_{\|}^{\,\prime\prime\prime}$, then
letting us replace $\vec{\kappa}_{\|,\bar{v}}$ by
$\vec{\kappa}_{\|,\bar{m}}+\vec{q}_{\|}^{\,\prime\prime\prime}+\vec{q}_{\|}^{\,\prime\prime}$,
which again let us replace $\vec{\kappa}_{\|,\bar{n}}$ by
$\vec{\kappa}_{\|,\bar{m}}+\vec{q}_{\|}^{\,\prime\prime\prime}+\vec{q}_{\|}^{\,\prime\prime}-\vec{q}_{\|}^{\,\prime}$.
(ii) In the second part of the sum, we find that
$\vec{\kappa}_{\|,\bar{l}}$ can be replaced by
$\vec{\kappa}_{\|,\bar{m}}+\vec{q}_{\|}^{\,\prime\prime\prime}$, then
letting us replace $\vec{\kappa}_{\|,\bar{v}}$ by
$\vec{\kappa}_{\|,\bar{m}}+\vec{q}_{\|}^{\,\prime\prime\prime}-\vec{q}_{\|}^{\,\prime}$,
which again let us replace $\vec{\kappa}_{\|,\bar{n}}$ by
$\vec{\kappa}_{\|,\bar{m}}+\vec{q}_{\|}^{\,\prime\prime\prime}+\vec{q}_{\|}^{\,\prime\prime}-\vec{q}_{\|}^{\,\prime}$.
(iii) In the third part of the sum, we find that $\vec{\kappa}_{\|,\bar{l}}$
can be replaced by $\vec{\kappa}_{\|,\bar{m}}-\vec{q}_{\|}^{\,\prime}$,
then letting us replace $\vec{\kappa}_{\|,\bar{v}}$ by
$\vec{\kappa}_{\|,\bar{m}}+\vec{q}_{\|}^{\,\prime\prime\prime}-\vec{q}_{\|}^{\,\prime}$,
which again let us replace $\vec{\kappa}_{\|,\bar{n}}$ by
$\vec{\kappa}_{\|,\bar{m}}+\vec{q}_{\|}^{\,\prime\prime\prime}+\vec{q}_{\|}^{\,\prime\prime}-\vec{q}_{\|}^{\,\prime}$.
Finally (iv), we observe that the substitution of
$\vec{\kappa}_{\|,\bar{n}}$ by
$\vec{\kappa}_{\|,\bar{m}}+\vec{q}_{\|}^{\,\prime\prime\prime}+\vec{q}_{\|}^{\,\prime\prime}-\vec{q}_{\|}^{\,\prime}$
is global, and we thus get the resulting current density element
\begin{eqnarray}
\lefteqn{
 J_{ijkh}^{\rm{G}}(z;\vec{q}_{\|})=
 -{1\over8\hbar^3}{2\over(2\pi)^6}
 \idotsint\sum_{nmvl}
 {1\over\tilde{\omega}_{nm}(\vec{\kappa}_{\|,\bar{m}}+\vec{q}_{\|}^{\,\prime\prime\prime}+\vec{q}_{\|}^{\,\prime\prime}
 -\vec{q}_{\|}^{\,\prime},\vec{\kappa}_{\|,\bar{m}})-\omega}
}\nonumber\\ &\quad&\times
\left\{
\left[
 \left({f_{l}(\vec{\kappa}_{\|,\bar{m}}+\vec{q}_{\|}^{\,\prime\prime\prime})
 -f_{m}(\vec{\kappa}_{\|,\bar{m}})\over\tilde{\omega}_{l{}m}(\vec{\kappa}_{\|,\bar{m}}
 +\vec{q}_{\|}^{\,\prime\prime\prime},\vec{\kappa}_{\|,\bar{m}})-\omega}
 +{f_{l}(\vec{\kappa}_{\|,\bar{m}}+\vec{q}_{\|}^{\,\prime\prime\prime})-f_{v}(\vec{\kappa}_{\|,\bar{m}}
 +\vec{q}_{\|}^{\,\prime\prime\prime}+\vec{q}_{\|}^{\,\prime\prime})\over
 \tilde{\omega}_{vl}(\vec{\kappa}_{\|,\bar{m}}+\vec{q}_{\|}^{\,\prime\prime\prime}+\vec{q}_{\|}^{\,\prime\prime},
 \vec{\kappa}_{\|,\bar{m}}+\vec{q}_{\|}^{\,\prime\prime\prime})-\omega}\right)
\right.\right.\nonumber\\ &&\times
 {1\over\tilde{\omega}_{v{}m}(\vec{\kappa}_{\|,\bar{m}}+\vec{q}_{\|}^{\,\prime\prime\prime}
 +\vec{q}_{\|}^{\,\prime\prime},\vec{\kappa}_{\|,\bar{m}})-2\omega}
 +\left({f_{l}(\vec{\kappa}_{\|,\bar{m}}+\vec{q}_{\|}^{\,\prime\prime\prime})
 -f_{v}(\vec{\kappa}_{\|,\bar{m}}+\vec{q}_{\|}^{\,\prime\prime\prime}+\vec{q}_{\|}^{\,\prime\prime})\over
 \tilde{\omega}_{vl}(\vec{\kappa}_{\|,\bar{m}}+\vec{q}_{\|}^{\,\prime\prime\prime}+\vec{q}_{\|}^{\,\prime\prime},
 \vec{\kappa}_{\|,\bar{m}}+\vec{q}_{\|}^{\,\prime\prime\prime})-\omega}
\right.\nonumber\\ &&\left.\!
 +{f_{n}(\vec{\kappa}_{\|,\bar{m}}+\vec{q}_{\|}^{\,\prime\prime\prime}+\vec{q}_{\|}^{\,\prime\prime}-\vec{q}_{\|}^{\,\prime})
 -f_{v}(\vec{\kappa}_{\|,\bar{m}}+\vec{q}_{\|}^{\,\prime\prime\prime}+\vec{q}_{\|}^{\,\prime\prime})\over
 \tilde{\omega}_{nv}(\vec{\kappa}_{\|,\bar{m}}+\vec{q}_{\|}^{\,\prime\prime\prime}+\vec{q}_{\|}^{\,\prime\prime}
 -\vec{q}_{\|}^{\,\prime},\vec{\kappa}_{\|,\bar{m}}+\vec{q}_{\|}^{\,\prime\prime\prime}+\vec{q}_{\|}^{\,\prime\prime})+\omega}
 \right)
\nonumber\\ &&\times\left.\!
 {1\over\tilde{\omega}_{nl}(\vec{\kappa}_{\|,\bar{m}}+\vec{q}_{\|}^{\,\prime\prime\prime}
 +\vec{q}_{\|}^{\,\prime\prime}-\vec{q}_{\|}^{\,\prime},\vec{\kappa}_{\|,\bar{m}}+\vec{q}_{\|}^{\,\prime\prime\prime})}
\right]
 j_{h,ml}(z''';2\vec{\kappa}_{\|,\bar{m}}+\vec{q}_{\|}^{\,\prime\prime\prime})
\nonumber\\ &&\times
 j_{k,lv}(z'';2\vec{\kappa}_{\|,\bar{m}}+2\vec{q}_{\|}^{\,\prime\prime\prime}+\vec{q}_{\|}^{\,\prime\prime})
 j_{j,v{}n}(z';2\vec{\kappa}_{\|,\bar{m}}+2\vec{q}_{\|}^{\,\prime\prime\prime}+2\vec{q}_{\|}^{\,\prime\prime}
 -\vec{q}_{\|}^{\,\prime})
\nonumber\\ &&
 +
\left[
 \left({f_{l}(\vec{\kappa}_{\|,\bar{m}}+\vec{q}_{\|}^{\,\prime\prime\prime})
 -f_{m}(\vec{\kappa}_{\|,\bar{m}})\over\tilde{\omega}_{l{}m}(\vec{\kappa}_{\|,\bar{m}}
 +\vec{q}_{\|}^{\,\prime\prime\prime},\vec{\kappa}_{\|,\bar{m}})-\omega}
 +{f_{l}(\vec{\kappa}_{\|,\bar{m}}+\vec{q}_{\|}^{\,\prime\prime\prime})-f_{v}(\vec{\kappa}_{\|,\bar{m}}
 +\vec{q}_{\|}^{\,\prime\prime\prime}-\vec{q}_{\|}^{\,\prime})\over
 \tilde{\omega}_{vl}(\vec{\kappa}_{\|,\bar{m}}+\vec{q}_{\|}^{\,\prime\prime\prime}-\vec{q}_{\|}^{\,\prime},
 \vec{\kappa}_{\|,\bar{m}}+\vec{q}_{\|}^{\,\prime\prime\prime})+\omega}\right)
\right.\nonumber\\ &&\times
 {1\over\tilde{\omega}_{v{}m}(\vec{\kappa}_{\|,\bar{m}}+\vec{q}_{\|}^{\,\prime\prime\prime}
 -\vec{q}_{\|}^{\,\prime},\vec{\kappa}_{\|,\bar{m}})}
 +\left({f_{l}(\vec{\kappa}_{\|,\bar{m}}+\vec{q}_{\|}^{\,\prime\prime\prime})
 -f_{v}(\vec{\kappa}_{\|,\bar{m}}+\vec{q}_{\|}^{\,\prime\prime\prime}-\vec{q}_{\|}^{\,\prime})\over
 \tilde{\omega}_{vl}(\vec{\kappa}_{\|,\bar{m}}+\vec{q}_{\|}^{\,\prime\prime\prime}-\vec{q}_{\|}^{\,\prime},
 \vec{\kappa}_{\|,\bar{m}}+\vec{q}_{\|}^{\,\prime\prime\prime})+\omega}
\right.\nonumber\\ &&\left.\!
 +{f_{n}(\vec{\kappa}_{\|,\bar{m}}+\vec{q}_{\|}^{\,\prime\prime\prime}+\vec{q}_{\|}^{\,\prime\prime}-\vec{q}_{\|}^{\,\prime})
 -f_{v}(\vec{\kappa}_{\|,\bar{m}}+\vec{q}_{\|}^{\,\prime\prime\prime}-\vec{q}_{\|}^{\,\prime})\over
 \tilde{\omega}_{nv}(\vec{\kappa}_{\|,\bar{m}}+\vec{q}_{\|}^{\,\prime\prime\prime}+\vec{q}_{\|}^{\,\prime\prime}
 -\vec{q}_{\|}^{\,\prime},\vec{\kappa}_{\|,\bar{m}}+\vec{q}_{\|}^{\,\prime\prime\prime}-\vec{q}_{\|}^{\,\prime})-\omega}
 \right)
\nonumber\\ &&\times\left.
 {1\over\tilde{\omega}_{nl}(\vec{\kappa}_{\|,\bar{m}}+\vec{q}_{\|}^{\,\prime\prime\prime}
 +\vec{q}_{\|}^{\,\prime\prime}-\vec{q}_{\|}^{\,\prime},\vec{\kappa}_{\|,\bar{m}}+\vec{q}_{\|}^{\,\prime\prime\prime})}
\right]
 j_{h,ml}(z''';2\vec{\kappa}_{\|,\bar{m}}+\vec{q}_{\|}^{\,\prime\prime\prime})
\nonumber\\ &&\times
 j_{k,v{}n}(z'';2\vec{\kappa}_{\|,\bar{m}}+2\vec{q}_{\|}^{\,\prime\prime\prime}+\vec{q}_{\|}^{\,\prime\prime}
 -2\vec{q}_{\|}^{\,\prime})
 j_{j,lv}(z';2\vec{\kappa}_{\|,\bar{m}}+2\vec{q}_{\|}^{\,\prime\prime\prime}-\vec{q}_{\|}^{\,\prime})
\nonumber\\ &&
 +
\left[
 \left({f_{l}(\vec{\kappa}_{\|,\bar{m}}-\vec{q}_{\|}^{\,\prime})-f_{m}(\vec{\kappa}_{\|,\bar{m}})
 \over\tilde{\omega}_{l{}m}(\vec{\kappa}_{\|,\bar{m}}-\vec{q}_{\|}^{\,\prime},
 \vec{\kappa}_{\|,\bar{m}})+\omega}
 +{f_{l}(\vec{\kappa}_{\|,\bar{m}}-\vec{q}_{\|}^{\,\prime})-f_{v}(\vec{\kappa}_{\|,\bar{m}}
 +\vec{q}_{\|}^{\,\prime\prime\prime}-\vec{q}_{\|}^{\,\prime})\over
 \tilde{\omega}_{vl}(\vec{\kappa}_{\|,\bar{m}}+\vec{q}_{\|}^{\,\prime\prime\prime}-\vec{q}_{\|}^{\,\prime},
 \vec{\kappa}_{\|,\bar{m}}-\vec{q}_{\|}^{\,\prime})-\omega}\right)
\right.\nonumber\\ &&\times
 {1\over\tilde{\omega}_{v{}m}(\vec{\kappa}_{\|,\bar{m}}+\vec{q}_{\|}^{\,\prime\prime\prime}
 -\vec{q}_{\|}^{\,\prime},\vec{\kappa}_{\|,\bar{m}})}
 +\left({f_{l}(\vec{\kappa}_{\|,\bar{m}}-\vec{q}_{\|}^{\,\prime})
 -f_{v}(\vec{\kappa}_{\|,\bar{m}}+\vec{q}_{\|}^{\,\prime\prime\prime}-\vec{q}_{\|}^{\,\prime})\over
 \tilde{\omega}_{vl}(\vec{\kappa}_{\|,\bar{m}}+\vec{q}_{\|}^{\,\prime\prime\prime}-\vec{q}_{\|}^{\,\prime},
 \vec{\kappa}_{\|,\bar{m}}-\vec{q}_{\|}^{\,\prime})-\omega}
\right.\nonumber\\ &&\left.\!
 +{f_{n}(\vec{\kappa}_{\|,\bar{m}}+\vec{q}_{\|}^{\,\prime\prime\prime}+\vec{q}_{\|}^{\,\prime\prime}-\vec{q}_{\|}^{\,\prime})
 -f_{v}(\vec{\kappa}_{\|,\bar{m}}+\vec{q}_{\|}^{\,\prime\prime\prime}-\vec{q}_{\|}^{\,\prime})\over
 \tilde{\omega}_{nv}(\vec{\kappa}_{\|,\bar{m}}+\vec{q}_{\|}^{\,\prime\prime\prime}+\vec{q}_{\|}^{\,\prime\prime}
 -\vec{q}_{\|}^{\,\prime},\vec{\kappa}_{\|,\bar{m}}+\vec{q}_{\|}^{\,\prime\prime\prime}-\vec{q}_{\|}^{\,\prime})-\omega}
 \right)
\nonumber\\ &&\times\left.\!
 {1\over\tilde{\omega}_{nl}(\vec{\kappa}_{\|,\bar{m}}+\vec{q}_{\|}^{\,\prime\prime\prime}
 +\vec{q}_{\|}^{\,\prime\prime}-\vec{q}_{\|}^{\,\prime},\vec{\kappa}_{\|,\bar{m}}-\vec{q}_{\|}^{\,\prime})-2\omega}
\right]
 j_{h,lv}(z''';2\vec{\kappa}_{\|,\bar{m}}+\vec{q}_{\|}^{\,\prime\prime\prime}-2\vec{q}_{\|}^{\,\prime})
\nonumber\\ &&\times\left.\!
 j_{k,v{}n}(z'';2\vec{\kappa}_{\|,\bar{m}}+2\vec{q}_{\|}^{\,\prime\prime\prime}+\vec{q}_{\|}^{\,\prime\prime}
 -2\vec{q}_{\|}^{\,\prime})
 j_{j,ml}(z';2\vec{\kappa}_{\|,\bar{m}}-\vec{q}_{\|}^{\,\prime})
\right\}
\nonumber\\ &&\times
 j_{i,nm}(z;2\vec{\kappa}_{\|,\bar{m}}+\vec{q}_{\|}^{\,\prime\prime\prime}+\vec{q}_{\|}^{\,\prime\prime}-\vec{q}_{\|}^{\,\prime})
 \delta(\vec{q}_{\|}^{\,\prime\prime\prime}+\vec{q}_{\|}^{\,\prime\prime}-\vec{q}_{\|}^{\,\prime}-\vec{q}_{\|})
 A_{h}(z''';\vec{q}_{\|}^{\,\prime\prime\prime})
 A_{k}(z'';\vec{q}_{\|}^{\,\prime\prime})
\nonumber\\ &&\times
 A_{j}^{*}(z';\vec{q}_{\|}^{\,\prime})
 d^2\kappa_{\|,\bar{m}}
 d^2q_{\|}'''d^2q_{\|}''d^2q_{\|}'
 dz'''dz''dz'.
\label{eq:J3G}
\end{eqnarray}
On the form of Eq.~(\ref{eq:J3zq}) we thus get part G of the
conductivity tensor as
\begin{eqnarray}
\lefteqn{
 {\Xi}_{ijkh}^{\rm{G}}(z,z',z'',z''';\vec{q}_{\|},\vec{q}_{\|}^{\,\prime},
 \vec{q}_{\|}^{\,\prime\prime},\vec{q}_{\|}^{\,\prime\prime\prime})=
 {2{\rm{i}}\over\omega^3}{1\over(2\pi)^2}{1\over8\hbar^3}
}\nonumber\\ &\quad&\times
 \sum_{nmvl}\int
 {1\over\tilde{\omega}_{nm}(\vec{\kappa}_{\|}+\vec{q}_{\|}^{\,\prime\prime\prime}+\vec{q}_{\|}^{\,\prime\prime}
 -\vec{q}_{\|}^{\,\prime},\vec{\kappa}_{\|})-\omega}
\left\{
\left[
 {1\over\tilde{\omega}_{v{}m}(\vec{\kappa}_{\|}+\vec{q}_{\|}^{\,\prime\prime\prime}
 +\vec{q}_{\|}^{\,\prime\prime},\vec{\kappa}_{\|})-2\omega}
\right.\right.\nonumber\\ &&\times
 \left({f_{l}(\vec{\kappa}_{\|}+\vec{q}_{\|}^{\,\prime\prime\prime})
 -f_{m}(\vec{\kappa}_{\|})\over\tilde{\omega}_{l{}m}(\vec{\kappa}_{\|}
 +\vec{q}_{\|}^{\,\prime\prime\prime},\vec{\kappa}_{\|})-\omega}
 +{f_{l}(\vec{\kappa}_{\|}+\vec{q}_{\|}^{\,\prime\prime\prime})-f_{v}(\vec{\kappa}_{\|}
 +\vec{q}_{\|}^{\,\prime\prime\prime}+\vec{q}_{\|}^{\,\prime\prime})\over
 \tilde{\omega}_{vl}(\vec{\kappa}_{\|}+\vec{q}_{\|}^{\,\prime\prime\prime}+\vec{q}_{\|}^{\,\prime\prime},
 \vec{\kappa}_{\|}+\vec{q}_{\|}^{\,\prime\prime\prime})-\omega}\right)
\nonumber\\ &&
 +
 {1\over\tilde{\omega}_{nl}(\vec{\kappa}_{\|}+\vec{q}_{\|}^{\,\prime\prime\prime}
 +\vec{q}_{\|}^{\,\prime\prime}-\vec{q}_{\|}^{\,\prime},\vec{\kappa}_{\|}+\vec{q}_{\|}^{\,\prime\prime\prime})}
 \left({f_{l}(\vec{\kappa}_{\|}+\vec{q}_{\|}^{\,\prime\prime\prime})
 -f_{v}(\vec{\kappa}_{\|}+\vec{q}_{\|}^{\,\prime\prime\prime}+\vec{q}_{\|}^{\,\prime\prime})\over
 \tilde{\omega}_{vl}(\vec{\kappa}_{\|}+\vec{q}_{\|}^{\,\prime\prime\prime}+\vec{q}_{\|}^{\,\prime\prime},
 \vec{\kappa}_{\|}+\vec{q}_{\|}^{\,\prime\prime\prime})-\omega}
\right.\nonumber\\ &&\left.\!\left.\!
 +{f_{n}(\vec{\kappa}_{\|}+\vec{q}_{\|}^{\,\prime\prime\prime}+\vec{q}_{\|}^{\,\prime\prime}-\vec{q}_{\|}^{\,\prime})
 -f_{v}(\vec{\kappa}_{\|}+\vec{q}_{\|}^{\,\prime\prime\prime}+\vec{q}_{\|}^{\,\prime\prime})\over
 \tilde{\omega}_{nv}(\vec{\kappa}_{\|}+\vec{q}_{\|}^{\,\prime\prime\prime}+\vec{q}_{\|}^{\,\prime\prime}
 -\vec{q}_{\|}^{\,\prime},\vec{\kappa}_{\|}+\vec{q}_{\|}^{\,\prime\prime\prime}+\vec{q}_{\|}^{\,\prime\prime})+\omega}
 \right)
\right]
 j_{h,ml}(z''';2\vec{\kappa}_{\|}+\vec{q}_{\|}^{\,\prime\prime\prime})
\nonumber\\ &&\times
 j_{k,lv}(z'';2\vec{\kappa}_{\|}+2\vec{q}_{\|}^{\,\prime\prime\prime}+\vec{q}_{\|}^{\,\prime\prime})
 j_{j,v{}n}(z';2\vec{\kappa}_{\|}+2\vec{q}_{\|}^{\,\prime\prime\prime}+2\vec{q}_{\|}^{\,\prime\prime}
 -\vec{q}_{\|}^{\,\prime})
\nonumber\\ &&
 +
\left[
 \left({f_{l}(\vec{\kappa}_{\|}+\vec{q}_{\|}^{\,\prime\prime\prime})
 -f_{m}(\vec{\kappa}_{\|})\over\tilde{\omega}_{l{}m}(\vec{\kappa}_{\|}
 +\vec{q}_{\|}^{\,\prime\prime\prime},\vec{\kappa}_{\|})-\omega}
 +{f_{l}(\vec{\kappa}_{\|}+\vec{q}_{\|}^{\,\prime\prime\prime})-f_{v}(\vec{\kappa}_{\|}
 +\vec{q}_{\|}^{\,\prime\prime\prime}-\vec{q}_{\|}^{\,\prime})\over
 \tilde{\omega}_{vl}(\vec{\kappa}_{\|}+\vec{q}_{\|}^{\,\prime\prime\prime}-\vec{q}_{\|}^{\,\prime},
 \vec{\kappa}_{\|}+\vec{q}_{\|}^{\,\prime\prime\prime})+\omega}\right)
\right.\nonumber\\ &&\times
 {1\over\tilde{\omega}_{v{}m}(\vec{\kappa}_{\|}+\vec{q}_{\|}^{\,\prime\prime\prime}
 -\vec{q}_{\|}^{\,\prime},\vec{\kappa}_{\|})}
 +\left({f_{l}(\vec{\kappa}_{\|}+\vec{q}_{\|}^{\,\prime\prime\prime})
 -f_{v}(\vec{\kappa}_{\|}+\vec{q}_{\|}^{\,\prime\prime\prime}-\vec{q}_{\|}^{\,\prime})\over
 \tilde{\omega}_{vl}(\vec{\kappa}_{\|}+\vec{q}_{\|}^{\,\prime\prime\prime}-\vec{q}_{\|}^{\,\prime},
 \vec{\kappa}_{\|}+\vec{q}_{\|}^{\,\prime\prime\prime})+\omega}
\right.\nonumber\\ &&\left.\!\left.\!
 +{f_{n}(\vec{\kappa}_{\|}+\vec{q}_{\|}^{\,\prime\prime\prime}+\vec{q}_{\|}^{\,\prime\prime}-\vec{q}_{\|}^{\,\prime})
 -f_{v}(\vec{\kappa}_{\|}+\vec{q}_{\|}^{\,\prime\prime\prime}-\vec{q}_{\|}^{\,\prime})\over
 \tilde{\omega}_{nv}(\vec{\kappa}_{\|}+\vec{q}_{\|}^{\,\prime\prime\prime}+\vec{q}_{\|}^{\,\prime\prime}
 -\vec{q}_{\|}^{\,\prime},\vec{\kappa}_{\|}+\vec{q}_{\|}^{\,\prime\prime\prime}-\vec{q}_{\|}^{\,\prime})-\omega}
 \right)
 {1\over\tilde{\omega}_{nl}(\vec{\kappa}_{\|}+\vec{q}_{\|}^{\,\prime\prime\prime}
 +\vec{q}_{\|}^{\,\prime\prime}-\vec{q}_{\|}^{\,\prime},\vec{\kappa}_{\|}+\vec{q}_{\|}^{\,\prime\prime\prime})}
\right]
\nonumber\\ &&\times
 j_{h,ml}(z''';2\vec{\kappa}_{\|}+\vec{q}_{\|}^{\,\prime\prime\prime})
 j_{k,v{}n}(z'';2\vec{\kappa}_{\|}+2\vec{q}_{\|}^{\,\prime\prime\prime}+\vec{q}_{\|}^{\,\prime\prime}
 -2\vec{q}_{\|}^{\,\prime})
 j_{j,lv}(z';2\vec{\kappa}_{\|}+2\vec{q}_{\|}^{\,\prime\prime\prime}-\vec{q}_{\|}^{\,\prime})
\nonumber\\ &&
 +
\left[
 \left({f_{l}(\vec{\kappa}_{\|}-\vec{q}_{\|}^{\,\prime})-f_{m}(\vec{\kappa}_{\|})
 \over\tilde{\omega}_{l{}m}(\vec{\kappa}_{\|}-\vec{q}_{\|}^{\,\prime},
 \vec{\kappa}_{\|})+\omega}
 +{f_{l}(\vec{\kappa}_{\|}-\vec{q}_{\|}^{\,\prime})-f_{v}(\vec{\kappa}_{\|}
 +\vec{q}_{\|}^{\,\prime\prime\prime}-\vec{q}_{\|}^{\,\prime})\over
 \tilde{\omega}_{vl}(\vec{\kappa}_{\|}+\vec{q}_{\|}^{\,\prime\prime\prime}-\vec{q}_{\|}^{\,\prime},
 \vec{\kappa}_{\|}-\vec{q}_{\|}^{\,\prime})-\omega}\right)
\right.\nonumber\\ &&\times
 {1\over\tilde{\omega}_{v{}m}(\vec{\kappa}_{\|}+\vec{q}_{\|}^{\,\prime\prime\prime}
 -\vec{q}_{\|}^{\,\prime},\vec{\kappa}_{\|})}
 +\left({f_{l}(\vec{\kappa}_{\|}-\vec{q}_{\|}^{\,\prime})
 -f_{v}(\vec{\kappa}_{\|}+\vec{q}_{\|}^{\,\prime\prime\prime}-\vec{q}_{\|}^{\,\prime})\over
 \tilde{\omega}_{vl}(\vec{\kappa}_{\|}+\vec{q}_{\|}^{\,\prime\prime\prime}-\vec{q}_{\|}^{\,\prime},
 \vec{\kappa}_{\|}-\vec{q}_{\|}^{\,\prime})-\omega}
\right.\nonumber\\ &&\left.\!
 +{f_{n}(\vec{\kappa}_{\|}+\vec{q}_{\|}^{\,\prime\prime\prime}+\vec{q}_{\|}^{\,\prime\prime}-\vec{q}_{\|}^{\,\prime})
 -f_{v}(\vec{\kappa}_{\|}+\vec{q}_{\|}^{\,\prime\prime\prime}-\vec{q}_{\|}^{\,\prime})\over
 \tilde{\omega}_{nv}(\vec{\kappa}_{\|}+\vec{q}_{\|}^{\,\prime\prime\prime}+\vec{q}_{\|}^{\,\prime\prime}
 -\vec{q}_{\|}^{\,\prime},\vec{\kappa}_{\|}+\vec{q}_{\|}^{\,\prime\prime\prime}-\vec{q}_{\|}^{\,\prime})-\omega}
 \right)
\nonumber\\ &&\times\left.
 {1\over\tilde{\omega}_{nl}(\vec{\kappa}_{\|}+\vec{q}_{\|}^{\,\prime\prime\prime}
 +\vec{q}_{\|}^{\,\prime\prime}-\vec{q}_{\|}^{\,\prime},\vec{\kappa}_{\|}-\vec{q}_{\|}^{\,\prime})-2\omega}
\right]
 j_{h,lv}(z''';2\vec{\kappa}_{\|}+\vec{q}_{\|}^{\,\prime\prime\prime}-2\vec{q}_{\|}^{\,\prime})
\nonumber\\ &&\times\left.
 j_{k,v{}n}(z'';2\vec{\kappa}_{\|}+2\vec{q}_{\|}^{\,\prime\prime\prime}+\vec{q}_{\|}^{\,\prime\prime}
 -2\vec{q}_{\|}^{\,\prime})
 j_{j,ml}(z';2\vec{\kappa}_{\|}-\vec{q}_{\|}^{\,\prime})
\right\}
\nonumber\\ &&\times
 j_{i,nm}(z;2\vec{\kappa}_{\|}+\vec{q}_{\|}^{\,\prime\prime\prime}+\vec{q}_{\|}^{\,\prime\prime}-\vec{q}_{\|}^{\,\prime})
 \delta(\vec{q}_{\|}^{\,\prime\prime\prime}+\vec{q}_{\|}^{\,\prime\prime}-\vec{q}_{\|}^{\,\prime}-\vec{q}_{\|})
 d^2\kappa_{\|},
\end{eqnarray}
where we have omitted the now superfluous index $m$ from the surface
states $\vec{\kappa}_{\|}$. As was the case with parts A--F, we are
particularly interested in finding the PCDFWM response tensor, and
thus we insert the DFWM pump fields given by Eqs.~(\ref{eq:Apump-1})
and (\ref{eq:Apump-2}) into Eq.~(\ref{eq:J3G}). This allows us to
carry out the integrals over $q_{\|}'''$ and $q_{\|}''$,
consequentally followed by solution to the integral over $q_{\|}'$.
The resulting expression is then split according to
Eq.~(\ref{eq:J3-wzq}), and the PCDFWM conductivity tensor part G
appear as Eq.~(\ref{eq:XiAG}).

\chapter[Principal analytic solution to the integrals over
$\vec{\kappa}_{\|}$]{Principal analytic solution \\ to the integrals
  over $\vec{\kappa}_{\|}$ in the low temperature
  limit}\label{ch:SolveQ}\label{ch:Solve-Q}\label{app:B}
\noindent
In this appendix we discuss the analytic solution to the integrals
over $\vec{\kappa}_{\|}$ appearing in the linear and nonlinear
conductivity tensor. The discussion is limited to cover the low
temperature limit, and it is presented as a principal solution to all
integrals over $\vec{\kappa}_{\|}$ appearing in
Eqs.~(\ref{eq:SigmaAA})--(\ref{eq:XiAG}).

\section{General type of integrals}
Every integral over $\vec{\kappa}_{\|}$ in both the linear
conductivity tensor, Eqs.~(\ref{eq:SigmaAA}) and (\ref{eq:SigmaAB}),
and the nonlinear conductivity tensor,
Eqs.~(\ref{eq:XiAA})--(\ref{eq:XiAG}), can when scattering takes place
in the $x$-$z$-plane be expressed as a sum over terms of the general
type
\begin{equation}
 {\cal{F}}_{pq}^{\beta}(n,\{a\},\{b\},s)
 =\int\int
 {{\kappa}_{x}^{p}\kappa_{y}^{q}f_{n}(\vec{\kappa}_{\|}+s\vec{e}_{x})
 \over\prod_{k=1}^{\beta}[a_k\kappa_{x}+b_k]}
 d\kappa_{x}d\kappa_{y},
\label{eq:FpqB-indef}
\end{equation}
where $p,k,\beta$ are nonnegative integers, and $q$ is an even
nonnegative integer. The functions in general depends on (i) the
quantum number $n$, which is a positive nonzero integer, (ii) a set of
real quantities, $\{a\}\equiv\{a_1,\dots,a_\beta\}$ appearing in front
of the integration variable $\kappa_x$ in the denominator, (iii) a set
of complex nonzero quantities, $\{b\}\equiv\{b_1,\dots,b_\beta\}$
apearing as the other quantity in each term of the denominator, and
(iv) the real quantity $s$ representing the displacement of the center
of the Fermi-Dirac distribution function from
$(\kappa_x,\kappa_y)=(0,0)$.  The quantity $s$ together with each
element in the set $\{a\}$ is in general a function of the parallel
components of the probe and pump wavevectors, $\vec{q}_{\|}$ and
$\vec{k}_{\|}$. Each element in the set $\{b\}$ is furthermore a
function of $\tau$, the relaxation time.

In the low temperature limit the Fermi-Dirac distribution function is
zero outside the Fermi sphere and equal to one inside, and it is
therefore advantageous to shift $\kappa_x$ by $-s$, followed by a
one-to-one mapping of the $x$-$y$-plane into polar coordinates
($r$-$\theta$-plane). Using in this way $\kappa_x=r\cos\theta$,
$\kappa_y=r\sin\theta$, and thus $d\kappa_xd\kappa_y=rd\theta{}dr$,
the indefinite integral in Eq.~(\ref{eq:FpqB-indef}) is turned into
the definite integral
\begin{equation}
 {\cal{F}}_{pq}^{\beta}(n,\{a\},\{b\},s)
 =\int_{0}^{{\alpha}(n)}\int_{0}^{2\pi}{r(r\cos\theta-s)^{p}(r\sin\theta)^{q}
  \over\prod_{k=1}^{\beta}[a_k(r\cos\theta-s)+b_k]}d\theta dr.
\label{eq:FpqB}
\end{equation}
The quantity $\alpha(n)=\sqrt{k_F^2-(\pi{}n/d)^2}$ is the radius of
the (two-dimensional) Fermi circle for state $n$, given by
Eq.~(\ref{eq:alpha(n)}). The Fermi wavenumber $k_F$ obeys the relation
$k_F>\pi{}n/d$, since the Fermi-Dirac distribution function is zero
for $k_F<\pi{}n/d$, and thus in that case the integral would vanish.

\begin{figure}[tb]
\setlength{\unitlength}{1mm}
\psset{unit=1mm}
\begin{center}
\begin{picture}(80,76)(0,4)
\newgray{mygray}{0.95}
\multiput(2,0)(24,0){3}{\multiput(0,0)(0,24){3}{
  \psline[linewidth=0.385]{-}(6,6)(4,6)(4,22)(6,22)
  \psline[linewidth=0.385]{-}(22,6)(24,6)(24,22)(22,22)
}}
{\small
 \put(2,70){\makebox(0,5)[l]{$[xxkh]$}}
 \put(2,46){\makebox(0,5)[l]{$[yxkh]$}}
 \put(2,22){\makebox(0,5)[l]{$[zxkh]$}}
 \put(40,70){\makebox(0,5)[c]{$[xykh]$}}
 \put(40,46){\makebox(0,5)[c]{$[yykh]$}}
 \put(40,22){\makebox(0,5)[c]{$[zykh]$}}
 \put(78,70){\makebox(0,5)[r]{$[xzkh]$}}
 \put(78,46){\makebox(0,5)[r]{$[yzkh]$}}
 \put(78,22){\makebox(0,5)[r]{$[zzkh]$}}
}
\put(70,8){\makebox(0,0)[c]{$\bigtriangleup$}}
\put(46,32){\makebox(0,0)[c]{$\otimes$}}
\put(70,38){\makebox(0,0)[c]{$\otimes$}}
\put(64,32){\makebox(0,0)[c]{$\otimes$}}
\put(46,14){\makebox(0,0)[c]{$\otimes$}}
\put(40,8){\makebox(0,0)[c]{$\otimes$}}
\put(64,14){\makebox(0,0)[c]{$\otimes$}}
\put(40,38){\makebox(0,0)[c]{$\star$}}
\put(70,56){\makebox(0,0)[c]{$+$}}
\put(22,8){\makebox(0,0)[c]{$+$}} 
\put(70,20){\makebox(0,0)[c]{$+$}}
\put(58,8){\makebox(0,0)[c]{$+$}} 
\put(46,62){\circle{2}}
\put(40,56){\circle{2}}
\put(64,62){\circle{2}}
\put(16,14){\circle{2}}
\put(40,20){\circle{2}}
\put(34,14){\circle{2}}
\put(22,38){\circle{2}}
\put(46,44){\circle{2}}
\put(34,32){\circle{2}}
\put(58,38){\circle{2}}
\put(16,32){\circle{2}}
\put(64,44){\circle{2}}
\put(22,56){\circle*{2}}
\put(70,68){\circle*{2}}
\put(58,56){\circle*{2}}
\put(22,20){\circle*{2}}
\put(10,8){\circle*{2}} 
\put(58,20){\circle*{2}}
\put(16,62){\makebox(0,0)[c]{$\times$}}
\put(40,68){\makebox(0,0)[c]{$\times$}}
\put(34,62){\makebox(0,0)[c]{$\times$}}
\put(16,44){\makebox(0,0)[c]{$\times$}}
\put(10,38){\makebox(0,0)[c]{$\times$}}
\put(34,44){\makebox(0,0)[c]{$\times$}}
\put(22,68){\makebox(0,0)[c]{$\oplus$}}
\put(10,56){\makebox(0,0)[c]{$\oplus$}}
\put(58,68){\makebox(0,0)[c]{$\oplus$}}
\put(10,20){\makebox(0,0)[c]{$\oplus$}}
\put(10,68){\makebox(0,0)[c]{$\bigtriangledown$}}
\put(10,14){\circle*{0.1}}
\put(10,32){\circle*{0.1}}
\put(10,44){\circle*{0.1}}
\put(10,62){\circle*{0.1}}
\put(16,8){\circle*{0.1}}
\put(16,20){\circle*{0.1}}
\put(16,38){\circle*{0.1}}
\put(16,56){\circle*{0.1}}
\put(16,68){\circle*{0.1}}
\put(22,14){\circle*{0.1}}
\put(22,32){\circle*{0.1}}
\put(22,44){\circle*{0.1}}
\put(22,62){\circle*{0.1}}
\put(34,8){\circle*{0.1}}
\put(34,20){\circle*{0.1}}
\put(34,38){\circle*{0.1}}
\put(34,56){\circle*{0.1}}
\put(34,68){\circle*{0.1}}
\put(40,14){\circle*{0.1}}
\put(40,32){\circle*{0.1}}
\put(40,44){\circle*{0.1}}
\put(40,62){\circle*{0.1}}
\put(46,8){\circle*{0.1}}
\put(46,20){\circle*{0.1}}
\put(46,38){\circle*{0.1}}
\put(46,56){\circle*{0.1}}
\put(46,68){\circle*{0.1}}
\put(58,14){\circle*{0.1}}
\put(58,32){\circle*{0.1}}
\put(58,44){\circle*{0.1}}
\put(58,62){\circle*{0.1}}
\put(64,8){\circle*{0.1}}
\put(64,20){\circle*{0.1}}
\put(64,38){\circle*{0.1}}
\put(64,56){\circle*{0.1}}
\put(64,68){\circle*{0.1}}
\put(70,14){\circle*{0.1}}
\put(70,32){\circle*{0.1}}
\put(70,44){\circle*{0.1}}
\put(70,62){\circle*{0.1}}
\end{picture}
\end{center}
\caption[Distribution of the elements of the nonlinear conductivity
tensor in terms of $p$ and $q$]{Distribution of the elements of the
  nonlinear conductivity tensor in terms of $p$ and $q$. In one
  element ($\bigtriangleup$) the terms appear with $(p,q)=(0,0)$. In
  six elements ($\otimes$) the terms appear with $(p,q)=(0,2)$. In one
  element ($\star$) the terms appear with $(p,q)=(0,4)$. In four
  elements ($+$) the terms appear with $(p,q)\in\{(1,0),(0,0)\}$. In
  twelve elements ($\circ$) the terms appear with
  $(p,q)\in\{(1,2),(0,2)\}$. In six elements ($\bullet$) the terms
  appear with $(p,q)\in\{(2,0),(1,0),(0,0)\}$. In six elements
  ($\times$) the terms appear with $(p,q)\in\{(2,2),(1,2),(0,2)\}$. In
  four elements ($\oplus$) the terms appear with
  $(p,q)\in\{(3,0),(2,0),(1,0),(0,0)\}$. The final element
  ($\bigtriangledown$) the terms appear with
  $(p,q)\{(4,0),(3,0),(2,0),(1,0),(0,0)\}$.  Elements labelled with a
  `$\cdot$' are zero.\label{fig:pq-distribution}}
\end{figure}

\section{Specific integrals to be solved}
The necessary combinations of $p$ and $q$ in Eq.~(\ref{eq:FpqB}) to be
calculated in order to solve the integrals over $\vec{\kappa}_{\|}$ in
the nonlinear conductivity tensor are summarized in
Fig.~\ref{fig:pq-distribution}. From Fig.~\ref{fig:pq-distribution} we
observe that a total of nine different combinations of $p$ and $q$
need to be calculated, namely when $(p,q)$ takes the values $(0,0)$,
$(0,2)$, $(0,4)$, $(1,0)$, $(1,2)$, $(2,0)$, $(2,2)$, $(3,0)$, or
$(4,0)$, and it is seen from Eqs.~(\ref{eq:SigmaAA})--(\ref{eq:XiAG})
that $\beta$ can take the values $\beta\in\{1,2,3\}$. However, the
complexity of the total solution can be reduced, since functions with
$\beta=2$ can be expressed in terms of functions with $\beta=1$ in the
following way
\begin{equation}
 {\cal{F}}_{pq}^{2}(n,a_1,a_2,b_1,b_2,s)=
 {a_1{\cal{F}}_{pq}^{1}(n,a_1,b_1,s)-a_2{\cal{F}}_{pq}^{1}(n,a_2,b_2,s)
  \over{}a_1b_2-a_2b_1}, 
\label{eq:Fpq2}
\end{equation}
$a_k\neq0, k\in\{1,2\}$.  In a similar fashion, the functions with
$\beta=3$ can be written in terms of functions with $\beta=1$, namely
\begin{eqnarray}
\lefteqn{
 {\cal{F}}_{pq}^{3}(n,a_1,a_2,a_3,b_1,b_2,b_3,s)=
 {a_1^2{\cal{F}}_{pq}^{1}(n,a_1,b_1,s)
  \over(a_2b_1-b_2a_1)(a_3b_1-b_3a_1)}
}\nonumber\\ &\quad&
 +{a_2^2{\cal{F}}_{pq}^{1}(n,a_2,b_2,s)
  \over(a_2b_1-b_2a_1)(a_3b_2-b_3a_2)}
 +{a_3^2{\cal{F}}_{pq}^{1}(n,a_3,b_3,s)
  \over(a_3b_1-b_3a_1)(a_3b_2-b_3a_2)},
\label{eq:Fpq3}
\end{eqnarray}
provided $a_k\neq0$, $k\in\{1,2,3\}$. If any $a_k$, for instance
$a_1$, becomes zero, we observe from Eq.~(\ref{eq:FpqB}) that the
order of the denominator becomes smaller by one. This implies in
Eq.~(\ref{eq:Fpq2}) that
${\cal{F}}_{pq}^{2}(n,0,a_2,b_1,b_2,s)={\cal{F}}_{pq}^{1}(n,a_2,b_2,s)/b_1$.
The similar conclusion with respect to Eq.~(\ref{eq:Fpq3}) is 
${\cal{F}}_{pq}^{3}(n,0,a_2,a_3,b_1,b_2,b_3,s)={\cal{F}}_{pq}^{2}(n,a_2,a_3,b_2,b_3,s)/b_1$.
A corresponding reduction applies for any other $a_k=0$ in
Eqs.~(\ref{eq:Fpq2}) and (\ref{eq:Fpq3}).

As a consequence of Eqs.~(\ref{eq:Fpq2}) and (\ref{eq:Fpq3}), the
integrals appearing in Eqs.~(\ref{eq:SigmaAA})--({\ref{eq:XiAG}}) can
now be written in terms of functions of the type
\begin{equation}
 {\cal{F}}_{pq}^{1}(n,a,b,s)
 =\int_{0}^{{\alpha}(n)}\int_{0}^{2\pi}{r(r\cos\theta-s)^{p}(r\sin\theta)^{q}
  \over{}b-as+ar\cos\theta}d\theta dr,
\label{eq:Fpq1}
\end{equation}
dropping the now superfluous index on $a$ and $b$. Since the following
treatment is a formal solution of Eq.~(\ref{eq:Fpq1}), we will also
drop the reference to $n$ for brevity, letting
$\alpha\equiv\alpha(n)$. 

\section{Solution when $a=0$}
Before carrying on with the solution to Eq.~(\ref{eq:Fpq1}) in the
appropriate cases, we take a look at it in the case where $a=0$ (as
would be the case in the local limit, for example). Then the only term
left in the denominator is $b$, which is constant with respect to the
integration variables. The solution to the remaining thus becomes
trivial, with the results
\begin{eqnarray}
\lefteqn{
 {\cal{F}}_{00}^{1}(n,0,b,s)={2\over{}b}\int_{0}^{\alpha}\int_{0}^{\pi}
 r d\theta dr
 ={\pi\alpha^2\over{}b},
}\label{eq:F-00-1-a=0}\\
\lefteqn{
 {\cal{F}}_{02}^{1}(n,0,b,s)={2\over{}b}\int_{0}^{\alpha}\int_{0}^{\pi}
 r^3\sin^2\theta d\theta dr
 ={\pi\alpha^4\over4b},
}\label{eq:F-02-1-a=0}\\
\lefteqn{
 {\cal{F}}_{04}^{1}(n,0,b,s)={2\over{}b}\int_{0}^{\alpha}\int_{0}^{\pi}
 r^5\sin^4\theta d\theta dr
 ={\pi\alpha^6\over8b},
}\label{eq:F-04-1-a=0}\\
\lefteqn{
 {\cal{F}}_{10}^{1}(n,0,b,s)={2\over{}b}\int_{0}^{\alpha}\int_{0}^{\pi}
 \left[r^2\cos\theta-rs\right]d\theta dr
 =-{\pi{}s\alpha^2\over{}b},
}\label{eq:F-10-1-a=0}\\
\lefteqn{
 {\cal{F}}_{12}^{1}(n,0,b,s)={2\over{}b}\int_{0}^{\alpha}\int_{0}^{\pi}
 \left[r^4\cos\theta\sin^2\theta-r^3s\sin^2\theta\right]d\theta dr
 =-{\pi{}s\alpha^4\over4b},
}\label{eq:F-12-1-a=0}\\
\lefteqn{
 {\cal{F}}_{20}^{1}(n,0,b,s)={2\over{}b}\int_{0}^{\alpha}\int_{0}^{\pi}
 \left[r^3\cos^2\theta+rs^2-2sr^2\cos\theta\right]d\theta dr
 ={\pi\alpha^4\over4b}+{\pi{}s^2\alpha^2\over{}b},
}\label{eq:F-20-1-a=0}\\
\lefteqn{
 {\cal{F}}_{22}^{1}(n,0,b,s)={2\over{}b}\int_{0}^{\alpha}\int_{0}^{\pi}
 \left[r^5\cos^2\theta\sin^2\theta+r^3s^2\sin^2\theta
 -2r^4s\cos\theta\sin^2\theta\right]d\theta dr
}\nonumber\\ &\quad&
 ={\pi\alpha^6\over24b}+{\pi{}s^2\alpha^4\over4b},
\label{eq:F-22-1-a=0}\\
\lefteqn{
 {\cal{F}}_{30}^{1}(n,0,b,s)={2\over{}b}\int_{0}^{\alpha}\int_{0}^{\pi}
 \left[r^4\cos^3\theta-3r^3s\cos^2\theta
 +3r^2s^2\cos\theta-rs^3\right]d\theta dr
}\nonumber\\ &&
 =-{3\pi{}s\alpha^4\over4b}-{\pi{}s^3\alpha^2\over{}b},
\label{eq:F-30-1-a=0}\\
\lefteqn{
 {\cal{F}}_{40}^{1}(n,0,b,s)={2\over{}b}\int_{0}^{\alpha}\int_{0}^{\pi}
 \left[r^5\cos^4\theta-4r^4s\cos^3\theta
 +6r^3s^2\cos^2\theta-4r^2s^3\cos\theta
\right.}\nonumber\\ &&\left.
 +rs^4\right]d\theta dr
 ={\pi\alpha^6\over8b}+{3\pi{}s^2\alpha^4\over2b}
  +{\pi{}s^4\alpha^2\over{}b},
\label{eq:F-40-1-a=0}
\end{eqnarray}
where we have made use of the facts that (i) $q$ is an even integer,
and (ii) $\cos\theta$ and $\sin^2\theta$ are symmetric around
$\theta=\pi$ in solving the angular integrals.

\section{General solution}
When $a\neq0$, we have to consider the full solution to
Eq.~(\ref{eq:Fpq1}) for the nine different combinations of $p$ and $q$
we need. To solve Eq.~(\ref{eq:Fpq1}), let us make the substitutions
\begin{equation}
 \eta\equiv{b-as\over{}a\alpha},\quad 
 r\equiv\alpha{}u,
\label{eq:Substitute}
\end{equation}
giving $dr=\alpha{}du$. Thereby Eq.~(\ref{eq:Fpq1}) is turned into the
nine functions
\begin{eqnarray}
\lefteqn{
 {\cal{F}}_{00}^{1}(n,a,b,s)={\alpha\over{}a}\int_{0}^{1}\int_{0}^{2\pi}
 {u\over\eta+u\cos\theta}d\theta du,
}\label{eq:F00}
\\
\lefteqn{
 {\cal{F}}_{02}^{1}(n,a,b,s)=
 {\alpha^3\over{}a}\int_{0}^{1}\int_{0}^{2\pi}
 \left[{u^3\over\eta+u\cos\theta}-{u^3\cos^2\theta\over\eta+u\cos\theta}\right]
 d\theta du,
}\label{eq:F02}
\\
\lefteqn{
 {\cal{F}}_{04}^{1}(n,a,b,s)={\alpha^5\over{}a}\int_{0}^{1}\int_{0}^{2\pi}
 \left[{u^5\over\eta+u\cos\theta}-{2u^5\cos^2\theta\over\eta+u\cos\theta}
 +{u^5\cos^4\theta\over\eta+u\cos\theta}\right]
 d\theta du,
}\label{eq:F04}
\\
\lefteqn{
 {\cal{F}}_{10}^{1}(n,a,b,s)={\alpha^2\over{}a}\int_{0}^{1}\int_{0}^{2\pi}
 \left[{u^2\cos\theta\over\eta+u\cos\theta}
 -{s\over\alpha}{u\over\eta+u\cos\theta}\right]
 d\theta du,
}\label{eq:F10}
\\
\lefteqn{
 {\cal{F}}_{12}^{1}(n,a,b,s)={\alpha^4\over{}a}\int_{0}^{1}\int_{0}^{2\pi}
 \left[{u^4\cos\theta\over\eta+u\cos\theta}
 -{u^4\cos^3\theta\over\eta+u\cos\theta}
 -{s\over\alpha}{u^3\over\eta+u\cos\theta}
\right.}\nonumber\\ &\quad&\left.\!
 +{s\over\alpha}{u^3\cos^2\theta\over\eta+u\cos\theta}
 \right]d\theta du,
\label{eq:F12}
\\
\lefteqn{
 {\cal{F}}_{20}^{1}(n,a,b,s)={\alpha^3\over{}a}\int_{0}^{1}\int_{0}^{2\pi}
 \left[{u^3\cos^2\theta\over\eta+u\cos\theta}
 -{2s\over\alpha}{u^2\cos\theta\over\eta+u\cos\theta}
 +{s^2\over\alpha^2}{u\over\eta+u\cos\theta}
 \right]d\theta du,
}\nonumber\\
\label{eq:F20}
\\
\lefteqn{
 {\cal{F}}_{22}^{1}(n,a,b,s)={\alpha^5\over{}a}\int_{0}^{1}\int_{0}^{2\pi}
 \left[{u^5\cos^2\theta\over\eta+u\cos\theta}
 -{u^5\cos^4\theta\over\eta+u\cos\theta}
 -{2s\over\alpha}{u^4\cos\theta\over\eta+u\cos\theta}
\right.}\nonumber\\ &&\left.\!
 +{2s\over\alpha}{u^4\cos^3\theta\over\eta+u\cos\theta}
 +{s^2\over\alpha^2}{u^3\over\eta+u\cos\theta}
 -{s^2\over\alpha^2}{u^3\cos^2\theta\over\eta+u\cos\theta}
 \right]d\theta du,
\label{eq:F22}
\\
\lefteqn{
 {\cal{F}}_{30}^{1}(n,a,b,s)={\alpha^4\over{}a}\int_{0}^{1}\int_{0}^{2\pi}
 \left[{u^4\cos^3\theta\over\eta+u\cos\theta}
 -{3s\over\alpha}{u^3\cos^2\theta\over\eta+u\cos\theta}
 +{3s^2\over\alpha^2}{u^2\cos\theta\over\eta+u\cos\theta}
\right.}\nonumber\\ &&\left.\!
 -{s^3\over\alpha^3}{u\over\eta+u\cos\theta}
 \right]d\theta du,
\label{eq:F30}
\\
\lefteqn{
 {\cal{F}}_{40}^{1}(n,a,b,s)={\alpha^5\over{}a}\int_{0}^{1}\int_{0}^{2\pi}
 \left[{u^5\cos^4\theta\over\eta+u\cos\theta}
 -{4s\over\alpha}{u^4\cos^3\theta\over\eta+u\cos\theta}
 +{6s^2\over\alpha^2}{u^3\cos^2\theta\over\eta+u\cos\theta}
\right.}\nonumber\\ &&\left.\!
 -{4s^3\over\alpha^3}{u^2\cos\theta\over\eta+u\cos\theta}
 +{s^4\over\alpha^4}{u\over\eta+u\cos\theta}
 \right]d\theta du,
\label{eq:F40}
\end{eqnarray}
where we have made use of the relations
\begin{eqnarray}
 \sin^2\theta&=&1-\cos^2\theta,
\label{eq:angle1}
\\
 \cos\theta\sin^2\theta&=&\cos\theta-\cos^3\theta,
\\
 \cos^2\theta\sin^2\theta&=&\cos^2\theta-\cos^4\theta,
\\
 \sin^4\theta&=&1-2\cos^2\theta+\cos^4\theta.
\label{eq:angle4}
\end{eqnarray}

\subsection{Solution to the angular integrals}
Next, to carry out the angular integrals, we put
\begin{equation}
 t=e^{{\rm{i}}\theta}
\end{equation}
so that these integrals become
\begin{equation}
 \int_0^{2\pi}{\cos^h\theta\over\eta+u\cos\theta}d\theta
 ={1\over{}2^h{\rm{i}}u}\oint{(1+t^2)^h\over{}t^h(t-t_+)(t-t_-)}dt,\!\!
\label{eq:t+t-h}
\end{equation}
where $h\in\{0,1,2,3,4\}$. In Eq.~(\ref{eq:t+t-h}), the poles at
$t_{\pm}$ in the $t$-plane are located at
\begin{equation}
 t_{\pm}=-{\eta\over{}u}\pm\sqrt{\left({\eta\over{}u}\right)^2-1},
\end{equation}

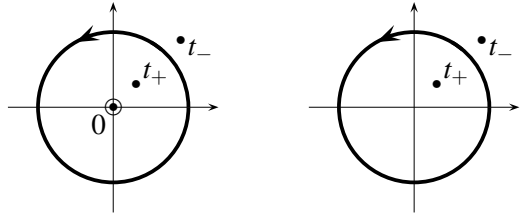
\begin{figure}[tb]
\setlength{\unitlength}{1mm}
\psset{unit=1mm}
\begin{center}
\begin{pspicture}(0,-5)(80,30)
\put(40,0){
\psline[linewidth=0.1]{->}(6,10)(34,10)
\psline[linewidth=0.1]{->}(20,-4)(20,24)
\psarc[linewidth=0.5]{-}(20,10){10}{0}{60}
\psarc[linewidth=0.5]{->}(20,10){10}{50}{120}
\psarc[linewidth=0.5]{-}(20,10){10}{110}{0}
\put(29,19){\circle*{1}}
\put(23,13){\circle*{1}}
\put(30,19){\makebox(0,0)[tl]{$t_-$}}
\put(24,13){\makebox(0,0)[bl]{$t_+$}}
}
\put(0,0){
\psline[linewidth=0.1]{->}(6,10)(34,10)
\psline[linewidth=0.1]{->}(20,-4)(20,24)
\psarc[linewidth=0.5]{-}(20,10){10}{0}{60}
\psarc[linewidth=0.5]{->}(20,10){10}{50}{120}
\psarc[linewidth=0.5]{-}(20,10){10}{110}{0}
\put(29,19){\circle*{1}}
\put(23,13){\circle*{1}}
\put(30,19){\makebox(0,0)[tl]{$t_-$}}
\put(24,13){\makebox(0,0)[bl]{$t_+$}}
\put(20,10){\circle*{1}}
\put(20,10){\circle{2}}
\put(19,9){\makebox(0,0)[tr]{$0$}}
}
\end{pspicture}
\end{center}
\caption[Poles and contours for solution of integrals]{The poles
  appearing in the complex $t$-plane in Eq.~(\ref{eq:t+t-h}) are of
  order $1$ at $t_{\pm}$ and of order $h$ at $t=0$, as shown to the
  left. To the right is shown the special case where $h=0$ and the
  pole at $t=0$ vanishes. The closed contour shown in each diagram is
  the integration path used.\label{fig:21}}
\end{figure}

\noindent
and the integration runs along the unit circle. Since we have
$t_+t_-=1$, one of these poles is inside the unit circle while the
other is outside. When $h>0$ there are an additional pole of order $h$
at $t=0$. Using the unit circles shown in Fig.~\ref{fig:21} as the
integration paths, we find by a residue calculation
\begin{eqnarray}
 \int_{0}^{2\pi}{1\over\eta+u\cos\theta}d\theta
 &=&{2\pi\over\sqrt{\eta^2-u^2}},
\label{eq:Ang-nocos}
\\
 \int_{0}^{2\pi}{\cos\theta\over\eta+u\cos\theta}d\theta
 &=&{2\pi\over{}u}\left[1-{\eta\over\sqrt{\eta^2-u^2}}\right],
\\
 \int_{0}^{2\pi}{\cos^2\theta\over\eta+u\cos\theta}d\theta
 &=&{2\pi\eta\over{}u^2}\left[{\eta\over\sqrt{\eta^2-u^2}}-1\right],
\\
 \int_{0}^{2\pi}{\cos^3\theta\over\eta+u\cos\theta}d\theta
 &=&{\pi\over{}u}+{2\pi\eta^2\over{}u^3}
 \left[1-{\eta\over\sqrt{\eta^2-u^2}}\right],
\\
 \int_{0}^{2\pi}{\cos^4\theta\over\eta+u\cos\theta}d\theta
 &=&{2\pi\eta^3\over{}u^4}\left[{\eta\over\sqrt{\eta^2-u^2}}-1\right]
 -{\pi\eta\over{}u^2}.
\end{eqnarray}
Inserting these results into Eqs.~(\ref{eq:F00})--(\ref{eq:F02}), we
get
\begin{eqnarray}
\lefteqn{
 {\cal{F}}_{00}^{1}(n,a,b,s)=
 {2\pi\alpha\over{}a}\int_{0}^{1}
 {u\over\sqrt{\eta^2-u^2}}du,
}
\label{eq:F00R}
\\
\lefteqn{
 {\cal{F}}_{02}^{1}(n,a,b,s)=
 {2\pi\alpha^3\over{}a}\int_{0}^{1}
 \left[{u^3\over\sqrt{\eta^2-u^2}}
 -{\eta^2u\over\sqrt{\eta^2-u^2}}+\eta{}u\right]
 du,
}\label{eq:F02R}
\\
\lefteqn{
 {\cal{F}}_{04}^{1}(n,a,b,s)=
 {\pi\alpha^5\over{}a}\int_{0}^{1}
 \left[{2u^5\over\sqrt{\eta^2-u^2}}
 -{4\eta^2u^3\over\sqrt{\eta^2-u^2}}
 +{2\eta^4u\over\sqrt{\eta^2-u^2}}
 +3\eta{}u^3
 -2\eta^3u
 \right] du,
}\nonumber\\
\label{eq:F04R}
\\
\lefteqn{
 {\cal{F}}_{10}^{1}(n,a,b,s)=
 {2\pi\alpha^2\over{}a}\int_{0}^{1}
 \left[u-\left(\eta+{s\over\alpha}\right){u\over\sqrt{\eta^2-u^2}}\right]
 du,
}\label{eq:F10R}
\\
\lefteqn{
 {\cal{F}}_{12}^{1}(n,a,b,s)=
 {\pi\alpha^4\over{}a}\int_{0}^{1}
 \left[{u^3}
 -\left({2\eta^2}+{2s\eta{}\over\alpha}\right)u
 -\left({2\eta}+{2s\over\alpha}\right){u^3\over\sqrt{\eta^2-u^2}}
\right.}\nonumber\\ &\quad&\left.\!
 +\left({2\eta^3}+{2s\eta^2\over\alpha}\right){u\over\sqrt{\eta^2-u^2}}
 \right] du,
\label{eq:F12R}
\\
\lefteqn{
 {\cal{F}}_{20}^{1}(n,a,b,s)=
 {2\pi\alpha^3\over{}a}\int_{0}^{1}
 \left[\left(\eta^2+{2s\eta\over\alpha}+{s^2\over\alpha^2}\right)
 {u\over\sqrt{\eta^2-u^2}}-\left(\eta+{2s\over\alpha}\right)u
 \right] du,
}\label{eq:F20R}
\\
\lefteqn{
 {\cal{F}}_{22}^{1}(n,a,b,s)=
 {\pi\alpha^5\over{}a}\int_{0}^{1}
 \left[\left({2\eta^2}+{4\eta{}s\over\alpha}
 +{2s^2\over\alpha^2}\right){u^3\over\sqrt{\eta^2-u^2}}
 -\left(\eta+{2s\over\alpha}\right)u^3
\right.}\nonumber\\ &&\left.\!
 -\left({2\eta^4}+{4\eta^3s\over\alpha}
 +{2\eta^2s^2\over\alpha^2}\right){u\over\sqrt{\eta^2-u^2}}
 +\left({2\eta^3}+{4s\eta^2\over\alpha}+{2s^2\eta{}\over\alpha^2}\right)u
 \right] du,
\label{eq:F22R}
\\
\lefteqn{
 {\cal{F}}_{30}^{1}(n,a,b,s)=
 {\pi\alpha^4\over{}a}\int_{0}^{1}
 \left[{u^3}
 +\left({2\eta^2}+{6s\eta\over\alpha}+{6s^2\over\alpha^2}\right)u
\right.}\nonumber\\ &&\left.\!
 -\left({2\eta^3}+{6s\eta^2\over\alpha}+{6s^2\eta\over\alpha^2}
 +{2s^3\over\alpha^3}\right){u\over\sqrt{\eta^2-u^2}}
 \right] du,
\label{eq:F30R}
\\
\lefteqn{
 {\cal{F}}_{40}^{1}(n,a,b,s)=
 {\pi\alpha^5\over{}a}\int_{0}^{1}
 \left[
 \left({2\eta^4}+{8s\eta^3\over\alpha}+{12s^2\eta^2\over\alpha^2}
 +{8s^3\eta\over\alpha^3}+{2s^4\over\alpha^4}\right){u\over\sqrt{\eta^2-u^2}}
\right.}\nonumber\\ &&\left.\!
 -\left(\eta+{4s\over\alpha}\right)u^3
 -\left(2\eta^3+{8s\eta^2\over\alpha}+{12s^2\eta\over\alpha^2}
 +{8s^3\over\alpha^3}\right)u
 \right] du.
\label{eq:F40R}
\end{eqnarray}

\subsection{Solution to the radial integrals}
This result leaves us with radial integrals of the type
\begin{equation}
 \int_0^1{u^h\over\sqrt{\eta^2-u^2}}du,\qquad h\in\{1,3,5\},
\label{eq:IntSq}
\end{equation}
apart from the trivial $u^n$-type of integrals. The three integrals in
Eq.~(\ref{eq:IntSq}) can be found for example in
\citeN{Gradshteyn:94:1} as Eqs.~(2.271.7), (2.272.7), and (2.273.8),
which by insertion of $u$ and $\eta$ becomes
\begin{eqnarray}
 \int{u\over\sqrt{\eta^2-u^2}}du&=&-\sqrt{\eta^2-u^2},
\\
 \int{u^3\over\sqrt{\eta^2-u^2}}du&=&-{1\over3}\Bigl[u^2+2\eta^2\Bigr]
 \sqrt{\eta^2-u^2},
\\
 \int{u^5\over\sqrt{\eta^2-u^2}}du&=&
  -{1\over15}\Bigl[3u^4+8\eta^4+4u^2\eta^2\Bigr]\sqrt{\eta^2-u^2},
\end{eqnarray}
in that order, verified by differentiation, and after correction of
the misprint appearing in Eq.~(2.273.8). By insertion of the limits we
get
\begin{eqnarray}
 \int_0^1{u\over\sqrt{\eta^2-u^2}}du&=&\eta-\sqrt{\eta^2-1},
\label{eq:2.271.7}
\\
 \int_0^1{u^3\over\sqrt{\eta^2-u^2}}du&=&
 {2\eta^3\over3}-{1+2\eta^2\over3}\sqrt{\eta^2-1},
\label{eq:2.272.7}
\\
 \int_{0}^{1}{u^5\over\sqrt{\eta^2-u^2}}du&=&
  {8\over15}\eta^5-{1\over15}\Bigl[3+8\eta^4+4\eta^2\Bigr]\sqrt{\eta^2-1}.
\label{eq:2.273.8}
\end{eqnarray}
By insertion of Eqs.~(\ref{eq:2.271.7})--(\ref{eq:2.273.8}) into
Eqs.~(\ref{eq:F00R})--(\ref{eq:F40R}) and solving the trivial
$u^n$-type of integrals, the resulting expressions for the nine
different cases of Eq.~(\ref{eq:Fpq1}) thus become
\begin{eqnarray}
\lefteqn{
 {\cal{F}}_{00}^{1}(n,a,b,s)=
 {2\pi\alpha\over{}a}
 \left[\eta-\sqrt{\eta^2-1}\right]
,}
\label{eq:F00Rs}
\\
\lefteqn{
 {\cal{F}}_{02}^{1}(n,a,b,s)=
 {2\pi\alpha^3\over{}a}
 \left[
 {\eta^2-1\over3}\sqrt{\eta^2-1}-{\eta^3\over3}+{\eta\over2}
 \right],
}\label{eq:F02Rs}
\\
\lefteqn{
 {\cal{F}}_{04}^{1}(n,a,b,s)=
 {\pi\alpha^5\over{}a}
 \left[
 \left({4\over5}\eta^2-{2\over5}\eta^4-{2\over5}\right)\sqrt{\eta^2-1}
 +{2\over5}\eta^5-\eta^3+{3\over4}\eta
 \right]
}\label{eq:F04Rs}
\\
\lefteqn{
 {\cal{F}}_{10}^{1}(n,a,b,s)=
 {2\pi\alpha^2\over{}a}
 \left[{1\over2}-\left(\eta+{s\over\alpha}\right)
 \left(\eta-\sqrt{\eta^2-1}\right)
 \right],
}\label{eq:F10Rs}
\\
\lefteqn{
 {\cal{F}}_{12}^{1}(n,a,b,s)=
 {\pi\alpha^4\over{}a}
 \left[
 \left({2s\over3\alpha}+{2\over3}\eta-{2s\over3\alpha}\eta^2-{2\over3}\eta^3
 \right)\sqrt{\eta^2-1}
 +{2\over3}\eta^4+{2s\over3\alpha}\eta^3
\right.}\nonumber\\ &\quad&\left.\!
 -\eta^2-{s\over\alpha}\eta+{1\over4}
 \right],
\label{eq:F12Rs}
\\
\lefteqn{
 {\cal{F}}_{20}^{1}(n,a,b,s)=
 {2\pi\alpha^3\over{}a}
 \left[\left(\eta^2+{2s\eta\over\alpha}+{s^2\over\alpha^2}\right)
 \left(\eta-\sqrt{\eta^2-1}\right)
 -{1\over2}\left(\eta+{2s\over\alpha}\right)
 \right],
}\label{eq:F20Rs}
\\
\lefteqn{
 {\cal{F}}_{22}^{1}(n,a,b,s)=
 {\pi\alpha^5\over{}a}
 \left[
 \left({2\over3}\eta^4+{4s\over3\alpha}\eta^3
 +\left({2s^2\over3\alpha^2}-{2\over3}\right)\eta^2-{4s\over3\alpha}\eta
 -{2s^2\over3\alpha^2}\right)\sqrt{\eta^2-1}
\right.}\nonumber\\ &&\left.
 -{2\over3}\eta^5-{4s\over3\alpha}\eta^4
 +\left(1-{2s^2\over3\alpha^2}\right)\eta^3
 +{2s\over\alpha}\eta^2+\left({s^2\over\alpha^2}-{1\over4}\right)\eta
 -{s\over2\alpha}
 \right],
\label{eq:F22Rs}
\\
\lefteqn{
 {\cal{F}}_{30}^{1}(n,a,b,s)=
 {\pi\alpha^4\over{}a}
 \left[{1\over4}
 +{1\over2}\left({2\eta^2}+{6s\eta\over\alpha}+{6s^2\over\alpha^2}\right)
\right.}\nonumber\\ &&\left.\!
 -\left({2\eta^3}+{6s\eta^2\over\alpha}+{6s^2\eta\over\alpha^2}
 +{2s^3\over\alpha^3}\right)
 \left(\eta-\sqrt{\eta^2-1}\right)
 \right],
\label{eq:F30Rs}
\\
\lefteqn{
 {\cal{F}}_{40}^{1}(n,a,b,s)=
 {\pi\alpha^5\over{}a}
 \left[
 \left({2\eta^4}+{8s\eta^3\over\alpha}+{12s^2\eta^2\over\alpha^2}
 +{8s^3\eta\over\alpha^3}+{2s^4\over\alpha^4}\right)
 \left(\eta-\sqrt{\eta^2-1}\right)
\right.}\nonumber\\ &&\left.\!
 -{1\over4}\left(\eta+{4s\over\alpha}\right)
 -{1\over2}\left(2\eta^3+{8s\eta^2\over\alpha}+{12s^2\eta\over\alpha^2}
 +{8s^3\over\alpha^3}\right)
 \right].
\label{eq:F40Rs}
\end{eqnarray}

\section{Verification}
In order to verify the result, we take the limit where
$a\rightarrow0$, we use the binomial series expansion
\begin{eqnarray}
\lefteqn{
 \sqrt{\eta^2-1}=
 \eta-{1\over2\eta}-\sum_{n=2}^{\infty}{(2n-3)!!\over(2n)!!\eta^{2n-1}}
}\nonumber\\ &\quad&=
 \eta
 -{1\over2\eta}
 -{1\over8\eta^3}
 -{1\over16\eta^5}
 -{5\over128\eta^7}
 -{7\over256\eta^9}
 -{21\over1024\eta^{11}}
 -{33\over2048\eta^{13}}
\nonumber\\ &&
 -{429\over32768\eta^{15}}
 -{715\over65536\eta^{17}}-\dots,
\label{eq:binomial}
\end{eqnarray}
where `$x!!$' is the `double factorial' operator taken for the integer
$x$, given by $x!!\equiv{}x(x-2)(x-4)\cdots(x-k)$ for $x>k$. Thus by
insertion of Eq.~(\ref{eq:binomial}) into
Eqs.~(\ref{eq:F00Rs})--(\ref{eq:F40Rs}) we get for small values of $a$
\begin{eqnarray}
\lefteqn{
 {\cal{F}}_{00}^{1}(n,a,b,s)=
 {2\pi\alpha\over{}a}
 \left[{1\over2\eta}+{1\over8\eta^3}+{1\over16\eta^5}+{5\over128\eta^7}+\cdots
 \right]
,}
\label{eq:F00Ra}
\\
\lefteqn{
 {\cal{F}}_{02}^{1}(n,a,b,s)=
 {2\pi\alpha^3\over{}a}
 \left[{1\over8\eta}+{1\over48\eta^3}+{3\over384\eta^5}+{3\over768\eta^7}
 +\cdots
 \right],
}\label{eq:F02Ra}
\\
\lefteqn{
 {\cal{F}}_{04}^{1}(n,a,b,s)=
 {\pi\alpha^5\over{}a}
 \left[
 {1\over8\eta}+{1\over320\eta^3}+{19\over640\eta^5}+{21\over2560\eta^7}
 +\cdots
 \right]
}\label{eq:F04Ra}
\\
\lefteqn{
 {\cal{F}}_{10}^{1}(n,a,b,s)=
 -{2\pi\alpha^2\over{}a}
 \left[{s\over2\alpha\eta}+{1\over8\eta^2}+{s\over8\alpha\eta^3}
 +{1\over16\eta^4}+\cdots
 \right],
}\label{eq:F10Ra}
\\
\lefteqn{
 {\cal{F}}_{12}^{1}(n,a,b,s)=
 -{\pi\alpha^4\over{}a}
 \left[{s\over4\alpha\eta}+{1\over24\eta^2}+{s\over24\alpha\eta^3}
 +{3\over192\eta^4}+\cdots
 \right],
}\label{eq:F12Ra}
\\
\lefteqn{
 {\cal{F}}_{20}^{1}(n,a,b,s)=
 {2\pi\alpha^3\over{}a}
 \left[\left({s^2\over2\alpha^2}+{1\over8}\right){1\over\eta}
 +{s\over4\alpha\eta^2}
 +\left({s^2\over8\alpha^2}+{1\over16}\right){1\over\eta^3}
 +{s\over8\alpha\eta^4}+\cdots
 \right],
}\nonumber\\
\label{eq:F20Ra}
\\
\lefteqn{
 {\cal{F}}_{22}^{1}(n,a,b,s)=
 {\pi\alpha^5\over{}a}\!
 \left[\!\left(\!{s^2\over4\alpha^2}+{1\over24}\!\right)\!{1\over\eta}
 +{s\over12\alpha\eta^2}
 +\!\left(\!{s^2\over24\alpha^2}+{7\over192}\!\right)\!{1\over\eta^3}
 +{7s\over96\alpha\eta^4}+\cdots
 \right]\!,
}\nonumber\\
\label{eq:F22Ra}
\\
\lefteqn{
 {\cal{F}}_{30}^{1}(n,a,b,s)=
 -{\pi\alpha^4\over{}a}
 \left[\left({s^3\over\alpha^3}+{3s\over4\alpha}\right){1\over\eta}
 +\left({3s^2\over4\alpha^2}+{1\over8}\right){1\over\eta^2}
 +\left({s^3\over4\alpha^3}+{3s\over8\alpha}\right){1\over\eta^3}
\right.}\nonumber\\ &\quad&\left.\!
 +\left({3s^2\over8\alpha^2}+{5\over64}\right){1\over\eta^4}+\cdots
 \right],
\label{eq:F30Ra}
\\
\lefteqn{
 {\cal{F}}_{40}^{1}(n,a,b,s)=
 {\pi\alpha^5\over{}a}
 \left[
 \left({s^4\over\alpha^4}+{3s^2\over2\alpha^2}+{1\over8}\right){1\over\eta}
 +\left({s^3\over\alpha^3}+{s\over2\alpha}\right){1\over\eta^2}
\right.}\nonumber\\ &\quad&\left.\!
 +\left({s^4\over4\alpha^4}+{3s^2\over4\alpha^2}+{5\over64}\right)
 {1\over\eta^3}
 +\left({s^3\over2\alpha^3}+{5s\over16\alpha}\right){1\over\eta^4}
 +\cdots
 \right],
\label{eq:F40Ra}
\end{eqnarray}
which by insertion of $\eta=(b-as)/(a\alpha)$ and subsequently letting
$a=0$ reduces to the results of
Eqs.~(\ref{eq:F-00-1-a=0})--(\ref{eq:F-40-1-a=0}).

\section{Result}
In terms of the original $a$ and $b$ quantities, the resulting
expressions for the specific integrals are:
\begin{eqnarray}
\lefteqn{
 {\cal{F}}_{00}^{1}(n,a,b,s)={2\pi\over{}a^2}
 \left[b-as-\sqrt{(b-as)^2-a^2\alpha^2}\right],
}\label{eq:F00s}
\\
\lefteqn{
 {\cal{F}}_{02}^{1}(n,a,b,s)=
 {2\pi\over3a^4}\left[
 \left((b-as)^2-{\alpha^2a^2}\right)^{3\over2}
 -(b-as)^3
 +{3\over2}{\alpha^2a^2(b-as)}\right].
}\\
\lefteqn{
 {\cal{F}}_{04}^{1}(n,a,b,s)=
 {2\pi\over5a^6}
 \left[
 (b-as)^5-{5\over2}a^2\alpha^2(b-as)^3+{15\over8}a^4\alpha^4(b-as)
\right.}\nonumber\\ &\quad&\left.\!
 -\left((b-as)^2-a^2\alpha^2\right)^{5\over2}
 \right]
\\
\lefteqn{
 {\cal{F}}_{10}^{1}(n,a,b,s)=
 {2\pi\alpha^2\over{}a}
 \left[{1\over2}-{b\over{}a^2\alpha^2}
 \left({b-as}-\sqrt{(b-as)^2-a^2\alpha^2}\right)
 \right],
}\\
\lefteqn{
 {\cal{F}}_{12}^{1}(n,a,b,s)=
 {\pi\over{}a^5}\left[{1\over4}a^4\alpha^4
 -b\left(a^2\alpha^2(b-as)-2(b-as)^2\right)
\right.}\nonumber\\ &&\left.\!
 -{2b\over3}\left((b-as)^2-a^2\alpha^2\right)^{3\over2}\right]
\\
\lefteqn{
 {\cal{F}}_{20}^{1}(n,a,b,s)=
 {2\pi\over{}a^4}
 \left[b^2\left(b-as-\sqrt{(b-as)^2-a^2\alpha^2}\right)
 -{1\over2}a^2\alpha^2(b+as)\right],
}\\
\lefteqn{
 {\cal{F}}_{22}^{1}(n,a,b,s)=
 {2\pi\over3a^6}\left[b^2\left(6a^2\alpha^2(b-as)-4a\alpha(b-as)^3
 +\left((b-as)^2-a^2\alpha^2\right)^{3\over2}\right)
\right.}\nonumber\\ &&\left.\!
 -{3\over8}a^4\alpha^4(b+as)\right]
\\
\lefteqn{
 {\cal{F}}_{30}^{1}(n,a,b,s)=
 {\pi\alpha^4\over{}a}
 \left[{1\over4}
 +{(b+as)^2\over{}a^2\alpha^2}
 -{2b^3\over{}a^4\alpha^4}
 \left(b-as-\sqrt{(b-as)^2-a^2\alpha^2}\right)
 \right],
}\\
\lefteqn{
 {\cal{F}}_{40}^{1}(n,a,b,s)=
 {\pi\alpha^5\over{}a}
 \left[
 {2b^4\over{}a^5\alpha^5}
 \left(b-as-\sqrt{(b-as)^2-a^2\alpha^2}\right)
 -{b\over4a\alpha}
 -{b^3\over{}a^3\alpha^3}
\right.}\nonumber\\ &&\left.\!
 -{s\over\alpha}\left({3\over4}+{(b+as)^2\over{}a^2\alpha^2}\right)
 \right].
\label{eq:F40s}
\end{eqnarray}
Using Eqs.~(\ref{eq:F00s})--(\ref{eq:F40s}) together with
Eqs.~(\ref{eq:Fpq2}) and (\ref{eq:Fpq3}) when they are needed, the
solution to the integrals appearing in
Eqs.~(\ref{eq:SigmaAA})--(\ref{eq:XiAG}), (\ref{eq:C}), (\ref{eq:D}),
(\ref{eq:N}), and (\ref{eq:A-xx})--(\ref{eq:A-zz}) are obtained in a
straightforward manner. They can be found on detailed form in
Appendix~\ref{app:C}.

\chapter[Analytic solution to the integrals over
$\vec{\kappa}_{\|}$]{Analytic solution to the integrals over
  $\vec{\kappa}_{\|}$ appearing in the conductivity tensors
  when scattering takes place in the $x$-$z$-plane}\label{app:C}
\noindent
Taking a close look at the expressions for the different nonlinear
conductivity tensor parts in Eqs.~(\ref{eq:XiAA})--(\ref{eq:XiAG}) we
observe by insertion of the transition current density given by
Eq.~(\ref{eq:Jm->n||}) that they can be separated into two independent
parts. One of these parts depends solely on $\vec{\kappa}_{\|}$ and
the other part depends only on $z$. In the remaining of this
Appendix they will be denoted by $\xi$ and ${\cal{Z}}$,
respectively. The quantity $\xi$ can be solved according to the
solution scheme given in Appendix~\ref{ch:Solve-Q}, and the explicit
solution will therefore in the following be given in terms of the
functions solved in Appendix~\ref{ch:Solve-Q}. Furthermore it is
possible to split the $z$-dependent part into independent functions of
each $z$-coordinate. Since the $z$-dependence involves only the wave
functions, we define the following three new quantities
\begin{eqnarray}
 {\cal{Z}}_{nm}^{x}(z)&=&{\cal{Z}}_{nm}^{\,y}(z)=\psi_m^*(z)\psi_n(z),
\label{eq:Z-nm-xy}\\
 {\cal{Z}}_{nm}^{z}(z)&=&\psi_m^*(z){\partial\psi_n(z)\over\partial{}z}
 -\psi_n(z){\partial\psi_m^*(z)\over\partial{}z},
\label{eq:Z-nm-z}
\end{eqnarray}
in order to reduce the expressions in the following.

As the first step in preparing the solutions to the integrals over
$\vec{\kappa}_{\|}$ we identify the transition frequencies occuring in
the nonlinear conductivity tensor parts. Each of these transition
frequencies can on general form be written
\begin{eqnarray}
 {\omega}_{nm}(\vec{\kappa}_{\|}+\vec{\beta},\vec{\kappa}_{\|}+\vec{\gamma})=
 {1\over\hbar}\left[\varepsilon_n-\varepsilon_m+{\hbar^2\over2m_e}
 \left(2\kappa_x(\beta_x-\gamma_x)+\beta_x^2-\gamma_x^2\right)\right].
\label{eq:transit-nm}
\end{eqnarray}
Looking at Eqs.~(\ref{eq:XiAA})--(\ref{eq:XiAG}), we observe that the
following transition frequencies appear in the nonlinear conductivity
tensor:
\begin{eqnarray}
 {\omega}_{nm}(\vec{\kappa}_{\|},\vec{\kappa}_{\|})&=&
 {\omega}_{nm}(\vec{\kappa}_{\|}+\vec{q}_{\|},\vec{\kappa}_{\|}+\vec{q}_{\|})=
 {1\over\hbar}\left[\varepsilon_n-\varepsilon_m\right],
\label{eq:omega-nm-1}\\
 \omega_{nm}(\vec{\kappa}_{\|}+\vec{k}_{\|},\vec{\kappa}_{\|})
 &=&{1\over\hbar}\left[\varepsilon_n-\varepsilon_m+{\hbar^2k_x\over2m_e}
  \left(2\kappa_x+k_x\right)\right],
\label{eq:omega-nm-2}\\
 \omega_{nm}(\vec{\kappa}_{\|},\vec{\kappa}_{\|}+\vec{k}_{\|})
 &=&{1\over\hbar}\left[\varepsilon_n-\varepsilon_m-{\hbar^2k_x\over2m_e}
  \left(2\kappa_x+k_x\right)\right],
\label{eq:omega-nm-3}\\
 {\omega}_{nm}(\vec{\kappa}_{\|}-\vec{k}_{\|},\vec{\kappa}_{\|})&=&
 {1\over\hbar}\left[\varepsilon_n-\varepsilon_m+{\hbar^2k_x\over2m_e}
 \left(k_x-2\kappa_x\right)\right],
\label{eq:omega-nm-4}\\
 {\omega}_{nm}(\vec{\kappa}_{\|},\vec{\kappa}_{\|}-\vec{k}_{\|})&=&
 {1\over\hbar}\left[\varepsilon_n-\varepsilon_m+{\hbar^2k_x\over2m_e}
 \left(2\kappa_x-k_x\right)\right],
\label{eq:omega-nm-5}\\
 {\omega}_{nm}(\vec{\kappa}_{\|}+\vec{q}_{\|},\vec{\kappa}_{\|})&=&
 {1\over\hbar}\left[\varepsilon_n-\varepsilon_m+{\hbar^2q_x\over2m_e}
 \left(2\kappa_x+q_x\right)\right],
\label{eq:omega-nm-6}\\
 \omega_{nm}(\vec{\kappa}_{\|}+\vec{q}_{\|}
 ,\vec{\kappa}_{\|}+\vec{k}_{\|})&=&
 {1\over\hbar}\left[\varepsilon_n-\varepsilon_m+{\hbar^2\over2m_e}
  \left(2\kappa_x(q_x-k_x)+q_x^2-k_x^2\right)\right],
\label{eq:omega-nm-7}\\
 {\omega}_{nm}(\vec{\kappa}_{\|}+\vec{q}_{\|},
 \vec{\kappa}_{\|}-\vec{k}_{\|})&=&
 {1\over\hbar}\left[\varepsilon_n-\varepsilon_m+{\hbar^2\over2m_e}
 \left(2\kappa_x(q_x+k_x)+q_x^2-k_x^2\right)\right],
\label{eq:omega-nm-8}\\
 {\omega}_{nm}(\vec{\kappa}_{\|}-\vec{k}_{\|}+\vec{q}_{\|},
 \vec{\kappa}_{\|})&=&
 {1\over\hbar}\left[\varepsilon_n-\varepsilon_m+{\hbar^2\over2m_e}(q_x-k_x)
 \left(2\kappa_x+(q_x-k_x)\right)\right],
\label{eq:omega-nm-9}\\
 \omega_{nm}(\vec{\kappa}_{\|}+\vec{k}_{\|}+\vec{q}_{\|}
 ,\vec{\kappa}_{\|})&=&
 {1\over\hbar}\left[\varepsilon_n-\varepsilon_m+{\hbar^2\over2m_e}
  (q_x+k_x)\left(2\kappa_x+(q_x+k_x)\right)\right],
\label{eq:omega-nm-10}\\
 \omega_{nm}(\vec{\kappa}_{\|}+\vec{k}_{\|}+\vec{q}_{\|}
 ,\vec{\kappa}_{\|}+\vec{k}_{\|})&=&
 {1\over\hbar}\left[\varepsilon_n-\varepsilon_m+{\hbar^2q_x\over2m_e}
  \left(2\kappa_x+q_x+2k_x\right)\right],
\label{eq:omega-nm-11}\\
 \omega_{nm}(\vec{\kappa}_{\|}+\vec{k}_{\|}+\vec{q}_{\|}
 ,\vec{\kappa}_{\|}+\vec{q}_{\|})&=&
 {1\over\hbar}\left[\varepsilon_n-\varepsilon_m+{\hbar^2k_x\over2m_e}
  \left(2\kappa_x+k_x+2q_x\right)\right],
\label{eq:omega-nm-12}\\
 {\omega}_{nm}(\vec{\kappa}_{\|}+\vec{q}_{\|},
 \vec{\kappa}_{\|}-\vec{k}_{\|}+\vec{q}_{\|})&=&
 {1\over\hbar}\left[\varepsilon_n-\varepsilon_m+{\hbar^2k_x\over2m_e}
 \left(2\kappa_x+2q_x-k_x\right)\right],
\label{eq:omega-nm-13}\\
 {\omega}_{nm}(\vec{\kappa}_{\|}-\vec{k}_{\|}+\vec{q}_{\|},
 \vec{\kappa}_{\|}-\vec{k}_{\|})&=&
 {1\over\hbar}\left[\varepsilon_n-\varepsilon_m+{\hbar^2q_x\over2m_e}
 \left(2\kappa_x+q_x-2k_x\right)\right],
\label{eq:omega-nm-14}\\
 {\omega}_{nm}(\vec{\kappa}_{\|}-\vec{k}_{\|}+\vec{q}_{\|},
 \vec{\kappa}_{\|}+\vec{q}_{\|})&=&
 {1\over\hbar}\left[\varepsilon_n-\varepsilon_m+{\hbar^2k_x\over2m_e}
 \left(k_x-2q_x-2\kappa_x\right)\right].
\label{eq:omega-nm-15}
\end{eqnarray}
We observe from Eqs.~(\ref{eq:XiAA})--(\ref{eq:XiAG}) that these
transition frequencies gives rise to a number of different $a$ and $b$
coefficients, which we will use in the later sections of this
Appendix. It turns out that there are a total of four different $a$'s
and thirteen different $b$'s. To present an overview and for the sake
of easy reference they are all listed together below, viz.
\begin{eqnarray}
 a_{1}&=&{\hbar{}k_x\over{}m_e},
\\
 a_{2}&=&{\hbar{}q_x\over{}m_e},
\\
 a_{3}&=&{\hbar\over{}m_e}(q_x+k_x),
\\
 a_{4}&=&{\hbar\over{}m_e}(q_x-k_x),
\\
 b_{nm}^{1}&=&{1\over\hbar}(\varepsilon_n-\varepsilon_m)
 +{\hbar{}k_x^2\over2m_e}-\omega-{\rm{i}}\tau_{nm}^{-1},
\\
 b_{nm}^{2}&=&{1\over\hbar}(\varepsilon_n-\varepsilon_m)
 -{\hbar{}k_x^2\over2m_e}-\omega-{\rm{i}}\tau_{nm}^{-1},
\\
 b_{nm}^{3}&=&{1\over\hbar}(\varepsilon_n-\varepsilon_m)
 +{\hbar{}q_x^2\over2m_e}+\omega-{\rm{i}}\tau_{nm}^{-1},
\\
 b_{nm}^{4}&=&{1\over\hbar}(\varepsilon_n-\varepsilon_m)
 +{\hbar{}q_x^2\over2m_e}-\omega-{\rm{i}}\tau_{nm}^{-1},
\\
 b_{nm}^{5}&=&{1\over\hbar}(\varepsilon_n-\varepsilon_m)
 +{\hbar\over2m_e}(q_x-k_x)^2-{\rm{i}}\tau_{nm}^{-1},
\\
 b_{nm}^{6}&=&{1\over\hbar}(\varepsilon_n-\varepsilon_m)
 -{\hbar\over2m_e}(q_x-k_x)^2-{\rm{i}}\tau_{nm}^{-1},
\\
 b_{nm}^{7}&=&{1\over\hbar}(\varepsilon_n-\varepsilon_m)
 +{\hbar\over2m_e}(q_x+k_x)^2-{\rm{i}}\tau_{nm}^{-1},
\\
 b_{nm}^{8}&=&{1\over\hbar}(\varepsilon_n-\varepsilon_m)
 +{\hbar\over2m_e}(q_x^2-k_x^2)-{\rm{i}}\tau_{nm}^{-1},
\\
 b_{nm}^{9}&=&{1\over\hbar}(\varepsilon_n-\varepsilon_m)
 +{\hbar{}q_x\over2m_e}(q_x+2k_x)+\omega-{\rm{i}}\tau_{nm}^{-1},
\\
 b_{nm}^{10}&=&{1\over\hbar}(\varepsilon_n-\varepsilon_m)
 +{\hbar{}k_x\over2m_e}(k_x+2q_x)-\omega-{\rm{i}}\tau_{nm}^{-1},
\\
 b_{nm}^{11}&=&{1\over\hbar}(\varepsilon_n-\varepsilon_m)
 +{\hbar{}q_x\over2m_e}(q_x-2k_x)+\omega-{\rm{i}}\tau_{nm}^{-1},
\\ 
 b_{nm}^{12}&=&{1\over\hbar}(\varepsilon_n-\varepsilon_m)
 +{\hbar{}k_x\over2m_e}(2q_x-k_x)-\omega-{\rm{i}}\tau_{nm}^{-1},
\\
 b_{nm}^{13}&=&{1\over\hbar}(\varepsilon_n-\varepsilon_m)
 +{\hbar{}k_x\over2m_e}(k_x-2q_x)-\omega-{\rm{i}}\tau_{nm}^{-1}.
\end{eqnarray}

\section{Nonlinear process A}
Starting with the pure interband term in Eq.~(\ref{eq:XiAA}) we
separate the $z$-dependent and the $\vec{\kappa}_{\|}$-dependent parts
in the following way:
\begin{eqnarray}
 {\Xi}_{xxxx}^{\rm{A}}(z,z',z'',z''';\vec{q}_{\|},\vec{k}_{\|})=
 {2\over({\rm{i}}\omega)^3}
 \sum_{nm}{\cal{Z}}_{nm}^{\rm{A}}(z,z',z'',z''')
 \xi_{nm}^{\rm{A}},
\end{eqnarray}
where the indices on the quantities ${\cal{Z}}$ and $\xi$ follows the
indices of the quantum numbers in the sum. The $z$-dependent part
above is
\begin{equation}
 {\cal{Z}}_{nm}^{\rm{A}}(z,z',z'',z''')=
 {\cal{Z}}_{mn}^{x}(z''){\cal{Z}}_{nm}^{x}(z)
 \delta(z-z')\delta(z''-z'''),
\end{equation}
and the solution to the integral over $\vec{\kappa}_{\|}$ in the
low-temperature limit 
\begin{equation}
 \xi_{nm}^{\rm{A}}=
 -{e^4\over32\pi{}m_e^2\hbar}
 {\alpha_n^2-\alpha_m^2\over(\varepsilon_n-\varepsilon_m)/\hbar
 -2\omega-{\rm{i}}\tau_{nm}^{-1}}
\end{equation}
readily appears by use of Eq.~(\ref{eq:Solve-Q-Beta=0}), since only
the Fermi-Dirac distribution functions depend on $\vec{\kappa}_{\|}$.

\section{Nonlinear process B}
In order to solve the integral over $\kappa_{\|}$ in
Eq.~(\ref{eq:XiAB}), we rewrite it into
\begin{eqnarray}
{\Xi}_{xxkh}^{\rm{B}}(z,z',z'',z''';\vec{q}_{\|},\vec{k}_{\|})=
 {2\over({\rm{i}}\omega)^3}\left({e\hbar\over2m_e}\right)^2\sum_{nmv}
 {\cal{Z}}_{kh,nmv}^{\rm{B}}(z,z',z'',z''')
 \xi_{kh,nmv}^{\rm{B}}(\vec{q}_{\|},\vec{k}_{\|}),
\nonumber\\
\label{eq:XiAB-M}
\end{eqnarray}
in which both quantities ${\cal{Z}}$ and $\xi$ are indexed according
to their dependence on the two Cartesian indices of $\Xi$ and the
quantum numbers in the sum. Above, the $z$-dependent part in general
is given by
\begin{eqnarray}
 {\cal{Z}}_{kh,nmv}^{\rm{B}}(z,z',z'',z''')=
 {\cal{Z}}_{v{}n}^{h}(z'''){\cal{Z}}_{mv}^{k}(z'')
 \delta(z-z'){\cal{Z}}_{nm}^{x}(z),
\end{eqnarray}
in terms of Eqs.~(\ref{eq:Z-nm-xy}) and (\ref{eq:Z-nm-z}). Of these,
the two with Cartesian indices $xx$ and $yy$ are equal. The solution
to the other quantity above,
$\xi_{kh,nmv}^{\rm{B}}(\vec{q}_{\|},\vec{k}_{\|})$, in terms of the $a$'s
and $b$'s and the functions solved in Appendix~\ref{ch:Solve-Q}, is
written
\begin{eqnarray}
\lefteqn{ 
 \xi_{kh,nmv}^{\rm{B}}(\vec{q}_{\|},\vec{k}_{\|})=
 -{e^2\over4m_e\hbar^2}{1\over(2\pi)^2}
 {1\over(\varepsilon_n-\varepsilon_m)/\hbar-2\omega-{\rm{i}}\tau_{nm}^{-1}}
 \left\{
 {\cal{F}}_{kh}^{\rm{B}}(m,a_1,b_{v{}m}^{1},0)
\right.}\nonumber\\ &&\left.
 -{\cal{F}}_{kh}^{\rm{B}}(v,a_1,b_{v{}m}^{1},k_x)
 +{\cal{F}}_{kh}^{\rm{B}}(n,-a_1,b_{nv}^{2},0)
 -{\cal{F}}_{kh}^{\rm{B}}(v,-a_1,b_{nv}^{2},k_x)
 \right\},
\end{eqnarray}
which is written in terms of a set of functions ${\cal{F}}$ that vary
from element to element. These functions are of the order $\beta=1$
because only one transition frequency appears inside each integral.
They are determined from the $\vec{\kappa}_{\|}$-dependent parts of
the microscopic current densities, and they become
\begin{eqnarray}
 {\cal{F}}_{zz}^{\rm{B}}&=&{\cal{F}}_{00}^{1},
\\
 {\cal{F}}_{yy}^{\rm{B}}&=&4{\cal{F}}_{02}^{1},
\\
 {\cal{F}}_{xz}^{\rm{B}}&=&{\cal{F}}_{zx}^{\rm{B}}=
 2{\cal{F}}_{10,n}^{1}+k_x{\cal{F}}_{00}^{1},
\\
 {\cal{F}}_{xx}^{\rm{B}}&=&4{\cal{F}}_{20}^{1}
 +4k_x{\cal{F}}_{10}^{1}+k_x^2{\cal{F}}_{00}^{1},
\end{eqnarray}
in short notation, since the functions at the right side of these
equations take the same arguments as the functions to the left.

\section{Nonlinear process C}
Separating Eq.~(\ref{eq:XiAC}) into its $z$-dependent and
$\vec{\kappa}_{\|}$-dependent parts, we write
\begin{eqnarray}
 {\Xi}_{xxxx}^{\rm{C}}(z,z',z'',z''';\vec{q}_{\|},\vec{k}_{\|})=
 {2\over({\rm{i}}\omega)^3}
 \sum_{nm}{\cal{Z}}_{nm}^{\rm{C}}(z,z',z'',z''')
 \xi_{nm}^{\rm{C}}(\vec{q}_{\|},\vec{k}_{\|})
\label{eq:XiAC-M}
\end{eqnarray}
where the indices on the new quantities follows the quantum numbers in
the sum. The $z$-independent part in this part of the conductivity
tensor is
\begin{equation}
{\cal{Z}}_{nm}^{\rm{C}}(z,z',z'',z''')=
 {\cal{Z}}_{mn}^{x}(z'){\cal{Z}}_{nm}^{x}(z)
 \delta(z'-z''')\delta(z-z''),
\end{equation}
in terms of Eq.~(\ref{eq:Z-nm-xy}). The solution to the quantity
$\xi_{nm}^{\rm{C}}(\vec{q}_{\|},\vec{k}_{\|})$ then appears as
\begin{eqnarray}
 \xi_{nm}^{\rm{C}}(\vec{q}_{\|},\vec{k}_{\|})=
 -{e^4\over16\pi^2m_e^2\hbar}
 \left\{ 
 {\cal{F}}_{00}^{1}(n,a_4,b_{nm}^{5},q_x-k_x)
 -{\cal{F}}_{00}^{1}(m,a_4,b_{nm}^{5},0)
 \right\}
\end{eqnarray}
in terms of the $a$'s and $b$'s and the functions solved in
Appendix~\ref{ch:Solve-Q}.

\section{Nonlinear process D}
Performing an adequate separation of variables in Eq.~(\ref{eq:XiAD}),
it is written
\begin{eqnarray}
\lefteqn{
 {\Xi}_{xjkx}^{\rm{D}}(z,z',z'',z''';\vec{q}_{\|},\vec{k}_{\|})=
 {2\over({\rm{i}}\omega)^3}\left({e\hbar\over2m_e}\right)^2
 \sum_{nmv}
 \left\{
 {\cal{Z}}_{jk,nmv}^{\rm{Da}}(z,z',z'',z''')
 \xi_{jk,nmv}^{\rm{Da}}(\vec{q}_{\|},\vec{k}_{\|})
\right.}\nonumber\\ &\quad&\left.
 +
 {\cal{Z}}_{jk,nmv}^{\rm{Db}}(z,z',z'',z''')
 \xi_{jk,nmv}^{\rm{Db}}(\vec{q}_{\|},\vec{k}_{\|})
 \right\},
\label{eq:XiAD-M}
\end{eqnarray}
where the four new quantities are indexed according to the varying
Cartesian coordinates of $\Xi$ and the quantum numbers in the sum.
The $z$-dependent terms in Eq.~(\ref{eq:XiAD-M}) are
\begin{eqnarray}
 {\cal{Z}}_{jk,nmv}^{\rm{Da}}(z,z',z'',z''')&=&\delta(z-z''')
 {\cal{Z}}_{mv}^{k}(z''){\cal{Z}}_{v{}n}^{j}(z'){\cal{Z}}_{nm}^{x}(z),
\\
 {\cal{Z}}_{jk,nmv}^{\rm{Db}}(z,z',z'',z''')&=&\delta(z-z''')
 {\cal{Z}}_{v{}n}^{k}(z''){\cal{Z}}_{mv}^{j}(z'){\cal{Z}}_{nm}^{x}(z)
\end{eqnarray}
in terms of Eqs.~(\ref{eq:Z-nm-xy}) and (\ref{eq:Z-nm-z}). In both
equations above, the $xx$ and $yy$ permutations are the same. The
solution to the $\xi$ quantities we write in terms of the $a$'s,
$b$'s, and the functions solved in Appendix~\ref{ch:Solve-Q}, the
result being
\begin{eqnarray}
\lefteqn{
 \xi_{jk,nmv}^{\rm{Da}}(\vec{q}_{\|},\vec{k}_{\|})=
 -{e^2\over16\pi^2m_e\hbar^2}
 \left\{
  {\cal{F}}_{jk}^{\rm{Da}}(m,\{a_3,a_1\},\{b_{nm}^{7},b_{v{}m}^{1}\},0)
\right.}\nonumber\\ &\quad&
 -{\cal{F}}_{jk}^{\rm{Da}}(v,\{a_3,a_1\},\{b_{nm}^{7},b_{v{}m}^{1}\},k_x)
 +{\cal{F}}_{jk}^{\rm{Da}}(n,\{a_3,a_2\},\{b_{nm}^{7},b_{nv}^{9}\},q_x+k_x)
\nonumber\\ &&\left.
 -{\cal{F}}_{jk}^{\rm{Da}}(v,\{a_3,a_2\},\{b_{nm}^{7},b_{nv}^{9}\},k_x)
 \right\},
\\
\lefteqn{
 \xi_{jk,nmv}^{\rm{Db}}(\vec{q}_{\|},\vec{k}_{\|})=
 -{e^2\over16\pi^2m_e\hbar^2}
 \left\{
  {\cal{F}}_{jk}^{\rm{Db}}(m,\{a_3,a_2\},\{b_{nm}^{7},b_{v{}m}^{3}\},0)
\right.}\nonumber\\ &&
 -{\cal{F}}_{jk}^{\rm{Db}}(v,\{a_3,a_2\},\{b_{nm}^{7},b_{v{}m}^{3}\},q_x)
 +{\cal{F}}_{jk}^{\rm{Db}}(n,\{a_3,a_1\},\{b_{nm}^{7},b_{nv}^{10}\},q_x+k_x)
\nonumber\\ &&\left.
 -{\cal{F}}_{jk}^{\rm{Db}}(v,\{a_3,a_1\},\{b_{nm}^{7},b_{nv}^{10}\},q_x)
 \right\},
\end{eqnarray}
which have been written in terms of a set of functions ${\cal{F}}$
that vary from element to element. These functions are again
determined from the $\vec{\kappa}_{\|}$-dependent part of the
microscopic current densities appearing in
${\Xi}_{xjkx}^{\rm{D}}(z,z',z'',z''';\vec{q}_{\|},\vec{k}_{\|})$. They
are
\begin{eqnarray}
 {\cal{F}}_{zz}^{\rm{Da}}&=&
 {\cal{F}}_{zz}^{\rm{Db}}=
 {\cal{F}}_{00}^{2},
\\
 {\cal{F}}_{yy}^{\rm{Da}}&=&
 {\cal{F}}_{yy}^{\rm{Db}}=
 4{\cal{F}}_{02}^{2},
\\
 {\cal{F}}_{xz}^{\rm{Da}}&=&2{\cal{F}}_{10}^{2}+(2k_x+q_x){\cal{F}}_{00,n}^{2},
\\
 {\cal{F}}_{xz}^{\rm{Db}}&=&2{\cal{F}}_{10}^{2}+q_x{\cal{F}}_{00}^{2},
\\
 {\cal{F}}_{zx}^{\rm{Da}}&=&2{\cal{F}}_{10}^{2}+k_x{\cal{F}}_{00}^{2},
\\
 {\cal{F}}_{zx}^{\rm{Db}}&=&2{\cal{F}}_{10}^{2}+(k_x+2q_x){\cal{F}}_{00}^{2},
\\
 {\cal{F}}_{xx}^{\rm{Da}}&=&4{\cal{F}}_{20}^{2}+2(q_x+3k_x){\cal{F}}_{10}^{2}
 +(2k_x^2+q_xk_x){\cal{F}}_{00}^{2},
\\
 {\cal{F}}_{xx}^{\rm{Db}}&=&4{\cal{F}}_{20}^{2}+2(3q_x+k_x){\cal{F}}_{10}^{2}
 +(2q_x^2+q_xk_x){\cal{F}}_{00}^{2},
\end{eqnarray}
again in short notation, and for the same reason as before.

\section{Nonlinear process E}
Separation of the $z$-dependent part and the
$\vec{\kappa}_{\|}$-dependent part in Eq.~(\ref{eq:XiAE}) yields
\begin{eqnarray}
\lefteqn{
 {\Xi}_{ijxx}^{\rm{E}}(z,z',z'',z''';\vec{q}_{\|},\vec{k}_{\|})=
 {2\over({\rm{i}}\omega)^3}\left({e\hbar\over2m_e}\right)^2\sum_{nmv}
\left\{
 {\cal{Z}}_{ij,nmv}^{\rm{Ea}}(z,z',z'',z''')
 \xi_{ij,nmv}^{\rm{Ea}}(\vec{q}_{\|},\vec{k}_{\|})
\right.}\nonumber\\ &\quad&\left.
 +
 {\cal{Z}}_{ij,nmv}^{\rm{Eb}}(z,z',z'',z''')
 \xi_{ij,nmv}^{\rm{Eb}}(\vec{q}_{\|},\vec{k}_{\|})
 \right\},
\label{eq:XiAE-M}
\end{eqnarray}
the new quantities being indexed according to their dependence on the
varying Cartesian coordinates in
${\Xi}_{ijxx}^{\rm{E}}(z,z',z'',z''';\vec{q}_{\|},\vec{k}_{\|})$ and
the quantum numbers in the sum. The $z$-dependent quantities are again
written in terms of Eqs.~(\ref{eq:Z-nm-xy})--(\ref{eq:Z-nm-z}), with
the result
\begin{eqnarray}
 {\cal{Z}}_{ij,nmv}^{\rm{Ea}}(z,z',z'',z''')&=&\delta(z''-z''')
 {\cal{Z}}_{mv}^{x}(z''){\cal{Z}}_{v{}n}^{j}(z'){\cal{Z}}_{nm}^{i}(z),
\\
 {\cal{Z}}_{ij,nmv}^{\rm{Eb}}(z,z',z'',z''')&=&\delta(z''-z''')
 {\cal{Z}}_{v{}n}^{x}(z''){\cal{Z}}_{mv}^{j}(z'){\cal{Z}}_{nm}^{i}(z),
\end{eqnarray}
and again it appears, the quantities with Cartesian indices $xx$ and
$yy$ are equal in each of the above equations. The solutions to the two
$\xi$ quantities are obtained in terms of the $a$'s and $b$'s and the
functions solved in Appendix~\ref{ch:Solve-Q}, and they become
\begin{eqnarray}
\lefteqn{
 \xi_{ij,nmv}^{\rm{Ea}}(\vec{q}_{\|},\vec{k}_{\|})=
 -{e^2\over2^6\pi^2m_e\hbar^2}
 \left\{
 {{\cal{F}}_{ij}^{\rm{E}1}(m,a_2,b_{nm}^{4},0)
  -{\cal{F}}_{ij}^{\rm{E}1}(v,a_2,b_{nm}^{4},0)
  \over(\varepsilon_{v}-\varepsilon_m)/\hbar-2\omega-{\rm{i}}\tau_{v{}m}^{-1}}
\right.}\nonumber\\ &\quad&
 +{\cal{F}}_{ij}^{\rm{E}2}(n,\{a_2,a_2\},\{b_{nm}^{4},b_{nv}^{3}\},q_x)
 -{\cal{F}}_{ij}^{\rm{E}2}(v,\{a_2,a_2\},\{b_{nm}^{4},b_{nv}^{3}\},0)
 \Bigr\},
\\
\lefteqn{
 \xi_{ij,nmv}^{\rm{Eb}}(\vec{q}_{\|},\vec{k}_{\|})=
 -{e^2\over2^6\pi^2m_e\hbar^2}
 \left\{
 {{\cal{F}}_{ij}^{\rm{E}1}(n,a_2,b_{nm}^{4},q_x)
  -{\cal{F}}_{ij}^{\rm{E}1}(v,a_2,b_{nm}^{4},q_x)
  \over(\varepsilon_n-\varepsilon_{v})/\hbar-2\omega-{\rm{i}}\tau_{nv}^{-1}}
\right.}\nonumber\\ &&
 +{\cal{F}}_{ij}^{\rm{E}2}(m,\{a_2,a_2\},\{b_{nm}^{4},b_{v{}m}^{3}\},0)
 -{\cal{F}}_{ij}^{\rm{E}2}(v,\{a_2,a_2\},\{b_{nm}^{4},b_{v{}m}^{3}\},q_x)
 \Bigr\},
\end{eqnarray}
which is written in terms of a set of functions ${\cal{F}}$ that vary
from element to element. Their structure is as before determined from
the $\vec{\kappa}_{\|}$-dependent parts of the transition current
densities, and we find for the pure interband transitions functions of
order $\beta=1$, since only one transition frequency occurs in each
integral. They are
\begin{eqnarray}
 {\cal{F}}_{zz}^{\rm{E}1}&=&{\cal{F}}_{00}^{1},
\\
 {\cal{F}}_{yy}^{\rm{E}1}&=&4{\cal{F}}_{02}^{1},
\\
 {\cal{F}}_{xz}^{\rm{E}1}&=&{\cal{F}}_{zx}^{\rm{E}1}=
 2{\cal{F}}_{10}^{1}+q_x{\cal{F}}_{00}^{1},
\\
 {\cal{F}}_{xx}^{\rm{E}1}&=&4{\cal{F}}_{20}^{1}+2q_x{\cal{F}}_{10}^{1}
 +2q_x^2{\cal{F}}_{00}^{1},
\end{eqnarray}
in short notation, and for the mixed interband/intraband transitions
functions of order $\beta=2$ because two transition frequencies occur
in each integral. They are
\begin{eqnarray}
 {\cal{F}}_{zz}^{\rm{E}2}&=&{\cal{F}}_{00}^{2},
\\
 {\cal{F}}_{yy}^{\rm{E}2}&=&4{\cal{F}}_{02}^{2},
\\
 {\cal{F}}_{xz}^{\rm{E}2}&=&{\cal{F}}_{zx}^{\rm{E}2}=
 2{\cal{F}}_{10}^{2}+q_x{\cal{F}}_{00}^{2},
\\
 {\cal{F}}_{xx}^{\rm{E}2}&=&4{\cal{F}}_{20}^{2}
 +2q_x{\cal{F}}_{10}^{2}+2q_x^2{\cal{F}}_{00}^{2},
\end{eqnarray}
again in short notation, since all arguments to the functions are of
the same type.

\section{Nonlinear process F}
The separation of variables of Eq.~(\ref{eq:XiAF}) into
$z$-independent and $\vec{\kappa}_{\|}$-independent terms gives
\begin{eqnarray}
\lefteqn{ {\Xi}_{ixxh}^{\rm{F}}(z,z',z'',z''';\vec{q}_{\|},\vec{k}_{\|})=
 {2\over({\rm{i}}\omega)^3}\left({e\hbar\over2m_e}\right)^2
 \sum_{nmv}
 \left\{
 {\cal{Z}}_{ih,nmv}^{\rm{Fa}}(z,z',z'',z''')
 \xi_{ih,nmv}^{\rm{Fa}}(\vec{q}_{\|},\vec{k}_{\|})
\right.}\nonumber\\ &\quad&\left.
 +
 {\cal{Z}}_{ih,nmv}^{\rm{Fb}}(z,z',z'',z''')
 \xi_{ih,nmv}^{\rm{Fb}}(\vec{q}_{\|},\vec{k}_{\|})
 \right\},
\label{eq:XiAF-M}
\end{eqnarray}
where the new quantities have been indexed according to their
dependence on the Cartesian indices of $\Xi$ and the quantum numbers
in the sum. In Eq.~(\ref{eq:XiAF-M}) above, the
$\vec{\kappa}_{\|}$-independent terms are
\begin{eqnarray}
 {\cal{Z}}_{ih,nmv}^{\rm{Fa}}(z,z',z'',z''')&=&\delta(z'-z''')
 {\cal{Z}}_{v{}n}^{h}(z''){\cal{Z}}_{mv}^{x}(z'){\cal{Z}}_{nm}^{i}(z),
\\
 {\cal{Z}}_{ih,nmv}^{\rm{Fb}}(z,z',z'',z''')&=&\delta(z'-z''')
 {\cal{Z}}_{mv}^{h}(z''){\cal{Z}}_{v{}n}^{x}(z'){\cal{Z}}_{nm}^{i}(z).
\end{eqnarray}
in terms of the three quantities defined in
Eqs.~(\ref{eq:Z-nm-xy})--(\ref{eq:Z-nm-z}), and again the $xx$ and
$yy$ elements in each of the two above quantities are equal. The
solutions to the $z$-independent terms appear in terms of the $a$'s
and $b$'s and the functions solved in Appendix~\ref{ch:Solve-Q} as
\begin{eqnarray}
\lefteqn{
 \xi_{ih,nmv}^{\rm{Fa}}(\vec{q}_{\|},\vec{k}_{\|})=
 -{e^2\over2^5\pi^2m_e\hbar^2}
\Bigl\{
 {\cal{F}}_{ih}^{\rm{Fa}}(m,\{a_2,a_4\},\{b_{nm}^{4},b_{v{}m}^{5}\},0)
}\nonumber\\ &\quad&
 -{\cal{F}}_{ih}^{\rm{Fa}}(v,\{a_2,a_4\},\{b_{nm}^{4},b_{v{}m}^{5}\},q_x-k_x)
 +{\cal{F}}_{ih}^{\rm{Fa}}(n,\{a_2,a_1\},\{b_{nm}^{4},b_{nv}^{12}\},q_x)
\nonumber\\ &&
 -{\cal{F}}_{ih}^{\rm{Fa}}(v,\{a_2,a_1\},\{b_{nm}^{4},b_{nv}^{12}\},q_x-k_x)
 \Bigr\},
\\
\lefteqn{
 \xi_{ih,nmv}^{\rm{Fb}}(\vec{q}_{\|},\vec{k}_{\|})=
 -{e^2\over2^5\pi^2m_e\hbar^2}
 \Bigl\{
 {\cal{F}}_{ih}^{\rm{Fb}}(m,\{a_2,a_1\},\{b_{nm}^{4},b_{v{}m}^{1}\},0)
}\nonumber\\ &&
 -{\cal{F}}_{ih}^{\rm{Fb}}(v,\{a_2,a_1\},\{b_{nm}^{4},b_{v{}m}^{1}\},k_x)
 +{\cal{F}}_{ih}^{\rm{Fb}}(n,\{a_2,a_4\},\{b_{nm}^{4},b_{nv}^{8}\},q_x)
\nonumber\\ &&
 -{\cal{F}}_{ih}^{\rm{Fb}}(v,\{a_2,a_4\},\{b_{nm}^{4},b_{nv}^{8}\},k_x)
 \Bigr\},
\end{eqnarray}
which are written in terms of a set of functions ${\cal{F}}$ that vary
from element to element. They are again determined from the
$\vec{\kappa}_{\|}$-dependent parts of the transition current
densities appearing, and thus they become
\begin{eqnarray}
 {\cal{F}}_{zz}^{\rm{Fa}}&=&{\cal{F}}_{zz}^{\rm{Fb}}={\cal{F}}_{00}^{2},
\\
 {\cal{F}}_{yy}^{\rm{Fa}}&=&{\cal{F}}_{yy}^{\rm{Fb}}=4{\cal{F}}_{02}^{2},
\\
 {\cal{F}}_{xz}^{\rm{Fa}}&=&{\cal{F}}_{xz}^{\rm{Fb}}=2{\cal{F}}_{00}^{2}
 +q_x{\cal{F}}_{00}^{2},
\\
 {\cal{F}}_{zx}^{\rm{Fa}}&=&2{\cal{F}}_{10}^{2}+(q_x-k_x){\cal{F}}_{00}^{2},
\\
 {\cal{F}}_{zx}^{\rm{Fb}}&=&2{\cal{F}}_{10}^{2}+k_x{\cal{F}}_{00}^{2},
\\
 {\cal{F}}_{xx}^{\rm{Fa}}&=&4{\cal{F}}_{20}^{2}
 +2(2q_x-k_x){\cal{F}}_{00}^{2}+(2q_x^2-q_xk_x){\cal{F}}_{00}^{2},
\\
 {\cal{F}}_{xx}^{\rm{Fb}}&=&4{\cal{F}}_{20}^{2}
 +2(q_x+k_x){\cal{F}}_{10}^{2}+q_xk_x{\cal{F}}_{00}^{2},
\end{eqnarray}
again in the abbreviated notation, where the functions in general take
arguments of the type $(n,\{a_1,a_2\},\{b_1,b_2\},s)$.

\section{Nonlinear process G}
Finally, Eq.~(\ref{eq:XiAG}) becomes in terms of $z$-independent and
$\vec{\kappa}_{\|}$-independent terms
\begin{eqnarray}
\lefteqn{{\Xi}_{ijkh}^{\rm{G}}(z,z',z'',z''';\vec{q}_{\|},\vec{k}_{\|})=
 {2\over({\rm{i}}\omega)^3}\left({e\hbar\over2m_e}\right)^4
 \sum_{nmvl}
\left\{
 {\cal{Z}}_{ijkh,nmvl}^{\rm{Ga}}(z,z',z'',z''')
 \xi_{ijkh,nmvl}^{\rm{Ga}}(\vec{q}_{\|},\vec{k}_{\|})
\right.}\nonumber\\ &\quad&\left.
 +{\cal{Z}}_{ijkh,nmvl}^{\rm{Gb}}(z,z',z'',z''')
 \xi_{ijkh,nmvl}^{\rm{Gb}}(\vec{q}_{\|},\vec{k}_{\|})
 +{\cal{Z}}_{ijkh,nmvl}^{\rm{Gc}}(z,z',z'',z''')
 \xi_{ijkh,nmvl}^{\rm{Gc}}(\vec{q}_{\|},\vec{k}_{\|})
\!\right\},
\nonumber\\
\label{eq:XiAG-M}
\end{eqnarray}
where again the new quantities have been indexed according to their
dependence on the various Cartesian indices of $\Xi$ and the quantum
numbers of the sum. The $z$-dependent terms in Eq.~(\ref{eq:XiAG-M})
are on general form
\begin{eqnarray}
 {\cal{Z}}_{ijkh,nmvl}^{\rm{Ga}}(z,z',z'',z''')=
 {\cal{Z}}_{ml}^{h}(z'''){\cal{Z}}_{lv}^{k}(z'')
 {\cal{Z}}_{v{}n}^{j}(z'){\cal{Z}}_{nm}^{i}(z),
\label{eq:Z-ijkh,nmvl-Ga}\\
 {\cal{Z}}_{ijkh,nmvl}^{\rm{Gb}}(z,z',z'',z''')=
 {\cal{Z}}_{ml}^{h}(z'''){\cal{Z}}_{v{}n}^{k}(z'')
 {\cal{Z}}_{lv}^{j}(z'){\cal{Z}}_{nm}^{i}(z),
\label{eq:Z-ijkh,nmvl-Gb}\\
 {\cal{Z}}_{ijkh,nmvl}^{\rm{Gc}}(z,z',z'',z''')=
 {\cal{Z}}_{lv}^{h}(z'''){\cal{Z}}_{v{}n}^{k}(z'')
 {\cal{Z}}_{ml}^{j}(z'){\cal{Z}}_{nm}^{i}(z),
\label{eq:Z-ijkh,nmvl-Gc}
\label{eq:Z-ijkh,nmvl-Gf}
\end{eqnarray}
in terms of the quantities defined in
Eqs.~(\ref{eq:Z-nm-xy})--(\ref{eq:Z-nm-z}), and as in the previous
cases we may observe that any element with a Cartesian index $x$ is
equal to the element with the Cartesian index $y$ on the same place,
the other Cartesian indices unchanged. The $z$-independent terms we
write using the $a$'s and $b$'s and the functions solved in
Appendix~\ref{ch:Solve-Q}, as before. They finally become
\begin{eqnarray}
\lefteqn{
 \xi_{ijkh,nmvl}^{\rm{Ga}}(\vec{q}_{\|},\vec{k}_{\|})=
 -{1\over8\hbar^3}{1\over(2\pi)^2}
 \left\{
 {1\over(\varepsilon_{v}-\varepsilon_m)/\hbar-2\omega-{\rm{i}}\tau_{v{}m}^{-1}}
\right.}\nonumber\\ &\quad&\times
 \Bigl[
 {\cal{F}}_{ijkh}^{\rm{Ga}1}
  \left(l,\{a_2,-a_1\},\{b_{nm}^{4},b_{{l}m}^{1}\},-k_x\right)
 -{\cal{F}}_{ijkh}^{\rm{Ga}1}
 \left(m,\{a_2,-a_1\},\{b_{nm}^{4},b_{{l}m}^{1}\},0\right)
\nonumber\\ &&
 +{\cal{F}}_{ijkh}^{\rm{Ga}1}
  \left(l,\{a_2,a_1\},\{b_{nm}^{4},b_{vl}^{2}\},-k_x\right)
 -{\cal{F}}_{ijkh}^{\rm{Ga}1}
  \left(v,\{a_2,a_1\},\{b_{nm}^{4},b_{vl}^{2}\},0\right)
 \Bigr]
\nonumber\\ &&
 +{\cal{F}}_{ijkh}^{\rm{Ga}2}
  \left(l,\{a_2,a_1,a_3\},\{b_{nm}^{4},b_{vl}^{2},b_{nl}^{8}\},-k_x
  \right)
\nonumber\\ &&
 -{\cal{F}}_{ijkh}^{\rm{Ga}2}
  \left(v,\{a_2,a_1,a_3\},\{b_{nm}^{4},b_{vl}^{2},b_{nl}^{8}\},0
  \right)
\nonumber\\ &&
 +{\cal{F}}_{ijkh}^{\rm{Ga}2}
  \left(n,\{a_2,a_2,a_3\},\{b_{nm}^{4},b_{nv}^{3},b_{nl}^{8}\},q_x\right)
\nonumber\\ &&
 -{\cal{F}}_{ijkh}^{\rm{Ga}2}
  \left(v,\{a_2,a_2,a_3\},\{b_{nm}^{4},b_{nv}^{3},b_{nl}^{8}\},0\right)
\Bigr\}
\\
\lefteqn{
 \xi_{ijkh,nmvl}^{\rm{Gb}}(\vec{q}_{\|},\vec{k}_{\|})=
 -{1\over8\hbar^3}{1\over(2\pi)^2}
 \Bigl\{
 {\cal{F}}_{ijkh}^{\rm{Gb}}
  \left(l,\{a_2,-a_1,a_4\},\{b_{nm}^{4},b_{{l}m}^{1},b_{{v}m}^{5}\},
  -k_x\right)
}\nonumber\\ &&
 -{\cal{F}}_{ijkh}^{\rm{Gb}}
  \left(m,\{a_2,-a_1,a_4\},\{b_{nm}^{4},b_{{l}m}^{1},b_{{v}m}^{5}\},0
  \right)
\nonumber\\ &&
 +{\cal{F}}_{ijkh}^{\rm{Gb}}
  \left(l,\{a_2,a_2,a_4\},\{b_{nm}^{4},b_{vl}^{11},b_{{v}m}^{5}\},
  -k_x\right)
\nonumber\\ &&
 -{\cal{F}}_{ijkh}^{\rm{Gb}}
  \left(v,\{a_2,a_2,a_4\},\{b_{nm}^{4},b_{vl}^{11},b_{{v}m}^{5}\},
  q_x-k_x\right)
\nonumber\\ &&
 +{\cal{F}}_{ijkh}^{\rm{Gb}}
  \left(l,\{a_2,a_2,a_3\},\{b_{nm}^{4},b_{vl}^{11},b_{nl}^{8}\},
  -k_x\right)
\nonumber\\ &&
 -{\cal{F}}_{ijkh}^{\rm{Gb}}
  \left(v,\{a_2,a_2,a_3\},\{b_{nm}^{4},b_{vl}^{11},b_{nl}^{8}\},
  q_x-k_x\right)
\nonumber\\ &&
 +{\cal{F}}_{ijkh}^{\rm{Gb}}
  \left(n,\{a_2,a_1,a_3\},\{b_{nm}^{4},b_{nv}^{12},b_{nl}^{8}\},q_x\right)
\nonumber\\ &&
 -{\cal{F}}_{ijkh}^{\rm{Gb}}
  \left(v,\{a_2,a_1,a_3\},\{b_{nm}^{4},b_{nv}^{12},b_{nl}^{8}\},
  q_x-k_x\right)
\Bigr\}
\\
\lefteqn{
 \xi_{ijkh,nmvl}^{\rm{Gc}}(\vec{q}_{\|},\vec{k}_{\|})=
 -{1\over8\hbar^3}{1\over(2\pi)^2}
 \Bigl\{
 {\cal{F}}_{ijkh}^{\rm{Gc}1}
  \left(l,\{a_2,a_2,a_4\},\{b_{nm}^{4},b_{{l}m}^{3},b_{{v}m}^{5}\},q_x
  \right)
}\nonumber\\ &&
 -{\cal{F}}_{ijkh}^{\rm{Gc}1}
  \left(m,\{a_2,a_2,a_4\},\{b_{nm}^{4},b_{{l}m}^{3},b_{{v}m}^{5}\},0
  \right)
\nonumber\\ &&
 +{\cal{F}}_{ijkh}^{\rm{Gc}1}
  \left(l,\{a_2,-a_1,a_4\},\{b_{nm}^{4},b_{vl}^{13},b_{{v}m}^{5}\},
  q_x\right)
\nonumber\\ &&
 -{\cal{F}}_{ijkh}^{\rm{Gc}1}
  \left(v,\{a_2,-a_1,a_4\},\{b_{nm}^{4},b_{vl}^{13},b_{{v}m}^{5}\},
  q_x-k_x\right)
 +{1\over(\varepsilon_n-\varepsilon_{l})/\hbar-2\omega-{\rm{i}}\tau_{nl}^{-1}}
\nonumber\\ &&\times
\Bigl[
  {\cal{F}}_{ijkh}^{\rm{Gc}2}
  \left(l,\{a_2,-a_1\},\{b_{nm}^{4},b_{vl}^{13}\},q_x\right)
 -{\cal{F}}_{ijkh}^{\rm{Gc}2}
  \left(v,\{a_2,-a_1\},\{b_{nm}^{4},b_{vl}^{13}\},q_x-k_x\right)
\nonumber\\ &&
 +{\cal{F}}_{ijkh}^{\rm{Gc}2}
  \left(n,\{a_2,a_1\},\{b_{nm}^{4},b_{nv}^{12}\},q_x\right)
 -{\cal{F}}_{ijkh}^{\rm{Gc}2}
  \left(v,\{a_2,a_1\},\{b_{nm}^{4},b_{nv}^{12}\},q_x-k_x\right)
\Bigr]\Bigr\},
\nonumber\\
\label{eq:GenericXi}
\end{eqnarray}
and again they are written in terms of a set of functions ${\cal{F}}$
that vary from element to element. As was the case in the previous
sections, these functions are determined from the $z$-independent
parts of the transition current densities appearing in $\Xi$.

In passing we should notice that parts $(Ga1)$ and $(Gc2)$ has
$\beta=2$ because of the pure interband transition appearing in one of
their denominators, while parts $(Ga2)$, $(Gb)$, and $(Gc1)$ has
$\beta=3$ since all their transitions are mixed interband/intraband
transitions, we observe that a lot of ${\cal{F}}$ functions are equal.
In the simplest case, we observe
\begin{eqnarray}
 {\cal{F}}_{zzzz}^{\rm{Ga}1}&=&{\cal{F}}_{zzzz}^{\rm{Gc}2}
 ={\cal{F}}_{00}^{2}
\\
 {\cal{F}}_{zzzz}^{\rm{Gb}}&=&{\cal{F}}_{zzzz}^{\rm{Gc}1}=
 {\cal{F}}_{zzzz}^{\rm{Ga}2}
 ={\cal{F}}_{00}^{3}.
\end{eqnarray}
At the second level of complexity we find
\begin{eqnarray}
\lefteqn{
 {\cal{F}}_{yyzz}^{\rm{Ga}1}={\cal{F}}_{yyzz}^{\rm{Gc}2}=
 {\cal{F}}_{yzyz}^{\rm{Ga}1}={\cal{F}}_{yzyz}^{\rm{Gc}2}=
 {\cal{F}}_{yzzy}^{\rm{Ga}1}={\cal{F}}_{yzzy}^{\rm{Gc}2}=
 {\cal{F}}_{zyyz}^{\rm{Ga}1}={\cal{F}}_{zyyz}^{\rm{Gc}2}=
 {\cal{F}}_{zyzy}^{\rm{Ga}1}={\cal{F}}_{zyzy}^{\rm{Gc}2}=
}\nonumber\\ &\quad&
 {\cal{F}}_{zzyy}^{\rm{Ga}1}={\cal{F}}_{zzyy}^{\rm{Gc}2}
 =4{\cal{F}}_{02}^{2},
\\
\lefteqn{
 {\cal{F}}_{yyzz}^{\rm{Gb}}={\cal{F}}_{yyzz}^{\rm{Gc}1}=
 {\cal{F}}_{yyzz}^{\rm{Ga}2}={\cal{F}}_{yzyz}^{\rm{Gb}}=
 {\cal{F}}_{yzyz}^{\rm{Gc}1}={\cal{F}}_{yzyz}^{\rm{Ga}2}=
 {\cal{F}}_{yzzy}^{\rm{Gb}}={\cal{F}}_{yzzy}^{\rm{Gc}1}=
 {\cal{F}}_{yzzy}^{\rm{Ga}2}={\cal{F}}_{zyyz}^{\rm{Gb}}=
}\nonumber\\ &&
 {\cal{F}}_{zyyz}^{\rm{Gc}1}={\cal{F}}_{zyyz}^{\rm{Ga}2}=
 {\cal{F}}_{zyzy}^{\rm{Gb}}={\cal{F}}_{zyzy}^{\rm{Gc}1}=
 {\cal{F}}_{zyzy}^{\rm{Ga}2}={\cal{F}}_{zzyy}^{\rm{Gb}}=
 {\cal{F}}_{zzyy}^{\rm{Gc}1}={\cal{F}}_{zzyy}^{\rm{Ga}2}
 =4{\cal{F}}_{02}^{3}.
\end{eqnarray}
The third level of complexity gives
\begin{eqnarray}
 {\cal{F}}_{yyyy}^{\rm{Ga}1}&=&{\cal{F}}_{yyyy}^{\rm{Gc}2}
 =16{\cal{F}}_{04}^{2},
\\
 {\cal{F}}_{yyyy}^{\rm{Gb}}&=&{\cal{F}}_{yyyy}^{\rm{Gc}1}=
 {\cal{F}}_{yyyy}^{\rm{Ga}2}
 =16{\cal{F}}_{04}^{3}.
\end{eqnarray}
At the fourth level of complexity we observe
\begin{eqnarray}
 {\cal{F}}_{xzzz}^{\rm{Ga}1}&=&{\cal{F}}_{xzzz}^{\rm{Gc}2}=
 {\cal{F}}_{zxzz}^{\rm{Ga}1}={\cal{F}}_{zxzz}^{\rm{Gc}2}
 =2{\rm{i}}{\cal{F}}_{10}^{2}+2{\rm{i}}q_x{\cal{F}}_{00}^{2},
\\
 {\cal{F}}_{xzzz}^{\rm{Gb}}&=&{\cal{F}}_{xzzz}^{\rm{Gc}1}=
 {\cal{F}}_{xzzz}^{\rm{Ga}2}={\cal{F}}_{zxzz}^{\rm{Gc}1}=
 {\cal{F}}_{zxzz}^{\rm{Ga}2}
 =2{\rm{i}}{\cal{F}}_{10}^{3}+2{\rm{i}}q_x{\cal{F}}_{00}^{3},
\\
 {\cal{F}}_{zzxz}^{\rm{Ga}1}&=&{\cal{F}}_{zzzx}^{\rm{Ga}1}
 =2{\rm{i}}{\cal{F}}_{10}^{2}-2{\rm{i}}k_x{\cal{F}}_{00}^{2},
\\
 {\cal{F}}_{zzxz}^{\rm{Ga}2}&=&{\cal{F}}_{zzzx}^{\rm{Gb}}=
 {\cal{F}}_{zzzx}^{\rm{Ga}2}
 =2{\rm{i}}{\cal{F}}_{10}^{3}-{\rm{i}}k_x{\cal{F}}_{00}^{3},
\\
 {\cal{F}}_{zzxz}^{\rm{Gc}2}&=&{\cal{F}}_{zzzx}^{\rm{Gc}2}
 =2{\rm{i}}{\cal{F}}_{10}^{2}+{\rm{i}}(2q_x-k_x){\cal{F}}_{00}^{2},
\\
 {\cal{F}}_{zzxz}^{\rm{Gb}}&=&{\cal{F}}_{zzxz}^{\rm{Gc}1}=
 {\cal{F}}_{zzzx}^{\rm{Gc}1}
 =2{\rm{i}}{\cal{F}}_{10}^{3}+{\rm{i}}(2q_x-k_x){\cal{F}}_{00}^{3},
\end{eqnarray}
and the independent element
\begin{equation}
 {\cal{F}}_{zxzz}^{\rm{Gb}}
 =2{\rm{i}}{\cal{F}}_{10}^{3}+2{\rm{i}}(q_x-2k_x){\cal{F}}_{00}^{3}.
\end{equation}
In the fifth case we find
\begin{eqnarray}
\lefteqn{
 {\cal{F}}_{xyyz}^{\rm{Ga}1}={\cal{F}}_{xyyz}^{\rm{Gc}2}=
 {\cal{F}}_{xyzy}^{\rm{Ga}1}={\cal{F}}_{xyzy}^{\rm{Gc}2}=
 {\cal{F}}_{xzyy}^{\rm{Ga}1}={\cal{F}}_{xzyy}^{\rm{Gc}2}=
 {\cal{F}}_{yxyz}^{\rm{Ga}1}={\cal{F}}_{yxyz}^{\rm{Gc}2}=
 {\cal{F}}_{yxzy}^{\rm{Ga}1}={\cal{F}}_{yxzy}^{\rm{Gc}2}=
}\nonumber\\ &\quad&
 {\cal{F}}_{zxyy}^{\rm{Ga}1}={\cal{F}}_{zxyy}^{\rm{Gc}2}
 =-8{\rm{i}}{\cal{F}}_{12}^{2}-4{\rm{i}}q_x{\cal{F}}_{02}^{2},
\\
\lefteqn{
 {\cal{F}}_{xyyz}^{\rm{Gb}}={\cal{F}}_{xyyz}^{\rm{Gc}1}=
 {\cal{F}}_{xyyz}^{\rm{Ga}2}={\cal{F}}_{xyzy}^{\rm{Gb}}=
 {\cal{F}}_{xyzy}^{\rm{Gc}1}={\cal{F}}_{xyzy}^{\rm{Ga}2}=
 {\cal{F}}_{xzyy}^{\rm{Gb}}={\cal{F}}_{xzyy}^{\rm{Gc}1}=
 {\cal{F}}_{xzyy}^{\rm{Ga}2}={\cal{F}}_{yxyz}^{\rm{Gc}1}=
}\nonumber\\ &&
 {\cal{F}}_{yxyz}^{\rm{Ga}2}={\cal{F}}_{yxzy}^{\rm{Gc}1}=
 {\cal{F}}_{yxzy}^{\rm{Ga}2}={\cal{F}}_{zxyy}^{\rm{Gc}1}=
 {\cal{F}}_{zxyy}^{\rm{Ga}2}
 =-8{\rm{i}}{\cal{F}}_{12}^{3}-4{\rm{i}}q_x{\cal{F}}_{02}^{3},
\\
\lefteqn{
 {\cal{F}}_{yxyz}^{\rm{Gb}}={\cal{F}}_{yxzy}^{\rm{Gb}}=
 {\cal{F}}_{zxyy}^{\rm{Gb}}
 =-8{\rm{i}}{\cal{F}}_{12}^{3}-4{\rm{i}}(q_x-2k_x){\cal{F}}_{02}^{3},
}\\
\lefteqn{
 {\cal{F}}_{yyxz}^{\rm{Ga}1}={\cal{F}}_{yzxy}^{\rm{Ga}1}=
 {\cal{F}}_{zyxy}^{\rm{Ga}1}={\cal{F}}_{yyzx}^{\rm{Ga}1}=
 {\cal{F}}_{yzyx}^{\rm{Ga}1}={\cal{F}}_{zyyx}^{\rm{Ga}1}
 =-8{\rm{i}}{\cal{F}}_{12}^{2}+4{\rm{i}}k_x{\cal{F}}_{02}^{2},
}\\
\lefteqn{
 {\cal{F}}_{yyxz}^{\rm{Ga}2}={\cal{F}}_{yzxy}^{\rm{Ga}2}=
 {\cal{F}}_{zyxy}^{\rm{Ga}2}={\cal{F}}_{yyzx}^{\rm{Gb}}=
 {\cal{F}}_{yyzx}^{\rm{Ga}2}={\cal{F}}_{yzyx}^{\rm{Gb}}=
 {\cal{F}}_{yzyx}^{\rm{Ga}2}={\cal{F}}_{zyyx}^{\rm{Gb}}=
 {\cal{F}}_{zyyx}^{\rm{Ga}2}=
}\nonumber\\ &&
 -8{\rm{i}}{\cal{F}}_{12}^{3}+4{\rm{i}}k_x{\cal{F}}_{02}^{3},
\\
\lefteqn{
 {\cal{F}}_{yyxz}^{\rm{Gc}2}={\cal{F}}_{yzxy}^{\rm{Gc}2}=
 {\cal{F}}_{zyxy}^{\rm{Gc}2}={\cal{F}}_{yyzx}^{\rm{Gc}2}=
 {\cal{F}}_{yzyx}^{\rm{Gc}2}={\cal{F}}_{zyyx}^{\rm{Gc}2}
 =-8{\rm{i}}{\cal{F}}_{12}^{2}+4{\rm{i}}(k_x-2q_x){\cal{F}}_{02}^{2},
}\\
\lefteqn{
 {\cal{F}}_{yyxz}^{\rm{Gb}}={\cal{F}}_{yyxz}^{\rm{Gc}1}=
 {\cal{F}}_{yzxy}^{\rm{Gb}}={\cal{F}}_{yzxy}^{\rm{Gc}1}=
 {\cal{F}}_{zyxy}^{\rm{Gb}}={\cal{F}}_{zyxy}^{\rm{Gc}1}=
 {\cal{F}}_{yyzx}^{\rm{Gc}1}={\cal{F}}_{yzyx}^{\rm{Gc}1}=
 {\cal{F}}_{zyyx}^{\rm{Gc}1}=
}\nonumber\\ &&
 -8{\rm{i}}{\cal{F}}_{12}^{3}+4{\rm{i}}(k_x-2q_x){\cal{F}}_{02}^{3},
\end{eqnarray}
In the sixth case we observe the related functions
\begin{eqnarray}
\lefteqn{
 {\cal{F}}_{xxzz}^{\rm{Ga}1}={\cal{F}}_{xxzz}^{\rm{Gc}2}
 =-4{\cal{F}}_{20}^{2}-4q_x{\cal{F}}_{10}^{2}-q_x^2{\cal{F}}_{00}^{2},
}\\
\lefteqn{
 {\cal{F}}_{xxzz}^{\rm{Gc}1}={\cal{F}}_{xxzz}^{\rm{Ga}2}
 =-4{\cal{F}}_{20}^{3}-4q_x{\cal{F}}_{10}^{3}-q_x^2{\cal{F}}_{00}^{3},
}\\
\lefteqn{
 {\cal{F}}_{xzxz}^{\rm{Ga}1}={\cal{F}}_{xzzx}^{\rm{Ga}1}=
 {\cal{F}}_{zxxz}^{\rm{Ga}1}={\cal{F}}_{zxzx}^{\rm{Ga}1}
 =-4{\cal{F}}_{20}^{2}-2(q_x-k_x){\cal{F}}_{10}^{2}
 +q_xk_x{\cal{F}}_{00}^{2},
}\\
\lefteqn{
 {\cal{F}}_{xzxz}^{\rm{Ga}2}={\cal{F}}_{xzzx}^{\rm{Gb}}=
 {\cal{F}}_{xzzx}^{\rm{Ga}2}={\cal{F}}_{zxxz}^{\rm{Ga}2}=
 {\cal{F}}_{zxzx}^{\rm{Ga}2}
 =-4{\cal{F}}_{20}^{3}-4(q_x-k_x){\cal{F}}_{10}^{3}
 +q_xk_x{\cal{F}}_{00}^{3},
}\\
\lefteqn{
 {\cal{F}}_{xzxz}^{\rm{Gb}}={\cal{F}}_{xzxz}^{\rm{Gc}1}=
 {\cal{F}}_{xzzx}^{\rm{Gc}1}={\cal{F}}_{zxxz}^{\rm{Gc}1}=
 {\cal{F}}_{zxzx}^{\rm{Gc}1}=
}\nonumber\\ &\quad&
 -4{\cal{F}}_{20}^{3}-2(3q_x-k_x){\cal{F}}_{10}^{3}
 -q_x(2q_x-k_x){\cal{F}}_{00}^{3},
\\
\lefteqn{
 {\cal{F}}_{xzxz}^{\rm{Gc}2}={\cal{F}}_{xzzx}^{\rm{Gc}2}=
 {\cal{F}}_{zxxz}^{\rm{Gc}2}={\cal{F}}_{zxzx}^{\rm{Gc}2}
 =-4{\cal{F}}_{20}^{2}-2(3q_x-k_x){\cal{F}}_{10}^{2}
 -q_x(2q_x-k_x){\cal{F}}_{00}^{2},
}
\end{eqnarray}
and the eight independent functions
\begin{eqnarray}
 {\cal{F}}_{xxzz}^{\rm{Gb}}&=&
 -4{\cal{F}}_{20}^{3}-4(q_x-k_x){\cal{F}}_{10}^{3}
 -q_x(q_x-2k_x){\cal{F}}_{00}^{3},
\\ 
 {\cal{F}}_{zxxz}^{\rm{Gb}}&=&
 -4{\cal{F}}_{20}^{3}-6(q_x-k_x){\cal{F}}_{10}^{3}
 -(2q_x^2+2k_x^2-5q_xk_x){\cal{F}}_{00}^{3},
\\
 {\cal{F}}_{zxzx}^{\rm{Gb}}&=&
 -4{\cal{F}}_{20}^{3}-2(q_x-3k_x){\cal{F}}_{10}^{3}
 -k_x(2k_x-q_x){\cal{F}}_{00}^{3},
\\
 {\cal{F}}_{zzxx}^{\rm{Ga}1}&=&
 -4{\cal{F}}_{20}^{2}+4k_x{\cal{F}}_{10}^{2}-k_x^2{\cal{F}}_{00}^{2},
\\
 {\cal{F}}_{zzxx}^{\rm{Ga}2}&=&
 -4{\cal{F}}_{20}^{3}+4k_x{\cal{F}}_{10}^{3}-k_x^2{\cal{F}}_{00}^{3},
\\
 {\cal{F}}_{zzxx}^{\rm{Gb}}&=&
 -4{\cal{F}}_{20}^{3}-4(q_x-k_x){\cal{F}}_{10}^{3}
 -k_x(k_x-2q_x){\cal{F}}_{00}^{3},
\\
 {\cal{F}}_{zzxx}^{\rm{Gc}1}&=&
 -4{\cal{F}}_{20}^{3}-2(2q_x-k_x){\cal{F}}_{10}^{3}
 -(2q_x-k_x)^2{\cal{F}}_{00}^{3},
\\
 {\cal{F}}_{zzxx}^{\rm{Gc}2}&=&
 -4{\cal{F}}_{20}^{2}-2(2q_x-k_x){\cal{F}}_{10}^{2}
 -(2q_x-k_x)^2{\cal{F}}_{00}^{2}.
\end{eqnarray}
The seventh case has the following related functions
\begin{eqnarray}
\lefteqn{
 {\cal{F}}_{xxyy}^{\rm{Ga}1}={\cal{F}}_{xxyy}^{\rm{Gc}2}
 =16{\cal{F}}_{22}^{2}+16q_x{\cal{F}}_{12}^{2}+4q_x^2{\cal{F}}_{02}^{2},
}\\
\lefteqn{
 {\cal{F}}_{xxyy}^{\rm{Gc}1}={\cal{F}}_{xxyy}^{\rm{Ga}2}
 =16{\cal{F}}_{22}^{3}+16q_x{\cal{F}}_{12}^{3}+4q_x^2{\cal{F}}_{02}^{3},
}\\
\lefteqn{
 {\cal{F}}_{xyxy}^{\rm{Ga}1}={\cal{F}}_{xyyx}^{\rm{Ga}1}=
 {\cal{F}}_{yxxy}^{\rm{Ga}1}={\cal{F}}_{yxyx}^{\rm{Ga}1}
 =16{\cal{F}}_{22}^{2}+8(q_x-k_x){\cal{F}}_{12}^{2}-4q_xk_x{\cal{F}}_{02}^{2},
}\\
\lefteqn{
 {\cal{F}}_{xyxy}^{\rm{Ga}2}={\cal{F}}_{xyyx}^{\rm{Gb}}=
 {\cal{F}}_{xyyx}^{\rm{Ga}2}={\cal{F}}_{yxxy}^{\rm{Ga}2}=
 {\cal{F}}_{yxyx}^{\rm{Ga}2}
 =16{\cal{F}}_{22}^{3}+8(q_x-k_x){\cal{F}}_{12}^{3}-4q_xk_x{\cal{F}}_{02}^{3},
}\\
\lefteqn{
 {\cal{F}}_{xyxy}^{\rm{Gb}}={\cal{F}}_{xyxy}^{\rm{Gc}1}=
 {\cal{F}}_{xyyx}^{\rm{Gc}1}={\cal{F}}_{yxxy}^{\rm{Gc}1}=
 {\cal{F}}_{yxyx}^{\rm{Gc}1}=
}\nonumber\\ &\quad&
 16{\cal{F}}_{22}^{3}+8(3q_x-k_x){\cal{F}}_{12}^{3}
 +4q_x(2q_x-k_x){\cal{F}}_{02}^{3},
\\
\lefteqn{
 {\cal{F}}_{xyxy}^{\rm{Gc}2}={\cal{F}}_{xyyx}^{\rm{Gc}2}=
 {\cal{F}}_{yxxy}^{\rm{Gc}2}={\cal{F}}_{yxyx}^{\rm{Gc}2}
 =16{\cal{F}}_{22}^{2}+8(3q_x-k_x){\cal{F}}_{12}^{2}
 +4q_x(2q_x-k_x){\cal{F}}_{02}^{2},
}
\end{eqnarray}
and the eight independent functions
\begin{eqnarray}
 {\cal{F}}_{xxyy}^{\rm{Gb}}&=&
 16{\cal{F}}_{22}^{3}+16(q_x-k_x){\cal{F}}_{12}^{3}
 +4q_x(q_x-2k_x){\cal{F}}_{02}^{3},
\\
 {\cal{F}}_{yxxy}^{\rm{Gb}}&=&
 16{\cal{F}}_{22}^{3}+24(q_x-k_x){\cal{F}}_{12}^{3}
 +4(2q_x^2+2k_x^2-5q_xk_x){\cal{F}}_{02}^{3},
\\
 {\cal{F}}_{yxyx}^{\rm{Gb}}&=&
 16{\cal{F}}_{22}^{3}+8(q_x-3k_x){\cal{F}}_{12}^{3}
 +4k_x(2k_x-q_x){\cal{F}}_{02}^{3},
\\
 {\cal{F}}_{yyxx}^{\rm{Ga}1}&=&
 16{\cal{F}}_{22}^{2}-16k_x{\cal{F}}_{12}^{2}+4k_x^2{\cal{F}}_{02}^{2},
\\
 {\cal{F}}_{yyxx}^{\rm{Ga}2}&=&
 16{\cal{F}}_{22}^{3}-16k_x{\cal{F}}_{12}^{3}+4k_x^2{\cal{F}}_{02}^{3},
\\
 {\cal{F}}_{yyxx}^{\rm{Gb}}&=&
 16{\cal{F}}_{22}^{3}+16(q_x-k_x){\cal{F}}_{12}^{3}
 +4k_x(k_x-2q_x){\cal{F}}_{02}^{3},
\\
 {\cal{F}}_{yyxx}^{\rm{Gc}1}&=&
 16{\cal{F}}_{22}^{3}+16(2q_x-k_x){\cal{F}}_{12}^{3}
 +4(2q_x-k_x)^2{\cal{F}}_{02}^{3},
\\
 {\cal{F}}_{yyxx}^{\rm{Gc}2}&=&
 16{\cal{F}}_{22}^{2}+16(2q_x-k_x){\cal{F}}_{12}^{2}
 +4(2q_x-k_x)^2{\cal{F}}_{02}^{2}.
\end{eqnarray}
The eighth case gives
\begin{eqnarray}
\lefteqn{
 {\cal{F}}_{xxxz}^{\rm{Ga}1}={\cal{F}}_{xxzx}^{\rm{Ga}1}
 =-8{\rm{i}}{\cal{F}}_{30}^{2}-4{\rm{i}}(2q_x-k_x){\cal{F}}_{20}^{2}
 -2{\rm{i}}q_x(q_x-2k_x){\cal{F}}_{10}^{2}+{\rm{i}}q_x^2k_x{\cal{F}}_{00}^{2},
}\\
\lefteqn{
 {\cal{F}}_{xxxz}^{\rm{Ga}2}={\cal{F}}_{xxzx}^{\rm{Ga}2}
 =-8{\rm{i}}{\cal{F}}_{30}^{3}-4{\rm{i}}(2q_x-k_x){\cal{F}}_{20}^{3}
 -2{\rm{i}}q_x(q_x-2k_x){\cal{F}}_{10}^{3}+{\rm{i}}q_x^2k_x{\cal{F}}_{00}^{3},
}\\
\lefteqn{
 {\cal{F}}_{xxxz}^{\rm{Gc}1}={\cal{F}}_{xxzx}^{\rm{Gc}1}
 =-8{\rm{i}}{\cal{F}}_{30}^{3}-4{\rm{i}}(4q_x-k_x){\cal{F}}_{20}^{3}
 -2{\rm{i}}q_x(5q_x-2k_x){\cal{F}}_{10}^{3}
}\nonumber\\ &\quad&
 -{\rm{i}}q_x^2(2q_x-k_x){\cal{F}}_{00}^{3},
\\
\lefteqn{
 {\cal{F}}_{xxxz}^{\rm{Gc}2}={\cal{F}}_{xxzx}^{\rm{Gc}2}
 =-8{\rm{i}}{\cal{F}}_{30}^{2}-4{\rm{i}}(4q_x-k_x){\cal{F}}_{20}^{2}
 -2{\rm{i}}q_x(5q_x-2k_x){\cal{F}}_{10}^{2}
}\nonumber\\ &&
 -{\rm{i}}q_x^2(2q_x-k_x){\cal{F}}_{00}^{2},
\\
\lefteqn{
 {\cal{F}}_{xzxx}^{\rm{Ga}1}={\cal{F}}_{zxxx}^{\rm{Ga}1}
 =-8{\rm{i}}{\cal{F}}_{30}^{2}-4{\rm{i}}(q_x-2k_x){\cal{F}}_{20}^{2}
 -2{\rm{i}}k_x(k_x-2q_x){\cal{F}}_{10}^{2}-{\rm{i}}q_xk_x^2{\cal{F}}_{00}^{2},
}\\
\lefteqn{
 {\cal{F}}_{xzxx}^{\rm{Ga}2}={\cal{F}}_{zxxx}^{\rm{Ga}2}
 =-8{\rm{i}}{\cal{F}}_{30}^{3}-4{\rm{i}}(q_x-2k_x){\cal{F}}_{20}^{3}
 -2{\rm{i}}k_x(k_x-2q_x){\cal{F}}_{10}^{3}-{\rm{i}}q_xk_x^2{\cal{F}}_{00}^{3},
}\\
\lefteqn{
 {\cal{F}}_{xzxx}^{\rm{Gc}1}={\cal{F}}_{zxxx}^{\rm{Gc}1}
 =-8{\rm{i}}{\cal{F}}_{30}^{3}-4{\rm{i}}(5q_x-2k_x){\cal{F}}_{20}^{3}
 -2{\rm{i}}(2q_x-k_x)(4q_x-k_x){\cal{F}}_{10}^{3}
}\nonumber\\ &&
 -{\rm{i}}q_x(2q_x-k_x)^2{\cal{F}}_{00}^{3},
\\
\lefteqn{
 {\cal{F}}_{xzxx}^{\rm{Gc}2}={\cal{F}}_{zxxx}^{\rm{Gc}2}
 =-8{\rm{i}}{\cal{F}}_{30}^{2}-4{\rm{i}}(5q_x-2k_x){\cal{F}}_{20}^{2}
 -2{\rm{i}}(2q_x-k_x)(4q_x-k_x){\cal{F}}_{10}^{2}
}\nonumber\\ &&
 -{\rm{i}}q_x(2q_x-k_x)^2{\cal{F}}_{00}^{2},
\end{eqnarray}
and the four independent elements
\begin{eqnarray}
\lefteqn{
 {\cal{F}}_{xxxz}^{\rm{Gb}}=
 -8{\rm{i}}{\cal{F}}_{30}^{3}-4{\rm{i}}(4q_x-3k_x){\cal{F}}_{20}^{3}
 -2{\rm{i}}(5q_x^2+2k_x^2-8q_xk_x){\cal{F}}_{10}^{3}
}\nonumber\\ &\quad&
 -{\rm{i}}q_x(2q_x^2+2k_x^2-5q_xk_x){\cal{F}}_{00}^{3},
\\
\lefteqn{
 {\cal{F}}_{xxzx}^{\rm{Gb}}=
 -8{\rm{i}}{\cal{F}}_{30}^{3}-4{\rm{i}}(2q_x-3k_x){\cal{F}}_{20}^{3}
 -2{\rm{i}}(q_x^2+2k_x^2-4q_xk_x){\cal{F}}_{10}^{3}
}\nonumber\\ &&
 +{\rm{i}}q_xk_x(q_x-2k_x){\cal{F}}_{00}^{3},
\\
\lefteqn{
 {\cal{F}}_{xzxx}^{\rm{Gb}}=
 -8{\rm{i}}{\cal{F}}_{30}^{3}-4{\rm{i}}(3q_x-2k_x){\cal{F}}_{20}^{3}
 -2{\rm{i}}(2q_x^2+k_x^2-4q_xk_x){\cal{F}}_{10}^{3}
}\nonumber\\ &&
 +{\rm{i}}q_xk_x(2q_x-k_x){\cal{F}}_{00}^{3},
\\
\lefteqn{
 {\cal{F}}_{zxxx}^{\rm{Gb}}=
 -8{\rm{i}}{\cal{F}}_{30}^{3}-4{\rm{i}}(3q_x-4k_x){\cal{F}}_{20}^{3}
 -2{\rm{i}}(2q_x^2+5k_x^2-8q_xk_x){\cal{F}}_{10}^{3}
}\nonumber\\ &&
 +{\rm{i}}k_x(2q_x^2+2k_x^2-5q_xk_x){\cal{F}}_{00}^{3}.
\end{eqnarray}
The most complex solution group of Cartesian indices gives the five
independent functions
\begin{eqnarray}
\lefteqn{
 {\cal{F}}_{xxxx}^{\rm{Ga}1}=
 16{\cal{F}}_{40}^{2}+16(q_x-k_x){\cal{F}}_{30}^{2}
 +4(q_x^2+k_x^2-4q_xk_x){\cal{F}}_{20}^{2}
 +4q_xk_x(k_x-q_x){\cal{F}}_{10}^{2}
}\nonumber\\ &\quad&
 +(q_xk_x)^2{\cal{F}}_{00}^{2},
\\
\lefteqn{
 {\cal{F}}_{xxxx}^{\rm{Ga}2}=
 16{\cal{F}}_{40}^{3}+16(q_x-k_x){\cal{F}}_{30}^{3}
 +4(q_x^2+k_x^2-4q_xk_x){\cal{F}}_{20}^{3}
 +4q_xk_x(k_x-q_x){\cal{F}}_{10}^{3}
}\nonumber\\ &&
 +(q_xk_x)^2{\cal{F}}_{00}^{3},
\\
\lefteqn{
 {\cal{F}}_{xxxx}^{\rm{Gb}}=
 16{\cal{F}}_{40}^{3}+32(q_x-k_x){\cal{F}}_{30}^{3}
 +4(5q_x^2+5k_x^2-12q_xk_x){\cal{F}}_{20}^{3}
}\nonumber\\ &&
 +2(q_x-k_x)(2q_x^2+2k_x^2-8q_xk_x){\cal{F}}_{10}^{3}
 -q_xk_x(2q_x^2+2k_x^2-5q_xk_x){\cal{F}}_{00}^{3},
\\
\lefteqn{
 {\cal{F}}_{xxxx}^{\rm{Gc}1}=
 16{\cal{F}}_{40}^{3}+16(3q_x-k_x){\cal{F}}_{30}^{3}
 +4((2q_x-k_x)^2+q_x(9q_x-4k_x)){\cal{F}}_{20}^{3}
}\nonumber\\ &&
 +4q_x(2q_x-k_x)(3q_x-k_x){\cal{F}}_{10}^{3}
 +q_x^2(2q_x-k_x)^2{\cal{F}}_{00}^{3},
\\
\lefteqn{
 {\cal{F}}_{xxxx}^{\rm{Gc}2}=
 16{\cal{F}}_{40}^{2}+16(3q_x-k_x){\cal{F}}_{30}^{2}
 +4((2q_x-k_x)^2+q_x(9q_x-4k_x)){\cal{F}}_{20}^{2}
}\nonumber\\ &&
 +4q_x(2q_x-k_x)(3q_x-k_x){\cal{F}}_{10}^{2}
 +q_x^2(2q_x-k_x)^2{\cal{F}}_{00}^{2}.
\end{eqnarray}
The immediate conclusion of these observations is that only $65$ of
the original $246$ possible functions are independent.

\section{The ${\cal{Z}}$ coefficients with uniform pump field
 amplitudes}\label{sec:Z}
If the pump fields have uniform amplitudes along the $z$-axis, we take
the local limit in the two coordinates $z'''$ and $z''$ in the
${\cal{Z}}$ coefficients, such that in general we may write,
\begin{equation}
 {\cal{Z}}(z,z')=\iint{\cal{Z}}(z,z',z'',z''')dz'''dz'',
\end{equation}
and thus
\begin{equation}
 \tensor{\Xi}(z,z';\vec{q}_{\|},\vec{k}_{\|})=
 \iint\tensor{\Xi}(z,z',z'',z''';\vec{q}_{\|},\vec{k}_{\|})dz'''dz''.
\end{equation}
Using an orthogonal set of wave functions, parity teaches that the
integrals over $x$- and $y$-components gives
\begin{equation}
 \int{\cal{Z}}_{nm}^{x}(z)dz=\int{\cal{Z}}_{nm}^{\,y}(z)dz=\delta_{nm},
\end{equation}
where $\delta_{nm}$ is the Kronecker delta. The only question left is
the $z$-components, which may be determined as soon as the wave
functions for the system is known. Then part $A$ of the nonlinear
conductivity tensor does not contribute to the phase conjugated
response because of the result of integration over $z''$. For the same
reason, part E vanish, since the pure interband terms vanish by
themselves, and the rest of part Ea becomes equal in magnitude to
the rest of part Eb, but with the opposite sign. All other terms
still contribute to the response.

\subsection{Infinite barrier quantum well}
If we choose a quantum well within the infinite barrier model with
boundaries at $0$ and $-d$ as the source, then we find
\begin{equation}
 {\cal{Z}}(z,z')=\int_{-d}^{0}\int_{-d}^{0}{\cal{Z}}(z,z',z'',z''')dz'''dz'',
\end{equation}
and thus
\begin{equation}
 \tensor{\Xi}(z,z';\vec{q}_{\|},\vec{k}_{\|})=
 \int_{-d}^{0}\int_{-d}^{0}
 \tensor{\Xi}(z,z',z'',z''';\vec{q}_{\|},\vec{k}_{\|})dz'''dz'',
\end{equation}
in general. Since the individual ${\cal{Z}}_{nm}^{x,y,z}(z)$ are
independent, the result is written as a product of these in the
coordinates $z$, $z'$, $z''$, and $z'''$. Then the integrals over
$z$-components gives
\begin{equation}
 \int_{-d}^{0}{\cal{Z}}_{nm}^{z}(z)dz=
 {4nm[1-(-1)^{n+m}]\over(n^2-m^2)d}.
\end{equation}

\subsection{Probe with a single wavevector}
If we take the probe field as
$\vec{E}(z';\vec{q}_{\|})=\vec{E}e^{{\rm{i}}q_{\perp}z'}$ and thus
$\vec{E}^{*}(z';\vec{q}_{\|})=\vec{E}^{*}e^{-{\rm{i}}q_{\perp}z'}$, then we
may further reduce the $z$-dependence, if we define
\begin{equation}
 {\mathfrak{Z}}(z)=\int_{-d}^{0}{\cal{Z}}(z,z')e^{-{\rm{i}}q_{\perp}z'}dz',
\end{equation}
where we have indicated the modification before integration of the
${\cal{Z}}(z,z')$ quantity by shifting the symbol from calligraphic
style to fraktur. Then, from the two integrals
\begin{eqnarray}
\lefteqn{
 \int_{-d}^{0}{\cal{Z}}_{nm}^{x}(z)e^{{\pm}{\rm{i}}q_{\perp}z}dz=
 \int_{-d}^{0}{\cal{Z}}_{nm}^{\,y}(z)e^{{\pm}{\rm{i}}q_{\perp}z}dz=
}\nonumber\\ &\quad&
 \mp{4\pi^2nm{\rm{i}}q_{\perp}d[e^{{\mp}{\rm{i}}q_{\perp}d}(-1)^{n+m}-1]
 \over[({\rm{i}}q_{\perp}d)^2+\pi^2(n+m)^2][({\rm{i}}q_{\perp}d)^2+\pi^2(n-m)^2]},
\\
\lefteqn{
 \int_{-d}^{0}{\cal{Z}}_{nm}^{z}(z)e^{{\pm}{\rm{i}}q_{\perp}z}dz=
 {4\pi^4(n^2-m^2)nm[e^{{\mp}{\rm{i}}q_{\perp}d}(-1)^{n+m}-1]
 \over{}d[({\rm{i}}q_{\perp}d)^2+\pi^2(n+m)^2][({\rm{i}}q_{\perp}d)^2+\pi^2(n-m)^2]},
}
\end{eqnarray}
we may deduce the reduced quantities.

\subsection{Combining ${\mathfrak{Z}}$ with $\Omega$}
If we combine the ${\mathfrak{Z}}$-quantities with the two different
exponentials appearing in the $\Omega$-quantities, we get new
quantities that are independent of $z$. Defining
\begin{equation}
 {\mathfrak{Z}}^{\pm}=\int_{-d}^{0}{\mathfrak{Z}}(z)e^{\pm{}{\rm{i}}q_{\perp}z}dz,
\end{equation}
we may now deduce the $z$-independent quantities that will finally
appear in the expressions for the $\Omega$-quantities. Even though we
have modified the ${\mathfrak{Z}}$ coefficients again before
integration, we keep the symbol, since the main purpose of selecting
another symbol is to indicate that it is modified from the original
${\cal{Z}}$ quantities.

\section{The nonlinear conductivity tensor in the $\Omega$ coefficients}\label{sec:Ps}
If we concatenate the definitions made in this compliment until now,
we may define the new quantity to be considered in the $\Omega$
coefficients [Eqs.~(\ref{eq:14.44})--(\ref{eq:14.46})]
\begin{equation}
 {\tensor{{\mathfrak{X}}}}^{\pm}(\vec{q}_{\|},\vec{k}_{\|})=
 \int_{-d}^{0}\int_{-d}^{0}\int_{-d}^{0}\int_{-d}^{0}
 {\tensor{\Xi}}(z,z',z'',z''';\vec{q}_{\|},\vec{k}_{\|})
 e^{-{\rm{i}}q_{\perp}z'}e^{\pm{}{\rm{i}}q_{\perp}z}
 dz'''dz''dz'dz.
\end{equation}
As we did with the ${\cal{Z}}$ coefficients, the modification before
integration have led us to define the new quantity in fraktur rather
as $\Xi$.

Then the nonlinear conductivity tensors combined with the
$z$-dependent parts of the probe and the $\Omega$ coefficient can be
written
\begin{eqnarray}
\lefteqn{
 {{\mathfrak{X}}}_{xxxx}^{\rm{A}\pm}(\vec{q}_{\|},\vec{k}_{\|})=
 {2\over({\rm{i}}\omega)^3}\sum_{nm}{\mathfrak{Z}}_{nm}^{\rm{A}\pm}\xi_{nm}^{\rm{A}},
}\\
\lefteqn{
 {{\mathfrak{X}}}_{xxkh}^{\rm{B}\pm}(\vec{q}_{\|},\vec{k}_{\|})=
 {2\over({\rm{i}}\omega)^3}\left({e\hbar\over2m_e}\right)^2\sum_{nmv}
 {\mathfrak{Z}}_{kh,nmv}^{\rm{B}\pm}\xi_{kh,nmv}^{\rm{B}}(\vec{q}_{\|},\vec{k}_{\|}),
}\\
\lefteqn{
 {{\mathfrak{X}}}_{xxxx}^{\rm{C}\pm}(\vec{q}_{\|},\vec{k}_{\|})=
 {2\over({\rm{i}}\omega)^3}
 \sum_{nm}{\mathfrak{Z}}_{nm}^{\rm{C}\pm}\xi_{nm}^{\rm{C}}(\vec{q}_{\|},\vec{k}_{\|}),
}\\
\lefteqn{
 {{\mathfrak{X}}}_{xjkx}^{\rm{D}\pm}(\vec{q}_{\|},\vec{k}_{\|})=
 {2\over({\rm{i}}\omega)^3}\left({e\hbar\over2m_e}\right)^2\sum_{nmv}
 \left\{
 {\mathfrak{Z}}_{jk,nmv}^{\rm{Da}\pm}
 \xi_{jk,nmv}^{\rm{Da}}(\vec{q}_{\|},\vec{k}_{\|})
 +{\mathfrak{Z}}_{jk,nmv}^{\rm{Db}\pm}
 \xi_{jk,nmv}^{\rm{Db}}(\vec{q}_{\|},\vec{k}_{\|})
 \right\},
}\nonumber\\ \\
\lefteqn{
 {{\mathfrak{X}}}_{ijxx}^{\rm{E}\pm}(\vec{q}_{\|},\vec{k}_{\|})=
 {2\over({\rm{i}}\omega)^3}\left({e\hbar\over2m_e}\right)^2\sum_{nmv}
 \left\{
 {\mathfrak{Z}}_{ij,nmv}^{\rm{Ea}\pm}
 \xi_{ij,nmv}^{\rm{Ea}}(\vec{q}_{\|},\vec{k}_{\|})
 +{\mathfrak{Z}}_{ij,nmv}^{\rm{Eb}\pm}
 \xi_{ij,nmv}^{\rm{Eb}}(\vec{q}_{\|},\vec{k}_{\|})
 \right\},
}\nonumber\\ \\
\lefteqn{
 {{\mathfrak{X}}}_{ixxh}^{\rm{F}\pm}(\vec{q}_{\|},\vec{k}_{\|})=
 {2\over({\rm{i}}\omega)^3}\left({e\hbar\over2m_e}\right)^2\sum_{nmv}
 \left\{
 {\mathfrak{Z}}_{ih,nmv}^{\rm{Fa}\pm}
 \xi_{ih,nmv}^{\rm{Fa}}(\vec{q}_{\|},\vec{k}_{\|})
 +{\mathfrak{Z}}_{ih,nmv}^{\rm{Fb}\pm}
 \xi_{ih,nmv}^{\rm{Fb}}(\vec{q}_{\|},\vec{k}_{\|})
 \right\},
}\nonumber\\ \\
\lefteqn{
 {{\mathfrak{X}}}_{ijkh}^{\rm{G}\pm}(\vec{q}_{\|},\vec{k}_{\|})=
 {2\over({\rm{i}}\omega)^3}\left({e\hbar\over2m_e}\right)^4\sum_{nmvl}
\left\{
 {\mathfrak{Z}}_{ijkh,nmvl}^{\rm{Ga}\pm}
 \xi_{ijkh,nmvl}^{\rm{Ga}}(\vec{q}_{\|},\vec{k}_{\|})
\right.}\nonumber\\ &\quad&\left.
 +{\mathfrak{Z}}_{ijkh,nmvl}^{\rm{Gb}\pm}
 \xi_{ijkh,nmvl}^{\rm{Gb}}(\vec{q}_{\|},\vec{k}_{\|})
 +{\mathfrak{Z}}_{ijkh,nmvl}^{\rm{Gc}\pm}
 \xi_{ijkh,nmvl}^{\rm{Gc}}(\vec{q}_{\|},\vec{k}_{\|})
\right\},
\end{eqnarray}
in terms of the ${\mathfrak{Z}}$ coefficients just calculated.

\section{The local limit in three coordinates}
In the local limit, the vector potentials (and electric fields) are
independent of their $z$-coordinate, and thus the nonlinear
conductivity tensor alone has to be integrated over the $z'''$, $z''$,
and $z'$ coordinates, viz.
\begin{equation}
 \tensor{\Xi}(z;\vec{q}_{\|},\vec{k}_{\|})=
 \iiint\tensor{\Xi}(z,z',z'',z''';\vec{q}_{\|},\vec{k}_{\|})dz'''dz''dz'.
\end{equation}

\section{Analytical expressions for ${\cal{C}}$, ${\cal{D}}$, and
  ${\cal{N}}$}\label{sec:CDN} 
In the quantity ${\cal{C}}(q_{\|}-k_{\|})$, given by Eq.~(\ref{eq:C}),
$\beta=1$ in Eq.~(\ref{eq:FpqB-indef}) and $p=0$, and in the quantity
${\cal{D}}(q_{\|},k_{\|})$ [Eq.~(\ref{eq:D})], $\beta=p=2$. In the
quantity ${\cal{N}}$ [Eq.~(\ref{eq:N})], $p=a_k=0$ and $b_k=1$, and it
can be solved immediately, with the result
\begin{equation}
 {\cal{N}}={4\over(2\pi)^2}\int_{0}^{{\alpha}}\int_{0}^{\pi}rd\theta dr
 ={{\alpha}^2\over2\pi}.
\label{eq:Solve-Q-Beta=0}
\end{equation}
Using the analysis in Appendix~\ref{app:B}, the quantities
${\cal{C}}(q_{\|}\pm{}k_{\|})$ and ${\cal{D}}(q_{\|},k_{\|})$ can be
expressed as
\begin{equation}
 {\cal{C}}(q_{\|}\pm{}k_{\|})=
  2\left[{\cal{F}}_{0}^{1}(a_1,b_1,q_{\|}\pm{}k_{\|})
  -{\cal{F}}_{0}^{1}(a_1,b_1,0)\right],
\label{eq:C-solved}
\end{equation}
and
\begin{eqnarray}
\lefteqn{ 
 {\cal{D}}(q_{\|},k_{\|})=2\left[{\cal{F}}_{2}^{2}(a_2,a_3,b_2,b_3,0)
 -{\cal{F}}_{2}^{2}(a_2,a_3,b_2,b_3,k_{\|})
\right.}\nonumber\\ &\quad&
 +{\cal{F}}_{2}^{2}(a_2,a_4,b_2,b_4,k_{\|}+q_{\|})
 -{\cal{F}}_{2}^{2}(a_2,a_4,b_2,b_4,k_{\|})
\nonumber\\ &&
 +{\cal{F}}_{2}^{2}(a_2,a_4,b_2,b_5,0)
 -{\cal{F}}_{2}^{2}(a_2,a_4,b_2,b_5,q_{\|})
\nonumber\\ &&\left.
 +{\cal{F}}_{2}^{2}(a_2,a_3,b_2,b_6,k_{\|}+q_{\|})
 -{\cal{F}}_{2}^{2}(a_2,a_3,b_2,b_6,q_{\|})\right],
\nonumber\\
\label{eq:D-solved}
\end{eqnarray}
where
\begin{eqnarray}
 a_1&=&\hbar(q_{\|}-k_{\|})/m_e, \\
 a_2&=&\hbar(q_{\|}+k_{\|})/m_e, \\
 a_3&=&\hbar{}k_{\|}/m_e, \\
 a_4&=&\hbar{}q_{\|}/m_e, \\
 b_1&=&\hbar(q_{\|}-k_{\|})^2/(2m_e)-i/\tau, \\
 b_2&=&\hbar(q_{\|}+k_{\|})^2/(2m_e)-i/\tau, \\
 b_3&=&\hbar{}k_{\|}^2/(2m_e)-i/\tau-\omega, \\
 b_4&=&\hbar{}q_{\|}(q_{\|}+2k_{\|})/(2m_e)-i/\tau+\omega, \\
 b_5&=&\hbar{}q_{\|}^2/(2m_e)-i/\tau+\omega, \\
 b_6&=&\hbar{}k_{\|}(k_{\|}+2q_{\|})/(2m_e)-i/\tau-\omega.
\end{eqnarray}

\section{Analytic expressions for the ${\cal{Q}}$
 quantities}\label{sec:Q}
Let us finish this appendix by giving the solutions to
Eqs.~(\ref{eq:A-xx})--(\ref{eq:A-zz}). They are
\begin{eqnarray}
\lefteqn{ {\cal{Q}}_{\,\,\,nm}^{xx}(\vec{q}_{\|},\omega)={2i\hbar\over(2\pi)^2}
 \left({e\hbar\over2m_{e}}\right)^2
}\nonumber\\ &\quad&\times
 \left[
 4{\cal{F}}_{20,n}^{2}(a_1,a_2,b_{1,nm},b_{2,nm},q_x)
 +4q_x{\cal{F}}_{10,n}^{2}(a_1,a_2,b_{1,nm},b_{2,nm},q_x)
\right.\nonumber\\ &&
 +q_x^2{\cal{F}}_{00,n}^{2}(a_1,a_2,b_{1,nm},b_{2,nm},q_x)
 -4{\cal{F}}_{20,m}^{2}(a_1,a_2,b_{1,nm},b_{2,nm},0)
\nonumber\\ &&\left.\!
 -4q_x{\cal{F}}_{10,m}^{2}(a_1,a_2,b_{1,nm},b_{2,nm},0)
 -q_x^2{\cal{F}}_{00,m}^{2}(a_1,a_2,b_{1,nm},b_{2,nm},0)
 \right],
\label{eq:Q1}
\\
\lefteqn{ {\cal{Q}}_{\,\,\,nm}^{xz}(\vec{q}_{\|},\omega)={2i\hbar\over(2\pi)^2}
 \left({e\hbar\over2m_{e}}\right)^2
}\nonumber\\ &&\times
 \left[
 2{\cal{F}}_{10,n}^{2}(a_1,a_2,b_{1,nm},b_{2,nm},q_x)
 +q_x{\cal{F}}_{00,n}^{2}(a_1,a_2,b_{1,nm},b_{2,nm},q_x)
\right.\nonumber\\ &&\left.
 -2{\cal{F}}_{10,m}^{2}(a_1,a_2,b_{1,nm},b_{2,nm},0)
 -q_x{\cal{F}}_{00,m}^{2}(a_1,a_2,b_{1,nm},b_{2,nm},0)
 \right],
\\
\lefteqn{
 {\cal{Q}}_{\,\,\,nm}^{yy}(\vec{q}_{\|},\omega)={2i\hbar\over(2\pi)^2}
 \left({e\hbar\over2m_{e}}\right)^2
}\nonumber\\ &&\times
 4\left[
 {\cal{F}}_{02,n}^{2}(a_1,a_2,b_{1,nm},b_{2,nm},q_x)
 -{\cal{F}}_{02,m}^{2}(a_1,a_2,b_{1,nm},b_{2,nm},0)
 \right],
\\
\lefteqn{
 {\cal{Q}}_{\,\,\,nm}^{zz}(\vec{q}_{\|},\omega)={2i\hbar\over(2\pi)^2}
 \left({e\hbar\over2m_{e}}\right)^2
}\nonumber\\ &&\times
 \left[
 {\cal{F}}_{00,n}^{2}(a_1,a_2,b_{1,nm},b_{2,nm},q_x)
 -{\cal{F}}_{00,m}^{2}(a_1,a_2,b_{1,nm},b_{2,nm},0)
 \right],
\label{eq:Q4}
\end{eqnarray}
according to the treatment of these types of integrals given in
Complement~\ref{ch:Solve-Q}. Above we have used
\begin{eqnarray}
 a_1&=&a_2={\hbar{}q_x\over{}m_{e}},
\\
 b_{1,nm}&=&
 \varepsilon_n-\varepsilon_m+{\hbar{}q_x^2\over2m_{e}}-{i\hbar\over\tau_{nm}},
\\
 b_{2,nm}&=&
 \varepsilon_n-\varepsilon_m+{\hbar{}q_x^2\over2m_{e}}-{i\hbar\over\tau_{nm}}
 -\hbar\omega.
\end{eqnarray}

\newpage
\thispagestyle{plain}
\newpage

\chapter{Fermi energy, quantum well thickness, and $\alpha(n)$}\label{app:D}
\noindent
The number of electrons $n(\vec{r})$ in a system where the spin
energies are degenerate can be written
\begin{equation}
 n(\vec{r})=2\sum_{N}|\Psi_{N}(\vec{r})|^2f_{N},
\label{eq:Number-r}
\end{equation}
where the number $2$ represents the degeneracy of the spin energies,
and the sum runs over all electron states in the system multiplied by
the probability of finding an electron in that state. This probability
is given as a Fermi-Dirac distribution
\begin{equation}
 f_{N}={1\over1+\exp(({\cal{E}}_{N}-\mu)/(k_BT))},
\end{equation}
where ${\cal{E}}_{N}$ is the energy of the electron in state $N$,
$\mu$ is the chemical potential, $k_B$ is Boltzmann's constant and $T$
is the absolute temperature.

We will now look at the case where we have two-dimensional
translational invariance along the $x$-$y$-plane. In this case the
wave function gives plane-wave solutions in the direction of the plane,
\begin{equation}
 \Psi_{N}(\vec{r})={1\over2\pi}\psi_n(z)e^{i\vec{\kappa}_{\|}\cdot\vec{r}}
\end{equation}
and the corresponding energy is
\begin{equation}
 {\cal{E}}_{N}=\varepsilon_{n}+{\hbar^2\kappa_{\|}^2\over2m_e},
\end{equation}
where $\kappa_{\|}=|\kappa_{\|}|$. By insertion into
Eq.~(\ref{eq:Number-r}), it is converted into
\begin{equation}
 n(z)=2\sum_{n}|\psi_{n}(z)|^2\int_{-\infty}^{\infty}\left({1+\exp\left[
 {\varepsilon_n+(\hbar^2\kappa_{\|}^2)/(2m_e)-\mu\over{}k_BT}
 \right]}\right)^{-1}
 {d^2\kappa_{\|}\over(2\pi)^2},
\end{equation}
taking into account that the sum over the plane-wave expansion parallel
to the surface can be converted into an integral, and the notation
$n(z)\equiv{}n(\vec{r})$ is introduced for consistency. Solving the
integral, we get
\begin{equation}
 n(z)={2m\over\hbar^2}\sum_{n}G_n|\psi_{n}(z)|^2,
\end{equation}
with
\begin{equation}
 G_n={k_{B}T\over2\pi}
 \ln\left[1+\exp\left({\mu-\varepsilon_n\over{}k_BT}\right)\right]
\end{equation}
as the number of electrons in the quantum well for any temperature
$T$. 

\section{Fermi energy in the low temperature limit}
In the low temperature limit, the chemical potential obeys,
\begin{equation}
 \lim_{T\rightarrow0}\mu={\cal{E}}_{F},
\end{equation}
where ${\cal{E}}_{F}$ is the Fermi energy. Then
\begin{equation}
 G_n=\left\{
 \begin{array}{lll}
 0 & \mbox{for} & \varepsilon_n>{\cal{E}}_{F}, \\
 ({\cal{E}}_{F}-\varepsilon_n)/(2\pi)&\mbox{for}&\varepsilon_n<{\cal{E}}_{F},
 \end{array}\right.
\end{equation}
for $T\rightarrow0$, and thus
\begin{equation}
 n(z)|_{T\rightarrow0}={m\over\pi\hbar^2}\sum_{n}
 ({\cal{E}}_{F}-\varepsilon_n)\Theta({\cal{E}}_{F}-\varepsilon_n)
 |\psi_n(z)|^2
\end{equation}
is the number of electrons (negative charges) in the system.

Additionally, the global neutrality condition teaches that if the net
electric charge should be zero, the number of positive charges should
be equal to the number of negative charges, that is
\begin{eqnarray}
 ZN_+d&=&\int{}n(z)dz
\nonumber\\ &=&
 {m\over\pi\hbar^2}\sum_{n}
 ({\cal{E}}_{F}-\varepsilon_n)\Theta({\cal{E}}_{F}-\varepsilon_n)
 \int|\psi_n(z)|^2dz
\nonumber\\ &=&
 {m\over\pi\hbar^2}\sum_{n}
 ({\cal{E}}_{F}-\varepsilon_n)\Theta({\cal{E}}_{F}-\varepsilon_n)
\end{eqnarray}
where $N_+$ is the number of positive ions per unit volume and $Z$ is
the valence of each of these ions.

Defining the quantity $N_F$ as the index of the highest occupied
level, this may be rewritten into
\begin{equation}
 ZN_+d={m\over\pi\hbar^2}\sum_{n=1}^{N_F}
 \left({\cal{E}}_{F}-\varepsilon_n\right),
\end{equation}
from which the Fermi energy easily is extracted as
\begin{equation}
 {\cal{E}}_{F}={1\over{}N_F}\left[{\pi\hbar^2\over{}m_e}ZN_+d
 +\sum_{n=1}^{N_F}\varepsilon_n\right].
\label{eq:EF-T=0}
\end{equation}

\section{Infinite barrier quantum well}
In the infinite barrier model for a quantum well extending from $0$ to
$-d$ in the $z$-direction and infinitely in the $x$-$y$-plane, we have
\begin{equation}
 \varepsilon_n={\pi^2\hbar^2n^2\over2m_ed^2},
\end{equation}
which inserted into Eq.~(\ref{eq:EF-T=0}) gives
\begin{eqnarray}
 {\cal{E}}_{F}&=&{1\over{}N_F}\left[{\pi\hbar^2\over{}m_e}ZN_+d
 +{\pi^2\hbar^2\over2m_ed^2}\sum_{n=1}^{N_F}n^2\right]
\nonumber\\ &=&
 {\pi\hbar^2\over{}N_Fm_e}\left[ZN_+d+{\pi\over2d^2}{N_F(N_F+1)(2N_F+1)\over6}
 \right].
\label{eq:EF-IB}
\end{eqnarray}
From this equation, the limits on the thickness of the quantum well
can be determined if we know the number of bound states we want below
the Fermi level, the minimal thickness for the quantum well to have
$n$ levels being determined from the simple relation
${\cal{E}}_{F}=\varepsilon_n$, and thus the maximal thickness can be
determined from ${\cal{E}}_{F}=\varepsilon_{n+1}$, since it has the
same limit value as the minimal thickness to obtain $n+1$ bound
states. Thus, for $n$ bound states below the Fermi level,
\begin{equation}
 {\pi\hbar^2\over{}nm_e}\left[ZN_+d+{\pi\over2d^2}{n(n+1)(2n+1)\over6}
 \right]={\pi^2\hbar^2n^2\over2m_ed^2},
\end{equation}
which gives the related minimal and maximal thicknesses to have these
$n$ bound states
\begin{eqnarray}
 d_{\rm{min}}^{n}&=&
 \sqrt[3]{{\pi{}n\over2ZN_+}\left[n^2-{(n+1)(2n+1)\over6}\right]},
\nonumber\\ 
 d_{\rm{max}}^{n}&=&
 \sqrt[3]{{\pi{}(n+1)\over2ZN_+}\left[(n+1)^2-{(n+2)(2n+3)\over6}\right]}.
\end{eqnarray}
For a quantum well with only a single bound state we thus get
\begin{eqnarray}
 d_{\rm{min}}^{(1)}&=&0,
\nonumber\\ 
 d_{\rm{max}}^{(1)}&=&\sqrt[3]{3\pi/2ZN_+},
\end{eqnarray}
for two bound states
\begin{eqnarray}
 d_{\rm{min}}^{(2)}&=&\sqrt[3]{3\pi/2ZN_+},
\nonumber\\ 
 d_{\rm{max}}^{(2)}&=&\sqrt[3]{39\pi/6ZN_+},
\end{eqnarray}
and so on.

Since $\hbar^2k_F^2=2m_e{\cal{E}}_{F}$, the radius of the
two-dimensional Fermi circle for state $n$, $\alpha(n)$, used as
integration boundary in Appendix~\ref{ch:Solve-Q} can be found using
Eq.~(\ref{eq:EF-IB}). It is
\begin{equation}
 \alpha(n)=\sqrt{{\pi{}ZN_+d\over{}N_F}+{\pi^2\over2d}{(N_F+1)(2N_F+1)\over6}
 -{n^2\pi^2\over{}d^2}}.
\label{eq:alpha(n)}
\end{equation}

\newpage
\thispagestyle{plain}
\newpage

\chapter{Solution to integrals over $z$ in Chapter~\ref{ch:14}}\label{app:E}
In this appendix we give some intermediate steps of the solution to
integrals in the quantity $\tensor{K}(\vec{q}_{\|},\omega)$ appering
in the description of the multilevel quantum well in
Chapter~\ref{ch:14}.

Inserting the expressions for the different $F$ quantities
[Eqs.~(\ref{eq:F-xx})--(\ref{eq:F-zz})] into
Eqs.~(\ref{eq:Kxx})--(\ref{eq:Kzz}), we get
\begin{eqnarray}
\lefteqn{K_{xx,mn}^{vl}(\vec{q}_{\|},\omega)=
 -{\rm{i}}\mu_0\omega
 \left\{ 
 {\cal{Q}}_{\,\,\,lv}^{xx}(\vec{q}_{\|},\omega)
 \iint{\cal{Z}}^{x}_{mn}(z)G_{xx}(z,z'';\vec{q}_{\|},\omega){\cal{Z}}^{x}_{lv}(z'')dz''dz
\right.}\nonumber\\ &\quad&\left.\!
 -i{\cal{Q}}_{\,\,\,lv}^{xz}(\vec{q}_{\|},\omega)
 \iint{\cal{Z}}^{x}_{mn}(z)G_{xz}(z,z'';\vec{q}_{\|},\omega){\cal{Z}}^{z}_{lv}(z'')dz''dz
 \right\},
\label{eq:K-xx-i}
\\
\lefteqn{K_{xz,mn}^{vl}(\vec{q}_{\|},\omega)=
 {\rm{i}}\mu_0\omega
 \left\{ 
 i{\cal{Q}}_{\,\,\,lv}^{xz}(\vec{q}_{\|},\omega)
 \iint{\cal{Z}}^{x}_{mn}(z)G_{xx}(z,z'';\vec{q}_{\|},\omega){\cal{Z}}^{x}_{lv}(z'')dz''dz
\right.}\nonumber\\ &&\left.\!
 +{\cal{Q}}_{\,\,\,lv}^{zz}(\vec{q}_{\|},\omega)
 \iint{\cal{Z}}^{x}_{mn}(z)G_{xz}(z,z'';\vec{q}_{\|},\omega){\cal{Z}}^{z}_{lv}(z'')dz''dz
 \right\},
\\
\lefteqn{K_{yy,mn}^{vl}(\vec{q}_{\|},\omega)=
 -{\rm{i}}\mu_0\omega {\cal{Q}}_{\,\,\,lv}^{yy}(\vec{q}_{\|},\omega)
 \iint{\cal{Z}}^{x}_{mn}(z)G_{yy}(z,z'';\vec{q}_{\|},\omega){\cal{Z}}^{x}_{lv}(z'')dz''dz,
}\\
\lefteqn{K_{zx,mn}^{vl}(\vec{q}_{\|},\omega)=
 -{\rm{i}}\mu_0\omega
 \left\{
 {\cal{Q}}_{\,\,\,lv}^{xx}(\vec{q}_{\|},\omega){q_{\|}\over{}q_{\perp}}
 \iint{\cal{Z}}^{z}_{mn}(z)G_{xx}(z,z'';\vec{q}_{\|},\omega){\cal{Z}}^{x}_{lv}(z'')dz''dz
\right.}\nonumber\\ &&\left.\!
 -i{\cal{Q}}_{\,\,\,lv}^{xz}(\vec{q}_{\|},\omega)
 {q_{\|}\over{}q_{\perp}}
 \iint{\cal{Z}}^{z}_{mn}(z)G_{xz}(z,z'';\vec{q}_{\|},\omega){\cal{Z}}^{z}_{lv}(z'')dz''dz
 \right\},
\\
\lefteqn{K_{zz,mn}^{vl}(\vec{q}_{\|},\omega)=
 {\rm{i}}\mu_0\omega
 \left\{
 i{\cal{Q}}_{\,\,\,lv}^{xz}(\vec{q}_{\|},\omega){q_{\|}\over{}q_{\perp}}
 \iint{\cal{Z}}^{z}_{mn}(z)G_{xx}(z,z'';\vec{q}_{\|},\omega){\cal{Z}}^{x}_{lv}(z'')dz''dz
\right.}\nonumber\\ &&\left.\!
 +{\cal{Q}}_{\,\,\,lv}^{zz}(\vec{q}_{\|},\omega)
 {q_{\|}\over{}q_{\perp}}
 \iint{\cal{Z}}^{z}_{mn}(z)G_{xz}(z,z'';\vec{q}_{\|},\omega){\cal{Z}}^{z}_{lv}(z'')dz''dz
 \right\}.
\label{eq:K-zz-i}
\end{eqnarray}
Using an infinite barrier potential along the $z$-direction of the
quantum well, the ${\cal{Z}}(z)$ quantities are described in
Eqs.~(\ref{eq:ZxIB}) and (\ref{eq:ZzIB}). Then by use of
\citeN{Gradshteyn:94:1}, Eqs.~2.663.1 and 2.663.3,
\begin{eqnarray}
 \int e^{ax}\sin(bx)dx&=&{e^{ax}[a\sin(bx)-b\cos(bx)]\over{}a^2+b^2},
\\
 \int e^{ax}\cos(bx)dx&=&{e^{ax}[a\cos(bx)+b\sin(bx)]\over{}a^2+b^2},
\end{eqnarray}
we find that integrals over the source region takes the form
\begin{eqnarray}
 \int_{-d}^{0}e^{ax}\sin\left({b\pi{}x\over{}d}\right)dx&=&
 {\pi{}bd[e^{-ad}(-1)^b-1]\over{}a^2d^2+\pi^2b^2},
\\
 \int_{-d}^{0}e^{ax}\cos\left({b\pi{}x\over{}d}\right)dx&=&
 {ad^2[1-e^{-ad}(-1)^b]\over{}a^2d^2+\pi^2b^2},
\end{eqnarray}
in which $b$ is an integer. This result leads to
\begin{eqnarray}
\lefteqn{
 \int G_{xx}(z,z';\vec{q}_{\|},\omega){\cal{Z}}^{x}_{nm}(z')dz'=
}\nonumber\\ &\quad&
 {2\pi^2nmc_0^2q_{\perp}^2d\left[
  1+r^p-\left(e^{-iq_{\perp}d}+r^pe^{iq_{\perp}d}\right)(-1)^{n+m}\right]
 \over\omega^2[(iq_{\perp}d)^2+\pi^2(n-m)^2][(iq_{\perp}d)^2+\pi^2(n+m)^2]}
 e^{-iq_{\perp}z},
\\
\lefteqn{
 \int G_{xz}(z,z';\vec{q}_{\|},\omega){\cal{Z}}^{x}_{nm}(z')dz'=
}\nonumber\\ &\quad&
 {2\pi^2nmc_0^2q_{\|}q_{\perp}d\left[
  1-r^p-\left(e^{-iq_{\perp}d}-r^pe^{iq_{\perp}d}\right)(-1)^{n+m}\right]
 \over\omega^2[(iq_{\perp}d)^2+\pi^2(n-m)^2][(iq_{\perp}d)^2+\pi^2(n+m)^2]}
 e^{-iq_{\perp}z},
\\
\lefteqn{
 \int G_{yy}(z,z';\vec{q}_{\|},\omega){\cal{Z}}^{x}_{nm}(z')dz'=
}\nonumber\\ &\quad&
 {2\pi^2nmd\left[
  1-r^s-\left(e^{-iq_{\perp}d}-r^se^{iq_{\perp}d}\right)(-1)^{n+m}\right]
 \over[(iq_{\perp}d)^2+\pi^2(n-m)^2][(iq_{\perp}d)^2+\pi^2(n+m)^2]}
 e^{-iq_{\perp}z},
\\
\lefteqn{
 \int G_{xz}(z,z';\vec{q}_{\|},\omega){\cal{Z}}^{z}_{nm}(z')dz'=
}\nonumber\\ &\quad&
 {2\pi^4nm(n^2-m^2)c_0^2q_{\|}
 [(e^{-iq_{\perp}d}+r^pe^{iq_{\perp}d})(-1)^{n+m}-1-r^p]
 \over{}i\omega^2d
 [(iq_{\perp}d)^2+\pi^2(n-m)^2][(iq_{\perp}d)^2+\pi^2(n+m)^2]}
 e^{-iq_{\perp}z},
\end{eqnarray}
and
\begin{eqnarray}
 \int {\cal{Z}}^{x}_{mn}(z)e^{-iq_{\perp}z}dz&=&
 {4\pi^2nmiq_{\perp}d[e^{iq_{\perp}d}(-1)^{n+m}-1]
 \over[(iq_{\perp}d)^2+\pi^2(n-m)^2][(iq_{\perp}d)^2+\pi^2(n+m)^2]},
\\
 \int {\cal{Z}}^{z}_{mn}(z)e^{-iq_{\perp}z}dz&=&
 {4\pi^4nm(n^2-m^2)[e^{iq_{\perp}d}(-1)^{n+m}-1]
 \over{}d[(iq_{\perp}d)^2+\pi^2(n-m)^2][(iq_{\perp}d)^2+\pi^2(n+m)^2]},
\end{eqnarray}
since $(-1)^{n-m}=(-1)^{n+m}$ for $n$ and $m$ integers. Then
\begin{eqnarray}
\lefteqn{ F_{nm}^{xx}(z;\vec{q}_{\|},\omega)=
 -{2\pi^2inm\over\epsilon_0\omega}
 e^{-iq_{\perp}z}
 \left\{ 
 {\cal{Q}}_{\,\,\,nm}^{xx}(\vec{q}_{\|},\omega)
 {q_{\perp}^2d}
 +{\cal{Q}}_{\,\,\,nm}^{xz}(\vec{q}_{\|},\omega)
 {\pi^2(n^2-m^2)q_{\|} \over{}d}
 \right\} 
}\nonumber\\ &\quad&\times
 {1+r^p-\left(e^{-iq_{\perp}d}+r^pe^{iq_{\perp}d}\right)(-1)^{n+m}
 \over[(iq_{\perp}d)^2+\pi^2(n-m)^2][(iq_{\perp}d)^2+\pi^2(n+m)^2]}
\label{eq:F-xx-i}\\
\lefteqn{ F_{nm}^{xz}(z;\vec{q}_{\|},\omega)=
 {2\pi^2inm\over\epsilon_0\omega}
 e^{-iq_{\perp}z}
 \left\{ 
 i{\cal{Q}}_{\,\,\,nm}^{xz}(\vec{q}_{\|},\omega)
 q_{\perp}^2d
 -{\cal{Q}}_{\,\,\,nm}^{zz}(\vec{q}_{\|},\omega)
 {\pi^2(n^2-m^2)q_{\|}\over{}id}
 \right\}
}\nonumber\\ &&\times
 {1+r^p-\left(e^{-iq_{\perp}d}+r^pe^{iq_{\perp}d}\right)(-1)^{n+m}
 \over[(iq_{\perp}d)^2+\pi^2(n-m)^2][(iq_{\perp}d)^2+\pi^2(n+m)^2]},
\label{eq:F-xz-i}\\
\lefteqn{ F_{nm}^{yy}(z;\vec{q}_{\|},\omega)=
 -2\pi^2{\rm{i}}\mu_0\omega nmd
 e^{-iq_{\perp}z}
 {\cal{Q}}_{\,\,\,nm}^{yy}(\vec{q}_{\|},\omega)
}\nonumber\\ &&\times
 {1-r^s-\left(e^{-iq_{\perp}d}-r^se^{iq_{\perp}d}\right)(-1)^{n+m}
 \over[(iq_{\perp}d)^2+\pi^2(n-m)^2][(iq_{\perp}d)^2+\pi^2(n+m)^2]},
\label{eq:F-yy-i}
\end{eqnarray}
If we now insert Eqs.~(\ref{eq:F-xx-i})--(\ref{eq:F-yy-i}),
(\ref{eq:F-zx}) and (\ref{eq:F-zz}) into
Eqs.~(\ref{eq:K-xx-i})--(\ref{eq:K-zz-i}) and perform the remaining
integration, we get Eqs.~(\ref{eq:Kxx})--(\ref{eq:Kzz}).

\newpage
\thispagestyle{plain}
\newpage
\chead[\fancyplain{}{\leftmark}]{\fancyplain{}{\leftmark}}
\bibliographystyle{ta-chicago}

\newpage
\thispagestyle{plain}
\noindent{\large\bf Dansk resum{\'e}}
\addcontentsline{toc}{chapter}{Dansk resum{\'e}}\par
\vspace{2.3ex plus.2ex}
\noindent 
For at studere den mulige fasekonjugation af optiske n{\ae}rfelter er
det n{\o}dvendigt at g{\aa} ud over den langsomtvarierende
indhyldningskurve approksimation samt den elektriske
dipoltiln{\ae}rmelse, der normalt anvendes i fasekonjugationsstudier
hvor rumligt ud{\ae}mpede (eller i det mindste svagt d{\ae}mpede)
svingninger blandes. I den foreliggende afhand\-ling pr{\ae}senteres
en vilk{\aa}rlig-fase tiln{\ae}rmet beregning af den uline{\ae}re og
ikke-lokale optiske responstensor der beskriver den uliner{\ae}re
str{\o}mt{\ae}thed af tredie orden, som genereres af
fireb{\o}lgeblanding i en uensartet elektrongas. Beskrivelsen er
baseret p{\aa} en halv\-klas\-sisk model, hvori det elektromagnetiske
felt antages at v{\ae}re en klassisk st{\o}rrelse og udgangspunktet er
bev{\ae}gelsesligningen for t{\ae}thedsmatrix-operatoren.
Vekselvirknings Hamil\-ton-operatoren anvendes i dens minimale
koblingsform, og den indeholder det led i
str{\o}m\-t{\ae}t\-heds\-operatoren, der er proportionalt med det
p{\aa}trykte vektorpotential. Ved brug af denne formalisme er den
rumlige struktur af systemets optiske respons beskrevet ved hj{\ae}lp
af mikroskopiske overgangsstr{\o}mt{\ae}theder. Beregningen inkluderer
derfor b{\aa}de bidrag fra $\vec{p}\cdot\vec{A}$ og
$\vec{A}\cdot\vec{A}$ leddene i vekselvirknings-Hamilton-operatoren.
Det er vist at der introduceres nogle vigtige f{\ae}nomener, som er
begrebs\-m{\ae}ssigt forskellige fra de der har deres oprindelse i
$\vec{p}\cdot\vec{A}$-leddet, ved at inkludere $\vec{A}\cdot\vec{A}$
leddet i vekselvirknings Hamilton-operatoren. For at fremh{\ae}ve den
fysiske mening af de forskellige processer er koblingerne mellem
observationspunkter for feltet og str{\o}m\-t{\ae}t\-heden pr{\ae}senteret
i form af diagrammer. Resultatet af en analyse af tensorsymmetrierne,
der er tilknyttet $\vec{p}\cdot\vec{A}$ og $\vec{A}\cdot\vec{A}$
vekselvirkningerne er summeret i form af symmetriskemaer for
fasekonjugationsprocessen. Den teoretiske model efterf{\o}lges af en
beregning af det fasekonjugerede respons fra en et-niveau metallisk
kvantebr{\o}nd. Et-niveau kvantebr{\o}nden repr{\ae}senterer den
simplest mulige konfiguration en kvantebr{\o}nds-fasekonju\-gator kan
have. Ydermere er den et interessant objekt, idet dens optiske respons
ikke indeholder noget dipol-led. Diskussionen af responset er baseret
p{\aa} stimulering af processen ved brug af lys, der er polariseret
enten i spredningsplanet eller vinkelret p{\aa} spredningsplanet. Det
vises, at fasekonjugationsprocessen er ekstremt effektiv i det
d{\ae}mpede omr{\aa}de af b{\o}lgevektor-spektret.  Dern{\ae}st
anskues problemet med at generere plane b{\o}lger til excitation i den
h{\o}je ende af det d{\ae}mpede spektrum, og vi diskuterer brugen af
en bredb{\aa}ndskilde (i vinkelspektret) til at stimulere
processen. En s{\aa}dan bredb{\aa}ndskilde kan v{\ae}re en
kvantetr{\aa}d, og det fasekonjugerede vinkelspektrum fra en
kvantetr{\aa}d pr{\ae}\-sen\-teres og diskuteres. Kvantetr{\aa}dens
subb{\o}lgel{\ae}ngde st{\o}rrelse g{\o}r den en mulig kandidat til en
diskussion af den mulige rumlige komprimering af lys, og rumlig
begr{\ae}ns\-ning af lys foran en et-niveau metallisk
kvantebr{\o}ndsfasekonjugator er diskuteret i to dimensioner.  Det
retf{\ae}rdigg{\o}res at man ved et passende valg af
str{\o}mt{\ae}thedens orientering i kvantetr{\aa}den kan opn{\aa} en
feltkomprimering, der er v{\ae}sentligt p{\aa} den anden side af
Rayleighs gr{\ae}nsev{\ae}rdi. Afhandlingen afsluttes med en kort
beskrivelse af det mere generelle tilf{\ae}lde, hvor kvantebr{\o}nden
tillades at have mere end en energi-egentilstand. Numeriske
resultater, der viser responset hvis en to-niveau kvantebr{\o}nd
anvendes som fasekonjugerende medium, er pr{\ae}senteret og
diskuteret.
\newpage
\thispagestyle{empty}
\setcounter{page}{0}
\vbox to2cm{ }
\newpage
\setcounter{page}{-1}
\setlength{\unitlength}{1mm}
\psset{unit=1mm}
\thispagestyle{empty}
\vfil
\begin{pspicture}(0,0)(141,204)
\put(0,204){\makebox(0,0)[tl]{%
 \begin{minipage}[tl]{141mm}
   \noindent 
   In order to study the possible phase conjugation of optical
   near-fields, it is necessary to go beyond the slowly varying
   envelope- and electric dipole approximations that are normally
   applied in phase conjugation studies where spatially non-decaying
   (or at least slowly decaying) modes are mixed.  In the present
   dissertation a random-phase-approximation calculation of the
   nonlocal nonlinear optical response tensor describing the third
   order nonlinear current density generated by degenerate four-wave
   mixing in an inhomogeneous electron gas is established. The
   description is based on a semi-classical approach, in which the
   electromagnetic field is considered as a classical quantity, and
   the starting point is the equation of motion for the density matrix
   operator. The interaction Hamiltonian is taken in its minimal
   coupling form, and it includes the term in the current density
   operator which is proportional to the prevailing vector potential.
   Using this formalism the spatial structure of the optical response
   of the system is described in terms of the microscopic transition
   current densities. The calculation thus includes contributions
   originating from both the $\vec{p}\cdot\vec{A}$ and
   $\vec{A}\cdot\vec{A}$ terms in the interaction Hamiltonian.  It is
   demonstrated that inclusion of the $\vec{A}\cdot\vec{A}$ term in
   the interaction Hamiltonian introduces some important phenomena
   that are conceptually different from those originating in the
   $\vec{p}\cdot\vec{A}$ part.  To emphasize the physical meaning of
   the various processes, the couplings between observation points for
   the field and the current density is presented in a diagrammatic
   form. The result of an analysis of the tensor symmetries associated
   with the $\vec{p}\cdot\vec{A}$ and $\vec{A}\cdot\vec{A}$
   interactions are summarized in terms of symmetry schemes for the
   phase conjugation process. The theoretical model is followed by a
   calculation of the phase conjugated response from a single-level
   metallic quantum well.  The single-level quantum well represents
   the simplest possible configuration of a quantum-well phase
   conjugator. Furthermore, it is an interesting object, since its
   optical response contains no dipole terms. The discussion of the
   response is based on the use of light that is polarized either in
   the scattering plane or perpendicular to the scattering plane to
   excite the process.  It is demonstrated that the phase conjugation
   process is extremely efficient in the evanescent regime of the
   wavevector spectrum. We address the problem of plane-wave
   excitation in the high wavenumber end of the evanescent regime and
   discuss the use of a broadband source to excite the process. One
   possible broad angular band source is a quantum wire, and the phase
   conjugated angular spectrum from a quantum wire is presented and
   discussed. The subwavelength size of the quantum wire makes it a
   possible candidate for discussion of confinement of light, and the
   confinement of light in two dimensions in front of a single-level
   metallic quantum-well phase conjugator is discussed. It is
   justified that by a proper choice of orientation of the current in
   the quantum wire a field compression substantially beyond the
   Rayleigh limit is obtained. The thesis is concluded with a short
   description of the more general case where the quantum well is
   allowed to have more than one energy eigenstate, and numerical
   results showing the response from a two-level quantum well as the
   phase conjugating medium are presented and discussed.
 \end{minipage}
}}
\put(0,0){\makebox(0,0)[bl]{\sc Med et resum{\'e} p{\aa} dansk.}}
\put(141,0){\makebox(0,0)[br]{ISBN 87-89195-16-7}}
\end{pspicture}
\end{document}